МІНІСТЕРСТВО ОСВІТИ І НАУКИ УКРАЇНИ

Одеський національний університет імені І. І. Мечникова

**В. М. Адамян, М. Я. Сушко**

# ВСТУП ДО МАТЕМАТИЧНОЇ ФІЗИКИ. ВАРІАЦІЙНЕ ЧИСЛЕННЯ ТА КРАЙОВІ ЗАДАЧІ

Навчальний посібник
для студентів фізичних
та інженерно-фізичних спеціальностей
вищих навчальних закладів

Рекомендовано
Міністерством освіти і науки України

У посібнику розглядаються постановка й методи розв'язування простих лінійних задач класичної математичної фізики. Коло розглянутих питань охоплює принципи варіаційного числення, одновимірні крайові задачі теорії коливань і теплопровідності з докладним аналізом крайової задачі Штурма — Ліувілля та обґрунтуванням методу Фур'є, приклади розв'язування відповідних задач у двох і трьох вимірах із необхідними елементами теорії спеціальних функцій.

Для студентів фізичних, інженерно-фізичних та математичних спеціальностей університетів.

# ЗМІСТ

*Розділ 7*

## ПРОСТОРОВІ КРАЙОВІ ЗАДАЧІ

# Передмова

Математична фізика як окрема галузь знань на стику фізики й математики сформувалася наприкінці XIX століття і спочатку ототожнювалася з теоретичною фізикою. Її предмет визначився як розвиток математичних підходів до формулювання фізичних теорій і розробка універсальних методів розв'язування фізичних проблем. Математичні поняття й моделі, які були започатковані при аналізі класичних задач гідродинаміки, акустики, теорії пружності, термодинаміки та електромагнетизму, виявилися настільки глибокими і плідними, що з відкриттям на початку XX століття квантових закономірностей і зміною понятійної бази фізичних уявлень не виникло потреби у створенні нового математичного апарату. Стрімкий розвиток квантової теорії протягом першого десятиліття її існування цілком відбувався на основі вже раніше розроблених методів математичної фізики. При цьому саме квантова теорія по-новому освітила і фактично матеріалізувала фундаментальні факти й положення, що зародилися в надрах математичної фізики при розв'язуванні задач класичної фізики. До них треба віднести ортогональні розклади за власними функціями крайових задач, асоційованих із лінійними диференціальними рівняннями математичної фізики, зображення розв'язків цих задач як векторів певних лінійних просторів із, взагалі кажучи, нескінченною розмірністю та спектральну теорію лінійних операторів у нескінченновимірних просторах.

В останні два десятиліття, завдяки повсюдному поширенню комп'ютерних інструментів автоматизації і багатократного прискорення рутинних математичних операцій, зміст і характер застосувань математики до розв'язування фізичних і технічних проблем кардинально змінилися. Стали доступними розрахунки, раніше немислимі в силу їх складності, комбіновані та чисто комп'ютерні експеримен-ти, зокрема, й такі, які неможливо здійснити в природі чи реальних лабораторних умовах. Це, звичайно, знизило практичну цінність і наукову актуальність математичних проблем, пов'язаних із пошуком і дослідженням специфічних класів точно розв'язуваних задач аналізу й математичної фізики, методів і прийомів «ручної» роботи при доведенні розв'язків наукових і прикладних задач «до числа». Однак принципи і підходи математичної фізики, які були закладені в дискусіях Ейлера і Д'Аламбера про коливання натягненої струни та в роботах Фур'є про поширення тепла і які до початку минулого століття перетворилися в універсальний інструмент дослідження й побудови математичних моделей фізичних (і не лише фізичних) явищ і процесів, повною мірою зберігають свою актуальність і сьогодні.

Цей посібник має на меті ознайомити студентів із проблемами, які можна віднести до фундаментального доробку класичної математичної фізики. Його основу складає друга частина лекційного курсу з методів математичної фізики, що читається студентам фізичних спеціальностей Одеського національного університету імені І. І. Мечникова в осінньому семестрі третього року навчання. Хоча посібник є продовженням нашого попереднього видання «Вступ до математичної фізики», який охоплює матеріал першої частини відповідного курсу і присвячений постановці й розв'язуванню задач для диференціальних рівнянь математичної фізики в необмеженому просторі, а також елементам теорії інтегрального перетворення Фур'є й теорії узагальнених функцій, читати його можна незалежно.

Перші розділи посібника присвячено стислому викладу елементів варіаційного числення; більш розгорнутий розгляд цих питань можна знайти в невеличкій книзі авторів «Варіаційне числення». Центральне місце в посібнику посідають крайові задачі на власні значення для рівняння Штурма — Ліувілля на скінченному інтервалі, півосі й осі та застосування отриманих результатів до розв'язування конкретних одновимірних крайових задач для рівняння коливань неоднорідної струни та рівнянь теплопровідності й дифузії. Відповідні розділи викладено, на наш погляд, з достатнім ступенем повноти. В останніх розділах посібника на простих прикладах окреслюються підходи до розв'язування крайових задач для хвильового рівняння та рівнянь теплопровідності й дифузії у двох і трьох вимірах.

Зважаючи на спрямованість посібника, ми не намагалися досягнути більшого ступеня загальності математичних положень і фактів, ніж той, який тепер потрібен для фізичних застосувань. Доведення окремих математичних пропозицій і теорем лише намічаються та роз'яснюються на прикладах, без ретельного з'ясування технічних деталей. Однак при всіх спрощеннях твердження теорем залишаються точними, а умови, за яких наведені формулювання й формули є коректними, обговорюються достатньо повно. Кожний розділ містить тренувальні завдання; деякі з них доповнюють, ілюструють і поглиблюють основний текст.

Сподіваємося, що книга буде корисною для студентів та аспірантів фізичних, інженерно-фізичних і математичних спеціальностей університетів.

***Вадим Адамян***
***Мирослав Сушко***

# Розділ 1

# НАЙПРОСТІША ЗАДАЧА ВАРІАЦІЙНОГО ЧИСЛЕННЯ

## 1.1. ОСНОВНІ ПОНЯТТЯ ТА ОЗНАЧЕННЯ

*Поняття функціонала* природно виникає при розв'язуванні широкого класу геометричних і фізичних задач. Для прикладу, розглянемо найпростішу задачу про обчислення площі фігури, обмеженої на площині $XOY$ графіком неперервної однозначної функції $y = y(x)$, прямими $x = x_1$, $x = x_2$ та віссю абсцис $OX$ (рис. 1.1). Ця площа, як відомо, визначається інтегралом

$$S = \int_{x_1}^{x_2} dx \left| y(x) \right| \tag{1.1}$$

і тому набуває того чи іншого значення в залежності від функції $y = y(x)$. Іншими словами, площа $S$ є функцією кривої $y = y(x)$, тобто числовою функцією від функції.

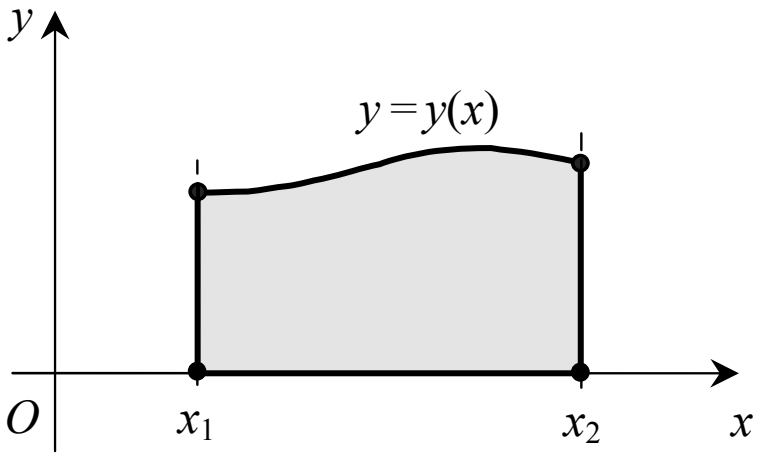

Рис. 1.1. Площа під графіком заданої функції

**Означення 1.1.1.** Кажуть, що на деякому класі функцій задано функціонал $J$, якщо вказано правило, за яким кожній функції $y(x)$ цього класу ставиться у відповідність певне число $J[y]$.

У наведеному прикладі функціонал площі $S = S[y]$ задається на класі $C([x_1, x_2])$ функцій, неперервних на відрізку $[x_1, x_2]$. Надалі під класом $C^k([x_1, x_2])$ розумітимемо множину функцій, неперервних на відрізку $[x_1, x_2]$ разом зі своїми першими $k$ похідними.

Клас функцій, на якому задано функціонал, називають *областю визначення функціонала*. Підкреслимо: щоб задати функціонал, треба вказати як правило відповідності, так і область визначення.

**Завдання 1.1.1.** Знайдіть явний вигляд та область визначення таких функціоналів:

а) довжина гладкої кривої $y = y(x)$;

б) площа поверхні, утвореної обертанням кривої $y = y(x)$ навколо осі $OX$;

в) об'єм тіла, обмеженого поверхнею обертання кривої $y = y(x)$ навколо осі $OX$ та площинами $x = x_1$, $x = x_2$;

г) координати центра мас однорідної нитки, форма якої описується рівнянням $y = y(x)$.

*Відповіді*: а) $J[y] = \int\limits_{x_1}^{x_2} dx\sqrt{1+y'^2(x)},\ C^1([x_1,x_2])$;

б) $J[y] = 2\pi\int\limits_{x_1}^{x_2} dx|y(x)|\sqrt{1+y'^2(x)},\ C^1([x_1,x_2])$;

в) $J[y] = \pi\int\limits_{x_1}^{x_2} dx\, y^2(x),\ C([x_1,x_2])$;

г) $X[y] = \dfrac{1}{M}\int\limits_{x_1}^{x_2} dx\, x\sqrt{1+y'^2(x)},\ Y[y] = \dfrac{1}{M}\int\limits_{x_1}^{x_2} dx\, y\sqrt{1+y'^2(x)}$,

де $M = \int\limits_{x_1}^{x_2} dx\sqrt{1+y'^2(x)},\ C^1([x_1,x_2])$.

*Основна задача варіаційного числення* полягає у відшуканні такої кривої (чи поверхні) $y_0(x)$, для якої значення $J[y_0]$ заданого функціонала є найменшим або найбільшим по відношенню до його значень $J[y]$ на всіх близьких до $y_0(x)$ кривих $y(x)$ із заданого класу функцій. Кривих (поверхонь) із цією властивістю може бути декілька; вони називаються *екстремалями* функціонала $J$.

Сформульована задача схожа на задачу диференціального числення про відшукання екстремумів функції $f(x)$, тобто тих значень $x_0$ змінної $x$, для яких величини $f(x_0)$ є найменшими або найбільшими в порівнянні зі значеннями $f(x)$ у достатньо близьких до $x_0$ точках.

Наведемо більш точні означення.

**Означення 1.1.2.** $\varepsilon$-Околом порядку $k$ кривої $y_0(x)$ на проміжку $[x_1,x_2]$ називають множину всіх кривих $y(x)$, для яких скрізь на цьому проміжку виконуються нерівності

$$|y(x)-y_0(x)| \le \varepsilon,\ |y'(x)-y_0'(x)| \le \varepsilon, \ldots,\ |y^{(k)}(x)-y_0^{(k)}(x)| \le \varepsilon.$$

Число $\varepsilon$ називають відстанню порядку $k$ між кривими $y(x)$ і $y_0(x)$.

**Означення 1.1.3.** Кажуть, що функціонал $J$ має на кривій $y_0(x)$ із заданої множини кривих класу $C^k([x_1, x_2])$ відносний екстремум, якщо нерівність $J[y] \le J[y_0]$ (або $J[y] \ge J[y_0]$) виконується для всіх кривих $y(x)$ цієї множини, належних ε-околу порядку $k$ кривої $y_0(x)$. Якщо ж ця нерівність справджується для всіх кривих $y(x)$ заданої множини, розміщених у деякій області $D$ площини $XOY$, то кажуть, що функціонал $J[y]$ набуває на кривій $y_0(x)$ абсолютного екстремуму в області $D$.

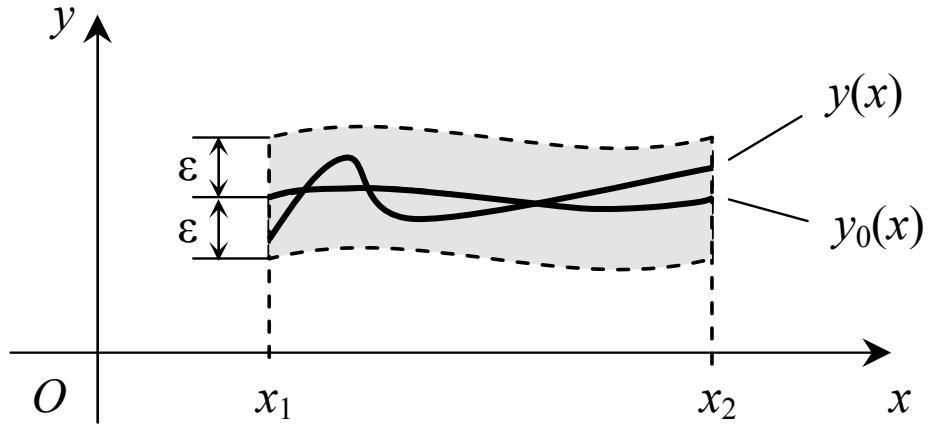

Рис. 1.2. ε-Окіл (щонайменше нульового порядку) кривої $y_0(x)$

**Завдання 1.1.2.** Серед усіх плоских гладких кривих, що з'єднують дві задані точки, укажіть ту, що має найменшу довжину. Який це тип екстремуму?

*Відповідь*: пряма, абсолютний.

Відносні екстремуми, які набуваються на кривих з ε-околів нульового та першого порядків кривої $y_0(x)$, називаються, відповідно, сильними та слабкими. Надалі ми зосереджуємося на вивченні слабких екстремумів, оскільки необхідні умови їх існування є одночасно необхідними умовами існування сильних та абсолютних екстремумів.

## 1.2. НЕОБХІДНА УМОВА ЕКСТРЕМУМУ ФУНКЦІОНАЛА. ТЕОРЕМА ЕЙЛЕРА — ЛАГРАНЖА

Розглянемо функціонал найпростішого типу

$$J[y] = \int_{x_1}^{x_2} dx F(x, y, y'), \tag{1.2}$$

де функція $F(x, y, y')$ є неперервною разом зі своїми першими та другими похідними за всіма аргументами в деякій області $D$ площини $XOY$ і при довільних значеннях похідної $y'(x)$. Для дослідження необхідних умов існування екстремуму такого функціонала використовується наступний факт.

**Лема 1.2.1.** Якщо неперервна функція $f(x)$ задовольняє співвідношення

$$\int_{x_1}^{x_2} dx f(x) h(x) = 0,$$

де $h(x)$ — довільна функція, що є гладкою на проміжку $[x_1, x_2]$ та дорівнює нулю на його кінцях $(h(x_1) = h(x_2) = 0)$, то скрізь на цьому проміжку $f(x) \equiv 0$.

*Доведення.* Припустимо, що твердження леми хибне, і в деякій точці $x_0$ проміжку $[x_1, x_2]$ значення $f(x_0) \neq 0$, наприклад, $f(x_0) > 0$. Унаслідок неперервності функція $f(x)$ буде додатною й у деякому околі $[x_0 - \delta, x_0 + \delta]$, $\delta > 0$, точки $x_0$, який повністю належить $[x_1, x_2]$. Скориставшись тепер довільністю $h(x)$, візьмемо її у вигляді

$$h(x) = \begin{cases} 0, & x_1 \le x < x_0 - \delta, \\ \eta^2(x), & x_0 - \delta \le x \le x_0 + \delta, \\ 0, & x_0 + \delta < x \le x_2, \end{cases}$$

де $\eta(x)$ — довільна гладка функція, яка відмінна від нуля при $x \in (x_0 - \delta, x_0 + \delta)$ та дорівнює нулю при $x = x_0 \pm \delta$. Дістаємо

$$\int_{x_1}^{x_2} dx f(x) h(x) = \int_{x_0-\delta}^{x_0+\delta} dx f(x) \eta^2(x) > 0,$$

що суперечить умові леми.

Зауважимо, що схожу лему можна довести й для кратних інтегралів. Її часто називають *основною лемою варіаційного числення.*

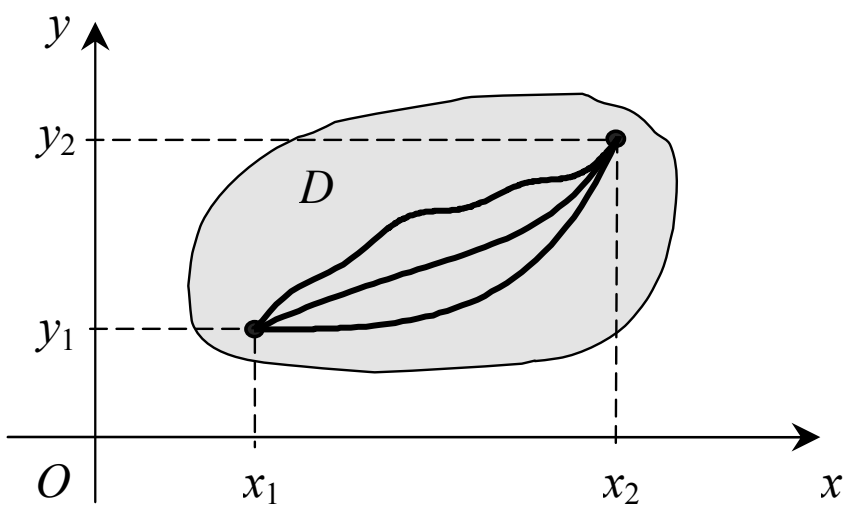

Рис. 1.3. Допустимі гладкі криві з жорстко закріпленими кінцями

**Лема 1.2.2.** Нехай крива $y_0(x)$ повністю лежить в області $D$ і є екстремаллю функціонала (1.2) у класі $C^2([x_1, x_2])$ кривих зі спільним початком і спільним кінцем (рис. 1.3):

$$y(x_1) = y_1, \quad y(x_2) = y_2. \qquad (1.3)$$

Тоді вона задовольняє диференціальне рівняння

$$\frac{\partial F}{\partial y} - \frac{d}{dx}\frac{\partial F}{\partial y'} = 0. \qquad (1.4)$$

*Доведення*. Нехай $y_0(x)$ — екстремаль функціонала (1.2). Розглянемо множину кривих $y(x) = y_0(x) + \delta y(x) = y_0(x) + \alpha h(x)$ з $\varepsilon$-околу другого порядку кривої $y_0(x)$ (див. рис. 1.3). Тут $\delta y$ — відхилення функції $y$ від екстремалі, $\alpha$ — близький до нуля параметр ($|\alpha| \le \alpha_0$), $h(x)$ — довільна гладка функція, для якої $\max\limits_{x_1 \le x \le x_2} \left\{ |\delta y(x)|, |\delta y'(x)|, |\delta y''(x)| \right\} < \varepsilon.$ На кінцях проміжку $y(x_i) = y_0(x_i) + \alpha h(x_i)$, тобто $y_i = y_i + \alpha h(x_i)$ ($i = 1,2$), звідки $h(x_1) = h(x_2) = 0$. Зауважимо, що криві $y(x)$ з області визначення функціонала називають *допустимими* кривими досліджуваної варіаційної задачі.

Функціонал (1.2) при підстановці в нього функції $y(x)$ являє собою функцію від параметра $\alpha$:

$$\varphi(\alpha) = J[y_0 + \alpha h] = \int\limits_{x_1}^{x_2} dx F(x, y_0 + \alpha h, y_0' + \alpha h'),$$

яка при $\alpha = 0$ має екстремум (оскільки $J[y_0]$ — екстремальне значення функціонала); тому $\varphi'(0) = 0$.

Умови, які задовольняє функція $F(x,\alpha) \equiv F(x, y_0 + \alpha h, y_0' + \alpha h')$, означають, що функції $F(x,\alpha)$ і $\partial F(x,\alpha)/\partial\alpha$ неперервні в прямокутнику $[x_1, x_2] \times [-\alpha_0, \alpha_0]$, $\alpha_0 \cdot \max\limits_{x_1 \le x \le x_2} \left\{ |h(x)|, |h'(x)| \right\} < \varepsilon$, і тому диференціювання за $\alpha$ під знаком інтеграла є законним. Маємо:

$$\varphi'(\alpha) = \int\limits_{x_1}^{x_2} dx \left\{ \left. \frac{\partial F(x,y,y')}{\partial y} \right|_{y = y_0 + \alpha h, y' = y_0' + \alpha h'} \cdot h(x) + \right.$$

$$\left. + \left. \frac{\partial F(x,y,y')}{\partial y'} \right|_{y = y_0 + \alpha h, y' = y_0' + \alpha h'} \cdot h'(x) \right\}. \qquad (1.5)$$

Підставляючи $\alpha = 0$ та інтегруючи другий доданок частинами, дістаємо:

$$\varphi'(0) = \left. \left. \frac{\partial F(x,y,y')}{\partial y'} \right|_{y = y_0, y' = y_0'} \cdot h(x) \right|_{x_1}^{x_2} +$$

$$+ \int\limits_{x_1}^{x_2} dx \left\{ \left. \frac{\partial F(x,y,y')}{\partial y} \right|_{y = y_0, y' = y_0'} - \frac{d}{dx} \left. \frac{\partial F(x,y,y')}{\partial y'} \right|_{y = y_0, y' = y_0'} \right\} h(x) = 0. \qquad (1.6)$$

Перший доданок у формулі (1.6) після підстановки меж інтегрування дорівнює нулю. Другий доданок містить під знаком інтеграла довільну гладку функцію $h(x)$, $h(x_1)=h(x_2)=0$. Згідно з лемою 1.2.1, він дорівнює нулю лише за умови, що дорівнює нулю вираз у фігурних дужках, тобто за умови, що функція $y_0(x)$ задовольняє рівняння (1.4).

Рівняння (1.4) називається *рівнянням Ейлера — Лагранжа* для екстремалі фунціонала (1.2). Зауважимо, що щойно наведене виведення цього рівняння базувалося на припущенні, що екстремаль $y_0(x)$ функціонала (1.2) є функцією класу $C^2([x_1,x_2])$. Покажемо тепер, що ця вимога є занадто сильною — рівняння (1.3) справджується й тоді, коли екстремум шукається на множині функцій класу $C^1([x_1,x_2])$. При цьому ми також доведемо, що похідна $\partial F/\partial y'$ має на екстремалі $y_0(x)$ повну похідну за $x$, і що в тих точках кривої $y_0(x)$, де $\partial^2 F/\partial y'^2 \neq 0$, існує неперервна друга похідна $y_0''(x)$.

**Лема 1.2.3.** Якщо неперервна функція $f(x)$ задовольняє співвідношення

$$\int_{x_1}^{x_2} dx\, f(x)g(x)=0,$$

де довільна функція $g(x)$ є неперервною на проміжку $[x_1,x_2]$ та задовольняє умову

$$\int_{x_1}^{x_2} dx\, g(x)=0,$$

то скрізь на цьому проміжку $f(x)$ є стала.

*Доведення.* Очевидно, що для довільної сталої $C$

$$\int_{x_1}^{x_2} dx C g(x)=0.$$

Виберемо $C$ із співвідношення

$$\int_{x_1}^{x_2} dx\, f(x)=C(x_2-x_1)=C\int_{x_1}^{x_2} dx.$$

Можемо записати:

$$\int_{x_1}^{x_2} dx\big[f(x)-C\big]g(x)=0.$$

Скориставшись тепер довільністю $g(x)$, візьмемо її у вигляді $g(x)=f(x)-C$. Дістаємо

$$\int_{x_1}^{x_2} dx\big[f(x)-C\big]^2=0,$$

тобто $f(x) \equiv C.$

Безпосереднім наслідком леми 1.2.3 є

**Лема 1.2.4 (дю Буа-Реймона)**. Якщо неперервна функція $f(x)$ задовольняє співвідношення

$$\int_{x_1}^{x_2} dx f(x) h'(x) = 0,$$

де $h(x)$ — довільна функція, що має неперервну похідну $h'(x)$ на проміжку $[x_1, x_2]$ і дорівнює нулю на його межах, то скрізь на цьому проміжку $f(x)$ є стала.

Повернімося тепер до формули (1.5). Означимо функцію $G(x)$ співвідношенням

$$G(x) = \int_{x_1}^{x} dx \left. \frac{\partial F(x,y,y')}{\partial y} \right|_{y=y_0, y'=y_0'},$$

звідки

$$\frac{dG(x)}{dx} = \left. \frac{\partial F(x,y,y')}{\partial y} \right|_{y=y_0, y'=y_0'}.$$

Інтегруючи частинами, можемо записати:

$$\int_{x_1}^{x_2} dx \left\{ \left. \frac{\partial F(x,y,y')}{\partial y} \right|_{y=y_0, y'=y_0'} \right\} h(x) = \int_{x_1}^{x_2} dx \frac{dG}{dx} h(x) = G(x) h(x) \Big|_{x_1}^{x_2} - \int_{x_1}^{x_2} dx\, G(x) h'(x).$$

Замість рівності (1.6) маємо:

$$\varphi'(0) = G(x) h(x) \Big|_{x_1}^{x_2} + \int_{x_1}^{x_2} dx \left\{ -G(x) + \left. \frac{\partial F(x,y,y')}{\partial y'} \right|_{y=y_0, y'=y_0'} \right\} h'(x) = 0. \quad (1.7)$$

Перший доданок у цій формулі, як і раніше, обертається в нуль після підстановки меж інтегрування. Звідси випливає, що другий доданок дорівнює нулю для довільної функції $h(x)$, що задовольняє умови леми дю Буа-Реймона. Отже, згідно з цією лемою, вираз у фігурних дужках у формулі (1.7) є сталою величиною:

$$-G(x) + \left. \frac{\partial F}{\partial y'} \right|_{y=y_0, y'=y_0'} = C. \quad (1.8)$$

Це співвідношення заміняє диференціальне рівняння Ейлера — Лагранжа (1.4); воно називається *інтегральною формою рівняння Ейлера — Лагранжа* для функції $y_0(x)$.

Узявши до уваги, що функція $G(x)$ має неперервну першу похідну, бачимо, що похідна $\partial F/\partial y'$ для екстремальної кривої $y_0(x)$ диференційовна. Диференціюючи (1.8) за змінною $x$, для $y_0(x)$ дістаємо рівняння (1.4):

$$\frac{d}{dx}\frac{\partial F}{\partial y'}\bigg|_{y=y_0,\,y'=y_0'} = \frac{dG}{dx} = \frac{\partial F}{\partial y}\bigg|_{y=y_0,\,y'=y_0'}.$$

Якщо тепер у деякій точці $(x,y)$ кривої $y_0(x)$ $\partial^2 F/\partial y'^2 \neq 0$, то можемо записати:

$$\frac{d}{dx}\frac{\partial F}{\partial y'} = \frac{\partial^2 F}{\partial x \partial y'} + \frac{\partial^2 F}{\partial y \partial y'}\cdot y_0' + \frac{\partial^2 F}{\partial y'^2}\cdot \lim_{\Delta x\to 0}\frac{\Delta y_0'}{\Delta x},$$

де $\Delta y_0'$ — приріст функції $y_0'(x)$ при зміщенні з точки з абсцисою $x$ у точку з абсцисою $x+\Delta x$. Звідси

$$\lim_{\Delta x\to 0}\frac{\Delta y_0'}{\Delta x} = \frac{\dfrac{d}{dx}\dfrac{\partial F}{\partial y'} - \dfrac{\partial^2 F}{\partial x \partial y'} - \dfrac{\partial^2 F}{\partial y \partial y'}\cdot y_0'}{\dfrac{\partial^2 F}{\partial y'^2}},$$

тобто друга похідна функції $y_0(x)$ існує скрізь, де $\partial^2 F/\partial y'^2 \neq 0$.

Отже, необхідну умову існування екстремуму (відносного, абсолютного) функціонала (1.2) у класі гладких кривих зі спільним початком і спільним кінцем можна сформулювати таким чином:

**Теорема 1.2.1 (Ейлера — Лагранжа).** Для того щоб функція $y_0(x)$ із класу $C^1([x_1,x_2])$ надавала екстремум функціоналу (1.2) за умов (1.3), необхідно, щоб вона задовольняла рівняння (1.4). Якщо для цієї функції $\partial^2 F/\partial y'^2 \neq 0$ скрізь на відрізку $[x_1,x_2]$, то $y_0(x) \in C^2([x_1,x_2])$.

У загальному випадку рівняння Ейлера — Лагранжа є *звичайним диференціальним рівнянням другого порядку* відносно функції $y(x)$. Серед двічі неперервно диференційовних функцій воно відбирає ті, що можуть бути екстремалями, але не обов’язково ними є. Відібрані функції, взагалі кажучи, залежать від двох довільних сталих інтегрування, тобто утворюють двопараметричну сім’ю кривих $y = y(x,C_1,C_2)$. Сталі $C_1$ і $C_2$ визначаються за допомогою умов (1.3) на межах проміжку $[x_1,x_2]$.

З упевненістю можна стверджувати, що функціонал (1.2) не має екстремалей, якщо рівняння Ейлера — Лагранжа: взагалі не має розв’язків; або має розв’язки, але вони не задовольняють умови (1.3);

або має розв'язки, що задовольняють умови (1.3), але не належать класу $C^1([x_1,x_2])$. У задачах прикладного характеру факт існування гладкого екстремуму визначається, як правило, самою постановкою задачі.

На завершення цього підрозділу введемо кілька нових понять та доведемо *інваріантність* рівняння Ейлера — Лагранжа відносно перетворення координат.

Функцію $\delta y(x)$ називають ***варіацією функції*** $y(x)$, а величину $\Delta J[y]=J[y+\delta y]-J[y]$ — приростом функціонала $J[y]$, зумовленим варіацією $\delta y$. Головну (лінійну) за $\delta y$ частину приросту функціонала називають ***варіацією функціонала*** $\delta J[y]$. Щоб знайти $\delta J[y]$ у нашому випадку, у різниці $\Delta J[y]$ достатньо виділити член, лінійний за $\alpha$:

$$\Delta J[y]=J[y+\delta y]-J[y]=J[y+\alpha h]-J[y]=\varphi(\alpha)-\varphi(0)=$$
$$=\varphi(0)+\varphi'(0)\alpha+O(\alpha^2)+\ldots-\varphi(0)=\varphi'(0)\alpha+O(\alpha^2)+\ldots.$$

Виокремлюючи перший доданок, маємо:

$$\delta J[y]=\int_{x_1}^{x_2}\left\{\frac{\partial F}{\partial y}\delta y+\frac{\partial F}{\partial y'}\delta y'\right\}dx.$$

Якщо $y(x)\in C^2([x_1,x_2])$, то, ураховуючи формулу (1.6), знаходимо:

$$\delta J[y]=\frac{\partial F}{\partial y'}\delta y(x)\bigg|_{x_1}^{x_2}+\int_{x_1}^{x_2}\left\{\frac{\partial F}{\partial y}-\frac{d}{dx}\frac{\partial F}{\partial y'}\right\}\delta y(x)dx.$$

Для кривих із закріпленими кінцями

$$\delta J[y]=\int_{x_1}^{x_2}\left\{\frac{\partial F}{\partial y}-\frac{d}{dx}\frac{\partial F}{\partial y'}\right\}\delta y(x)dx. \qquad (1.9)$$

Для екстремальної кривої $y=y_0(x)$ маємо: $\delta J[y_0]=0$. Це співвідношення нагадує теорему Ферма для точок екстремуму $x_0$ функції $f(x)$: $df(x_0)=0$. Аналогія між повним диференціалом функції та варіацією функціонала дозволяє шукати останній за звичайними правилами знаходження диференціала $df(x)=f'(x)dx$, тільки тепер у ролі приросту аргументу $dx$ виступає варіація функції $\delta y$, і підінтегральний вираз у формулі для $\Delta J[y]=J[y+\delta y]-J[y]$ формально розвивається за $\delta y$. Варіацію $\delta y$ можна диференціювати один або декілька разів, причому $\dfrac{d^k\delta y}{dx^k}=\delta\dfrac{d^k y}{dx^k}$.

Підкреслимо: якщо існує варіація функціонала як диференціал за параметром (див. доведення леми 1.2.2), то існує й варіація функціо-нала як головна лінійна частина приросту функціонала, і ці два означення рівносильні.

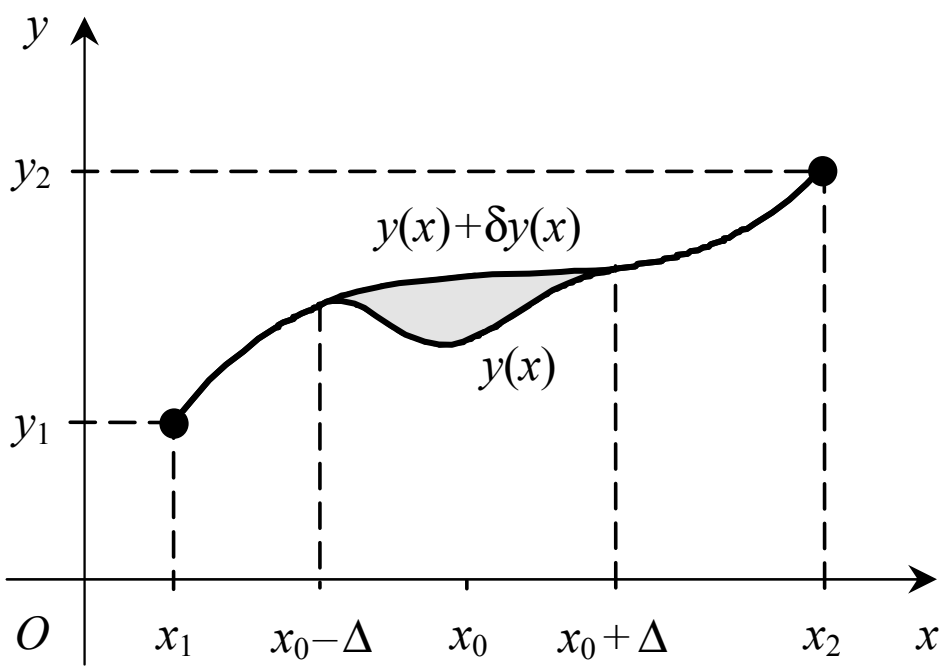

Рис. 1.4. Площа Δ-околу точки $x_0$

Внески в $\Delta J[y]$, пропорційні другому та більш високим степеням $\delta y$, називають другою варіацією ($\delta^2 J[y]$) та варіаціями вищих порядків.

Аналогом похідної у варіаційному численні виступає функціональна похідна. Виберемо варіацію $\delta y$ так, щоб вона відрізнялася від нуля в малому $\Delta$-околі точки $x_0$, тобто щоб криві $y$ та $y+\delta y$ збігалися скрізь на проміжку $[x_1,x_2]$, за винятком інтервалу $(x_0-\Delta,x_0+\Delta)\subset[x_1,x_2]$ (рис. 1.4). Користуючись теоремою про середнє, інтеграл (1.9) перепишемо у вигляді

$$\delta J[y]=\int\limits_{x_0-\Delta}^{x_0+\Delta}\left\{\frac{\partial F}{\partial y}-\frac{d}{dx}\frac{\partial F}{\partial y'}\right\}\delta y(x)dx=\left\{\frac{\partial F}{\partial y}-\frac{d}{dx}\frac{\partial F}{\partial y'}\right\}\Bigg|_{x=\tilde{x}_0}\cdot S,$$

де $S=\int\limits_{x_0-\Delta}^{x_0+\Delta}\delta y(x)dx$ — з точністю до знака площа (затемненої області) між кривими $y$ та $y+\delta y$ (називатимемо її площею $\Delta$-околу), $\tilde{x}_0$ — деяка точка з $\Delta$-околу. *Функціональною похідною* за кривою $y$ у точці $x_0$ називають границю відношення

$$\frac{\delta J[y]}{\delta y}\equiv\lim_{\Delta\to 0}\frac{\delta J[y]}{S}=\left\{\frac{\partial F}{\partial y}-\frac{d}{dx}\frac{\partial F}{\partial y'}\right\}\Bigg|_{x=x_0}. \qquad (1.10)$$

*Функціональна похідна вздовж екстремальної кривої дорівнює нулю в кожній точці цієї кривої.*

Поняття функціональної похідної дозволяє досить просто довести той факт, що властивість кривої бути екстремаллю є інваріантна відносно перетворення координат.

**Теорема 1.2.2.** Нехай крива $y=y_0(x)$ є екстремаллю функціонала (1.2) на множині гладких кривих, що з'єднують точки $(x_1,y_1)$ і $(x_2,y_2)$. Крім того, нехай після переходу від координат $(x,y)$ до криволінійних координат $(t,q)$ за формулами

$$x = \varphi(t,q), \quad y = \psi(t,q),$$

з відмінним від нуля якобіаном переходу

$$I \equiv \begin{vmatrix} \dfrac{\partial x}{\partial t} & \dfrac{\partial x}{\partial q} \\ \dfrac{\partial y}{\partial t} & \dfrac{\partial y}{\partial q} \end{vmatrix} \neq 0,$$

її рівняння набирає вигляду $q = q_0(t)$, а функціонал $J[y]$ перетворюється на функціонал $J_1[q]$,

$$J[y] = J_1[q] = \int_{t_1}^{t_2} dt L(t,q,q'), \tag{1.11}$$

від кривих $q = q(t)$, що з'єднують точки $(t_1,q_1)$ і $(t_2,q_2)$ на площині $TOQ$, такі, що $x_i = \varphi(t_i,q_i)$, $y_i = \psi(t_i,q_i)$, $i = 1,2$. Тоді екстремум функціонала $J_1[q]$ досягається саме на кривій $q = q_0(t)$, яка буде розв'язком рівняння Ейлера — Лагранжа для $J_1[q]$:

$$\frac{\partial L}{\partial q} - \frac{d}{dt}\frac{\partial L}{\partial q'} = 0. \tag{1.12}$$

Іншими словами, якщо загальний розв'язок рівняння (1.4) має вигляд $y = y_0(x,C_1,C_2)$, то загальний розв'язок рівняння (1.12) визначається як неявна функція $q = q_0(t)$, що задовольняє рівняння $\psi(t,q_0(t)) = y_0(\varphi(t,q_0(t)),C_1,C_2)$.

*Доведення.* Справді, якщо в $\Delta$-околі кожної точки $x_0$ з інтервалу $(x_1,x_2)$ для екстремальної кривої виконується співвідношення

$$\lim_{\Delta\to 0} \frac{\delta J[y]}{S} = 0,$$

то і в $\Delta_1$-околі кожної відповідної точки $t_0$ з інтервалу $(t_1,t_2)$ функціональна похідна від $J_1[q]$ дорівнює нулю:

$$\lim_{\Delta_1\to 0} \frac{\delta J_1[q]}{S_1} = \lim_{\Delta\to 0} \frac{\delta J[y]}{S}\frac{S}{S_1} = 0,$$

оскільки відношення площ $S$ і $S_1$ старого й нового околів прямує до детермінанта $I$, який не дорівнює нулю.

Інваріантність рівняння Ейлера — Лагранжа відносно перетворення координат дозволяє при розв'язуванні конкретних геометричних та фізичних задач в однаковій мірі користуватися різними криволінійними системами координат, лише б вони були взаємно однозначно пов'язані з декартовими. Перехід до криволінійних координат, наприклад, сферичних, стає особливо ефективним у тому випадку, коли система має відповідну симетрію, зокрема, сферичну, оскільки тоді рівняння Ейлера — Лагранжа стає помітно простішим. Зауважимо, що перехід до нових координат можна здійснювати безпосередньо в підінтегральному виразі функціонала (1.2), а потім для нового інтеграла писати рівняння Ейлера — Лагранжа — це й буде початкове рівняння в декартових координатах, віднесене до нових змінних.

## 1.3. ПОНИЖЕННЯ ПОРЯДКУ РІВНЯННЯ ЕЙЛЕРА — ЛАГРАНЖА

Як уже зазначалося, рівняння Ейлера — Лагранжа є, взагалі кажучи, звичайним диференціальним рівнянням другого порядку, і тому знаходження його розв'язків є в більшості випадків значно складнішою задачею, ніж розв'язування диференціальних рівнянь першого порядку. У зв'язку з цим особливий інтерес привертають умови, за яких його порядок можна понизити чи, навіть, це рівняння можна аналітично розв'язати. Це можливо в тих випадках, коли *інтегрант* (підінтегральна функція) $F(x,y,y')$ функціонала (1.2) не залежить від одного чи двох своїх аргументів. У фізичних задачах відсутність такої залежності є проявом певних симетрій, наприклад, простору та часу, що призводять до виконання відповідних законів збереження.

Зупинимося на аналізі зазначених випадків докладніше. Нагадаємо, що під $x$ і $y$ можна розуміти довільні криволінійні координати.

1) Якщо $F = F(y')$, тобто залежить лише від похідної $y'$, то $\partial F/\partial y = 0$ і $\dfrac{d}{dx}\dfrac{\partial F}{\partial y'} = \left(\dfrac{\partial}{\partial y'}\dfrac{\partial F}{\partial y'}\right)\dfrac{dy'}{dx}$. Рівняння (1.4) зводиться до рівності

$$y''\frac{\partial^2 F(y')}{\partial y'^2} = 0, \tag{1.13}$$

і хоча б один із співмножників повинен дорівнювати нулю. Нехай $y''(x) = 0$, тоді $y'(x) = C_1$, $y(x) = C_1 x + C_2$. Ці криві утворюють двопараметричну сім'ю прямих. У другому можливому випадку, коли

$$\frac{\partial^2 F(y')}{\partial y'^2} \equiv f(y') = 0,$$

рівняння $f(y')=0$ є функціональним рівнянням відносно похідної $y'$. Нехай $y_i'(x)\equiv k_i$ — усі його дійсні корені (їх може й не бути). Тоді $y_i(x)=k_i x+C$. Ці криві утворюють більш вузьку однопараметричну ($k_i$ — фіксовані числа) сім'ю прямих, що входить до зазначеної вище двопараметричної. Отже, у випадку, коли $F=F(y')$, екстремалями функціонала (1.2) можуть бути лише прямі

$$y(x)=C_1 x+C_2. \tag{1.14}$$

2) Якщо $F=F(x,y)$, тобто не залежить від похідної $y'$, то $\partial F/\partial y'=0$ і рівняння (1.4) вироджується у функціональне відносно $y(x)$:

$$\frac{\partial F(x,y)}{\partial y}\equiv g(x,y)=0. \tag{1.15}$$

Якщо розв'язки $y_i=y_i(x)$ останнього рівняння існують, то, на відміну від розв'язків диференціального рівняння, вони не містять довільних сталих, підбором яких задовольняють умови (1.3). Можна очікувати, що функції $y_i(x)$ підкорятимуться умовам (1.3) лише у виключних ситуаціях, тобто в загальному випадку відповідний функціонал не матиме екстремалей.

3) Якщо $F$ залежить від $y'$ лінійно, тобто має вигляд $F=a(x,y)+$ $+b(x,y)y'$, де $a(x,y)$, $b(x,y)$ — деякі функції, то рівняння Ейлера — Лагранжа знову вироджується у функціональне:

$$\frac{\partial a(x,y)}{\partial y}-\frac{\partial b(x,y)}{\partial x}\equiv g(x,y)=0.$$

Як щойно було зазначено, розв'язок подібного рівняння в загальному випадку не задовольняє умови в крайніх точках, і тому функціонал (1.2), найімовірніше, екстремалей не має. Більше того, варіаційна задача взагалі втрачає зміст, якщо

$$\frac{\partial a(x,y)}{\partial y}-\frac{\partial b(x,y)}{\partial x}\equiv 0,$$

бо тоді лінійна форма $a(x,y)dx+b(x,y)dy$ є повним диференціалом, і значення функціонала (1.2) залежать лише від початкової та кінцевої точок, а не від вибору кривої $y=y(x)$.

4) $F=F(x,y')$, тобто не залежить явно від $y$ (змінну $y$ у цьому випадку називають *циклічною*). Тоді $\partial F/\partial y=0$ і рівняння (1.4) набирає вигляду

$$\frac{d}{dx}\frac{\partial F(x,y')}{\partial y'}=0.$$

Його відразу можна один раз зінтегрувати — рівняння Ейлера — Лагранжа має перший інтеграл

$$\frac{\partial F(x,y')}{\partial y'} = C_1. \tag{1.16}$$

Це є диференціальне рівняння першого порядку, яке не залежить явно від функції $y(x)$. Розв'язавши його відносно $y'$, отримаємо рівняння або кілька рівнянь типу $y' = f(x,C_1)$, звідки $y(x)$ знаходимо інтегруванням:

$$y(x) = \int dx f(x,C_1) + C_2. \tag{1.17}$$

5) $F = F(y,y')$, тобто не залежить явно від $x$. Помноживши обидві частини виразу

$$\frac{\partial F}{\partial y} - \frac{d}{dx}\frac{\partial F}{\partial y'} = \frac{\partial F}{\partial y} - \frac{\partial^2 F}{\partial y'^2}y'' - \frac{\partial^2 F}{\partial y \partial y'}y' = 0$$

на $y'$, перепишемо його як повну похідну

$$\frac{d}{dx}\left(F - y'\frac{\partial F}{\partial y'}\right) = 0$$

(перевірте!). Відповідно, рівняння Ейлера — Лагранжа має перший інтеграл

$$F - y'\frac{\partial F}{\partial y'} = C_1. \tag{1.18}$$

Це є диференціальне рівняння першого порядку, яке не залежить явно від $x$. Розв'язавши його відносно $y'$, дістанемо рівняння або кілька рівнянь типу $y' = f(y,C_1)$. Відокремлюючи в ньому змінні та інтегруючи, функцію $y$ знаходимо із співвідношення

$$\int \frac{dy}{f(y,C_1)} = x + C_2\ . \tag{1.19}$$

**Завдання 1.3.1.** Серед усіх плоских гладких кривих, що з'єднують задані точки $(x_1,y_1)$ і $(x_2,y_2)$, знайдіть ту, що має найменшу довжину.

*Вказівка*: див. завдання 1.1.1 а).

*Відповідь*: пряма $y(x) = y_1 + \dfrac{x - x_1}{x_2 - x_1}(y_2 - y_1)$.

**Завдання 1.3.2.** Доведіть, що функціонал площі (1.1) не має екстремалей у класі гладких кривих, що з'єднують задані точки $(x_1,y_1)$ і $(x_2,y_2)$. Відповідь поясніть.

*Вказівка*: рівняння Ейлера — Лагранжа вироджується в неправильну тотожність.

**Завдання 1.3.3.** Знайдіть *геодезичну* поверхні сфери радіусом $R$ (лінію найменшої довжини між двома заданими точками на цій поверхні).

*Розв'язання.* Декартові координати $(x, y, z)$ точки на поверхні сфери пов'язані з азимутальним $\varphi$ та полярним $\theta$ кутами сферичної системи координат з початком у центрі сфери співвідношеннями

$$x = R\sin\theta\cos\varphi, \quad y = R\sin\theta\sin\varphi, \quad z = R\cos\theta, \quad 0 \le \varphi < 2\pi, \quad 0 \le \theta \le \pi. \tag{1.20}$$

Відповідно, квадрат елемента довжини на поверхні сфери

$$dl^2 = dx^2 + dy^2 + dz^2 = R^2 d\theta^2 + R^2 \sin^2\theta d\varphi^2,$$

і шукана крива $\varphi = \varphi(\theta)$ є екстремаллю функціонала

$$L[\varphi] = R\int_{\theta_1}^{\theta_2} d\theta\sqrt{1 + \sin^2\theta\varphi'^2}, \quad \varphi' \equiv \frac{d\varphi}{d\theta},$$

яка з'єднує задані точки з координатами, скажімо, $(R, \theta_1, \varphi_1)$ та $(R, \theta_2, \varphi_2)$. Підінтегральна функція в ньому не залежить від змінної $\varphi$, тому можемо скористатися першим інтегралом виду (1.16) відповідного рівняння Ейлера — Лагранжа:

$$\frac{\sin^2\theta\varphi'}{\sqrt{1 + \sin^2\theta\varphi'^2}} = C_1.$$

Звідси

$$\varphi' = \frac{C_1}{\sin^2\theta\sqrt{1 - \dfrac{C_1^2}{\sin^2\theta}}}.$$

Відокремивши змінні $\varphi$ і $\theta$ та перейшовши до нової змінної $u = \operatorname{ctg}\theta$, дістаємо:

$$d\varphi = -\frac{C_1 du}{\sqrt{1 - C_1^2 - C_1^2 u^2}},$$

звідки

$$\varphi + C_2 = \arccos\left(\frac{C_1 u}{\sqrt{1 - C_1^2}}\right),$$

$C_2$ — ще одна стала інтегрування. Позначивши $C_1^* \equiv C_1 \Big/ \sqrt{1 - C_1^2}$, знаходимо рівняння геодезичної у вигляді

$$C_1^* \operatorname{ctg}\theta = \cos(\varphi + C_2). \tag{1.21}$$

Сталі $C_1^*$ і $C_2$ знаходимо з умови, що крива (1.21) починається в точці $(R, \theta_1, \varphi_1)$ і закінчується в точці $(R, \theta_2, \varphi_2)$. Дістаємо систему

$$C_1^* \operatorname{ctg}\theta_1 = \cos(\varphi_1 + C_2),$$
$$C_1^* \operatorname{ctg}\theta_2 = \cos(\varphi_2 + C_2).$$

Якщо $C_1^* = 0$, то $\varphi = \text{const}$. У цьому випадку геодезична є відрізком меридіана сфери, тобто великого кола, що проходить через полюси сфери. Якщо ж $C_1^* \neq 0$, то перепишемо формулу (1.21) у вигляді

$$\operatorname{ctg}\theta = A\cos\varphi + B\sin\varphi,$$

де $A$ і $B$ — деякі комбінації сталих. Повертаючись за допомогою формул (1.20) до декартових координат, отримуємо:

$$z = Ax + By.$$

Бачимо, що й у цьому випадку геодезична виявляється відрізком великого кола, яке є перерізом сфери $x^2 + y^2 + z^2 = R^2$ та площини $z = Ax + By$, що проходить через її центр.

**Завдання 1.3.4.** Серед усіх кривих, що з'єднують задані точки $A$ і $B$ у вертикальній площині (точки не лежать на одній вертикалі), віднайдіть ту, рухаючись по якій під впливом сили тяжіння матеріальна точка пройде шлях від $A$ до $B$ за найкоротший час. Початкова швидкість точки дорівнює нулю, тертя та опір середовища нехтовно малі. (Задача про брахістохрону І. Бернуллі[1].)

---

[1] Ця задача була сформульована І. Бернуллі в 1696 році в замітці «Нова задача, до розв'язування якої запрошуються математики», де він написав: «У вертикальній площині дано дві точки *A* і *B*. Визначити шлях *AMB*, спускаючись по якому під впливом власної ваги, тіло *M*, почавши рухатися з точки *A*, дійде до точки *B* за найкоротший час. Щоб зацікавити аматорів подібних запитань та заохотити їх до більш активних спроб розв'язати вказану задачу, повідомляю, що вона не зводиться до марної розумової вправи, позбавленої будь-якого практичного значення — як це може комусь здатися. Насправді ця задача має великий практичний інтерес, при цьому, крім механіки, ще й для інших дисциплін, що може всім здатися неправдоподібним» (Johann Bernoulli. Problema novum ad cujus solutionem Mathematici invitantur, Acta Eruditorum Lipsiae, Junii A.MDCXCVI, p. 269). Розв'язки цієї задачі, отримані самим І. Бернуллі, а також Лейбніцом, Лопіталем, Я. Бернуллі та Ньютоном, і поклали початок розвитку варіаційного числення та, взагалі, дали значний поштовх розвитку аналізу. Загальні підходи до розв'язання задач варіаційного числення були розроблені Ейлером і Лагранжем у XVIII столітті.

*Розв'язання*. Нехай маса матеріальної точки $m$, прискорення вільного падіння $g$. Напрямимо вісь $OX$ горизонтально, вісь $OY$ — вертикально вниз, початок координат помістимо в точку $A$, а координати точки $B$ позначимо через $(l,h)$ (див. рис. 1.5). Щоб знайти час $T[y]$, потрібний матеріальній точці для руху вздовж шуканої кривої $y(x)$, візьмемо до уваги, що ділянку $dl$ з координатами кінців $(x,y)$, $(x+dx, y+dy)$ та довжиною $dl=\sqrt{dx^2+dy^2}=\sqrt{1+y'^2(x)}\,dx$ точка проходить за час $dt = dl/v(y)$, де $v(y)$ — швидкість точки на цій ділянці. Останню знаходимо із закону збереження механічної енергії: $\frac{1}{2}mv^2(y)=mgy$ (початкова швидкість матеріальної точки дорівнює нулю, потенціальну енергію відраховуємо від осі $OX$). Дістаємо:

$$T[y]=\int_0^l dx\frac{\sqrt{1+y'^2}}{\sqrt{2gy}}. \quad (1.22)$$

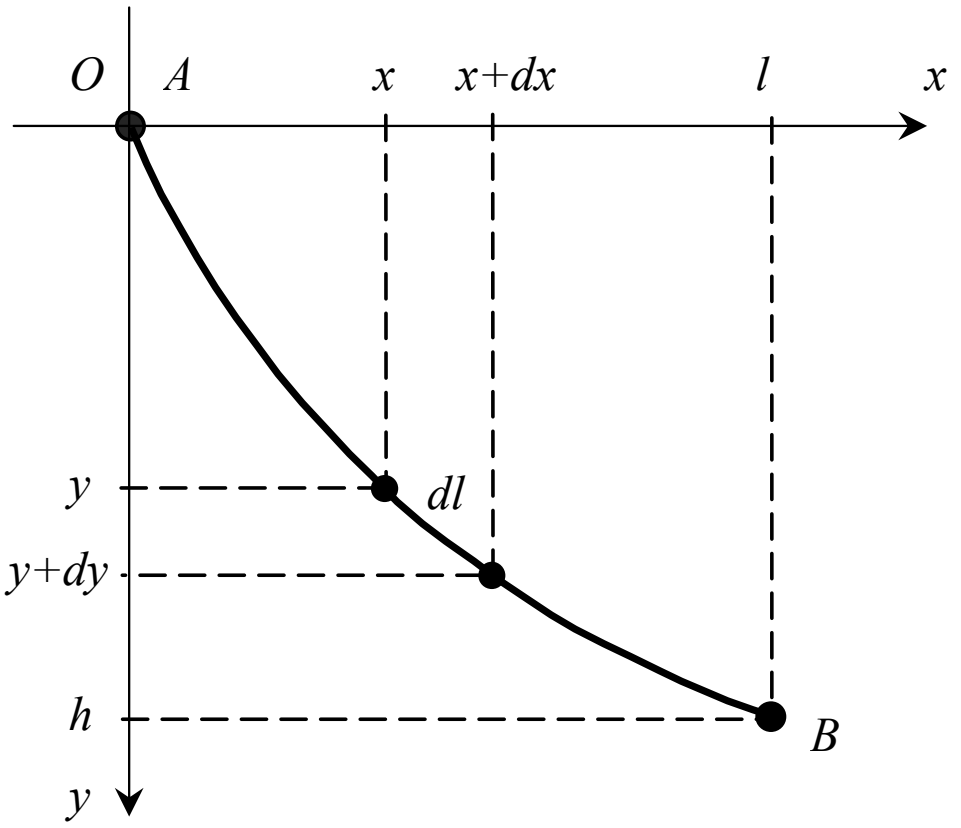

Рис. 1.5. Геометрія задачі про брахістохрону

Знайдемо екстремалі функціонала (1.22), скориставшись першим інтегралом (1.18). Останній набирає вигляду (множник $\sqrt{2g}$ заносимо в сталу інтегрування $C$)

$$\frac{1}{\sqrt{y}\sqrt{1+y'^2}}=C. \quad (1.23)$$

Розв'язок цього рівняння зручно подати в параметричному вигляді $x=x(\varphi)$, $y=y(\varphi)$, де $\varphi_1\le\varphi\le\varphi_2$, при цьому точці $A$ зручно віднести значення параметра $\varphi_1=0$: $x(0)=0$, $y(0)=0$. Для точки $B$, де $\varphi=\varphi_2$, можемо записати: $x(\varphi_2)=l$, $y(\varphi_2)=h$. Підставивши в рівняння (1.23) $y'=\operatorname{ctg}\frac{\varphi}{2}$, дістаємо:

$$y(\varphi)=\frac{1}{C^2}\sin^2\frac{\varphi}{2}=C_1(1-\cos\varphi), \quad (1.24)$$

де для зручності ми ввели нову сталу $C_1\equiv 1/(2C^2)$. Відповідно,

$$dx(\varphi)=\operatorname{tg}\frac{\varphi}{2}dy(\varphi)=C_1\operatorname{tg}\frac{\varphi}{2}\sin\varphi\, d\varphi=C_1(1-\cos\varphi)d\varphi,$$

звідки

$$x(\varphi) = C_1(\varphi - \sin\varphi) + C_2. \qquad (1.25)$$

З умов у точці $A$ випливає, що стала інтегрування $C_2 = 0$. З умов у точці $B$ для сталих $\varphi_2$ і $C_1$ знаходимо співвідношення

$$\frac{1-\cos\varphi_2}{\varphi_2 - \sin\varphi_2} = \frac{h}{l}, \quad C_1 = \frac{h}{1-\cos\varphi_2}. \qquad (1.26)$$

Очевидно, що в загальному випадку значення цих сталих можна відновити лише числовими методами.

Отже, шукана екстремаль функціонала часу $T[y]$ дається формулами (1.24)–(1.26). Вона називається *брахістохроною* і є відрізком *циклоїди* — однієї з кривих, що описуються формулами (1.24) і (1.25). Зауважимо, що стала $C_1$ дорівнює радіусу $R$ кола, що творить циклоїду (кожна точка кола описує циклоїду, коли воно котиться без ковзання вздовж осі $OX$), а стала $\varphi_2$ — куту, на який треба повернути коло, щоб його верхня в початковий момент точка перемістилася з положення $A$ в положення $B$ (рис. 1.6).

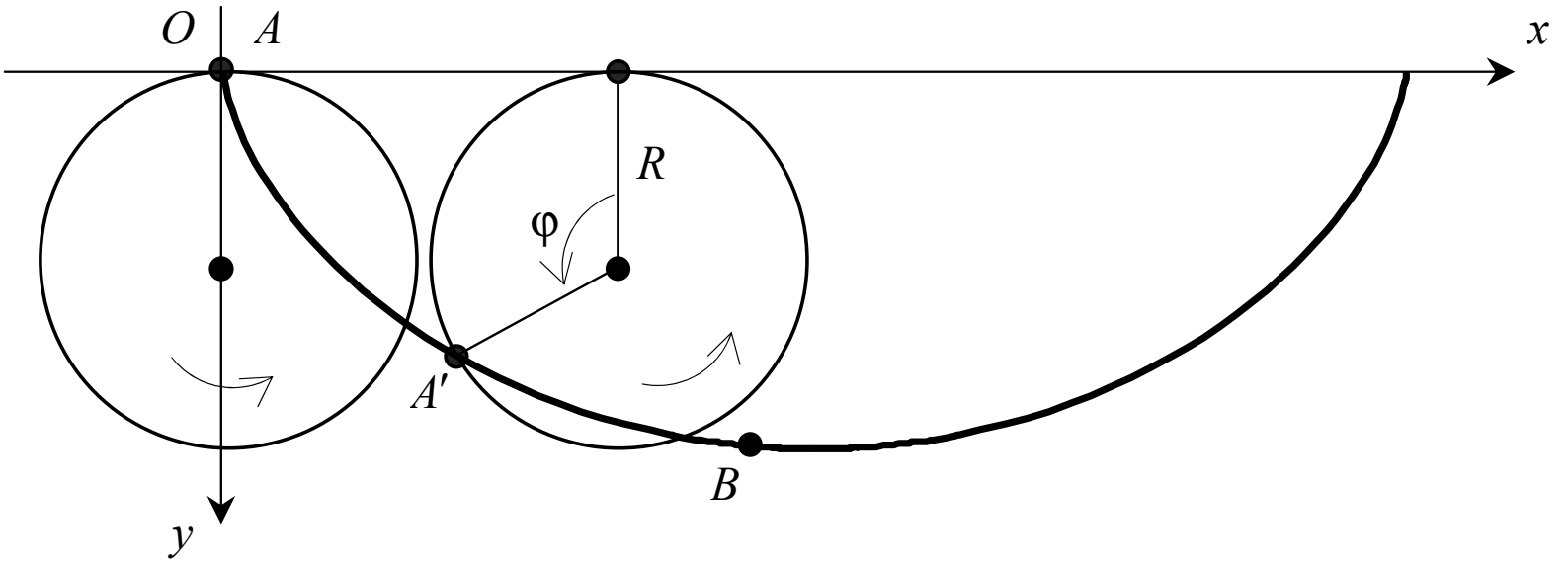

Рис. 1.6. Циклоїда, утворена точкою $A$ твірного кола. Відрізок $AB$ — брахістохрона

Природно виникає питання про характер часової залежності координат матеріальної точки при русі вздовж брахістохрони. Відповідь на нього можна знайти за допомогою загального принципу найменшої дії Гамільтона, який на сьогодні є одним з найбільш важливих екстремальних принципів у фізиці[1]. Опишемо, як на його основі будується лагранжевий формалізм класичної механіки.

[1] Як загальний принцип механіки та оптики, принцип найменшої дії вперше був сформульований у словесні формі П. Мопертюї в 1746 р. Згодом був розвинутий Ейлером, Лагранжем та Гамільтоном.

## 1.4. ЕКСТРЕМАЛЬНІ ПРИНЦИПИ У ФІЗИЦІ

У рамках формалізму Лагранжа кожній механічній системі, підкореній ідеальним в'язям, ставиться у відповідність певна функція $L(q_k, \dot{q}_k, t)$ *узагальнених координат* $q_k(t)$ та *узагальнених швидкостей* $\dot{q}_k(t)$ точок системи, а також часу $t$ ($k = 1, 2, ..., s$). Нагадаємо, що узагальненими координатами називають будь-які величини (не обов'язково декартові координати точок системи), за допомогою яких можна однозначно задати положення системи у просторі. Мінімально необхідну кількість $s \geq 1$ таких величин називають кількістю ступенів вільності системи. Узагальнені швидкості означаються як похідні узагальнених координат за часом: $\dot{q}_k \equiv dq_k(t)/dt$.

Функцію $L(q_k, \dot{q}_k, t)$ називають *функцією Лагранжа* системи; вона визначається як різниця кінетичної $K$ та потенціальної $\Pi$ енергій системи як функцій змінних $q_k(t)$ і $\dot{q}_k(t)$. Її похідні

$$P_k = \frac{\partial L}{\partial \dot{q}_k}, \quad F_k = \frac{\partial L}{\partial q_k} \tag{1.27}$$

називаються відповідно *узагальненими імпульсами* та *узагальненими силами*. Повна енергія механічної системи[1]

$$E = K + \Pi = \sum_k \frac{\partial L}{\partial \dot{q}_k} \dot{q}_k - L. \tag{1.28}$$

Згідно з *принципом найменшої дії* (*принципом Гамільтона*), рух системи на проміжку часу $[t_1, t_2]$ між двома фіксованими положеннями, які характеризуються значеннями $q_k(t_1)$ і $q_k(t_2)$ узагальнених координат точок системи, відбувається таким чином, що *функціонал дії*

$$S = \int_{t_1}^{t_2} dt L(q_k, \dot{q}_k, t) \tag{1.29}$$

набуває екстремального (зазвичай кажуть мінімального) значення:

$$\delta S = \delta \int_{t_1}^{t_2} dt L(q_k, \dot{q}_k, t) = 0. \tag{1.30}$$

[1] За аналогією, перші інтеграли виду (1.16) і (1.18) також інколи називають інтегралами імпульсу та енергії.

Очевидно, що для системи з одним ступенем вільності $q(t)$ необхідною умовою для цього є виконання рівняння

$$\frac{\partial L}{\partial q} - \frac{d}{dt}\frac{\partial L}{\partial \dot{q}} = 0. \qquad (1.31)$$

Як буде показано далі, при наявності $s$ ступенів вільності $q_k(t)$ еволюція останніх від початкового положення $q_k(t_1)$ до кінцевого $q_k(t_2)$ визначається системою диференціальних рівнянь Ейлера — Лагранжа

$$\frac{\partial L}{\partial q_k} - \frac{d}{dt}\frac{\partial L}{\partial \dot{q}_k} = 0, \quad k = 1,2,...,s. \qquad (1.32)$$

Рівняння (1.31) та система рівнянь (1.32) залишаються інваріантними відносно вибору узагальнених координат $q_k(t)$. Для знаходження сталих інтегрування в розв'язках цих рівнянь замість умов типу (1.3) (тобто значень координат точок системи в початковий та кінцевий моменти часу: $q_k(t_1) = q_k^{(1)}$ і $q_k(t_2) = q_k^{(2)}$) використовують так звані *початкові умови* — значення координат і швидкостей у початковий момент часу $t_1$: $q_k(t_1) = q_k^{(1)}$, $\dot{q}_k(t_1) = v_k^{(1)}$ (як правило, уважають, що $t_1 = 0$). У фізичному плані це означає, що рух механічної системи жорстко детермінований: при відомому рівнянні руху він повністю визначається початковими координатами та початковими швидкостями точок системи[1].

З огляду на означення (1.27) рівняння руху (1.32) набирають вигляду рівнянь Ньютона, записаних для узагальнених імпульсів та узагальнених сил:

$$\dot{P}_k = F_k, \quad k = 1,2,...,s. \qquad (1.33)$$

У порівнянні з рівняннями Ньютона, записаними для кожної матеріальної точки системи, рівняння (1.32), (1.33) мають ту перевагу, що їх кількість дорівнює кількості ступенів вільності системи, і при наявності в'язей, що обмежують рух системи, є меншою від $3N$, тобто є меншою від кількості рівнянь Ньютона, потрібних для опису системи $N$ матеріальних точок. Крім того, у рівняння Ейлера — Лагранжа не входять сили реакції в'язей, які заздалегідь невідомі.

---

[1] Для диференціальних рівнянь другого порядку початкові умови гарантують єдиність розв'язку при довільних значеннях $(q_k^{(1)}, v_k^{(1)})$ (окрім спеціальних випадків, коли $t_1$ — особлива точка). Умови ж на кінцях проміжку $[t_1, t_2]$ не гарантують узагалі існування відповідної траєкторії. Наприклад, для умов $q(0) = a$, $q(2\pi/\omega) = b$ розв'язок рівняння $\ddot{q} + \omega^2 q = 0$ на проміжку $[0, 2\pi/\omega]$ існує лише при $a = b$, і при виконанні цієї умови існує нескінченно багато розв'язків цього рівняння.

Скористаємося окресленим підходом для аналізу руху циклоїдального та математичного маятників.

**Завдання 1.4.1.** Матеріальна точка рухається в однорідному полі тяжіння по гладкій циклоїді, розміщеній у вертикальній площині. Знайдіть закон руху точки, якщо радіус твірного кола циклоїди дорівнює $R$, і рух починається з найвищої точки циклоїди (точка $A$ на рис. 1.6). (Циклоїдальний маятник.)

*Розв'язання.* Нехай миттєве положення $A'$ точки на циклоїді в момент часу $t$ задається кутом повороту $\varphi = \varphi(t)$ радіуса твірного кола відносно від'ємного напряму осі $OY$ (див. рис. 1.6). Декартові координати точки в цей момент

$$x(t) = R(\varphi - \sin\varphi), \quad y(t) = R(1 - \cos\varphi).$$

Записавши $dx = R(1 - \cos\varphi)d\varphi$, $dy = R\sin\varphi d\varphi$, знаходимо декартові компоненти швидкості:

$$\dot{x}(t) = R(1 - \cos\varphi)\dot{\varphi}, \quad \dot{y}(t) = R\sin\varphi\dot{\varphi},$$

та довжину $l = l(\varphi)$ дуги $AA'$ циклоїди:

$$l = \int\limits_{AA'} dl = 2R\int\limits_0^{\varphi} d\varphi \sin\frac{\varphi}{2} = 4R\left(1 - \cos\frac{\varphi}{2}\right).$$

Функція Лагранжа точки $L = K - \Pi$, де

$$K = \frac{m}{2}\left(\dot{x}^2 + \dot{y}^2\right) = 2mR^2\sin^2\frac{\varphi}{2}\dot{\varphi}^2,$$

$$\Pi = -mgy = -2mgR\sin^2\frac{\varphi}{2} = 2mgR\cos^2\frac{\varphi}{2} - 2mgR.$$

Відповідне рівняння Ейлера — Лагранжа

$$\frac{\partial L}{\partial\varphi} - \frac{d}{dt}\frac{\partial L}{\partial\dot{\varphi}} = 0$$

досить складне. Зручніше скористатися його інваріантністю відносно перетворень координат та перейти до нової узагальненої координати $s = l - 4R = -4R\cos\frac{\varphi}{2}$, яка описує зміщення матеріальної точки від найнижчої точки циклоїди. Опустивши несуттєву сталу в потенціальній енергії, маємо:

$$K = \frac{m}{2}\dot{s}^2, \quad \Pi = \frac{mg}{8R}s^2.$$

Рівняння Ейлера — Лагранжа в нових координатах

$$\frac{\partial L}{\partial s} - \frac{d}{dt}\frac{\partial L}{\partial \dot{s}} = 0$$

зводиться до рівняння руху гармонічного осцилятора

$$\ddot{s} + \omega_0^2 s = 0 \qquad (1.34)$$

з частотою коливань $\omega_0 = \sqrt{g/(4R)}$. Його загальний розв'язок має вигляд

$$s(t) = C_1 \cos\omega_0 t + C_2 \sin\omega_0 t. \qquad (1.35)$$

Сталі $C_1$ і $C_2$ знаходимо з початкових умов $l(0) = 0,\ \dot{l}(0) = 0,$ тобто $s(0) = -4R,\ \dot{s}(0) = 0.$ Дістаємо: $C_1 = -4R,\ C_2 = 0.$

Рух точки вздовж циклоїди *ізохронний* — при довільних початкових відхиленнях $|s(0)| \le 4R$ точка досягає найнижчого положення за однакові проміжки часу, які дорівнюють чверті її періоду коливань.

Час $t^*$, який точка витрачає на проходження брахістохрони $AB$ (див. завдання 1.3.4), знаходимо із співвідношення

$$s(t^*) = l_{AB} - 4R = -4R\cos\frac{\varphi_2}{2},$$

звідки $t^* = \varphi_2/(2\omega_0)$.

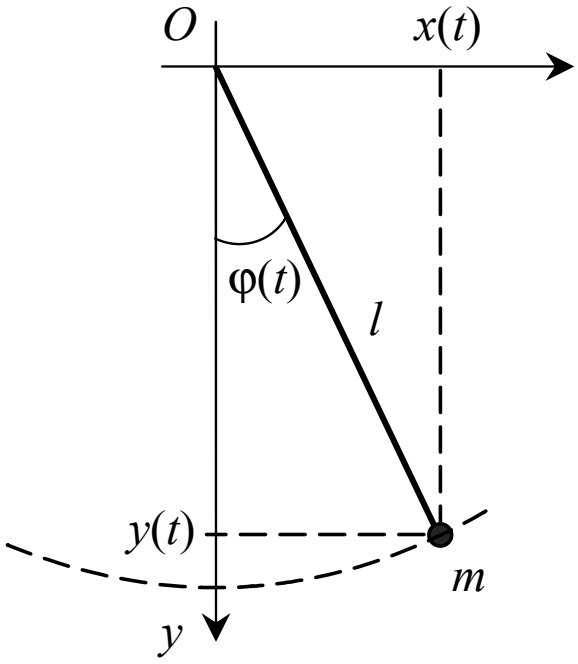

Рис. 1.7. Математичний маятник

**Завдання 1.4.2.** Матеріальну точку підвісили на нерозтяжній невагомій нитці довжиною $l$ та, відхиливши на кут $\varphi_0$, плавно відпустили. Знайдіть рівняння коливань точки та виясніть умови, за яких вони стають гармонічними. (Математичний маятник.)

*Розв'язання.* Найбільш зручною узагальненою координатою є кут $\varphi = \varphi(t)$, який описує відхилення маятника від положення рівноваги (рис. 1.7). Тоді:

$$x(t)=l\sin\varphi,\quad y(t)=l\cos\varphi.$$

Кінетична енергія та потенціальна енергія (за нульовий рівень якої вибираємо пряму $y=0$) дорівнюють:

$$K=\frac{m}{2}\left(\dot{x}^2+\dot{y}^2\right)=\frac{1}{2}ml^2\dot{\varphi}^2,\quad \Pi=-mgy=-mgl\cos\varphi.$$

Рівняння руху математичного маятника має вигляд

$$\ddot{\varphi}+\omega_0^2\sin\varphi=0,\quad \omega_0^2\equiv\frac{g}{l}.$$

Воно зводиться до рівняння гармонічних коливань (1.34) лише у випадку малих кутів відхилення $(\varphi<<1)$ від положення рівноваги. На практиці ці кути не повинні перевищувати $15^0$ $(\approx 0{,}26$ рад$)$[1].

Як уже зазначалося, принцип найменшої дії Гамільтона для механічних систем матеріальних точок є одним з найбільш загальних формулювань законів їх руху. Його перевага перед іншими полягає в тому, що він легко поширюється не лише на механічні системи з неперервним розподілом маси, але й на системи іншої фізичної природи — наприклад, електромагнітне поле, поля піонів чи кварк-глюонні поля. А саме, кожній польовій системі ставиться у відповідність функціонал дії $S$, який залежить від відповідних польових характеристик (векторів напруженості електричної $\mathbf{E}(\mathbf{r},t)$ та магнітної $\mathbf{H}(\mathbf{r},t)$ компонент електромагнітного поля або його тензора $F_{\mu\nu}(\mathbf{r},t)$, (псевдо) скалярної хвильової функції $\phi(\mathbf{r},t)$ мезонного поля, біспінорів $\psi(\mathbf{r},t)$ діраківського поля тощо) та їх похідних, які самі є функціями координат точок 4-вимірного простору — часу. Закономірності зміни стану такої системи у просторі та часі між двома фіксованими моментами $t_1$ і $t_2$ описуються певними диференціальними рівняннями, які випливають з умови екстремальності $S$ на відрізку $[t_1,t_2]$.

Застосувавши принцип найменшої дії до рівноважних систем, параметри яких не змінюються з часом, приходимо до видозмінених, але еквівалентних його формулювань у вигляді так званих *умов рівноваги*. Наведемо дві з них, до яких ми звертатимемося дещо пізніше:

1) стійкому стану рівноваги механічної системи, що знаходиться в полі потенціальних сил і підкорена голономним (накладеним тільки на координати) ідеальним стаціонарним в'язям, відповідає таке роз-ташування її частин, при якому повна потенціальна енергія системи є мінімальною (теорема Лагранжа — Діріхле);

[1] Докладний аналіз руху математичного маятника при довільних кутах відхилення можна знайти, наприклад, у § 1.5 посібника авторів Варіаційне числення. — Одеса : Астропринт, 2005. — 128 с.

2) енергія електростатичного поля, створюваного зарядженими провідниками, є мінімальною по відношенню до енергії полів, які створювалися б усіма іншими розподілами зарядів по об'єму провідників (теорема Томсона). Аналогічне твердження справджується і для магнітостатичних полів, спричинених стаціонарними струмами.

При аналізі систем, що складаються з величезної кількості мікрочастинок, ми переходимо від опису індивідуальної поведінки частинок системи до опису системи як цілого за допомогою відносно невеликої кількості параметрів. До цих параметрів належать як механічні (об'єм, тиск) чи електричні (дипольний момент) за своєю природою, так і ті, що мають виключно *статистичну* природу (температура, ентропія, функції розподілу), оскільки з'являються внаслідок статистичного усереднення процесів у системі за ймовірностями реалізації різних мікростанів системи. Як результат, закономірності поведінки таких систем лежать поза межами застосовності принципу найменшої дії.

З іншого боку, для таких систем також вдається встановити низку варіаційних принципів, зокрема, принцип максимуму ентропії для процесів у замкненій системі. Обмежившись розглядом рівноважної (термодинамічної) системи, можна від її опису за допомогою ентропії перейти до її опису за допомогою термодинамічного потенціалу, що відображає характер накладених на систему зовнішніх обмежень, та сформулювати для цього потенціалу варіаційний принцип, яким установлюється критерій рівноважного стану системи. Одне з формулювань цього принципу таке: у стані теплової рівноваги вільна енергія $F(T,V)$ термодинамічної системи (тобто робота, яку система здатна здійснити при оборотному ізотермічному процесі) є мінімальною в порівнянні з усіма іншими змінами стану при сталих температурі $T$ та об'ємі $V$. Аналогічні теореми справджуються і для інших термодинамічних потенціалів як функцій власних змінних — термодинамічного потенціалу Гіббса $\Phi(T,P)$, внутрішньої енергії $E(S,V)$ і теплової функції $W(S,P)$. Тут $S$ — ентропія системи, $P$ — тиск у системі.

**Завдання 1.4.3.** Між двома дротяними кільцями радіусами $R$ натягнуто мильну плівку. Відрізок, що з'єднує центри кілець, перпендикулярний до площин кілець і має довжину $l$ (рис. 1.8). Уважаючи товщину плівки сталою та малою, знайдіть рівняння поверхні, утвореної плівкою. (Задача Плато про мінімальну поверхню обертання.)

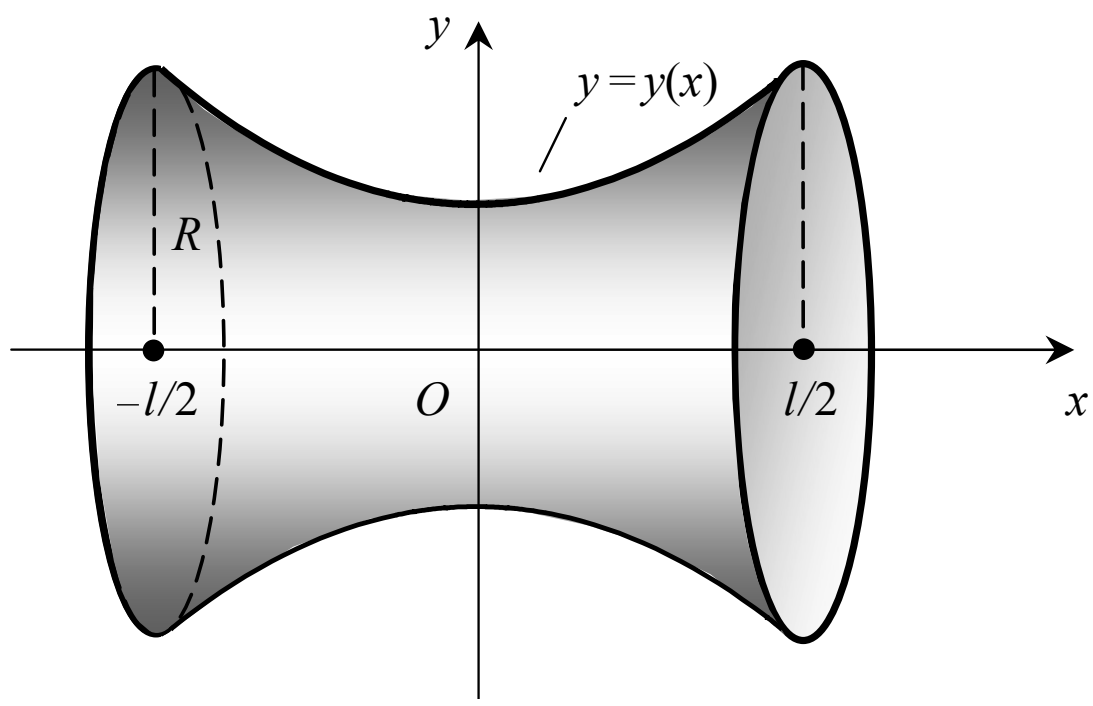

Рис. 1.8. Плівка, яка має форму поверхні, утвореної обертанням гладкої кривої $y(x)$ навколо осі $OX$

*Розв'язання*. При фіксованих температурі та густині мильного розчину плівка набирає такої форми, для якої функціонал надлишкової вільної енергії $\Delta F$ системи, пов'язаний з утворенням плівки, мінімальний. Останній дорівнює енергії поверхневого натягу плівки: $\Delta F = 2\sigma S,$ де $\sigma$ — коефіцієнт поверхневого натягу, $S$ — площа бічної поверхні плівки, коефіцієнт 2 враховує наявність у плівки внутрішньої та зовнішньої поверхонь. З міркувань симетрії зрозуміло, що поверхня плівки збігається з поверхнею, утвореною обертанням гладкої кривої $y(x) > 0$ навколо осі $OX$. Тому

$$\Delta F[y] = 4\pi\sigma \int_{-l/2}^{l/2} dxy\sqrt{1+y'^2}, \tag{1.36}$$

і задача зводиться до відшукання кривої, що мінімізує функціонал $\Delta F[y]$ за умов

$$y\left(-\frac{l}{2}\right) = y\left(\frac{l}{2}\right) = R. \tag{1.37}$$

Скориставшись першим інтегралом (1.18), можемо записати

$$y = C_1\sqrt{1+y'^2}, \;\; y' = \pm\frac{1}{C_1}\sqrt{y^2 - C_1^2},$$

звідки, відокремивши змінні, дістаємо:

$$\pm\int \frac{C_1 dy}{\sqrt{y^2 - C_1^2}} = x + C_2.$$

Інтеграл справа легко обчислюється за допомогою підстановки $y = C_1 \operatorname{ch} t$ й дорівнює $C_1 t$. Маємо $t = \pm(x + C_2)/C_1$, звідки

$$y(x) = C_1 \operatorname{ch}\frac{x + C_2}{C_1}. \quad (1.38)$$

Шукана крива (1.38) є *ланцюговою лінією*, що проходить через точки (1.37). Поверхня, утворена обертанням ланцюгової лінії, називається *катеноїдом*.

З умов (1.37) випливає, що стала $C_2 = 0$ (крива симетрична відносно осі $OY$), а стала $C_1$ задовольняє трансцендентне рівняння $C_1 \operatorname{ch}\frac{l}{2C_1} = R.$

Як завжди в подібних випадках, аналізуємо це рівняння, перейшовши до безрозмірних величин. Дістаємо:

$$\operatorname{ch} u = \alpha u, \quad (1.39)$$

де $\alpha = 2R/l$, $u = l/(2C_1)$. При $\alpha << 1$ $\operatorname{ch} u >> \alpha u$, і рівняння (1.39) розв'язків не має (див. рис. 1.9). При зростанні $\alpha$ до певного значення $\alpha^*$ пряма $\alpha u$ стає дотичною до графіка $\operatorname{ch} u$ — маємо єдиний розв'язок $u^*$. При $\alpha > \alpha^*$ існують два розв'язки $u_1$ і $u_2$; їх числові значення залежать від $\alpha$.

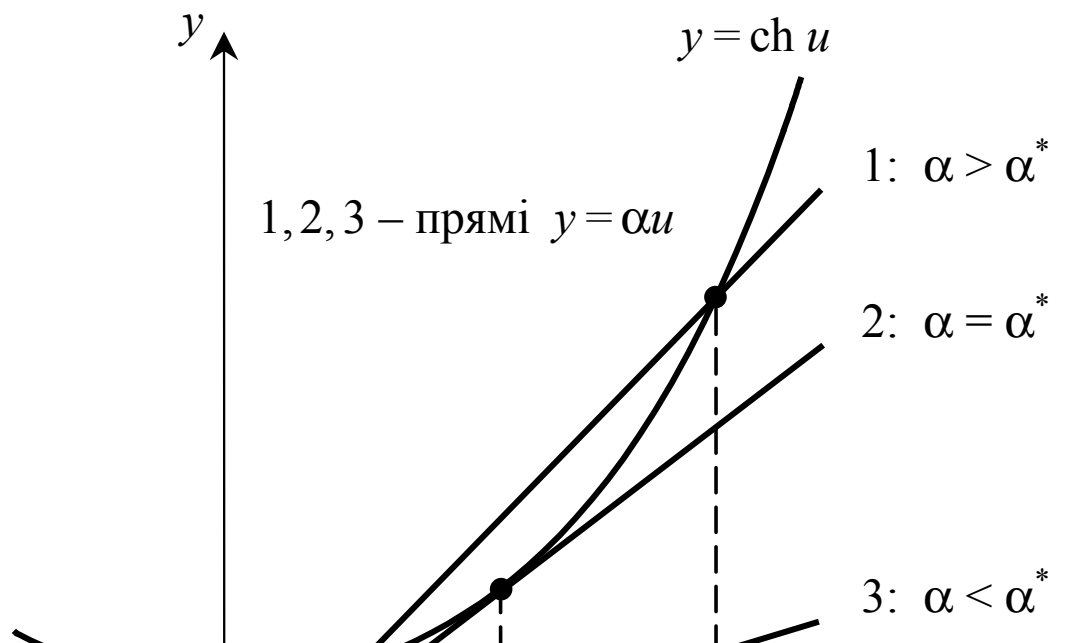

Рис. 1.9. Розв'язки рівняння ch $u = \alpha u$ при різних значеннях параметра $\alpha$

Тангенс кута нахилу дотичної до функції $\operatorname{ch} u$ в точці $u^*$ дорівнює $\operatorname{sh} u^*$, тому для $u^*$ маємо рівняння $\operatorname{ch} u^* = u^* \operatorname{sh} u^*$. Числові обчислення дають: $\alpha^* = \operatorname{sh} u^* \approx 1{,}509 \approx 3/2$.

Отже, при $l > 4R/3$ задача не має розв'язків у класі $C^1([-l/2, l/2])$: серед гладких поверхонь обертання нема такої, яка б реалізовувала мінімум функціонала (1.36), проходячи че-рез точки (1.37). Цю ситуацію можна інтерпретувати наступним чином. Якщо відстань між кільцями досить велика в порівнянні з їх радіусами, то площа поверхні, утвореної двома кругами радіусами $R$ та відрізком осі $OX$ між ними, є меншою, ніж площа довільної гладкої поверхні, натягненої на кільця (схожу ситуацію маємо й для функціонала площі (1.1)). З фізичного погляду це означає, що при $l > 4R/3$ плівка стає нестійкою й розривається.

З двох розв'язків, що маємо при $l < 4R/3$, реалізується лише один. Щоб його визначити, обчислюємо значення вільної енергії (1.36) для функції (1.38) при двох значеннях сталої $C_1$ ($l/(2u_1)$ і $l/(2u_2)$). Шуканому кореню відповідає менше значення вільної енергії.

Поява «зайвого» кореня відображає той факт, що виконання рівняння Ейлера — Лагранжа (1.4) є необхідною, але не достатньою умовою існування мінімуму функціонала.

Завершимо цей підрозділ прикладом на застосування відомого з геометричної оптики екстремального *принципу Ферма*[1]: серед допустимих шляхів між двома фіксованими точками в оптично неоднорідному середовищі світло поширюється (за стаціонарних умов) тим, на який воно (або, точніше, фронт світлової хвилі) витрачає найменше часу.

**Завдання 1.4.4.** У прозорому середовищі зі змінним показником заломлення $n(x,y)$ світло поширюється з точки $A(0,0)$ у точку $B(l,h)$. Знайдіть траєкторію променя світла, якщо $n(x,y) = n_0(1+\alpha y)$. Проаналізуйте її форму для слабко неоднорідного ($\alpha << 1$) та однорідного ($\alpha = 0$) середовищ.

*Вказівка*. Задача зводиться до відшукання екстремуму функціонала

$$T[y] = \int_A^B \frac{dl}{u(x,y)} = \frac{1}{c}\int_0^l dx\, n(x,y)\sqrt{1+y'^2} \qquad (1.40)$$

за додаткових умов $y(0) = 0$ та $y(l) = h$. Тут $c$ — швидкість світла у вакуумі, $u(x,y) = c/n(x,y)$ — швидкість світла в середовищі.

[1] П. Ферма постулював свій принцип у 1662 р., ще до появи основ класичної механіки. Показавши, що заломленому на поверхні розділу двох середовищ променю світла відповідає шлях, на який світло витрачає найменше часу, він узагальнив цей результат, посилаючись на міркування, що «природа завжди діє найкоротшим та найпростішим способом». Принцип Ферма мав зміст найбільш загального закону, з якого випливали відомі на той час закони геометричної оптики (закон прямолінійного поширення світла в однорідному середовищі, закони відбивання та заломлення світла на поверхні розділу двох середовищ із різними оптичними густинами). Сучасне виведення цього принципу базується на аналізі рівнянь хвильової оптики у граничному випадку малих довжин хвиль.

*Відповідь*. Траєкторією є відрізок ланцюгової лінії

$$y(x)=\frac{C_1}{\alpha}\operatorname{ch}\left[\frac{\alpha}{C_1}\left(x+C_2\right)\right]-\frac{1}{\alpha}, \tag{1.41}$$

де сталі $C_1$ та $C_2$ задовольняють рівняння

$$\frac{C_1}{\alpha}\operatorname{ch}\left[\frac{\alpha}{C_1}C_2\right]-\frac{1}{\alpha}=0,$$

$$\frac{C_1}{\alpha}\operatorname{ch}\left[\frac{\alpha}{C_1}\left(l+C_2\right)\right]-\frac{1}{\alpha}=h. \tag{1.42}$$

Урахувавши тотожності $\operatorname{ch}(x+y)=\operatorname{ch}x\operatorname{ch}y+\operatorname{sh}x\operatorname{sh}y,\ \operatorname{ch}^2x-\operatorname{sh}^2x=1$ та першу з формул (1.42), з формули (1.41) можемо виключити $C_2$:

$$y(x)=\frac{1}{\alpha}\left[-1+\operatorname{ch}\frac{\alpha x}{C_1}\right]+\frac{1}{\alpha}\sqrt{1-C_1^2}\operatorname{sh}\frac{\alpha x}{C_1}.$$

Сталу $C_1$ знаходимо з умови $y(l)=h.$

При $\alpha\ll 1$ з точністю до квадратичних за $\alpha$ членів маємо:

$$y(x)\approx\sqrt{1-C_1^2}\left[\frac{x}{C_1}+\frac{1}{6}\alpha^2\left(\frac{x}{C_1}\right)^3\right]+\frac{1}{2}\alpha\left(\frac{x}{C_1}\right)^2.$$

Для однорідного середовища

$$y(x)=\sqrt{1-C_1^2}\,\frac{x}{C_1}=\frac{h}{l}x,$$

тобто світло поширюється в ньому по прямій. Останній результат можна отримати відразу ж, якщо врахувати, що при $\alpha=0$ підінтегральний вираз у функціоналі (1.40) залежить лише від похідної $y'$.

## 1.5. ДОСТАТНЯ УМОВА СЛАБКОГО МІНІМУМУ

Зупинимося коротко на достатніх умовах існування мінімуму (максимуму) функціонала найпростішого типу (1.2). Обмежимося розглядом мінімуму, який реалізується на множині гладких кривих зі спільним початком і спільним кінцем.

Нехай $y_0(x)$ — крива, яка надає мінімум функціоналу (1.2). Тоді для довільної гладкої функції $h(x)$, що задовольняє умови $h(x_1)=h(x_2)=0,$ і достатньо малих значень параметра $\alpha$ маємо:

$$\varphi_h(\alpha) = J[y_0 + \alpha h] = \int\limits_{x_1}^{x_2} dx F(x, y_0 + \alpha h, y_0' + \alpha h') >$$

$$> \varphi(0) = \int\limits_{x_1}^{x_2} dx F(x, y_0, y_0'), \ \ |\alpha| > 0,$$

тобто

$$\varphi_h'(0) = 0, \ \ \varphi_h''(0) \geq 0. \tag{1.43}$$

Як було встановлено раніше (див. лему 1.2.2 та теорему Ейлера — Лагранжа), перша з цих умов веде до рівняння Ейлера — Лагранжа (1.4). Розв'язавши його та скориставшись умовами жорсткого закріплення (1.3), знаходимо $y_0(x)$.

Позначимо через $C_0^1([x_1, x_2])$ множину гладких функцій на відрізку $[x_1, x_2]$, які дорівнюють нулю при $x = x_1$ і $x = x_2$. Друге із співвідношень (1.43) потребує проаналізувати умови, за яких для довільної функції $h(x) \in C_0^1([x_1, x_2])$ справджується нерівність

$$\varphi_h''(0) = \int\limits_{x_1}^{x_2} dx \left( S h^2 + R h'^2 \right) \geq 0, \tag{1.44}$$

де $S$ та $R$ виражаються через $y_0(x)$ за формулами

$$S(x) \equiv \left( \frac{\partial^2 F}{\partial y^2} - \frac{d}{dx} \frac{\partial^2 F}{\partial y \partial y'} \right) \Bigg|_{y = y_0}, \ \ R(x) \equiv \frac{\partial^2 F}{\partial y'^2} \Bigg|_{y = y_0}. \tag{1.45}$$

Далі вважаємо, що $S(x)$ і $R(x)$ — неперервні функції на відрізку $[x_1, x_2]$, і що $R(x)$ ще має на $[x_1, x_2]$ неперервну похідну.

**Теорема 1.5.1**. Нерівність (1.44) виконується для довільної функції $h(x) \in C_0^1([x_1, x_2])$ лише тоді, коли

$$\mu \equiv \min_{x \in [x_1, x_2]} R(x) \geq 0. \tag{1.46}$$

*Доведення*. Припустимо, що $R(x)$ досягає мінімуму в точці $x_0 \in (x_1, x_2)$, і що $R(x_0) = \mu < 0$. Оскільки $R(x)$ — неперервна функція, то існує таке число $\Delta > 0$, що $R(x) < \mu/2$, якщо $|x - x_0| \leq \Delta$.

Розглянемо послідовність функцій

$$h_n(x) = \begin{cases} \dfrac{1}{\sqrt{\Delta}} \left[ \sin \dfrac{\pi n}{\Delta}(x - x_0) - \dfrac{1}{2} \sin \dfrac{2\pi n}{\Delta}(x - x_0) \right], & |x - x_0| < \Delta, \quad n = 1, 2, \ldots, \\ 0, & |x - x_0| > \Delta. \end{cases}$$

Усі ці функції належать множині $C_0^1([x_1, x_2])$, і для них

$$\int_{x_1}^{x_2} dx\, h_n^2(x) = \int_{x_0-\Delta}^{x_0+\Delta} dx\, h_n^2(x) = \frac{5}{4},$$
$$\int_{x_1}^{x_2} dx\, h_n'^2(x) = \int_{x_0-\Delta}^{x_0+\Delta} dx\, h_n'^2(x) = \frac{2\pi^2 n^2}{\Delta^2}. \qquad (1.47)$$

Нехай $M \equiv \max\limits_{x\in[x_1,x_2]} S(x)$. Тоді, згідно з формулами (1.47),

$$\varphi_{h_n}''(0) = \int_{x_1}^{x_2} dx\left[S(x)h_n^2(x) + R(x)h_n'^2(x)\right] = \int_{x_0-\Delta}^{x_0+\Delta} dx\left[S(x)h_n^2(x) + R(x)h_n'^2(x)\right] \le$$
$$\le M\int_{x_0-\Delta}^{x_0+\Delta} dx\, h_n^2(x) + \frac{\mu}{2}\int_{x_0-\Delta}^{x_0+\Delta} dx\, h_n'^2(x) = \frac{5}{4}M + \frac{\pi^2 n^2 \mu}{\Delta^2}.$$

Якщо $\mu < 0$, то $\varphi_{h_n}''(0) < 0$ при $n > \sqrt{5\Delta^2 M/(4\pi^2|\mu|)}$. Отже, умова (1.46) є необхідною для того, щоб функція $y_0(x)$ надавала мінімум функціоналу (1.2).

Далі вважатимемо, що $R(x) > 0$, тобто $\mu > 0$. Ця умова називається *посиленою умовою Лежандра*.

**Завдання 1.5.1**. Нехай функціонал (1.2) породжується функціями виду

$$F(x, y, y') = \sqrt{1 + y'^2}\, H(y), \qquad (1.48)$$

де за змістом задачі $\min H(y) > 0$. Переконайтеся, що для мінімуму такого функціонала посилена умова Лежандра виконується.

**Теорема 1.5.2**. Якщо посилена умова Лежандра виконується, то нерівність

$$S(x) = \left.\left(\frac{\partial^2 F}{\partial y^2} - \frac{d}{dx}\frac{\partial^2 F}{\partial y \partial y'}\right)\right|_{y=y_0} \ge 0 \qquad (1.49)$$

є достатньою для того, щоб функціонал (1.2) мав мінімум на функції $y_0(x) \in C_0^1([x_1, x_2])$.

*Доведення*. За умов теореми для будь-якої функції $h(x) \in C_0^1([x_1, x_2])$ маємо:

$$\varphi_h''(0) = \int_{x_1}^{x_2} dx\left[S(x)h^2(x) + R(x)h'^2(x)\right] \ge \int_{x_1}^{x_2} dx\, R(x)h'^2(x) \ge \mu\int_{x_1}^{x_2} dx\, h'^2(x) \ge 0.$$

Знак рівності в останній нерівності треба писати лише тоді, коли неперервна функція $h'(x)$ є тотожним нулем на відрізку $[x_1, x_2]$,

а з огляду на умови $h(x_1)=h(x_2)=0$ — лише коли $h(x)\equiv 0$. Отже, $\varphi''_h(0)>0$ для ненульових $h(x)\in C_0^1([x_1,x_2])$.

**Завдання 1.5.2**. Доведіть, що розв'язки рівняння Ейлера — Лагранжа для функцій виду

$$F(x,y,y')=H(y)G(y') \tag{1.50}$$

надають мінімальні значення функціоналу (1.2) з-поміж усіх кривих зі спільним початком і спільним кінцем, якщо двічі диференційовні функції $H(y)$ і $G(y')$ задовольняють, наприклад, умови

$$H(y)>0,\ G''(y')>0,\ G(y')-y'G'(y')\geq 0,\ H''(y)H(y)-H'^2(y)\geq 0. \tag{1.51}$$

**Завдання 1.5.3**. Перевірте виконання умов теорем 1.5.1, 1.5.2 та умов (1.51) для функціоналів часу в завданнях 1.3.4, 1.4.4.

Якщо $\min\limits_{x\in[x_1,\,x_2]} S(x)<0,$ то при виконанні посиленої умови Лежандра квадратичний функціонал (1.44) має або лише додатні значення для всіх ненульових функцій $h(x)\in C_0^1([x_1,x_2])$, або набуває від'ємних значень для деяких ненульових функцій цього класу. У першому випадку функція $y_0(x)$ є кривою, що справді надає мінімум функціоналу (1.2), тоді як у другому вона такою не є.

Щоб з'ясувати, коли реалізується саме перший випадок, візьмемо довільну неперервну функцію $\rho(x)$, для якої

$$\min_{x\in[x_1,\,x_2]} \rho(x)>0, \tag{1.52}$$

та розглянемо на множині $C_0^1([x_1,x_2])$ функціонал більш складної структури

$$Q[h]=\frac{\int\limits_{x_1}^{x_2} dx\left[S(x)h^2(x)+R(x)h'^2(x)\right]}{\int\limits_{x_1}^{x_2} dx\,\rho(x)h^2(x)}. \tag{1.53}$$

Уведення функціонала (1.53) диктується тією обставиною, що квадратичний функціонал (1.44) в загальному випадку може набувати необмежених за модулем від'ємних значень, якщо $\min\limits_{x\in[x_1,\,x_2]} S(x)<0.$ Справді, нехай остання умова виконується, і при цьому $\varphi''_{\tilde h}(0)=a<0$ для деякої функції $\tilde h(x)\in C_0^1([x_1,x_2])$. Тоді на функції $N\tilde h(x)$ функціонал (1.44) набуває значення $aN^2<0$, при цьому число $N$ можна взяти як завгодно великим.

Що ж стосується функціонала (1.53), то він задовольняє умову

$$Q[h] > \Lambda_0 \equiv \min_{x\in[x_1,x_2]} \frac{S(x)}{\rho(x)} > -\infty. \tag{1.54}$$

Дійсно, якщо $R(x) > 0$, то при ненульових $h(x) \in C_0^1([x_1,x_2])$ внесок доданка $\int_{x_1}^{x_2} dx\, R(x) h'^2(x)$ у $Q[u]$ ненульовий і додатний, оскільки рівність $\int_{x_1}^{x_2} dx\, R(x) h'^2(x) = 0$ справджується лише при $h'(x) \equiv 0$, що для функцій класу $C_0^1([x_1,x_2])$ є рівносильним умові $h(x) \equiv 0$ (див. доведення теореми 1.5.2). Тому можемо записати:

$$Q[h] > \frac{\int_{x_1}^{x_2} dx\, S(x) h^2(x)}{\int_{x_1}^{x_2} dx\, \rho(x) h^2(x)} = \frac{\int_{x_1}^{x_2} dx \frac{S(x)}{\rho(x)} \rho(x) h^2(x)}{\int_{x_1}^{x_2} dx\, \rho(x) h^2(x)} \geq \Lambda_0.$$

Позначивши

$$\lambda_0 \equiv \inf_{h(x)\in C_0^1[x_1,\, x_2]} Q[h], \tag{1.55}$$

для кожної $h(x) \in C_0^1([x_1,x_2])$ маємо:

$$\int_{x_1}^{x_2} dx \left[ S(x) h^2(x) + R(x) h'^2(x) \right] \geq \lambda_0 \int_{x_1}^{x_2} dx\, \rho(x) h^2(x). \tag{1.56}$$

Звідси випливає, що твердження теорем 1.5.1, 1.5.2 можна посилити наступним чином.

**Теорема 1.5.3**. Серед гладких кривих зі спільним початком і спільним кінцем крива $y_0(x)$ надає мінімум функціоналу (1.2), якщо $\lambda_0 > 0$, і не надає, якщо $\lambda_0 < 0$. Випадок $\lambda_0 = 0$ вимагає додаткового дослідження.

Щоб знайти число $\lambda_0$, зазначимо, що згідно з нерівністю (1.56) точна нижня межа значень функціонала

$$\Phi[h] \equiv \int_{x_1}^{x_2} dx \left[ S(x) h^2(x) + R(x) h'^2(x) \right] - \lambda_0 \int_{x_1}^{x_2} dx\, \rho(x) h^2(x)$$

на множині функцій $C_0^1([x_1,x_2])$ дорівнює нулю. Нехай ця межа досягається на ненульовій функції $u(x) \in C_0^1([x_1,x_2])$, тобто $\Phi[u] = 0$. Тоді для будь-якої функції $h(x) \in C_0^1([x_1,x_2])$ і дійсних значень параметра $\varepsilon$ функція

$$f(\varepsilon) \equiv \Phi[u + \varepsilon h]$$

має абсолютний мінімум при $\varepsilon = 0$. Звідси знаходимо, що функція $u(x)$ задовольняє рівняння Ейлера — Лагранжа

$$-\frac{d}{dx}\left( R(x)\frac{du(x)}{dx} \right) + S(x)u(x) = \lambda_0 \rho(x)u(x). \qquad (1.57)$$

Умови закріплення мають вигляд

$$u(x_1) = 0, \quad u(x_2) = 0. \qquad (1.58)$$

Для подальшого аналізу замінимо $\lambda_0$ у формулі (1.57) комплексним параметром $z$. Дістаємо систему

$$\begin{gathered} -\frac{d}{dx}\left( R(x)\frac{du(x)}{dx} \right) + S(x)u(x) = z\rho(x)u(x), \\ u(x_1) = 0, \quad u(x_2) = 0, \end{gathered} \qquad (1.59)$$

яка називається *крайовою задачею Штурма — Ліувілля*. Диференціальне рівняння в ній називається *рівнянням Штурма — Ліувілля*, а додаткові умови в крайніх точках — *крайовими умовами*.

Система (1.59) лінійна та однорідна відносно функції $u(x)$, і тому має тривіальний розв'язок $u(x) \equiv 0$. При довільних значеннях параметра $z$ він є, взагалі кажучи, єдиним розв'язком, що задовольняє як рівняння Штурма — Ліувілля, так і крайові умови. З другого боку, можуть існувати виняткові значення параметра $z$, для яких рівняння Штурма — Ліувілля має нетривіальні розв'язки, що задовольняють крайові умови. Ці виняткові значення параметра $z$ називаються *власними значеннями* крайової задачі Штурма — Ліувілля, а відповідні нетривіальні розв'язки системи (1.59) при цих значеннях $z$ — *власними функціями*, що відповідають цим власним значенням.

**Завдання 1.5.4.** Знайдіть власні значення і відповідні власні функції крайової задачі (1.59) для випадку, коли всі коефіцієнтні функції є тотожними сталими: $R(x) \equiv \mu$, $S(x) \equiv q$, $\rho(x) \equiv \rho$.

*Розв'язання*. При заданих значеннях коефіцієнтів загальний розв'язок рівняння Штурма — Ліувілля має вигляд

$$u(x,z) = A\sin\sqrt{\frac{\rho z - q}{\mu}}(x - x_1) + B\cos\sqrt{\frac{\rho z - q}{\mu}}(x - x_1). \qquad (1.60)$$

Крайова умова зліва дає $B = 0$, а крайова умова справа веде до співвідношення

$$A\sin\sqrt{\frac{\rho z-q}{\mu}}(x_2-x_1)=0,$$

з якого випливає, що нетривіальні розв'язки ( $A\neq 0$ ) існують лише при таких значеннях параметра $z$:

$$z_n=\frac{1}{\rho}\left[\frac{\mu\pi^2n^2}{(x_2-x_1)^2}+q\right],\quad n=1,2,.... \tag{1.61}$$

Відповідні власні функції мають вигляд

$$u_n(x,z)=A\sin\frac{\pi n(x-x_1)}{x_2-x_1}. \tag{1.62}$$

Повертаючись до рівняння Штурма — Ліувілля в (1.59), позначимо через $\psi(x,z)$ той його єдиний розв'язок, який зліва задовольняє умови

$$\psi(x_1,z)=0,\quad \psi'(x_1,z)=1. \tag{1.63}$$

Застосовуючи стандартні методи теорії звичайних лінійних диференціальних рівнянь, можна показати (див. підрозділ 5.2), що при фіксованих значеннях змінної $x$ функція $\psi(x,z)$ є *цілою* відносно параметра $z$, тобто аналітичною на відкритій комплексній площині. Порівнюючи перші крайові умови в (1.59) і (1.63), бачимо, що кожна власна функція крайової задачі (1.59) пропорційна функції $\psi(x,z)$, якщо $z$ збігається з власним значенням. З другої крайової умови в (1.59) знаходимо, що власні значення задачі (1.59) є нулями функції $\psi(x_2,z)$, тобто коренями рівняння $\psi(x_2,z)=0.$ Оскільки однозначна аналітична функція в області аналітичності може мати лише ізольовані нулі, то в кожному крузі скінченного радіуса на комплексній площині може знаходитися лише скінченна кількість власних значень крайової задачі (1.59). Тому ці власні значення утворюють послідовність точок $\{z_n\}$ із єдиною точкою скупчення на нескінченності.

Нехай $z_n$ — власне значення крайової задачі (1.59), а $u_n(x)$ — власна функція, яка йому відповідає. Помноживши ліву і праву частини рівняння Штурма — Ліувілля для $u_n(x)$ на комплексно спряжену функцію $\overline{u_n(x)}$ та зінтегрувавши обидві частини здобутої рівності за $x$ у межах від $x_1$ до $x_2$, дістаємо:

$$-\int_{x_1}^{x_2}\overline{u_n(x)}\frac{d}{dx}\left(R(x)u_n'(x)\right)dx+\int_{x_1}^{x_2}S(x)\left|u_n(x)\right|^2dx=z_n\int_{x_1}^{x_2}\rho(x)\left|u_n(x)\right|^2dx. \tag{1.64}$$

Обчислюючи перший інтеграл в (1.64) частинами, з урахуванням крайових умов (1.59) далі знаходимо:

$$z_n = \frac{\int_{x_1}^{x_2} R(x)\left|u_n'(x)\right|^2 dx + \int_{x_1}^{x_2} S(x)\left|u_n(x)\right|^2 dx}{\int_{x_1}^{x_2} \rho(x)\left|u_n(x)\right|^2 dx}. \tag{1.65}$$

Вираз у правій частині рівності (1.65) — це дійсна величина. Звідси бачимо, що *власні значення крайової задачі Штурма — Ліувілля (1.59) є дійсними числами*.

Для дійсних чисел $z_n$ коефіцієнти в рівнянні (1.59) для функції $u_n(x)$ стають дійсними, тому *дійсними є функція* $\psi(x, z_n)$ *та (при дійсних коефіцієнтах* $C_n$*) власні функції* $u_n(x) = C_n \psi(x, z_n)$. Згадавши означення (1.53) функціонала $Q[h]$ і нерівність (1.54), дістаємо:

$$z_n = Q[u_n] > \Lambda_0 = \min_{x \in [x_1, x_2]} \frac{S(x)}{\rho(x)}. \tag{1.66}$$

Тим самим ми довели, що правильна

**Теорема 1.5.4**. Власні значення крайової задачі Штурма — Ліувілля (1.59) є дійсними числами з інтервалу $(\Lambda_0, +\infty)$ з єдиною точкою скупчення на $+\infty$.

З означення (1.55) та системи (1.57), (1.58) бачимо, що $\lambda_0$ є власним значенням, причому *найменшим*. Тому теорему 1.5.3 можна сформулювати так:

**Теорема 1.5.5**. Серед гладких кривих зі спільним початком і спільним кінцем крива $y_0(x)$ надає мінімум функціоналу (1.2), якщо найменше власне значення $\lambda_0$ крайової задачі Штурма — Ліувілля (1.59) додатне, і не надає, якщо воно від'ємне.

**Завдання 1.5.5.** Нехай $\min_{x \in [x_1, x_2]} R(x) = \mu$ і $\min_{x \in [x_1, x_2]} S(x) = q$. Доведіть, що $\lambda_0 > 0$, якщо

$$\frac{\mu \pi^2}{(x_2 - x_1)^2} + q > 0. \tag{1.67}$$

*Вказівка*. Скористайтеся нерівністю

$$\int_{x_1}^{x_2} dx\left[S(x)h^2(x) + R(x)h'^2(x)\right] \geq \int_{x_1}^{x_2} dx\left[qh^2(x) + \mu h'^2(x)\right]$$

та результатом завдання 1.5.4.

Завершимо цей підрозділ, навівши ще одну достатню умову мінімуму функціонала (1.2) при виконанні посиленої умови Лежандра. Для цього розглянемо введений вище розв'язок $\psi(x,z)$ рівняння Штурма — Ліувілля (1.59), підкорений умовам (1.63). З теорії рівняння Штурма — Ліувілля відомо, що кількість нулів функції $\psi(x,0)$ в інтервалі $(x_1,x_2)$ точно збігається з кількістю від'ємних власних значень крайової задачі (1.59). Отже, крива $y_0(x)$ надає функціоналу (1.2) мінімум, якщо функція $\psi(x,0)$ не має нулів на інтервалі $(x_1,x_2]$. Ця умова відома як *посилена умова Якобі.*

*Виконання посилених умов Лежандра та Якобі є достатньою умовою того, щоб екстремаль $y_0(x)$ надавала мінімум функціоналу найпростішого типу (1.2) на класі гладких кривих. Достатня умова максимуму формулюється аналогічно, лише тепер $R<0$.*

**Завдання 1.5.6.** Серед гладких кривих, які проходять через точки $A(0,0)$ і $B(1,1)$, знайдіть ту, яка надає екстремум функціоналу

$$J[y]=\int_0^1 dx\, y'^3.$$

*Розв'язання.* Загальний розв'язок рівняння Ейлера — Лагранжа для цього функціонала $y(x)=C_1x+C_2$. З урахуванням умов закріплення маємо $y_0(x)=x$.

Посилена умова Лежандра для функції $y_0(x)$ виконується скрізь на інтервалі $(0,1)$: $R=\left.\frac{\partial^2 F}{\partial y'^2}\right|_{y=y_0}=6y_0'=6>0$. Перевіряємо виконання посиленої умови Якобі. Потрібно знайти розв'язок рівняння (1.59) при $R=6$, $S=0$ і $z=0$. Воно набирає вигляду $\psi''(x,0)=0$, звідки $\psi(x,0)=D_1x+D_2$, $D_1$ і $D_2$ — сталі інтегрування. Умови зліва $\psi(0,0)=0$, $\psi'(0,0)=1$ остаточно дають: $\psi(x,0)=x$. Ця функція не має жодних нулів на інтервалі $(0,1]$.

Отже, екстремаль $y_0(x)=x$ надає мінімум заданому функціоналу. Цей результат можна було б отримати відразу, скориставшись теоремою 1.5.2.

Вивчення достатніх умов екстремуму для більш складних типів функціоналів є значно складнішим і тому в цьому тексті не проводиться. Як уже зазначалося, у фізичних та прикладних задачах існування таких екстремумів випливає із самої постановки задачі.

## *КОНТРОЛЬНІ ПИТАННЯ ДО РОЗДІЛУ 1*

1. *Як визначаються функціонали на множині гладких функцій на відрізку* $[x_1, x_2]$*? Наведіть приклади таких функціоналів. Чи є відмінність між поняттями функції і функціонала?*
2. *У чому полягає основна задача варіаційного числення? Сформулюйте леми, які постійно використовуються при відшуканні необхідних умов існування розв'язку цієї задачі.*
3. *Яке рівняння повинна задовольняти крива* $y_0(x)$, *яка надає екстремум функціоналу* $J[y] = \int_{x_1}^{x_2} F(x, y, y')dx$ *у класі функцій* $C^1([x_1, x_2])$, *де* $F$ *— двічі диференційовна за всіма аргументами? Яка додаткова умова гарантує диференційовність двічі екстремальної кривої* $y_0(x)$*?*
4. *У яких випадках порядок рівняння Ейлера — Лагранжа можна понизити? Які криві можуть бути екстремалями функціонала найпростішого типу, якщо його інтегрант* $F(x, y, y')$ *залежить лише від* $y'$*? Знайдіть перший інтеграл рівняння Ейлера — Лагранжа для випадку, коли інтегрант функціонала найпростішого типу не залежить від* $x$.
5. *Яка умова є необхідною для того, щоб екстремальна крива* $y_0(x)$ *надавала слабкий мінімум функціоналу найпростішого типу? Яку вимогу треба додати до неї, щоб уже отримати критерій слабкого мінімуму відповідного функціонала?*

# Розділ 2

# СКЛАДНІШІ ТИПИ ВАРІАЦІЙНИХ ЗАДАЧ

## 2.1. ЗАДАЧІ З РУХОМИМИ КІНЦЯМИ

До цього часу при вивченні функціонала (1.2) розглядалися лише ті гладкі криві, кінці яких жорстко закріплено в нерухомих точках з абсцисами $x_1$ та $x_2$. Ординати цих точок задавалися співвідношеннями (1.3).

**Означення 2.1.1.** Додаткові умови, яким підкоряються значення допустимих функцій та їх похідних на межах досліджуваного інтервалу, називаються крайовими.

Крайові умови (1.3) задавалися ззовні і ніяким чином не були пов'язані з явним видом функціонала (1.2). Інша ситуація зустрічається у випадках, коли екстремум функціонала (1.2) вивчається на більш широкому класі гладких функцій, наприклад, коли кінці допустимих кривих можуть вільно ковзати вздовж деяких кривих. Тепер явний вигляд крайових умов залежить від конкретного виду інтегранта $F(x,y,y')$, а необхідна умова існування екстремуму вимагає, щоб екстремальна крива задовольняла як відповідне рівняння Ейлера — Лагранжа, так і ці нові умови.

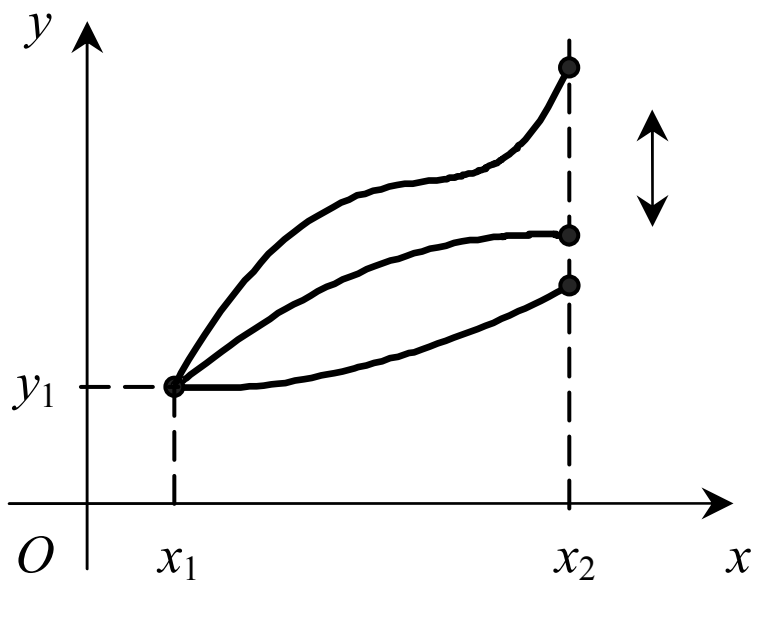

Рис. 2.1. Допустимі гладкі криві, які закінчуються в довільних точках прямої $x = x_2$ («праві кінці кривих вільно ковзають по вертикалі»)

Знайдемо необхідні умови існування екстремуму функціонала (1.2) у класі $C^1([x_1,x_2])$ кривих, ліві кінці яких жорстко закріплено в точці $x = x_1$, а праві можуть вільно ковзати вздовж вертикальної прямої $x = x_2$ (рис. 2.1).

При фіксованому значенні зліва

$$y(x_1) = y_1 \tag{2.1}$$

необхідна умова екстремуму набирає вигляду (див. формулу (1.6))

$$\left.\frac{\partial F}{\partial y'}\cdot h(x)\right|_{x=x_2}+\int_{x_1}^{x_2}dx\left\{\frac{\partial F}{\partial y}-\frac{d}{dx}\frac{\partial F}{\partial y'}\right\}h(x)=0. \qquad (2.2)$$

Те ж саме можемо записати в термінах варіації $\delta y(x)$ функції $y(x)$:

$$\left.\frac{\partial F}{\partial y'}\cdot \delta y(x)\right|_{x=x_2}+\int_{x_1}^{x_2}dx\left\{\frac{\partial F}{\partial y}-\frac{d}{dx}\frac{\partial F}{\partial y'}\right\}\delta y(x)=0. \qquad (2.3)$$

На інтервалі $(x_1,x_2)$, а також у точці $x=x_2$ значення функції $h(x)$ (варіації $\delta y(x)$) довільні. Користуючись цим, розглянемо спершу окремий випадок, коли $h(x_2)=0$ $(\delta y(x_2)=0)$. Тоді перший доданок у (2.2) (і (2.3)) дорівнює нулю. У силу ж довільності функції $h(x)$ (варіації $\delta y(x)$) на інтервалі $(x_1,x_2)$ другий доданок дорівнює нулю тоді, коли дорівнює нулю вираз у фігурних дужках, тобто коли справджується рівняння Ейлера — Лагранжа (1.4).

Нехай тепер значення $h(x_2)$ (чи $\delta y(x_2)$) довільні. Другий доданок у (2.2) (і (2.3)) дорівнює нулю, бо шукана екстремаль, належачи до більш широкого класу допустимих кривих, має задовольняти необхідну умову екстремуму для більш вузького класу кривих із закріпленими кінцями, тобто має справджуватися рівняння (1.4). Тоді перший доданок у (2.2) (і (2.3)) дорівнює нулю лише за умови

$$\left.\frac{\partial F}{\partial y'}\right|_{x=x_2}=0 \qquad (2.4)$$

– дістаємо співвідношення, якому повинні підкорятися значення екстремалі та її похідної в точці $x=x_2$.

Таким чином, необхідна умова існування екстремуму функціонала (1.2) у класі гладких кривих, ліві кінці яких жорстко закріплено, а праві можуть вільно ковзати вздовж вертикальної прямої $x=x_2$, складається з: а) рівняння Ейлера — Лагранжа (1.4); б) крайової умови (2.1); в) крайової умови (2.4). Додатково треба перевірити належність знайдених розв'язків екстремальної задачі саме до класу гладких кривих.

**Зауваження 2.1.1.** Якщо й ліві кінці допустимих кривих можуть вільно ковзати по вертикалі $x=x_1$, то умова (2.1) заміняється умовою (2.4), записаною для $x=x_1$. У загальному ж випадку крайові умови є різними комбінаціями умов (1.3), (2.1), (2.4) та їм подібних. *Конкретний вигляд крайової умови в заданій точці визначається характером обмеження на поведінку допустимої кривої у цій точці.*

**Завдання 2.1.1.** По якій кривій, розташованій у вертикальній площині $XOY$ (вісь $OX$ напрямлена горизонтально, вісь $OY$ — вертикально вниз), має рухатися матеріальна точка під дією сили тяжіння, щоб, відправившись із початку координат з нульовою початковою швидкістю, перетнути вертикальну пряму $x = l$ за найкоротший час? (Модифікована задача про брахістохрону І. Бернуллі.)

*Вказівка*: $y'(l) = 0$; див. також завдання 1.3.4.

*Відповідь*: $x(\varphi) = \frac{l}{\pi}(\varphi - \sin\varphi)$, $y(\varphi) = \frac{l}{\pi}(1 - \cos\varphi)$, $\varphi \in [0, \pi]$.

**Завдання 2.1.2.** Горизонтальний тонкий стержень довжиною $L$ жорстко закріплено в точці $A$. До його другого кінця $B$ підвішено кульку масою $m$. Нехтуючи масою стержня, знайдіть його рівноважний профіль (рис. 2.2).

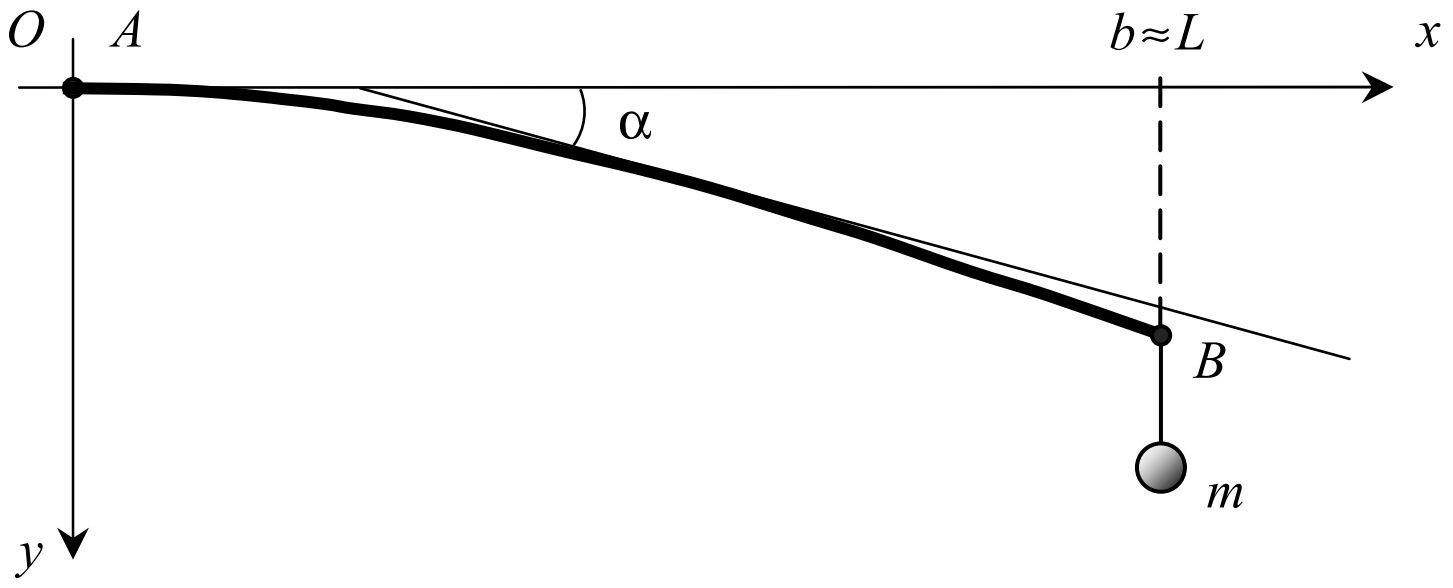

Рис. 2.2. Прогин стержня під вагою кульки. Для наочності величину прогину перебільшено

*Розв'язання*. Треба знайти функцію $y = y(x)$, яка реалізує мінімум функціонала потенціальної енергії системи $П = П_1 + П_2$. Тут: $П_1$ — функціонал потенціальної енергії системи в полі сили тяжіння; він дорівнює потенціальній енергії кульки:

$$П_1 = -mgY, \qquad (2.5)$$

де $Y$ — ордината точки $B$ (енергію вимірюємо від рівня $y = 0$, несуттєву адитивну сталу $-mga$, $a$ — довжина нитки, опускаємо); $П_2$ — функціонал потенціальної енергії пружної деформації стержня; він обчислюється за формулою

$$П_2 = \frac{1}{2} EJ \int_0^L dl \left( \frac{d\alpha}{dl} \right)^2, \qquad (2.6)$$

де $\alpha$ — кут між дотичною до профілю $y(x)$ і віссю $OX$, $d\alpha/dl = 1/R$ — кривина стержня ($R$ — радіус кривини), $E$ — модуль розтягу Юнга, $J$ — головний момент інерції поперечного перерізу стержня.

Зауважимо, що головні моменти поперечного перерізу $S$ відносно головних осей $\xi$ і $\eta$ визначаються подібно до звичайних моментів інерції, лише замість елемента маси у відповідних формулах стоїть елемент поверхні $dS$:

$$J_\xi = \int_S dS\,\eta^2, \quad J_\eta = \int_S dS\,\xi^2.$$

Для кругового перерізу радіусом $R$, у якого центр інерції знаходиться в центрі круга, а напрями головних осей довільні, $J = J_\xi = J_\eta = \pi R^4/4$.

Виразивши ординату $Y$ через кут $\alpha$,

$$Y = \int_{AB} dy = \int_0^L dl \sin\alpha,$$

для $\Pi$ дістаємо:

$$\Pi = \int_0^L dl\left(\frac{1}{2}EJ\alpha'^2 - mg\sin\alpha\right).$$

Мінімум функціонала $\Pi$ реалізується для функції $\alpha = \alpha(l)$, яка задовольняє рівняння

$$EJ\alpha'' + mg\cos\alpha = 0$$

та крайові умови $\alpha(0) = 0$, $\alpha'(L) = 0$.

Уважаючи, що прогин стержня малий ($\alpha << 1$), та нехтуючи величинами другого порядку малості, можемо записати: $\cos\alpha \approx 1$, $y'(x) = \mathrm{tg}\alpha \approx \alpha$. Бачимо також, що довжина стержня в цьому наближенні залишається незмінною:

$$L = \int_0^L dl = \int_0^b dx\sqrt{1 + y'^2} \approx b,$$

де $b$ — абсциса точки $B$. Тому $\alpha' \approx y''$, $\alpha'' \approx y'''$, а рівняння для профілю стержня та крайові умови набирають вигляду

$$EJy''' + mg = 0, \quad y(0) = y'(0) = y''(L) = 0, \tag{2.7}$$

звідки

$$y(x) = -\frac{mg}{6EJ}x^3 + C_2x^2 + C_1x + C_0.$$

Сталі інтегрування дорівнюють: $C_0 = 0$, $C_1 = 0$, $C_2 = mgL/(2EJ)$. Отже, остаточна відповідь така:

$$y(x)=\frac{mg}{6EJ}x^2(3L-x).$$

Задачі про форму деформованих стержнів, балок та поверхонь за тих чи інших крайових умов складають предмет вивчення теорії пружності та її спрощеного варіанта — опору матеріалів. Як правило, для їх розв'язання доводиться аналізувати функціонали, що залежать від функції та її вищих похідних. Такі функціонали вивчаються в підрозділі 2.2.

Подальшим узагальненням функціоналів найпростішого типу (1.2) виступають функціонали виду

$$J[y]=\int_{\Gamma} dxF(x,y,y'), \tag{2.8}$$

що задаються на множині гладких кривих $\Gamma$, кінці яких можуть ковзати по відомих гладких лініях $y=\varphi(x)$ та $y=\psi(x)$ (не обов'язково вертикальних). Крім нових крайових умов, новим елементом відповідної екстремальної задачі є те, що абсциси початку та кінця екстремальної кривої теж треба відшукувати.

Знайдемо екстремум функціонала типу (2.8) у класі гладких кривих, ліві кінці яких жорстко закріплено в точці $x=0$, а праві можуть ковзати вздовж гладкої кривої $y=\psi(x)$. Для цього функціонал подамо у вигляді інтеграла зі змінною верхньою межею:

$$J[y]=\int_0^{\tilde{x}} dxF(x,y,y'). \tag{2.9}$$

Нехай його екстремум досягається на кривій $y=y(x)$, правий кінець якої має абсцису $x_2$. Оскільки останній лежить на кривій $y=\psi(x)$, можемо записати (див. рис. 2.3):

$$y(x_2)=\psi(x_2). \tag{2.10}$$

Зліва ж маємо:

$$y(0)=0. \tag{2.11}$$

Розглянемо тепер множину допустимих кривих $y(x)+\alpha h(x)$, де $\alpha$ — малий параметр, $h(x)$ — довільна гладка функція. Праві кінці допустимих кривих лежать на кривій $y=\psi(x)$, тому їх абсциси $\tilde{x}$ задовольняють співвідношення

$$y(\tilde{x})+\alpha h(\tilde{x})=\psi(\tilde{x}). \tag{2.12}$$

Зауважимо, що абсциси $\tilde{x}$ залежать від параметра $\alpha$, причому при $\alpha=0$ (для екстремальної кривої) $\tilde{x}(0)=x_2$. Зліва й далі маємо $y(0)+\alpha h(0)=0$, тобто $h(0)=0$.

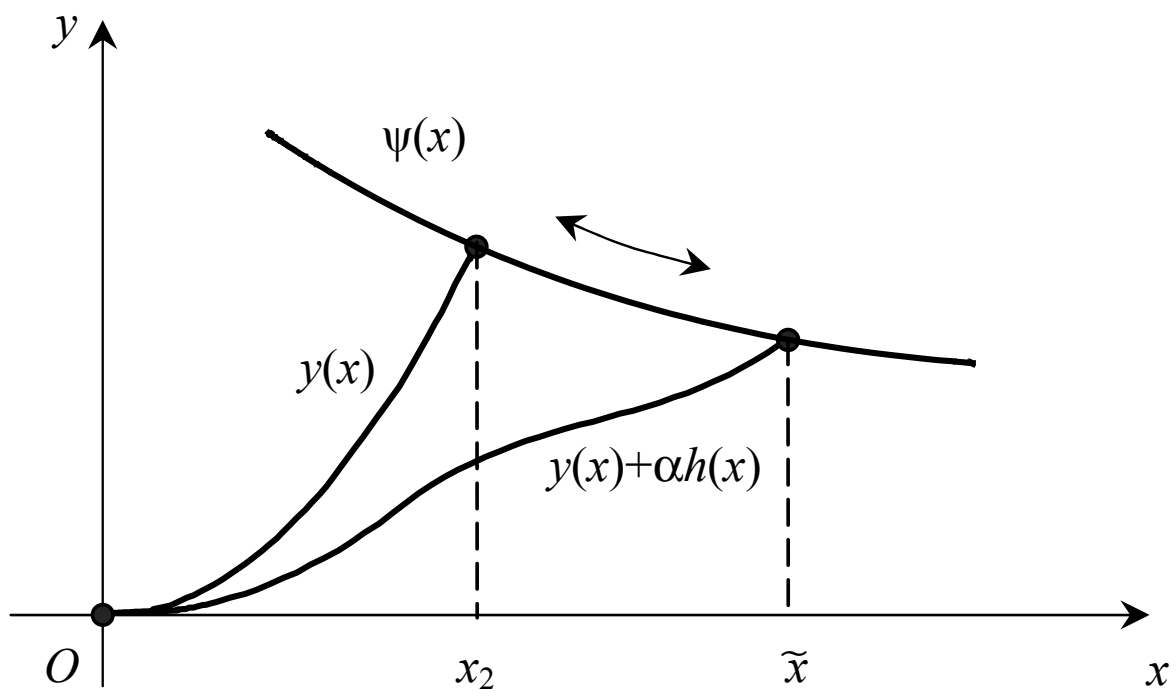

Рис. 2.3. Вільне ковзання правих кінців допустимих кривих по заданій кривій $\psi(x)$

Для екстремалі $y(x)$ функція $\varphi(\alpha) \equiv J[y + \alpha h]$ набуває екстремуму. Можемо записати:

$$\varphi'(0) = F\big|_{x=x_2} \cdot \tilde{x}'(0) + \int_0^{\tilde{x}(0)} dx \left\{ \frac{\partial F}{\partial y} \cdot h(x) + \frac{\partial F}{\partial y'} \cdot h'(x) \right\} =$$

$$= F\big|_{x=x_2} \cdot \tilde{x}'(0) + \frac{\partial F}{\partial y'}\bigg|_{x=x_2} \cdot h(x_2) + \int_0^{x_2} dx \left\{ \frac{\partial F}{\partial y} - \frac{d}{dx}\frac{\partial F}{\partial y'} \right\} h(x) = 0. \quad (2.13)$$

Число $\tilde{x}'(0)$ знаходимо, диференціюючи (2.12) за $\alpha$:

$$y'(\tilde{x})\tilde{x}'(\alpha) + h(\tilde{x}) + \alpha h'(\tilde{x})\tilde{x}'(\alpha) = \psi'(\tilde{x})\tilde{x}'(\alpha),$$

$$\tilde{x}'(0) = \frac{h(x_2)}{\psi'(x_2) - y'(x_2)}.$$

Дістаємо:

$$\left[ \frac{F}{\psi' - y'} + \frac{\partial F}{\partial y'} \right]_{x=x_2} \cdot h(x_2) + \int_0^{x_2} dx \left\{ \frac{\partial F}{\partial y} - \frac{d}{dx}\frac{\partial F}{\partial y'} \right\} h(x) = 0.$$

Звідси, з огляду на довільність функції $h(x)$ на проміжку $(0, x_2]$, робимо висновок, що екстремаль $y(x)$ має задовольняти як рівняння Ейлера — Лагранжа (1.4), так і таку крайову умову справа:

$$\left[ F + (\psi' - y')\frac{\partial F}{\partial y'} \right]_{x=x_2} = 0. \quad (2.14)$$

Крім того, повинні справджуватися рівності (2.11) (крайова умова для жорстко закріпленого лівого кінця) та (2.10) (правий кінець екстремалі лежить на заданій кривій).

Співвідношення (2.14) називається *умовою трансверсальності*. У граничному випадку вертикальної прямої ($\psi' \to \infty$) воно переходить у крайову умову (2.4). Граничний випадок горизонтальної прямої дістаємо, поклавши $\psi' = 0$.

Користуючись попередніми результатами, можемо тепер сформулювати загальну схему розв'язування екстремальної задачі для функціонала (2.8). Спершу розв'язуємо рівняння Ейлера — Лагранжа (1.4) та знаходимо двопараметричну сім'ю кривих $y = y(x, C_1, C_2)$. Потім серед них відшукуємо ті, які на лівих (з абсцисою $x_1$) і правих (з абсцисою $x_2$) кінцях задовольняють умови трансверсальності

$$\left[F + (\varphi' - y')\frac{\partial F}{\partial y'}\right]_{x=x_1} = 0, \quad \left[F + (\psi' - y')\frac{\partial F}{\partial y'}\right]_{x=x_2} = 0 \qquad (2.15)$$

та рівняння

$$y(x_1, C_1, C_2) = \varphi(x_1), \quad y(x_2, C_1, C_2) = \psi(x_2). \qquad (2.16)$$

За допомогою цих чотирьох рівнянь знаходимо $x_1$, $x_2$, $C_1$ та $C_2$.

Підкреслимо, що умова трансверсальності повинна виконуватися лише для того кінця екстремальної кривої, який може вільно рухатися по певній кривій (чи поверхні).

**Завдання 2.1.3.** Знайдіть найкоротшу відстань від точки $A(0,0)$ до кривої $y = \psi(x)$.

*Розв'язання.* Треба знайти довжину $L$ екстремалі $y = y(x)$ функціонала

$$J[y] = \int_0^{\tilde{x}} dx\sqrt{1 + y'^2},$$

яка з'єднує точку $A$ з деякою точкою $B$ (з поки що невідомою абсцисою $x_2$) на лінії $y = \psi(x)$: $y(x_2) = \psi(x_2)$. Крайова умова в точці $A$ має вигляд $y(0) = 0$. Умова трансверсальності в точці $B$ записується як $y'(x_2)\psi'(x_2) = -1$, тобто вироджується в умову ортогональності екстремалі та заданої кривої.

Загальний розв'язок рівняння Ейлера — Лагранжа в нашому випадку відомий: $y(x) = C_1 x + C_2$. З урахуванням крайових умов для екстремалі дістаємо: $y(x) = -\dfrac{1}{\psi'(x_2)}x$. Точку $x_2$ знаходимо як корінь рівняння $x_2 = -\psi'(x_2)\psi(x_2)$. Шукана відстань $L = \sqrt{x_2^{\,2} + \psi^2(x_2)}$.

**Завдання 2.1.4.** Нехай інтегрант у функціоналі (2.8) $F(x, y, y') = f(x, y)\sqrt{1 + y'^2}$. Покажіть, що якщо значення функції $f$ на ліво-му кінці не дорівнює нулю, то умова трансверсальності на ньому вироджується в умову ортогональності екстремалі $y = y(x)$ та лінії $y = \varphi(x)$: $y'(x_1)\varphi'(x_1) = -1$.

Нехай тепер $y(x)$ — екстремаль функціонала типу (2.9) у класі гладких кривих, ліві кінці яких жорстко закріплено в початку координат, а праві можуть вільно ковзати у площині $XOY$. З'ясуємо, яку крайову умову задовольняє $y(x)$ на рухомому кінці.

Візьмемо до уваги, що співвідношення (2.14) повинно тепер справджуватися для довільної функції $\psi(x)$. Таке можливо, якщо одночасно виконуються умови

$$F\Big|_{x=x_2} = 0, \quad \frac{\partial F}{\partial y'}\Big|_{x=x_2} = 0. \qquad (2.17)$$

Система цих рівнянь дозволяє знайти координати $(x_2, y_2)$ правого кінця екстремалі ($y_2 = y(x_2)$). Доповнивши (2.17) рівнянням Ейлера — Лагранжа (1.4) та крайовою умовою (2.11), дістаємо необхідні умови для $y(x)$.

Поява двох умов (2.17) відображає той факт, що вільне ковзання правих кінців допустимих кривих у площині $XOY$ є суперпозицією двох незалежних рухів — вільних ковзань уздовж горизонтального та вертикального напрямів. Першому відповідає незалежна варіація абсциси $\delta x_2$, а другому — незалежна варіація ординати $\delta y_2$ правих кінців. Аналогічні умови повинні задовольнятися й на лівому кінці екстремальної кривої, якщо йому також дозволено вільно ковзати у площині $XOY$.

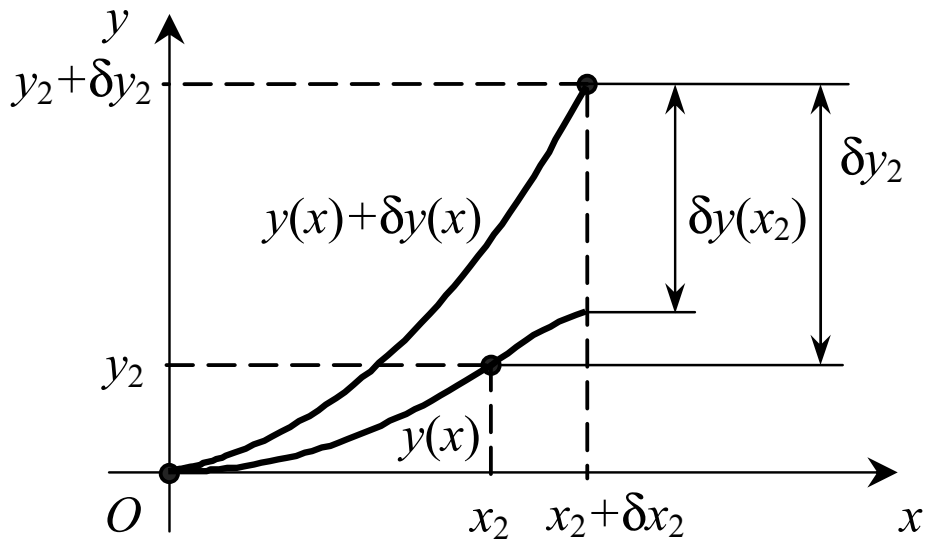

Рис. 2.4. Варіація форми кривої та координат її кінця при вільному русі останнього у площині

Цей самий результат можна отримати й так. Нехай $(x_2, y_2)$ — координати правого кінця екстремалі $y(x)$, а $(x_2 + \delta x_2, y_2 + \delta y_2)$ — координати правих кінців допустимих кривих $y(x) + \delta y(x)$ (рис. 2.4). Подамо першу варіацію функціонала (2.9) у вигляді (див. формулу (2.13))

$$\delta J[y] = F\Big|_{x=x_2} \cdot \delta x_2 + \frac{\partial F}{\partial y'}\Big|_{x=x_2} \cdot \delta y(x_2) + \int_0^{x_2} dx \left\{ \frac{\partial F}{\partial y} - \frac{d}{dx}\frac{\partial F}{\partial y'} \right\} \delta y(x). \qquad (2.18)$$

У другому доданку цієї формули треба, однак, перейти від $\delta y(x_2)$ (варіації форми допустимих функцій у точці $x = x_2$) до незалежних варіацій $\delta x_2$ і $\delta y_2$ координат правих кінців допустимих функцій. У лінійному наближенні маємо:

$$\delta y_2 = y(x_2 + \delta x_2) + \delta y(x_2 + \delta x_2) - y(x_2) \approx y'(x_2)\delta x_2 + \delta y(x_2).$$

З урахуванням цієї формули необхідну умову екстремуму функціонала (2.9) можемо записати у вигляді

$$\delta J[y] = \left( F - y' \frac{\partial F}{\partial y'} \right)\bigg|_{x=x_2} \cdot \delta x_2 + \frac{\partial F}{\partial y'}\bigg|_{x=x_2} \cdot \delta y_2 +$$

$$+ \int_0^{x_2} dx \left\{ \frac{\partial F}{\partial y} - \frac{d}{dx}\frac{\partial F}{\partial y'} \right\} \delta y(x) = 0. \tag{2.19}$$

Розглядаючи по черзі кожну з варіацій $\delta y(x)$, $x \in (0, x_2)$, $\delta x_2$ і $\delta y_2$, дістаємо рівняння (1.4) та крайові умови (2.17).

Окрім уже розглянутих типів обмежень на поведінку кінців допустимих кривих — жорсткого закріплення та вільного ковзання вздовж певних кривих — можна уявити собі й інші типи в'язей. Прикладом може бути так зване пружне закріплення кінців за допомогою пружинок малої (у граничному випадку — нульової) довжини. Вплив таких пружинок на рух кінців допустимих кривих можна врахувати за допомогою деякої функції $\Phi\big(y(x_1), y(x_2)\big)$, яка залежить лише від значень $y(x_1)$ та $y(x_2)$ шуканої функції на кінцях відрізка $[x_1, x_2]$. Відповідні функціонали мають вигляд

$$J[y] = \int_{x_1}^{x_2} dx F(x, y, y') + \Phi\big(y(x_1), y(x_2)\big) \tag{2.20}$$

і називаються *функціоналами Больца*, а задача про відшукання їх екстремалей — *задачею Больца*.

**Завдання 2.1.5.** Покажіть, що екстремаль функціонала (2.20) з диференційовною функцією $\Phi\big(y(x_1), y(x_2)\big)$ повинна задовольняти рівняння Ейлера — Лагранжа (1.4) та крайові умови

$$\frac{\partial F}{\partial y'}\bigg|_{x=x_1} - \frac{\partial \Phi}{\partial y(x_1)} = 0, \quad \frac{\partial F}{\partial y'}\bigg|_{x=x_2} + \frac{\partial \Phi}{\partial y(x_2)} = 0. \tag{2.21}$$

*Вказівка.* Варіація функції $\Phi\big(y(x_1), y(x_2)\big)$ для кривих $y(x) + \delta y(x)$ з $\varepsilon$-околу першого порядку кривої $y(x)$ дорівнює

$$\delta\Phi\big(y(x_1),y(x_2)\big)=\Phi\big(y(x_1)+\delta y(x_1),y(x_2)+\delta y(x_2)\big)-\Phi\big(y(x_1),y(x_2)\big)\approx$$
$$\approx\frac{\partial\Phi\big(y(x_1),y(x_2)\big)}{\partial y(x_1)}\delta y(x_1)+\frac{\partial\Phi\big(y(x_1),y(x_2)\big)}{\partial y(x_2)}\delta y(x_2).$$

Перша варіація функціонала (2.20) набирає вигляду

$$\delta J[y]=\left[-\frac{\partial F}{\partial y'}\bigg|_{x=x_1}+\frac{\partial\Phi}{\partial y(x_1)}\right]\delta y(x_1)+\left[\frac{\partial F}{\partial y'}\bigg|_{x=x_2}+\frac{\partial\Phi}{\partial y(x_2)}\right]\delta y(x_2)+$$
$$+\int_{x_1}^{x_2}dx\left\{\frac{\partial F}{\partial y}-\frac{d}{dx}\frac{\partial F}{\partial y'}\right\}\delta y(x).$$

Уважаючи значення $\delta y(x_1)$ та $\delta y(x_2)$ довільними (чому?), знайдіть умови, за яких $\delta J[y]=0$.

**Завдання 2.1.6.** Покажіть, що гладка екстремаль функціонала

$$\Pi[y]=\frac{1}{2}\int_0^l\big[p(x)y'^2(x)+q(x)y^2(x)\big]dx+\frac{k_1}{2}y^2(0)+\frac{k_2}{2}y^2(l)$$

задовольняє рівняння Штурма — Ліувілля

$$-\frac{d}{dx}\big(p(x)y'(x)\big)+q(x)y(x)=0,\ \ 0<x<l,$$

та однорідні крайові умови

$$y'(0)-h_1y(0)=0,\ \ y'(l)+h_2y(l)=0,$$

де $h_1\equiv k_1/p(0),\ \ h_2\equiv k_2/p(l)$.

## 2.2. ФУНКЦІОНАЛИ ВІД ФУНКЦІЇ ТА ЇЇ ВИЩИХ ПОХІДНИХ

Тепер уважатимемо, що при $x\in[x_1,x_2]$ функція $F$ залежить ще й від другої похідної $y''$ та є неперервною разом зі всіма своїми частинними похідними до третього порядку включно при всіх можливих значеннях своїх аргументів.

**Теорема 2.2.1.** Нехай крива $y_0(x)$ є екстремаллю функціонала

$$J[y]=\int_{x_1}^{x_2}dxF(x,y,y',y'') \qquad (2.22)$$

у класі $C^2([x_1,x_2])$ допустимих кривих, які разом зі своїми першими похідними набувають фіксованих значень на межах відрізка $[x_1,x_2]$:

$$y(x_1)=y_{10},\ \ y'(x_1)=y_{11},\ \ y(x_2)=y_{20},\ \ y'(x_2)=y_{21}. \qquad (2.23)$$

Припустимо, що $y_0(x)\in C^4([x_1,x_2])$. Тоді $y_0(x)$ задовольняє *рівняння Ейлера — Пуассона*

$$\frac{\partial F}{\partial y}-\frac{d}{dx}\frac{\partial F}{\partial y'}+\frac{d^2}{dx^2}\frac{\partial F}{\partial y''}=0. \qquad (2.24)$$

*Доведення.* Розглянемо множину кривих $y(x)=y_0(x)+\delta y(x)$ з ε-околу другого порядку кривої $y_0(x)$. Інтегруючи частинами, першу варіацію функціонала (2.22) подамо у вигляді

$$\delta J[y]=\left(\frac{\partial F}{\partial y'}-\frac{d}{dx}\frac{\partial F}{\partial y''}\right)\delta y(x)\Bigg|_{x_1}^{x_2}+\frac{\partial F}{\partial y''}\delta y'(x)\Bigg|_{x_1}^{x_2}+$$

$$+\int_{x_1}^{x_2}\left\{\frac{\partial F}{\partial y}-\frac{d}{dx}\frac{\partial F}{\partial y'}+\frac{d^2}{dx^2}\frac{\partial F}{\partial y''}\right\}\delta y(x)\,dx. \qquad (2.25)$$

Унаслідок крайових умов $\delta y(x_1)=\delta y'(x_1)=\delta y(x_2)=\delta y'(x_2)=0$ подвійні підстановки у формулі (2.25) дорівнюють нулю, тому перша варіація збігається з інтегралом у правій частині цієї формули. Останній дорівнює нулю для довільної варіації $\delta y$ (яка на кінцях відрізка $[x_1,x_2]$ дорівнює нулю разом із першою похідною), якщо справджується рівність (2.24).

**Зауваження 2.2.1.** Інтегруючи частинами у формулі (2.25), ми припустили, що шукана функція має неперервні похідні третього та четвертого порядків. Можна довести, що для екстремальної кривої це припущення справджується скрізь, де $\partial^2 F/\partial y''^2\neq 0$.

**Зауваження 2.2.2.** У загальному випадку рівняння Ейлера — Пуассона (2.24) є звичайним диференціальним рівнянням четвертого порядку, тому його загальний розв'язок містить, узагалі кажучи, чотири сталі інтегрування. Останні визначаються за допомогою крайових умов, що відповідають досліджуваному класу кривих.

**Завдання 2.2.1.** Знайдіть необхідні умови існування екстремуму функціонала (2.22) у класі $C^2([x_1,x_2])$ кривих, праві кінці яких жорстко закріплено під фіксованим кутом до осі $OX$, а ліві:

а) можуть вільно ковзати по вертикальній прямій $x=x_1$, утворюючи при цьому з нею фіксований кут (маючи в точці $x_1$ заданий кут нахилу дотичної);

б) закріплено в точці $x_1$ шарнірно (тобто в точці $x_1$ вони можуть утворювати довільні кути з віссю $OX$);

в) можуть вільно ковзати по вертикальній прямій $x = x_1$, утворюючи з нею довільні кути (маючи в точці $x_1$ дотичні довільних напрямів).

*Вказівка.* Узявши до уваги характер поведінки допустимих кривих у крайових точках, проаналізуйте умови, за яких перша варіація (2.25) дорівнює нулю. Наприклад, у випадку б) рівність $\delta J = 0$ має справджуватися при $\delta y(x_1) = \delta y(x_2) = \delta y'(x_2) = 0$ і довільних значеннях $\delta y'(x_1)$, оскільки тангенс кута нахилу дотичних зліва є довільним.

*Відповіді.* Екстремальні криві задовольняють рівняння Ейлера — Пуассона (2.24) та такі крайові умови:

а) $$\left.\left(\frac{\partial F}{\partial y'} - \frac{d}{dx}\frac{\partial F}{\partial y''}\right)\right|_{x=x_1} = 0,\ \ y'(x_1) = y_{11},\ \ y(x_2) = y_{20},\ \ y'(x_2) = y_{21}; \quad (2.26)$$

б) $$\left.\frac{\partial F}{\partial y''}\right|_{x=x_1} = 0,\ \ y(x_1) = y_{10},\ \ y(x_2) = y_{20},\ \ y'(x_2) = y_{21}; \quad (2.27)$$

в) $$\left.\left(\frac{\partial F}{\partial y'} - \frac{d}{dx}\frac{\partial F}{\partial y''}\right)\right|_{x=x_1} = 0,\ \ \left.\frac{\partial F}{\partial y''}\right|_{x=x_1} = 0,\ \ y(x_2) = y_{20},\ \ y'(x_2) = y_{21}. \quad (2.28)$$

Невиконання хоча б однієї з цих умов означає, що функціонал не має екстремуму серед кривих класу $C^2([x_1, x_2])$.

**Завдання 2.2.2.** Тонкий стержень довжиною $L$ замуровано в точках $A$ та $B$, що лежать на одній горизонталі (рис. 2.5). Знайдіть форму профілю стержня внаслідок провисання під власною вагою. Погонна густина стержня $\rho$.

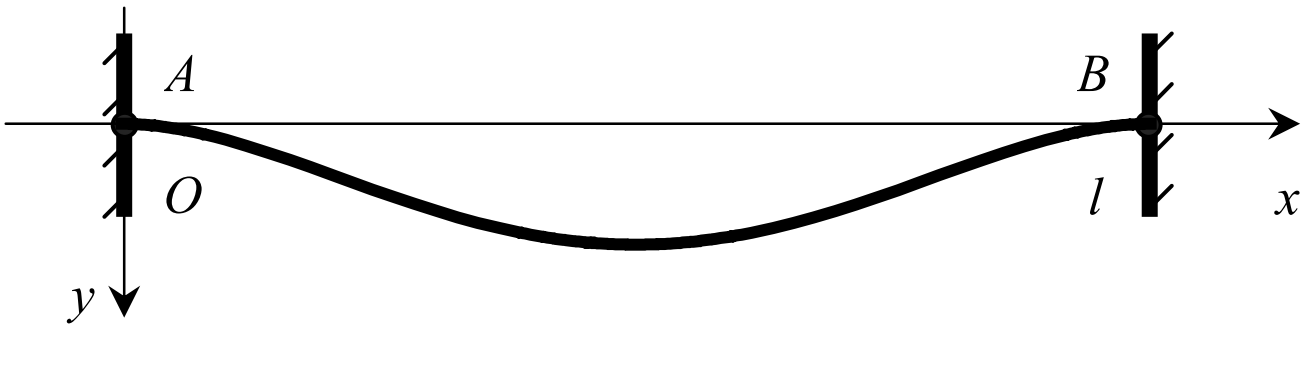

Рис. 2.5. Прогин стержня під власною вагою. Для наочності величину прогину значно перебільшено

*Вказівки.* Ураховуючи співвідношення $\operatorname{tg}\alpha = y'(x)$ і $dl = dx\sqrt{1 + y'^2(x)}$, для кривини стержня можемо записати: $d\alpha/dl =$

$= y''\big/\left(1+y'^2\right)^{3/2}$. Функціонал повної потенціальної енергії деформованого стержня в полі тяжіння

$$\Pi[y] = \int_0^b dx \left\{ \frac{1}{2} EJ \frac{y''^2}{\left(1+y'^2\right)^{5/2}} - \rho g y \sqrt{1+y'^2} \right\},$$

де $b$ — абсциса точки $B$ (див. також завдання 2.1.2). Задача зводиться до відшукання екстремуму функціонала $\Pi[y]$ у класі двічі неперервно диференційовних функцій, що підкоряються крайовим умовам $y(0) = y'(0) = y(b) = y'(b) = 0$. Відповідне рівняння Ейлера — Пуассона досить складне, але його можна спростити, якщо природно припустити, що провисання стержня є малим: $y'^2 \ll 1$. Тоді рівняння набирає вигляду $y^{(4)} = \dfrac{\rho g}{EJ}$, а в крайових умовах справа можна покласти $b \approx L$.

*Відповідь*: $y(x) = \dfrac{\rho g}{24EJ} x^2 (L-x)^2$. Максимальне провисання спостерігається посередині стержня: $y_{\max} = y\left(\dfrac{L}{2}\right) = \dfrac{mgL^4}{384EJ}$.

**Завдання 2.2.3.** Знайдіть форму профілю стержня внаслідок провисання під власною вагою, якщо:

а) обидва кінці стержня закріплено шарнірно в точках $x = 0$ та $x = L$ горизонтальної осі;

б) кінець $x = 0$ спирається на стінку, а кінець $x = L$ замуровано в стінку;

в) кінець $x = 0$ замуровано в стінку, а кінець $x = L$ вільний.

*Вказівки*. У наближенні $y'^2(x) \ll 1$ формули (2.26)–(2.28) дають такі (на додаток до умов жорсткого закріплення) крайові умови: якщо кінець $x = x_0$ ($= 0$ або $L$) закріплено шарнірно або спирається на стінку, то $y(x_0) = y''(x_0) = 0$; якщо кінець $x = x_0$ вільний, то $y''(x_0) = y'''(x_0) = 0$.

*Відповіді*: а) $y(x) = \dfrac{\rho g}{24EJ} x\left(x^3 - 2Lx^2 + L^3\right)$; б) $y(x) = \dfrac{\rho g}{48EJ} x(2x^3 - {}$ $-3Lx^2 + L^3)$; в) $y(x) = \dfrac{\rho g}{24EJ} x^2(x^2 - 4Lx + 6L^2)$.

**Завдання 2.2.4.** Нехай $y(x)$ — екстремаль функціонала (2.22) у класі двічі неперервно диференційовних кривих, які разом зі своїми першими похідними набувають фіксованих значень на лівих кінцях, а на правих можуть вільно ковзати під довільними кутами у площині $XOY$. Знайдіть умови трансверсальності для рухомого кінця кривої $y(x)$.

*Відповідь*:

$$\left.\frac{\partial F}{\partial y''}\right|_{x=x_2}=0,\ \left.\left(\frac{\partial F}{\partial y'}-\frac{d}{dx}\frac{\partial F}{\partial y''}\right)\right|_{x=x_2}=0,\ \left.F\right|_{x=x_2}=0. \qquad (2.29)$$

Система рівнянь (2.29) дозволяє знайти абсцису $x_2$ та ординату $y_2=y(x_2)$ правого кінця екстремалі, а також тангенс кута нахилу дотичної $y'(x_2)$ до екстремалі в цій точці. Разом із рівнянням Ейлера — Пуассона (2.24) та крайовими умовами жорсткого закріплення типу (2.23) дістаємо необхідні умови, які повинна задовольняти $y(x)$.

Результати теореми 2.2.1 та завдання 2.2.1 досить легко узагальнити. Справджується така

**Теорема 2.2.2 (Ейлера — Пуассона)**. Нехай функція $F$ неперервна разом зі всіма своїми частинними похідними до порядку $n+1$ включно та при всіх можливих значеннях аргументів $x,y,y',y'',...,y^{(n)}$. Для того, щоб деяка крива була екстремаллю функціонала

$$J[y]=\int_{x_1}^{x_2}dxF(x,y,y',y'',...,y^{(n)}) \qquad (2.30)$$

у класі $C^n([x_1,x_2])$ кривих, підкорених крайовим умовам

$$y(x_1)=y_{10},\ y'(x_1)=y_{11},...,\ y^{(n-1)}(x_1)=y_{1,n-1},$$
$$y(x_2)=y_{20},\ y'(x_2)=y_{21},...,\ y^{(n-1)}(x_2)=y_{2,n-1}, \qquad (2.31)$$

необхідно, щоб вона задовольняла рівняння

$$\frac{\partial F}{\partial y}-\frac{d}{dx}\frac{\partial F}{\partial y'}+\frac{d^2}{dx^2}\frac{\partial F}{\partial y''}+...+(-1)^n\frac{d^n}{dx^n}\frac{\partial F}{\partial y^{(n)}}=0. \qquad (2.32)$$

Порядок рівняння (2.32) в певних випадках можна понизити.

1) Якщо функція $F$ не залежить явно від $y$, то рівняння (2.32) має перший інтеграл:

$$\frac{\partial F}{\partial y'}-\frac{d}{dx}\frac{\partial F}{\partial y''}+...+(-1)^{n-1}\frac{d^{n-1}}{dx^{n-1}}\frac{\partial F}{\partial y^{(n)}}=C.$$

2) Якщо функція $F$ не залежить явно від $x$,

$$J[y]=\int_{x_1}^{x_2} dxF(y,y',y'',...,y^{(n)}),$$

то за незалежну змінну зручно взяти $y$ і далі шукати $x$ як функцію від неї: $x=x(y)$. У нових змінних $J[y]$ набирає вигляду

$$J[x]=\int_{y(x_1)}^{y(x_2)} dy\, x'F\left(y,\frac{1}{x'},-\frac{x''}{x'^3},...\right),$$

де $x^{(n)}\equiv\dfrac{d^n x}{dy^n}$. Позначивши нову підінтегральну функцію через

$$\Phi\left(y,x',x'',...,x^{(n)}\right)\equiv x'F\left(y,\frac{1}{x'},-\frac{x''}{x'^3},...\right),$$

знаходимо перший інтеграл рівняння Ейлера — Пуассона у вигляді

$$\frac{\partial\Phi}{\partial x'}-\frac{d}{dy}\frac{\partial\Phi}{\partial x''}+...+(-1)^{n-1}\frac{d^{n-1}}{dy^{n-1}}\frac{\partial\Phi}{\partial x^{(n)}}=C.$$

3) Якщо $F$ залежить лише від похідних $y^{(k)}$, $y^{(k+1)}$, ..., $y^{(n)}$, $k\geq 2$, то відповідне рівняння Ейлера — Пуассона

$$(-1)^k\frac{d^k}{dx^k}\frac{\partial F}{\partial y^{(k)}}+(-1)^{k+1}\frac{d^{k+1}}{dx^{k+1}}\frac{\partial F}{\partial y^{(k+1)}}+...+(-1)^n\frac{d^n}{dx^n}\frac{\partial F}{\partial y^{(n)}}=0$$

після інтегрування $k$ разів веде до інтеграла

$$\frac{\partial F}{\partial y^{(k)}}-\frac{d}{dx}\frac{\partial F}{\partial y^{(k+1)}}+...+(-1)^{n-k}\frac{d^{n-k}}{dx^{n-k}}\frac{\partial F}{\partial y^{(n)}}=Q_{k-1}(x),$$

де $Q_{k-1}(x)=C_1x^{k-1}+C_2x^{k-2}+...+C_{k-1}x+C_k$ — поліном степеня $k-1$, $C_1,...,C_k$ — довільні сталі.

4) Якщо $F$ залежить лише від $y^{(n)}$, то (2.32) зводиться до рівняння

$$\frac{d^n}{dx^n}\frac{\partial F}{\partial y^{(n)}}=0,$$

звідки

$$\frac{\partial F}{\partial y^{(n)}}=P_{n-1}(x),$$

де $P_{n-1}(x)$ — поліном степеня $n-1$. Ліва частина цього рівняння містить лише $y^{(n)}$. Розв'язавши його відносно $y^{(n)}$, дістанемо $y^{(n)}$ як функцію від $x$:

$$y^{(n)}=f\left(P_{n-1}(x)\right).$$

Функція $y$ відновлюється далі $n$-разовим інтегруванням за $x$ обох частин останнього рівняння:

$$y(x)=\int\ldots\int(dx)^n f\left(P_{n-1}(x)\right)+Q_{n-1}(x),$$

де $Q_{n-1}(x)$ — довільний поліном степеня $n-1$.

## 2.3. ЗАДАЧІ ДЛЯ ПРОСТОРОВИХ КРИВИХ

До цього часу розглядалися задачі, у яких функціонали залежали від плоских кривих. При подальшому узагальненні цих задач приходимо до функціоналів, визначених на множинах просторових ліній.

Розглянемо, наприклад, задачу про поширення світла в неоднорідному середовищі між заданими точками $A(x_1,y_1,z_1)$ і $B(x_2,y_2,z_2)$. Уважаючи швидкість світла в середовищі відомою функцією $u=u(x,y,z)$ просторових координат та користуючись принципом Ферма (див. підрозділ 1.4 та завдання 1.4.4), бачимо, що задача зводиться до відшукання у тривимірному просторі кривої $\gamma:\ y=y(x),\ \ z=z(x)$ із фіксованими кінцями, уздовж якої промінь світла доходить із точки $A$ в точку $B$ за мінімальний час, тобто для якої функціонал

$$T[y,z]=\int_{x_1}^{x_2}dx\frac{\sqrt{1+y'^2+z'^2}}{u(x,y,z)} \tag{2.33}$$

набуває найменшого значення при додаткових умовах

$$y(x_1)=y_1,\ \ y(x_2)=y_2,\ \ z(x_1)=z_1,\ \ z(x_2)=z_2. \tag{2.34}$$

Зауважимо, що функціонал (2.33) залежить від двох функцій та їх перших похідних. У випадку $(n+1)$-вимірного простору, у якому криві задаються за допомогою $n$ функцій $y_i(x)$, $x\in[x_1,x_2]$, доводиться аналізувати фунціонали виду

$$J[y_1,y_2,\ldots,y_n]=\int_{x_1}^{x_2}dxF(x;y_1,y_2,\ldots,y_n;y_1',y_2',\ldots,y_n'). \tag{2.35}$$

Уточнивши для просторової кривої поняття $\varepsilon$-околу, для них теж можемо користуватися (з відповідними змінами) означеннями відносного та абсолютного екстремумів, наведеними в підрозділі 1.1.

**Означення 2.3.1.** $\varepsilon$-Околом порядку $k$ просторової кривої $\gamma_0:\ y_{0i}=y_{0i}(x),\ \ i=1,2,\ldots,n,$ на проміжку $[x_1,x_2]$ називають множину всіх просторових кривих $\gamma:\ y_i=y_i(x),\ \ i=1,2,\ldots,n,$ для яких скрізь на цьому проміжку і при всіх $i$ виконуються нерівності

$$\left|y_i(x)-y_{0i}(x)\right|\le\varepsilon,\ \left|y_i'(x)-y_{0i}'(x)\right|\le\varepsilon,\ldots,\ \left|y_i^{(k)}(x)-y_{0i}^{(k)}(x)\right|\le\varepsilon.$$

Число $\varepsilon$ називають відстанню порядку $k$ між кривими $\gamma$ і $\gamma_0$.

Нехай функції $y_0(x)$ та $z_0(x)$ задають просторову криву $\gamma_0$, яка реалізує екстремум функціонала

$$J[y,z]=\int_{x_1}^{x_2}dxF(x,y,z,y',z') \tag{2.36}$$

на функціях класу $C^2([x_1,x_2])$ з фіксованими значеннями (2.34) на кінцях, а підінтегральна функція $F$ неперервна разом зі своїми частинними похідними до другого порядку включно. Розглянемо множину допустимих просторових кривих $\gamma$: $y(x)=y_0(x)+\delta y(x)$, $z(x)=z_0(x)+\delta z(x)$ з $\varepsilon$-околу першого порядку кривої $\gamma_0$ та проаналізуємо умови, за яких перша варіація функціонала (2.36)

$$\delta J[y,z]=\frac{\partial F}{\partial y'}\delta y(x)\Bigg|_{x_1}^{x_2}+\int_{x_1}^{x_2}\left\{\frac{\partial F}{\partial y}-\frac{d}{dx}\frac{\partial F}{\partial y'}\right\}\delta y(x)dx+$$
$$+\frac{\partial F}{\partial z'}\delta z(x)\Bigg|_{x_1}^{x_2}+\int_{x_1}^{x_2}\left\{\frac{\partial F}{\partial z}-\frac{d}{dx}\frac{\partial F}{\partial z'}\right\}\delta z(x)dx \tag{2.37}$$

дорівнює нулю при довільних значеннях варіацій $\delta y(x)$ і $\delta z(x)$ на внутрішніх точках проміжку $[x_1,x_2]$ та нульових значеннях на його межах: $\delta y(x_1)=\delta y(x_2)=0$, $\delta z(x_1)=\delta z(x_2)=0$. Оскільки варіації $\delta y(x)$ і $\delta z(x)$ незалежні, відразу бачимо, що функції $y_0(x)$ та $z_0(x)$ задовольняють систему диференціальних рівнянь Ейлера — Лагранжа

$$\begin{aligned}&\frac{\partial F}{\partial y}-\frac{d}{dx}\frac{\partial F}{\partial y'}=0,\\&\frac{\partial F}{\partial z}-\frac{d}{dx}\frac{\partial F}{\partial z'}=0.\end{aligned} \tag{2.38}$$

**Зауваження 2.3.1.** У загальному випадку *узагальнена теорема Ейлера — Лагранжа* стверджує: якщо крива $\gamma: y_1=y_1(x), y_2=y_2(x),\ldots,y_n= =y_n(x)$ — екстремаль функціонала (2.35) серед кривих класу $C^1([x_1,x_2])$ із закріпленими кінцями, то функції $y_i(x)$ задовольняють систему диференціальних рівнянь

$$\frac{\partial F}{\partial y_i}-\frac{d}{dx}\frac{\partial F}{\partial y_i'}=0,\ i=1,2,\ldots,n. \tag{2.39}$$

Якщо при цьому вздовж $\gamma$ функціональний визначник

$$\Delta \equiv \left| \frac{\partial^2 F}{\partial y_i' \partial y_k'} \right| = \begin{vmatrix} \frac{\partial^2 F}{\partial y_1' \partial y_1'} & \frac{\partial^2 F}{\partial y_1' \partial y_2'} & \cdots & \frac{\partial^2 F}{\partial y_1' \partial y_n'} \\ \cdots & \cdots & \cdots & \cdots \\ \cdots & \cdots & \cdots & \cdots \\ \frac{\partial^2 F}{\partial y_n' \partial y_1'} & \frac{\partial^2 F}{\partial y_n' \partial y_2'} & \cdots & \frac{\partial^2 F}{\partial y_n' \partial y_n'} \end{vmatrix} \neq 0, \qquad (2.40)$$

то функції $y_i'(x)$ мають неперервні похідні, тобто екстремаль $\gamma$ належить класу $C^2([x_1, x_2])$ (ця обставина використовувалася при знаходженні варіації (2.37)).

Рівняння (2.39) інваріантні відносно перетворень координат.

**Завдання 2.3.1.** Випишіть рівняння Ейлера — Лагранжа для функціонала (2.33) та знайдіть їх загальний розв'язок для однорідного середовища.

*Розв'язання*. Безпосереднє диференціювання дає:

$$\frac{\partial u}{\partial y}\frac{\sqrt{1+y'^2+z'^2}}{u^2} + \frac{d}{dx}\frac{y'}{u\sqrt{1+y'^2+z'^2}} = 0,$$

$$\frac{\partial u}{\partial z}\frac{\sqrt{1+y'^2+z'^2}}{u^2} + \frac{d}{dx}\frac{z'}{u\sqrt{1+y'^2+z'^2}} = 0.$$

Для однорідного середовища $u = \text{const}$, і ці рівняння можна один раз зінтегрувати:

$$\frac{y'}{\sqrt{1+y'^2+z'^2}} = C_1, \quad \frac{z'}{\sqrt{1+y'^2+z'^2}} = C_2,$$

де $C_1$ та $C_2$ — сталі. Розв'язуючи останні рівняння відносно $y'$ та $z'$, знаходимо $y'(x) = a_1$, $z'(x) = a_2$, звідки

$$y(x) = a_1 x + b_1, \quad z(x) = a_2 x + b_2.$$

Тут $a_i$ та $b_i$, $i = 1,2$ — деякі сталі, чиї значення визначаються через координати початкової та кінцевих точок.

Таким чином, в однорідному середовищі світло поширюється по прямій.

**Завдання 2.3.2.** Доведіть правильність таких тверджень.

1) Якщо підінтегральна функція $F$ у функціоналі (2.35) не залежить явно від змінної (кількох змінних) $y_k$, то система рівнянь

Ейлера — Лагранжа (2.39) має перший інтеграл (кілька перших інтегралів)

$$\frac{\partial F}{\partial y_k'} = C_k. \tag{2.41}$$

Усього можуть існувати $n$ перших інтегралів такого типу.

2) Якщо функція $F$ не залежить явно від змінної $x$, то система рівнянь (2.39) має перший інтеграл

$$F - \sum_{i=1}^{n} y_i' \frac{\partial F}{\partial y_i'} = C. \tag{2.42}$$

**Завдання 2.3.3.** Виходячи з принципу найменшої дії Гамільтона (1.30), виведіть рівняння руху (1.32) механічної системи з $s$ ступенями вільності.

Розширення класу допустимих кривих, на якому шукається екстремум функціонала типу (2.35), відповідним чином змінює крайові умови, що їх (як і систему рівнянь (2.39)) повинна задовольняти екстремаль. Нехай, наприклад, екстремаль функціонала (2.35) шукається у класі гладких просторових кривих, ліві кінці яких жорстко закріплено, а праві можуть вільно ковзати у площині $x = x_2$. Якщо при ковзанні правого кінця допустимої кривої $\gamma$: $y_i = y_i(x)$, $i = 1,2,...,n$, у площині $x = x_2$ можуть змінюватися значення однієї або кількох функцій $y_k = y_k(x)$, то для кожної з них маємо:

$$\left.\frac{\partial F}{\partial y_k'}\right|_{x=x_2} = 0. \tag{2.43}$$

Умова (2.43) є узагальненням крайової умови (2.4) на випадок функціонала, що залежить від просторової кривої. Умова трансверсальності (2.14) також досить легко узагальнюється на цей випадок.

**Завдання 2.3.4.** Доведіть таку теорему: якщо крива $\gamma$: $y = y(x)$, $z = z(x)$ надає екстремум функціоналу

$$J[\Gamma] = \int_{\Gamma} dx F(x, y, z, y', z') \tag{2.44}$$

на множині гладких кривих Г, що з'єднують фіксовану точку $A(x_1, y_1, z_1)$ з точками заданої гладкої кривої $C$: $y = \varphi(x)$, $z = \psi(x)$, то в точці $B(x_2, y_2, z_2)$, яка належить обом кривим $\gamma$ та $C$, виконується умова трансверсальності

$$\left[F + (\varphi' - y')\frac{\partial F}{\partial y'} + (\psi' - z')\frac{\partial F}{\partial z'}\right]_{x=x_2} = 0. \tag{2.45}$$

Співвідношення (2.45) разом із рівняннями

$$y(x_2)=\varphi(x_2),\quad z(x_2)=\psi(x_2) \tag{2.46}$$

та умовами жорсткого закріплення в точці $A$ дозволяють знайти значення всіх сталих інтегрування в загальному розв'язку системи рівнянь Ейлера — Лагранжа (2.38) та значення абсциси $x_2$.

**Завдання 2.3.5.** Нехай у функціоналі (2.44)

$$F(x,y,z,y',z')=f(x,y,z)\sqrt{1+y'^2+z'^2}. \tag{2.47}$$

Доведіть, що екстремаль $\gamma$ цього функціонала ортогональна до кривої $C$.

*Вказівка*. Координати одиничного вектора дотичної $\mathbf{n}_l(t)$ до гладкої кривої $l$: $x=\xi(t)$, $y=\eta(t)$, $z=\zeta(t)$, $t_1\le t\le t_2$, знаходяться за формулою

$$\mathbf{n}_l(t)=\frac{1}{\sqrt{\xi'^2(t)+\eta'^2(t)+\zeta'^2(t)}}\big(\xi'(t),\eta'(t),\zeta'(t)\big).$$

Тому умова ортогональності $\mathbf{n}_\gamma(x_2)\cdot\mathbf{n}_C(x_2)=0$ кривих $\gamma$: $x=t$, $y=y(x)$, $z=z(x)$ і $C$: $x=t$, $y=\varphi(x)$, $z=\psi(x)$ у точці $x_2$ має вигляд

$$1+y'(x_2)\varphi'(x_2)+z'(x_2)\psi'(x_2)=0.$$

Зауважимо, що правильне й обернене твердження: якщо для екстремалі $\gamma$ функціонала (2.44) і будь-якої гладкої кривої $C$ умова трансверсальності збігається з умовою ортогональності, то функція $F$ має структуру (2.47), де $f$ — довільна диференційовна функція змінних $x$, $y$ та $z$.

**Завдання 2.3.6.** Нехай із точки $A$ випущено світловий сигнал. Доведіть, що промінь $AB$, уздовж якого сигнал уперше досягає заданої гладкої поверхні $S$, ортогональний до неї.

*Вказівка*. Згідно з принципом Ферма, треба знайти екстремаль функціонала

$$T[\Gamma]=\int_\Gamma\frac{\sqrt{1+y'^2+z'^2}}{u(x,y,z)}dx$$

на множині гладких кривих Г, що з'єднують точку $A$ з точками поверхні $S$. З попереднього завдання випливає, що промінь $AB$ є ортогональним до будь-якої кривої, що лежить на поверхні $S$ і проходить через точку $B$.

Важливим фізичним наслідком завдання 2.3.6 є *принцип Малюса* в геометричній оптиці: фронт світлової хвилі (геометричне місце точок, які досягаються світловим сигналом за рівні проміжки часу) є ортогональним до всіх променів світлового потоку. Цей принцип справджується й тоді, коли промені заломлюються та відбиваються довільну кількість разів.

До відшукання екстремалей функціоналів типу (2.35) веде й задача про *геодезичні* — криві найбільшої або найменшої довжини між двома заданими точками на двовимірній поверхні (див. завдання 1.3.3) або в багатовимірному рімановому просторі. Поняття геодезичної є узагальненням поняття прямої лінії евклідового простору — відрізки останньої є, як ми вже бачили, найкоротшими серед усіх спрямлюваних ліній, що з'єднують дві довільно задані точки евклідового простору.

Нехай положення точки на деякій поверхні визначається координатами $(u,v)$, а квадрат елемента довжини на ній

$$dl^2 = E(u,v)du^2 + 2F(u,v)dudv + G(u,v)dv^2. \tag{2.48}$$

Тоді геодезичними цієї поверхні є ті лінії $u = u(\tau)$, $v = v(\tau)$, $\tau \in [\tau_1, \tau_2]$, які надають екстремум функціоналу довжини

$$L[u,v] = \int_{\tau_1}^{\tau_2} d\tau \sqrt{Eu'^2 + 2Fu'v' + Gv'^2}, \tag{2.49}$$

де $u' \equiv du/d\tau$, $v' \equiv dv/d\tau$.

**Завдання 2.3.7.** Знайдіть геодезичні лінії поверхні кругового циліндра радіусом $R$.

*Розв'язання.* Рівняння циліндричної поверхні радіусом $R$ має вигляд

$$x = R\cos\varphi,\ y = R\sin\varphi,\ 0 \le \varphi < 2\pi,\ -\infty < z < +\infty,$$

тому квадрат елемента довжини на ній $dl^2 = R^2 d\varphi^2 + dz^2$. Якщо рівняння геодезичної шукати в параметричній формі $\varphi = \varphi(\tau)$, $z = z(\tau)$, $\tau_1 \le \tau \le \tau_2$, то функціонал довжини набирає вигляду функціонала, що залежить від двох функцій та їх перших похідних:

$$L[\varphi, z] = \int_{\tau_1}^{\tau_2} d\tau \sqrt{R^2\varphi'^2 + z'^2}. \tag{2.50}$$

Підінтегральна функція у формулі (2.50) залежить лише від похідних $\varphi'$ та $z'$, тому система рівнянь Ейлера — Лагранжа має перші інтеграли

$$\frac{R^2\varphi'}{\sqrt{R^2\varphi'^2 + z'^2}} = C_1,\quad \frac{z'}{\sqrt{R^2\varphi'^2 + z'^2}} = C_2.$$

Звідси випливає, що шукані функції є лінійними:

$$\varphi(\tau) = a_1\tau + b_1,\ \ z(\tau) = a_2\tau + b_2, \qquad (2.51)$$

де $a_i$ і $b_i$ — сталі, які визначаються за допомогою координат початкової та кінцевої точок.

Геодезичні (2.51) — це *гвинтові лінії*. Якщо кінцеві точки знаходяться на одній вертикалі $(\varphi = \text{const})$, геодезичні вироджуються у відрізки прямих; якщо ж на одній висоті $(z = \text{const})$ — у дуги кола.

**Завдання 2.3.8.** Те саме для кругового конуса

$$x = az\cos\varphi,\ \ y = az\sin\varphi,\ \ 0 \le \varphi < 2\pi,\ \ 0 \le z \le H.$$

*Вказівка*. Квадрат елемента довжини на поверхні конуса

$$dl^2 = a^2z^2d\varphi^2 + (1+a^2)dz^2.$$

*Відповідь*: $\varphi = C_2 + \sqrt{1+a^{-2}}\arcsin\left(C_1/(az)\right)$, де сталі інтегрування визначаються через координати $(\varphi_1, z_1)$ та $(\varphi_2, z_2)$ початкової та кінцевої точок геодезичної. При $\varphi_1 = \varphi_2$, $z_1 \ne z_2$ маємо $C_1 = 0$, тобто геодезична — це відрізок твірної конуса. При $\varphi_1 \ne \varphi_2$ розв'язок зручно подати у вигляді

$$z = z_1 \sin\frac{\varphi_1 - C_2}{\sqrt{1+a^{-2}}} \Bigg/ \sin\frac{\varphi - C_2}{\sqrt{1+a^{-2}}}.$$

Проаналізуйте його поведінку для випадку $z_1 = z_2$ (кінці геодезичної знаходяться на одній висоті).

На завершення виведемо диференціальне рівняння геодезичної для загального випадку $n$-вимірного ріманового простору.

**Означення 2.3.2.** $n$-Вимірний простір з довільним базисом $\mathbf{e}_i$ і координатами $x^i$, $i = 1,2,\ldots,n$, називається рімановим, якщо його метрика задається співвідношенням

$$dl^2 = \sum_{i,k=1}^{n} g_{ik}(x^1, x^2, \ldots, x^n)dx^i dx^k \equiv g_{ik}dx^i dx^k \ge 0, \qquad (2.52)$$

де $g_{ik}(x^1, x^2, \ldots, x^n) = \mathbf{e}_i \cdot \mathbf{e}_k$ — компоненти метричного тензора. За нижнім і верхнім індексами, що повторюються, тут і далі розуміємо підсумовування.

З означення (2.52) випливає, що геодезичними ріманового простору є екстремалі функціонала

$$L[x^i] = \int_{\tau_1}^{\tau_2} d\tau \sqrt{g_{ik}(x^1, x^2, \ldots, x^n)\frac{dx^i}{d\tau}\frac{dx^k}{d\tau}}, \qquad (2.53)$$

який виражає довжину відрізка кривої $x^i = x^i(\tau),\ i = 1,2,...,n,\ \tau \in [\tau_1, \tau_2]$, між заданими точками. Більше того, будь-яка достатньо мала ділянка геодезичної є найкоротшою серед усіх спрямлюваних ліній, що з'єднують ці точки.

Запишемо систему рівнянь Ейлера — Лагранжа для функціонала (2.53). Вони набирають найбільш компактного вигляду, якщо в якості $\tau$ скористатися так званим канонічним параметром. За означенням, це будь-який параметр, який можна подати у вигляді $\tau = al + b$, де $a \neq 0$ і $b$ — сталі, а $l$ — довжина відрізка геодезичної. Для канонічного параметра величина

$$g_{ik}\frac{dx^i}{d\tau}\frac{dx^k}{d\tau} = g_{ik}\frac{dx^i}{adl}\frac{dx^k}{adl} = \frac{1}{a^2} = \text{const},$$

і тому екстремум функціонала (2.53) досягається на тій самій кривій, що й екстремум функціонала

$$M[x^i] = \int_{\tau_1}^{\tau_2} d\tau\, g_{ik}\dot{x}^i\dot{x}^k, \tag{2.54}$$

де $\dot{x}^i \equiv dx^i/d\tau$. Справді:

$$\delta M = \delta\int_{\tau_1}^{\tau_2} d\tau\left(\sqrt{g_{ik}\dot{x}^i\dot{x}^k}\right)^2 = 2\int_{\tau_1}^{\tau_2} d\tau\frac{1}{\sqrt{a^2}}\delta\sqrt{g_{ik}\dot{x}^i\dot{x}^k} = \frac{2}{|a|}\delta\int_{\tau_1}^{\tau_2} d\tau\sqrt{g_{ik}\dot{x}^i\dot{x}^k}.$$

Оскільки $\partial\dot{x}^i/\partial\dot{x}^k = \delta^i_k$ — символ Кронекера, $g_{ik}$ — симетричний тензор, то, перепозначаючи індекси підсумовування, для похідних підінтегральної функції $g_{ik}\dot{x}^i\dot{x}^k \equiv F$ у формулі (2.54) маємо:

$$\frac{\partial F}{\partial x^l} = \frac{\partial g_{ik}}{\partial x^l}\dot{x}^i\dot{x}^k,\quad \frac{\partial F}{\partial \dot{x}^l} = g_{ik}\delta^i_l\dot{x}^k + g_{ik}\dot{x}^i\delta^k_l = g_{lk}\dot{x}^k + g_{li}\dot{x}^i,$$

$$\frac{d}{d\tau}\frac{\partial F}{\partial \dot{x}^l} = \frac{\partial g_{lk}}{\partial x^i}\dot{x}^i\dot{x}^k + g_{lk}\ddot{x}^k + \frac{\partial g_{li}}{\partial x^k}\dot{x}^i\dot{x}^k + g_{li}\ddot{x}^i = \left(\frac{\partial g_{lk}}{\partial x^i} + \frac{\partial g_{li}}{\partial x^k}\right)\dot{x}^i\dot{x}^k + 2g_{li}\ddot{x}^i.$$

Позначивши тепер

$$\Gamma_{l,ik} = \frac{1}{2}\left(\frac{\partial g_{li}}{\partial x^k} + \frac{\partial g_{lk}}{\partial x^i} - \frac{\partial g_{ik}}{\partial x^l}\right),$$

для екстремалей функціонала (2.54), а, отже, і функціонала (2.53), дістаємо:

$$g_{li}\ddot{x}^i + \Gamma_{l,ik}\dot{x}^i\dot{x}^k = 0.$$

Величини $\Gamma_{l,ik}$ називаються символами Кристоффеля першого роду. Зручніше, однак, перейти до символів Кристоффеля другого роду

$$\Gamma_{ik}^{m} = g^{ml}\Gamma_{l,ik},$$

де тензор $g^{ml}$ визначається із співвідношення

$$g^{ml}g_{li} = \delta_i^m.$$

Тоді остаточна відповідь така:

$$\frac{d^2x^m}{d\tau^2} + \Gamma_{ik}^{m}\frac{dx^i}{d\tau}\frac{dx^k}{d\tau} = 0. \qquad (2.55)$$

Зауважимо, що з довільної точки ріманового простору в заданому напрямі виходить тільки одна геодезична.

## 2.4. ЗАДАЧІ НА УМОВНИЙ ЕКСТРЕМУМ

У задачах на умовний екстремум екстремаль функціонала відшукується серед функцій заданого класу, які, крім крайових, задовольняють певні додаткові умови. У багатьох випадках ці задачі можна розв'язати, звівши їх до звичайних задач варіаційного числення для функціоналів більш загального виду.

Зупинимося спершу на *ізопериметричній задачі*. У найпростішому варіанті вона формулюється наступним чином: серед гладких функцій із жорстко закріпленими кінцями знайти екстремаль функціонала

$$J[y] = \int_{x_1}^{x_2} dxF(x, y, y'), \qquad (2.56)$$

для якої інший функціонал набуває заданого сталого значення:

$$L[y] = \int_{x_1}^{x_2} dx\, G(x, y, y') = L_0. \qquad (2.57)$$

Як і раніше, уважатимемо, що функції $F(x, y, y')$ і $G(x, y, y')$ є неперервними разом зі своїми першими та другими похідними за всіма аргументами в деякій області $D$ площини $XOY$ і при довільних значеннях похідної $y'(x)$.

**Теорема 2.4.1 (Ейлера)**. Якщо крива $y_0(x) \in C^2([x_1, x_2])$ надає екстремум функціоналу (2.56) у класі $C^1([x_1, x_2])$ за умови (2.57) та звичайних умов жорсткого закріплення

$$y(x_1) = y_1, \quad y(x_2) = y_2, \qquad (2.58)$$

і якщо $y_0(x)$ не є екстремаллю функціонала (2.57) (тобто число $L_0$ не є екстремальним значенням функціонала (2.57)), то існує така стала $\lambda$, що крива $y_0(x)$ надає екстремум функціоналу

$$J^*[y] = J[y] - \lambda L[y] = \int_{x_1}^{x_2} dx\big[F(x,y,y') - \lambda G(x,y,y')\big]. \qquad (2.59)$$

Стала (параметр) $\lambda$ називається невизначеним множником Лагранжа.

*Доведення.* Називатимемо допустимими такі варіації $\delta y(x) \in \in C_0^1([x_1,x_2])$ кривої $y_0(x)$, для яких

$$L[y_0 + \delta y] = L_0. \qquad (2.60)$$

Для допустимих варіацій маємо

$$\delta L[y_0] = \int_{x_1}^{x_2} \left\{ \frac{\partial G}{\partial y} - \frac{d}{dx}\frac{\partial G}{\partial y'} \right\} \delta y(x) dx = 0, \qquad (2.61)$$

а також, як і у випадку функціонала найпростішого типу (див. підрозділ 1.2),

$$\delta J[y_0] = \int_{x_1}^{x_2} \left\{ \frac{\partial F}{\partial y} - \frac{d}{dx}\frac{\partial F}{\partial y'} \right\} \delta y(x) dx = 0. \qquad (2.62)$$

Умова (2.61) істотно звужує клас можливих варіацій екстремальної кривої. Зокрема, не є допустимими варіації, які відрізняються від нуля в малому околі лише однієї точки $x_0 \in (x_1, x_2)$, для якої

$$\left[ \frac{\partial G(x, y_0(x), y_0'(x))}{\partial y} - \frac{d}{dx} \frac{\partial G(x, y_0(x), y_0'(x))}{\partial y'} \right]_{x=x_0} \neq 0. \qquad (2.63)$$

Точки, для яких справджується умова (2.63), існують, бо інакше функція $y_0(x)$ задовольняла б на інтервалі $(x_1, x_2)$ рівняння

$$\frac{\partial G}{\partial y} - \frac{d}{dx}\frac{\partial G}{\partial y'} = 0$$

для екстремалі функціонала (2.57), а це суперечить умові теореми.

Розглянемо тепер таку допустиму варіацію, яка відмінна від нуля в як завгодно малих $\delta$-околах точки $x_0$ та іншої (довільної) точки $x' \in (x_1, x_2)$. Позначивши

$$S_0 \equiv \int_{x_0-\delta}^{x_0+\delta} \delta y(x) dx, \quad S' \equiv \int_{x'-\delta}^{x'+\delta} \delta y(x) dx,$$

замість (2.61) і (2.62) можемо записати:

$$\delta L[y_0] \approx \left[\frac{\partial G}{\partial y} - \frac{d}{dx}\frac{\partial G}{\partial y'}\right]_{x=x_0} \cdot S_0 + \left[\frac{\partial G}{\partial y} - \frac{d}{dx}\frac{\partial G}{\partial y'}\right]_{x=x'} \cdot S' = 0, \qquad (2.64)$$

$$\delta J[y_0] \approx \left[\frac{\partial F}{\partial y} - \frac{d}{dx}\frac{\partial F}{\partial y'}\right]_{x=x_0} \cdot S_0 + \left[\frac{\partial F}{\partial y} - \frac{d}{dx}\frac{\partial F}{\partial y'}\right]_{x=x'} \cdot S' = 0. \qquad (2.65)$$

З рівняння (2.64) знаходимо:

$$S_0 = -\frac{\left[\frac{\partial G}{\partial y} - \frac{d}{dx}\frac{\partial G}{\partial y'}\right]_{x=x'}}{\left[\frac{\partial G}{\partial y} - \frac{d}{dx}\frac{\partial G}{\partial y'}\right]_{x=x_0}} \cdot S'.$$

Поклавши тепер

$$\lambda = \frac{\left[\frac{\partial F}{\partial y} - \frac{d}{dx}\frac{\partial F}{\partial y'}\right]_{x=x_0}}{\left[\frac{\partial G}{\partial y} - \frac{d}{dx}\frac{\partial G}{\partial y'}\right]_{x=x_0}},$$

рівняння (2.65) можемо подати у вигляді

$$\left\{\frac{\partial F}{\partial y} - \frac{d}{dx}\frac{\partial F}{\partial y'} - \lambda\left[\frac{\partial G}{\partial y} - \frac{d}{dx}\frac{\partial G}{\partial y'}\right]\right\}\Bigg|_{x=x'} \cdot S' = 0.$$

Оскільки $S'$ — мале, але довільне число, а $x'$ — довільна точка інтервалу $(x_1, x_2)$, бачимо, що екстремальна функція задовольняє рівняння

$$\frac{\partial}{\partial y}(F - \lambda G) - \frac{d}{dx}\frac{\partial}{\partial y'}(F - \lambda G) = 0. \qquad (2.66)$$

Таким чином, щоб розв'язати ізопериметричну задачу (2.56)–(2.58), спершу за допомогою заданих функціоналів $J[y]$ та $L[y]$ будуємо функціонал $J^*[y]$ (2.59) з підінтегральною функцією $F^*(x,y,y') = F(x,y,y') - \lambda G(x,y,y')$ та виписуємо для нього відповідне диференціальне рівняння Ейлера — Лагранжа (2.66). Розв'язавши це рівняння, отримуємо в загальному випадку трипараметричну сім'ю кривих $y = y(x;\lambda,C_1,C_2)$, серед яких, можливо, знаходиться функція, що є екстремаллю функціонала (2.56) і надає фіксованого (не екстремального) значення функціоналу (2.57). Щоб її виокремити, за допомогою крайових умов (2.58) та умови (2.57) знаходимо значення сталих інтегрування $C_1$, $C_2$ та параметра $\lambda$:

$$y(x_1;\lambda,C_1,C_2) = y_1,\ \ y(x_2;\lambda,C_1,C_2) = y_2,\ \ L[y(x;\lambda,C_1,C_2)] = L_0.$$

**Завдання 2.4.1.** Знайдіть рівноважний профіль важкої однорідної гнучкої нерозтяжної нитки довжиною $l$, яка провисає під дією сили тяжіння. Кінці нитки закріплено на одному рівні. (Задача про ланцюгову лінію.)

*Розв'язання.* Нехай кінці нитки закріплено в точках $x=\pm a$ горизонтальної осі $OX$, а вісь $OY$ напрямлена вертикально вгору (рис. 2.6). У стані рівноваги система має мінімальну потенціальну енергію, тобто її центр ваги $Y$ знаходиться в найнижчій точці. Треба знайти мінімум функціонала

$$Y[y]=\int_{-a}^{+a} dx\, y\sqrt{1+y'^2}$$

за додаткової умови, що довжина нитки залишається сталою:

$$L[y]=\int_{-a}^{+a} dx\sqrt{1+y'^2}=l. \qquad (2.67)$$

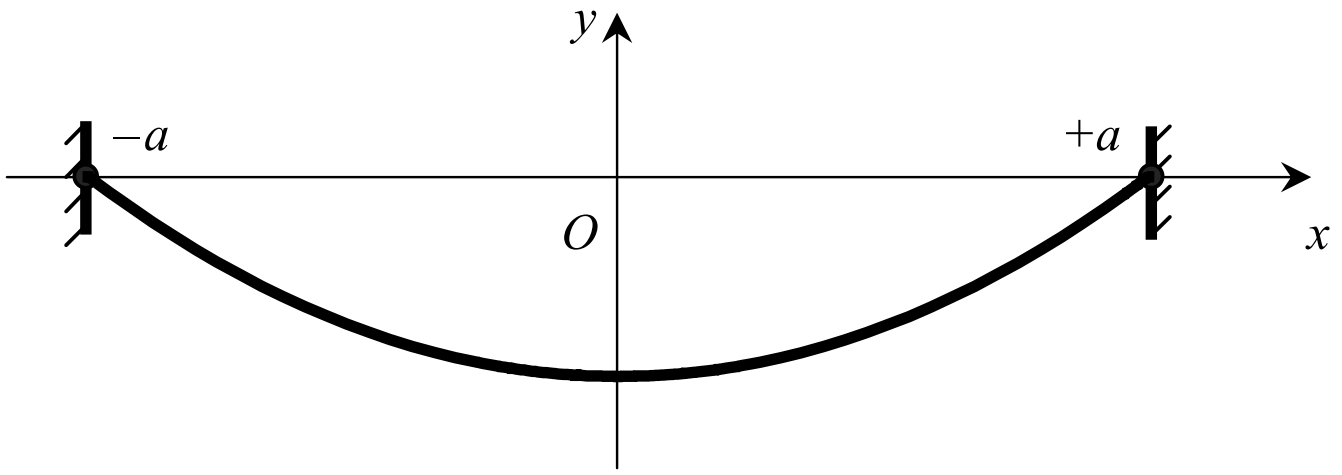

Рис. 2.6. Провисання нитки під власною вагою

Розв'язком цієї ізопериметричної задачі є екстремаль функціонала

$$Y^*[y]=\int_{-a}^{+a} dx(y-\lambda)\sqrt{1+y'^2},$$

яка задовольняє крайові умови $y(-a)=y(a)=0$ та умову (2.67). Перший інтеграл відповідного рівняння Ейлера — Лагранжа має вигляд

$$\frac{y-\lambda}{\sqrt{1+y'^2}}=C_1,$$

де $\lambda$ — невизначений множник Лагранжа, $C_1$ — стала інтегрування. Підставивши $y'=\operatorname{sh} t$, знаходимо:

$$y(x)=\lambda+C_1\operatorname{ch}\left(\frac{x}{C_1}+C_2\right).$$

Крайові умови дають: $C_2 = 0$, $\lambda = -C_1 \operatorname{ch}(a/C_1)$. З умови (2.67) для сталої $C_1$ дістаємо рівняння $\operatorname{sh}(a/C_1) = l/(2C_1)$. У безрозмірному вигляді маємо: $\operatorname{sh} u = \alpha u$, де $u \equiv a/C_1$, $\alpha \equiv l/(2a)$. Тривіальний розв'язок $u = 0$ цього рівняння відповідає випадку, коли $C_1 = \infty$ і $y = 0$. З фізичного погляду це можливо лише за умови, що довжина нитки $l = 2a$. Нетривіальний розв'язок з'являється тоді, коли тангенс кута нахилу прямої $y = \alpha u$ є більшим за тангенс кута нахилу дотичної до функції $y = \operatorname{sh} u$ в нулі (див. рис. 2.7), тобто коли $\alpha > 1$, або $l > 2a$. Отже, задача має розв'язок за умови, що $l \geq 2a$.

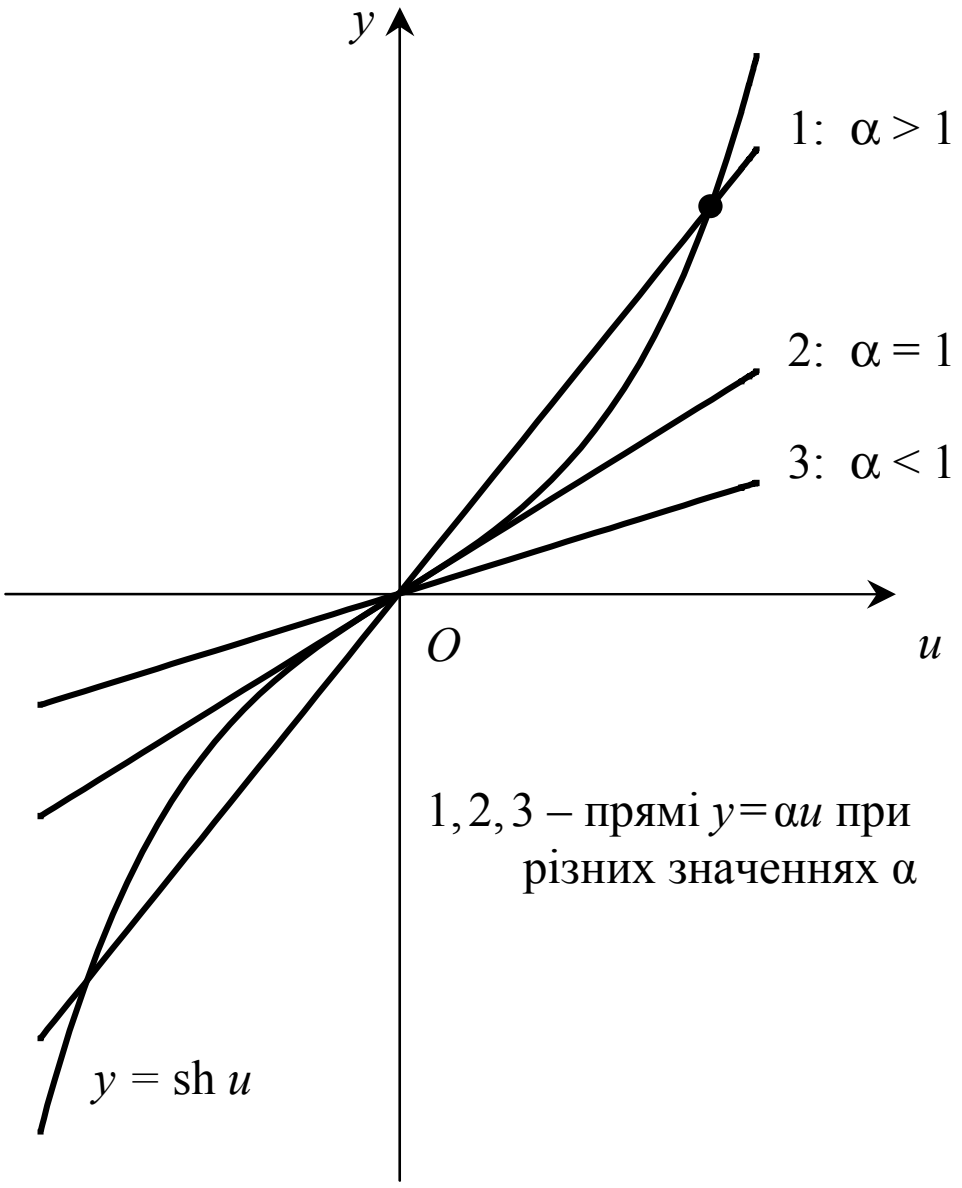

Рис. 2.7. Графічний аналіз рівняння $\operatorname{sh} u = \alpha u$

**Завдання 2.4.2.** Серед усіх кривих довжиною $l$, що з'єднують фіксовані точки $A$ та $B$ ($l \geq AB$), знайдіть ту, яка разом з відрізком $AB$ обмежує плоску замкнену область із найбільшою площею.

*Розв'язання.* Нехай відстань $AB = 2a$. Виберемо за вісь $OX$ пряму, що проходить через задані точки $A$ та $B$, і помістимо початок координат посередині відрізка $AB$, тобто $A(-a, 0)$ та $B(a, 0)$ (див. рис. 2.8). Уважатимемо також, що шукана крива $\gamma$ знаходиться над віссю $OX$. Тоді задача зводиться до відшукання максимуму функціонала площі

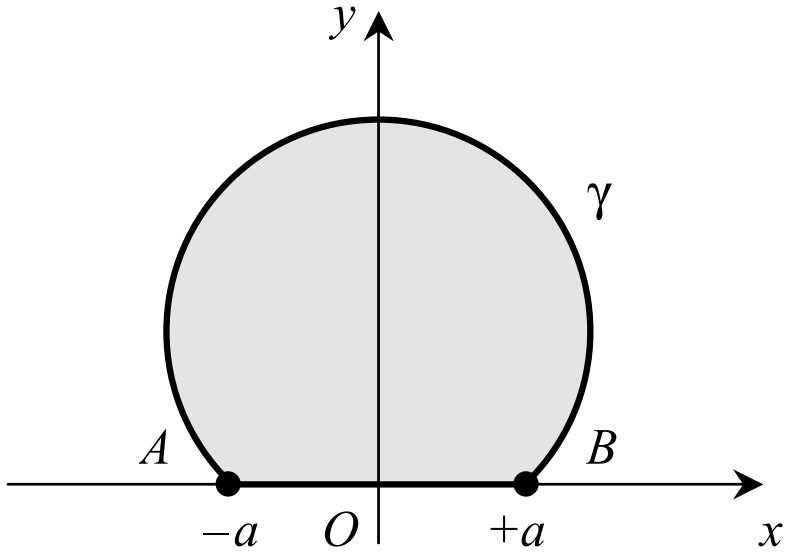

Рис. 2.8. Площа, обмежена відрізком $AB$ та кривою $\gamma$ із заданою довжиною

$$S[y]=\int_{-a}^{+a}dx\,y$$

за умови, що довжина кривої $\gamma$ залишається сталою:

$$L[y]=\int_{-a}^{+a}dx\sqrt{1+y'^2}=l.$$

Згідно з теоремою Ейлера, треба знайти екстремаль функціонала

$$S^*[y]=\int_{-a}^{+a}dx\left(y-\lambda\sqrt{1+y'^2}\right),$$

яка задовольняє крайові умови $y(-a)=y(a)=0$ та умову (2.67). Перший інтеграл відповідного рівняння Ейлера — Лагранжа має вигляд

$$y=C_1+\frac{\lambda}{\sqrt{1+y'^2}},$$

де $\lambda$ — невизначений множник Лагранжа, $C_1$ — стала інтегрування. Це рівняння легко інтегрується в параметричному вигляді за допомогою підстановки $y'=\text{tg}\,\varphi.$ Маємо:

$$y=C_1+\lambda\cos\varphi,\quad x=C_2-\lambda\sin\varphi,$$

де $C_2$ — нова стала інтегрування. Виключивши параметр $\varphi$ за допомогою тотожності $\sin^2\varphi+\cos^2\varphi=1,$ бачимо, що крива $\gamma$ є дугою кола

$$(y-C_1)^2+(x-C_2)^2=\lambda^2$$

з центром у точці $(C_2,C_1)$ і радіусом $R=|\lambda|$.

Крайові умови дають $C_1=\pm\sqrt{\lambda^2-a^2},$ $C_2=0.$ Для сталої $R>0$ з умови (2.67) дістаємо рівняння $2R\arcsin(a/R)=l,$ яке зводимо до безрозмірного вигляду $\arcsin u=\alpha u,$ де $u\equiv a/R,$ $\alpha\equiv l/(2a).$ Тривіальний розв'язок цього рівняння $u=0$ відповідає значенню $R=\infty.$ У цьому випадку крива $\gamma$ лежить на осі $OX,$ а відповідна площа набуває нульового (найменшого) значення. Нетривіальний розв'язок існує, якщо тангенс кута нахилу прямої $y=\alpha u$ є більшим за тангенс кута нахилу дотичної до функції $y=\arcsin u$ в нулі (див. рис. 2.9), тобто коли $\alpha>1,$ або $l>2a.$ При цьому функція, яка описує дугу $\gamma,$ є однозначною лише за умови $\alpha\le\pi/2,$ тобто $l\le\pi a,$ та має вигляд $y(x)=\sqrt{R^2-x^2}-\sqrt{R^2-a^2},$ $x\in[-a,a].$ Площа, обмежена нею та відрізком $AB,$ дорівнює

$$S_{l\le\pi a}(R)=\int_{-a}^{a}dx\left(\sqrt{R^2-x^2}-\sqrt{R^2-a^2}\right)=$$

$$=-a\sqrt{R^2-a^2}+R^2\arcsin\frac{a}{R}=\frac{Rl}{2}-a\sqrt{R^2-a^2}.$$

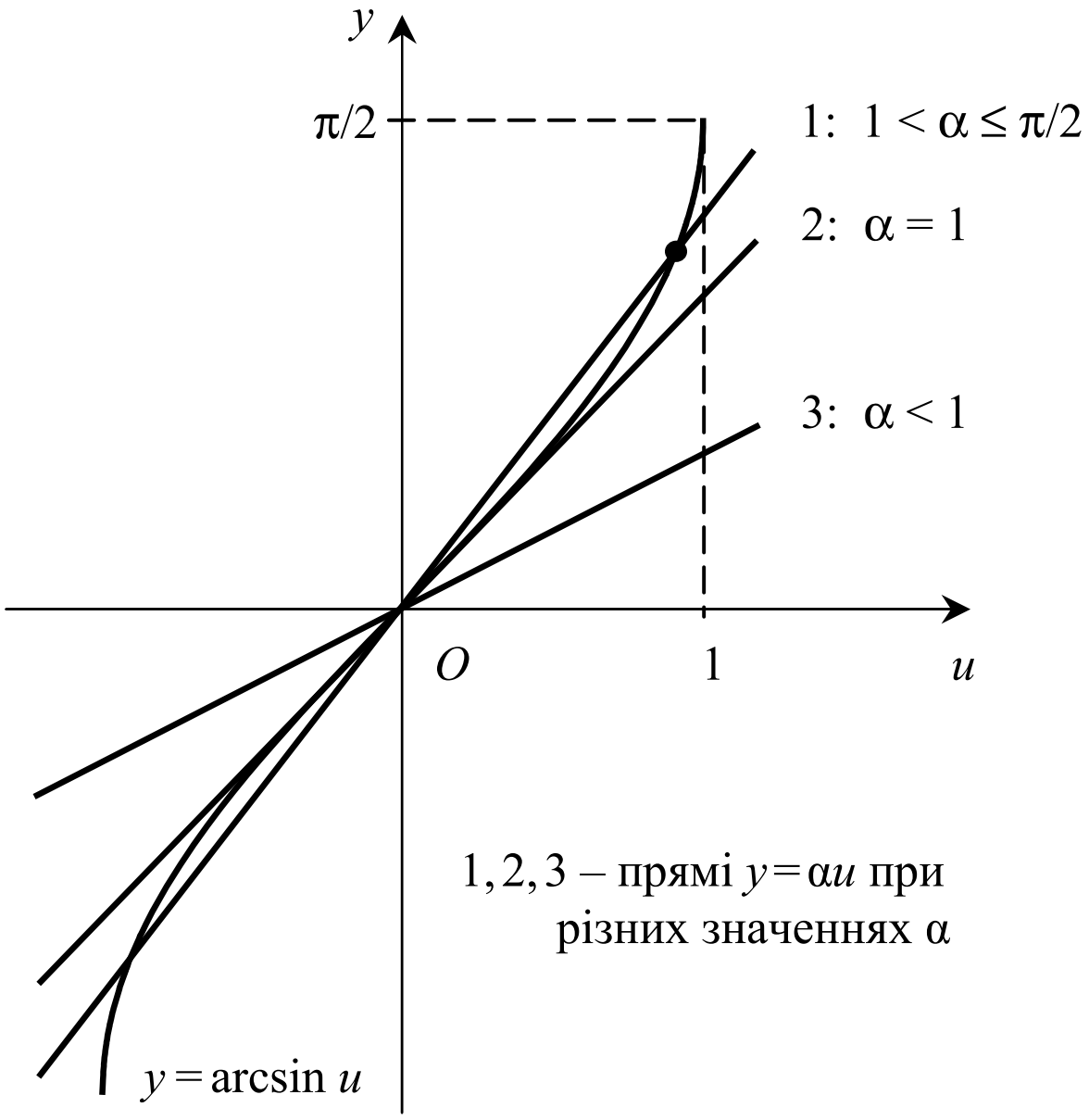

Рис. 2.9. Графічний аналіз рівняння arcsin $u = \alpha u$

Зокрема, при $l=\pi a$ радіус $R=a$ і площа $S_{l=\pi a}(R)=\pi a^2/2$.

Якщо ж $l>\pi a$, проекцією дуги $\gamma$ на вісь $OX$ є відрізок $[-R,R]$, а обмежена нею та відрізком $AB$ площа $S_{l>\pi a}(R)=\pi R^2-S_{l\le\pi a}(R)$.

Крайові умови жорсткого закріплення не є обов'язковими при формулюванні ізопериметричної задачі. Крайові умови інших типів також можливі.

**Завдання 2.4.3.** Покажіть, що гладка екстремаль функціонала

$$\Pi[X]=\frac{1}{2}\int_0^l\left[p(x)X'^2(x)+q(x)X^2(x)\right]dx+\frac{k_1}{2}X^2(0)+\frac{k_2}{2}X^2(l) \quad (2.68)$$

за умови, що інший функціонал

$$K[X]=\frac{1}{2}\int_0^l \rho(x)X^2(x)dx \qquad (2.69)$$

набуває сталого значення, є розв'язком *крайової задачі Штурма — Ліувілля*, що складається з однорідного диференціального рівняння

$$-\frac{d}{dx}\big(p(x)X'(x)\big)+q(x)X(x)=\lambda\rho(x)X(x),\ \ 0<x<l, \qquad (2.70)$$

та однорідних крайових умов

$$X'(0)-h_1X(0)=0,\ \ X'(l)+h_2X(l)=0,\ \ h_1\equiv k_1/p(0),\ \ h_2\equiv k_2/p(l). \qquad (2.71)$$

Узагальненням найпростішої ізопериметричної задачі (2.56)–(2.58) є задача про відшукання гладкої екстремалі функціонала

$$J[y_1,y_2,...,y_n]=\int_{x_1}^{x_2} dxF(x,y_1,y_2,...,y_n;y_1',y_2',...,y_n') \qquad (2.72)$$

за додаткової умови, що залишаються сталими значення $k$ інших функціоналів

$$L_j[y_1,y_2,...,y_n]=\int_{x_1}^{x_2} dx\, G_j(x,y_1,y_2,...,y_n;y_1',y_2',...,y_n')=L_j,\ \ j=1,2,...,k, \ (2.73)$$

та при фіксованих значеннях функцій $y_i(x)$ на кінцях відрізка $[x_1,x_2]$. Аналогічно до задачі (2.56)–(2.58), вона розв'язується шляхом уведення $k$ невизначених множників Лагранжа $\lambda_j$ та побудови нового функціонала

$$J^*[y_1,y_2,...,y_n]=J[y_1,y_2,...,y_n]-\sum_{j=1}^{k}\lambda_jL_j[y_1,y_2,...,y_n] \qquad (2.74)$$

з підінтегральною функцією

$$F^*=F-\sum_{j=1}^{k}\lambda_jG_j. \qquad (2.75)$$

Шукана екстремаль є розв'язком системи $n$ диференціальних рівнянь Ейлера — Лагранжа для функціонала $J^*[y_1,y_2,...,y_n]$,

$$\frac{\partial F^*}{\partial y_i}-\frac{d}{dx}\frac{\partial F^*}{\partial y_i'}=0,\ \ i=1,2,...,n, \qquad (2.76)$$

який задовольняє умови (2.73) та умови жорсткого закріплення.

Інший тип задач на умовний екстремум дістаємо, вимагаючи, щоб екстремаль функціонала (2.72) замість умови сталості функціоналів (2.73) задовольняла *в'язі*

$$f_j(x, y_1, y_2, ..., y_n; y_1', y_2', ..., y_n') = 0, \quad j = 1, 2, ..., m < n. \tag{2.77}$$

Ця задача називається *задачею Лагранжа*. Нагадаємо, що в'язь $f_j$ називається *голономною*, якщо вона не містить (або зводиться до вигляду, що не містить) похідні $y_i'(x), \ i = 1, 2, ..., n.$ У противному разі вона називається *неголономною*.

Прикладами задачі Лагранжа з голономною в'яззю є розглянуті вище задачі про геодезичну (див. завдання 2.3.7 та 2.3.8), яка з'єднує задані точки на певній поверхні

$$f(x, y, z) = 0. \tag{2.78}$$

Отже, у найпростішому варіанті задачу Лагранжа можна сформулювати так: знайти гладку екстремаль функціонала виду

$$J[y, z] = \int_{x_1}^{x_2} dx F(x, y, z; y', z'), \tag{2.79}$$

яка задовольняє голономну в'язь (2.78) та крайові умови жорсткого закріплення

$$y(x_1) = y_1, \ z(x_1) = z_1, \ y(x_2) = y_2, \ z(x_2) = z_2. \tag{2.80}$$

**Теорема 2.4.2.** Нехай функції $F(x, y, z; y', z')$ та $f(x, y, z)$ неперервні разом зі своїми частинними похідними відповідно до другого та першого порядків включно. Тоді гладка екстремаль функціонала (2.79) за умов (2.78), (2.80) задовольняє систему диференціальних рівнянь Ейлера — Лагранжа

$$\frac{\partial \tilde{F}}{\partial y} - \frac{d}{dx}\frac{\partial \tilde{F}}{\partial y'} = 0, \quad \frac{\partial \tilde{F}}{\partial z} - \frac{d}{dx}\frac{\partial \tilde{F}}{\partial z'} = 0, \tag{2.81}$$

де

$$\tilde{F}(x, y, z, \lambda; y', z') = F(x, y, z; y', z') - \lambda(x) f(x, y, z), \tag{2.82}$$

а $\lambda(x)$ — невизначена функція класу $C^1([x_1, x_2])$.

Сталі інтегрування і функція $\lambda(x)$ у загальному розв'язку системи (2.81) знаходяться за допомогою крайових умов (2.80) і в'язі (2.78).

*Доведення*. Як і у випадку ізопериметричної задачі, спробуємо задачу (2.78)–(2.80) на умовний екстремум функціонала (2.79) звести до задачі про звичайний екстремум функціонала більш загального виду з підінтегральною функцією (2.82):

$$\tilde{J}[y, z, \lambda] = \int_{x_1}^{x_2} dx \tilde{F}(x, y, z, \lambda; y', z'). \tag{2.83}$$

Якщо розглядати $\lambda(x)$ як ще одну функцію, від якої залежить функціонал $\tilde{J}[y,z,\lambda]$, то його екстремаль повинна задовольняти систему трьох диференціальних рівнянь Ейлера — Лагранжа, а саме, два рівняння (2.81) та рівняння

$$\frac{\partial \tilde{F}}{\partial \lambda}-\frac{d}{dx}\frac{\partial \tilde{F}}{\partial \lambda'}=0. \qquad (2.84)$$

А це є не що інше, як рівняння (2.78).

**Завдання 2.4.4.** Доведіть, що головна нормаль до геодезичної лінії поверхні збігається в кожній точці з нормаллю до поверхні. (Основна властивість геодезичної.)

*Розв'язання.* Потрібно проаналізувати задачу Лагранжа (2.78)–(2.80), де у випадку геодезичної лінії підінтегральна функція у функціоналі (2.79) має вигляд $F(y',z')=\sqrt{1+y'^2+z'^2}$. З рівнянь Ейлера — Лагранжа (2.81) для допоміжної функції

$$\tilde{F}=\sqrt{1+y'^2+z'^2}-\lambda(x)f(x,y,z)$$

випливає, що

$$\frac{d}{dx}\left(\frac{y'}{F}\right)+\lambda f_y=0,\ \ \frac{d}{dx}\left(\frac{z'}{F}\right)+\lambda f_z=0. \qquad (2.85)$$

Ще одне рівняння дістанемо, диференціюючи (2.78) за змінною $x$:

$$f_x+f_y y'+f_z z'=0.$$

Тепер помножимо обидві частини останнього рівняння на $\lambda$ та підставимо замість $\lambda f_y$ і $\lambda f_z$ їх значення з формул (2.85):

$$-y'\frac{d}{dx}\left(\frac{y'}{F}\right)-z'\frac{d}{dx}\left(\frac{z'}{F}\right)+\lambda f_x=0.$$

Розкриваючи похідні за $x$ та враховуючи співвідношення

$$F^2=1+y'^2+z'^2,\ \ F\frac{dF}{dx}=y'y''+z'z'',$$

знаходимо:

$$\frac{d}{dx}\left(\frac{1}{F}\right)+\lambda f_x=0. \qquad (2.86)$$

Вирази в дужках під знаком похідної дорівнюють напрямним косинусам дотичної до геодезичної, тому рівняння (2.85), (2.86) можна переписати так:

$$\frac{d\cos\alpha}{dx}+\lambda f_x=0,\quad \frac{d\cos\beta}{dx}+\lambda f_y=0,\quad \frac{d\cos\gamma}{dx}+\lambda f_z=0. \tag{2.87}$$

Користуючись формулою $dx/dl=\cos\alpha$, операцію диференціювання за змінною $x$ у формулах (2.87) замінимо операцією диференціювання за довжиною геодезичної: $d/dx=(\cos\alpha)^{-1}\,d/dl$. Отримуємо:

$$\frac{d\cos\alpha}{dl}=\mu f_x,\quad \frac{d\cos\beta}{dl}=\mu f_y,\quad \frac{d\cos\gamma}{dl}=\mu f_z, \tag{2.88}$$

де $\mu\equiv-\lambda\cos\alpha$.

Як відомо з диференціальної геометрії, ліві частини формул (2.88) пропорційні напрямним косинусам головної нормалі до кривої, а праві — напрямним косинусам нормалі до поверхні. Звідси й випливає, що вздовж геодезичної лінії головна нормаль до неї є одночасно й нормаллю до поверхні.

Загальна задача Лагранжа про відшукання екстремуму функціонала (2.72) за умови, що шукані функції $y_i(x)$ належать класу $C^1([x_1,x_2])$, задовольняють $m$ диференціальних співвідношень (неголономних в'язей) (2.77) та певні крайові умови кількістю $2n+m$, розв'язується аналогічно до попередньої.

Нехай функції в (2.77) визначені та мають неперервні частинні похідні другого порядку за всіма своїми аргументами, а матриця $\left\|\partial f_j/\partial y_i'\right\|$ має ранг $m$ в усіх точках $(x,y_1,y_2,...,y_n)$ деякої просторової області, коли похідні $y_1',y_2',...,y_n'$ пробігають довільні значеннях на кінцях інтервалу $[x_1,x_2]$. Тоді шуканою екстремаллю є інтегральна крива системи $(n+m)$ диференціальних рівнянь

$$\frac{\partial\tilde{F}}{\partial y_i}-\frac{d}{dx}\frac{\partial\tilde{F}}{\partial y_i'}=0,\quad i=1,2,...,n, \tag{2.89}$$

$$f_j=0,\ j=1,2,...,m<n, \tag{2.90}$$

де допоміжна функція

$$\tilde{F}=F-\sum_{j=1}^{m}\lambda_j(x)f_j, \tag{2.91}$$

а невизначені функції Лагранжа $\lambda_j(x)$ належать класу $C^1([x_1,x_2])$.

Зауважимо, що співвідношення (2.90) можна розглядати як рівняння Ейлера — Лагранжа, у яких функція (2.91) диференціюється за невідомими функціями $\lambda_j$.

**Завдання 2.4.5.** По якій замкненій кривій у горизонтальній площині повинен рухатися літак зі сталою за величиною власною швидкістю $v$, щоб за фіксований проміжок часу облетіти найбільшу площу? Під час руху літака дме вітер, швидкість якого відносно землі стала за напрямом та величиною $w < v$. (Задача Чаплигіна про максимальну площу обльоту.)

*Розв'язання*. Вивчатимемо рух літака відносно нерухомої системи відліку, пов'язаної із землею. Нехай рух відбувається в горизонтальній площині $XOY$, а вісь $OX$ напрямлена вздовж швидкості вітру $\mathbf{w}$ (рис. 2.10). Позначимо через $\alpha(t)$ кут між напрямом власної швидкості літака $\mathbf{v}$ та віссю $OX$, а через $x(t)$ і $y(t)$ — компоненти радіус-вектора центра мас літака $\mathbf{r}$ у момент часу $t$. Тоді швидкість літака відносно землі $\mathbf{u} = \mathbf{v} + \mathbf{w}$ має компоненти

$$\dot{x} = v\cos\alpha + w, \quad \dot{y} = v\sin\alpha. \tag{2.92}$$

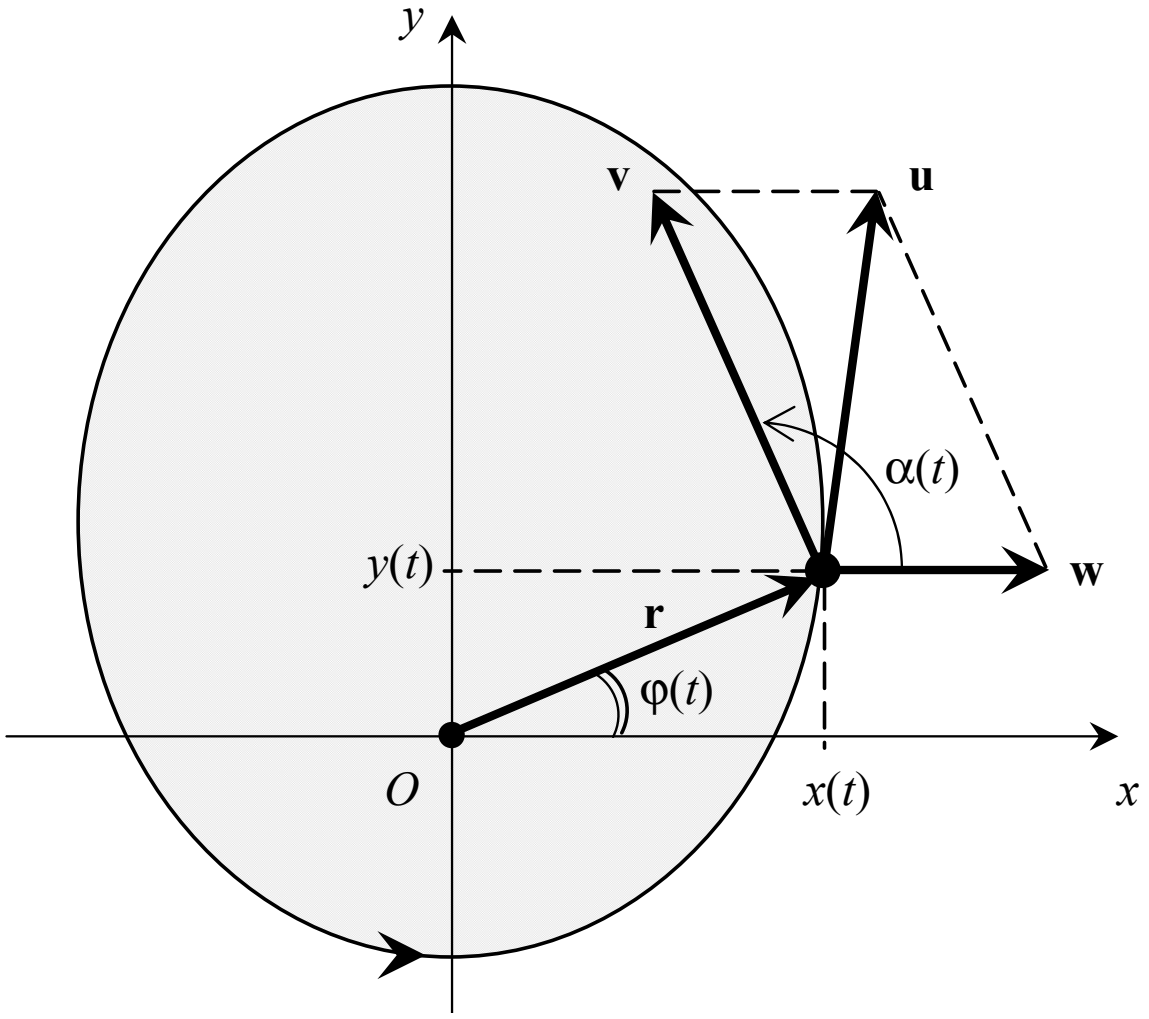

Рис. 2.10. Кінематика задачі Чаплигіна

Площа, яку описує літак у горизонтальній площині за час $dt$, дорівнює половині модуля (фактично половині $z$-компоненти) векторного добутку $[\mathbf{r}, \mathbf{u}\,dt]$. Тому площа, яку літак облітає за проміжок часу $T$, виражається інтегралом

$$S = \frac{1}{2}\int_0^T dt(x\dot{y} - y\dot{x}). \tag{2.93}$$

Задача, таким чином, зводиться до відшукання трьох функцій $\alpha(t)$, $x(t)$ і $y(t)$, що надають максимум функціоналу (2.93) та задовольняють дві неголономні в'язі (2.92).

Побудуємо систему рівнянь (2.89). Оскільки функція (2.91) у нашому випадку має вигляд

$$\tilde{F}=\frac{1}{2}(x\dot{y}-y\dot{x})-\lambda_1(t)(\dot{x}-v\cos\alpha-w)-\lambda_2(t)(\dot{y}-v\sin\alpha), \qquad (2.94)$$

дістаємо:

$$\dot{\lambda}_1+\dot{y}=0,\ \ \dot{\lambda}_2-\dot{x}=0,\ \ -\lambda_1\sin\alpha+\lambda_2\cos\alpha=0. \qquad (2.95)$$

З перших двох рівнянь (2.95)

$$\lambda_1=-y+C_1,\ \ \lambda_2=x+C_2. \qquad (2.96)$$

Паралельним перенесенням осей координат можна домогтися, щоб сталі $C_1$ та $C_2$ дорівнювали нулю. Тоді третє рівняння руху набирає вигляду

$$x\cos\alpha+y\sin\alpha=0. \qquad (2.97)$$

Разом із двома рівняннями в'язей (2.92) воно утворює замкнену систему для знаходження функцій $\alpha(t)$, $x(t)$ та $y(t)$.

Перейдемо до полярних координат

$$x=r\cos\varphi,\quad y=r\sin\varphi.$$

Тоді рівняння (2.97) зводиться до

$$\cos(\alpha-\varphi)=0,$$

звідки

$$\alpha=\varphi+\frac{\pi}{2}. \qquad (2.98)$$

З урахуванням цього співвідношення рівняння (2.92) набирають вигляду

$$\begin{aligned}\dot{r}\cos\varphi-r\sin\varphi\dot{\varphi}&=-v\sin\varphi+w,\\ \dot{r}\sin\varphi+r\cos\varphi\dot{\varphi}&=v\cos\varphi.\end{aligned} \qquad (2.99)$$

Помножимо перше з рівнянь (2.99) на $\cos\varphi$, друге — на $\sin\varphi$, та додамо їх ліві та праві частини. Знаходимо:

$$\frac{dr}{dt}=w\cos\varphi=\frac{w}{v}\frac{dy}{dt}.$$

Обидві частини цього рівняння легко інтегруються:

$$r=\frac{w}{v}y+C=\frac{w}{v}r\sin\varphi+C,$$

де $C$ — стала інтегрування. Звідси

$$r = \frac{C}{1 - \frac{w}{v}\sin\varphi} = \frac{C}{1 + \frac{w}{v}\cos\alpha}. \tag{2.100}$$

Бачимо, що траєкторія літака з максимальною площею обльоту — це конічний переріз із фокусом у початку координат і ексцентриситетом $\varepsilon = w/v$. Оскільки швидкість вітру вважається меншою за швидкість літака, то $\varepsilon < 1$, тобто цей переріз є еліпсом з великою віссю вздовж осі $OY$. Рухаючись по такому еліпсу, пілот повинен витримувати кут (2.98) для напряму власної швидкості літака відносно землі.

## 2.5. ФУНКЦІОНАЛИ ВІД ФУНКЦІЇ БАГАТЬОХ ЗМІННИХ

У цьому підрозділі вивчається основна задача варіаційного числення для функціоналів, що залежать від функцій кількох змінних та їх частинних похідних. Найпростіший представник такого типу функціоналів залежить від функції двох змінних та її перших похідних:

$$J[z] = \iint_D dxdyF\left(x, y, z(x,y), \frac{\partial z(x,y)}{\partial x}, \frac{\partial z(x,y)}{\partial y}\right). \tag{2.101}$$

Уважатимемо, що допустимі функції $z = z(x,y)$ визначені в деякій області $D$ площини $XOY$ з кусково-гладкою межею Г, неперервні разом зі своїми першими частинними похідними, а множина їх значень для точок $(x,y) \in D$ утворює деяку просторову область $G$ (рис. 2.11).

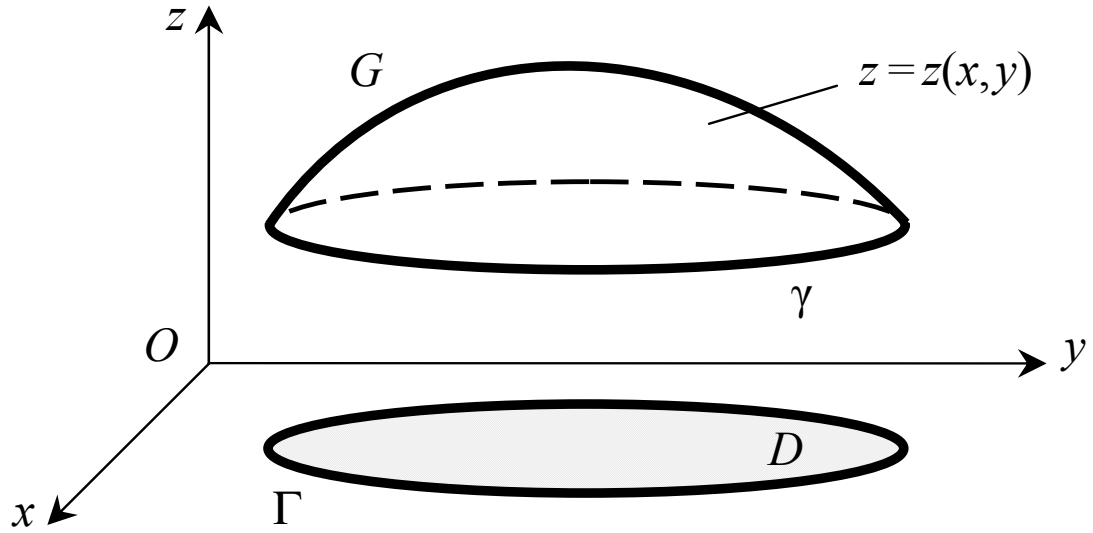

Рис. 2.11. Поверхня у тривимірному просторі, задана функцією $z = z(x,y)$. Крива $\gamma$ — край поверхні

**Теорема 2.5.1.** Нехай при довільних скінченних значеннях $z_x \equiv \partial z/\partial x$ та $z_y \equiv \partial z/\partial y$ підінтегральна функція $F(x,y,z,z_x,z_y)$ у функціоналі (2.101) є в області $G$ двічі неперервно диференційовною за своїми аргументами. Якщо функція $z_0(x,y)$, яка надає екстремум функціоналу (2.101) на множині допустимих функцій з однаковими фіксованими значеннями $z(x,y)\big|_\Gamma$ на контурі Г, має неперервні другі похідні, то вона задовольняє *рівняння Ейлера — Остроградського*

$$\frac{\partial F}{\partial z} - \frac{\partial}{\partial x}\frac{\partial F}{\partial z_x} - \frac{\partial}{\partial y}\frac{\partial F}{\partial z_y} = 0. \qquad (2.102)$$

*Доведення.* Розглянемо множину допустимих поверхонь

$$z(x,y) = z_0(x,y) + \delta z(x,y) = z_0(x,y) + \alpha h(x,y),$$

для яких на межі Г $\delta z(x,y)\big|_\Gamma = h(x,y)\big|_\Gamma = 0$. Варіацію функціонала (2.101)

$$\delta J[z] = \alpha \iint\limits_D dxdy \left( \frac{\partial F}{\partial z} h + \frac{\partial F}{\partial z_x}\frac{\partial h}{\partial x} + \frac{\partial F}{\partial z_y}\frac{\partial h}{\partial y} \right)$$

за допомогою двовимірних векторів $\mathbf{A} = \left( \frac{\partial F}{\partial z_x}, \frac{\partial F}{\partial z_y} \right)$, $\nabla h = \left( \frac{\partial h}{\partial x}, \frac{\partial h}{\partial y} \right)$ та тотожності

$$\mathbf{A}\cdot\nabla h = \nabla\left(h\mathbf{A}\right) - h\nabla\mathbf{A}$$

подамо у вигляді

$$\delta J[z] = \alpha \iint\limits_D dxdy \left( \frac{\partial F}{\partial z} h + \mathbf{A}\cdot\nabla h \right) = \alpha \iint\limits_D dxdy \left\{ \left( \frac{\partial F}{\partial z} - \nabla\mathbf{A} \right) h + \nabla\left(h\mathbf{A}\right) \right\}.$$

Застосовуючи до останнього члена в $\delta J[z]$ двовимірний варіант теореми Остроградського — Гаусса

$$\iint\limits_D dxdy\, \nabla\left(h\mathbf{A}\right) = \oint\limits_\Gamma dl\, hA_n,$$

де $A_n$ — перпендикулярна до контура Г компонента вектора $\mathbf{A}$, дістаємо:

$$\delta J[z] = \iint\limits_D dxdy \left( \frac{\partial F}{\partial z} - \nabla\mathbf{A} \right) \delta z(x,y) + \oint\limits_\Gamma dl \delta z(x,y) A_n. \qquad (2.103)$$

Другий доданок у формулі (2.103) дорівнює нулю, оскільки значення допустимих поверхонь на контурі Г фіксовані. Рівність $\delta J[z_0] = 0$ справджуватиметься для довільних значень варіації $\delta z(x,y)\big|_D$ всере-дині області $D$ лише тоді, коли скрізь в області $D$ функція $z_0(x, y)$ задовольнятиме рівняння

$$\frac{\partial F}{\partial z} - \nabla \mathbf{A} = 0,$$

тобто рівняння (2.102).

**Завдання 2.5.1.** Випишіть рівняння Ейлера — Остроградського для задачі про знаходження мінімальної площі, натягненої на контур $\Gamma$.

*Вказівка.* Треба знайти мінімум функціонала площі поверхні, що натягнена на контур $\Gamma$ і проектується на область $D$:

$$J[z] = \iint_D dxdy \sqrt{1 + \left(\frac{\partial z}{\partial x}\right)^2 + \left(\frac{\partial z}{\partial y}\right)^2}.$$

Відповідне рівняння Ейлера — Остроградського

$$\frac{\partial}{\partial x} \frac{z_x}{\sqrt{1 + z_x^2 + z_y^2}} + \frac{\partial}{\partial y} \frac{z_y}{\sqrt{1 + z_x^2 + z_y^2}} = 0$$

означає, що середня кривина шуканої поверхні для всіх точок $D$ дорівнює нулю.

**Завдання 2.5.2.** Узагальніть рівняння Ейлера — Остроградського на випадок функціонала виду

$$J[z] = \int\int ... \int dx_1 dx_2 ... \ dx_n F(x_1, x_2, ..., x_n; z, z_{x_1}, z_{x_2}, ..., z_{x_n}),$$

де $z_{x_i} \equiv \partial z(x_1, x_2, ..., x_n) / \partial x_i$.

*Відповідь*:

$$\frac{\partial F}{\partial z} - \sum_{i=1}^{n} \frac{\partial}{\partial x_i} \frac{\partial F}{\partial z_{x_i}} = 0. \qquad (2.104)$$

**Завдання 2.5.3.** Покажіть, що потенціал електростатичного поля у вакуумі задовольняє рівняння Лапласа

$$\Delta \varphi(\mathbf{r}) = 0.$$

*Вказівки.* Об'ємна густина енергії електростатичного поля у вакуумі

$$w(\mathbf{r}) = \frac{1}{8\pi} \mathbf{E}^2(\mathbf{r}) = \frac{1}{8\pi} \mathbf{E}^2(x, y, z).$$

Узявши до уваги зв'язок між напруженістю поля $\mathbf{E}(\mathbf{r})$ та його потенціалом $\varphi(\mathbf{r})$,

$$\mathbf{E}(\mathbf{r}) = -\nabla\varphi(\mathbf{r}),$$

виразіть повну енергію електростатичного поля

$$W = \int_V d\mathbf{r}\, w(\mathbf{r}) = \frac{1}{8\pi}\int_V d\mathbf{r}\, \mathbf{E}^2(\mathbf{r}) = \frac{1}{8\pi}\iiint_V dxdydz\, \mathbf{E}^2(x,y,z)$$

через потенціал $\varphi(\mathbf{r})$ та скористайтеся теоремою Томсона (див. підрозділ 1.4) для енергії електростатичного поля, створюваного зарядженими провідниками.

Оскільки $W$ належить до класу функціоналів, що залежать від функції кількох змінних та її перших похідних, його екстремальна поверхня має задовольняти рівняння Ейлера — Остроградського (2.104):

$$\frac{\partial F}{\partial\varphi} - \frac{\partial}{\partial x}\frac{\partial F}{\partial\varphi_x} - \frac{\partial}{\partial y}\frac{\partial F}{\partial\varphi_y} - \frac{\partial}{\partial z}\frac{\partial F}{\partial\varphi_z} = 0,$$

де в нашому випадку

$$F = \frac{1}{8\pi}\left(\nabla\varphi\right)^2 = \frac{1}{8\pi}\left(\varphi_x^2 + \varphi_y^2 + \varphi_z^2\right).$$

**Завдання 2.5.4.** Інтегралом Діріхле від функції $z = z(x_1, x_2, ..., x_n)$ по області $D$ $n$-вимірного простору називається інтеграл

$$J[z] = \int_D ... \int dx_1 ...\, dx_n \sum_{i=1}^{n} (z_{x_i})^2.$$

Уважаючи, що функція $z$ набуває заданих значень на границі області $D$, знайдіть умови, за яких досягається мінімум інтеграла Діріхле.

*Відповідь.* Функція $z$ має задовольняти $n$-вимірне рівняння Лапласа

$$\Delta z \equiv \frac{\partial^2 z}{\partial x_1^2} + \frac{\partial^2 z}{\partial x_2^2} + ... + \frac{\partial^2 z}{\partial x_n^2} = 0.$$

**Завдання 2.5.5.** Те саме для функціонала

$$J[z] = \int_D ... \int dx_1 ...\, dx_n \left[\sum_{i=1}^{n} (z_{x_i})^2 + 2zf(x_1, ..., x_n)\right].$$

*Відповідь.* Функція $z$ повинна задовольняти $n$-вимірне рівняння Пуассона

$$\Delta z = f.$$

**Завдання 2.5.6.** Нехай екстремум функціонала (2.101) шукається серед поверхонь, краї яких можуть вільно ковзати в напрямі, перпендикулярному до площини $XOY$, залишаючись при цьому над контуром $\Gamma$ (рис. 2.12). Уважаючи решту умов теореми 2.5.1 на підінтегральну функцію $F(x,y,z,z_x,z_y)$ та на допустимі функції виконаними, знайдіть необхідні умови, які тепер повинна задовольняти екстремальна поверхня $z_0(x,y)$ функціонала (2.101).

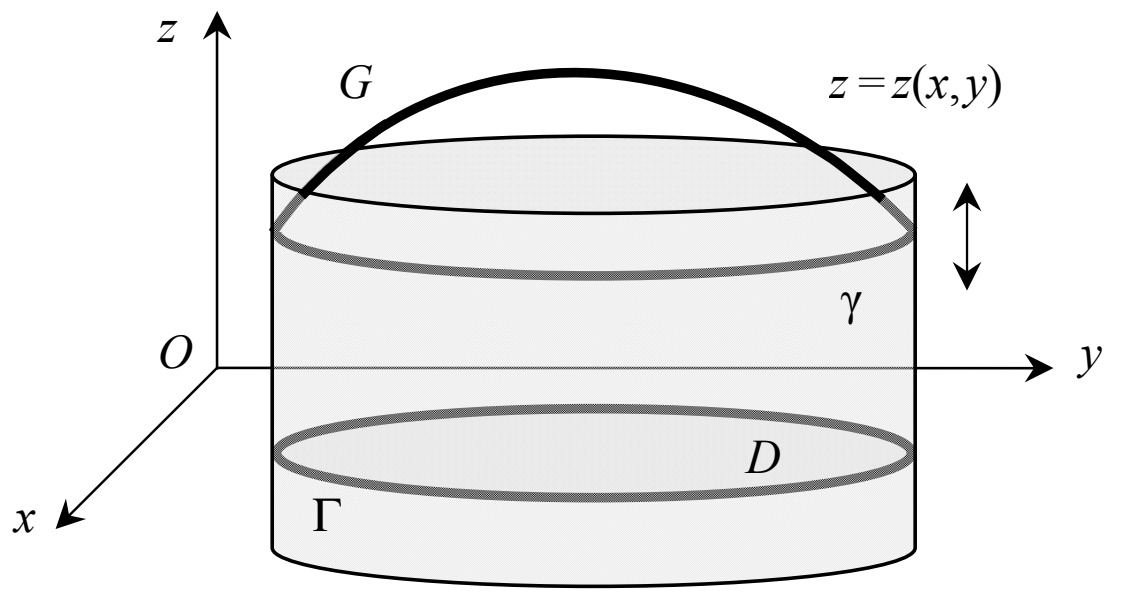

Рис. 2.12. Допустимі поверхні, краї яких вільно ковзають у вертикальному напрямі по заданій поверхні

*Вказівка*. Проаналізуйте умови, за яких справджується рівність $\delta J[z_0]=0$ для варіації (2.103) при довільних значеннях варіацій $\delta z(x,y)\big|_D$ та $\delta z(x,y)\big|_\Gamma$.

*Відповідь.* Екстремальна поверхня $z_0(x,y)$ задовольняє рівняння Ейлера — Остроградського (2.102) та крайову умову

$$A_n\big|_\Gamma = 0, \tag{2.105}$$

де $\mathbf{A}=\left(\dfrac{\partial F}{\partial z_x},\dfrac{\partial F}{\partial z_y}\right)$.

**Завдання 2.5.7.** Нехай задано функціонал типу (2.101), але з додатковим внеском на межі $\Gamma$:

$$J[z]=\iint\limits_D dxdyF\big(x,y,z(x,y),z_x(x,y),z_y(x,y)\big)+\oint\limits_\Gamma dlf\big(z(l)\big), \tag{2.106}$$

де гладка функція $f\big(z(l)\big)$, $z(l)\equiv z(x,y)\big|_\Gamma$, залежить лише від значень допустимих функцій $z(x,y)$ на контурі $\Gamma$. Уважаючи решту умов теореми 2.5.1 на підінтегральну функцію $F(x,y,z,z_x,z_y)$ та допустимі функції виконаними, знайдіть необхідні умови, які повинна задовольняти екстремальна поверхня $z_0(x,y)$ функціонала (2.106).

*Відповідь.* Екстремальна поверхня задовольняє рівняння Ейлера — Остроградського (2.102) та крайову умову

$$\left.\left(A_n + \frac{\partial f}{\partial z}\right)\right|_{\Gamma} = 0. \qquad (2.107)$$

**Завдання 2.5.8.** Покажіть, що екстремальна поверхня функціонала

$$J[z] = \iint_D dxdyF\left(x, y, z, \frac{\partial z}{\partial x}, \frac{\partial z}{\partial y}, \frac{\partial^2 z}{\partial x^2}, \frac{\partial^2 z}{\partial x \partial y}, \frac{\partial^2 z}{\partial y^2}\right)$$

задовольняє рівняння

$$\frac{\partial F}{\partial z} - \frac{\partial}{\partial x}\frac{\partial F}{\partial z_x} - \frac{\partial}{\partial y}\frac{\partial F}{\partial z_y} + \frac{\partial^2}{\partial x^2}\frac{\partial F}{\partial z_{xx}} + \frac{\partial^2}{\partial x \partial y}\frac{\partial F}{\partial z_{xy}} + \frac{\partial^2}{\partial y^2}\frac{\partial F}{\partial z_{yy}} = 0.$$

**Завдання 2.5.9.** Покажіть, що екстремальна поверхня функціонала

$$J[z] = \iint_D dxdy\left[z_{xx}^{\,2} + z_{yy}^{\,2} + 2z_{xy}^{\,2} - 2zf(x, y)\right]$$

задовольняє неоднорідне бігармонічне рівняння

$$\Delta\Delta z = f(x, y).$$

## *КОНТРОЛЬНІ ПИТАННЯ ДО РОЗДІЛУ 2*

1. *Які крайові умови задовольняє екстремаль класу $C^1([x_1, x_2])$ функціонала найпростішого типу, якщо її лівий кінець виходить із точки $(x_1, y_1)$, а правий вільно ковзає по прямій $x = x_2$?*
2. *Які крайові умови задовольняє гладка екстремаль функціонала $\int_\Gamma F(x, y, y')dx$, якщо її лівий кінець виходить із точки $(x_1, y_1)$, а правий вільно ковзає по кривій $y = \varphi(x)$? До якої зводиться умова на правому кінці, якщо $F(x, y, y') = f(x, y)\sqrt{1 + y'^2}$?*
3. *Яке рівняння повинна задовольняти екстремаль $y_0(x)$ функціонала $J[y] = \int_{x_1}^{x_2} F(x, y, y', y'')dx$ у класі кривих $C^2([x_1, x_2])$ з кінцями, закріпленими під фіксованими кутами? Функція $F$ уважається двічі диференційовною за своїми аргументами.*
4. *Як відрізняються крайові умови для рівняння Ейлера — Пуассона в задачах про екстремальні криві, кінці яких можуть або вільно ковза-*

*ти вздовж вертикальних прямих під фіксованими кутами, або перетинатися з ними у фіксованих точках під довільними кутами до осі абсцис?*

5. *Як узагальнюється необхідна умова екстремальності кривої у випадку функціоналів від гладких просторових кривих, що з'єднують дві фіксовані точки у тривимірному просторі?*
6. *Як формулюється найпростіший варіант ізопериметричної задачі для функціонала* $J[y]=\int_{x_1}^{x_2} F(x,y,y')dx$ *у класі гладких кривих із фіксованими кінцями? Користуючись загальним алгоритмом відшукання екстремумів функціоналів такого типу, як побудувати алгоритм розв'язування відповідної ізопериметричної задачі?*
7. *Які умови задовольняє двічі диференційовна функція* $z(x,y)$, *яка надає екстремум функціоналу*

$$J[z]=\iint_D F\left(x,y,z(x,y),\frac{\partial z(x,y)}{\partial x},\frac{\partial z(x,y)}{\partial y}\right)dxdy$$

*у класі функцій, що визначені й диференційовні в області* $D$ *площини* $XOY$ *та набувають заданих значень на її кусково-гладкій межі* $\Gamma$*?*

8. *Яку крайову умову на межі* $\Gamma$ *області* $D$ *площини* $XOY$ *має задовольняти функція* $z(x,y)$, *що реалізує екстремум функціонала*

$$J[z]=\iint_D F\left(x,y,z(x,y),\frac{\partial z(x,y)}{\partial x},\frac{\partial z(x,y)}{\partial y}\right)dxdy+\oint_\Gamma f\big(z(x,y)\big)dl,$$

*де* $f(z)$ *— гладка функція?*

# Розділ 3

# МАЛІ КОЛИВАННЯ СИСТЕМ ІЗ РОЗПОДІЛЕНИМИ ПАРАМЕТРАМИ

## 3.1. РІВНЯННЯ КОЛИВАНЬ ТА КРАЙОВІ УМОВИ

Згідно з принципом найменшої дії, механічний рух системи дискретних точок з відомою функцією Лагранжа $L$ при наявності ідеальних в'язей можна описати за допомогою рівнянь Ейлера — Лагранжа, які є наслідком екстремальності функціонала дії для цієї системи (див. підрозділ 1.4). Для системи з неперервним розподілом маси, що заповнює область з об'ємом $V$, виникає потреба перейти до об'ємної густини (*лагранжіана*) $\mathcal{L}$ функції Лагранжа, визначеної співвідношенням

$$L = \int\limits_V dV \mathcal{L}. \tag{3.1}$$

Дія для такої системи набирає вигляду

$$S = \int\limits_{t_1}^{t_2} dt \int\limits_V dV \mathcal{L}, \tag{3.2}$$

тобто є функціоналом, що залежить від функції кількох змінних. Вимагаючи далі, щоб для істинного руху він набував екстремального значення, дістаємо рівняння руху у вигляді рівняння Ейлера — Остроградського для лагранжіана $\mathcal{L}$.

Почнемо розгляд таких рівнянь із випадку рівняння малих вільних поперечних коливань струни (пружної гнучкої нитки).

Нехай у стані рівноваги струна збігається з відрізком $[0,l]$ осі $OX$, а її коливання відбуваються у вертикальній площині $XOU$ та характеризуються деякою функцією $u(x,t)$, яка має зміст поперечного відхилення від стану рівноваги точки $x$ струни в момент часу $t$ (див. рис. 3.1). Оскільки струна не має розривів, функція $u(x,t)$ неперервна.

Частинна похідна $u_t(x,t) \equiv \partial u(x,t)/\partial t$ від зміщення за часом має зміст поперечної швидкості точки $x$ струни в момент часу $t$. Оскільки струна не рветься, її сусідні точки рухаються з близькими швидкостями, тобто функція $u_t(x,t)$ теж неперервна.

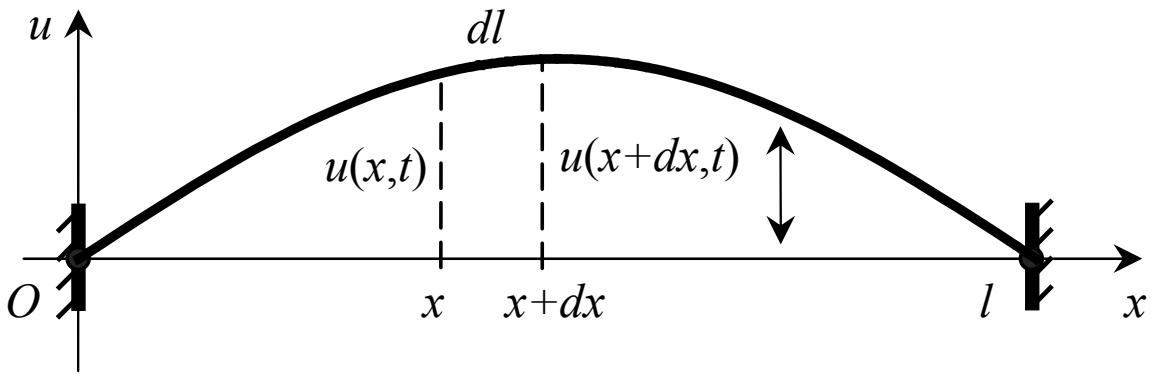

Рис. 3.1. Миттєві поперечні зміщення точок струни

Частинна похідна $u_x(x,t) \equiv \partial u(x,t)/\partial x$ від зміщення за координатою пов'язана з деформацією струни: ділянка струни $(x, x+dx)$ має в деформованому стані, з точністю до нескінченно малих більш високого порядку, довжину $dl = \sqrt{1+u_x^2(x,t)}\,dx$. Функція $u_x(x,t)$ також є неперервною, бо профіль струни внаслідок її гнучкості не має гострих кутів.

За означенням, коливання вважаються *малими*, якщо виконується співвідношення

$$u_x^2(x,t) << 1. \tag{3.3}$$

Довжина струни $l'$ зазнає при малих коливаннях незначних змін, залишаючись майже рівною довжині $l$ недеформованої струни:

$$l' = \int_0^l dx\sqrt{1+u_x^2(x,t)} \simeq \int_0^l dx = l.$$

Тому при малих коливаннях сильно натягненої струни можна знехтувати змінами сили її натягу внаслідок деформації в порівняння з силою натягу $T_0$ струни в стані спокою.

Виведемо рівняння малих поперечних коливань струни за допомогою принципу найменшої дії. Функціонал дії для струни записується у вигляді

$$S = \int_{t_1}^{t_2} dt\left[K(t) - \Pi(t)\right] = \int_{t_1}^{t_2} dt \int_0^l dx\mathcal{L}, \tag{3.4}$$

де $K(t)$, $\Pi(t)$ та $\mathcal{L}$ — відповідно кінетична енергія, потенціальна енергії та лагранжіан струни, обчислені в наближенні (3.3).

Нехай $\rho(x)$ — погонна густина струни в точці $x$. Тоді маса елементарної ділянки $(x, x+dx)$ струни $dm(x) = \rho(x)dx$, а кінетична енергія цієї ділянки в момент часу $t$

$$dK(x,t) = \frac{1}{2}dm(x)u_t^2(x,t) = \frac{1}{2}\rho(x)u_t^2(x,t)\,dx.$$

Кінетична енергія всієї струни

$$K(t)=\frac{1}{2}\int_0^l dx\rho(x)u_t^2(x,t).$$

Потенціальна енергія $d\Pi(x,t)$, яку запасає ділянка $(x,x+dx)$ у процесі коливань, дорівнює роботі $dA$, яку треба виконати проти сили $T_0$, щоб розтягнути цю ділянку на величину

$$\delta l=dl-dx=\left(\sqrt{1+u_x^2(x,t)}-1\right)dx\simeq$$

$$\simeq\left(1+\frac{1}{2}u_x^2(x,t)-\frac{1}{8}u_x^4(x,t)+\ldots-1\right)dx\simeq\frac{1}{2}u_x^2(x,t)dx.$$

(Прокоментуйте ту обставину, що довжина струни при малих коливаннях уважається незмінною, $l'\simeq l$, тоді як при обчисленні потенціальної енергії ми вважаємо, що $l'-l\neq 0$). Маємо:

$$d\Pi(x,t)=dA=T_0\delta l\simeq\frac{1}{2}T_0u_x^2(x,t)\,dx,$$

$$\Pi(t)=\frac{1}{2}\int_0^l dxT_0u_x^2(x,t).$$

Отже, функція Лагранжа струни має вигляд

$$L=\frac{1}{2}\int_0^l dx\left[\rho(x)u_t^2(x,t)-T_0u_x^2(x,t)\right],\qquad(3.5)$$

а функціонал дії струни —

$$S=\frac{1}{2}\int_{t_1}^{t_2}dt\int_0^l dx\left[\rho(x)u_t^2(x,t)-T_0u_x^2(x,t)\right].\qquad(3.6)$$

Задача, таким чином, зводиться до відшукання екстремалей функціонала (3.6), що залежить від функції двох змінних та її перших частинних похідних: $\mathcal{L}=\mathcal{L}\left(u(x,t),u_x(x,t),u_t(x,t)\right)$. Варіювання здійснюється за функцією $u(x,t)$, значення якої в точках $t=t_1$, $t=t_2$ (оскільки вивчається рух струни між двома фіксованими моментами часу) та в точках $x=0$, $x=l$ (оскільки кінці струни жорстко закріплено) фіксовані. Іншими словами, $\delta u(0,t)=\delta u(l,t)=\delta u(x,t_1)=\delta u(x,t_2)=0$.

Згідно з теоремою 2.5.1, така екстремаль обов'язково задовольняє рівняння

$$\frac{\partial\mathcal{L}}{\partial u}-\frac{\partial}{\partial t}\frac{\partial\mathcal{L}}{\partial u_t}-\frac{\partial}{\partial x}\frac{\partial\mathcal{L}}{\partial u_x}=0.$$

У нашому випадку

$$\frac{\partial \mathcal{L}}{\partial u}=0,\quad \frac{\partial \mathcal{L}}{\partial u_t}=\rho(x)u_t(x,t),\quad \frac{\partial \mathcal{L}}{\partial u_x}=-T_0u_x(x,t),$$

і, отже, *рівняння малих вільних поперечних коливань струни* має вигляд

$$\rho(x)u_{tt}(x,t)=\frac{\partial}{\partial x}\big(T_0u_x(x,t)\big). \tag{3.7}$$

Для однорідної струни густина $\rho(x)$ є сталою, тому рівняння (3.7) зводиться до диференціального рівняння в частинних похідних другого порядку зі сталими коефіцієнтами:

$$u_{tt}(x,t)=a^2u_{xx}(x,t), \tag{3.8}$$

де величина $a^2\equiv T_0/\rho$ має розмірність квадрата швидкості.

**Завдання 3.1.1.** Виведіть рівняння малих поперечних коливань струни, якщо в процесі коливань на неї ще діє зовнішня сила, перпендикулярна до рівноважного профілю струни. Погонна густина сили (тобто сила, яка діє на одиницю довжини струни) дорівнює $F(x,t)$.

*Вказівка*. Узагальнена сила, яка діє на матеріальну точку, дорівнює частинній похідній функції Лагранжа точки за узагальненою координатою точки: $F=\partial L(q,\dot q,t)/\partial q$ (див. підрозділ 1.4). У нашому випадку на елементарну ділянку $(x,x+dx)$ струни діє сила $F(x,t)dx=\partial\Delta L_{\text{ext}}(u,u_x,u_t,t)/\partial u$, де $\Delta L_{\text{ext}}$ — внесок цієї ділянки у функцію Лагранжа струни, зумовлений зовнішньою силою:

$$\Delta L_{\text{ext}}=\int_x^{x+dx}\mathcal{L}_{\text{ext}}dx\approx\mathcal{L}_{\text{ext}}dx.$$

Дістаємо: $F(x,t)=\partial\mathcal{L}_{\text{ext}}(u,u_x,u_t,t)/\partial u$. Узявши, зокрема, лагранжіан $\mathcal{L}_{\text{ext}}(u,u_x,u_t,t)=F(x,t)u(x,t)$, наявність зовнішньої сили можемо врахувати за допомогою додаткового внеску

$$L_{\text{ext}}=\int_0^l dx\,F(x,t)u(x,t)$$

у функцію Лагранжа (3.5) (або $\Pi_{\text{ext}}=-L_{\text{ext}}$ у потенціальну енергію, якщо $F(x,t)$ — потенціальна сила).

*Відповідь*:

$$\rho(x)u_{tt}(x,t)=\frac{\partial}{\partial x}\big(T_0u_x(x,t)\big)+F(x,t). \tag{3.9}$$

**Завдання 3.1.2.** Виведіть *рівняння малих поперечних коливань мембрани* із закріпленим краєм при наявності поперечних потенціальних зовнішніх сил.

*Вказівки.* Нехай рівноважний профіль мембрани утворює у площині $XOY$ деяку область $D$, відхилення мембрани від стану рівноваги описується функцією $u = u(x,y,t)$, і на одиницю площі мембрани діє сила $F(x,y,t)$, перпендикулярна до площини $XOY$. Крім того, нехай функції $\rho(x,y)$ і $k(x,y)$ характеризують відповідно поверхневу густину та натяг мембрани. Тоді її кінетична енергія, потенціальна енергія пружної деформації та потенціальна енергія в полі зовнішніх сил дорівнюють:

$$K(t) = \frac{1}{2}\iint\limits_D dxdy\rho(x,y)u_t^2(x,y,t),$$

$$\Pi(t) = \iint\limits_D dxdy\,k(x,y)\left(\sqrt{1+u_x^2(x,y,t)+u_y^2(x,y,t)}-1\right) \approx$$

$$\approx \frac{1}{2}\iint\limits_D dxdy\,k(x,y)\left(u_x^2(x,y,t)+u_y^2(x,y,t)\right),$$

$$\Pi_{\text{ext}}(t) = -\iint\limits_D dxdyF(x,y,t)u(x,y,t).$$

*Відповідь*:

$$\rho u_{tt} = \frac{\partial}{\partial x}\left(ku_x\right) + \frac{\partial}{\partial y}\left(ku_y\right) + F. \qquad (3.10)$$

**Завдання 3.1.3.** Виведіть *рівняння малих поздовжніх коливань пружного стержня* та крайові умови на його кінцях, якщо вони: а) закріплені пружно, тобто на кожний кінець діє з боку кріплення поздовжня сила, пропорційна зміщенню та напрямлена протилежно до нього; б) закріплені жорстко; в) вільні, тобто не закріплені. Покажіть, що б) та в) є граничними випадками а).

*Розв'язання.* а) Скористаємося принципом найменшої дії. Функціонал дії для самого стержня записується у вигляді (3.4), де лагранжіан залежить від поздовжнього зміщення $u(x,t)$ точок стержня та його перших похідних: $\mathcal{L} = \mathcal{L}(u,u_x,u_t)$. Функція $u(x,t)$ описує відхилення точок стержня в момент часу $t$ від тих положень $x$, які вони мали в недеформованому стержні (див. рис. 3.2).

Нехай $\rho(x)$ та $S(x)$ — відповідно густина та площа поперечного перерізу стержня в точці $x$. Знайдемо кінетичну енергію $dK(x,t)$ елементарної ділянки $(x, x+dx)$ стержня, обмеженої двома близькими поперечними перерізами $x$ та $x+dx$, а також потенціальну енергію

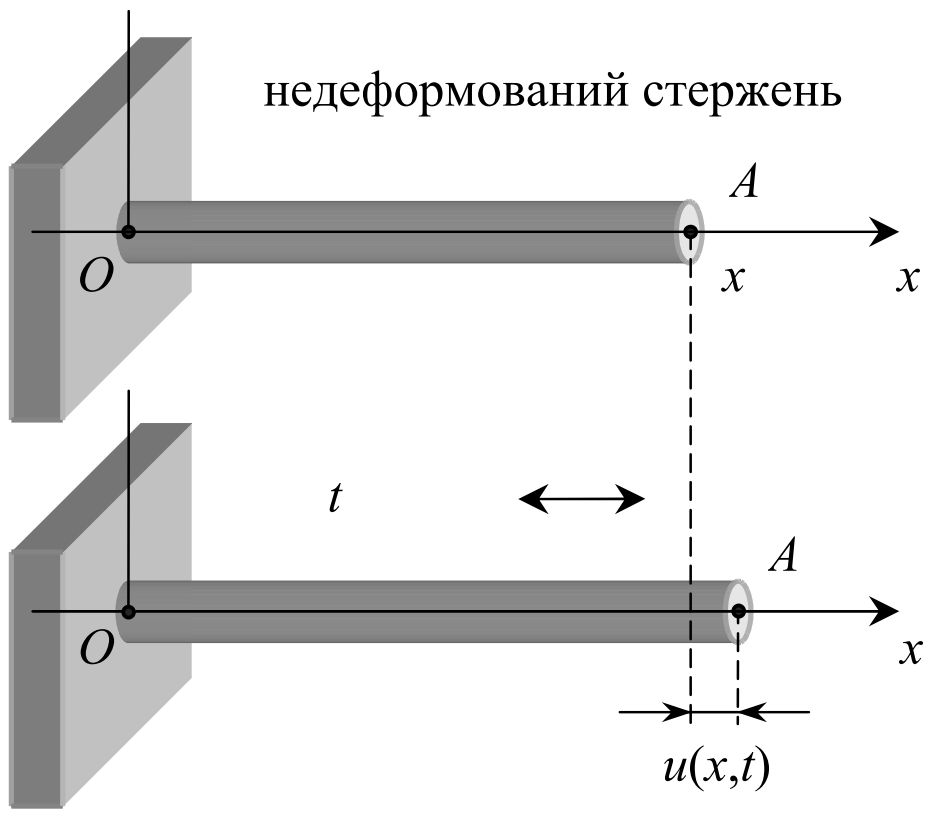

Рис. 3.2. Зміщення точок стержня при поздовжніх коливаннях деформації $d\Pi(x,t)$ цієї ділянки при поздовжніх коливаннях. При аналізі малих коливань достатньо обмежитися членами другого порядку малості за функцією $u(x,t)$ та її похідними.

Оскільки маса ділянки $(x, x+dx)$ дорівнює $\rho(x)S(x)dx$, і у згаданому наближенні швидкості всіх точок ділянки можна вважати однаковими й рівними $u_t(x,t)$, то

$$dK(x,t) = \frac{1}{2}\rho(x)S(x)u_t^2(x,t)dx.$$

Потенціальну енергію $d\Pi(x,t)$ знайдемо за допомогою закону Гука, який стверджує, що механічне напруження $\sigma(x,t)$ в перерізі $x$ у момент часу $t$ прямо пропорційне відносному видовженню $\varepsilon(x,t)$ стержня в цьому перерізі:

$$\sigma(x,t) = \frac{|\mathbf{F}(x,t)|}{S(x)} = E(x)|\varepsilon(x,t)|,$$

де $\mathbf{F}(x,t)$ — сила пружності, яка діє в перерізі $x$ у момент часу $t$, $E(x)$ — модуль Юнга матеріалу стержня. Оскільки $\varepsilon(x,t) = \delta l/l$, де $l = dx$ — рівноважна довжина ділянки $(x, x+dx)$, $\delta l = u(x+dx,t) - u(x,t) \simeq u_x(x,t)dx$ — її видовження, то

$$\varepsilon(x,t) \simeq \frac{u_x(x,t)dx}{dx} = u_x(x,t),$$

тобто похідна $u_x(x,t)$ має зміст відносного видовження при поздовжніх коливаннях. Проекція сили $\mathbf{F}(x,t)$ на вісь $OX$ дорівнює

$$F(x,t) = -E(x)S(x)u_x(x,t),$$

де знак «–» ураховує напрям цієї сили. Зокрема, якщо стержень видовжується вздовж осі $OX$, то $u_x(x,t) > 0$ і $F(x,t) < 0$.

Потенціальна енергія $d\Pi(x,t)$, яку ділянка $(x, x+dx)$ запасає у процесі коливань, дорівнює роботі $dA$, що треба виконати проти сили $\mathbf{F}(x,t)$, щоб цю ділянку розтягнути. Оскільки при розтягуванні стержня $F(x,t)$ змінюється за лінійним законом, то її середнє значення

$$\overline{F(x,t)} = -\frac{0 + E(x)S(x)u_x(x,t)}{2} = -\frac{1}{2}E(x)S(x)u_x(x,t).$$

Тоді

$$d\Pi(x,t) = dA = -\overline{F(x,t)}\delta l = \frac{1}{2}E(x)S(x)u_x^2(x,t)dx.$$

Інтегруючи за всіма елементарними ділянками, дістаємо кінетичну та потенціальну енергії стержня:

$$K(t) = \frac{1}{2}\int_0^l dx\, \rho(x)S(x)u_t^2(x,t),$$

$$\Pi(t) = \frac{1}{2}\int_0^l dx E(x)S(x)u_x^2(x,t).$$

Урахуємо тепер ще той факт, що кінці стержня закріплено пружно, наприклад, за допомогою пружинок із дуже малими масами та довжинами. Кінетична енергія таких пружинок практично дорівнює нулю, а їх потенціальна енергія визначається зміщеннями лівого $x = 0$ та правого $x = l$ кінців стержня:

$$\Pi_1(t) = \frac{1}{2}k_1 u^2(0,t), \quad \Pi_2(t) = \frac{1}{2}k_2 u^2(l,t),$$

де $k_1$ та $k_2$ — коефіцієнти жорсткості відповідно лівої та правої пружинок. Отже, пружне закріплення кінців стержня враховується за допомогою додаткових внесків у точках $x = 0$ та $x = l$ до потенціальної енергії системи.

Для функціонала дії всієї системи остаточно маємо:

$$S = \frac{1}{2}\int_{t_1}^{t_2} dt \left\{ \int_0^l dx \left[ \rho(x)S(x)u_t^2(x,t) - E(x)S(x)u_x^2(x,t) \right] - \right.$$

$$\left. -k_1 u^2(0,t) - k_2 u^2(l,t) \right\}. \tag{3.11}$$

Задача зводиться до відшукання екстремалей функціонала з додатковими внесками на краях. Варіювання проводиться за функцією $u(x,t)$, значення якої в точках $t = t_1$, $t = t_2$ у всіх випадках а) — в) фіксовані (вивчається рух системи між фіксованими моментами часу): $\delta u(x,t_1) = \delta u(x,t_2) = 0$. Значення ж функції $u(x,t)$ в точках $x = 0$ та $x = l$, а, отже, і варіацій $\delta u(0,t)$ та $\delta u(l,t)$ є, взагалі кажучи, довільними.

Для варіації дії (3.11) маємо:

$$\delta S = \int_{t_1}^{t_2} dt \left\{ \int_0^l dx \left[ \rho S u_t \delta u_t - E S u_x \delta u_x \right] - k_1 u(0,t) \delta u(0,t) - k_2 u(l,t) \delta u(l,t) \right\}.$$

Інтегруючи частинами та прирівнюючи $\delta S$ до нуля, отримуємо

$$\int_{t_1}^{t_2} dt \left\{ \int_0^l dx \left[ -\rho S u_{tt} + \frac{\partial}{\partial x}\left( E S u_x \right) \right] \delta u(x,t) + \right.$$
$$+ \left[ E S u_x(x,t) - k_1 u(x,t) \right]\Big|_{x=0} \delta u(0,t) -$$
$$\left. - \left[ E S u_x(x,t) + k_2 u(x,t) \right]\Big|_{x=l} \delta u(l,t) \right\} = 0, \qquad (3.12)$$

звідки, унаслідок довільності варіацій $\delta u(0,t)$ та $\delta u(l,t)$, знаходимо як рівняння руху стержня, так і крайові умови:

$$\rho S u_{tt} = \frac{\partial}{\partial x}\left( E S u_x \right), \qquad (3.13)$$

$$u_x(0,t) - h_1 u(0,t) = 0, \;\; u_x(l,t) + h_2 u(l,t) = 0, \qquad (3.14)$$

де $h_1 \equiv k_1 / E(0) S(0), \;\; h_2 \equiv k_2 / E(l) S(l)$.

*Відповіді*: б) Потрібно знайти варіацію функціонала

$$S = \frac{1}{2} \int_{t_1}^{t_2} dt \int_0^l dx \left[ \rho(x) S(x) u_t^2(x,t) - E(x) S(x) u_x^2(x,t) \right] \qquad (3.15)$$

за додаткової умови жорсткого закріплення: $\delta u(0,t) = \delta u(l,t) = 0$. Дістаємо рівняння (3.13) та крайові умови

$$u(0,t) = 0, \;\; u(l,t) = 0. \qquad (3.16)$$

Це є граничний випадок задачі а) при $k_{1,2} \to \infty$ (або $h_{1,2} \to \infty$).

в) Треба знайти варіацію функціонала (3.15) за додаткової умови, що кінці стержня вільні, тобто варіації $\delta u(0,t)$ і $\delta u(l,t)$ довільні. Отримуємо рівняння (3.13) та крайові умови

$$u_x(0,t) = 0, \;\; u_x(l,t) = 0. \qquad (3.17)$$

Це є граничний випадок задачі а) при $k_{1,2} \to 0$ (або $h_{1,2} \to 0$).

**Завдання 3.1.4.** Виведіть рівняння та крайові умови, які описують *малі згинальні коливання пружного стержня* із закріпленим лівим кінцем, якщо його правий кінець: а) закріплено жорстко; б) закріплено шарнірно; в) вільний. Окремо розгляньте випадок однорідного стержня.

*Розв'язання*. Потенціальна енергія пружної деформації згину (див. завдання 2.1.2 та 2.2.2) дорівнює

$$\Pi(t)=\frac{1}{2}\int_0^l E(x)J(x)u_{xx}^2(x,t)dx,$$

де функція $u(x,t)$ описує поперечне відхилення точки $x$ стержня в момент часу $t$.

Функціонал дії

$$S=\frac{1}{2}\int_{t_1}^{t_2}dt\int_0^l dx\left[\rho(x)S(x)u_t^2(x,t)-E(x)J(x)u_{xx}^2(x,t)\right]. \qquad (3.18)$$

Після дворазового інтегрування частинами та з урахуванням умов $\delta u(x,t_1)=\delta u(x,t_2)=0$ його перша варіація набирає вигляду

$$\delta S=-\int_{t_1}^{t_2}dt\int_0^l dx\left[\rho Su_{tt}+\frac{\partial^2}{\partial x^2}\left(ESu_{xx}\right)\right]\delta u(x,t)-$$

$$-\int_{t_1}^{t_2}dt\left[EJu_{xx}(x,t)\delta u_x(x,t)\right]\Big|_{x=0}^{x=l}+$$

$$+\int_{t_1}^{t_2}dt\left[\left(\frac{\partial}{\partial x}EJu_{xx}(x,t)\right)\delta u(x,t)\right]\Bigg|_{x=0}^{x=l}. \qquad (3.19)$$

Прирівнюючи її до нуля та враховуючи довільність варіації $\delta u(x,t)$, знаходимо:

$$\rho Su_{tt}+\frac{\partial^2}{\partial x^2}\left(EJu_{xx}\right)=0. \qquad (3.20)$$

Це є рівняння четвертого порядку відносно змінної $x$, тому його потрібно доповнити чотирма крайовими умовами.

У випадку жорсткого закріплення крайня точка стержня не зміщується, а дотична до стержня в ній залишається горизонтальною (варіації $\delta u(x,t)$ та $\delta u_x(x,t)$ у крайній точці дорівнюють нулю). Тому зліва маємо:

$$u(0,t)=0,\ \ u_x(0,t)=0. \qquad (3.21)$$

Крайові умови справа у випадку а) такі самі:

$$u(l,t)=0,\ \ u_x(l,t)=0. \qquad (3.22)$$

У випадку б) крайня права точка не зміщується $(\delta u(l,t)=0)$, а дотична до стержня в ній змінює, взагалі кажучи, свій напрям ($\delta u_x(l,t)$ — довільна). Маємо:

$$u(l,t)=0,\ \ u_{xx}(l,t)=0. \tag{3.23}$$

У випадку в) варіації $\delta u(l,t)$ та $\delta u_x(l,t)$ довільні, тому дістаємо

$$u_{xx}(l,t)=0,\ \ \frac{\partial}{\partial x}\left(EJu_{xx}(x,t)\right)\Big|_{x=l}=0. \tag{3.24}$$

Зокрема, для однорідного стержня

$$u_{xx}(l,t)=0,\ \ u_{xxx}(l,t)=0. \tag{3.25}$$

**Завдання 3.1.5.** Виведіть рівняння, яке описує малі згинальні коливання пружного стержня при наявності зовнішньої сили, перпендикулярної до його рівноважного профілю. Погонна густина сили $F(x,t)$.

*Відповідь*:

$$\rho S u_{tt}+\frac{\partial^2}{\partial x^2}\left(EJu_{xx}\right)=F.$$

## 3.2. ПОСТАНОВКА ОДНОВИМІРНИХ КРАЙОВИХ ЗАДАЧ. ЄДИНІСТЬ РОЗВ'ЯЗКУ

З математичного погляду вивчення процесу малих одновимірних коливань просторово-обмеженої системи зводиться до відшукання функції із заданими аналітичними властивостями, що задовольняє рівняння коливань та певні додаткові умови, достатні для того, щоб шукана функція була єдиною та однозначною.

Щоб з'ясувати ці умови, сформулюємо (спираючись на задачі про поперечні коливання неоднорідної струни у фіксованій площині та поздовжні коливання неоднорідного стержня вздовж власної осі) узагальнюючу математичну модель для опису одновимірних коливань *механічних систем із розподіленими параметрами*, тобто систем, геометричні і фізичні параметри яких змінюються, взагалі кажучи, від точки до точки. Як і раніше (див. підрозділ 3.1), коливальний рух таких систем будемо описувати за допомогою скалярної функції $u(x,t)$, визначеної на півсмузі $D:\{0\le x\le l,\ t\ge 0\}$, $l$ — довжина системи. У кожний момент часу $t>0$ функція $u(x,t)$ дорівнює зміщенню поперечного перерізу системи з координатою $x$ зі свого положення рівноваги; у недеформованому стані $u(x,t)=0$. Ми вважатимемо функцію $u(x,t)$ неперервною разом із першими та другими похідними.

Припустимо тепер, що система з розподіленими параметрами коливається, взагалі кажучи, у середовищі, ефективний вплив якого на систему та її кінці може розглядатися як пружний. Щоб змоделювати параметри такої системи, уведемо на відрізку $[0,l]$ функції $\rho(x)$, $p(x)$, $q(x)$ та сталі $h_1$, $h_2$. Для струни $\rho(x)$ має зміст погонної густини струни в точці $x$, а $p(x)$ дорівнює (сталій) силі натягу струни $T_0$. У випадку стержня під $\rho(x)$ і $p(x)$ слід розуміти добутки $\rho_V(x)S(x)$ і $E(x)S(x)$, де $\rho_V(x)$, $S(x)$ та $E(x)$ — відповідно об'ємна густина, площа поперечного перерізу і модуль Юнга матеріалу стержня в точці $x$. Функція $q(x)$ визначає силу пружного зв'язку внутрішніх точок системи з середовищем, обчислену на одиницю довжини системи; моделюватимемо цю силу у вигляді $-q(x)u(x,t)$. І, нарешті, сталі $h_1$ і $h_2$ характеризують пружні сили, чисельно рівні $-h_1 p(0)u(0,t)$ і $-h_2 p(l)u(l,t)$, що діють на лівий $(x=0)$ та правий $(x=l)$ кінці коливальної системи з боку тіл, до яких ці кінці прикріплені. Раніше (див. підрозділ 3.1) уже наводився модельний приклад такої системи — горизонтальний стержень, кінці якого прикріплені до безмасових горизонтальних пружинок з коефіцієнтами жорсткості $k_1, k_2 \geq 0$. Вплив пружинок на поздовжні коливання стержня моделювався через додатковий внесок

$$\Delta\Pi(t) = \frac{1}{2}k_1 u^2(0,t) + \frac{1}{2}k_2 u^2(l,t) \tag{3.26}$$

у потенціальну енергію системи; відповідно, $h_1 = k_1 / E(0)S(0)$, $h_2 = k_2 / E(l)S(l)$.

Уявімо тепер собі струну, кінці якої прикріплені до малих безмасових кілець, які можуть ковзати без тертя по гладких прямолінійних дротинах, що проходять через кінці струни перпендикулярно до її рівноважного профілю (див. рис. 3.3). Насадивши на дротини безмасові гладкі пружинки та прикріпивши їх за одні кінці до кілець, а за другі — до дротин у точках, розташованих від рівноважного профілю струни на відстанях, що дорівнюють довжинам пружинок у недеформованому стані, приходимо до модельної задачі про

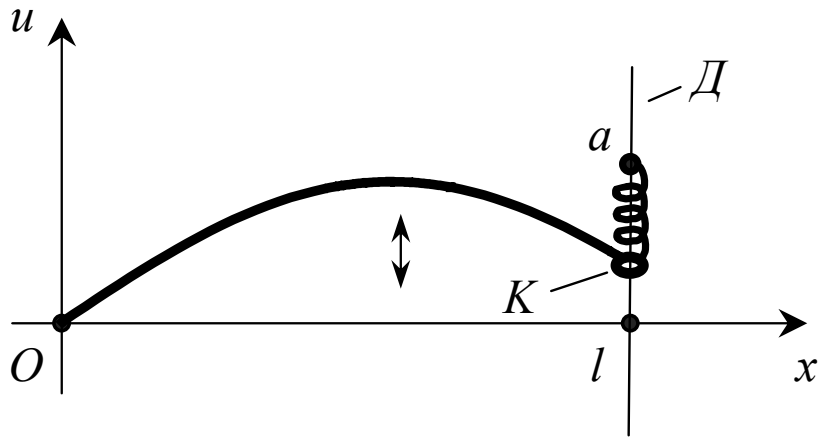

Рис. 3.3. Струна з лівим жорстко та правим пружно закріпленими кінцями. Позначено: *К* — кільце, *Д* — дротина, *a* — точка кріплення пружини до дротини. Відстань *al* дорівнює довжині недеформованої пружини. Точки струни рухаються паралельно дротині поперечні коливання струни з пружно закріпленими кінцями. Якщо наявність пружинок знову враховувати за допомогою внеску виду (3.26) у потенціальну енергію системи, то тепер $h_i = k_i / T_0$ ($i = 1,2$), $T_0$ — сила натягу струни.

З фізичного змісту функцій $\rho(x)$, $p(x)$ і $q(x)$ випливає, що у випадку гомогенної системи (утвореної з одного й того самого матеріалу та з однаковими характеристиками в усіх точках) їх можна вважати гладкими, а у випадку гетерогенної системи (складеної з двох чи більше гомогенних областей) — гладкими майже скрізь на $[0,l]$, за винятком обмеженої кількості точок розриву першого роду. Для реальних фізичних систем із розподіленими параметрами ці вимоги можна дещо послабити. Надалі вважатимемо, якщо не обумовлено супротивне, що функції $\rho(x)$ і $q(x)$ — неперервні, а $p(x)$ — неперервно диференційовна на $[0,l]$, причому всюди

$$\rho(x) > 0, \;\; p(x) > 0, \;\; q(x) \geq 0. \tag{3.27}$$

Відповідно, сталі

$$h_1, h_2 \geq 0, \tag{3.28}$$

при цьому при $h_1, h_2 \to \infty$ приходимо до граничного випадку жорсткого закріплення кінців, а при $h_1, h_2 = 0$ — граничного випадку вільних кінців. У залежності від значень цих сталих, тобто від фізичних умов, підтримуваних на кінцях коливальної системи, функція $u(x,t)$ має задовольняти в цих точках певні співвідношення — *крайові умови*.

Рівняння руху та крайові умови для функції $u(x,t)$ знайдемо, скориставшись принципом найменшої дії. Для цього на множині неперервно диференційовних функцій, заданих на півсмузі $D$, розглянемо квадратичні функціонали

$$K[u;t] = \frac{1}{2}\int_0^l dx\, \rho(x) u_t^2(x,t), \tag{3.29}$$

$$\begin{aligned} \Pi[u;t] = \frac{1}{2}\int_0^l dx \Big[ p(x) u_x^2(x,t) + q(x) u^2(x,t) \Big] + \\ + \frac{1}{2} h_1 p(0) u^2(0,t) + \frac{1}{2} h_2 p(l) u^2(l,t). \end{aligned} \tag{3.30}$$

Функціонал $K[u;t]$ має зміст кінетичної енергії системи, а функціонал $\Pi[u;t]$ — потенціальної енергії системи з урахуванням її пружного зв'язку з середовищем, включаючи тіла, до яких вона прикріплена.

Якщо зміщення точок системи в моменти часу $t_1$ і $t_2$, $0<t_1<t_2$, визначаються відповідно функціями $u_1(x)$ і $u_2(x)$, і жодні інші зовнішні сили на систему не діють, то зміщення цих самих точок у проміжні моменти часу $t \in (t_1,t_2)$ визначаються, згідно з принципом найменшої дії, функцією $u(x,t)$, що надає мінімум функціоналу дії

$$S[u] = \int_{t_1}^{t_2} dt \left\{ K[u;t] - \Pi[u;t] \right\}. \qquad (3.31)$$

З необхідної умови екстремуму функціонала (3.31) випливає, що функція $u(x,t)$ , яка описує рух системи, має справджувати диференціальне рівняння

$$\rho(x)\frac{\partial^2 u}{\partial t^2} = \frac{\partial}{\partial x}\left( p(x)\frac{\partial u}{\partial x} \right) - q(x)u, \ 0<x<l, \ t>0, \qquad (3.32)$$

та задовольняти крайові умови

$$u_x(0,t) - h_1 u(0,t) = 0, \ u_x(l,t) + h_2 u(l,t) = 0, \ h_1,h_2 \geq 0, \ t \geq 0. \qquad (3.33)$$

Рівнянням (3.32) визначається загальний закон руху системи з розподіленими параметрами при малих коливаннях у пружному середовищі. З усіх функцій воно відбирає певну підмножину двічі диференційовних функцій, визначених у досліджуваній області координат і часу і в загальному випадку залежних від чотирьох сталих інтегрування. При розв'язуванні конкретної фізичної задачі з цих функцій виокремлюється більш вузька двопараметрична підмножина функцій, що задовольняють крайові умови (3.33). Щоб знайти серед останніх ту функцію (при повній і коректній постановці задачі — єдину), яка правильно описує конкретний процес коливань, треба, очевидно, мати ще дві умови. Ними виступають *початкові умови* — співвідношення, які відображають механічний стан точок системи в момент часу $t_0$, з якого починають спостерігати за їх коливаннями[1].

---

[1] Треба нагадати, що при використанні принципу найменшої дії для аналізу руху механічної системи природно виникає питання, чи можна відновити проміжні значення функції $u(x,t)$, $t \in (t_1,t_2)$, що задовольняє лінійну систему (3.32), (3.33), за її значеннями $u_1(x)$ і $u_2(x)$ у моменти часу $t=t_1$ і $t=t_2$. Однак така задача, взагалі кажучи, не завжди має розв'язок — унаслідок можливої несумісності заданих початкового і кінцевого положень точок системи і проміжку часу, що їх розділяє. Наприклад, з подальшого (див. завдання 4.6.2) зрозуміло, що однорідна струна із закріпленими кінцями, форма якої в момент часу $t_0=0$ описується функцією $u_0(x) = A\sin(\pi x/l)$, $A \neq 0$, не може при вільних коливаннях набувати в момент часу $t_1 = l/a$ форми, яка описується функцією $u_1(x) = B\sin(2\pi x/l)$, $B \neq 0$.

Як правило, початкові умови задають зміщення $u_0(x)$ та швидкості $v_0(x)$ точок системи в початковий момент $t_0$. Поклавши, за традицією, $t_0 = 0$, можемо записати їх у вигляді

$$u(x,0) = u_0(x), \;\; u_t(x,0) = v_0(x), \;\; 0 \le x \le l. \tag{3.34}$$

Надалі, крім окремих випадків, уважатимемо, що початкові функції $u_0(x)$ та $v_0(x)$ є відповідно двічі та один раз неперервно диференційовними в області визначення.

Якщо на внутрішні точки системи діють, крім пружних сил з боку навколишнього середовища, й інші зовнішні сили, то до правої частини рівняння (3.32) треба додати їх погонну густину $F(x,t)$:

$$\rho(x)\frac{\partial^2 u}{\partial t^2} = \frac{\partial}{\partial x}\left( p(x)\frac{\partial u}{\partial x}\right) - q(x)u + F(x,t), \;\; 0 < x < l, \;\; t > 0. \tag{3.35}$$

У більшості задач функцію $F(x,t)$ можна вважати неперервною. Виняток складають окремі ідеалізовані випадки, коли на систему діють зосереджені сили, ударні навантаження тощо.

*Крайовою задачею для рівняння коливань (3.32) (або (3.35)) називатимемо задачу про відшукання в області $D$ такого його двічі неперервно диференційовного розв'язку, який задовольняє крайові умови (3.33) та початкові умови (3.34).* Окремими випадками цієї загальної задачі виступають крайові задачі про коливання систем із жорстко закріпленими та вільними кінцями; відповідні крайові умови

$$u(0,t) = 0, \;\; u(l,t) = 0 \tag{3.36}$$

та

$$u_x(0,t) = 0, \;\; u_x(l,t) = 0 \tag{3.37}$$

знаходимо з умов (3.33), перейшовши до границь $h_1, h_2 \to \infty$ (умови жорсткого закріплення кінців) та $h_1, h_2 \to 0$ (умови на вільних кінцях).

**Зауваження 3.2.1**. На практиці типовими є ситуації, коли крайові умови на кінцях системи різняться (див. рис. 3.3). Більше того, вони можуть і не зводитися до умов виду (3.33). Якщо, наприклад, на лівий кінець стержня діє сила $F(t)$, а правий кінець закріплено жорстко, то крайові умови мають вигляд

$$u_x(0,t) = -\frac{F(t)}{E(0)S(0)}, \;\; u(l,t) = 0.$$

Покажемо тепер, що розв'язки крайових задач (3.32)–(3.34) та (3.33)–(3.35) єдині, і що цей факт можна розглядати як наслідок закону збереження енергії. Для цього зауважимо, що коли зовнішня сила $F(x,t)$ відсутня, то наша система є консервативною, і рівняння (3.32) має перший інтеграл енергії $E = \text{const}$. Справді, за допомогою рівняння (3.35) для похідної кінетичної енергії системи $K(t)$ за часом знаходимо

$$\frac{dK}{dt} = \int_0^l dx\, \rho(x) u_t(x,t) u_{tt}(x,t) =$$

$$= \int_0^l dx u_t(x,t) \left[ \frac{\partial}{\partial x}\left( p(x) \frac{\partial u(x,t)}{\partial x} \right) - q(x) u(x,t) \right] + \int_0^l dx\, F(x,t) u_t(x,t). \quad (3.38)$$

Останній інтеграл у цій формулі дорівнює загальній потужності зовнішніх сил (за винятком сил пружного зв'язку системи з середовищем). Інтегруючи в передостанньому інтегралі частинами та враховуючи крайові умови (3.33), дістанемо:

$$\frac{dK}{dt} = -\frac{d\Pi}{dt} + \int_0^l dx\, F(x,t) u_t(x,t), \quad (3.39)$$

де

$$\Pi(t) = \frac{1}{2} \int_0^l dx\, p(x) u_x^2(x,t) + \frac{1}{2} \int_0^l dx q(x) u^2(x,t) +$$

$$+ \frac{1}{2} h_1 p(0) u^2(0,t) + \frac{1}{2} h_2 p(l) u^2(l,t) \quad (3.40)$$

– потенціальна енергія системи з урахуванням умов закріплення та пружного зв'язку її внутрішніх точок з середовищем. Формула (3.39) еквівалентна співвідношенню

$$\frac{dE}{dt} = \int_0^l dx\, F(x,t) u_t(x,t), \quad (3.41)$$

звідки при $F(x,t) = 0$ дістаємо закон збереження повної механічної енергії системи: $E = K(t) + \Pi(t) = \text{const}$.

**Теорема 3.2.1 (єдиності)**. Нехай функції $\rho(x) > 0$, $p(x) > 0$, $q(x) \geq 0$ задовольняють перелічені вище умови, сталі $h_1, h_2 \geq 0$. Тоді в області $D$ може існувати лише одна двічі неперервно диференційовна функція $u(x,t)$, яка задовольняє рівняння (3.35), крайові умови (3.33) та початкові умови (3.34).

*Доведення.* Доводимо від супротивного. Припустимо, що існують два розв'язки $u_1(x,t)$ і $u_2(x,t)$ розглядуваної крайової задачі. Їх різниця $U(x,t) = u_1(x,t) - u_2(x,t)$ задовольняє однорідне рівняння

$$\rho(x)\frac{\partial^2 U}{\partial t^2} = \frac{\partial}{\partial x}\left(p(x)\frac{\partial U}{\partial x}\right) - q(x)U, \qquad (3.42)$$

крайові умови

$$U_x(0,t) - h_1 U(0,t) = 0,\ U_x(l,t) + h_2 U(l,t) = 0,\ h_1, h_2 \geq 0, \qquad (3.43)$$

та нульові початкові умови

$$U(x,0) = 0,\ U_t(x,0) = 0. \qquad (3.44)$$

Згідно з формулою (3.41), повна механічна енергія системи, коливання якої описуються функцією $U(x,t)$, зберігається $(F(x,t) = 0)$, а тому її значення $E(t)$ в довільний момент часу $t$ дорівнює значенню $E(0)$ в початковий момент часу $t = 0$:

$$E(t) = \frac{1}{2}\int_0^l dx\left[\rho(x)U_t^2(x,t) + p(x)U_x^2(x,t) + q(x)U^2(x,t)\right] +$$

$$+\frac{1}{2}h_1 p(0)U^2(0,t) + \frac{1}{2}h_2 p(l)U^2(l,t) = E(0). \qquad (3.45)$$

З початкових умов (3.44) випливає, що $E(0) = 0.$ Оскільки всі доданки у формулі (3.45) за умовами теореми невід'ємні, то їх сума дорівнює нулю лише за умови, що кожний з них дорівнює нулю. При $q(x) = 0$ звідси випливає, що дорівнюють нулю частинні похідні функції $U(x,t)$: $U_t(x,t) = U_x(x,t) = 0.$ Тому $U(x,t) = \text{const},$ і з початкових умов (3.44) знаходимо, що $U(x,t) \equiv 0.$ При $q(x) > 0$ разом із похідними $U_t(x,t)$ і $U_x(x,t)$ дорівнює нулю і сама функція $U(x,t)$, тобто знову маємо $U(x,t) \equiv 0.$

**Зауваження 3.2.2**. Якщо функція $q(x,t) \equiv 0$ і при цьому також $h_1 = h_2 = 0,$ то енергія $E(t)$ може дорівнювати нулю і при ненульовому значенні $U(x,t) = \text{const},$ тобто коли незакріплена недеформована система паралельно зсувається як ціле в інше місце.

При формулюванні крайових задач, що описують малі згинальні коливання стержня (див. завдання 3.1.4, 3.1.5), слід ураховувати, що ці коливання описуються диференціальним рівнянням у частинних похідних четвертого порядку

$$\rho(x)\frac{\partial^2 u}{\partial t^2} + \frac{\partial^2}{\partial x^2}\left(p(x)\frac{\partial^2 u}{\partial x^2}\right) = F(x,t), \qquad (3.46)$$

де шукана функція $u(x,t)$ описує поперечне відхилення точки стержня з координатою $x$ у момент часу $t$ від того положення, яке ця точка займає в недеформованому стержні, під функціями $\rho(x)$ і $p(x)$ слід розуміти добутки $\rho_V(x)S(x)$ і $E(x)J(x)$, де $\rho_V(x)$, $S(x)$, $E(x)$ та $J(x)$ — відповідно об'ємна густина, площа поперечного перерізу, модуль Юнга та головний момент інерції поперечного перерізу стержня в точці $x$, і функція $F(x,t)$ описує погонну густину всіх зовнішніх сил, що діють на внутрішні точки стержня. Розв'язок рівняння (3.46) повинен задовольняти чотири крайові та дві початкові умови. Крайові умови для цього рівняння були розглянуті в завданні 3.1.4 (формули (3.21)–(3.25)). Початкові умови й далі мають вигляд (3.34).

**Завдання 3.2.1.** Доведіть, що в області $D$ може існувати лише одна функція $u(x,t)$, двічі неперервно диференційовна за часом і чотири рази за координатою, яка задовольняє рівняння (3.46), крайові умови (3.21)–(3.25) і початкові умови (3.34). Укажіть обмеження, які при цьому треба накласти на функції $\rho(x)$, $p(x)$, $u_0(x)$, $v_0(x)$ і $F(x,t)$, та сформулюйте відповідну теорему єдиності.

## *КОНТРОЛЬНІ ПИТАННЯ ДО РОЗДІЛУ 3*

*1. Як визначаються кінетична і потенціальна енергії та функція Лагранжа натягненої струни з пружно закріпленими кінцями?*

*2. Як визначаються функції Лагранжа неоднорідного стержня із закріпленими кінцями в задачах про його малі поздовжні та поперечні коливання?*

*3. Як із варіаційних принципів механіки можна вивести рівняння руху для пружних коливань стержнів і струн із пружно закріпленими кінцями та крайові умови для відповідних функцій зміщень?*

*4. Як ставляться крайові задачі для вільних і вимушених коливань струн і стержнів із пружно закріпленими кінцями?*

*5. Який закон збереження справджується при вільних коливаннях струн і стержнів із пружно закріпленими кінцями? Як за його допомогою довести єдиність розв'язків крайових задач для вільних і вимушених коливань відповідних механічних систем?*

# Розділ 4
# КОЛИВАННЯ ОДНОРІДНИХ СИСТЕМ

## 4.1. ЗАДАЧА КОШІ ДЛЯ НЕОБМЕЖЕНОЇ СТРУНИ. ВІЛЬНІ КОЛИВАННЯ

Уявімо тепер ситуацію, коли кінці системи (струни чи стержня) знаходяться настільки далеко від області початкового збурення, що час, за який збурення дійде до них, набагато перевищує час спостереження за системою. Очевидно, що за такий проміжок часу крайові умови на кінцях системи не встигнуть вплинути на характер процесів у ній, і формально можна вважати, що кінці системи знаходяться на нескінченності. Зміщення точок такої необмеженої системи з рівноважного стану визначатимуться лише початковим умовами та фізичними властивостями внутрішніх областей системи. При наявності ще й зовнішньої сили (з погонною густиною $F(x,t)$), прикладеної до внутрішніх точок системи, відповідна задача для знаходження зміщення $u(x,t)$ набирає вигляду

$$\rho(x)\frac{\partial^2 u}{\partial t^2}=\frac{\partial}{\partial x}\left(p(x)\frac{\partial u}{\partial x}\right)+F(x,t),\ -\infty<x<\infty,\ t>0, \qquad (4.1)$$

$$u(x,0)=u_0(x),\ u_t(x,0)=v_0(x), \qquad (4.2)$$

де функції $u_0(x)$ та $v_0(x)$ описують відповідно початкові зміщення та початкові швидкості точок системи. Задача (4.1), (4.2) називається *задачею Коші для необмеженої струни (стержня) або задачею Коші на необмеженій прямій*. Для *однорідної* струни вона набирає вигляду

$$\frac{\partial^2 u}{\partial t^2}=a^2\frac{\partial^2 u}{\partial x^2}+f(x,t),\ -\infty<x<\infty,\ t>0, \qquad (4.3)$$

$$u(x,0)=u_0(x),\ u_t(x,0)=v_0(x), \qquad (4.4)$$

де $a^2\equiv T_0/\rho,\ f(x,t)\equiv F(x,t)/\rho$.

Очевидно, що доведену в попередньому підрозділі теорему єдиності для обмеженої струни (стержня) можна, з відповідними модифікаціями, перенести й на задачу (4.1), (4.2) та її простіший варіант (4.3), (4.4). Звідси випливає, що якщо функції $\rho(x)$ і $p(x)$ — строго додатні та, відповідно, неперервна і гладка, $u_0(x)$ — двічі диференційовна, $v_0(x)$ — диференційовна, $F(x,t)$ — неперервна та обмежена, то задача Коші для необмеженої прямої має єдиний двічі неперервно диференційовний розв'язок. Виявляється, що для однорідних систем його можна побудувати в загальному випадку. Перейдемо до його відшукання.

Почнемо з випадку, коли зовнішня сила відсутня. Маємо таку *задачу Коші для вільних коливань необмеженої однорідної струни (стержня):*

$$\frac{\partial^2 u}{\partial t^2} = a^2 \frac{\partial^2 u}{\partial x^2}, \quad -\infty < x < \infty, \quad t > 0, \tag{4.5}$$

$$u(x,0) = u_0(x), \quad u_t(x,0) = v_0(x). \tag{4.6}$$

Перейдемо в рівнянні (4.5) до нових змінних $\alpha = x + at$, $\beta = x - at$. Диференціюючи $u = u\big(\alpha(x,t), \beta(x,t)\big)$ як складену функцію, дістаємо:

$$\frac{\partial u}{\partial x} = \frac{\partial u}{\partial \alpha}\frac{\partial \alpha}{\partial x} + \frac{\partial u}{\partial \beta}\frac{\partial \beta}{\partial x} = \frac{\partial u}{\partial \alpha} + \frac{\partial u}{\partial \beta},$$

$$\frac{\partial^2 u}{\partial x^2} = \frac{\partial}{\partial x}\left(\frac{\partial u}{\partial \alpha}\right) + \frac{\partial}{\partial x}\left(\frac{\partial u}{\partial \beta}\right) = \frac{\partial^2 u}{\partial \alpha^2} + 2\frac{\partial^2 u}{\partial \alpha \partial \beta} + \frac{\partial^2 u}{\partial \beta^2}.$$

Аналогічним чином знаходимо:

$$\frac{\partial^2 u}{\partial t^2} = a^2\left(\frac{\partial^2 u}{\partial \alpha^2} - 2\frac{\partial^2 u}{\partial \alpha \partial \beta} + \frac{\partial^2 u}{\partial \beta^2}\right).$$

Рівняння (4.5) набирає вигляду

$$\frac{\partial^2 u}{\partial \alpha \partial \beta} = 0. \tag{4.7}$$

Похідна за змінною $\alpha$ від функції $\partial u / \partial \beta$ дорівнює нулю, якщо остання не залежить від $\alpha$, тобто є, у загальному випадку, функцією змінної $\beta$:

$$\frac{\partial u}{\partial \beta} = f(\beta).$$

Інтегруючи це співвідношення за $\beta$ та враховуючи, що замість довільної сталої інтегрування (як це має місце при інтегруванні функції однієї змінної) виникає, взагалі кажучи, довільна функція другої незалежної змінної, маємо:

$$u(\alpha,\beta) = f_1(\alpha) + \int f(\beta)\,d\beta \equiv f_1(\alpha) + f_2(\beta). \tag{4.8}$$

Повертаючись до вихідних змінних, приходимо до висновку, що розв’язок рівняння (4.5) має вигляд

$$u(x,t) = f_1(x+at) + f_2(x-at), \tag{4.9}$$

тобто є сумою двох функцій $f_1$ та $f_2$, кожна з яких залежить лише від певної комбінації змінних $x$ і $t$.

Явний вигляд $f_1$ та $f_2$ знаходимо за допомогою початкових умов (4.6). Після підстановки в них виразу (4.9) дістаємо систему рівнянь

$$\begin{aligned} f_1(x) + f_2(x) &= u_0(x), \\ f_1'(x) - f_2'(x) &= \frac{1}{a} v_0(x), \end{aligned} \tag{4.10}$$

де штрих означає диференціювання. Інтегруючи друге рівняння, можемо записати

$$\begin{aligned} f_1(x) + f_2(x) &= u_0(x), \\ f_1(x) - f_2(x) &= \frac{1}{a} \int_{x_0}^{x} v_0(x')dx' + C, \end{aligned} \tag{4.11}$$

де $C$ — невідома стала інтегрування, яка дорівнює різниці $f_1(x_0) - f_2(x_0)$ в довільно вибраній точці $x_0$. Для подальшого розгляду значення $x_0$ та $C$ виявляються несуттєвими.

Розв’язок системи (4.11) має вигляд

$$\begin{aligned} f_1(x) &= \frac{1}{2} u_0(x) + \frac{1}{2a} \int_{x_0}^{x} v_0(x')dx' + \frac{1}{2} C, \\ f_2(x) &= \frac{1}{2} u_0(x) - \frac{1}{2a} \int_{x_0}^{x} v_0(x')dx' - \frac{1}{2} C. \end{aligned} \tag{4.12}$$

Замінивши $x$ у першій із цих формул на $x+at$, а у другій — на $x-at$, та підставивши здобуті вирази у формулу (4.9), знаходимо розв’язок задачі (4.5), (4.6):

$$u(x,t) = \frac{u_0(x+at) + u_0(x-at)}{2} + \frac{1}{2a} \int_{x-at}^{x+at} v_0(x')dx'. \tag{4.13}$$

Співвідношення (4.13) називається *формулою Д’Аламбера*.

**Завдання 4.1.1.** Безпосередньою підстановкою переконайтеся, що формула (4.13) задовольняє рівняння (4.5) і початкові умови (4.6).

Нехай, наприклад, у початковий момент часу $t = 0$ необмежена однорідна струна була деформована таким чином, що її профіль опи-сувався парною функцією $u_0(x)$, що швидко спадає на нескінченності, скажімо $u_0(x) = A\exp(-x^2/l^2)$. Після того як струну відпустили з цього положення, її профіль у будь-який інший момент часу $t$ є суперпозицією двох симетричних контурів з удвічі меншими амплітудами (див. рис. 4.1):

$$u(x,t) = \frac{A}{2} e^{-(x+at)^2/l^2} + \frac{A}{2} e^{-(x-at)^2/l^2}.$$

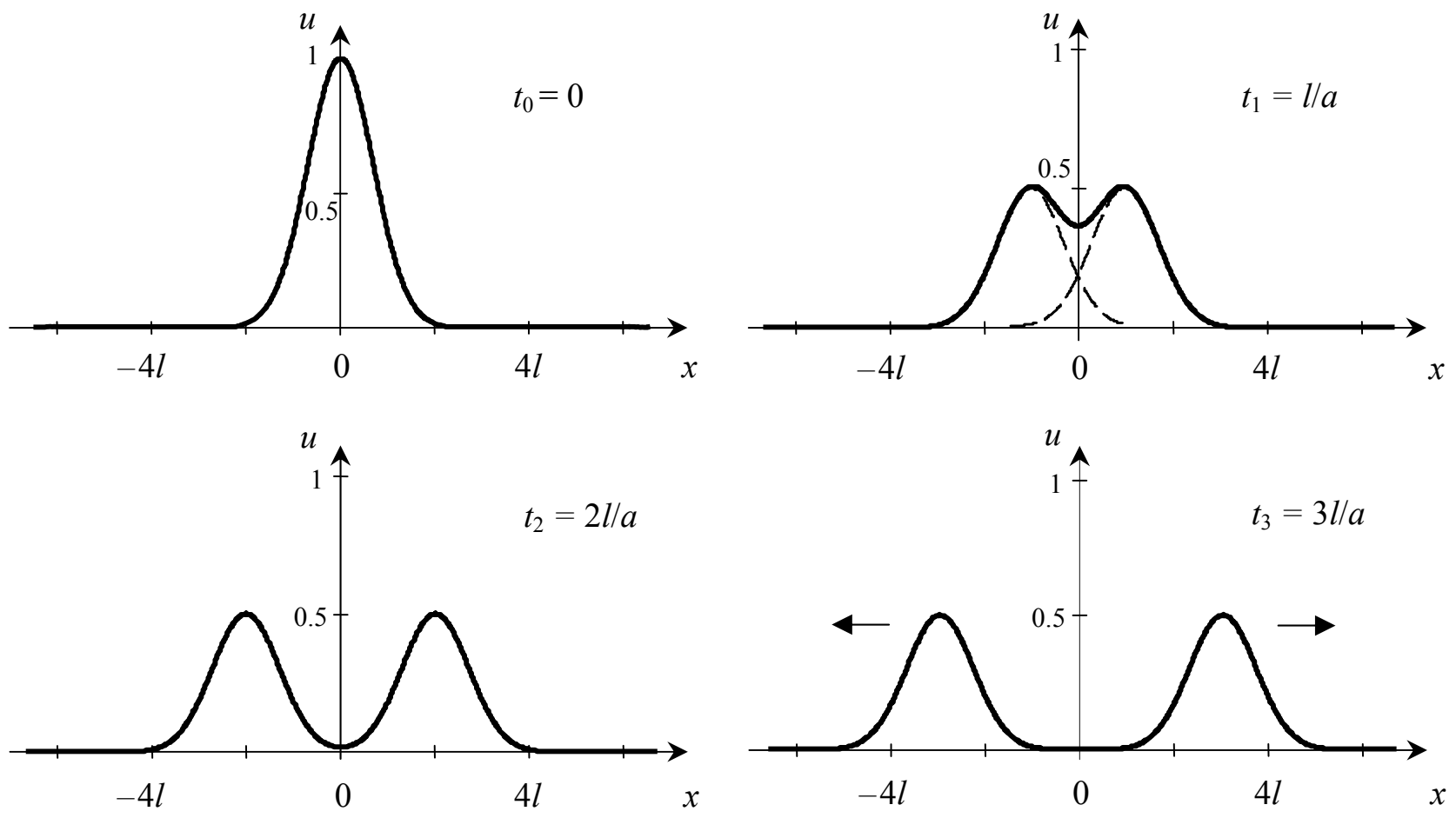

Рис. 4.1. Формування біжучих хвиль уздовж необмеженої струни внаслідок початкової деформації $u_0(x) = A\exp(-x^2/l^2)$, $A = 1$. Зображено профілі струни в моменти часу $t_k = kl/a$, $k = 0 \div 3$

Координати центрів цих контурів змінюються з часом за законами $x_0 - at$ та $x_0 + at$ $(x_0 = 0)$, тобто рухаються вздовж осі $x$ відповідно вліво та вправо з однаковими сталими швидкостями $a$. Переносячи цей результат на загальний випадок, можемо сказати, що розв'язок задачі Коші для необмеженої однорідної струни $-\infty < x < \infty$ є суперпозицією двох збурень $f_1(x+at)$ та $f_2(x-at)$, що поширюються по осі $x$ зі сталою швидкістю $a$ відповідно вліво та вправо. Ці рухомі збурення називаються *біжучими хвилями*. Для спостерігача, що рухається зі швидкістю $a$ разом з однією з цих хвиль, її профіль, очевидно, залишатиметься незмінним.

Розв'язок (4.13) задачі Коші (4.5), (4.6) можна перенести й на більш широкі класи початкових функцій $u_0(x)$ та $v_0(x)$. Нехай, наприклад, останні відмінні від нуля лише на скінченних відрізках та неперервні, а функція $u_0(x)$ додатково має першу похідну. Тоді функції $u_0(x)$ та $v_0(x)$ можна рівномірно апроксимувати нескінченними послідовностями диференційовних функцій $\{\varphi_n(x)\}$ та $\{\psi_n(x)\}$, що мають відповідно перші дві та першу похідні. При фіксованому $n$ кожна пара початкових функцій $\varphi_n(x)$ та $\psi_n(x)$ визначає єдиний розв'язок $u_n(x,t)$ задачі Коші для необмеженої струни. Беручи до уваги неперервну залежність розв'язку задачі Коші від початкових функцій, можна показати, що послідовність $\{u_n(x,t)\}$ рівномірно збігається до деякої функції $u(x,t)=\lim\limits_{n\to\infty}u_n(x,t)$. Ця функція називається *узагальненим розв'язком* задачі Коші (4.5), (4.6) та дається формулою (4.13), де $u_0(x)=\lim\limits_{n\to\infty}\varphi_n(x)$, $v_0(x)=\lim\limits_{n\to\infty}\psi_n(x)$. Аналогічний висновок справджується і для розривних початкових швидкостей.

**Завдання 4.1.2.** Нехай у початковий момент часу всім точкам необмеженої недеформованої струни з проміжку $[-l,l]$ надали однакової поперечної швидкості $v_0$. Зобразіть профіль струни в послідовні моменти часу $t_k=kl/(2a)$, $k=0,1,2,...$. Чим біжучі хвилі у цьому випадку відрізняються від усамітнених біжучих хвиль із наведеного вище прикладу?

*Відповідь*: $u(x,t)=\dfrac{1}{2a}\left[\psi(x+at)-\psi(x-at)\right]$, де

$$\psi(z)=\int_0^z v_0(x')dx'=\begin{cases}-v_0 l, & \text{якщо } z<-l,\\ v_0 z, & \text{якщо } -l\le z\le l,\\ v_0 l, & \text{якщо } z>l.\end{cases}$$

Профілі струни для вказаних моментів часу зображено на рис. 4.2. Починаючи з моменту часу $t=l/a$, формується плато висотою $u_{\max}=$ $=lv_0/a$ зі зростаючою шириною.

Із завдання 4.1.2 випливає важливий висновок: біжучі хвилі, що поширюються вздовж необмеженої струни, у загальному випадку не мають чітко вираженого заднього фронту. Таким чином, *для одновимірних задач принцип Гюйгенса не справджується*.

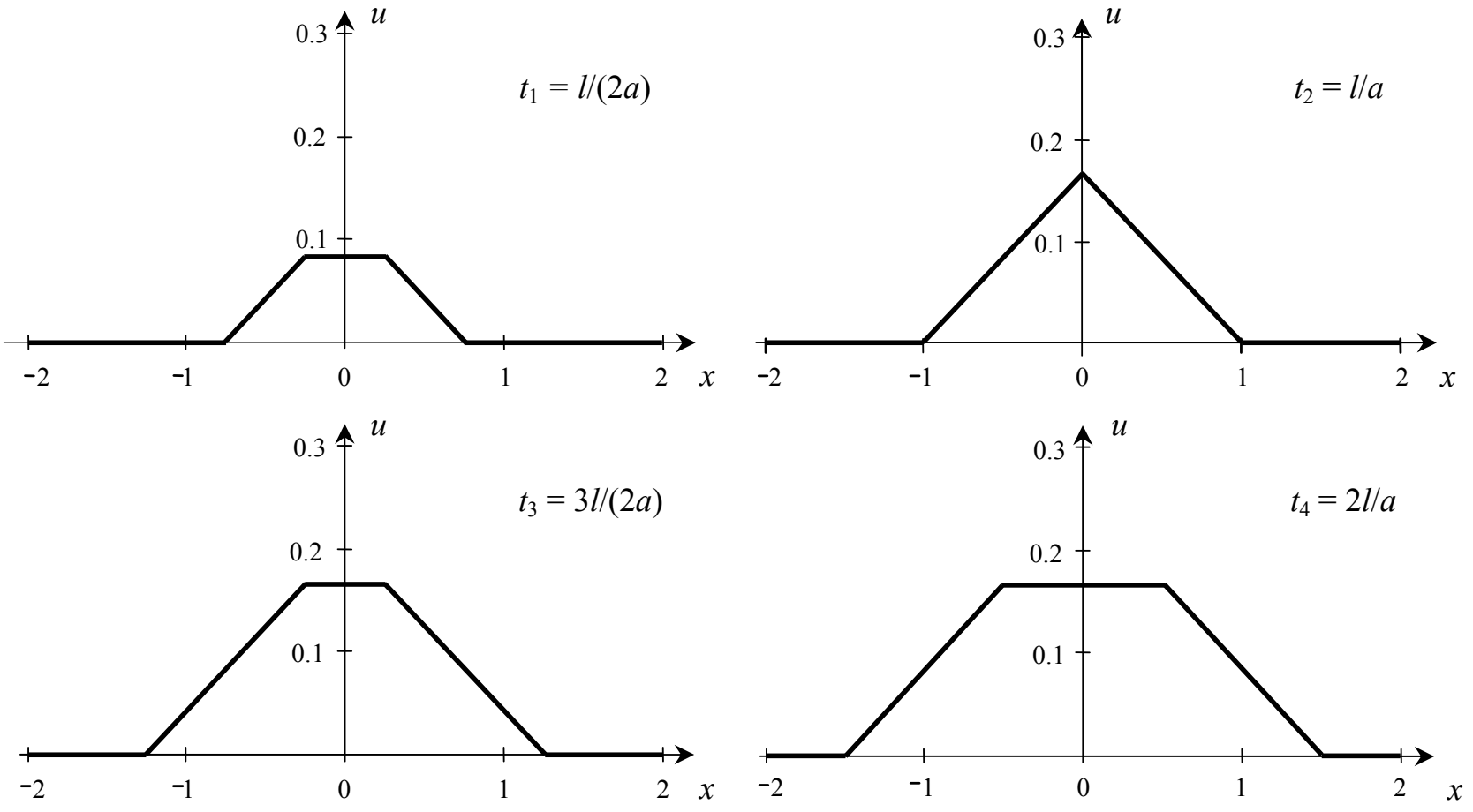

Рис. 4.2. Профілі необмеженої струни в моменти часу $t_k = kl/(2a)$, $k = 1 \div 4$, після того, як у момент $t = 0$ всім точкам ділянки $[-l, l]$ надали поперечної швидкості $v_0$. Профілі зображено для значень $l = 0.5$ м, $a = 3$ м/с, $v_0 = 1$ м/с

## 4.2. ВИМУШЕНІ КОЛИВАННЯ НЕОБМЕЖЕНОЇ СТРУНИ

Перейдемо тепер до розв’язування задачі Коші про вимушені коливання необмеженої однорідної струни, тобто до відшукання функції, яка є розв’язком рівняння (4.3) з $f(x,t) \neq 0$ та задовольняє початкові умови (4.4). Почнемо з наступного зауваження. Параметр $t$ у формулі (4.13) має зміст проміжку часу, який минув з початкового моменту $t_0 = 0$. Якщо ж початкові функції задані для ненульового моменту $t_0 = t_1$,

$$u(x,t_1) = u_0(x), \; u_t(x,t_1) = v_0(x), \tag{4.14}$$

то зміщення точок струни в довільний момент часу $t \geq t_1$ описуватимуться формулою

$$u(x,t) = \frac{u_0\big(x + a(t-t_1)\big) + u_0\big(x - a(t-t_1)\big)}{2} + \frac{1}{2a}\int\limits_{x-a(t-t_1)}^{x+a(t-t_1)} v_0(x')dx'. \tag{4.15}$$

Нехай, наприклад, струна до моменту часу $t_1 \geq 0$ перебувала в стані спокою, а в момент часу $t_1$ кожній її малій ділянці $(x, x+dx)$ масою $dm(x) = \rho dx$ миттєвим поштовхом надали імпульсу $dm(x)v_1(x) = \rho v_1(x)dx,$ де $v_1(x)$ — швидкість ділянки. Згідно з формулою (4.15), рух такої струни описується виразом

$$u(x,t) = \begin{cases} 0, & t < t_1, \\ \dfrac{1}{2a} \displaystyle\int_{x-a(t-t_1)}^{x+a(t-t_1)} v_1(x')dx', & t > t_1. \end{cases} \tag{4.16}$$

Оскільки імпульс $\rho v_1(x)dx$ та поперечна сила $F(x,t)dx,$ що його надала, пов'язані співвідношенням

$$\rho v_1(x)dx = \int_{t_1-\varepsilon}^{t_1+\varepsilon} \big(F(x,t)dx\big)dt,$$

де $\varepsilon > 0$ — скінченне (хоч і мале) число, а $F(x,t) \neq 0$ лише при $t = t_1,$ то зазначений поштовх можна моделювати як дію поперечної сили з погонною густиною

$$F(x,t) = \rho v_1(x)\delta(t - t_1). \tag{4.17}$$

Відповідно, формулу (4.16) можна інтерпретувати як розв'язок задачі Коші (4.3), (4.4) для спеціального випадку, коли $f(x,t) = v_1(x)\delta(t - t_1),$ $u_0(x) = 0$ та $v_0(x) = 0.$

Розглянемо тепер випадок, коли коливання струни збуджуються поперечною силою, що складається з послідовності подібних миттєвих поштовхів, прикладених до струни в моменти часу $t_1 < t_2 < ... < t_n$. Погонну густину такої сили можна подати у вигляді

$$F(x,t) = \rho \sum_{k=1}^{n} v_k(x)\delta(t - t_k). \tag{4.18}$$

Скориставшись принципом суперпозиції, можемо далі стверджувати, що розв'язок задачі Коші (4.3), (4.4) для такої сили при нульових початкових умовах дається формулою

$$u(x,t) = \begin{cases} 0, & t < t_1, \\ \dfrac{1}{2a} \displaystyle\int_{x-a(t-t_1)}^{x+a(t-t_1)} v_1(x')dx', & t_1 < t < t_2, \\ \dots\dots\dots\dots\dots\dots\dots\dots\dots\dots, \\ \dfrac{1}{2a} \displaystyle\int_{x-a(t-t_1)}^{x+a(t-t_1)} v_1(x')dx' + ... + \dfrac{1}{2a} \displaystyle\int_{x-a(t-t_n)}^{x+a(t-t_n)} v_n(x')dx', & t > t_n. \end{cases} \tag{4.19}$$

Беручи до уваги явний вигляд густини (4.18), безпосереднім інтегруванням переконуємося, що формулі (4.19) можна надати вигляду $(f(x,t) = F(x,t)/\rho)$

$$u(x,t)=\frac{1}{2a}\int_0^t dt'\int_{x-a(t-t')}^{x+a(t-t')}dx'f(x',t'). \qquad (4.20)$$

Очевидно, що результат (4.20) справджується й у випадку, коли погонна густина сили $F(x,t)$ є кусково-неперервною функцією на проміжку часу від $t_0=0$ до $t_n$. Справді, розіб'ємо цей проміжок на нескінченно малі інтервали $[0,t_1)$, $[t_1,t_2)$, ... , $[t_{n-1},t_n)$. Оскільки значення густини сили на них практично не змінюються, виберемо їх рівними значенням $F(x,t_i^*)$ у деякі моменти часу $t_i^*$ $(i=1,2,...,n)$ із цих інтервалів. Розглянемо тепер малу ділянку $(x,x+dx)$ струни. За проміжок часу $dt_i=t_i-t_{i-1}$ на неї діє імпульс сили $F(x,t_i^*)dxdt_i$, унаслідок чого швидкість ділянки змінюється на величину $F(x,t_i^*)dxdt_i/(\rho dx)=f(x,t_i^*)dt_i$. Оскільки сила $F(x,t_i^*)dx$ перестає діяти в момент $t_i$, можна вважати, що її дія зводиться до того, щоб у момент $t_i$ надати точкам струни додаткової швидкості $v(x,t_i)=f(x,t_i^*)dt_i$. Остання вестиме, згідно з формулою (4.15), до зміщення точок струни в моменти часу $t>t_i$ на величину

$$u_i(x,t)=\frac{1}{2a}\int_{x-a(t-t_i)}^{x+a(t-t_i)}v(x',t_i)dx'=\frac{1}{2a}\int_{x-a(t-t_i)}^{x+a(t-t_i)}f(x',t_i^*)dx'dt_i. \qquad (4.21)$$

Підсумовуючи, згідно з принципом суперпозиції, за всіма такими зміщеннями та переходячи до границі $dt_i\to 0$, дістаємо формулу (4.20).

**Завдання 4.2.1.** Згадавши властивості дельта-функції Дірака, переконайтеся, що формула (4.19) випливає з формули (4.20), якщо погонна густина зовнішньої сили описується виразом (4.18).

**Завдання 4.2.2.** Сформулюйте задачу Коші, розв'язком якої є функція (4.21).

Повторюючи дослівно аргументацію, використану при виведенні формули (4.20), можна встановити просте загальне правило, за яким розв'язки задач Коші для неоднорідних лінійних диференціальних рівнянь у частинних похідних (хвильового рівняння, рівняння коливань струни, теплопровідності тощо) можна будувати через загальні розв'язки задач Коші для відповідних однорідних рівнянь. Це правило є аналогом методу варіації довільних сталих у теорії звичайних лінійних диференціальних рівнянь і називається *принципом Дюамеля*.

Сформулюємо принцип Дюамеля для лінійних диференціальних рівнянь виду

$$\frac{\partial^2 u}{\partial t^2} - \hat{L}[u] = f(x_1,...,x_n,t), \tag{4.22}$$

де $\hat{L}[u]$ — довільний лінійний диференціальний оператор, що не містить других похідних функції $u = u(x_1,...,x_n,t)$ за часом. Наприклад,

$$\hat{L}[u] = \sum_{i,j=1}^{n} \frac{\partial}{\partial x_i}\left( D_{ij}(x_1,...,x_n)\frac{\partial u(x_1,...,x_n,t)}{\partial x_j}\right) + V(x_1,...,x_n)u(x_1,...,x_n,t),$$

або, у випадку однієї просторової змінної $(n=1)$,

$$\hat{L}[u] = \frac{\partial}{\partial x}\left( p(x)\frac{\partial u(x,t)}{\partial x}\right) + q(x)u(x,t) - 2\eta\frac{\partial u(x,t)}{\partial t}.$$

Нехай для фіксованого значення параметра $\tau$ функція $\varphi(x_1,...,x_n,t\,|\,\tau)$ задовольняє однорідне рівняння

$$\frac{\partial^2 \varphi}{\partial t^2} - \hat{L}[\varphi] = 0, \;\; t > \tau, \tag{4.23}$$

та початкові умови

$$\varphi(x_1,...,x_n,t\,|\,\tau)\big|_{t=\tau} = 0, \;\; \left.\frac{\partial \varphi(x_1,...,x_n,t\,|\,\tau)}{\partial t}\right|_{t=\tau} = f(x_1,...,x_n,\tau). \tag{4.24}$$

Тоді, як стверджує принцип Дюамеля, функція

$$u(x_1,...,x_n,t) = \int_0^t \varphi(x_1,...,x_n,t\,|\,\tau)d\tau \tag{4.25}$$

є розв'язком рівняння (4.22), який задовольняє нульові початкові умови

$$u(x_1,...,x_n,0) = 0, \;\; \left.\frac{\partial u(x_1,...,x_n,t)}{\partial t}\right|_{t=0} = 0. \tag{4.26}$$

Це твердження легко перевірити безпосередньо. Справді, згідно з формулами (4.24), (4.25) маємо:

$$\frac{\partial u(x_1,...,x_n,t)}{\partial t} = \varphi(x_1,...,x_n,t\,|\,\tau)\big|_{\tau=t} +$$

$$+\int_0^t \frac{\partial \varphi(x_1,...,x_n,t\,|\,\tau)}{\partial t}d\tau = \int_0^t \frac{\partial \varphi(x_1,...,x_n,t\,|\,\tau)}{\partial t}d\tau,$$

$$\frac{\partial^2 u(x_1,...,x_n,t)}{\partial t^2} = \left.\frac{\partial \varphi(x_1,...,x_n,t\,|\,\tau)}{\partial t}\right|_{\tau=t} + \int_0^t \frac{\partial^2 \varphi(x_1,...,x_n,t\,|\,\tau)}{\partial t^2}d\tau =$$

$$= f(x_1,...,x_n,t) + \int_0^t \frac{\partial^2 \varphi(x_1,...,x_n,t\,|\,\tau)}{\partial t^2} d\tau,$$

$$\hat{L}[u(x_1,...,x_n,t)] = \int_0^t \hat{L}[\varphi(x_1,...,x_n,t\,|\,\tau)]d\tau.$$

За допомогою двох останніх співвідношень та формули (4.23) бачимо, що функція (4.25) задовольняє рівняння (4.22). Узявши ж до уваги перше співвідношення та формулу (4.25), відразу переконуємося, що справджуються й початкові умови (4.26).

**Завдання 4.2.3.** Сформулюйте й обґрунтуйте принцип Дюамеля для рівняння типу

$$\frac{\partial u}{\partial t} - \hat{L}[u] = f(x_1,...,x_n,t),$$

де $\hat{L}[u]$ — лінійний диференціальний оператор, який не містить похідних функції $u$ за часом.

Повернімося тепер до задачі Коші (4.3), (4.4). Її загальний розв'язок знайдемо, знову скориставшись принципом суперпозиції. Маємо:

$$u(x,t) = \frac{u_0(x+at) + u_0(x-at)}{2} +$$

$$+ \frac{1}{2a} \int_{x-at}^{x+at} v_0(x')dx' + \frac{1}{2a} \int_0^t dt' \int_{x-a(t-t')}^{x+a(t-t')} dx' f(x',t'). \qquad (4.27)$$

**Завдання 4.2.4.** Перевірте, що функція (4.27) справді задовольняє рівняння (4.3) та початкові умови (4.4).

З формули (4.27) випливає, що якщо початкові функції $u_0(x)$ та $v_0(x)$ швидко спадають на нескінченності, то з часом розв'язок задачі Коші $u(x,t)$ в будь-якій точці буде визначитися, з точністю до сталої

$$C_0 = \frac{1}{2a} \int_{-\infty}^{\infty} v_0(x')dx',$$

лише зовнішньою силою, тобто останнім доданком у формулі (4.27). А саме: якщо сила починає діяти в момент часу $t = 0,$ то при $t \to \infty$ зміщення точок необмеженої струни зі стану рівноваги описуватимуться формулою

$$u(x,t) = C_0 + \frac{1}{2a} \int_0^t dt' \int_{x-a(t-t')}^{x+a(t-t')} dx' f(x',t'). \qquad (4.28)$$

Розглянемо тепер важливий випадок, коли вимушені коливання відбуваються під дією сили, густина потужності якої в кожній точці змінюється за гармонічним законом із частотою $\omega$:

$$f(x,t) = f_0(x)\sin\big(\omega t + \varphi(x)\big). \tag{4.29}$$

Можна очікувати, що з часом кожна точка струни теж почне коливатися гармонічно з тією самою частотою $\omega$. Щоб у цьому переконатися, спочатку замість функції джерел (4.29) підставимо у формулу (4.28) комплекснозначну функцію

$$\tilde{f}(x,t) = \tilde{f}_0(x)e^{-i\omega t},\ \ \tilde{f}_0(x) = f_0(x)e^{-i\varphi(x)}, \tag{4.30}$$

та обчислимо при $t \to \infty$ частинний комплекснозначний розв'язок

$$\tilde{u}(x,t) = \frac{1}{2a}\int_0^t dt' \int_{x-a(t-t')}^{x+a(t-t')} dx'\,\tilde{f}(x',t');$$

далі за допомогою формули

$$u(x,t) = C_0 - \operatorname{Im}\tilde{u}(x,t) \tag{4.31}$$

знайдемо й саму функцію $u(x,t)$.

Почнемо з окремого випадку, коли гармонічну силу прикладено до точки $x_1$, тобто функція джерел має вигляд

$$\tilde{f}(x,t) = \tilde{f}_1\delta(x-x_1)e^{-i\omega t}, \tag{4.32}$$

де $\tilde{f}_1 = f_0(x_1)e^{-i\varphi(x_1)}$. Згадавши, що похідна функції Хевісайда $\theta'(x) = \delta(x)$, можемо записати

$$\int_a^b \delta(x'-x_1)\,dx' = \theta(b-x_1) - \theta(a-x_1).$$

За допомогою цієї рівності обчислюємо $\tilde{u}(x,t)$ з підінтегральною функцією (4.32). При $x > x_1$ і $t \to \infty$ маємо:

$$\tilde{u}(x,t) = \frac{1}{2a}\int_0^t dt' \int_{x-a(t-t')}^{x+a(t-t')} dx'\,\tilde{f}_1\delta(x'-x_1)e^{-i\omega t'} =$$

$$= \frac{1}{2a}\tilde{f}_1\int_0^t dt'\big\{1-\theta(x-x_1-a(t-t'))\big\}e^{-i\omega t'}.$$

Якщо $x - x_1 - a(t-t') > 0$, то підінтегральний вираз в останньому інтегралі дорівнює нулю. Якщо ж $x - x_1 - a(t-t') < 0$, то після переходу до нової змінної інтегрування $\tau = t - t'$ маємо

$$\tilde{u}(x,t) = \frac{1}{2a}\tilde{f}_1\int_{(x-x_1)/a}^{t} d\tau\, e^{-i\omega(t-\tau)} = \frac{1}{2ia\omega}\tilde{f}_1 - \frac{1}{2ia\omega}\tilde{f}_1 e^{i\frac{\omega}{a}(x-x_1)}e^{-i\omega t},$$

де було враховано, що $\tau > (x - x_1)/a$. Аналогічним чином знаходимо, що при $x < x_1$ і $t \to \infty$

$$\tilde{u}(x,t) = \frac{1}{2a}\tilde{f}_1 \int_{(x_1-x)/a}^{t} d\tau e^{-i\omega(t-\tau)} = \frac{1}{2ia\omega}\tilde{f}_1 - \frac{1}{2ia\omega}\tilde{f}_1 e^{i\frac{\omega}{a}(x_1-x)} e^{-i\omega t}.$$

Два останні вирази можна об'єднати однією формулою:

$$\tilde{u}(x,t) = \frac{1}{2ia\omega}\tilde{f}_1 + \frac{i}{2a\omega}\tilde{f}_1 e^{i\frac{\omega}{a}|x-x_1|} e^{-i\omega t}. \tag{4.33}$$

Таким чином, під дією гармонічної сили (4.32) всі точки необмеженої однорідної струни з часом починають коливатися гармонічно з частотою $\omega$, при цьому комплекснозначне зміщення точки з координатою $x$ у момент часу $t$ описується формулою (4.33).

З лінійності задачі Коші відносно сили, прикладеної до струни, випливає, що аналогічний висновок справджується і для випадку, коли на струну діє комплекснозначна сила виду

$$\tilde{f}(x,t) = \sum_{j=1}^{n} \tilde{f}_j \delta(x - x_j) e^{-i\omega t}. \tag{4.34}$$

Тепер зміщення точки струни з координатою $x$ у момент часу $t$ визначається виразом

$$\tilde{u}(x,t) = \frac{1}{2ia\omega}\sum_{j=1}^{n}\tilde{f}_j + \left(\frac{i}{2a\omega}\sum_{j=1}^{n}\tilde{f}_j e^{i\frac{\omega}{a}|x-x_j|}\right) e^{-i\omega t}. \tag{4.35}$$

Користуючись формулою (4.35), не важко зрозуміти і структуру функції, яка описує вимушені коливання необмеженої струни під дією гармонічної сили з розподіленою амплітудою. Подавши відповідну комплекснозначну функцію джерел у вигляді

$$\tilde{f}(x,t) = \tilde{f}(x) e^{-i\omega t}, \tag{4.36}$$

де $\tilde{f}(x)$ — кусково-неперервна і принаймні інтегровна функція,

$$\int_{-\infty}^{\infty} \left|\tilde{f}(x')\right| dx' < \infty,$$

можемо стверджувати, що під дією сили (4.36) усі точки струни з часом починають коливатися з частотою $\omega$ за законом

$$\tilde{u}(x,t) = \frac{1}{2ia\omega}\int_{-\infty}^{\infty} \tilde{f}(x')dx' + \tilde{A}_\omega(x) e^{-i\omega t}, \tag{4.37}$$

де

$$\tilde{A}_\omega(x) = \int_{-\infty}^{\infty} G_\omega(|x - x'|)\tilde{f}(x')dx', \quad (4.38)$$

$$G_\omega(|x|) = \frac{i}{2a\omega} e^{i\frac{\omega}{a}|x|}. \quad (4.39)$$

Функція $\tilde{A}_\omega(x)$ має зміст *комплекснозначної амплітуди коливань* точок з координатами $x$ навколо стаціонарного профілю струни; останній визначається сталою $C_0$ та першим доданком у формулі (4.37). Функція $G_\omega(|x|)$ називається *частотною функцією Гріна для одновимірного рівняння коливань на прямій.*

**Завдання 4.2.5.** Доведіть, що для кусково-неперервної інтегровної функції $\tilde{f}(x)$ амплітуда $\tilde{A}_\omega(x)$ задовольняє неоднорідне рівняння

$$-\tilde{A}''_\omega(x) - k^2\tilde{A}_\omega(x) = \frac{1}{a^2}\tilde{f}(x), \; k^2 \equiv \frac{\omega^2}{a^2}. \quad (4.40)$$

Зауважимо, що при додатних значеннях $k^2$ розв'язок рівняння (4.40) не є єдиний.

**Завдання 4.2.6.** Перевірте, що для неперервної фінітної функції $\tilde{f}(x)$ амплітуда (4.38) задовольняє *умови випромінювання* на нескінченності:

$$\lim_{x\to\pm\infty}\left[\tilde{A}'_\omega(x) \mp ik\tilde{A}_\omega(x)\right] = 0. \quad (4.41)$$

Неважко переконатися, що умови (4.41) виконуються й у випадку, коли $\tilde{f}(x)$ — довільна інтегровна функція.

Таким чином, серед множини обмежених розв'язків рівняння (4.40) формула (4.38) виокремлює єдиний розв'язок, який задовольняє умови випромінювання (4.41). Цей результат є наслідком того фізичного факту, що усталені коливання в кожній точці струни встановлюються під впливом зовнішньої сили поступово, після того як сила почала діяти в певний момент часу в минулому.

**Завдання 4.2.7.** Скориставшись формулами (4.32), (4.33), (4.37) і (4.38), визначте фізичний зміст частотної функції Гріна (4.39).

**Завдання 4.2.8.** На однорідну необмежену струну діє гармонічна сила (4.29). Виходячи з формул (4.31) та (4.37)–(4.39), знайдіть: а) закон $u(x,t)$, який описує зміщення точок струни в довільний момент часу; б) фізичні амплітуди коливань точок струни; в) зсув фаз між коливаннями точок струни і коливаннями сили.

## 4.3. КОЛИВАННЯ НАПІВОБМЕЖЕНОЇ СТРУНИ. МЕТОД ПРОДОВЖЕННЯ

Розглянемо тепер випадок, коли область початкового збурення $\Omega$ знаходиться достатньо близько до одного з кінців струни, скажімо, лівого $x=0$. Очевидно, що за проміжок часу $\tau = L/a$, де $L$ — найкоротша відстань від області $\Omega$ до точки $x=0$, $a$ — швидкість поширення хвиль уздовж струни, збурення дійде й до нього. З цього моменту часу коливання струни суттєвим чином залежатимуть від умов, які підтримуються на лівому кінці. Останні, як уже зазначалося, записуються у вигляді крайових (граничних) умов (див. формули (3.33), (3.36) і (3.37)). Уважаючи також, що другий кінець струни знаходиться настільки далеко, що за час спостереження за струною збурення не встигає його досягти, приходимо до *задачі Коші про коливання напівобмеженої струни*, або, іншими словами, *задачі Коші для рівняння коливань на півосі* $0 \le x < \infty$.

Почнемо розгляд цього типу задач із випадку вільних поперечних коливань однорідної напівобмеженої струни з жорстко закріпленим лівим кінцем; у стані спокою струна збігається з піввіссю $x \ge 0$. Для знаходження зміщення $u(x,t)$ точок струни маємо задачу:

$$\frac{\partial^2 u}{\partial t^2} = a^2 \frac{\partial^2 u}{\partial x^2}, \quad 0 < x < \infty, \quad t > 0, \tag{4.42}$$

$$u(0,t) = 0, \tag{4.43}$$

$$u(x,0) = u_0(x), \quad u_t(x,0) = v_0(x). \tag{4.44}$$

Припустивши, що початкові функції $u_0(x)$ та $v_0(x)$ належать до того самого класу функцій, що й початкові функції в задачі про вільні коливання необмеженої струни, шукатимемо розв'язок задачі (4.42)–(4.44) у класі функцій, двічі неперервно диференційовних в області $x > 0$, $t > 0$.

Розв'яжемо задачу (4.42)–(4.44), застосувавши так званий *метод продовження*. Ідея методу полягає в тому, щоб скористатися вже відомим розв'язком Д'Аламбера (4.13) задачі про вільні коливання необмеженої струни, доозначивши шукану функцію $u(x,t)$ та початкові функції $u_0(x)$, $v_0(x)$ на всю вісь таким чином, щоб крайова умова (4.43) автоматично задовольнялася. Покажемо, що для цього початкові функції можна продовжити на всю вісь непарним чином відносно точки $x=0$.

Позначимо доозначені на область $x<0$ функції через відповідно $U(x,t)$, $U_0(x)$ та $V_0(x)$. За означенням, функція $U(x,t)$ є розв'язком задачі Коші на необмеженій прямій з початковими функціями $U_0(x)$ та $V_0(x)$,

$$\frac{\partial^2 U}{\partial t^2}=a^2\frac{\partial^2 U}{\partial x^2},\quad -\infty<x<\infty,\quad t>0, \tag{4.45}$$

$$U(x,0)=U_0(x),\quad U_t(x,0)=V_0(x), \tag{4.46}$$

а тому можемо записати

$$U(x,t)=\frac{U_0(x+at)+U_0(x-at)}{2}+\frac{1}{2a}\int\limits_{x-at}^{x+at}V_0(x')\,dx'. \tag{4.47}$$

Вимагаючи далі, щоб у точці $x=0$ виконувалася б умова

$$U(0,t)=0, \tag{4.48}$$

дістаємо співвідношення

$$U_0(at)+U_0(-at)+\frac{1}{a}\int\limits_{-at}^{at}V_0(x')\,dx'=0,$$

яке можна автоматично задовольнити, вибравши функції $U_0(x)$ та $V_0(x)$ непарними відносно точки $x=0$ (нагадаємо, що інтеграл від непарної функції в симетричних межах дорівнює нулю):

$$U_0(-x)=-U_0(x),\quad V_0(-x)=-V_0(x). \tag{4.49}$$

Такі непарні (відносно точки $x=0$) продовження початкових функцій на всю вісь мають вигляд

$$U_0(x)=\begin{cases}u_0(x), & \text{якщо } x>0,\\ -u_0(-x), & \text{якщо } x<0,\end{cases}\quad V_0(x)=\begin{cases}v_0(x), & \text{якщо } x>0,\\ -v_0(-x), & \text{якщо } x<0.\end{cases} \tag{4.50}$$

Справді, легко перевірити, що функції (4.50) задовольняють співвідношення (4.49).

Таким чином, формулами (4.47) та (4.50) дається розв'язок формальної задачі Коші (4.45), (4.46) для необмеженої прямої, який завжди задовольняє умову (4.48). Крім того, при $x>0$ і $t=0$ функція (4.47) задовольняє початкові умови (4.44) вихідної задачі для напівобмеженої струни:

$$U(x,0)=U_0(x)=u_0(x),\quad U_t(x,0)=V_0(x)=v_0(x),\quad x>0. \tag{4.51}$$

Отже, при $x\ge 0$ і $t\ge 0$ вираз (4.47) задовольняє всі умови задачі (4.42)–(4.44) про вільні коливання напівобмеженої струни із закрі-пленим лівим кінцем. Згідно з теоремою єдиності робимо висновок, що тим самим побудовано розв'язок останньої.

Залишається у формулі (4.47) перейти від параметрів допоміжної задачі (4.45), (4.46) і (4.50) до параметрів вихідної задачі (4.42)–(4.44). Для цього скористаємося формулами (4.50) та перепишемо розв'язок (4.47) для області фізично допустимих значень змінних $x$ і $t$, що відповідають коливанням напівобмеженої струни, через початкові функції $u_0(x)$ і $v_0(x)$.

При $x > at > 0$ всі можливі значення аргументів функцій у правій частині формули (4.47) додатні. Згідно з формулами (4.50) можемо записати:

$$u(x,t) = \frac{u_0(x+at)+u_0(x-at)}{2} + \frac{1}{2a}\int_{x-at}^{x+at} v_0(x')dx', \quad x>0, \ \ 0<t<x/a. \qquad (4.52)$$

Очевидно, що за проміжок часу $0<t<x/a$ збурення від точки з координатою $x$ не встигає дійти до закріпленого кінця, а тому його вплив на коливання струни ще не проявляється. За цих умов розв'язок (4.52) для напівобмеженої струни збігається з розв'язком (4.13) для необмеженої струни.

Вплив закріпленого кінця на коливання напівобмеженої струни стає помітним у тих точках, що задовольняють умову $0<x<at$, тобто в моменти часу $t>x/a>0$. У цьому випадку праву частину формули (4.47) за допомогою співвідношень (4.50) можна переписати наступним чином:

$$u(x,t) = \frac{U_0(x+at)+U_0(x-at)}{2} + \frac{1}{2a}\int_{x-at}^{x+at} V_0(x')dx' =$$

$$= \frac{u_0(x+at)-u_0(at-x)}{2} + \frac{1}{2a}\int_{x-at}^{0} V_0(x')dx' +$$

$$+\frac{1}{2a}\int_{0}^{x+at} V_0(x')dx' = \{x' \to -x' \text{ у першому інтегралі}\} =$$

$$= \frac{u_0(x+at)-u_0(at-x)}{2} - \frac{1}{2a}\int_{at-x}^{0} V_0(-x')dx' + \frac{1}{2a}\int_{0}^{x+at} V_0(x')dx' =$$

$$= \frac{u_0(x+at)-u_0(at-x)}{2} + \frac{1}{2a}\int_{at-x}^{0} v_0(x')dx' + \frac{1}{2a}\int_{0}^{x+at} v_0(x')dx',$$

тобто

$$u(x,t)=\frac{u_0(x+at)-u_0(at-x)}{2}+\frac{1}{2a}\int_{at-x}^{x+at}v_0(x')dx',\quad x>0,\ t>x/a. \qquad (4.53)$$

Зазначимо, що обидві формули (4.52) і (4.53) можна об'єднати в одну:

$$u(x,t)=\frac{u_0(x+at)+\operatorname{sign}(x-at)u_0(|x-at|)}{2}+\frac{1}{2a}\int_{|x-at|}^{x+at}v_0(x')dx',$$
$$x>0,\ t>0, \qquad (4.54)$$

де $\operatorname{sign}x=\begin{cases}1, & x>0,\\ -1, & x<0,\end{cases}$ — знакова функція.

**Завдання 4.3.1.** Переконайтеся, що розв'язок (4.52), (4.53) задовольняє крайову умову (4.43).

*Вказівка.* Розгляньте поведінку функції $u(x,t)$ при $x\to 0$ і $t>0$.

**Завдання 4.3.2.** Виходячи з розв'язку (4.52), (4.53), знайдіть поперечну швидкість довільної точки $x$ напівобмеженої струни в довільний момент часу $t$. Переконайтеся, що початкові умови (4.44) задовольняються.

*Вказівка.* Знайдіть $u_t(x,t)$ та розгляньте поведінку функцій $u(x,t)$ і $u_t(x,t)$ при $t\to 0$ і $x>0$.

**Зауваження 4.3.1.** Початкова умова $u(x,0)=u_0(x)$ має узгоджуватися з крайовою умовою жорсткого закріплення $u(0,t)=0$, тобто початкове зміщення має задовольняти умову $u_0(0)=0$. Якщо ця вимога не виконується, то формулою (4.54) дається розв'язок задачі (4.42)–(4.44), який має розрив на лінії $x=at$ площини $(x,t)$.

**Завдання 4.3.3.** Нехай $u_0(x)$ і $v_0(x)$ — узагальнені функції. Безпосередньою підстановкою переконайтеся, що формулами (4.52) і (4.53) дається узагальнений розв'язок відповідної узагальненої задачі (4.42)–(4.44).

*Вказівка.* Узагальнені функції мають узагальнені похідні всіх порядків.

Проаналізуємо вплив закріпленого кінця на поведінку профілю напівобмеженої однорідної струни. Нехай, для прикладу, у початковий момент часу $t=0$ всі точки струни $x\geq 0$ нерухомі, а вона сама деформована за законом $u_0(x)=A\exp[-(x-x_0)^2/l^2]$. Унаслідок швидкого спадання функції $u_0(x)$ можна наближено вважати, що струна деформована лише в околі півшириною порядку $l$ точки $x_0$. Через час $\tau_1\approx l/a$ після того, як струну в початковий момент часу $t=0$ від-пускають із цього положення, і протягом інтервалу часу $\tau_2 \approx (x_0 - 2l)/a$ її профіль є суперпозицією двох симетричних контурів з удвічі меншими амплітудами (див. рис. 4.3):

$$u(x,t) = \frac{A}{2} e^{-(x-x_0+at)^2/l^2} + \frac{A}{2} e^{-(x-x_0-at)^2/l^2};$$

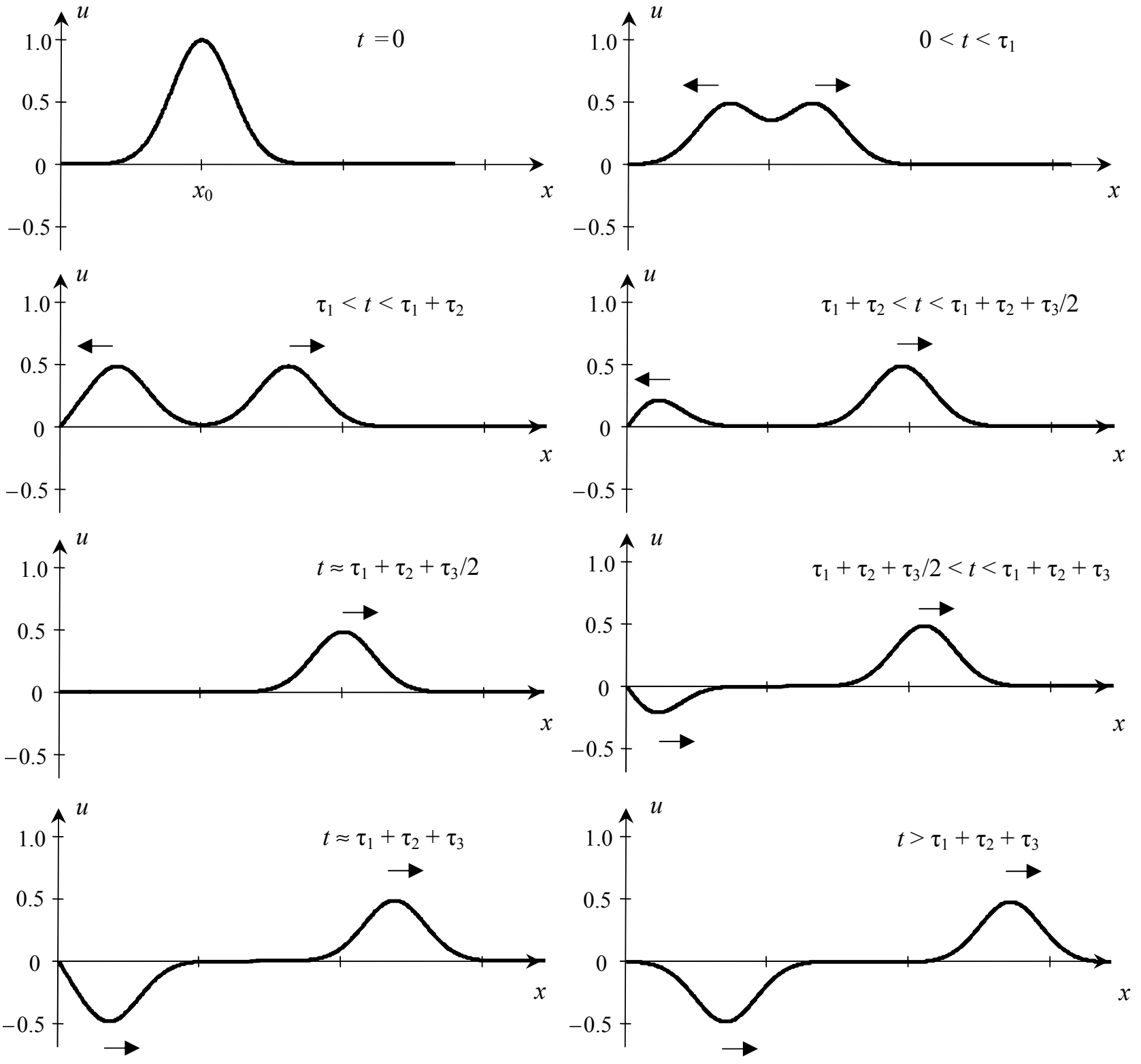

Рис. 4.3. Процес відбивання хвилі від жорстко закріпленого кінця напівобмеженої струни. Зображено профілі струни при $A = 1$ в послідовні моменти часу, який зростає зліва направо та зверху вниз центри контурів рухаються вздовж струни відповідно вліво та вправо з однаковими сталими швидкостями $a$. Ситуація повністю аналогічна тій, яку маємо у випадку необмеженої струни. Однак з моменту $t \approx \tau_1 + \tau_2$ лівий контур починає деформуватися під впливом закріпленого кінця. Цей вплив ефективно триває протягом часу $\tau_3 \approx 2l/a$ і призводить до того, що лівий контур спочатку звужується, потім зникає, потім знову формується, набуваючи своєї попередньої амплітуди, але протилежного напряму (протилежної фази), та зміщуючись управо від закріпленого кінця. У моменти часу $t \gtrsim \tau_1 + \tau_2 + \tau_3$ новоутворений контур є повністю сформованим, а профіль струни є суперпозицією двох усамітнених контурів (хвиль), що рухаються вправо з однаковими сталими швидкостями $a$, але протилежними фазами.

Докладніше простежити за процесом формування профілю відбитої хвилі можна за допомогою формули (4.47), зобразивши графічно непарне продовження функції $u_0(x)$ на всю вісь (див. рис. 4.4). З плином часу (при $t > 0$) ліве (в області $x < 0$) та праве $(x > 0)$ початкові збурення поступово розпадаються на пари усамітнених хвиль, що поширюються в обидва боки. Процес розпочинається в той момент, коли в точці $x = 0$ зустрічаються усамітнені хвилі, що надходять з областей $x < 0$ та $x > 0$. Відбиту хвилю отримуємо, геометрично додаючи ці хвилі для послідовних моментів часу та залишаючи лише ту частину отриманого інтерференційного контура, що знаходиться у фізичній області $x \geq 0$.

**Завдання 4.3.4.** Нехай у початковий момент часу всім точкам напівобмеженої недеформованої струни $x \geq 0$ з проміжку $[l, 2l]$ надали однакової поперечної швидкості. Зобразіть профіль струни, який формується під впливом закріпленого лівого кінця в послідовні моменти часу. Покажіть, що при $t \to \infty$ він має форму рівнобедреної трапеції, що рухається вправо зі сталою швидкістю.

*Вказівка.* Скористайтеся результатами завдання 4.1.2.

Розглянемо тепер вільні поперечні коливання напівобмеженої однорідної струни з незакріпленим лівим кінцем, коли замість умови (4.43) справджується крайова умова

$$\left. \frac{\partial u(x,t)}{\partial x} \right|_{x=0} = 0. \tag{4.55}$$

Розв'яжемо відповідну задачу (4.42), (4.55) і (4.44), знову скориставшись методом продовження. Нагадаємо, що в рамках цього методу розв'язування задачі для напівобмеженої (чи обмеженої) області зводиться до розв'язування задачі для необмеженої області з початковими функціями, які є спеціальними продовженнями початкових функцій вихідної задачі на весь простір, побудованими так, щоб автоматично справджувалися крайові умови на межі області визначення вихідної задачі.

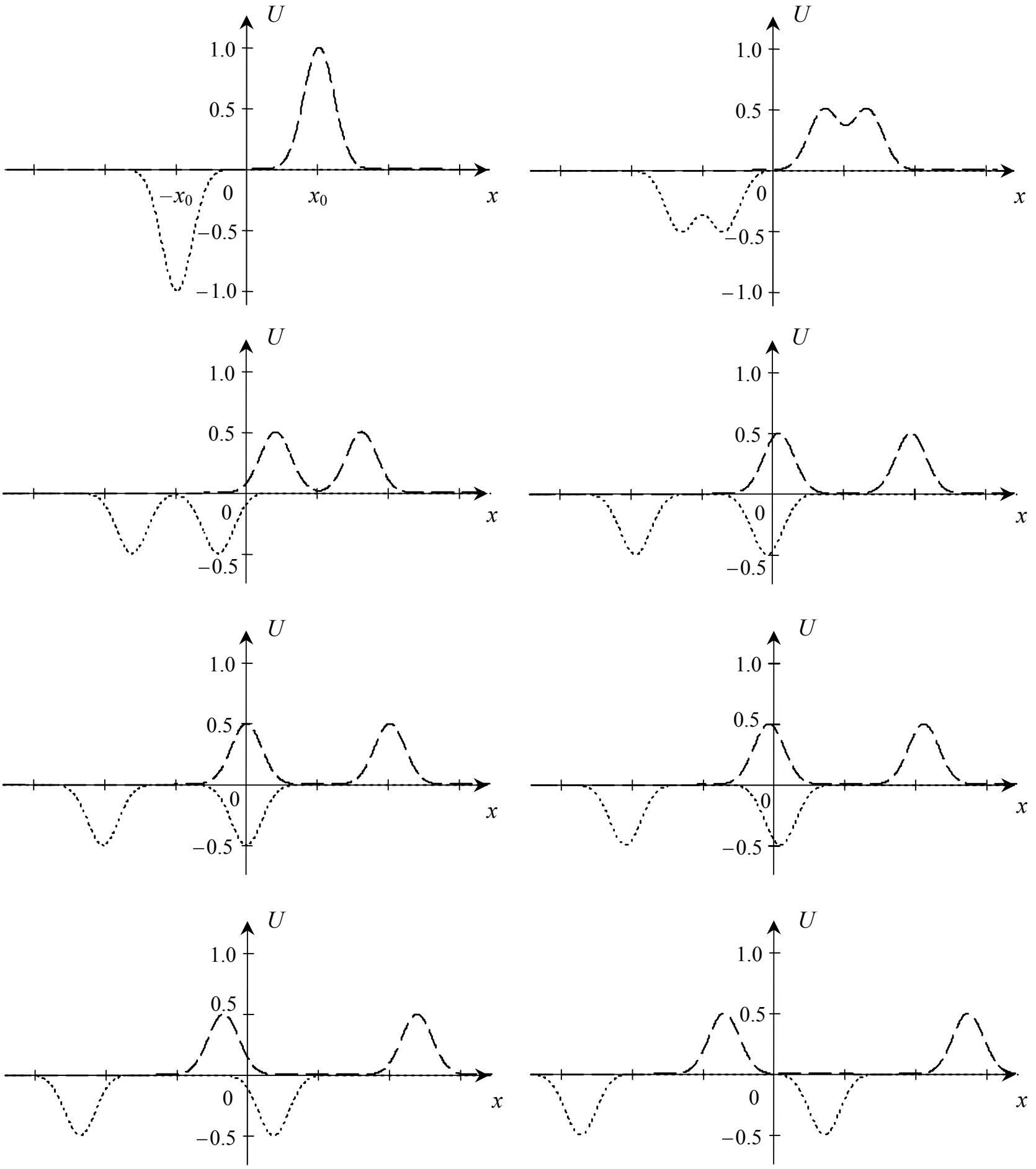

Рис. 4.4. Геометричні побудови для знаходження профілів напівобмеженої струни із закріпленим лівим кінцем методом продовження. Графіки на рис. 4.3 є суперпозицією в області $x \geq 0$ відповідних графіків на цьому рисунку

Нехай $U_0(x)$ та $V_0(x)$ — відповідні продовження початкових функцій вихідної задачі (4.42), (4.55) і (4.44) на всю вісь $-\infty < x < \infty$, $U(x,t)$ — розв'язок задачі Коші (4.45), (4.46) із цими функціями. Останній, очевидно, має вигляд (4.47). Функції $U_0(x)$ та $V_0(x)$ будуємо так, щоб у точці $x = 0$ справджувалася б умова

$$\left.\frac{\partial U(x,t)}{\partial x}\right|_{x=0} = 0. \tag{4.56}$$

Підставивши (4.47) у (4.56), дістаємо таке обмеження на $U_0(x)$ та $V_0(x)$:

$$U_0'(at) + U_0'(-at) + \frac{1}{a}\left[V_0(at) - V_0(-at)\right] = 0, \tag{4.57}$$

де штрих означає похідну відповідної функції. Отже, початкові функції $u_0(x)$ та $v_0(x)$ слід продовжувати на від'ємну піввісь так, щоби виконувалися співвідношення

$$U_0'(-x) = -U_0'(x), \;\; V_0(-x) = V_0(x), \tag{4.58}$$

тобто продовжені функції були парними відносно точки $x = 0$ (похідна від парної функції є непарна функція):

$$U_0(-x) = U_0(x), \;\; V_0(-x) = V_0(x). \tag{4.59}$$

**Завдання 4.3.5.** Парні (відносно точки $x = 0$) продовження початкових функцій $u_0(x)$ та $v_0(x)$ на всю вісь мають вигляд

$$U_0(x) = \begin{cases} u_0(x), \text{ якщо } x > 0, \\ u_0(-x), \text{ якщо } x < 0, \end{cases} \quad V_0(x) = \begin{cases} v_0(x), \text{ якщо } x > 0, \\ v_0(-x), \text{ якщо } x < 0. \end{cases} \tag{4.60}$$

Покажіть, що загальний розв'язок задачі (4.42), (4.55) і (4.44) для напівобмеженої однорідної струни з вільним лівим кінцем дається формулою

$$u(x,t) = \frac{u_0(x+at) + u_0(|x-at|)}{2} +$$

$$+\frac{1}{2a}\int_0^{x+at} v_0(x')dx' - \text{sign}(x-at)\frac{1}{2a}\int_0^{|x-at|} v_0(x')dx', \;\; x > 0, \;\; t > 0. \tag{4.61}$$

**Завдання 4.3.6.** Нехай у початковий момент часу $t = 0$ всі точки струни $x \geq 0$ з незакріпленим лівим кінцем були нерухомими, а вона сама була деформована за законом $u_0(x) = A\exp[-(x-x_0)^2/l^2]$. Опишіть профіль струни в послідовні моменти часу $t > 0$ та покажіть, що після відбиття хвилі від незакріпленого кінця він є суперпозицією двох усамітнених контурів (хвиль) з удвічі меншими амплітудами, що рухаються вправо з однаковими сталими швидкостями $a$ та однаковими фазами. Побудуйте відповідні графіки.

*Відповідь*. Див. рис. 4.5.

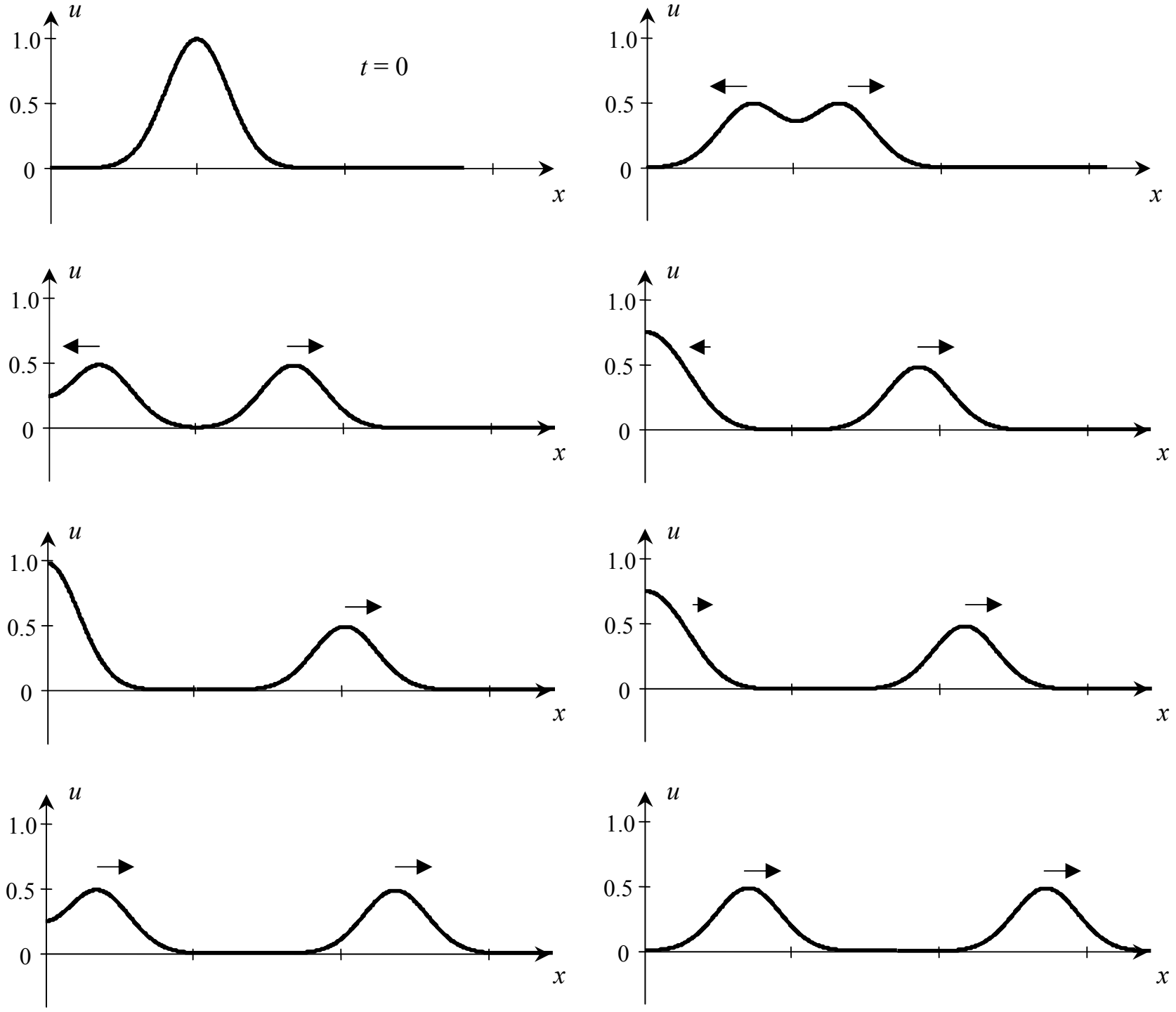

Рис. 4.5. Відбивання хвилі від вільного кінця напівобмеженої струни. Час зростає зліва направо та зверху вниз, $A = 1$

На завершення цього підрозділу розглянемо задачу Коші про вимушені коливання напівобмеженої струни. Для аналізу достатньо обмежитися випадком, коли вона складається з неоднорідного рівняння

$$\frac{\partial^2 u}{\partial t^2} = a^2 \frac{\partial^2 u}{\partial x^2} + f(x,t), \quad 0 < x < \infty, \quad t > 0, \tag{4.62}$$

крайової умови типу (4.43) або (4.55), а також нульових початкових умов

$$u(x,0) = 0, \quad u_t(x,0) = 0. \tag{4.63}$$

**Завдання 4.3.7.** Скориставшись формулою (4.54) та принципом Дюамеля, покажіть, що розв'язок задачі Коші (4.62), (4.43) та (4.63) для напівобмеженої струни із закріпленим лівим кінцем має вигляд

$$u(x,t)=\frac{1}{2a}\int_0^t d\tau \int_{|x-a(t-\tau)|}^{x+a(t-\tau)} f(x',\tau)dx'. \tag{4.64}$$

**Завдання 4.3.8.** Скориставшись формулою (4.61) та принципом Дюамеля, покажіть, що розв'язок задачі Коші (4.62), (4.55) та (4.63) для напівобмеженої струни з незакріпленим лівим кінцем має вигляд

$$u(x,t)=\frac{1}{2a}\int_0^t d\tau \left\{ \int_0^{x+a(t-\tau)} f(x',\tau)dx' + \operatorname{sign}\big(a(t-\tau)-x\big)\int_0^{|x-a(t-\tau)|} f(x',\tau)dx' \right\}. \tag{4.65}$$

**Завдання 4.3.9.** Покажіть, що формули (4.64) та (4.65) можна отримати й методом продовження, доозначивши $f(x,t)$ на від'ємну піввісь як відповідно непарну та парну функції.

Як і у випадку необмеженої струни, можна також довести, що під дією гармонічної сили $\tilde f(x,t)=\tilde f(x)e^{-i\omega t}$ всі точки напівобмеженої струни з часом починають коливатися за гармонічним законом $\tilde A_\omega(x)e^{-i\omega t}$ з комплекснозначною амплітудою

$$\tilde A_\omega(x)=\int_0^\infty G_\omega(|x-x'|)\tilde f(x')dx', \tag{4.66}$$

де $G_\omega(x,x')$ — частотна функція Гріна для напівобмеженої струни. Явний вигляд функції $G_\omega(x,x')$ залежить від способу закріплення кінця струни.

**Завдання 4.3.10.** Покажіть, що для напівобмеженої струни з жорстко закріпленим лівим кінцем

$$G_\omega(x,x')=\frac{i}{2a\omega}\left(e^{i\frac{\omega}{a}|x-x'|}-e^{i\frac{\omega}{a}(x+x')}\right),\ \ x,x'>0, \tag{4.67}$$

а для напівобмеженої струни з незакріпленим лівим кінцем

$$G_\omega(x,x')=\frac{i}{2a\omega}\left(e^{i\frac{\omega}{a}|x-x'|}+e^{i\frac{\omega}{a}(x+x')}\right),\ \ x,x'>0. \tag{4.68}$$

*Вказівка*. Скористайтеся методом продовження і формулами (4.38), (4.39), (4.66).

## 4.4. ВІЛЬНІ КОЛИВАННЯ ОБМЕЖЕНОЇ СТРУНИ ІЗ ЗАКРІПЛЕНИМИ КІНЦЯМИ

Повернімося до ситуації, коли вплив обох кінців струни на її коливання суттєвий. Як і раніше (див. підрозділ 3.2), уважатимемо, що рівноважний профіль натягненої струни збігається з відрізком $[0,l]$ дійсної осі. Якщо струна однорідна, а її кінці жорстко закріплені, то відповідна *крайова задача* про вільні коливання такої обмеженої струни полягає у відшуканні двічі неперервно диференційовної функції $u(x,t)$, яка разом із рівнянням

$$\frac{\partial^2 u}{\partial t^2} = a^2 \frac{\partial^2 u}{\partial x^2}, \quad 0 < x < l, \quad t > 0, \tag{4.69}$$

задовольняє крайові та початкові умови

$$u(0,t) = u(l,t) = 0, \quad t \geq 0, \tag{4.70}$$

$$u(x,0) = u_0(x), \quad u_t(x,0) = v_0(x), \quad 0 \leq x \leq l, \tag{4.71}$$

де функції $u_0(x)$ і $v_0(x)$ належать певному класу та узгоджуються з умовами (4.70), тобто дорівнюють нулю на кінцях проміжку $[0,l]$.

Припустимо спочатку, що $v_0(x)$ — неперервно диференційовна, а $u_0(x)$ — двічі неперервно диференційовна на відрізку $[0,l]$, та побудуємо розв'язок задачі (4.69)–(4.71) за допомогою методу продовження. Для цього перейдемо до допоміжної задачі Коші для однорідного рівняння коливань на всій осі $(-\infty,\infty)$:

$$\frac{\partial^2 U}{\partial t^2} = a^2 \frac{\partial^2 U}{\partial x^2}, \quad -\infty < x < \infty, \quad t > 0, \tag{4.72}$$

$$U(x,0) = U_0(x), \quad U_t(x,0) = V_0(x), \tag{4.73}$$

де початкові функції $U_0(x)$ та $V_0(x)$ є спеціальними продовженнями функцій $u_0(x)$ та $v_0(x)$ із відрізка $[0,l]$ на всю вісь. Ці продовження будуються таким чином, щоб розв'язок Д'Аламбера

$$U(x,t) = \frac{U_0(x+at) + U_0(x-at)}{2} + \frac{1}{2a} \int_{x-at}^{x+at} V_0(x')dx' \tag{4.74}$$

задачі (4.72), (4.73) автоматично задовольняв би обидві умови (4.70).

Як було показано в попередньому підрозділі, умова $U(0,t) = 0$ виконується, якщо продовжені функції $U_0(x)$ та $V_0(x)$ непарні відносно точки $x = 0$. Доведемо тепер, що друга умова $U(l,t) = 0$ задовольняється, якщо $U_0(x)$ та $V_0(x)$ непарні й відносно точки $x = l$. Справді, з формули (4.74) випливає, що при $x=l$ і довільного $t>0$ повинно справджуватися співвідношення

$$U_0(l+at)+U_0(l-at)+\frac{1}{a}\int_{l-at}^{l+at}V_0(x')dx'=0. \qquad (4.75)$$

Останнє задовольняється тотожно, якщо для будь-якого $x\in(-\infty,\infty)$

$$U_0(l+x)=-U_0(l-x),\ \ V_0(l+x)=-V_0(l-x), \qquad (4.76)$$

що і є умовами непарності функцій $U_0(x)$ та $V_0(x)$ відносно точки $x=l$.

Продовжені функції $U_0(x)$ та $V_0(x)$ із зазначеними властивостями можна побудувати за допомогою наступного покрокового алгоритму. Спочатку функції $u_0(x)$ та $v_0(x)$ із відрізка $[0,l]$ продовжуємо на відрізок $[-l,l]$ як непарні відносно точки $x=0$. Здобуті таким чином функції $u_0^{(1)}(x)$ та $v_0^{(1)}(x)$ із відрізка $[-l,l]$ продовжуємо на відрізок $[-l,3l]$ як непарні відносно точки $x=l$. Отримуємо нові функції $u_0^{(2)}(x)$ та $v_0^{(2)}(x)$, які далі з відрізка $[-l,3l]$ продовжуємо на відрізок $[-3l,3l]$ як непарні відносно точки $x=0$. Дістаємо деякі функції $u_0^{(3)}(x)$ та $v_0^{(3)}(x)$, які з відрізка $[-3l,3l]$ продовжуємо на відрізок $[-3l,5l]$ як непарні відносно точки $x=l$. Повторивши цю процедуру скінченну кількість разів $n$, можемо, у принципі, знайти певні апроксимації $u_0^{(n)}(x)$ та $v_0^{(n)}(x)$ до функцій $U_0(x)$ та $V_0(x)$, а далі — за допомогою формули (4.74) і цих апроксимацій — відновити значення зміщення $u(x,t)$ струни в точках $x\in(0,l)$ у довільний момент часу $t>0$ із заданою точністю.

Очевидно, що цей алгоритм дозволяє розвинути відносно простий комп'ютерний код для числового розв'язування задачі. Однак він не дає можливості простежити за загальними якісними рисами руху струни із закріпленими кінцями як лінійної коливальної системи.

Для якісного аналізу побудованого розв'язку крайової (4.69)–(4.71) подамо його в більш зручному аналітичному вигляді. Щоб просунутися у цьому напрямі, перш за все звернемо увагу на такий факт: функція $\varphi(x)$, яка непарна на всій осі, $\varphi(-x)=-\varphi(x)$, буде непарна ще й відносно точки $x=l$, $\varphi(l-x)=-\varphi(l+x)$, тоді й лише тоді, коли вона періодична з періодом $2l$, тобто коли

$$\varphi(x+2l)=\varphi(x).$$

Справді, якщо $\varphi(x)$ — непарна на всій осі та непарна відносно точки $x=l$, то для довільної точки $x\in(-\infty,\infty)$ маємо:

$$\varphi(x+2l)=\varphi(l+(l+x))=-\varphi(l-(l+x))=-\varphi(-x)=\varphi(x).$$

З іншого боку, якщо $\varphi(x)$ — непарна на всій осі та періодична з періодом $2l$, то

$$\varphi(l+x)=\varphi(2l+(x-l))=\varphi(x-l)=-\varphi(l-x).$$

Тепер нагадаємо, що кожній неперервній періодичній функції $\varphi(x)$, $x\in(-\infty,\infty)$, з періодом $2l$ відповідає *ряд Фур'є*

$$\varphi(x)=\frac{a_0}{2}+\sum_{n=1}^{\infty}\left[a_n\cos\frac{\pi nx}{l}+b_n\sin\frac{\pi nx}{l}\right], \tag{4.77}$$

де

$$a_0=\frac{1}{l}\int_{-l}^{+l}\varphi(x)\,dx, \tag{4.78}$$

$$a_n=\frac{1}{l}\int_{-l}^{+l}\varphi(x)\cos\frac{\pi nx}{l}dx,\quad n=1,2,3,..., \tag{4.79}$$

$$b_n=\frac{1}{l}\int_{-l}^{+l}\varphi(x)\sin\frac{\pi nx}{l}dx,\quad n=1,2,3,..., \tag{4.80}$$

який збігається до $\varphi(x)$ принаймні «в середньому квадратичному», тобто

$$\int_{-l}^{+l}\left|\varphi(x)-\frac{a_0}{2}-\sum_{n=1}^{N}\left[a_n\cos\frac{\pi nx}{l}+b_n\sin\frac{\pi nx}{l}\right]\right|^2dx\underset{N\to\infty}{\to}0. \tag{4.81}$$

При цьому $a_n,b_n\to 0$ при $n\to\infty$ щонайменше так, що

$$\sum_{n=1}^{\infty}\left\{|a_n|^2+|b_n|^2\right\}<\infty.$$

Підкреслимо, що *зображення довільної неперервної функції* $\varphi(x)$ *на проміжку* $[0,l]$ *у вигляді ряду (4.77) єдине.* Справді, нехай існує ще один ряд

$$\varphi(x)=\frac{\tilde{a}_0}{2}+\sum_{n=1}^{\infty}\left[\tilde{a}_n\cos\frac{\pi nx}{l}+\tilde{b}_n\sin\frac{\pi nx}{l}\right], \tag{4.82}$$

який збігається до $\varphi(x)$ «у середньому квадратичному», тобто для послідовності часткових сум

$$\tilde{S}_N(x)=\frac{\tilde{a}_0}{2}+\sum_{n=1}^{N}\left[\tilde{a}_n\cos\frac{\pi nx}{l}+\tilde{b}_n\sin\frac{\pi nx}{l}\right]$$

виконується співвідношення

$$\int_{-l}^{l}\left|\varphi(x)-\tilde{S}_N(x)\right|^2dx\underset{N\to\infty}{\to}0. \tag{4.83}$$

Подамо коефіцієнти Фур'є $a_n$ та $b_n$ ряду (4.77) для функції $\varphi(x)$ у вигляді ($N \ge n$)

$$a_0 = \frac{1}{l}\int_{-l}^{l}\varphi(x)dx = \frac{1}{l}\int_{-l}^{l}\tilde{S}_N(x)dx + \frac{1}{l}\int_{-l}^{l}\left[\varphi(x) - \tilde{S}_N(x)\right]dx, \tag{4.84}$$

$$a_n = \frac{1}{l}\int_{-l}^{l}\varphi(x)\cos\frac{\pi n x}{l}dx =$$

$$= \frac{1}{l}\int_{-l}^{l}\tilde{S}_N(x)\cos\frac{\pi n x}{l}dx + \frac{1}{l}\int_{-l}^{l}\left[\varphi(x) - \tilde{S}_N(x)\right]\cos\frac{\pi n x}{l}dx, \tag{4.85}$$

$$b_n = \frac{1}{l}\int_{-l}^{l}\varphi(x)\sin\frac{\pi n x}{l}dx =$$

$$= \frac{1}{l}\int_{-l}^{l}\tilde{S}_N(x)\sin\frac{\pi n x}{l}dx + \frac{1}{l}\int_{-l}^{l}\left[\varphi(x) - \tilde{S}_N(x)\right]\sin\frac{\pi n x}{l}dx. \tag{4.86}$$

Оскільки

$$\int_{-l}^{l}\cos\frac{\pi n x}{l}dx = 0, \quad \int_{-l}^{l}\sin\frac{\pi n x}{l}dx = 0, \quad n = 1,2,3,..., \tag{4.87}$$

то перший інтеграл у правій частині (4.84) для будь-якого $N \ge 0$ дорівнює $\tilde{a}_0$. Для другого інтеграла, скориставшись співвідношенням (4.83), маємо:

$$\left|\frac{1}{l}\int_{-l}^{l}\left[\varphi(x) - \tilde{S}_N(x)\right]dx\right| \le \sqrt{\frac{2}{l}}\left\{\int_{-l}^{l}\left|\varphi(x) - \tilde{S}_N(x)\right|^2 dx\right\}^{1/2} \underset{N\to\infty}{\to} 0.$$

Далі, за допомогою (4.87) та рівностей

$$\frac{1}{l}\int_{-l}^{l}\cos\frac{\pi m x}{l}\cos\frac{\pi n x}{l}dx = \delta_{mn}, \quad \frac{1}{l}\int_{-l}^{l}\sin\frac{\pi m x}{l}\sin\frac{\pi n x}{l}dx = \delta_{mn},$$
$$\int_{-l}^{l}\cos\frac{\pi m x}{l}\sin\frac{\pi n x}{l}dx = 0, \quad m,n = 1,2,3,..., \tag{4.88}$$

де $\delta_{mn}$ — символ Кронекера, знаходимо, що перші інтеграли у правих частинах формул (4.85) та (4.86) для будь-якого $N \ge n \ge 1$ дорівнюють відповідно $\tilde{a}_n$ та $\tilde{b}_n$. Для решти інтегралів за допомогою нерівності Коші — Буняковського дістаємо:

$$\left|\frac{1}{l}\int_{-l}^{l}\left[\varphi(x) - \tilde{S}_N(x)\right]\cos\frac{\pi n x}{l}dx\right| \le \frac{1}{l}\left\{\int_{-l}^{l}\cos^2\frac{\pi n x}{l}dx \cdot \int_{-l}^{l}\left|\varphi(x) - \tilde{S}_N(x)\right|^2 dx\right\}^{1/2} \underset{N\to\infty}{\to} 0,$$

$$\left|\frac{1}{l}\int_{-l}^{l}\left[\varphi(x)-\tilde{S}_N(x)\right]\sin\frac{\pi nx}{l}dx\right|\le\frac{1}{l}\left\{\int_{-l}^{l}\sin^2\frac{\pi nx}{l}dx\cdot\int_{-l}^{l}\left|\varphi(x)-\tilde{S}_N(x)\right|^2dx\right\}^{1/2}\underset{N\to\infty}{\to}0.$$

Отже, за умови (4.83) коефіцієнти $\tilde{a}_n$ та $\tilde{b}_n$ ряду (4.82) збігаються з відповідними коефіцієнтами $a_n$ та $b_n$ ряду (4.77).

Слід зазначити, що класичний алгоритм відновлення періодичної функції $\varphi(x)$ із періодом $2l$ за її відомими коефіцієнтами Фур'є $\{a_n, b_n\}$ як границі при $N\to\infty$ послідовності часткових тригонометричних сум

$$S_N(x)=\frac{a_0}{2}+\sum_{n=1}^{N}\left[a_n\cos\frac{\pi nx}{l}+b_n\sin\frac{\pi nx}{l}\right]$$

не завжди є ефективним чи, більше того, застосовним — навіть тоді, коли заздалегідь відомо, що функція $\varphi(x)$, яку треба відновити, неперервна на відрізку $[-l,l]$. Це пов'язано з тим, що, взагалі кажучи, ряд Фур'є (4.77) для довільної неперервної функції $\varphi(x)$ може бути розбіжним як в окремих точках, так і на щільних множинах точок відрізка $[-l,l]$. Однак можна запропонувати більш ефективні алгоритми відновлення функції за її коефіцієнтами Фур'є. Наприклад, усереднена послідовність

$$\sigma_N(x)=\frac{1}{N}\left[S_0+S_1(x)+\ldots+S_{N-1}(x)\right]=$$
$$=\frac{a_0}{2}+\sum_{n=1}^{N-1}\left(1-\frac{n}{N}\right)\left[a_n\cos\frac{\pi nx}{l}+b_n\sin\frac{\pi nx}{l}\right] \tag{4.89}$$

часткових тригонометричних сум $S_N(x)$ збігається до $\varphi(x)$ у кожній точці, де $\varphi(x)$ неперервна. Для неперервної ж функції $\varphi(x)$ послідовність (4.89) збігається до $\varphi(x)$ рівномірно на відрізку $[-l,l]$. У правильності останнього твердження можна переконатися, помітивши, що, згідно з формулами (4.78)–(4.80) та (4.89),

$$\sigma_N(x)=\frac{1}{2lN}\int_{-l}^{l}\frac{\sin^2\frac{\pi N(x-x')}{2l}}{\sin^2\frac{\pi(x-x')}{2l}}\varphi(x')dx' \tag{4.90}$$

і що послідовність невід'ємних функцій

$$f_N(x)=\frac{1}{2lN}\frac{\sin^2\frac{\pi Nx}{2l}}{\sin^2\frac{\pi x}{2l}} \tag{4.91}$$

прямує при $N\to\infty$ до дельта-функції Дірака $\delta(x)$.

Коли ж $\varphi(x)$ — неперервно диференційовна періодична функція з періодом $2l$, то ряд (4.77) збігається до неї абсолютно й рівномірно на всій осі. І, нарешті, якщо така функція $\varphi(x)$ має $k \geq 1$ неперервних похідних, то її коефіцієнти Фур'є ведуть себе із зростанням номера $n$ як $a_n = o\left(\frac{1}{n^k}\right)$, $b_n = o\left(\frac{1}{n^k}\right)$, при цьому ряди $\sum_{n=1}^{\infty} n^{k-1} |a_n|$, $\sum_{n=1}^{\infty} n^{k-1} |b_n|$, $\sum_{n=1}^{\infty} n^{2k} |a_n|^2$ та $\sum_{n=1}^{\infty} n^{2k} |b_n|^2$ збігаються.

**Завдання 4.4.1.** Доведіть рівності (4.87) і (4.88). Також покажіть, що для системи функцій $\left\{\sin\frac{\pi n x}{l}\right\}$, $n = 1,2,3,...$, справджуються рівності

$$\frac{2}{l}\int_0^l \sin\frac{\pi m x}{l}\sin\frac{\pi n x}{l}dx = \delta_{mn}. \tag{4.92}$$

**Завдання 4.4.2.** Доведіть, що непарній неперервній періодичній функції $\varphi(x)$ із періодом $2l$ відповідає ряд Фур'є

$$\varphi(x) = \sum_{n=1}^{\infty} b_n \sin\frac{\pi n x}{l}, \tag{4.93}$$

у якому, з огляду на формули (4.78)–(4.80) та непарність $\varphi(x)$, коефіцієнти $a_n = 0$, $n = 0,1,2,...$, і

$$b_n = \frac{2}{l}\int_0^l \varphi(x)\sin\frac{\pi n x}{l}dx, \quad n = 1,2,3,.... \tag{4.94}$$

**Завдання 4.4.3.** Доведіть формулу (4.90).

*Вказівка.* Скориставшись формулами (4.78)–(4.80), формулу (4.89) можна подати у вигляді

$$\sigma_N(x) = \frac{1}{2l}\int_{-l}^{l} dx'\,\varphi(x') +$$

$$+\sum_{n=1}^{N-1}\left(1-\frac{n}{N}\right)\frac{1}{l}\int_{-l}^{l} dx'\,\varphi(x')\left[\cos\frac{\pi n x'}{l}\cos\frac{\pi n x}{l} + \sin\frac{\pi n x'}{l}\sin\frac{\pi n x}{l}\right] =$$

$$= \frac{1}{l}\int_{-l}^{l} dx'\,\varphi(x')\left[\frac{1}{2} + \sum_{n=1}^{N-1}\left(1-\frac{n}{N}\right)\cos\frac{\pi n(x-x')}{l}\right].$$

Обчислення функціональних сум у цьому виразі за допомогою співвідношень

$$\frac{1}{2}+\sum_{n=1}^{N-1}\cos\alpha n=\frac{1}{2}+\mathrm{Re}\sum_{n=1}^{N-1}e^{i\alpha n},$$

$$\sum_{n=1}^{N-1}n\cos\alpha n=\frac{\partial}{\partial\alpha}\sum_{n=1}^{N-1}\sin\alpha n=\frac{\partial}{\partial\alpha}\mathrm{Im}\sum_{n=1}^{N-1}e^{i\alpha n},\ \ \alpha\equiv\frac{\pi(x-x')}{l},$$

зводиться до знаходження суми перших $N-1$ членів геометричної прогресії з першим членом $e^{i\alpha}$ та знаменником $e^{i\alpha}$:

$$\sum_{n=1}^{N-1}e^{i\alpha n}=\frac{e^{i\alpha}\left[1-e^{i\alpha(N-1)}\right]}{1-e^{i\alpha}}=\frac{e^{i\alpha}\left[1-e^{i\alpha(N-1)}\right]\left(1-e^{-i\alpha}\right)}{\left(1-e^{i\alpha}\right)\left(1-e^{-i\alpha}\right)}=\frac{e^{i\alpha}-1+e^{i\alpha(N-1)}-e^{i\alpha N}}{4\sin^2(\alpha/2)}.$$

**Завдання 4.4.4.** Доведіть, що послідовність функцій (4.91) прямує при $N\to\infty$ до дельта-функції Дірака $\delta(x)$.

*Вказівка.* Скориставшись співвідношенням $\lim\limits_{N\to\infty}\dfrac{\sin^2 Nx}{\pi Nx^2}=\delta(x)$, перевірте, що для довільної основної функції $\varphi(x)$

$$\lim_{N\to\infty}\int_{-l}^{l}f_N(x)\varphi(x)\,dx=\varphi(0).$$

Перейдемо тепер до аналізу загального розв’язку крайової задачі (4.69)–(4.71) про вільні поперечні коливання однорідної струни із закріпленими кінцями. Як було показано, він дається розв’язком (4.74) допоміжної задачі (4.72), (4.73), записаним для точок $x\in(0,l)$ та довільного моменту часу $t>0$. Функції $U_0(x)$ та $V_0(x)$ у формулі (4.74) — це непарні і періодичні, з періодом $2l$, продовження початкових функцій $u_0(x)$ та $v_0(x)$ (4.71) з відрізка $[0,l]$ на всю вісь. Підкреслимо, що внаслідок умов $u_0(0)=u_0(l)=0$ і $v_0(0)=v_0(l)=0$ продовження $U_0(x)$ та $V_0(x)$ є неперервними функціями на осі. Скориставшись цим фактом та подавши $U_0(x)$ у вигляді ряду Фур’є (див. завдання 4.4.2)

$$U_0(x)=\sum_{n=1}^{\infty}A_n\sin\frac{\pi nx}{l},\tag{4.95}$$

$$A_n=\frac{2}{l}\int_0^l U_0(x)\sin\frac{\pi nx}{l}dx=\frac{2}{l}\int_0^l u_0(x)\sin\frac{\pi nx}{l}dx,\tag{4.96}$$

знаходимо:

$$\begin{aligned}\frac{U_0(x+at)+U_0(x-at)}{2}&=\sum_{n=1}^{\infty}A_n\frac{1}{2}\left[\sin\frac{\pi n(x+at)}{l}+\sin\frac{\pi n(x-at)}{l}\right]=\\&=\sum_{n=1}^{\infty}A_n\cos\frac{\pi nat}{l}\sin\frac{\pi nx}{l}.\end{aligned}\tag{4.97}$$

Аналогічним чином, подавши $V_0(x)$ у вигляді ряду Фур'є

$$V_0(x) = \sum_{n=1}^{\infty} B_n \sin\frac{\pi n x}{l}, \tag{4.98}$$

$$B_n = \frac{2}{l}\int_0^l V_0(x)\sin\frac{\pi n x}{l}dx = \frac{2}{l}\int_0^l v_0(x)\sin\frac{\pi n x}{l}dx, \tag{4.99}$$

почленним інтегруванням ряду (4.98) дістанемо:

$$\frac{1}{2a}\int_{x-at}^{x+at} V_0(x')dx' = \sum_{n=1}^{\infty} B_n \frac{l}{\pi n a}\sin\frac{\pi n a t}{l}\sin\frac{\pi n x}{l}. \tag{4.100}$$

Підставивши вирази (4.97) та (4.100) у формулу (4.74), приходимо до висновку, що *для початкових функцій* $u_0(x)$ *та* $v_0(x)$, *що є відповідно двічі та один раз неперервно диференційовними, розв'язок крайової задачі (4.69)–(4.71) про вільні коливання однорідної струни із закріпленими кінцями має вигляд*

$$u(x,t) = \sum_{n=1}^{\infty} q_n(t)\sin\frac{\pi n x}{l}, \tag{4.101}$$

де

$$q_n(t) = A_n \cos\omega_n t + B_n \frac{\sin\omega_n t}{\omega_n}, \quad \omega_n = \frac{\pi n a}{l}, \tag{4.102}$$

а коефіцієнти $A_n$ та $B_n$ даються формулами (4.96) та (4.99). Неважко бачити, що цей розв'язок *неперервний*. Справді, з огляду на формули (4.102), (4.96), (4.99) та властивості початкових функцій, для всіх членів $u_n(x,t)$ функціонального ряду (4.101) маємо оцінку

$$|u_n(x,t)| \le |A_n| + \frac{l}{\pi a n}|B_n|, \quad 0 < x < l, \quad t > 0,$$

де, згідно із загальними властивостями рядів Фур'є (див. текст перед завданням 4.4.1), коефіцієнти $A_n = o\left(\frac{1}{n^2}\right)$, $B_n = o\left(\frac{1}{n}\right)$. Оскільки мажорантний числовий ряд (з додатними членами)

$$\sum_{n=1}^{\infty}\left(|A_n| + \frac{l}{\pi a n}|B_n|\right)$$

збігається, то за ознакою Вейєрштрасса ряд (4.101) при $t > 0$ збігається абсолютно й рівномірно до деякої неперервної функції на проміжку $[0,l]$. Очевидно, що ця функція задовольняє крайові умови (4.70).

Строго кажучи, після того, як ми скористалися рядами Фур'є, щоб подати розв'язок (4.74) крайової задачі (4.69)–(4.71) у вигляді ряду (4.101), вимоги до початкових функцій треба уточнити, оскільки розв'язок (4.101) теж має задовольняти рівняння (4.69), тобто бути двічі диференційовним за змінними $x$ і $t$, та неперервно переходити, зі своїми відповідними похідними, у початкові функції при $t \downarrow 0$. Щоб з'ясувати ці питання, спершу зазначимо, що достатньою умовою можливості почленного диференціювання ряду (4.101) певну кількість разів є рівномірна збіжність функціонального ряду, який формально одержуємо після такого диференціювання, а достатньою умовою рівномірної збіжності функціонального ряду — уже згадувана ознака Вейєрштрасса. Для функціональних рядів, що відповідають першим $(k=1)$ і другим $(k=2)$ похідним ряду (4.101) за змінними $x$ і $t$, мажорантними числовими рядами виступають лінійні комбінації рядів виду

$$\sum_{n=1}^{\infty} n^k\,|A_n|,\quad \sum_{n=1}^{\infty} n^{k-1}\,|B_n|,\quad k=1,2,$$

де коефіцієнти $A_n$ і $B_n$ визначаються формулами (4.96) та (4.99). Ці ряди гарантовано збігаються, якщо початкові функції $u_0(x)$ і $v_0(x)$ є відповідно *тричі і двічі неперервно диференційовними* та задовольняють умови

$$u_0(0)=u_0(l)=0,\quad u_0''(0)=u_0''(l)=0,\quad v_0(0)=v_0(l)=0. \qquad (4.103)$$

За цих умов ряд (4.101) і ряди, члени яких є першими та другими частинними похідними членів ряду (4.101), збігаються абсолютно й рівномірно на відрізку $[0,l]$ і при $t \geq 0$, а тому сума ряду (4.101) є двічі неперервно диференційовною функцією. Оскільки кожний член ряду (4.101) задовольняє рівняння (4.69) і крайові умови (4.70), то сума ряду (4.101) теж їх задовольняє. Зазначимо також, що умови (4.103) є природними, бо виражають рівність нулеві зміщень, швидкостей і прискорень на закріплених кінцях струни.

Нагадаємо, що для побудови двічі неперервно диференційовного розв'язку (4.74) цієї ж задачі методом продовження достатньо було обмежитися припущенням, що початкові функції $u_0(x)$ і $v_0(x)$ є відповідно *двічі та один раз неперервно диференційовними* та допускають неперервні непарні періодичні продовження з інтервалу $(0,l)$ на всю числову вісь. Очевидно, що при такому підході умови (4.103) теж виконуються[1].

---

[1] Неперервність перших похідних цих продовжень у точках $x=0,l$ виконується автоматично, оскільки вони — парні функції.

Таким чином, для позитивного розв'язання методами аналізу питань збіжності рядів виду (4.101), можливості їх почленного диференціювання тощо доводиться, як правило, висувати додаткові вимоги (які в загальному випадку залежать від обраного способу розв'язування крайової задачі) до початкових функцій, як це було продемонстровано вище на прикладі задачі про вільні коливання однорідної струни із закріпленими кінцями. Однак при розгляді фізичних та прикладних задач ці питання є радше технічними і зазвичай опускаються — обмежуються побудовою лише самих розв'язків крайових задач у вигляді рядів Фур'є, не обов'язково двічі неперервно диференційовних для заданих початкових функцій. Такі розв'язки називаються *узагальненими*. Зокрема, для функцій $u_0(x)$ і $v_0(x)$, що є відповідно двічі та один раз неперервно диференційовними і задовольняють крайові умови (4.70), ряд (4.101) дає узагальнений розв'язок крайової задачі (4.69)–(4.71) про вільні коливання однорідної струни із закріпленими кінцями, такий, що: для довільного $n$ його часткова сума $S_n(x,t)$ строго задовольняє рівняння коливань (4.69) і крайові умови (4.70), а при достатньо великому $n$ наближається до початкових функцій (4.71) *із похибкою, як завгодно малою рівномірно на всьому інтервалі* $(0,l)$ *і при* $t>0$; повна сума ряду (4.101) відрізняється від $S_n(x,t)$ рівномірно мало на відрізку $[0,l]$ і при $t \geq 0$.

Подальшому аналізу властивостей розкладів типу (4.101) присвячено підрозділ 4.11.

## 4.5. РОЗРИВНІ РОЗВ'ЯЗКИ ХВИЛЬОВОГО РІВНЯННЯ. ЯВИЩЕ ГІББСА

Припустимо, що початкові функції $u_0(x)$ і $v_0(x)$ задачі Коші для однорідного рівняння коливань на всій осі є двічі та один раз неперервно диференційовними скрізь, за винятком множини ізольованих точок $\{x_m\}$, де одна або обидві функції мають розриви першого роду. Поширивши формулу Д'Аламбера (4.13) на цей випадок, дістанемо функцію $u(x,t)$, яка задовольняє однорідне хвильове рівняння на всій півплощині $-\infty < x < \infty$, $t>0$, крім множини прямих

$$x \pm at = x_m,$$

і при цьому всюди, крім точок $\{x_m\}$,

$$\lim_{t\to 0} u(x,t) = u_0(x), \quad \lim_{t\to 0} v(x,t) = v_0(x).$$

Як уже зазначалося, такі узагальнені розв'язки хвильового рівняння не мають змісту для задач про пружні коливання струн чи стержнів. Однак існують інші, не менш важливі, фізичні та технічні задачі для хвильового рівняння, для яких розв'язки з такими властивостями є цілком природними.

Розглянемо, наприклад, задачу про локальні коливання електричного струму $I(x,t)$ і потенціалу $V(x,t)$ в тонкому провідному дроті; для простоти вважатимемо, що опором дроту та втратами заряду крізь його поверхню можна знехтувати. За цих умов зміна заряду на ділянці дроту $(x,x+dx)$ за інтервал часу $(t,t+dt)$ дорівнює

$$\left[I(x,t)-I(x+dx,t)\right]dt \approx -\frac{\partial I(x,t)}{\partial x}\,dx\,dt$$

і має збігатися зі зміною заряду

$$Cdx\left[V(x,t+dt)-V(x,t)\right] \approx C\frac{\partial V(x,t)}{\partial t}dx\,dt,$$

зумовленою коливанням потенціалу на цій ділянці за той самий інтервал часу. В останній формулі множник $C$ (умовно) визначається як електрична ємність одиниці довжини однорідного дроту. Прирівнюючи виписані вирази, отримуємо рівняння

$$C\frac{\partial V(x,t)}{\partial t}+\frac{\partial I(x,t)}{\partial x}=0. \qquad (4.104)$$

З іншого боку, із закону Ома для ділянки дроту $(x,x+dx)$ випливає, що падіння напруги на ній має дорівнювати електрорушійній силі самоіндукції. Маємо:

$$\left[V(x+dx,t)-V(x,t)\right] \approx \frac{\partial V(x,t)}{\partial x}dx=-Ldx\frac{\partial I(x,t)}{\partial t},$$

де стала $L$ (умовно) визначається як коефіцієнт самоіндукції одиниці довжини однорідного дроту. Отже, дістаємо ще одне рівняння

$$\frac{\partial V(x,t)}{\partial x}+L\frac{\partial I(x,t)}{\partial t}=0. \qquad (4.105)$$

Треба зауважити, що з фізичного погляду міркування, які були використані, щоб одержати рівняння (4.104) і (4.105), є нестрогими, а самі ці рівняння — наближеними. Однак при певних обмеженнях[1] на конфігурації провідників та частоти електромагнітних коливань рівняння (4.104) і (4.105) цілком придатні для інженерних застосувань.

---

[1] Зокрема, радіуси кривини провідників мають бути значно більшими за характерний лінійний розмір (діаметр) системи $l$, а характерний час зміни зовнішнього поля (наприклад, період) $T$ — достатньо великим, щоб значення струму та напруги, викликаних ним у провідниках, устигали за час $l/a=l\sqrt{LC}$ підстроїтися під нього: $T>>l\sqrt{LC}$.

З рівнянь (4.104) і (4.105) легко вивести замкнені рівняння для величин $I(x,t)$ та $V(x,t)$. Здиференціювавши, наприклад, (4.104) за змінною $x$ та виразивши з (4.105) похідну $\partial V(x,t)/\partial x$ через похідну $\partial I(x,t)/\partial t$, знаходимо, що змінний струм у дроті задовольняє хвильове рівняння

$$\frac{\partial^2 I(x,t)}{\partial t^2} = a^2 \frac{\partial^2 I(x,t)}{\partial x^2}, \quad a^2 = \frac{1}{LC}. \tag{4.106}$$

Аналогічне рівняння справджується й для потенціалу $V(x,t)$.

Якщо дріт має скінченну довжину $l$, а його кінці $x = 0$ та $x = l$ є ізольованими, то рівняння (4.106) треба доповнити крайовими умовами

$$I(0,t) = 0, \quad I(l,t) = 0. \tag{4.107}$$

Розглянемо тепер питання про передачу електричних сигналів по провідній лінії. На практиці такі сигнали можуть генеруватися кусково-гладкими імпульсами струму (наприклад, П- або пилкоподібними), тому, на відміну від задач про коливання струни, у цих задачах допустимими є й кусково-гладкі початкові функції, а, отже, і розривні розв'язки рівняння (4.106), що породжуються такими функціями та для нескінченної лінії даються формулою Д'Аламбера (4.13). Якщо ж провідна лінія має скінченну довжину, то, як і при аналізі коливань струни із закріпленими кінцями, розв'язки відповідних крайових задач типу (4.106), (4.107) для відрізка $[0,l]$ можна шукати, продовжуючи початкові функції з відрізка $[0,l]$ на всю вісь як непарні й періодичні з періодом $2l$ та переходячи до задачі Коші на всій осі. Підкреслимо, що навіть якщо всередині проміжку $[0,l]$ початкові функції не мають розривів, але не дорівнюють нулю на його кінцях, то зазначені їх продовження на всю вісь матимуть розриви в точках $2ml$ та (або) $(2m+1)l$, $m = 0, \pm 1, \pm 2, \ldots$, а тому будуть розривними і розв'язки, породжувані цими продовженнями.

При побудові розривних розв'язків хвильового рівняння на основі формул Д'Аламбера та методу продовження (коли це можливо) ніяких принципових труднощів не виникає. Однак вони з'являються при намаганні подати ці розривні розв'язки у вигляді рядів Фур'є, як це робилося в задачах про коливання обмеженої струни. Справа в тому, що для розривних розв'язків відповідні ряди Фур'є не збігаються рівномірно, що веде до небажаних ефектів при передачі електричних сигналів по провідних лініях. Тут треба взяти до уваги той факт, що реальні провідні лінії мають обмежену частотну смугу пропускання. Унаслідок цього замість електричного сигналу, що зображається у вигляді нескінченного ряду Фур'є, по лінії проходить спотворений сигнал, якому відповідає лише часткова сума цього ряду. При цьому, як ми зараз покажемо, відхилення розривного електричного сигналу, який збуджується в лінії, від того, що пройде по ній, узагалі кажучи, не зменшується при розширенні смуги пропускання лінії, тобто при збільшенні номера часткової суми відповідного ряду Фур'є.

Нехай $u_0(x)$ та $v_0(x)$ — двічі та один раз неперервно диференційовні функції на відрізку $[0,l]$, а $U_0(x)$ та $V_0(x)$ — їх непарні періодичні продовження на всю вісь. Якщо функції $u_0(x)$ та $v_0(x)$ не дорівнюють нулю на краях $[0,l]$, то вказані їх продовження є кусково-гладкими періодичними функціями з розривами першого роду (стрибками) в точках $2ml$ та (чи) $(2m+1)l$, $m = 0, \pm 1, \pm 2, \ldots$.

Нагадаємо, що для кусково-гладкої періодичної функції $\varphi(x)$ послідовність часткових сум $S_N(x)$ її ряду Фур'є збігається до $\varphi(x)$ у кожній точці, де $\varphi(x)$ є неперервною. Якщо ж $x_0$ — точка розриву функції $\varphi(x)$, то

$$\lim_{N\to\infty} S_N(x_0) = \frac{1}{2}\left[\varphi(x_0+0) + \varphi(x_0-0)\right], \tag{4.108}$$

при цьому збіжність ряду Фур'є до $\varphi(x)$ уже не є рівномірною на відрізку $[-l,l]$. Часткові суми $S_N(x)$ осцилюють навколо точки розриву $x_0$ з амплітудами, що перевищують значення $\left|\varphi(x_0-0) + \varphi(x_0-0)\right|/2$ на скінченну величину (до 18 % від указаного значення). Зі зростанням номера $N$ максимально можливе значення амплітуди осциляцій сум $S_N(x)$ не зменшується; звужується лише окіл точки $x_0$, у якому вони відбуваються. Цей феномен називається *явищем Гіббса.*

Проаналізуємо явище Гіббса докладніше на прикладі непарного періодичного продовження $U_0(x)$ функції

$$u_0(x) = d\left(1 - \frac{x}{l}\right), \quad 0 \le x \le l, \quad d > 0. \tag{4.109}$$

Це продовження має розриви першого роду в точках $2ml$, $m = 0, \pm 1, \pm 2, \ldots$, при цьому величина стрибка функції $U_0(x)$ у цих точках

$$U_0(2ml+0) - U_0(2ml-0) = 2d. \tag{4.110}$$

Продовженню $U_0(x)$ відповідає ряд Фур'є

$$U_0(x) = \frac{2d}{\pi}\sum_{n=1}^{\infty}\frac{1}{n}\sin\frac{\pi nx}{l}, \tag{4.111}$$

який, зокрема, збігається до $u_0(x)$ у кожній точці інтервалу $(0,l]$ і до нуля в точці $x=0$. Користуючись формулою

$$\sum_{n=1}^{N}\cos\frac{\pi nx}{l} = \sum_{n=1}^{N}\frac{1}{\sin\frac{\pi x}{2l}}\cos\frac{\pi nx}{l}\sin\frac{\pi x}{2l} =$$

$$= \frac{1}{2\sin\frac{\pi x}{2l}}\sum_{n=1}^{N}\left[\sin\frac{\pi(n+1/2)x}{l} - \sin\frac{\pi(n-1/2)x}{l}\right] = \frac{\sin\frac{\pi(N+1/2)x}{l}}{2\sin\frac{\pi x}{2l}} - \frac{1}{2},$$

часткові суми ряду (4.111) в околі точки $x=0$ можна подати у вигляді

$$S_N(x) = \frac{2d}{\pi}\sum_{n=1}^{N}\frac{1}{n}\sin\frac{\pi nx}{l} = \frac{2d}{l}\sum_{n=1}^{N}\int_0^x\cos\frac{\pi nx'}{l}dx' =$$

$$= \frac{2d}{l}\int_0^x dx'\left(\sum_{n=1}^{N}\cos\frac{\pi nx'}{l}\right) = \frac{2d}{l}\int_0^x dx'\frac{\sin\frac{\pi(N+1/2)x'}{l}}{2\sin\frac{\pi x'}{2l}} - \frac{d}{l}x. \tag{4.112}$$

Перейшовши під знаком інтеграла до безрозмірної змінної $\theta = \pi x'/l$, можемо далі записати:

$$S_N(x) = \frac{2d}{\pi}\int_0^{\pi x/l}\frac{\sin(N+1/2)\theta}{\theta}d\theta +$$

$$+\frac{2d}{\pi}\int_0^{\pi x/l}\sin(N+1/2)\theta\left[\frac{1}{2\sin(\theta/2)} - \frac{1}{\theta}\right]d\theta - \frac{d}{l}x. \tag{4.113}$$

Оскільки при $\theta \to 0$ підінтегральна функція у другому інтегралі неперервна та рівномірно обмежена,

$$\left|\sin(N+1/2)\theta\left[\frac{1}{2\sin(\theta/2)} - \frac{1}{\theta}\right]\right| \le \frac{\theta}{24} + O(\theta^2),$$

то при $x \to 0$ другий і третій доданки у правій частині формули (4.113) прямують до нуля рівномірно відносно $N$. Перший же доданок залежить лише від добутку $(N+1/2)x$, оскільки

$$\int_0^{\pi x/l}\frac{\sin(N+1/2)\theta}{\theta}d\theta = \int_0^{\pi(N+1/2)x/l}\frac{\sin\tau}{\tau}d\tau. \tag{4.114}$$

Інтеграл

$$\int_0^z \frac{\sin\tau}{\tau} d\tau \equiv \mathrm{Si}(z) \tag{4.115}$$

визначає функцію, яка називається *інтегральним синусом*. Графік цієї функції для невід'ємних дійсних значень змінної $z$ зображено на рис. 4.6. Неважко перевірити, що при $z > 0$ функція (4.115) набуває лише додатних значень, має максимуми при $z = (2m-1)\pi$ і мінімуми при $z = 2\pi m$ ($m$ — натуральне число), та прямує до $\pi/2$ при $z \to \infty$. Її абсолютний максимум при $z > 0$, який інколи називають *сталою Гіббса*, досягається при $z = \pi$ і дорівнює

$$G \equiv \int_0^\pi \frac{\sin\tau}{\tau} d\tau \approx 1{,}852 > \frac{\pi}{2} \approx 1{,}571. \tag{4.116}$$

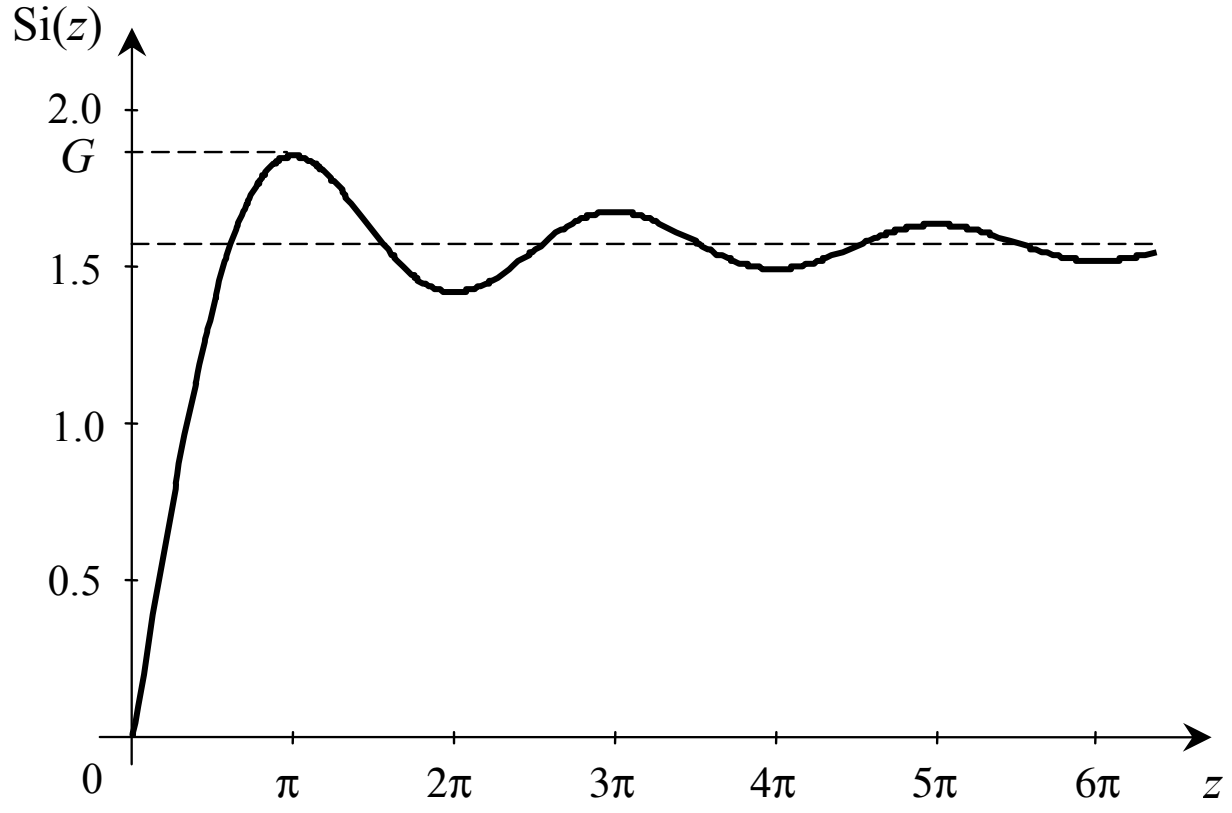

Рис. 4.6. Графік інтегрального синуса для невід'ємних дійсних значень аргументу $z$. Нижня штрихова лінія — асимптота $\pi/2$ при $z >> 1$

Звернемо увагу, що $2G/\pi \approx 1{,}18$.

З огляду на формули (4.113), (4.114) і (4.116), для значень часткових сум (4.113) у точках $x = l/(N+1/2)$ при $N >> 1$ дістаємо:

$$S_N\left(\frac{l}{N+1/2}\right) = \frac{2d}{\pi} G + O\left(\frac{1}{N+1/2}\right).$$

Бачимо, що

$$\lim_{N\to\infty} S_N\left(\frac{l}{N+1/2}\right) = \frac{2d}{\pi} G \approx 1{,}18d > d = U_0(+0). \tag{4.117}$$

Поведінку послідовності часткових сум ряду Фур'є (4.111) на відрізку $[-l, 3l]$ показано на рис. 4.7.

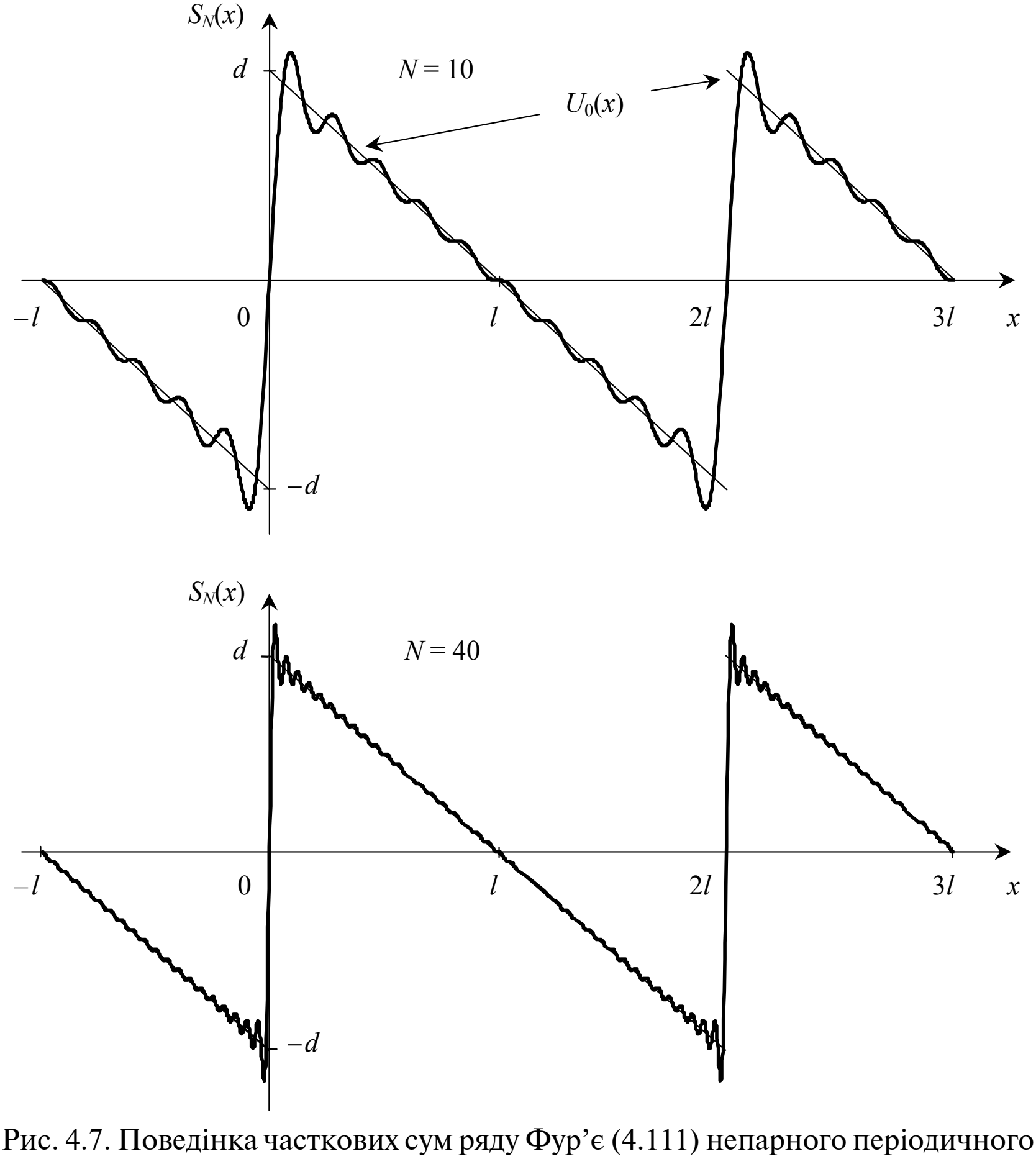

Рис. 4.7. Поведінка часткових сум ряду Фур'є (4.111) непарного періодичного продовження $U_0(x)$ функції (4.109). Графік $U_0(x)$ — сукупність нахилених паралельних відрізків

Отримані результати легко узагальнити на випадок довільної кусково-гладкої періодичної функції $\varphi(x)$ із періодом $2l$, що має розрив першого роду в точці $x_0$, і при цьому граничні точки $x_0 - 0$ та $x_0 + 0$ гілок її графіка не є симетричними відносно осі абсцис $OX$. Справді, перейдемо від такої функції до функції

$$\tilde{\varphi}(x)=\varphi(x)-\frac{1}{2}\left[\varphi(x_0+0)+\varphi(x_0-0)\right].$$

Очевидно, що для $\tilde{\varphi}(x)$ справджуються формули (4.113) і (4.114), якщо під $S_N(x)$ розуміти часткові суми ряду Фур'є функції $\tilde{\varphi}(x)$, а під $x$ на верхніх межах інтегралів — різницю $x-x_0$. Оскільки стрибки функцій $\tilde{\varphi}(x)$ і $\varphi(x)$ у точці $x_0$ мають однакове значення $2d=$ $=\varphi(x_0+0)-\varphi(x_0-0)$, можемо далі стверджувати: якщо $x_N=x_0+h_N$ — деяка послідовність точок, така, що $Nh_N \underset{N\to\infty}{\to} al$, то для часткових сум $S_N(x)$ ряду Фур'є функції $\varphi(x)$ справджується співвідношення

$$\lim_{N\to\infty} S_N(x_0+h_N)=\frac{1}{2}\left[\varphi(x_0+0)+\varphi(x_0-0)\right]+$$

$$+\frac{1}{\pi}\left[\varphi(x_0+0)-\varphi(x_0-0)\right]\int_0^{\pi a}\frac{\sin\tau}{\tau}d\tau, \tag{4.118}$$

при цьому в самій точці $x_0$ справджується співвідношення (4.108).

**Завдання 4.5.1**. Дослідіть та продемонструйте графічно явище Гіббса для непарного періодичного продовження функції $u_0(x)=A$, $0\le x\le l$, $A>0$.

## 4.6. ОБМЕЖЕНА СТРУНА ЯК КОЛИВАЛЬНА МЕХАНІЧНА СИСТЕМА

Оскільки довільна функція $\varphi(x)$, яка неперервна на відрізку $[0,l]$, однозначно визначається послідовністю коефіцієнтів ряду Фур'є в розкладі (4.93), то функції $q_n(t)$ в розкладі (4.101) утворюють набір незалежних параметрів, що дозволяють повністю описати стан струни із закріпленими кінцями при вільних коливаннях, тобто знайти миттєве положення її точок у кожний момент часу $t>0$. За означенням з аналітичної механіки, будь-які параметри з такими властивостями називаються *узагальненими координатами*. Відповідно, можемо сказати, що коефіцієнти $q_n(t)$ є узагальненими координатами струни при її вільних коливаннях, а зміщення струни $u(x,t)$ в кожній точці є лінійною сумою координат $q_n(t)$. У загальному випадку ця сума нескінченна, але за певних умов вона зводиться до одного або кількох доданків.

**Завдання 4.6.1.** Переконайтеся, що не лише повна сума ряду (4.101) для функції $u(x,t)$, а й кожний окремий член цього ряду задовольняє рівняння (4.69) та крайові умови (4.70).

**Завдання 4.6.2.** Нехай початкові функції $u_0(x)$ та $v_0(x)$ мають вигляд

$$u_0(x) = A_0 \sin\frac{\pi n_0 x}{l}, \quad v_0(x) = B_0 \sin\frac{\pi n_0 x}{l}. \tag{4.119}$$

Покажіть, що в цьому і лише в цьому випадку ряд (4.101) зводиться до одного доданка

$$u_{n_0}(x,t) = \left[ A_0 \cos\omega_{n_0} t + B_0 \frac{\sin\omega_{n_0} t}{\omega_{n_0}} \right] \sin\frac{\pi n_0 x}{l}, \quad \omega_{n_0} = \frac{\pi n_0 a}{l}. \tag{4.120}$$

*Вказівка.* Скористайтеся єдиністю зображення довільної неперервної функції на відрізку $[0,l]$ у вигляді ряду Фур'є (4.93).

Функція $u_{n_0}(x,t)$ описує спеціальний тип вільного руху струни, при якому всі її точки гармонічно коливаються з однаковою частотою $\omega_{n_0}$, але різними амплітудами. У точках $x_{n_0}^{(M)} = (2M+1)l/(2n_0)$, $M = 0,1,\ldots,n_0 - 1$, які є коренями рівняння $|\sin(\pi n_0 x/l)| = 1$, амплітуда досягає максимального значення. Ці точки називаються *пучностями* коливання (4.120). Точки $x_{n_0}^{(m)} = ml/n_0$, $m = 0,1,\ldots,n_0$, де $\sin(\pi n_0 x/l) = 0$ і, відповідно, амплітуда дорівнює нулю, називаються *вузлами* коливання (4.120). Коливання (4.120) можна уявляти собі як одночасний рух $n_0$ ділянок струни, обмежених вузлами, коли точки сусідніх ділянок рухаються в протилежних напрямах (протифазах, рис. 4.8).

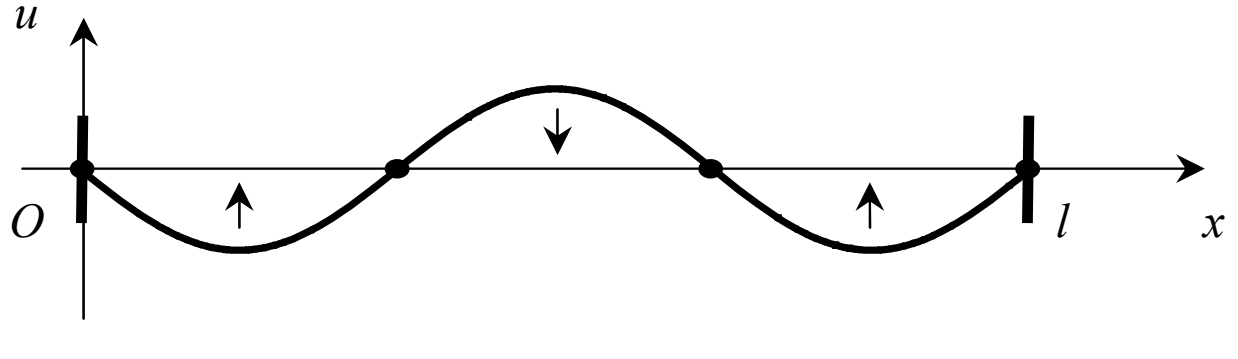

Рис. 4.8. Миттєвий профіль струни із закріпленими кінцями, яка коливається за законом (4.120) з $n_0 = 3$. Точки вказують положення вузлів, стрілки — пучностей (на осі $OX$), а також напрями руху точок струни на ділянках, обмежених вузлами. Досягнувши своїх крайніх положень, точки струни починають рухатися у зворотних напрямах. Такий коливальний рух є результатом інтерференції хвиль, спричинених початковим збуренням, які поширюються по струні в обох напрямах, багаторазово відбиваючись від її кінців

Рух коливальної системи, при якому за відсутності зовнішніх сил усі її матеріальні точки здійснюють гармонічні коливання з однаковою частотою, називається *власним коливанням* (або *стоячою хвилею*), сама ця частота — *власною частотою*, а функція, що описує (з точністю до сталого множника) залежність амплітуди коливань точок від їх положень, — *власною функцією*. Таким чином, $u_{n_0}(x,t)$, $\omega_{n_0}$ та $\sin(\pi n_0 x/l)$ у формулі (4.120) — відповідно власне коливання струни, збуджене спеціальними початковими умовами (4.119), його власна частота та власна функція. У загальному ж випадку (при довільних початкових умовах) зміщення точок струни $u(x,t)$ при її вільних коливаннях є лінійною сумою — *суперпозицією* — власних коливань з відповідними власними частотами та власними функціями.

Підкреслимо, що власні частоти і власні функції струни чи інших лінійних коливальних систем не залежать від початкових функцій, тобто від способу, яким збуджуються коливання, а визначаються лише фізичними і геометричними параметрами самої коливальної системи.

**Зауваження 4.6.1.** Однорідна струна із закріпленими кінцями не має інших власних коливань, крім тих, які представлені в загальному розв'язку (4.101) крайової задачі (4.69)–(4.71). Справді, нехай функція $u_\omega(x,t)$ описує власне коливання такої струни, що має власну частоту $\omega$. Згідно з означенням власного коливання цю функцію можна подати у вигляді

$$u_\omega(x,t) = A_\omega(x)\cos\omega t + B_\omega(x)\sin\omega t, \qquad (4.121)$$

де функції $A_\omega(x)$ та $B_\omega(x)$ визначають співвідношення амплітуд і фаз гармонічних коливань із частотою $\omega$ в різних точках струни. Очевидно, що функція $u_\omega(x,t)$ має задовольняти рівняння коливань (4.69) і крайові умови (4.70). Підставивши в них вираз (4.121) та взявши до уваги, що функції $\sin\omega t$ і $\cos\omega t$ — лінійно незалежні, знаходимо, що кожна з функцій $A_\omega(x)$ та $B_\omega(x)$ задовольняє рівняння

$$-y''(x) - \frac{\omega^2}{a^2}y(x) = 0 \qquad (4.122)$$

та крайові умови

$$y(0) = 0, \;\; y(l) = 0. \qquad (4.123)$$

Загальний розв'язок рівняння (4.122) має вигляд

$$y(x) = C_1\cos\frac{\omega}{a}x + C_2\sin\frac{\omega}{a}x, \qquad (4.124)$$

де $C_1$ і $C_2$ — сталі. Перша з умов (4.123) дає $C_1 = 0$. З другої ж умови знаходимо, що $C_2\sin(\omega l/a) = 0$. Бачимо, що нетривіальний розв'язок системи (4.122), (4.123) існує за умови $\sin(\omega l/a)=0,$ тобто коли частота $\omega$ набуває одного із значень $\omega_n=\pi na/l,\ \ n=1,2,3,\ldots$ .

Неважко перевірити, що рівняння руху для узагальнених координат $q_n(t)$ незалежні. У теорії лінійних коливальних систем координати з такою властивістю називають *нормальними*, тому власні коливання струни також називають її *нормальними модами*. Таким чином, узагальнені координати $q_n(t),\ \ n=1,2,3,\ldots,$ у формулі (4.101) визначають внески окремих нормальних мод у загальний вільний рух струни.

**Завдання 4.6.3.** Скориставшись формулами (4.102), (4.96) і (4.99), покажіть, що при довільних початкових збудженнях $u_0(x)$ та $v_0(x)$ однорідної струни із закріпленими кінцями рівняння руху для узагальнених координат $q_n(t),$ що описують її вільні коливання, розпадаються на нескінченну систему незалежних лінійних рівнянь[1]

$$\ddot{q}_n(t)+\omega_n^2 q_n(t)=0,\ \ \omega_n=\frac{\pi na}{l}=\frac{\pi n}{l}\sqrt{\frac{T_0}{\rho}},\ \ n=1,2,3,\ldots, \tag{4.125}$$

та початкових умов

$$q_n(0)=\frac{2}{l}\int_0^l u_0(x)\sin\frac{\pi nx}{l}dx,\ \ \dot{q}_n(0)=\frac{2}{l}\int_0^l v_0(x)\sin\frac{\pi nx}{l}dx. \tag{4.126}$$

У нормальних координатах функція Лагранжа $L(t)$ для вільних коливань струни розпадається на суму незалежних виразів, кожний з яких має вигляд функції Лагранжа для одновимірного лінійного осцилятора з масою $M=\rho l/2$ та частотою, яка дорівнює одній із власних частот $\omega_n$:

$$L(t)=\sum_{n=1}^{\infty}\frac{1}{2}M\left[\dot{q}_n^2-\omega_n^2 q_n^2\right]. \tag{4.127}$$

**Завдання 4.6.4.** Скориставшись розкладом (4.101), рівностями (4.92) і

$$\frac{2}{l}\int_0^l \cos\frac{\pi mx}{l}\cos\frac{\pi nx}{l}dx=\delta_{mn},\ \ m,n=1,2,3,\ldots, \tag{4.128}$$

покажіть, що в координатах $q_n(t)$ кінетична $K(t)$ та потенціальна $\Pi(t)$ енергії вільних коливань однорідної струни із закріпленими кінцями мають вигляд

[1] Тут і далі для похідних узагальнених координат за часом використовуються загальноприйняті в аналітичній механіці позначення: $\dot{q}_n(t)\equiv\dfrac{dq_n(t)}{dt},\ \ \ddot{q}_n(t)\equiv\dfrac{d^2q_n(t)}{dt^2}.$

$$K(t)=\int_0^l \frac{1}{2}\rho u_t^2(x,t)dx=\frac{1}{4}\rho l\sum_{n=1}^{\infty}\dot{q}_n^2, \qquad (4.129)$$

$$\Pi(t)=\int_0^l \frac{1}{2}T_0 u_x^2(x,t)dx=\frac{1}{4}T_0 l\sum_{n=1}^{\infty}\left(\frac{\pi n}{l}\right)^2 q_n^2, \qquad (4.130)$$

а функція Лагранжа $L(t)=K(t)-\Pi(t)$ такої струни дається формулою (4.127).

Раніше (див. підрозділ 3.2) уже було показано, що повна енергія $E=K(t)+\Pi(t)$ вільних коливань струни із закріпленими кінцями зберігається. Згідно з формулами (4.129) і (4.130) вона дається виразом

$$E=\sum_{n=1}^{\infty}\frac{1}{2}M\left[\dot{q}_n^2+\omega_n^2 q_n^2\right], \qquad (4.131)$$

тобто дорівнює сумі внесків

$$E_n=\frac{1}{2}M\left[\dot{q}_n^2+\omega_n^2 q_n^2\right] \qquad (4.132)$$

від окремих власних (нормальних) коливань. З незалежності останніх відразу ж випливає, що зберігається й енергія, що припадає на кожне окреме власне коливання.

**Завдання 4.6.5.** Скориставшись рівняннями руху (4.125), доведіть попереднє твердження безпосередньо.

Із закону збереження повної енергії вільних коливань струни випливає, що для вільних коливань функціонал енергії струни $E=K(t)+\Pi(t)=K(0)+\Pi(0)$ залежить лише від початкових функцій $u_0(x)$ та $v_0(x)$, а саме

$$E=\frac{1}{2}\int_0^l\left[\rho v_0^2(x)+T_0\left(\frac{\partial u_0(x)}{\partial x}\right)^2\right]dx. \qquad (4.133)$$

Скориставшись фактом збереження енергії $E_n$ кожного окремого власного коливання, маємо $E_n=\frac{1}{2}M\left[\dot{q}_n^2(0)+\omega_n^2 q_n^2(0)\right]$, звідки за допомогою формул (4.126) можемо виразити $E_n$ через початкові функції:

$$E_n=\frac{1}{2}M\left[\left(\frac{2}{l}\int_0^l v_0(x)\sin\frac{\pi n x}{l}dx\right)^2+\omega_n^2\left(\frac{2}{l}\int_0^l u_0(x)\sin\frac{\pi n x}{l}dx\right)^2\right]. \qquad (4.134)$$

Формула (4.134) дозволяє проаналізувати, як енергія початкового збудження струни перерозподіляється між її окремими власними коливаннями. Виявляється, що при звичайних способах збудження вільних коливань натягненої струни (щипком, смичком тощо) основна частка енергії коливань припадає на власне коливання з найменшою власною частотою

$$\omega_1 = \frac{\pi}{l}\sqrt{\frac{T_0}{\rho}}. \qquad (4.135)$$

Унаслідок такого розподілу енергії між власними коливаннями саме ця частота, яку називають основною, визначає *основний тон* звуку, випромінюваного струною при коливаннях у повітрі. Іншим власним коливанням відповідають *обертони*. Комбінація основного тону та обертонів, а також співвідношення між їх енергіями визначають *тембр* звуку.

**Завдання 4.6.6.** Початкове збудження однорідної струни із закріпленими кінцями описується функціями

$$u_0(x) = hx\left(1 - \frac{x}{l}\right), \quad v_0(x) = 0.$$

Знайдіть частку енергії вільних коливань струни, яка припадає на основну частоту.

*Відповідь*: $96/\pi^4 \approx 0{,}986$.

## 4.7. ПОЗДОВЖНІ КОЛИВАННЯ СТЕРЖНЯ СКІНЧЕННОЇ ДОВЖИНИ

Застосуємо тепер метод продовження для аналізу вільних поздовжніх коливань однорідного стержня. У випадку, коли недеформований профіль стержня збігається з відрізком $[0,l]$ дійсної осі, а обидва кінці стержня закріплені жорстко, ми знову повертаємося до крайової задачі (4.69)–(4.71), де функція $u(x,t)$ тепер дорівнює зміщенню точок стержня в довільний момент часу $t$ з тих положень $x$, які вони займали в недеформованому стержні, а параметр $a^2 = E/\rho_V$ ($E$ і $\rho_V$ — відповідно модуль Юнга та об'ємна густина речовини стержня). При двічі неперервно диференційовній початковій функції $u_0(x)$ та неперервно диференційовній початковій функції $v_0(x)$ узагальнений розв'язок цієї задачі дається формулами (4.101), (4.102), (4.96) і (4.99). При певних додаткових обмеженнях на функції $u_0(x)$ та $v_0(x)$ (див. підрозділ 4.4) він стає двічі неперервно диференційовним.

Інтуїтивно зрозуміло, що характеристики поздовжніх коливань повинні змінитися при переході до інших типів закріплення одного чи обох кінців стержня. Справді, нові крайові умови можна врахувати через нові вимоги до продовжень початкових функцій (див. підрозділ 4.3), що, очевидно, вестиме до появи нових власних функцій та власних частот коливальної системи. Продемонструємо цей факт на прикладі вільних поздовжніх коливань стержня із закріпленим лівим і вільним правим кінцями. Відповідна крайова задача має вигляд

$$\frac{\partial^2 u}{\partial t^2} = a^2 \frac{\partial^2 u}{\partial x^2}, \quad 0 < x < l, \quad t > 0, \tag{4.136}$$

$$u(0,t) = 0, \quad u_x(l,t) = 0, \quad t \geq 0, \tag{4.137}$$

$$u(x,0) = u_0(x), \quad u_t(x,0) = v_0(x), \quad 0 \leq x \leq l. \tag{4.138}$$

Як і раніше, функції $u_0(x)$ та $v_0(x)$ уважаємо відповідно двічі та один раз неперервно диференційовними.

Нехай $U_0(x)$ та $V_0(x)$ — продовження функцій $u_0(x)$ та $v_0(x)$ із відрізка $[0,l]$ на всю вісь, які будуємо в такий спосіб, щоб розв'язок допоміжної задачі

$$\frac{\partial^2 U}{\partial t^2} = a^2 \frac{\partial^2 U}{\partial x^2}, \quad -\infty < x < \infty, \quad t > 0, \tag{4.139}$$

$$U(x,0) = U_0(x), \quad U_t(x,0) = V_0(x), \tag{4.140}$$

який має вигляд

$$U(x,t) = \frac{U_0(x+at) + U_0(x-at)}{2} + \frac{1}{2a} \int_{x-at}^{x+at} V_0(x')dx', \tag{4.141}$$

автоматично задовольняв обидві умови (4.137).

Уже неодноразово зазначалося, що умова $U(0,t) = 0$ справджується, якщо продовжені функції $U_0(x)$ та $V_0(x)$ непарні відносно точки $x = 0$. Друга умова $U_x(l,t) = 0$ задовольняється, якщо при $x = l$ і для довільного $t > 0$ виконується співвідношення

$$U_0'(l+at) + U_0'(l-at) + \frac{1}{a}\left[V_0(l+at) - V_0(l-at)\right] = 0. \tag{4.142}$$

Останнє справджується тотожно, якщо для будь-якого $x \in (-\infty, \infty)$

$$U_0(l+x)=U_0(l-x),\ \ V_0(l+x)=V_0(l-x), \tag{4.143}$$

тобто функції $U_0(x)$ та $V_0(x)$ є парними продовженнями функцій $u_0(x)$ та $v_0(x)$ відносно точки $x=l.$

Нехай функція $\varphi(x)$ є непарною на всій осі та парною відносно точки $x=l,$ тобто

$$\varphi(-x)=-\varphi(x),\ \ \varphi(l-x)=\varphi(l+x).$$

Тоді для довільної точки $x\in(-\infty,\infty)$ можемо записати

$$\varphi(x+4l)=\varphi(l+(3l+x))=\varphi(l-(3l+x))=\varphi(-2l-x)=-\varphi(x+2l)=$$
$$=-\varphi(l+(x+l))=-\varphi(l-(x+l))=-\varphi(-x)=\varphi(x),$$

звідки бачимо, що $\varphi(x)$ — періодична функція з періодом $4l$:

$$\varphi(x+4l)=\varphi(x).$$

Аналогічним чином доводимо й обернене твердження: якщо $\varphi(x)$ — непарна на всій осі та періодична з періодом $4l,$ то вона парна відносно точки $x=l.$

Неперервно диференційовній функції $\varphi(x),\ \ x\in(-\infty,\infty),$ з періодом $4l$ відповідає ряд Фур'є

$$\varphi(x)=\frac{a_0}{2}+\sum_{m=1}^{\infty}\left[a_m\cos\frac{\pi m x}{2l}+b_m\sin\frac{\pi m x}{2l}\right], \tag{4.144}$$

де

$$a_m=\frac{1}{2l}\int_{-2l}^{+2l}\varphi(x)\cos\frac{\pi m x}{2l}dx,\ \ m=0,1,2,..., \tag{4.145}$$

$$b_m=\frac{1}{2l}\int_{-2l}^{+2l}\varphi(x)\sin\frac{\pi m x}{2l}dx,\ \ m=1,2,3,..., \tag{4.146}$$

який збігається до $\varphi(x)$ абсолютно й рівномірно. Для непарної функції всі коефіцієнти $a_m=0,$ тому ряд (4.144) набирає вигляду

$$\varphi(x)=\sum_{m=1}^{\infty}b_m\sin\frac{\pi m x}{2l}, \tag{4.147}$$

де

$$b_m=\frac{1}{l}\int_{0}^{2l}\varphi(x)\sin\frac{\pi m x}{2l}dx,\ m=1,2,3,.... \tag{4.148}$$

Якщо функція $\varphi(x)$ ще й парна відносно точки $x=l,$ то коефіцієнти $b_m$ можемо звести до інтегралів по «фізичній області» $[0,l]$:

$$b_m=\frac{1}{l}\int_0^{2l}\varphi(x)\sin\frac{\pi mx}{2l}dx=\frac{1}{l}\int_0^{l}\varphi(x)\sin\frac{\pi mx}{2l}dx+\frac{1}{l}\int_l^{2l}\varphi(x)\sin\frac{\pi mx}{2l}dx=$$

$$=\{x'=2l-x\}=\frac{1}{l}\int_0^{l}\varphi(x)\sin\frac{\pi mx}{2l}dx-\frac{1}{l}\int_l^{0}\varphi(2l-x')\sin\left(\pi m-\frac{\pi mx'}{2l}\right)dx'=$$

$$=\{\varphi(2l-x')=\varphi(x')\}=\frac{1}{l}\int_0^{l}\varphi(x)\sin\frac{\pi mx}{2l}dx-\frac{\cos\pi m}{l}\int_0^{l}\varphi(x')\sin\frac{\pi mx'}{2l}dx'=$$

$$=\frac{1-(-1)^m}{l}\int_0^{l}\varphi(x)\sin\frac{\pi mx}{2l}dx.$$

Очевидно, що коефіцієнти $b_m$ з парними номерами $m=2n$, $n=1,2,...,$ обертаються в нуль, а коефіцієнти з непарними номерами $m=2n+1$ дорівнюють

$$b_{2n+1}\equiv c_n=\frac{2}{l}\int_0^{l}\varphi(x)\sin\frac{\pi(n+1/2)x}{l}dx,\quad n=0,1,2,..., \tag{4.149}$$

де ми для зручності нумерації перейшли до індексу $n$ та ввели для коефіцієнтів нові позначення $c_n$. Відповідно, ряд Фур’є (4.147) набирає вигляду

$$\varphi(x)=\sum_{n=0}^{\infty}c_n\sin\frac{\pi(n+1/2)x}{l}. \tag{4.150}$$

Формули (4.149) та (4.150) дозволяють повернутися від допоміжної задачі (4.139), (4.140) до вихідної крайової задачі (4.136)–(4.138) та побудувати розв’язок останньої. Узявши до уваги властивості продовжених функцій $U_0(x)$ та $V_0(x)$, подамо їх у вигляді рядів Фур’є (4.150) для відрізка $[0,l]$:

$$U_0(x)=\sum_{n=0}^{\infty}A_n\sin\frac{\pi(n+1/2)x}{l}, \tag{4.151}$$

$$A_n=\frac{2}{l}\int_0^{l}U_0(x)\sin\frac{\pi(n+1/2)x}{l}dx=\frac{2}{l}\int_0^{l}u_0(x)\sin\frac{\pi(n+1/2)x}{l}dx, \tag{4.152}$$

$$V_0(x)=\sum_{n=0}^{\infty}B_n\sin\frac{\pi(n+1/2)x}{l}, \tag{4.153}$$

$$B_n=\frac{2}{l}\int_0^{l}V_0(x)\sin\frac{\pi(n+1/2)x}{l}dx=\frac{2}{l}\int_0^{l}v_0(x)\sin\frac{\pi(n+1/2)x}{l}dx. \tag{4.154}$$

Далі знаходимо:

$$\frac{U_0(x+at)+U_0(x-at)}{2}=$$

$$=\sum_{n=0}^{\infty}A_n\frac{1}{2}\left[\sin\frac{\pi(n+1/2)(x+at)}{l}+\sin\frac{\pi(n+1/2)(x-at)}{l}\right]= \quad (4.155)$$

$$=\sum_{n=0}^{\infty}A_n\cos\frac{\pi(n+1/2)at}{l}\sin\frac{\pi(n+1/2)x}{l},$$

$$\frac{1}{2a}\int_{x-at}^{x+at}V_0(x')dx'=$$

$$=\sum_{n=0}^{\infty}B_n\frac{l}{\pi(n+1/2)a}\sin\frac{\pi(n+1/2)at}{l}\sin\frac{\pi(n+1/2)x}{l}. \quad (4.156)$$

Підставивши ряди (4.155) і (4.156) у формулу (4.141), знаходимо *узагальнений розв'язок крайової задачі (4.136)–(4.138) про вільні поздовжні коливання однорідного стержня із закріпленим лівим і вільним правим кінцями*:

$$u(x,t)=\sum_{n=0}^{\infty}q_n(t)\sin\frac{\pi(n+1/2)x}{l}, \quad (4.157)$$

де

$$q_n(t)=A_n\cos\omega_n t+B_n\frac{\sin\omega_n t}{\omega_n},\quad \omega_n=\frac{\pi(n+1/2)a}{l}, \quad (4.158)$$

а коефіцієнти $A_n$ та $B_n$ даються формулами (4.152) та (4.154). Нагадаємо, що при виведенні цих формул *початкові функції* $u_0(x)$ *та* $v_0(x)$ *уважалися відповідно двічі неперервно диференційовною та неперервно диференційовною.* При додаткових обмеженнях на ці функції, що були обговорені в підрозділі 4.4, знайдений розв'язок є двічі неперервно диференційовною функцією за своїми аргументами.

Порівнюючи отриманий результат з відповідним результатом для однорідної обмеженої струни чи однорідного стержня із закріпленими кінцями (див. формули (4.101) і (4.102)), легко побачити, що в загальному випадку зміщення точок стержня $u(x,t)$ при вільних поздовжніх коливаннях також є суперпозицією власних коливань

$$u_n(x,t)=\left[A_n\cos\omega_n t+B_n\frac{\sin\omega_n t}{\omega_n}\right]\sin\frac{\pi(n+1/2)x}{l}, \quad (4.159)$$

з власними частотами $\omega_n = \frac{\pi(n+1/2)a}{l}$ та власними функціями $\sin\frac{\pi(n+1/2)x}{l}$, а механічний стан стержня описується набором узагальнених нормальних координат $q_n(t)$ (4.158). Звернемо, однак, увагу на той факт, що при цих спільних загальних властивостях розв'язків обох задач *власні функції та власні частоти коливань у них різні, що спричинено різними типами крайових умов*.

**Завдання 4.7.1.** Перевірте, що для систем функцій $\left\{\sin\frac{\pi(n+1/2)x}{l}\right\}$ і $\left\{\cos\frac{\pi(n+1/2)x}{l}\right\}$, $n = 0,1,2,...$, справджуються рівності

$$\frac{2}{l}\int_0^l \sin\frac{\pi(m+1/2)x}{l}\sin\frac{\pi(n+1/2)x}{l}dx = \delta_{mn}, \quad m,n = 0,1,2,..., \qquad (4.160)$$

$$\frac{2}{l}\int_0^l \cos\frac{\pi(m+1/2)x}{l}\cos\frac{\pi(n+1/2)x}{l}dx = \delta_{mn}, \quad m,n = 0,1,2,.... \qquad (4.161)$$

**Завдання 4.7.2.** Запишіть функцію Лагранжа і рівняння руху в нормальних координатах $q_n(t)$ для вільних поздовжніх коливань однорідного стержня із закріпленим лівим та вільним правим кінцями, виразіть початкові значення координат і швидкостей через початкові функції (4.138). Порівняйте здобуті результати з формулами (4.125)–(4.127).

**Завдання 4.7.3.** Обчисліть повну енергію вільних поздовжніх коливань стержня із закріпленим лівим та вільним правим кінцями. Порівняйте отриманий результат з формулами (4.131) і (4.132).

**Завдання 4.7.4.** Методом продовження знайдіть власні частоти і власні функції вільних поздовжніх коливань однорідного стержня $0 \le x \le l$ з вільним лівим та закріпленим правим кінцями.

*Відповідь*. Відповідно $\frac{\pi(n+1/2)a}{l}$ і $\cos\frac{\pi(n+1/2)x}{l}$, де $n = 0,1,2,....$

## 4.8. КОЛИВАННЯ ОБМЕЖЕНОЇ СТРУНИ ПРИ НАЯВНОСТІ ТЕРТЯ

Вільні коливання струни відповідають ідеалізованій ситуації, коли тертя між струною і середовищем, у якому знаходиться струна, відсутнє. На практиці навколишнє середовище має скінченну в'язкість, і тому будь-які коливання струни в ньому, викликані початковим збудженням, поступового згасають — унаслідок тертя струна втрачає свою енергію, віддаючи її середовищу у формі тепла.

При наявності тертя струна є відкритою системою, тому її рівняння руху в середовищі не можна вивести за допомогою лише законів механіки. Але якщо процес коливань струни набагато повільніший за дисипативні процеси в середовищі, то вплив останнього на струну можна врахувати, увівши в рівняння механічних коливань модельну дисипативну силу (силу тертя). Експеримент показує, що дисипативна сила $\Delta F_{\rm d}(x,t)$, яка діє з боку в'язкого середовища на малу ділянку $(x, x+\Delta x)$ однорідної струни, у першому наближенні пропорційна швидкості цієї ділянки та її довжині. Таку силу можна подати як

$$\Delta F_{\rm d}(x,t) = -2\eta\rho \frac{\partial u(x,t)}{\partial t}\Delta x, \qquad (4.162)$$

де $\eta$ — коефіцієнт тертя, $\eta > 0$.

Таким чином, рівняння «вільних» коливань однорідної струни у в'язкому середовищі набирає вигляду

$$\frac{\partial^2 u}{\partial t^2} = a^2 \frac{\partial^2 u}{\partial x^2} - 2\eta \frac{\partial u}{\partial t}, \quad 0 < x < l, \quad t > 0. \qquad (4.163)$$

Слово «вільні» тут і далі вказує на те, що на струну не діють ніякі зовнішні сили, за винятком дисипативної сили.

Крайова задача про «вільні» коливання однорідної струни із закріпленими кінцями у в'язкому середовищі полягає в тому, щоб знайти двічі диференційовну функцію $u(x,t)$, яка задовольняє рівняння (4.163), крайові умови (4.70) та початкові умови (4.71).

**Завдання 4.8.1.** Покажіть, що при дії на однорідну струну сили в'язкого тертя (4.162) її повна енергія $E(t)$ та кінетична енергія $T(t)$ пов'язані співвідношенням

$$\frac{dE}{dt} = -4\eta T. \qquad (4.164)$$

**Завдання 4.8.2.** Скориставшись співвідношенням (4.164), доведіть, що крайова задача про «вільні» коливання однорідної струни із закріпленими кінцями має єдиний розв'язок при $\eta \geq 0$.

Перейдемо до розв'язування крайової задачі про «вільні» коливання однорідної струни із закріпленими кінцями у в'язкому середовищі. Як і випадку істинно вільних (без тертя) коливань, подамо шукану функцію у вигляді

$$u(x,t) = \sum_{n=1}^{\infty} q_n(t) \sin \frac{\pi n x}{l}. \tag{4.165}$$

Припустимо, що ряд (4.165) можна двічі диференціювати, і при цьому отримувані ряди збігаються на відрізку $[0,l]$ абсолютно й рівномірно; крайові умови (4.70) виконуються автоматично. З умов (4.71) та єдиності розкладу неперервно диференційовної функції в ряд Фур'є знаходимо початкові значення узагальнених координат і узагальнених швидкостей. Так само як і для істинно вільних коливань струни,

$$q_n(0) = \frac{2}{l}\int_0^l u_0(x) \sin \frac{\pi n x}{l} dx, \quad \dot{q}_n(0) = \frac{2}{l}\int_0^l v_0(x) \sin \frac{\pi n x}{l} dx. \tag{4.166}$$

Узявши далі до уваги єдиність розкладу (4.165) для функції $u(x,t)$, підставляємо його в рівняння (4.163) та бачимо, що це рівняння еквівалентне такій системі незалежних рівнянь для узагальнених координат $q_n(t)$:

$$\ddot{q}_n(t) + 2\eta \dot{q}_n(t) + \omega_n^2 q_n(t) = 0, \quad n = 1,2,3,.... \tag{4.167}$$

Загальний розв'язок кожного з рівнянь (4.167) є лінійною комбінацією його частинних розв'язків $e^{-\eta t} \cos \Omega_n t$ і $e^{-\eta t} \sin \Omega_n t$, де $\Omega_n = \sqrt{\omega_n^2 - \eta^2}$ :

$$q_n(t) = A_n e^{-\eta t} \cos \Omega_n t + B_n e^{-\eta t} \sin \Omega_n t.$$

Виразивши коефіцієнти $A_n$ і $B_n$ через початкові значення (4.166), дістаємо:

$$q_n(t) = e^{-\eta t}\left[ q_n(0) \cos \Omega_n t + \left(\dot{q}_n(0) + \eta q_n(0)\right) \frac{\sin \Omega_n t}{\Omega_n} \right], \quad n = 1,2,3,.... \tag{4.168}$$

Отже, розв'язок крайової задачі (4.163), (4.70), (4.71) про «вільні» коливання однорідної закріпленої струни у в'язкому середовищі дається формулою (4.165) з узагальненими координатами (4.168) та початковими значеннями (4.166) — принаймні, коли непарні періо-дичні продовження $U_0(x)$, $V_0(x)$ початкових функцій $u_0(x)$, $v_0(x)$ є відповідно двічі та один раз неперервно диференційовними.

**Зауваження 4.8.1.** Якщо відлік часу починається з моменту $t=\tau$, треба аргумент $t$ у формулах (4.165) і (4.168) замінити на різницю $t-\tau$, а замість $q_n(0)$ і $\dot{q}_n(0)$ у формулах (4.166) і (4.168) писати відповідно $q_n(\tau)$ і $\dot{q}_n(\tau)$.

Неважко побачити, що з плином часу коливання струни у в'язкому середовищі поступово згасають. Характер згасання залежить від відносної величини коефіцієнта тертя $\eta$ між струною і середовищем. При малих значеннях $\eta \ll \pi a/l$ можна наближено говорити про майже періодичні коливання струни з частотою $\Omega_n$ та амплітудою, що поволі спадає за законом $e^{-\eta t}$. У протилежному граничному випадку струна аперіодично наближається до свого рівноважного стану, здійснюючи відносно нього кілька коливальних рухів або взагалі не здійснюючи жодного.

**Завдання 4.8.3.** Поясніть, як змінюються основний тон $\Omega_1$ та обертони $\Omega_n$, $n \geq 2$, коливань обмеженої струни в середовищі з відносно невеликою в'язкістю по відношенню до основного тону $\omega_1$ та обертонів $\omega_n$, $n \geq 2$, коливань вільної струни.

**Завдання 4.8.4.** За допомогою формул (4.165), (4.166) і (4.168) проаналізуйте та опишіть рух струни в сильно в'язкому середовищі з $\eta \geq \pi a/l$.

Нехай тепер на однорідну струну із закріпленими кінцями, що знаходиться у в'язкому середовищі, додатково діє деяка сила з погонною густиною $F(x,t)$. Рівняння руху струни набирає вигляду

$$\frac{\partial^2 u}{\partial t^2} = a^2 \frac{\partial^2 u}{\partial x^2} - 2\eta \frac{\partial u}{\partial t} + f(x,t), \quad 0 < x < l, \quad t > 0, \tag{4.169}$$

де $f(x,t) = F(x,t)/\rho$, і ми приходимо до *крайової задачі про вимушені коливання струни у в'язкому середовищі*: знайти такий двічі диференційовний розв'язок неоднорідного рівняння (4.169), який задовольняє крайові умови (4.70) та початкові умови (4.71). У силу лінійності задачі цей розв'язок можна подати у вигляді суми $u(x,t) = u_1(x,t) + u_2(x,t)$, де функція $u_1(x,t)$ задовольняє однорідне рівняння (4.163), крайові умови (4.70) та початкові умови (4.71), а функція $u_2(x,t)$ задовольняє неоднорідне рівняння (4.169), крайові умови (4.70) та нульові початкові умови

$$u_2(x,0)=0,\ \left.\frac{\partial u_2(x,t)}{\partial t}\right|_{t=0}=0. \qquad (4.170)$$

Очевидно, що функція $u_1(x,t)$ описує «вільні» коливання струни у в'язкому середовищі, збуджені ненульовими початковими функціями $u_0(x)$ та $v_0(x)$. Як щойно було показано, вона визначається формулами (4.165), (4.166) і (4.168). Другий доданок $u_2(x,t)$ описує вимушені коливання струни у в'язкому середовищі, які відбуваються лише завдяки дії прикладеної сили $F(x,t)$. Відшукаємо його, скориставшись принципом Дюамеля.

Нехай $\varphi(x,t\,|\,\tau)$ — розв'язок крайової задачі для однорідного рівняння (4.163) при $t>\tau$, який задовольняє крайові умови (4.70) і початкові умови

$$\varphi(x,\tau\,|\,\tau)=0,\ \varphi_t(x,\tau\,|\,\tau)=f(x,\tau),$$

де $f(x,\tau)$ — довільна функція, неперервно диференційовна при $x\in[0,l]$, $\tau>0$. Згідно з формулами (4.165), (4.166), (4.168) та зауваженням 4.8.1, можемо записати

$$\varphi(x,t\,|\,\tau)=\sum_{n=1}^{\infty}\dot{q}_n(\tau)e^{-\eta(t-\tau)}\frac{\sin\Omega_n(t-\tau)}{\Omega_n}\sin\frac{\pi nx}{l}=$$

$$=\sum_{n=1}^{\infty}\left(\frac{2}{l}\int_0^l f(x',\tau)\sin\frac{\pi nx'}{l}dx'\right)e^{-\eta(t-\tau)}\frac{\sin\Omega_n(t-\tau)}{\Omega_n}\sin\frac{\pi nx}{l},$$

або

$$\varphi(x,t\,|\,\tau)=\int_0^l G(x,x';t-\tau)F(x',\tau)\,dx', \qquad (4.171)$$

де введено (часову) функцію Гріна $G(x,x';t)$, формально визначену при $t>0$ як суму (умовно) збіжного ряду

$$G(x,x';t)=\frac{2}{\rho l}\sum_{n=1}^{\infty}e^{-\eta t}\frac{\sin\Omega_n t}{\Omega_n}\sin\frac{\pi nx}{l}\sin\frac{\pi nx'}{l},\quad 0\le x,x'\le l,\ t>0. \qquad (4.172)$$

Тоді за принципом Дюамеля шуканий доданок $u_2(x,t)$ дається виразом

$$u_2(x,t)=\int_0^t d\tau\int_0^l dx'\,G(x,x';t-\tau)F(x',\tau). \qquad (4.173)$$

З огляду на явний вигляд функції $G(x,x';t)$ остаточно знаходимо:

$$u_2(x,t)=\frac{2}{l}\sum_{n=1}^{\infty}\left\{\int_0^t e^{-\eta(t-\tau)}\frac{\sin\Omega_n(t-\tau)}{\Omega_n}\left[\int_0^l f(x',\tau)\sin\frac{\pi nx'}{l}dx'\right]d\tau\right\}\sin\frac{\pi nx}{l}. \qquad (4.174)$$

Зауважимо, що для неперервних чи лише обмежених при $x \in [0,l]$, $\tau > 0$ функцій $f(x,\tau)$ ряд (4.174) збігається абсолютно і рівномірно, а його сума є щонайменше неперервним узагальненим розв'язком задачі (4.169), (4.70), (4.170).

Повертаючись до задачі (4.169), (4.70) та (4.71) про вимушені коливання струни у в'язкому середовищі, бачимо, що при вказаних застереженнях її розв'язок зберігає структуру розв'язку (4.165) крайової задачі (4.163), (4.70), (4.71) про «вільні» коливання струни у цьому середовищі, тобто його можна подати у вигляді ряду

$$u(x,t) = \sum_{n=1}^{\infty} Q_n(t) \sin\frac{\pi n x}{l}, \tag{4.175}$$

коефіцієнти $Q_n(t)$ якого відіграють роль нормальних координат. Згідно з попередніми викладками,

$$\begin{aligned} Q_n(t) = e^{-\eta t}\left[ q_n(0)\left( \cos\Omega_n t + \frac{\eta}{\Omega_n}\sin\Omega_n t \right) + \dot{q}_n(0)\frac{\sin\Omega_n t}{\Omega_n} \right] + \\ + \frac{2}{l}\int_0^t d\tau e^{-\eta(t-\tau)}\frac{\sin\Omega_n(t-\tau)}{\Omega_n}\int_0^l f(x',\tau)\sin\frac{\pi n x'}{l}dx', \end{aligned} \tag{4.176}$$

де $q_n(0)$ і $\dot{q}_n(0)$ даються формулами (4.166).

Розглянемо окремо спеціальний випадок, коли прикладена сила змінюється з часом за гармонічним законом виду (4.29). Подавши її погонну густину в комплекснозначній формі

$$\tilde{F}(x,t) = \tilde{F}_0(x)e^{-i\omega t} \tag{4.177}$$

(див. формули (4.29) і (4.30)) та взявши до уваги співвідношення

$$\frac{1}{\Omega_n}\int_0^t e^{-\eta(t-\tau)}\sin\Omega_n(t-\tau)e^{-i\omega\tau}d\tau \underset{t\to\infty}{=} \frac{1}{\Omega_n^2 - (\omega + i\eta)^2}e^{-i\omega t} + O\left(e^{-\eta t}\right),$$

знаходимо, що викликані такою силою зміщення точок струни описуються при $t \to \infty$ комплекснозначною функцією

$$\tilde{u}_2(x,t) = \tilde{A}_\omega(x)e^{-i\omega t} + O\left(e^{-\eta t}\right), \tag{4.178}$$

де

$$\tilde{A}_\omega(x) = \int_0^l G_\omega(x,x')\tilde{F}_0(x')dx', \tag{4.179}$$

$$G_\omega(x,x') = \frac{2}{\rho l}\sum_{n=1}^{\infty}\frac{1}{\Omega_n^2 - (\omega + i\eta)^2}\sin\frac{\pi n x}{l}\sin\frac{\pi n x'}{l}. \tag{4.180}$$

Отже, під дією гармонічної сили з погонною густиною (4.177) усі точки струни з часом починають гармонічно коливатися з частотою сили $\omega$ та усталеною (комплекснозначною) амплітудою $\tilde{A}_\omega(x)$, що визначається формулами (4.179), (4.180). Функція $G_\omega(x,x')$ у цих формулах називається *частотною функцією Гріна* для однорідної струни із закріпленими кінцями. З рівняння (4.179) бачимо, що ця функція має наступний фізичний зміст: вона дорівнює комплекснозначній амплітуді усталених коливань струни в точці з координатою $x$, які виникають унаслідок дії на струну зосередженої гармонічної сили, що має одиничну амплітуду та прикладена до точки з координатою $x'$. Погонна густина такої сили описується виразом $\tilde{F}(x,t) = \delta(x - x')e^{-i\omega t}$.

**Завдання 4.8.5.** Проаналізуйте фізичний зміст частотної функції Гріна струни для випадку, коли тертям між струною і навколишнім середовищем можна знехтувати.

**Завдання 4.8.6.** На однорідну струну із закріпленими кінцями діє зосереджена сила $F(t) = F_0 \sin\omega t$. Силу прикладено до точки, що віддалена на чверть довжини струни від правого кінця. Виразіть через частотну функцію Гріна струни амплітуду коливань струни та зсув фаз між коливаннями струни та сили в точці, віддаленій на таку саму відстань від лівого кінця.

Окремо розгляньте випадок, коли тертям між струною і навколишнім середовищем можна знехтувати.

*Відповідь*: $A_\omega(l/4) = F_0\left|G_\omega(l/4, 3l/4)\right|$, $\operatorname{tg}\psi(l/4) = \dfrac{\operatorname{Im} G_\omega(l/4, 3l/4)}{\operatorname{Re} G_\omega(l/4, 3l/4)}$,
де $\psi(l/4)$ — фаза, на яку коливання струни в точці $x = l/4$ відстають від прикладеної сили. Якщо тертя відсутнє, то $A_\omega(l/4) = F_0 G_\omega(l/4, 3l/4)$, $\psi(l/4) = 0$, оскільки у цьому випадку $\operatorname{Im} G_\omega(x,x') = 0$ для будь-яких точок $x$ і $x'$ струни.

**Завдання 4.8.7.** Те саме для сили $F(t) = F_0 \cos\omega t$, прикладеної до середини струни.

## 4.9. ЧАСТОТНА ФУНКЦІЯ ГРІНА

Перейдемо до більш докладного аналізу властивостей частотної функції Гріна (4.180). Перш за все зазначимо, що для членів функціонального ряду в (4.180) справджується оцінка

$$\left|\frac{1}{\Omega_n^2-(\omega+i\eta)^2}\sin\frac{\pi n x}{l}\sin\frac{\pi n x'}{l}\right|\le\frac{1}{\left|\Omega_n^2-(\omega+i\eta)^2\right|}\underset{n\to\infty}{=}\frac{l^2}{\pi^2 n^2 a^2}\left[1+O\left(\frac{1}{n^2}\right)\right].$$

Тому цей ряд збігається абсолютно й рівномірно відносно змінних $x$ та $x'$. Звідси випливає, що функція Гріна $G_\omega(x,x')$, яка визначається через суму цього ряду, неперервна в області $0\le x,x'\le l$. З фізичного погляду ця властивість функції $G_\omega(x,x')$ узгоджується з вимогою, щоб струна при коливаннях не рвалася.

Щоб подати функцію $G_\omega(x,x')$ у скінченному вигляді, спершу покажемо, що вона є розв'язком певної крайової задачі. Для цього скористаємося такими евристичними міркуваннями. Оскільки в кожний момент часу загальний розв'язок (4.175) крайової задачі про вимушені коливання однорідної струни задовольняє рівняння (4.169), то, очевидно, його задовольняє і граничний вираз, який описує вимушені коливання струни при $t\to\infty$ під дією джерела (4.177). Звідси випливає, що головний член $\tilde{A}_\omega(x)e^{-i\omega t}$ у формулі (4.178), який при $t\to\infty$ описує усталені коливання струни, також має задовольняти рівняння (4.169), а разом з ним і крайові умови жорсткого закріплення (4.70). Підставивши цей член у рівняння (4.169) та умови (4.70), бачимо, що *комплекснозначна амплітуда усталених коливань* $\tilde{A}_\omega(x)$ *при довільній амплітуді функції джерел* $\tilde{f}_0(x)$ *задовольняє неоднорідне диференціальне рівняння*

$$-\frac{d^2y(x)}{dx^2}-\lambda y(x)=\frac{1}{a^2}\tilde{f}_0(x),\quad \lambda=\frac{\omega^2}{a^2}+\frac{2i\eta\omega}{a^2},\quad 0<x<l, \tag{4.181}$$

*та крайові умови*

$$y(0)=0,\;\; y(l)=0. \tag{4.182}$$

Згадавши, що у спеціальному випадку $\tilde{f}_0(x)=\delta(x-x')/\rho$ амплітуда $\tilde{A}_\omega(x)$ збігається з функцією $G_\omega(x,x')$, приходимо до висновку, що *частотна функція Гріна для однорідної струни із закріпленими кінцями є неперервним у точці* $x=x'$ *розв'язком крайової задачі*

$$-\frac{\partial^2 G_\omega(x,x')}{\partial x^2}-\lambda G_\omega(x,x')=\frac{1}{\rho a^2}\delta(x-x'),\quad 0<x,x'<l, \tag{4.183}$$

$$G_\omega(0,x')=0,\;\; G_\omega(l,x')=0. \tag{4.184}$$

Перейдемо тепер до побудови розв'язку задачі (4.183), (4.184). Для цього розіб'ємо відрізок $[0,l]$ на два проміжки $0\le x<x'$ і $x'<x\le l$, які назвемо відповідно проміжок I та проміжок II. На кожному з цих проміжків функція $G_\omega(x,x')$ задовольняє однорідне рівняння

$$\frac{\partial^2 G_\omega(x,x')}{\partial x^2} + \lambda G_\omega(x,x') = 0 \qquad (4.185)$$

у внутрішніх точках проміжку та одну з умов (4.184) у крайній точці: $G_\omega(0,x') = 0$ на проміжку I та $G_\omega(l,x') = 0$ на проміжку II. Загальний розв'язок рівняння (4.185) має вигляд

$$G_\omega(x,x') = A\sin\sqrt{\lambda}x + B\cos\sqrt{\lambda}x,$$

тому для проміжку I з урахуванням крайової умови $G_\omega(0,x') = 0$ дістаємо

$$G_\omega^{(I)}(x,x') = A\sin\sqrt{\lambda}x, \quad 0 \le x < x'. \qquad (4.186)$$

Якщо ж загальний розв'язок рівняння (4.185) подати у вигляді

$$G_\omega(x,x') = C\sin\sqrt{\lambda}(l-x) + D\cos\sqrt{\lambda}(l-x)$$

і скористатися умовою $G_\omega(l,x') = 0$, то для проміжку II знаходимо

$$G_\omega^{(II)}(x,x') = C\sin\sqrt{\lambda}(l-x), \quad x' < x \le l. \qquad (4.187)$$

Таким чином, для шуканої функції Гріна можемо записати

$$G_\omega(x,x') = \begin{cases} G_\omega^{(I)}(x,x'), & 0 \le x < x', \\ G_\omega^{(II)}(x,x'), & x' < x \le l. \end{cases} \qquad (4.188)$$

У цій формулі залишаються невідомими дві сталі інтегрування. Щоб їх відшукати, потрібно мати дві додаткові умови. Ними виступають *умови зшивання в точці* $x = x'$.

Перша умова зшивання випливає з факту неперервності частотної функції Гріна, який було встановлено на початку цього підрозділу. Згідно з цією властивістю функції $G_\omega(x,x')$ та означенням неперервності, для будь-якої точки $x' \in (0,l)$ виконується рівність

$$\lim_{\varepsilon\to 0} G_\omega(x'-\varepsilon, x') = \lim_{\varepsilon\to 0} G_\omega(x'+\varepsilon, x'),$$

або, у більш компактних позначеннях,

$$G_\omega(x'-0,x') = G_\omega(x'+0,x'). \qquad (4.189)$$

Другу умову зшивання знайдемо, зінтегрувавши обидві частини рівняння (4.183) за змінною $x$ у межах від $x'-\varepsilon$ до $x'+\varepsilon$, $\varepsilon > 0$, та перейшовши у здобутій рівності до границі $\varepsilon \to 0$. Маємо:

$$\lim_{\varepsilon\to 0}\left[-\int\limits_{x'-\varepsilon}^{x'+\varepsilon}\frac{\partial^2 G_\omega(x,x')}{\partial x^2}dx-\lambda\int\limits_{x'-\varepsilon}^{x'+\varepsilon}G_\omega(x,x')dx\right]=\frac{1}{\rho a^2}\lim_{\varepsilon\to 0}\int\limits_{x'-\varepsilon}^{x'+\varepsilon}\delta(x-x')dx,$$

звідки, з огляду на властивості дельта-функції та неперервність функції $G_\omega(x,x')$, дістаємо

$$\lim_{\varepsilon\to 0}\left[-\frac{\partial G_\omega(x,x')}{\partial x}\bigg|_{x=x'+\varepsilon}+\frac{\partial G_\omega(x,x')}{\partial x}\bigg|_{x=x'-\varepsilon}\right]=\frac{1}{\rho a^2},$$

або

$$G'_\omega(x'-0,x')-G'_\omega(x'+0,x')=\frac{1}{\rho a^2}. \tag{4.190}$$

Таким чином, на відміну від самої функції $G_\omega(x,x')$, яка залишається неперервною в точці $x=x'$ (умова (4.189)), її похідна $\partial G_\omega(x,x')/\partial x$ має в цій точці скінченний стрибок (умова (4.190)).

За допомогою введених нами позначень умови зшивання (4.189) і (4.190) набирають вигляду

$$\begin{gathered}G_\omega^{(\mathrm{I})}(x'-0,x')=G_\omega^{(\mathrm{II})}(x'+0,x'),\\ G_\omega^{(\mathrm{I})\prime}(x'-0,x')-G_\omega^{(\mathrm{II})\prime}(x'+0,x')=\frac{1}{\rho a^2}.\end{gathered} \tag{4.191}$$

Підставивши в них вирази (4.186) і (4.187), отримуємо систему лінійних рівнянь для відшукання сталих $A$ і $C$:

$$\begin{gathered}A\sin\sqrt{\lambda}x'-C\sin\sqrt{\lambda}(l-x')=0,\\ A\cos\sqrt{\lambda}x'+C\cos\sqrt{\lambda}(l-x')=\frac{1}{\rho a^2\sqrt{\lambda}}.\end{gathered} \tag{4.192}$$

Розв’язок системи (4.192) знаходимо за формулами Крамера $A=\Delta_A/\Delta$, $C=\Delta_C/\Delta$, де детермінант системи

$$\Delta=\begin{vmatrix}\sin\sqrt{\lambda}x' & -\sin\sqrt{\lambda}(l-x')\\ \cos\sqrt{\lambda}x' & \cos\sqrt{\lambda}(l-x')\end{vmatrix}=\sin\sqrt{\lambda}l,$$

а детермінанти $\Delta_A$ та $\Delta_C$ отримуємо, замінивши в $\Delta$ відповідно перший та другий стовпчики на стовпчик у правій частині системи (4.192):

$$\Delta_A=\begin{vmatrix}0 & -\sin\sqrt{\lambda}(l-x')\\ \dfrac{1}{\rho a^2\sqrt{\lambda}} & \cos\sqrt{\lambda}(l-x')\end{vmatrix}=\frac{\sin\sqrt{\lambda}(l-x')}{\rho a^2\sqrt{\lambda}},$$

Маємо:

$$\Delta_C = \begin{vmatrix} \sin\sqrt{\lambda}x' & 0 \\ \cos\sqrt{\lambda}x' & \dfrac{1}{\rho a^2\sqrt{\lambda}} \end{vmatrix} = \frac{\sin\sqrt{\lambda}x'}{\rho a^2\sqrt{\lambda}}.$$

$$A = \frac{\sin\sqrt{\lambda}(l-x')}{\rho a^2\sqrt{\lambda}\sin\sqrt{\lambda}l}, \quad C = \frac{\sin\sqrt{\lambda}x'}{\rho a^2\sqrt{\lambda}\sin\sqrt{\lambda}l}. \tag{4.193}$$

З формул (4.186)–(4.188) і (4.193) остаточно дістаємо:

$$G_\omega(x,x') = \frac{1}{\rho a^2\sqrt{\lambda}\sin\sqrt{\lambda}l}\begin{cases} \sin\sqrt{\lambda}x\sin\sqrt{\lambda}(l-x'), & 0\le x\le x', \\ \sin\sqrt{\lambda}(l-x)\sin\sqrt{\lambda}x', & x'\le x\le l. \end{cases} \tag{4.194}$$

Зауважимо, що виведення формули (4.194) може здатися не зовсім строгим, або принаймні таким, що потребує подальшого з'ясування технічних деталей. Однак у правильності формули (4.194) можна переконатися безпосередньо.

**Завдання 4.9.1.** Перевірте, що при фіксованому значенні $x'\in(0,l)$ ряд Фур'є для непарної неперервної функції на проміжку $(-l,l)$, яка при $x>0$ визначається виразом (4.194), збігається з рядом (4.180).

**Завдання 4.9.2.** Скориставшись явним виглядом власних частот $\Omega_n = \sqrt{\omega_n^2 - \eta^2}$ та формулою

$$\sum_{n=1}^{\infty}\frac{1}{n^2-b^2}\cos ny = \frac{1}{2b^2} - \frac{\pi}{2b\sin(\pi b)}\cos[(\pi-y)b], \quad 0\le y\le 2\pi, \quad b\ne n,$$

обчисліть суму ряду (4.180) безпосередньо та переконайтеся, що вона збігається з функцією (4.194).

*Вказівка*: $\sin\alpha\sin\beta = [\cos(\alpha-\beta)-\cos(\alpha+\beta)]/2.$

**Зауваження 4.9.1.** При $\eta>0$ крайова задача

$$\begin{gathered} \frac{\partial^2\tilde{u}}{\partial t^2} = a^2\frac{\partial^2\tilde{u}}{\partial x^2} - 2\eta\frac{\partial\tilde{u}}{\partial t} + \tilde{f}_0(x)e^{-i\omega t}, \quad 0<x<l, \quad t>0, \\ \tilde{u}(0,t) = \tilde{u}(l,t) = 0 \end{gathered} \tag{4.195}$$

має *єдиний розв'язок виду* $\tilde{A}_\omega(x)e^{-i\omega t}$, *що відповідає усталеним коливанням*. Справді, нехай існують два розв'язки задачі (4.195), $\tilde{A}_\omega^{(1)}(x)e^{-i\omega t}$ і $\tilde{A}_\omega^{(2)}(x)e^{-i\omega t}$. Записавши їх різницю у вигляді

$$\tilde{A}_\omega^{(2)}(x)e^{-i\omega t} - \tilde{A}_\omega^{(1)}(x)e^{-i\omega t} = g(x)e^{-i\omega t},$$

для функції $g(x)$ дістанемо крайову задачу

$$g'' + \lambda g = 0, \quad \lambda = \frac{\omega^2}{a^2} + \frac{2i\eta\omega}{a^2}, \quad 0 < x < l, \\ g(0) = g(l) = 0. \tag{4.196}$$

Як ми вже бачили, ця задача має нетривіальний розв'язок лише за умови, що $\sin\sqrt{\lambda}l = 0$, тобто коли $\sqrt{\lambda}l = \pi n$, або

$$\frac{\pi^2 n^2}{l^2} = \frac{\omega^2}{a^2} + \frac{2i\eta\omega}{a^2}, \quad n = 1,2,3,....$$

Однак при дійсних значеннях $\eta > 0$ ці рівності не можуть виконуватися для дійсних значень частоти $\omega$. Приходимо до висновку, що $g(x) \equiv 0$.

З єдиності розв'язку крайової задачі (4.195) про усталені коливання випливає, без посилань на розклад у ряд Фур'є, що вирази для функції Гріна (4.180) і (4.194) мають збігатися, оскільки за їх допомогою при $\eta > 0$ і довільної інтегровної функції джерел знаходиться єдиний розв'язок задачі (4.195) виду $\tilde{A}_\omega(x)e^{-i\omega t}$.

**Завдання 4.9.3.** Перевірте, що для довільної неперервної функції $\tilde{f}_0(x)$ функція

$$\tilde{u}(x,t) = e^{-i\omega t}\frac{1}{a^2\sqrt{\lambda}\sin\sqrt{\lambda}l}\left[\sin\sqrt{\lambda}(l-x)\int_0^x \sin\sqrt{\lambda}x'\tilde{f}_0(x')\,dx' + \right. \\ \left. + \sin\sqrt{\lambda}x\int_x^l \sin\sqrt{\lambda}(l-x')\tilde{f}_0(x')\,dx'\right] \tag{4.197}$$

є розв'язком крайової задачі (4.195).

На завершення цього підрозділу нагадаємо, що для того, щоб знайти частотну функцію Гріна для однорідної струни із закріпленими кінцями у скінченному вигляді, ми розглянули спеціальну крайову задачу (4.183), (4.184) та знайшли її єдиний розв'язок, побудувавши два розв'язки рівняння (4.183) в областях, що лежать зліва і справа від точки $x'$, та відновивши чотири невідомі сталі інтегрування в них (по дві в кожному) за допомогою крайових умов (4.184) та умов зшивання (4.191). Цей алгоритм побудови частотної функції Гріна можна застосувати і для коливальних систем з іншими типами крайових умов.

**Завдання 4.9.4.** Знайдіть частотну функцію Гріна для поздовжніх коливань однорідного стержня $0 \le x \le l$, лівий кінець якого закріплений жорстко, а правий — вільний.

*Вказівка.* Відповідна крайова задача складається з рівняння (4.183) та крайових умов $G_\omega(0,x')=0,\ G'_\omega(l,x')=0.$

*Відповідь:*

$$G_\omega(x,x')=\frac{1}{\rho_V S a^2\sqrt{\lambda}\cos\sqrt{\lambda}l}\begin{cases}\sin\sqrt{\lambda}x\cos\sqrt{\lambda}(l-x'), & 0\le x\le x',\\ \cos\sqrt{\lambda}(l-x)\sin\sqrt{\lambda}x', & x'\le x\le l.\end{cases}\qquad(4.198)$$

**Завдання 4.9.5.** Те саме, якщо правий кінець стержня закріплений пружно.

*Вказівка.* Крайова умова справа має вигляд $G'_\omega(l,x')+hG_\omega(l,x')=0.$

## 4.10. СИСТЕМИ З ПРУЖНО ЗАКРІПЛЕНИМИ КІНЦЯМИ. МЕТОД ВІДОКРЕМЛЕННЯ ЗМІННИХ

Підходи, які були розвинуті вище для дослідження коливань струни чи стержня із закріпленими або вільними кінцями, можна поширити й на випадок, коли кінці коливальної системи є рухомими, але для їх зміщення має бути виконана певна робота. Раніше (див. підрозділи 3.1, 3.2) уже наводилися модельні приклади таких систем — струна та стержень, до кінців яких прикріплено безмасові пружинки (з коефіцієнтами жорсткості $k_1,k_2\ge 0$). Уважаючи ці системи однорідними та користуючись результатами підрозділу 3.2, приходимо до *крайових задач про коливання однорідної струни (стержня) з пружно закріпленими кінцями*, які в загальному випадку включають неоднорідне рівняння коливань

$$\frac{\partial^2 u}{\partial t^2}=a^2\frac{\partial^2 u}{\partial x^2}+f(x,t),\quad 0<x<l,\quad t>0,\qquad(4.199)$$

крайові умови

$$u_x(0,t)-h_1u(0,t)=0,\quad u_x(l,t)+h_2u(l,t)=0,\quad t\ge 0,\qquad(3.33)$$

та початкові умови

$$u(x,0)=u_0(x),\quad u_t(x,0)=v_0(x),\quad 0\le x\le l.\qquad(3.34)$$

Нагадаємо, що для однорідної струни $h_i=k_i/T_0$, $T_0$ — сила натягу струни, і для однорідного стержня $h_i=k_i/ES$, $E$ і $S$ — модуль Юнга речовини та площа поперечного перерізу стержня $(i=1,2)$. У граничному випадку $h_1,h_2\to\infty$ умови (3.33) зводяться до умов жорсткого закріплення кінців (3.36), а при $h_1,h_2=0$ — умов на вільних кінцях (3.37).

Очевидно, що безпосереднє застосування методу продовження для розв'язування задачі (4.199), (3.33), (3.34) стає громіздким і малопрозорим, бо внаслідок більш складних крайових умов значно ускладнюються як побудова продовжених функцій, так і подальший їх аналіз. З іншого боку, інтуїтивно зрозуміло, що й для цього класу задач структура розв'язків у вигляді розкладів за власними коливаннями повинна зберігатися, оскільки формування стоячих хвиль є спільною рисою обмежених коливальних систем. Фактично ця ідея лежить в основі загального підходу до лінійних крайових задач, який називається *методом відокремлення змінних або методом Фур'є*[1]. Перейдемо до викладу основ цього методу та його застосувань до систем із пружно закріпленими кінцями.

Метод відокремлення змінних можна застосовувати до задач, поставлених для лінійних диференціальних рівнянь виду

$$\left[\Gamma(\tau_1,\ldots,\tau_s)-\Lambda(x_1,\ldots,x_n)\right]u(\tau_1,\ldots,\tau_s;x_1,\ldots,x_n)=0, \qquad (4.200)$$

де $\Gamma(\tau_1,\ldots,\tau_s)$ і $\Lambda(x_1,\ldots,x_n)$ — лінійні диференціальні оператори, що містять похідні лише за змінними $\tau_1,\ldots,\tau_s$ та $x_1,\ldots,x_n$; коефіцієнти в цих операторах можуть залежати також лише від змінних відповідно $\tau_1,\ldots,\tau_s$ та $x_1,\ldots,x_n$. Якщо, наприклад, множина змінних $\tau_1,\ldots,\tau_s$ зводиться до однієї часової змінної $t$, а множина змінних $x_1,\ldots,x_n$ — до трьох просторових координат $x$, $y$ і $z$, то вигляду (4.200) набирає тривимірне хвильове рівняння Д'Аламбера

$$\frac{\partial^2 u}{\partial t^2}-a^2\Delta u=0,$$

якщо покласти $\Gamma(t)=\partial^2/\partial t^2$, $\Lambda(x,y,z)=a^2\Delta$, та тривимірне рівняння теплопровідності для неоднорідного середовища

$$\frac{\partial u}{\partial t}-\frac{1}{\rho(x,y,z)c(x,y,z)}\operatorname{div}\left[\kappa(x,y,z)\operatorname{grad}u\right]=0,$$

якщо $\Gamma(t)=\partial/\partial t$, $\Lambda(x,y,z)=\left[\rho(x,y,z)c(x,y,z)\right]^{-1}\operatorname{div}\left[\kappa(x,y,z)\operatorname{grad}\right]$ і виконується умова, що густина $\rho(x,y,z)$, питома теплоємність $c(x,y,z)$ та коефіцієнт теплопровідності $\kappa(x,y,z)$ середовища не змінюються з часом.

Вихідною точкою методу відокремлення змінних є відшукання нетривіальних частинних розв'язків рівняння (4.200) у вигляді добутків

[1] У літературі також широко вживається термін «метод розділення змінних».

$$u(\tau_1,...,\tau_s;x_1,...,x_n)=T(\tau_1,...,\tau_s)X(x_1,...,x_n). \qquad (4.201)$$

Підставимо добуток (4.201) у рівняння (4.200) та поділимо обидві частини здобутого рівняння на шукану функцію, тобто на добуток (4.201). Дістаємо співвідношення

$$\frac{1}{T(\tau_1,...,\tau_s)}\Gamma(\tau_1,...,\tau_s)T(\tau_1,...,\tau_s)=$$
$$=\frac{1}{X(x_1,...,x_n)}\Lambda(x_1,...,x_n)X(x_1,...,x_n). \qquad (4.202)$$

Якщо множини змінних $\tau_1,...,\tau_s$ та $x_1,...,x_n$ незалежні, то рівність (4.202) може справджуватися лише за умови, що обидві її частини дорівнюють деякій сталій, яку позначимо через $-\lambda$. Звідси бачимо, що добуток (4.201) задовольняє рівняння (4.200) тоді й лише тоді, коли самі множники $T(\tau_1,...,\tau_s)$ і $X(x_1,...,x_n)$ задовольняють рівняння

$$\Gamma(\tau_1,...,\tau_s)T(\tau_1,...,\tau_s)+\lambda T(\tau_1,...,\tau_s)=0, \qquad (4.203)$$

$$\Lambda(x_1,...,x_n)X(x_1,...,x_n)+\lambda X(x_1,...,x_n)=0, \qquad (4.204)$$

з тією самою сталою $\lambda$.

У багатьох випадках область визначення $\Omega$ функцій $u(\tau_1,...,\tau_s;x_1,...,x_n)$ за змінними $x_1,...,x_n$ є обмеженою, а на шукані розв'язки рівняння (4.200) накладається додаткова вимога, щоб на поверхні $\Sigma$, що обмежує область $\Omega$, вони задовольняли крайову умову

$$\gamma(x_1,...,x_n)u(\tau_1,...,\tau_s;x_1,...,x_n)\big|_{\Sigma}=0, \qquad (4.205)$$

де $\gamma(x_1,...,x_n)$ — лінійний диференціальний оператор, який містить лише похідні за змінними $x_1,...,x_n$ та коефіцієнти якого можуть залежати лише від змінних $x_1,...,x_n$. Тоді умова (4.205) природно переноситься й на множник $X(x_1,...,x_n)$, тобто до рівняння (4.204) додається крайова умова

$$\gamma(x_1,...,x_n)X(x_1,...,x_n)\big|_{\Sigma}=0. \qquad (4.206)$$

У загальному випадку нетривіальні розв'язки рівняння (4.204), що додатково задовольняють умову (4.206), існують лише при певних значеннях $\lambda_\nu$ параметра $\lambda$. Ці особливі значення $\lambda_\nu$ називаються ***власними значеннями***, а нетривіальні розв'язки $X_\nu(x_1,...,x_n)$, що їм відповідають, — ***власними функціями*** крайової задачі (4.204), (4.206). Знайшовши $\lambda_\nu$ та $X_\nu(x_1,...,x_n)$, а також нетривіальні розв'язки $T_\nu(\tau_1,...,\tau_s)$ рівняння (4.203) при $\lambda=\lambda_\nu$, дістаємо нетривіальні частинні розв'язки $u_\nu(\tau_1,...,\tau_s;x_1,...,x_n)=T_\nu(\tau_1,...,\tau_s)X_\nu(x_1,...,x_n)$ рівняння (4.200), що також задовольняють крайову умову (4.205).

За допомогою функцій $u_\nu(\tau_1,...,\tau_s;x_1,...,x_n)$ розв'язок крайової задачі (4.200), (4.205) у рамках методу відокремлення змінних подається у вигляді функціонального ряду (узагальненого ряду Фур'є)

$$u(\tau_1,...,\tau_s;x_1,...,x_n)=\sum_\nu C_\nu T_\nu(\tau_1,...,\tau_s)X_\nu(x_1,...,x_n), \qquad (4.207)$$

де $C_\nu$ — деякі коефіцієнти.

Скористаємося щойно описаним методом, щоб побудувати загальний розв'язок крайової задачі про вільні коливання натягненої однорідної струни з пружно закріпленими кінцями. Оскільки рівняння руху струни в такій задачі має вигляд

$$\frac{\partial^2 u}{\partial t^2}-a^2\frac{\partial^2 u}{\partial x^2}=0, \qquad (4.208)$$

то тепер множина змінних $\tau_1,...,\tau_s$ зводиться до однієї часової змінної $t$, множина змінних $x_1,...,x_n$ — до однієї просторої змінної $x$, а оператори $\Gamma=\partial^2/\partial t^2$ і $\Lambda=a^2\partial^2/\partial x^2$. Виходячи із загальної схеми, частинні розв'язки рівняння (4.208) шукаємо у вигляді добутку $u(x,t)=T(t)X(x)$. Підставивши його в рівняння (4.208), дістаємо співвідношення

$$\frac{T''(t)}{a^2T(t)}=\frac{X''(x)}{X(x)},$$

яке може виконуватися для довільних значень $x$ і $t$ лише за умови, що обидві його частини дорівнюють деякій сталій $-\lambda$. Вимагаючи також, щоб частинні розв'язки рівняння (4.208) задовольняли крайові умови (3.33), знаходимо, що функція $T(t)$ задовольняє звичайне диференціальне рівняння

$$T''(t)+\lambda a^2T(t)=0, \qquad (4.209)$$

а функція $X(x)$ — звичайне диференціальне рівняння

$$X''(x)+\lambda X(x)=0 \qquad (4.210)$$

та крайові умови

$$X'(0)-h_1X(0)=0,\quad X'(l)+h_2X(l)=0. \qquad (4.211)$$

Загальні розв'язки рівнянь (4.209) і (4.210) описуються виразами

$$T(t) = A\cos\sqrt{\lambda}at + B\sin\sqrt{\lambda}at, \tag{4.212}$$

$$X(x) = C\cos\sqrt{\lambda}x + D\sin\sqrt{\lambda}x. \tag{4.213}$$

Крайова умова в точці $x = 0$ (перша формула в (4.211)) дає співвідношення $\sqrt{\lambda}D - h_1 C = 0,$ тому функцію (4.213) можемо подати у вигляді

$$X(x) = CX_\lambda(x), \tag{4.214}$$

де

$$X_\lambda(x) = \cos\sqrt{\lambda}x + \frac{h_1}{\sqrt{\lambda}}\sin\sqrt{\lambda}x. \tag{4.215}$$

Крайова умова в точці $x = l$ (друга формула в (4.211)) далі дає рівняння $Cm(\lambda) = 0,$ де позначено

$$m(\lambda) = X'_\lambda(l) + h_2 X_\lambda(l) = (h_1 + h_2)\cos\sqrt{\lambda}l + \left(-\sqrt{\lambda} + \frac{h_1 h_2}{\sqrt{\lambda}}\right)\sin\sqrt{\lambda}l. \tag{4.216}$$

Оскільки $C \neq 0$ (інакше матимемо $X(x) \equiv 0$), то робимо висновок, що нетривіальні розв'язки однорідної системи (4.210), (4.211) для координатної частини $X(x)$ частинних розв'язків задачі (4.208), (3.33) існують для тих і лише тих значень параметра $\lambda,$ що є нулями функції $m(\lambda),$ тобто коренями рівняння

$$m(\lambda) = 0. \tag{4.217}$$

Згідно із сказаним вище, ці особливі значення параметра $\lambda$ називаються власними значеннями крайової задачі (4.210), (4.211), а нетривіальні розв'язки системи (4.210), (4.211) при цих значеннях $\lambda$ — власними функціями цієї задачі.

З явного виразу (4.216) для функції $m(\lambda)$ випливає, що вона є однозначною аналітичною функцією параметра $\lambda$ у скінченній частині комплексної площини, а *всі її нулі лежать на дійсній півосі* $[0,\infty).$ Останнє твердження можна довести і без посилання на явний вигляд виразу (4.216). Справді, нехай $\lambda_\nu$ і $X_\nu(x)$ — довільне власне значення та відповідна власна функція крайової задачі (4.210), (4.211). Тоді справджується рівність

$$X''_\nu(x) + \lambda_\nu X_\nu(x) = 0.$$

Помножимо обидві її частини на комплексно спряжену функцію $\overline{X_\nu(x)}$ та зінтегруємо отримані вирази за змінною $x$ у межах від 0 до $l.$ Дістаємо:

$$-\int_0^l \overline{X_\nu(x)} X''_\nu(x)\,dx = \lambda_\nu \int_0^l |X_\nu(x)|^2\,dx.$$

Застосувавши інтегрування частинами та скориставшись умовами (4.211), для інтеграла в лівій частині маємо:

$$-\int_0^l \overline{X_\nu(x)} X_\nu''(x)\,dx = \overline{X_\nu(0)} X_\nu'(0) - \overline{X_\nu(l)} X_\nu'(l) + \int_0^l \left|X_\nu'(x)\right|^2 dx =$$

$$= h_1 \left|X_\nu(0)\right|^2 + h_2 \left|X_\nu(l)\right|^2 + \int_0^l \left|X_\nu'(x)\right|^2 dx.$$

Отже,

$$\lambda_\nu = \frac{h_1 \left|X_\nu(0)\right|^2 + h_2 \left|X_\nu(l)\right|^2 + \int_0^l \left|X_\nu'(x)\right|^2 dx}{\int_0^l \left|X_\nu(x)\right|^2 dx}. \qquad (4.218)$$

Оскільки $h_1, h_2 \geq 0$ і всі інші вирази в чисельнику цієї формули також невід'ємні, приходимо до висновку, що $\lambda_\nu \geq 0$.

Покажемо тепер, що на додатній півосі функція $m(\lambda)$ має нескінченно багато нулів. Для цього в рівнянні (4.217) при $\lambda > 0$ перейдемо до безрозмірних змінних

$$\xi = \sqrt{\lambda}\, l > 0, \quad \eta_1 = h_1 l \geq 0, \quad \eta_2 = h_2 l \geq 0 \qquad (4.219)$$

та подамо його у вигляді

$$(\eta_1 + \eta_2)\cos\xi - \left(\xi - \frac{\eta_1\eta_2}{\xi}\right)\sin\xi = 0. \qquad (4.220)$$

Увівши далі величину $\delta = \delta(\xi)$ за формулами

$$\cos\delta = \frac{\eta_1 + \eta_2}{\sqrt{(\eta_1 + \eta_2)^2 + \left(\xi - \frac{\eta_1\eta_2}{\xi}\right)^2}},$$
$$\sin\delta = \frac{\xi - \frac{\eta_1\eta_2}{\xi}}{\sqrt{(\eta_1 + \eta_2)^2 + \left(\xi - \frac{\eta_1\eta_2}{\xi}\right)^2}}, \qquad (4.221)$$

перепишемо рівняння (4.220) як

$$\sqrt{(\eta_1 + \eta_2)^2 + \left(\xi - \frac{\eta_1\eta_2}{\xi}\right)^2}\,\cos\left(\xi + \delta(\xi)\right) = 0. \qquad (4.222)$$

З огляду на формули (4.221) та допустимі значення змінних (4.219) маємо також нерівність

$$-\frac{\pi}{2} \le \delta(\xi) = \operatorname{arctg}\frac{\xi - \dfrac{\eta_1\eta_2}{\xi}}{\eta_1 + \eta_2} \le \frac{\pi}{2}. \quad (4.223)$$

З неї випливає, що аргумент косинуса $\xi + \delta(\xi)$ у формулі (4.222) — це монотонно зростаюча функція параметра $\xi$, значення якої пробігають піввісь $(-\pi/2, \infty)$, коли змінна $\xi$ пробігає піввісь $(0, \infty)$.

Таким чином, на півосі $(0, \infty)$ існує зростаюча послідовність чисел $\xi_1, \dots, \xi_n, \dots$, таких, що

$$\xi_n + \delta(\xi_n) = \frac{\pi}{2} + \pi(n-1), \quad n = 1, 2, \dots. \quad (4.224)$$

Оскільки числа $\xi_n$ є коренями рівняння (4.220), то числа

$$\lambda_n = \frac{\xi_n^2}{l^2}, \quad n = 1, 2, \dots, \quad (4.225)$$

є власними значеннями крайової задачі (4.210), (4.211). Їм відповідають власні функції (див. формули (4.214) і (4.215))

$$X_n(x) = C_n X_{\lambda_n}(x) = C_n\left(\cos\frac{\xi_n x}{l} + \frac{h_1 l}{\xi_n}\sin\frac{\xi_n x}{l}\right). \quad (4.226)$$

Беручи до уваги формули (4.212) при $\lambda = \lambda_n$, (4.225) і (4.226), бачимо, що частинні розв'язки лінійної системи (4.208), (3.33) мають вигляд

$$T_n(x)X_n(x) = C_n\left(A\cos\omega_n t + B\sin\omega_n t\right)\left(\cos\frac{\xi_n x}{l} + \frac{h_1 l}{\xi_n}\sin\frac{\xi_n x}{l}\right), \; \omega_n = \frac{\xi_n a}{l}. \quad (4.227)$$

Вони описують власні коливання струни (стержня) з пружно закріпленими кінцями, при яких точки струни (стержня) гармонічно коливаються з власними частотами $\omega_n = \xi_n a/l$, $n = 1, 2, \dots$, та амплітудами, пропорційними власним функціям (4.226) відповідної крайової задачі (4.210), (4.211).

**Завдання 4.10.1.** Проаналізуйте поведінку коренів рівняння (4.224) у граничних випадках $h_1, h_2 \to \infty$ і $h_1, h_2 \to 0$. Випишіть власні частоти, власні функції та власні коливання однорідної струни (стержня) для цих випадків.

**Завдання 4.10.2.** Застосуйте метод відокремлення змінних безпосередньо до задачі про вільні поперечні коливання однорідної струни із закріпленими кінцями. Порівняйте отримані результати з відповідними результатами завдання 4.10.1.

**Завдання 4.10.3.** Те саме для задачі про вільні поздовжні коливання однорідного стержня з вільними кінцями.

Власні функції (4.226) визначені з точністю до сталого множника $C_n$. У багатьох випадках його зручно вибирати таким чином, щоб виконувалася *умова нормування*

$$\int_0^l |X_n(x)|^2 \, dx = 1. \tag{4.228}$$

Власні функції, що задовольняють це співвідношення, називаються *нормованими*.

З огляду на формулу (4.214) умова нормування (4.228) набирає вигляду

$$\int_0^l |X_{\lambda_n}(x)|^2 \, dx = \frac{1}{C_n^2}. \tag{4.229}$$

Підставивши сюди явний вираз (4.215) для функції $X_{\lambda_n}(x)$, можемо безпосередньо обчислити інтеграл (4.229) та відновити сталу $C_n$. Однак, маючи на увазі отримати не лише значення $C_n$, але й низку додаткових співвідношень, ми розглянемо інтеграл

$$J_{\lambda,\lambda_n} = \int_0^l X_\lambda(x) X_{\lambda_n}(x) dx, \tag{4.230}$$

де $\lambda$ — довільне число, яке в загальному випадку не дорівнює власному значенню $\lambda_n$.

Нагадаємо, що функції $X_\lambda(x)$ і $X_{\lambda_n}(x)$ задовольняють рівняння (4.210) при значеннях сталої відповідно $\lambda$ і $\lambda = \lambda_n$. Випишемо ці рівняння, помножимо їх на, відповідно, $X_{\lambda_n}(x)$ і $X_\lambda(x)$, а потім віднімемо. У такий спосіб знаходимо, що підінтегральний вираз в інтегралі (4.230) можна подати у вигляді

$$X_\lambda(x) X_{\lambda_n}(x) = \frac{1}{\lambda - \lambda_n}\left[X_\lambda(x) X''_{\lambda_n}(x) - X''_\lambda(x) X_{\lambda_n}(x)\right] =$$

$$= \frac{1}{\lambda - \lambda_n} \frac{d}{dx}\left[X_\lambda(x) X'_{\lambda_n}(x) - X'_\lambda(x) X_{\lambda_n}(x)\right],$$

звідки бачимо, що

$$J_{\lambda,\lambda_n} = \frac{1}{\lambda-\lambda_n}\Big[X_\lambda(x)X'_{\lambda_n}(x) - X'_\lambda(x)X_{\lambda_n}(x)\Big]\Big|_0^l.$$

Оскільки обидві функції $X_\lambda(x)$ і $X_{\lambda_n}(x)$ задовольняють першу з крайових умов (4.211), далі знаходимо:

$$J_{\lambda,\lambda_n} = \frac{1}{\lambda-\lambda_n}\Big[X_\lambda(l)X'_{\lambda_n}(l) - X'_\lambda(l)X_{\lambda_n}(l)\Big].$$

Цей інтеграл зручно виразити через функцію (4.216), для чого в квадратних дужках додамо та віднімемо вираз $h_2X_\lambda(l)X_{\lambda_n}(l)$. Маємо:

$$J_{\lambda,\lambda_n} = \frac{1}{\lambda-\lambda_n}\Big[X_\lambda(l)X'_{\lambda_n}(l) + h_2X_\lambda(l)X_{\lambda_n}(l) - X'_\lambda(l)X_{\lambda_n}(l) - h_2X_\lambda(l)X_{\lambda_n}(l)\Big] =$$

$$= \frac{1}{\lambda-\lambda_n}\Big[m(\lambda_n)X_\lambda(l) - m(\lambda)X_{\lambda_n}(l)\Big].$$

Узявши до уваги, що $m(\lambda_n)=0$, остаточно отримуємо:

$$J_{\lambda,\lambda_n} = -\frac{m(\lambda)-m(\lambda_n)}{\lambda-\lambda_n}X_{\lambda_n}(l). \tag{4.231}$$

Проаналізуємо поведінку виразу (4.231). Перейшовши до границі при $\lambda\to\lambda_n$, дістаємо співвідношення

$$\int_0^l X_{\lambda_n}^2(x)\,dx = -m'(\lambda_n)X_{\lambda_n}(l), \tag{4.232}$$

де символ $m'(\lambda)$ означає похідну функції $m(\lambda)$ за аргументом $\lambda$. Оскільки, згідно з формулами (4.215)–(4.217) і (4.225),

$$m'(\lambda_n) = \frac{l}{2}\left\{-\frac{1}{\xi_n}\left[1+(h_1+h_2)l+\frac{h_1h_2l^2}{\xi_n^2}\right]\sin\xi_n + \left(-1+\frac{h_1h_2l^2}{\xi_n^2}\right)\cos\xi_n\right\},$$

$$X_{\lambda_n}(l) = \cos\xi_n + \frac{h_1l}{\xi_n}\sin\xi_n,$$

$$m(\lambda_n) = \big(h_1+h_2\big)\cos\xi_n + \frac{\xi_n}{l}\left(-1+\frac{h_1h_2l^2}{\xi_n^2}\right)\sin\xi_n = 0,$$

після дещо громіздких перетворень[1] знаходимо, що при $0<h_1,h_2<\infty$

---

[1] Помножте виписані вирази для $m'(\lambda_n)$ і $X_{\lambda_n}(l)$, винесіть за дужки множник $\cos^2\xi_n = \left(1+\mathrm{tg}^2\,\xi_n\right)^{-1}$, усюди замініть $\mathrm{tg}\,\xi_n$ виразом, знайденим із рівняння $m(\lambda_n)=0$, та спростіть отримані дроби.

$$C_n^{-2}=\frac{l}{2}\left(1+\frac{h_1^2l^2}{\xi_n^2}\right)+\frac{l^2}{2}\frac{\left(h_1+h_2\right)\left(1+\dfrac{h_1^2l^2}{\xi_n^2}\right)\left(1+\dfrac{h_1h_2l^2}{\xi_n^2}\right)}{\left(h_1^2+h_2^2\right)l^2+\xi_n^2+\dfrac{h_1^2h_2^2l^4}{\xi_n^2}}. \qquad (4.233)$$

З другого боку, переходячи у виразі (4.231) до границі $\lambda\to\lambda_m\neq\lambda_n$ та враховуючи співвідношення $m(\lambda_m)=0,$ знаходимо

$$\int_0^l X_{\lambda_m}(x)X_{\lambda_n}(x)\,dx=-\frac{m(\lambda_m)-m(\lambda_n)}{\lambda_m-\lambda_n}X_{\lambda_n}(l)=0. \qquad (4.234)$$

Звідси відразу бачимо, що власні функції (4.226) крайової задачі (4.210), (4.211), які відповідають різним власним значенням, задовольняють *умову ортогональності* (для однорідних систем)

$$\int_0^l \overline{X_m(x)}X_n(x)\,dx=0, \quad m\neq n. \qquad (4.235)$$

Функції з такою властивістю називаються *ортогональними*.

Зауважимо, що знак комплексного спряження у формулах (4.228), (4.235) (та їх аналогах (4.246), (4.247) для неоднорідних систем) можна опустити, якщо, як і в нашому випадку, власні функції можна вибрати дійсними.

Нехай $\lambda_1<\lambda_2<\ldots<\lambda_n<\ldots$ — зростаюча послідовність власних значень крайової задачі (4.210), (4.211), а $\left\{X_n(x)=C_nX_{\lambda_n}(x)\right\}_1^\infty$ — відповідна послідовність власних функцій. Якщо останні є нормованими, то згідно з формулами (4.228) і (4.235) усі члени послідовності $\left\{X_n(x)\right\}_1^\infty$ мають властивість

$$\int_0^l \overline{X_m(x)}X_n(x)\,dx=\delta_{mn}=\begin{cases}1, & \text{якщо } m=n,\\ 0, & \text{якщо } m\neq n,\end{cases} \qquad (4.236)$$

тобто є *ортонормованими*. Звернемо увагу, що таку саму властивість мають і послідовності нормованих власних функцій

$$\left\{X_n(x)=\sqrt{\frac{2}{l}}\sin\frac{\pi nx}{l}\right\}_1^\infty \text{ та } \left\{X_n(x)=\sqrt{\frac{2}{l}}\sin\frac{\pi(n+1/2)x}{l}\right\}_0^\infty,$$

що з'являються в задачах про коливання відповідно струни із закріпленими кінцями (див. формулу (4.92)) та стержня з лівим закріпленим і правим вільним кінцями (див. формулу (4.160)).

На підставі співвідношень (4.236) можемо стверджувати: якщо неперервна функція $\varphi(x)$ допускає зображення у вигляді рівномірно збіжного функціонального ряду

$$\varphi(x)=\sum_{n=1}^{\infty}c_n X_n(x), \tag{4.237}$$

то таке зображення єдине. Справді, помноживши обидві частини рівності (4.237) на функцію $X_m(x)$ та почленно зінтегрувавши рівномірно збіжний ряд у правій частині здобутої рівності, за допомогою формули (4.236) знаходимо однозначні вирази для коефіцієнтів ряду (4.237):

$$\int_0^l \varphi(x)X_m(x)\,dx=\sum_{n=1}^{\infty}c_n\int_0^l X_n(x)X_m(x)\,dx=\sum_{n=1}^{\infty}c_n\delta_{nm}=c_m. \tag{4.238}$$

Ряди виду (4.237) називаються *узагальненими рядами Фур'є* функції $\varphi(x)$ за (ортонормованими) власними функціями $X_n(x)$, а коефіцієнти (4.238) — відповідними *узагальненими коефіцієнтами Фур'є* чи просто *коефіцієнтами Фур'є* функції $\varphi(x)$. У наступному підрозділі буде показано, що послідовність часткових сум $S_N(x)$ ряду (4.237) для будь-якої функції $\varphi(x)$, неперервної на відрізку $[0,l]$, збігається до $\varphi(x)$ принаймні в середньому квадратичному, тобто

$$\lim_{N\to\infty}\int_0^l \left|\varphi(x)-S_N(x)\right|^2 dx=0;$$

якщо ж $\varphi(x)$ додатково є неперервно диференційовною на $[0,l]$ та задовольняє умови $\varphi(0)=\varphi(l)=0$, то ця послідовність збігається до $\varphi(x)$ рівномірно на $[0,l]$.

Той факт, що функції достатньо широкого класу можна розвинути на відрізку $[0,l]$ в узагальнені ряди Фур'є, дозволяє сформулювати простий алгоритм побудови розв'язку крайової задачі

$$\begin{gathered}\frac{\partial^2 u}{\partial t^2}=a^2\frac{\partial^2 u}{\partial x^2},\quad 0<x<l,\quad t>0,\\ u_x(0,t)-h_1u(0,t)=0,\quad u_x(l,t)+h_2u(l,t)=0,\quad t\ge 0,\\ u(x,0)=u_0(x),\quad u_t(x,0)=v_0(x),\quad 0\le x\le l,\end{gathered} \tag{4.239}$$

про вільні коливання струни з пружно закріпленими кінцями у вигляді суперпозиції власних коливань такої струни. Якщо початкові функції $u_0(x)$ і $v_0(x)$ допускають на $[0,l]$ зображення у вигляді абсолютно і рівномірно збіжних рядів виду (4.237), які можна почленно диференціювати відповідно двічі та один раз, і при цьому отримувані ряди теж є абсолютно і рівномірно збіжними, то (єдиний) двічі неперервно диференційовний розв'язок задачі (4.239) можна подати у вигляді

$$u(x,t)=\sum_{n=1}^{\infty}q_n(t)X_n(x), \tag{4.240}$$

де

$$q_n(t)=a_n\cos\omega_n t+\frac{b_n}{\omega_n}\sin\omega_n t,\quad \omega_n=a\sqrt{\lambda_n}, \tag{4.241}$$

$$a_n=\int_0^l u_0(x)X_n(x)dx,\quad b_n=\int_0^l v_0(x)X_n(x)dx. \tag{4.242}$$

Зокрема, якщо $u_0(x)\equiv 0$, то розв'язок задачі (4.239) набирає вигляду

$$u(x,t)=\rho\int_0^l G(x,x';t)v_0(x')dx', \tag{4.243}$$

де

$$G(x,x';t)=\frac{1}{\rho}\sum_{n=1}^{\infty}\frac{\sin\omega_n t}{\omega_n}X_n(x)X_n(x') \tag{4.244}$$

— (часова) функція Гріна для задачі (4.239). При наявності сили тертя з погонною густиною $-2\rho\eta u_t(x,t)$ вираз для функції Гріна змінюється, стаючи схожим за своєю структурою на вираз (4.172):

$$G(x,x';t)=\frac{1}{\rho}e^{-\eta t}\sum_{n=1}^{\infty}\frac{\sin\Omega_n t}{\Omega_n}X_n(x)X_n(x'),\quad \Omega_n=\sqrt{\omega_n^2-\eta^2}. \tag{4.245}$$

**Зауваження 4.10.1.** Ми ввели «зайві» множники $\rho$ у формулі (4.243) і $1/\rho$ у формулах (4.244) і (4.245) для того, щоб не змінювати фізичний зміст і розмірність функцій Гріна при переході до крайових задач про коливання неоднорідних струн і стержнів зі змінним розподілом маси $\rho(x)$. Як стане зрозуміло з подальшого викладу, для таких систем природними умовами нормування та ортогональності виступають співвідношення

$$\int_0^l \rho(x)|X_n(x)|^2dx=1, \tag{4.246}$$

$$\int_0^l \rho(x)\overline{X_m(x)}X_n(x)dx=0,\quad m\neq n, \tag{4.247}$$

а не (4.228) і (4.235), котрі зазвичай використовуються для однорідних систем.

Як і у випадку однорідної струни із закріпленими кінцями, розв'язок задачі про вимушені коливання однорідної струни з пруж-но закріпленими кінцями, що відбуваються у в'язкому середовищі під дією зовнішньої сили з погонною густиною $F(x,t)$ і при нульових початкових умовах $u_0(x)=0,\ \ v_0(x)=0,$ тобто крайової задачі

$$\begin{gathered}\frac{\partial^2 u}{\partial t^2}=a^2\frac{\partial^2 u}{\partial x^2}-2\eta\frac{\partial u}{\partial t}+f(x,t),\ \ 0<x<l,\ \ t>0,\\ u_x(0,t)-h_1u(0,t)=0,\ \ u_x(l,t)+h_2u(l,t)=0,\ \ t\ge 0,\\ u(x,0)=0,\ \ u_t(x,0)=0,\ \ 0\le x\le l,\end{gathered}\tag{4.248}$$

де $f(x,t)=F(x,t)/\rho,$ дається, згідно з принципом Дюамеля, формулою

$$u(x,t)=\int_0^t d\tau\int_0^l dx' G(x,x';t-\tau)F(x',\tau).\tag{4.249}$$

**Завдання 4.10.4.** Розв'яжіть крайову задачу (4.248) методом Фур'є. Переконайтеся, що знайдений розв'язок описується виразом (4.249) з функцією Гріна (4.245).

*Розв'язання.* Уважаючи власні значення $\lambda_n$ і власні функції $X_n(x)$ відповідної задачі про вільні коливання струни відомими (див. формули (4.225), (4.226) і (4.233)), розвиваємо функцію $f(x,t)$ в ряд Фур'є

$$f(x,t)=\sum_n f_n(t)X_n(x)\tag{4.250}$$

з коефіцієнтами

$$f_n(t)=\int_0^l f(x',t)X_n(x')dx'\tag{4.251}$$

та шукаємо розв'язок задачі (4.248) у вигляді ряду

$$u(x,t)=\sum_n q_n(t)X_n(x).\tag{4.252}$$

Підставимо ряди (4.250) та (4.252) в задачу (4.248). З огляду на рівняння $X_n''(x)=-\lambda_n X_n(x),$ крайові умови для функцій $X_n(x),$ а також повноту й ортонормованість останніх, для функцій $q_n(t)$ дістаємо звичайне неоднорідне диференціальне рівняння

$$\ddot q_n(t)+2\eta\,\dot q_n(t)+\omega_n^2 q_n(t)=f_n(t),\ \ t>0,\tag{4.253}$$

де $\omega_n=a\sqrt{\lambda_n}$ — частоти вільних коливань струни, та початкові умови

$$q_n(0)=0,\quad \dot q_n(0)=0.\tag{4.254}$$

Розв'язуємо рівняння (4.253) методом варіації сталих. Оскільки загальний розв'язок відповідного однорідного рівняння (коли $f_n(t)=0$) має вигляд

$$q_n(t) = A_n e^{-\eta t}\cos\Omega_n t + B_n e^{-\eta t}\sin\Omega_n t, \tag{4.255}$$

де $A_n$ і $B_n$ — сталі, $\Omega_n = \sqrt{\omega_n^2-\eta^2}$ — частоти коливань струни у в'язкому середовищі (припускаємо, що $\omega_n^2 > \eta^2$), то розв'язок неоднорідного рівняння (4.253) теж треба шукати в такому вигляді, але з коефіцієнтами $A_n$ і $B_n$, що є функціями аргументу $t$. Згідно із загальною схемою для рівнянь другого порядку, на ці функції накладаємо умову

$$\dot{A}_n\cos\Omega_n t + \dot{B}_n\sin\Omega_n t = 0. \tag{4.256}$$

Друге рівняння для $A_n$ і $B_n$ знаходимо, підставивши вираз (4.255) у рівняння (4.253) та виконавши відповідні скорочення:

$$-\dot{A}_n\sin\Omega_n t + \dot{B}_n\cos\Omega_n t = \frac{1}{\Omega_n}e^{\eta t}f_n(t). \tag{4.257}$$

Помноживши обидві частини рівняння (4.256) на $\cos\Omega_n t$, рівняння (4.257) — на $\sin\Omega_n t$, та віднявши здобуті рівняння, дістаємо

$$\dot{A}_n = -\frac{1}{\Omega_n}e^{\eta t}\sin\Omega_n t\, f_n(t),$$

звідки

$$A_n(t) = -\frac{1}{\Omega_n}\int_0^t e^{\eta\tau}\sin\Omega_n\tau\, f_n(\tau)\,d\tau,$$

де сталу інтегрування ми прирівняли до нуля в силу першої початкової умови (4.254): $q_n(0) = A_n(0) = 0$.

Далі множимо обидві частини рівняння (4.256) на $\sin\Omega_n t$, рівняння (4.257) — на $\cos\Omega_n t$, та додаємо здобуті рівняння. Дістаємо

$$\dot{B}_n = \frac{1}{\Omega_n}e^{\eta t}\cos\Omega_n t\, f_n(t),$$

$$B_n(t) = \frac{1}{\Omega_n}\int_0^t e^{\eta\tau}\cos\Omega_n\tau\, f_n(\tau)\,d\tau,$$

де, з огляду на другу умову (4.254) та явний вигляд функції $A_n(t)$, сталу інтегрування теж прирівняли до нуля: $\dot{q}_n(0) = \dot{A}_n(0) + \Omega_n B_n(0) = = \Omega_n B_n(0) = 0$, тобто $B_n(0) = 0$.

Підставивши знайдені функції $A_n$ і $B_n$ у праву частину виразу (4.255), після простих тригонометричних перетворень знаходимо:

$$q_n(t) = \int_0^t e^{-\eta(t-\tau)} \frac{\sin\Omega_n(t-\tau)}{\Omega_n} f_n(\tau)d\tau. \qquad (4.258)$$

Відповідь (4.249) отримуємо, підставивши вираз (4.258) у формулу (4.252) та записавши коефіцієнти $f_n(\tau)$ у вигляді (4.251).

Повторюючи майже дослівно аналіз, виконаний раніше для струни із закріпленими кінцями, знаходимо, що під дією комплекснозначної гармонічної сили з погонною густиною $\tilde{F}(x,t) = \tilde{F}_0(x)e^{-i\omega t}$ всі точки струни з пружно закріпленими кінцями поступово починають коливатися з частотою сили $\omega$ і комплекснозначною амплітудою

$$\tilde{A}_\omega(x) = \int_0^l G_\omega(x,x')\tilde{F}_0(x')dx', \qquad (4.259)$$

де

$$G_\omega(x,x') = \frac{1}{\rho}\sum_{n=1}^{\infty}\frac{1}{\Omega_n^2 - (\omega+i\eta)^2} X_n(x)X_n(x') \qquad (4.260)$$

– частотна функція Гріна струни. При $\eta \to 0$ амплітуда усталених коливань (4.259) є, очевидно, розв'язком крайової задачі

$$\begin{gathered} -y'' - \frac{\omega^2}{a^2} y = \frac{1}{\rho a^2}\tilde{F}_0(x), \\ y'(0) - h_1 y(0) = y'(l) + h_2 y(l) = 0. \end{gathered} \qquad (4.261)$$

Розв'язок цієї задачі при $\tilde{F}_0(x) = \delta(x - x')$ має зміст частотної функції Гріна для однорідної струни (стержня) з пружно закріпленими кінцями, що коливається без тертя.

**Завдання 4.10.5.** Покажіть, що частотна функція Гріна однорідної струни з пружно закріпленими кінцями, що коливається в нев'язкому середовищі, описується виразом

$$G_\omega(x,x') = \frac{1}{\rho a^2 m\left(\omega^2/a^2\right)} \times$$

$$\times \begin{cases} \left[\cos\frac{\omega}{a}x + \frac{h_1 a}{\omega}\sin\frac{\omega}{a}x\right]\left[\cos\frac{\omega}{a}(l-x') + \frac{h_2 a}{\omega}\sin\frac{\omega}{a}(l-x')\right], & x < x', \\ \left[\cos\frac{\omega}{a}(l-x) + \frac{h_2 a}{\omega}\sin\frac{\omega}{a}(l-x)\right]\left[\cos\frac{\omega}{a}x' + \frac{h_1 a}{\omega}\sin\frac{\omega}{a}x'\right], & x > x', \end{cases} \qquad (4.262)$$

де функція $m$ дається формулою (4.216) і $0 \le x, x' \le l$. Проаналізуйте граничні випадки жорстко закріплених та вільних кінців.

*Вказівка.* Покажіть, що розв'язок крайової задачі (4.261) при $\tilde{F}_0(x) = \delta(x - x')$ має структуру

$$G_\omega(x,x') = \begin{cases} A\left[\cos\frac{\omega}{a}x + \frac{h_1 a}{\omega}\sin\frac{\omega}{a}x\right], & x < x', \\ B\left[\cos\frac{\omega}{a}(l-x) + \frac{h_2 a}{\omega}\sin\frac{\omega}{a}(l-x)\right], & x > x', \end{cases}$$

та знайдіть коефіцієнти $A$ і $B$, скориставшись умовами зшивання (4.189), (4.190).

**Завдання 4.10.6.** Перевірте, що при $\eta \to 0$ ряд (4.260) є рядом Фур'є функції (4.262).

*Вказівка.* Обчисліть узагальнені коефіцієнти Фур'є $\int_0^l G_\omega(x,x')X_n(x)dx$, скориставшись рівністю

$$\int_0^l G_\omega(x,x')X_n(x)dx = \int_0^{x'} G_\omega(x,x')X_n(x)dx + \int_{x'}^l G_\omega(x,x')X_n(x)dx =$$

$$= \frac{a^2}{\omega_n^2 - \omega^2}\left\{\int_0^{x'}\left[G''_\omega(x,x')X_n(x) - G_\omega(x,x')X''_n(x)\right]dx + \right.$$

$$\left. + \int_{x'}^l \left[G''_\omega(x,x')X_n(x) - G_\omega(x,x')X''_n(x)\right]dx\right\},$$

крайовими умовами (3.33) та умовами зшивання (4.189), (4.190). Цю рівність легко довести, якщо для кожної з областей $x < x'$ і $x > x'$ записати рівняння для функцій $G_\omega(x,x')$ і $X_n(x)$, помножити ці рівняння відповідно на $X_n(x)$ і $G_\omega(x,x')$, почленно відняти та зінтегрувати за змінною $x$ у межах, що відповідають кожній області.

## 4.11. РОЗКЛАДИ ЗА ВЛАСНИМИ ФУНКЦІЯМИ КРАЙОВИХ ЗАДАЧ

Щоб побудувати та проаналізувати загальні розв'язки задач про вільні та вимушені коливання однорідних струн та стержнів при різних крайових умовах, ми явно скористалися тим фактом, що довільну неперервну періодичну функцію $\varphi(x)$ можна подати у вигляді ряду Фур'є, який збігається до $\varphi(x)$ принаймні «в середньому квадратичному». Ряди Фур'є, що з'являлися в розглянутих нами задачах, кожного разу зводилися до суперпозиції власних функцій, якими визначалася форма стоячих хвиль, що виникали в однорідній коливальній системі при заданих умовах закріплення кінців. Іншими словами, ці ряди Фур'є мали вигляд *розкладів за власними функціями* відповідних крайових задач.

Зараз, на прикладі задачі про вільні коливання однорідної струни із закріпленими кінцями, ми продемонструємо загальний метод, який дозволяє пересвідчитися безпосередньо, що власні функції струни (стержня) при крайових умовах виду (3.33) мають властивість *повноти*: довільні неперервні функції (зокрема, функції, що визначають миттєві положення точок однорідної струни при коливаннях) справді можна подати у вигляді розкладів за власними функціями струни. Цей метод не спирається на результати теорії рядів Фур'є, але здобуті на основі нього розклади мають ті ж самі властивості збіжності, що й раніше розглянуті ряди Фур'є.

Метод, про який йтиме мова, ґрунтується на врахуванні аналітичних властивостей частотної функції Гріна коливальної системи як функції комплексного параметра $\lambda = (\omega^2 + 2i\eta\omega)/a^2$. Далі вважатимемо, що цей параметр може набувати довільних комплексних значень.

Нагадаємо, що частотна функція Гріна однорідної струни із закріпленими кінцями дається формулою (4.194). Опускаючи для зручності сталий множник $(\rho a^2)^{-1}$, перепишемо цю функцію у вигляді

$$g_\lambda(x,x') = \frac{1}{\chi_\lambda(l)} \begin{cases} \chi_\lambda(x)\chi_\lambda(l-x'), & x \le x', \\ \chi_\lambda(x')\chi_\lambda(l-x), & x \ge x', \end{cases} \tag{4.263}$$

де

$$\chi_\lambda(x) = \frac{\sin\sqrt{\lambda}x}{\sqrt{\lambda}}, \quad 0 \le x \le l. \tag{4.264}$$

При фіксованому значенні аргументу $x$ функцію $\chi_\lambda(x)$ можна подати у вигляді ряду

$$\chi_\lambda(x) = \sum_{n=0}^{\infty} (-1)^n \frac{x^{2n+1}}{(2n+1)!} \lambda^n, \tag{4.265}$$

який має нескінченний радіус збіжності. Тому, згідно з означенням аналізу, функція $\chi_\lambda(x)$ при кожному значенні $x \in [0,l]$ є однознач-ною аналітичною функцією в кожній обмеженій області комплексної $\lambda$-площини[1]. Очевидно, що функція $g_\lambda(x,x')$, яка дорівнює добутку однозначних аналітичних функцій, поділеному на однозначну аналітичну функцію, є при $0 < x, x' < l$ однозначною аналітичною функцією у всій $\lambda$-площині, за винятком точок $\lambda_n = \pi^2 n^2/l^2$, $n = 1,2,...$, де її знаменник $\chi_\lambda(l)$ має прості нулі, а вона сама має прості полюси. Зауважимо, що за відсутності тертя, коли $\eta \downarrow 0$, полюси частотної функції Гріна як функції параметра $\lambda$ з точністю до множника $a^2$ збігаються з квадратами власних частот коливальної системи.

Нехай $\varphi(x)$ — довільна неперервно диференційовна функція на відрізку $[0,l]$, яка задовольняє умови $\varphi(0) = \varphi(l) = 0$. Розглянемо функцію

$$\Phi_\lambda(x) = \int_0^l g_\lambda(x,x')\varphi(x')\,dx', \quad 0 \le x \le l. \tag{4.266}$$

Якщо $\mathcal{L}$ — будь-який кусково-гладкий контур у комплексній $\lambda$-площині, який охоплює однозв'язну область, що не містить, як і сам контур, полюсів функції $g_\lambda(x,x')$, то, згідно з теоремою Коші з теорії функцій комплексної змінної, при $x \in [0,l]$ можемо записати

$$\oint_{\mathcal{L}} \Phi_\lambda(x)\,d\lambda = \int_0^l \left\{ \oint_{\mathcal{L}} g_\lambda(x,x')\,d\lambda \right\} \varphi(x')\,dx' = 0.$$

Звідси випливає, що й функція $\Phi_\lambda(x)$ при $x \in [0,l]$ є однозначною аналітичною функцією в скінченній частині комплексної площини, за винятком полюсів функції Гріна $g_\lambda(x,x')$.

Розглянемо тепер контури

$$\mathcal{L}_N = \left\{ \lambda:\ \lambda = R_N e^{i\theta},\ R_N = \frac{\pi^2}{l^2}(N+1/2)^2,\ 0 \le \theta < 2\pi \right\}, \quad N = 1,2,..., \tag{4.267}$$

та проаналізуємо послідовність контурних інтегралів

$$S_N(x) = -\frac{1}{2\pi i}\oint_{\mathcal{L}_N} \Phi_\lambda(x)\,d\lambda = -\int_0^l \left\{ \frac{1}{2\pi i}\oint_{\mathcal{L}_N} g_\lambda(x,x')\,d\lambda \right\} \varphi(x')\,dx' \tag{4.268}$$

при $N \to \infty$. За теоремою про лишки кожний інтеграл $S_N(x)$ дорівнює сумі (узятій з від'ємним знаком) лишків функції $\Phi_\lambda(x)$ у точках $\pi^2 n^2/l^2$, $n = 1,2,...,N$. Ця сума, у свою чергу, виражається через суму лишків функції $g_\lambda(x,x')$ у цих точках.

---

[1] Такі функції називаються цілими.

Згідно з формулою (4.263), лишок функції $g_\lambda(x,x')$ у точці $\lambda_n = \pi^2 n^2 / l^2$ при $x > x'$ дорівнює:

$$\lim_{\lambda\to\frac{\pi^2 n^2}{l^2}} g_\lambda(x,x')\left(\lambda - \frac{\pi^2 n^2}{l^2}\right) =$$

$$= \frac{\sin\frac{\pi n x'}{l}\sin\frac{\pi n(l-x)}{l}}{\frac{\pi n}{l}\left[\frac{d}{d\lambda}\sin\sqrt{\lambda} l\right]_{\lambda=\pi^2 n^2/l^2}} =$$

$$= -2\frac{(-1)^n \sin\frac{\pi n x'}{l}\sin\frac{\pi n x}{l}}{l(-1)^n} = -\frac{2}{l}\sin\frac{\pi n x'}{l}\sin\frac{\pi n x}{l}. \qquad (4.269)$$

Такий самий вираз дістаємо й при $x < x'$. Бачимо, на прикладі натягненої однорідної струни з жорстко закріпленими кінцями, що *для лишків частотної функції Гріна* $g_\lambda(x,x')$ *у її полюсах справджуються співвідношення*

$$\lim_{\lambda\to\lambda_n} g_\lambda(x,x')(\lambda-\lambda_n) = -X_n(x)X_n(x'), \qquad (4.270)$$

*де* $X_n(x)$ *— нормована власна функція, що відповідає власному значенню* $\lambda_n$. Це твердження носить загальний характер: воно залишається правильним не лише для однорідних струн (стержнів) з іншими умовами виду (3.33) на кінцях, але й для неоднорідних струн (стержнів).

**Завдання 4.11.1.** Скориставшись явними виразами для частотної функції Гріна стержня з а) лівим жорстко закріпленим і правим вільним кінцями і б) обома вільними кінцями, перевірте співвідношення (4.270) для цих випадків.

Повертаючись до інтегралів (4.268), на підставі формули (4.269) знаходимо, що

$$S_N(x) = \sum_{n=1}^{N}\left\{\frac{2}{l}\int_0^l \sin\frac{\pi n x'}{l}\varphi(x')dx'\right\}\sin\frac{\pi n x}{l}, \qquad (4.271)$$

тобто ці інтеграли дорівнюють частковим сумам ряду (4.93) для функції $\varphi(x)$.

З іншого боку, границю інтегралів (4.268) при $N\to\infty$ можна обчислити безпосередньо. Для цього підінтегральні функції $\sin\sqrt{\lambda}(l-x')$ і $\sin\sqrt{\lambda}x'$ у виразі для функції $\Phi_\lambda(x)$,

$$\Phi_\lambda(x)=\frac{\sin\sqrt{\lambda}(l-x)}{\sqrt{\lambda}\sin\sqrt{\lambda}l}\int_0^x\sin\sqrt{\lambda}x'\,\varphi(x')\,dx'+$$

$$+\frac{\sin\sqrt{\lambda}x}{\sqrt{\lambda}\sin\sqrt{\lambda}l}\int_x^l\sin\sqrt{\lambda}(l-x')\,\varphi(x')\,dx', \qquad (4.272)$$

перепишемо як похідні

$$\sin\sqrt{\lambda}x'=-\frac{1}{\sqrt{\lambda}}\frac{d}{dx'}\cos\sqrt{\lambda}x',\quad \sin\sqrt{\lambda}(l-x')=\frac{1}{\sqrt{\lambda}}\frac{d}{dx'}\cos\sqrt{\lambda}(l-x')$$

та виконаємо інтегрування частинами. Маємо $(\varphi'(x')\equiv d\varphi(x')/dx')$:

$$\Phi_\lambda(x)=\frac{\sin\sqrt{\lambda}(l-x)}{\sqrt{\lambda}\sin\sqrt{\lambda}l}\left[-\frac{1}{\sqrt{\lambda}}\cos\sqrt{\lambda}x'\,\varphi(x')\Big|_0^x\right]+$$

$$+\frac{\sin\sqrt{\lambda}x}{\sqrt{\lambda}\sin\sqrt{\lambda}l}\left[\frac{1}{\sqrt{\lambda}}\cos\sqrt{\lambda}(l-x')\,\varphi(x')\Big|_x^l\right]+$$

$$+\frac{\sin\sqrt{\lambda}(l-x)}{\lambda\sin\sqrt{\lambda}l}\int_0^x\cos\sqrt{\lambda}x'\,\varphi'(x')\,dx'-$$

$$-\frac{\sin\sqrt{\lambda}x}{\lambda\sin\sqrt{\lambda}l}\int_x^l\cos\sqrt{\lambda}(l-x')\,\varphi'(x')\,dx'. \qquad (4.273)$$

Беручи до уваги умови $\varphi(0)=\varphi(l)=0,$ бачимо, що підстановки верхніх та нижніх меж інтегрування у формулі (4.273) дають

$$\frac{-\sin\sqrt{\lambda}(l-x)\cos\sqrt{\lambda}x-\sin\sqrt{\lambda}x\cos\sqrt{\lambda}(l-x)}{\lambda\sin\sqrt{\lambda}l}\varphi(x)=-\frac{1}{\lambda}\varphi(x).$$

Отже,

$$\Phi_\lambda(x)=-\frac{1}{\lambda}\varphi(x)+\frac{1}{\lambda}\int_0^l Q(x,x';\lambda)\,\varphi'(x')\,dx', \qquad (4.274)$$

де

$$Q(x,x';\lambda)=\frac{1}{\sin\sqrt{\lambda}l}\begin{cases}\sin\sqrt{\lambda}(l-x)\cos\sqrt{\lambda}x', & x\ge x',\\ -\cos\sqrt{\lambda}(l-x')\sin\sqrt{\lambda}x, & x\le x'.\end{cases} \qquad (4.275)$$

Звідси знаходимо:

$$-\frac{1}{2\pi i}\oint_{\mathcal{L}_N}\Phi_\lambda(x)\,d\lambda=\varphi(x)-\frac{1}{2\pi i}\oint_{\mathcal{L}_N}\frac{d\lambda}{\lambda}\int_0^l Q(x,x';\lambda)\varphi'(x')\,dx'. \qquad (4.276)$$

Залишається довести, що інтеграли

$$J_N(x) \equiv -\frac{1}{2\pi i}\oint\limits_{\mathcal{L}_N}\frac{d\lambda}{\lambda}\int\limits_0^l Q(x,x';\lambda)\varphi'(x')\,dx' =$$

$$= -\frac{1}{2\pi}\int\limits_0^{2\pi} d\theta \int\limits_0^l Q(x,x';R_N e^{i\theta})\varphi'(x')\,dx' \qquad (4.277)$$

у правій частині формули (4.276), що беруться по колах $\mathcal{L}_N$, спадають до нуля при $N \to \infty$. Для цього знайдемо оцінки зверху та знизу для функцій, що стоять відповідно в чисельнику та знаменнику формули (4.275).

Перш за все зазначимо, що для будь-яких дійсних чисел $x$ і $y$

$$\left|1 \mp e^{ix}e^{-y}\right| = \sqrt{(1 \mp \cos x e^{-y})^2 + \sin^2 x e^{-2y}} = \sqrt{1 \mp 2\cos x e^{-y} + e^{-2y}}.$$

Узявши до уваги, що $-1 \le \cos x \le 1$ і $e^{-y} > 0$, звідси, зокрема, маємо нерівності

$$\left|1 - e^{ix}e^{-y}\right| \le \sqrt{1 + 2e^{-y} + e^{-2y}} = 1 + e^{-y}, \qquad (4.278)$$

$$\left|1 \mp e^{ix}e^{-y}\right| \ge \left|1 \mp \cos x e^{-y}\right| \ge \left|1 - e^{-y}\right|. \qquad (4.279)$$

Крім того, для будь-якого кола $\mathcal{L}_N$, на якому, нагадаємо, $\lambda = R_N e^{i\theta}$, $0 \le \theta < 2\pi$, і довільного дійсного параметра $u \ge 0$ можемо записати:

$$i\sqrt{\lambda}u = -\sqrt{R_N}\,u\sin(\theta/2) + i\sqrt{R_N}\,u\cos(\theta/2). \qquad (4.280)$$

Зокрема,

$$i\sqrt{\lambda}\,l = -\pi(N+1/2)\sin(\theta/2) + i\pi(N+1/2)\cos(\theta/2). \qquad (4.281)$$

Скориставшись співвідношеннями (4.280) і (4.278) та згадавши формулу $|e^{ix}| = 1$, для довільного кола $\mathcal{L}_N$ знаходимо:

$$\left|\sin\sqrt{\lambda}u\right| = \left|\frac{1}{2i}\left(e^{i\sqrt{\lambda}u} - e^{-i\sqrt{\lambda}u}\right)\right| =$$

$$= \frac{1}{2}e^{\sqrt{R_N}\,u\sin(\theta/2)}\left|1 - e^{2i\sqrt{R_N}\,u\cos(\theta/2)}e^{-2\sqrt{R_N}\,u\sin(\theta/2)}\right| \le$$

$$\le \frac{1}{2}e^{\sqrt{R_N}\,u\sin(\theta/2)}\left(1 + e^{-2\sqrt{R_N}\,u\sin(\theta/2)}\right) \le e^{\sqrt{R_N}\,u\sin(\theta/2)}. \qquad (4.282)$$

Аналогічно,

$$\left|\cos\sqrt{\lambda}u\right| \le e^{\sqrt{R_N}\,u\sin(\theta/2)}. \qquad (4.283)$$

Перейдемо тепер до аналізу поведінки функції $\sin\sqrt{\lambda}\,l$ на $\mathcal{L}_N$. Оскільки при $0 \le x \le \pi/2$ $\sin x \ge 2x/\pi$, то при $0 \le \theta \le \pi$ $\sin(\theta/2) \ge \theta/\pi$ і, відповідно, $1 - e^{-2\sqrt{R_N}\,l\sin(\theta/2)} \ge 1 - e^{-2\sqrt{R_N}\,l\theta/\pi}$. Ураховуючи цю нерівність і нерівність (4.279), на дузі $\pi\ln 2\big/\left(2\sqrt{R_N}\,l\right) < \theta \le \pi$ кола $\mathcal{L}_N$ маємо:

$$\left|\sin\sqrt{\lambda}\,l\right| = \frac{1}{2}e^{\sqrt{R_N}\,l\sin(\theta/2)}\left|1 - e^{2i\sqrt{R_N}\,l\cos(\theta/2)}e^{-2\sqrt{R_N}\,l\sin(\theta/2)}\right| \ge$$

$$\ge \frac{1}{2}e^{\sqrt{R_N}\,l\sin(\theta/2)}\left(1 - e^{-2\sqrt{R_N}\,l\sin(\theta/2)}\right) \ge$$

$$\ge \frac{1}{2}e^{\sqrt{R_N}\,l\sin(\theta/2)}\left(1 - e^{-2\sqrt{R_N}\,l\theta/\pi}\right) >$$

$$> \frac{1}{2}e^{\sqrt{R_N}\,l\sin(\theta/2)}\left(1 - e^{-\ln 2}\right) = \frac{1}{4}e^{\sqrt{R_N}\,l\sin(\theta/2)}. \qquad (4.284)$$

З іншого боку, при $\pi/2 \le x \le \pi$ $\sin x \ge 2(\pi - x)/\pi$, тому при $\pi \le \theta \le 2\pi$ $\sin(\theta/2) \ge (2\pi - \theta)/\pi$ і, відповідно, $1 - e^{-2\sqrt{R_N}\,l\sin(\theta/2)} \ge 1 - e^{-2\sqrt{R_N}\,l(2\pi-\theta)/\pi}$. Тоді на дузі $\pi \le \theta < 2\pi - \pi\ln 2\big/\left(2\sqrt{R_N}\,l\right)$ кола $\mathcal{L}_N$

$$\left|\sin\sqrt{\lambda}\,l\right| \ge \frac{1}{2}e^{\sqrt{R_N}\,l\sin(\theta/2)}\left(1 - e^{-2\sqrt{R_N}\,l\sin(\theta/2)}\right) \ge$$

$$\ge \frac{1}{2}e^{\sqrt{R_N}\,l\sin(\theta/2)}\left(1 - e^{-2\sqrt{R_N}\,l(2\pi-\theta)/\pi}\right) >$$

$$> \frac{1}{2}e^{\sqrt{R_N}\,l\sin(\theta/2)}\left(1 - e^{-\ln 2}\right) = \frac{1}{4}e^{\sqrt{R_N}\,l\sin(\theta/2)}. \qquad (4.285)$$

Таким чином, скрізь на дузі $\pi\ln 2\big/\left(2\sqrt{R_N}\,l\right) < \theta < 2\pi - \pi\ln 2\big/\left(2\sqrt{R_N}\,l\right)$ кола $\mathcal{L}_N$ справджується оцінка

$$\left|\sin\sqrt{\lambda}\,l\right| \ge \frac{1}{4}e^{\sqrt{R_N}\,l\sin(\theta/2)}. \qquad (4.286)$$

Для достатньо малих значень $\theta$, що відповідають дузі $0 \le \theta \le$ $\le \pi\ln 2\big/\left(2\sqrt{R_N}\,l\right) = \ln 2/(2N+1)$, оцінки носять більш тонкий характер. Перепишемо перший рядок формули (4.284) у вигляді

$$\left|\sin\sqrt{\lambda}\,l\right| = \frac{1}{2}e^{\sqrt{R_N}\,l\sin(\theta/2)}\left|1 - e^{2i\sqrt{R_N}\,l}e^{-2i\sqrt{R_N}\,l[1-\cos(\theta/2)]}e^{-2\sqrt{R_N}\,l\sin(\theta/2)}\right|.$$

Ураховуючи співвідношення $e^{2i\sqrt{R_N}\,l} = e^{i(2\pi N+\pi)} = -1$ та нерівність (4.279), знаходимо:

$$\left|\sin\sqrt{\lambda}\,l\right| = \frac{1}{2}e^{\sqrt{R_N}l\sin(\theta/2)}\left|1+e^{-2i\sqrt{R_N}l[1-\cos(\theta/2)]}e^{-2\sqrt{R_N}l\sin(\theta/2)}\right| \geq$$

$$\geq \frac{1}{2}e^{\sqrt{R_N}l\sin(\theta/2)}\left|1+\cos\left[2\sqrt{R_N}l\left(1-\cos\frac{\theta}{2}\right)\right]e^{-2\sqrt{R_N}l\sin(\theta/2)}\right|. \qquad (4.287)$$

Проаналізуємо поведінку аргументу косинуса в цій формулі при $N\to\infty$, коли $2\sqrt{R_N}l = \pi(2N+1)\to\infty$, а $\theta\to 0$. Оскільки при $0\leq x\leq \pi/2$ $\cos x \geq 1 - x^2/2$, то при $0\leq\theta\leq\pi$ маємо нерівність $1-\cos(\theta/2)\leq\theta^2/8$. З неї випливає, що для дуги $0\leq\theta\leq\ln 2/(2N+1)$

$$1-\cos\frac{\theta}{2}\leq\frac{\ln^2 2}{8(2N+1)^2},\quad 2\sqrt{R_N}l\left(1-\cos\frac{\theta}{2}\right)\leq\frac{\pi\ln^2 2}{8(2N+1)},$$

тобто при $N\to\infty$ аргумент косинуса монотонно спадає до нуля і, отже,

$$\cos\left[2\sqrt{R_N}l\left(1-\cos\frac{\theta}{2}\right)\right]\geq 1-\frac{\pi^2\ln^4 2}{128(2N+1)^2}.$$

Беручи далі до уваги нерівність $0\leq\sin x\leq x$ при $x\geq 0$, бачимо, що при $0\leq\theta\leq\ln 2/(2N+1)$ і $N\to\infty$ виконується нерівність

$$1\geq e^{-2\sqrt{R_N}l\sin(\theta/2)}\geq e^{-\sqrt{R_N}l\theta}\geq e^{-\pi\ln 2/2}=\left(\frac{1}{2}\right)^{\pi/2}>\frac{1}{4},$$

тому

$$\left|\sin\sqrt{\lambda}\,l\right|\geq\frac{1}{2}e^{\sqrt{R_N}l\sin(\theta/2)}\left\{1+e^{-2\sqrt{R_N}l\sin(\theta/2)}\left[1-\frac{\pi^2\ln^4 2}{128(2N+1)^2}\right]\right\}\geq$$

$$\geq\frac{1}{2}e^{\sqrt{R_N}l\sin(\theta/2)}\left\{1+\frac{1}{4}\left[1-\frac{\pi^2\ln^4 2}{128(2N+1)^2}\right]\right\}>\frac{5}{8}e^{\sqrt{R_N}l\sin(\theta/2)}. \qquad (4.288)$$

Таку саму нерівність дістаємо й для дуги $2\pi-\ln 2/(2N+1)<\theta<2\pi$. Справді, рівняння цієї дуги можна подати у вигляді $2\pi-\alpha<\theta<2\pi$, де $0<\alpha\leq\ln 2/(2N+1)$. Переписавши формулу (4.287) у вигляді

$$\left|\sin\sqrt{\lambda}\,l\right|\geq\frac{1}{2}e^{\sqrt{R_N}l\sin(\theta/2)}\left|1-\cos\left(2\sqrt{R_N}l\cos\frac{\theta}{2}\right)e^{-2\sqrt{R_N}l\sin(\theta/2)}\right|$$

та скориставшись формулами $\sin(\pi-\alpha/2)=\sin(\alpha/2)$ і $\cos(\pi-\alpha/2)=$ $=-\cos(\alpha/2)$, повертаємося до щойно розглянутого випадку.

Таким чином, з формул (4.286), (4.288) випливає, що скрізь на колах $\mathcal{L}_N$ при $N>>1$ справджується нерівність

$$\left|\sin\sqrt{\lambda}\,l\right| \ge \frac{1}{4}e^{\pi(N+1/2)\sin(\theta/2)}. \tag{4.289}$$

З нерівностей (4.282), (4.283), (4.289) і формули (4.275) знаходимо, що скрізь на колах $\mathcal{L}_N$ при $N \gg 1$

$$\left|Q(x,x';\lambda)\right| \le 4e^{-\pi(N+1/2)|x-x'|\sin(\theta/2)/l}. \tag{4.290}$$

Щоб оцінити $\left|J_N(x)\right|$, розіб'ємо область інтегрування в подвійному інтегралі у формулі (4.277) на дві частини, $|x'-x| \le \delta$ і $|x'-x| > \delta$, $\delta > 0$. Скориставшись нерівністю (4.290), для інтеграла $\left|J_N^{(1)}(x)\right|$ по першій області маємо оцінку:

$$\left|J_N^{(1)}(x)\right| = \frac{1}{2\pi}\left|\int_0^{2\pi} d\theta \int_{x-\delta}^{x+\delta} Q(x,x';R_N e^{i\theta})\varphi'(x')\,dx'\right| \le$$

$$\le \frac{1}{2\pi}\int_0^{2\pi} d\theta \int_{x-\delta}^{x+\delta} \left|Q(x,x';R_N e^{i\theta})\right|\left|\varphi'(x')\right| dx' \le 4\int_{x-\delta}^{x+\delta}\left|\varphi'(x')\right| dx'. \tag{4.291}$$

В інтегралі $\left|J_N^{(2)}(x)\right|$ по другій області для $\left|Q(x,x';R_N e^{i\theta})\right|$ скористаємося оцінками (див. формулу (4.290) і нерівності, що використовувалися при виведенні формул (4.284) і (4.285))

$$\left|Q(x,x';R_N e^{i\theta})\right| \le 4e^{-(N+1/2)\delta\theta/l}, \quad 0 \le \theta \le \pi,$$

$$\left|Q(x,x';R_N e^{i\theta})\right| \le 4e^{-(N+1/2)\delta(2\pi-\theta)/l}, \quad \pi \le \theta \le 2\pi.$$

Розбивши область інтегрування за кутом $\theta$ на проміжки $0 \le \theta \le \pi$ і $\pi \le \theta \le 2\pi$, маємо:

$$\left|J_N^{(2)}(x)\right| = \frac{1}{2\pi}\left|\int_0^{2\pi} d\theta \left[\int_0^{x-\delta} Q(x,x';R_N e^{i\theta})\varphi'(x')\,dx' + \right.\right.$$

$$\left.\left. + \int_{x+\delta}^{l} Q(x,x';R_N e^{i\theta})\varphi'(x')\,dx'\right]\right| \le$$

$$\le \frac{1}{2\pi}\int_0^{\pi} d\theta\, 4e^{-(N+1/2)\delta\theta/l}\left[\int_0^{x-\delta}\left|\varphi'(x')\right| dx' + \int_{x+\delta}^{l}\left|\varphi'(x')\right| dx'\right] +$$

$$+ \frac{1}{2\pi}\int_{\pi}^{2\pi} d\theta\, 4e^{-(N+1/2)\delta(2\pi-\theta)/l}\left[\int_0^{x-\delta}\left|\varphi'(x')\right| dx' + \int_{x+\delta}^{l}\left|\varphi'(x')\right| dx'\right] =$$

$$= \frac{4l}{\pi(N+1/2)\delta}\left(1-e^{-\pi(N+1/2)\delta/l}\right)\left[\int_0^{x-\delta}\left|\varphi'(x')\right|dx' + \int_{x+\delta}^{l}\left|\varphi'(x')\right|dx'\right] <$$

$$< \frac{4l}{\pi(N+1/2)\delta}\int_0^l\left|\varphi'(x')\right|dx'. \qquad (4.292)$$

Об'єднавши формули (4.291) і (4.292) разом, дістаємо:

$$\left|J_N(x)\right| < 4\int_{x-\delta}^{x+\delta}\left|\varphi'(x')\right|dx' + \frac{4l}{\pi(N+1/2)\delta}\int_0^l\left|\varphi'(x')\right|dx'. \qquad (4.293)$$

Для інтегровної похідної $\varphi'(x)$ перший доданок у цій формулі можна зробити як завгодно малим за рахунок вибору достатньо малого $\delta > 0.$ При будь-якому малому скінченному $\delta > 0$ другий доданок у цій формулі також спадає до нуля при $N \to \infty.$ Отже,

$$\lim_{N\to\infty}\left|J_N(x)\right| = 0. \qquad (4.294)$$

Зауважимо, що оцінка (4.294) для функції $\left|J_N(x)\right|$ не залежить від значення змінної $x.$ Тому послідовність (4.268) часткових сум $S_N(x)$ збігається до неперервно диференційовної функції $\varphi(x),$ що задовольняє умови $\varphi(0)=\varphi(l)=0,$ рівномірно на відрізку $[0,l].$

Скориставшись, як і вище, методом контурного інтегрування функції Гріна, нерівностями (4.282), (4.283), (4.289) та аргументами, наведеними після них, можемо довести таку теорему.

**Теорема 4.11.1**. Нехай $\left\{X_n(x)\right\}_1^\infty,\ \ 0 \le x \le l,$ — послідовність нормованих власних функцій, що відповідають власним коливанням однорідної струни при крайових умовах виду (3.33) (жорстке або пружне закріплення кінців, граничний випадок вільних кінців, комбінації цих типів закріплення). Тоді для будь-якої неперервно диференційовної функції $\varphi(x)$ на відрізку $[0,l],$ що задовольняє умови $\varphi(0)=\varphi(l)=0,$ ряд

$$\sum_{n=1}^{\infty}\left\{\int_0^l X_n(x')\varphi(x')dx'\right\}X_n(x) \qquad (4.295)$$

збігається до $\varphi(x)$ рівномірно на $[0,l].$

Зауважимо, що ми довели теорему 4.11.1 для послідовності власних функцій однорідної струни з жорстко закріпленими кінцями. Замінюючи в попередніх міркуваннях частотну функцію Гріна такої струни на частотні функції Гріна однорідної струни, що відповідають іншим видам закріплення її кінців, та користуючись оцінками (4.282), (4.283) і (4.289), можемо переконатися, що теорема 4.11.1 справджується і для решти перелічених крайових умов. Підкреслимо також, що у випадках, коли хоча б один кінець струни не є жорстко закріпленим або вільним, сума абсолютно збіжного ряду (4.295) не продовжується з відрізка $[0,l]$ на всю вісь як періодична функція, оскільки періоди власних функцій $X_n(x)$ у таких випадках є несумірними, тобто відношення періодів функцій $X_n(x)$ і $X_m(x)$ не є раціональними числами.

**Завдання 4.11.2.** Доведіть теорему 4.11.1 для розкладу (4.295) за власними функціями задачі про поздовжні коливання стержня з жорстко закріпленим лівим та вільним правим кінцями.

*Вказівка.* Скористайтеся явним виразом (4.198) для частотної функції Гріна стержня в цій задачі.

**Завдання 4.11.3.** Те саме для розкладу (4.295) за власними функціями задачі про поздовжні коливання стержня з вільними кінцями.

**Зауваження 4.11.1.** Для будь-якої функції $\varphi(x)$ на $[0,l]$, що задовольняє умову

$$\int_0^l |\varphi(x)|^2\, dx < \infty,$$

завжди можна знайти таку послідовність неперервно диференційовних функцій $\varphi_n(x)$ на $[0,l]$, що задовольняють умови $\varphi_n(0) = \varphi_n(l) = 0$, для якої

$$\lim_{n\to\infty} \int_0^l |\varphi(x) - \varphi_n(x)|^2\, dx = 0.$$

Спираючись на цей факт, за допомогою стандартних прийомів аналізу неважко довести, що *для будь-якої інтегровної з квадратом функції* $\varphi(x)$ *на відрізку* $[0,l]$ *послідовність часткових сум* $S_N(x)$ *ряду (4.295) збігається до* $\varphi(x)$ *у середньому квадратичному, тобто*

$$\lim_{N\to\infty} \int_0^l |\varphi(x) - S_N(x)|^2\, dx = 0.$$

## 4.12. ЗГИНАЛЬНІ КОЛИВАННЯ СТЕРЖНІВ

Уже неявно зазначалося, що метод відокремлення змінних можна застосовувати і для розв'язування крайових задач, поставлених для лінійних диференціальних рівнянь більш високого порядку, ніж рівняння поперечних коливань струни чи поздовжніх коливань стержня. Типовими прикладами, що мають важливе практичне значення, виступають задачі про згинальні коливання стержнів, зокрема, камертона. Рух останнього можна розглядати як малі поперечні коливання тонкого стержня, один кінець якого затиснутий у лещатах, а другий вільний. До цього ж класу задач ведуть і дослідження про поперечні коливання та механічну стійкість балок. Застосуванню та особливостям практичної реалізації методу Фур'є для аналізу зазначених задач і присвячено цей підрозділ.

Перш за все нагадаємо (див. підрозділ 3.2), що крайові задачі про малі згинальні коливання стержнів складаються з *диференціального рівняння другого порядку за часом та четвертого — за координатою*. Якщо стержень однорідний і в недеформованому стані збігається з відрізком $0 \le x \le l$, рівняння для функції $u(x,t)$, яка описує поперечне відхилення точки стержня з координатою $x$ у момент часу $t$ від її положення в недеформованому стержні, можна подати у вигляді

$$\frac{\partial^2 u}{\partial t^2} + c^2 \frac{\partial^4 u}{\partial x^4} = f(x,t), \quad 0 < x < l, \quad t > 0, \qquad (4.296)$$

де позначено $c^2 = EJ/\rho_V S$, $f(x,t) = F(x,t)/\rho_V S$, сталі $\rho_V$, $S$, $E$ та $J$ мають зміст, відповідно, об'ємної густини, площі поперечного перерізу, модуля Юнга та головного моменту інерції поперечного перерізу стержня, а (неперервна) функція $F(x,t)$ — погонної густини поперечних зовнішніх сил, що діють на стержень. *Початкові умови* вибираємо у стандартному вигляді:

$$u(x,0) = u_0(x), \quad u_t(x,0) = v_0(x), \quad 0 \le x \le l, \qquad (3.34)$$

де тепер початкові функції $u_0(x)$ та $v_0(x)$ є відповідно чотири рази та тричі неперервно диференційовними. Очевидно, що *кількість крайових умов для однозначного виокремлення розв'язку рівняння (4.296) за умов (3.34) становить чотири* — як правило, по дві на кожному кінці стержня. Згадавши формули (3.21)–(3.25) та позначивши координату кінця стержня через $x_0$ $(= 0$ або $l)$, можемо виписати найбільш прості з них. Маємо:

1) на жорстко затиснутому (замурованому) кінці:

$$u(x_0,t)=0,\ \ u_x(x_0,t)=0,\ \ t\geq 0; \tag{4.297}$$

2) на шарнірно закріпленому (опертому) кінці:

$$u(x_0,t)=0,\ \ u_{xx}(x_0,t)=0,\ \ t\geq 0; \tag{4.298}$$

3) на вільному кінці:

$$u_{xx}(x_0,t)=0,\ \ u_{xxx}(x_0,t)=0,\ \ t\geq 0. \tag{4.299}$$

Типова крайова задача про згинальні коливання однорідного стержня полягає в тому, щоб у півсмузі $D:\{0\leq x\leq l, t\geq 0\}$ знайти функцію $u(x,t)$, яка, маючи неперервні похідні до четвертого порядку за координатою та до другого — за часом, задовольняє рівняння (4.296) у внутрішніх точках півсмуги $D$ та крайові (типу (4.297)–(4.299)) і початкові умови (3.34) на її межі.

При вказаних властивостях функцій $u_0(x)$, $v_0(x)$ і $F(x,t)$ розв'язок цієї задачі єдиний (див. завдання 3.2.1). З огляду на її лінійність, можемо, як звичайно, обмежитися розглядом задачі про *вільні згинальні коливання* $(F(x,t)=0)$ та задачі про *вимушені згинальні коливання* (під дією сили $F(x,t)$ і при нульових початкових умовах). Сума розв'язків цих редукованих задач і дає шуканий розв'язок.

Користуючись загальною схемою методу Фур'є, почнемо аналіз із задачі про вільні коливання. Для цього шукаємо розв'язок відповідного рівняння

$$\frac{\partial^2 u}{\partial t^2}+c^2\frac{\partial^4 u}{\partial x^4}=0,\ \ 0<x<l,\ \ t>0, \tag{4.300}$$

у вигляді добутку $u(x,t)=T(t)X(x)$. Підставивши цей добуток у рівняння (4.300) та відокремивши змінні, для часової частини дістаємо рівняння другого порядку

$$T''+\lambda c^2 T=0,\ \ t>0, \tag{4.301}$$

а для координатної — рівняння четвертого порядку

$$X^{(4)}-\lambda X=0,\ \ 0<x<l. \tag{4.302}$$

Далі з умов (4.297)–(4.299) знаходимо крайові умови для функції $X(x)$:

1) на жорстко затиснутому кінці:

$$X(x_0)=0,\ \ X'(x_0)=0; \tag{4.303}$$

2) на шарнірно закріпленому (опертому) кінці:

$$X(x_0)=0,\quad X''(x_0)=0; \tag{4.304}$$

3) на вільному кінці:

$$X''(x_0)=0,\quad X'''(x_0)=0. \tag{4.305}$$

Отже, для координатної частини $X(x)$ шуканої функції $u(x,t)$ дістаємо крайову задачу, що складається з рівняння (4.302) та чотирьох умов типу (4.303)–(4.305) (зрозуміло, що можемо мати різні комбінації цих умов — у залежності від способів закріплення лівого та правого кінців стержня). Ця задача відрізняється від крайової задачі (4.210), (4.211) для струни з пружно закріпленими кінцями, але й для неї зберігаються означення *власних значень і власних функцій* як, відповідно, певних значень $\lambda_\nu$ параметра $\lambda$, для яких вона має нетривіальні розв'язки, та самих цих розв'язків $X_\nu$. Неважко бачити, що власні значення $\lambda_\nu$ *дійсні і невід'ємні*, а власні функції $X_\nu$, що відповідають різним власним значенням, — *ортогональні*.

Щоб довести ці твердження, скористаємося способом, випробуваним при аналізі крайової задачі (4.210), (4.211). А саме, помножимо рівняння (4.302) для функції $X_\nu(x)$, що відповідає власному значенню $\lambda_\nu$, на комплексно спряжену функцію $\overline{X_\nu(x)}$, та зінтегруємо обидві частини здобутої рівності за $x$ у межах від 0 до $l$. Маємо:

$$\int_0^l \overline{X_\nu(x)}X_\nu^{(4)}(x)\,dx=\lambda_\nu\int_0^l |X_\nu(x)|^2\,dx.$$

Застосовуючи дворазове інтегрування частинами до інтеграла зліва, знаходимо:

$$\lambda_\nu=\frac{\left[\overline{X_\nu(x)}X_\nu'''(x)-\overline{X_\nu'(x)}X_\nu''(x)\right]\Big|_0^l+\int_0^l |X_\nu''(x)|^2\,dx}{\int_0^l |X_\nu(x)|^2\,dx}. \tag{4.306}$$

Для будь-яких комбінацій крайових умов (4.303)–(4.305) вираз у квадратних дужках дорівнює нулю. Інтеграли, що залишаються у правій частині цієї рівності, дійсні та невід'ємні, а, отже, значення $\lambda_\nu$ також дійсні та невід'ємні.

Тепер, ураховуючи дійсність власних значень $\lambda_\nu$, запишемо рівняння (4.302) окремо для кожної з власних функцій $X_\nu(x)$ і $\overline{X_\mu(x)}$, що відповідають різним власним значенням $\lambda_\nu\neq\lambda_\mu$, помножимо перше рівняння на $\overline{X_\mu(x)}$, друге — на $X_\nu(x)$, віднімемо здобуті рівняння та зінтегруємо результат за $x$ у межах від 0 до $l$. Знаходимо:

$$(\lambda_\nu - \lambda_\mu)\int_0^l \overline{X_\mu(x)} X_\nu(x)\,dx = \int_0^l \left[\overline{X_\mu(x)} X_\nu^{(4)}(x) - \overline{X_\mu^{(4)}(x)} X_\nu(x)\right] dx.$$

Двічі інтегруючи частинами справа, дістаємо рівність

$$(\lambda_\nu - \lambda_\mu)\int_0^l \overline{X_\mu(x)} X_\nu(x)\,dx = \left[\overline{X_\mu(x)} X_\nu'''(x) - \overline{X_\mu'''(x)} X_\nu(x)\right]\Bigg|_0^l - \left[\overline{X_\mu'(x)} X_\nu''(x) - \overline{X_\mu''(x)} X_\nu'(x)\right]\Bigg|_0^l, \quad (4.307)$$

права частина якої дорівнює нулю для довільних комбінацій умов (4.303)–(4.305). Отже, власні функції задач (4.301)–(4.305) задовольняють співвідношення (4.235), тобто є ортогональними. Додатково вимагаючи, щоб власні функції $X_\nu(x)$ були нормованими (задовольняли співвідношення (4.228)), можемо записати для них умову *ортонормованості* (див. формулу (4.236)):

$$\int_0^l \overline{X_\mu(x)} X_\nu(x)\,dx = \delta_{\mu\nu}. \quad (4.308)$$

Перейдемо до відшукання власних значень і власних функцій конкретних коливальних систем, що здійснюють малі згинальні коливання. Для прикладу розглянемо вільні коливання стержня (балки) з жорстко закріпленими кінцями.

Спочатку звернемо увагу на те, що власні значення крайової задачі для рівняння (4.302) з умовами (4.303) на обох кінцях додатні. Справді, загальний розв'язок рівняння (4.302) при $\lambda = 0$ має вигляд

$$X(x) = K + Lx + Mx^2 + Nx^3, \quad (4.309)$$

де $K$, $L$, $M$ і $N$ — сталі інтегрування. Крайові умови (4.303), записані для точок $x = 0$ і $x = l$, дають співвідношення

$$K = 0, \quad L = 0, \quad Ml^2 + Nl^3 = 0, \quad 2Ml + 3Nl^2 = 0. \quad (4.310)$$

З двох останніх рівнянь знаходимо, що $M = 0$, $N = 0$. Отже, при $\lambda = 0$ задача (4.302), (4.303) має лише тривіальний розв'язок $X(x) \equiv 0$.

Загальні розв'язки рівнянь (4.301) і (4.302) при $\lambda > 0$ мають вигляд

$$T(t) = A\cos\sqrt{\lambda}ct + B\sin\sqrt{\lambda}ct, \quad (4.311)$$

$$X(x) = K \operatorname{ch} \sqrt[4]{\lambda} x + L \operatorname{sh} \sqrt[4]{\lambda} x + M \cos \sqrt[4]{\lambda} x + N \sin \sqrt[4]{\lambda} x. \qquad (4.312)$$

Крайові умови (4.303) тепер дають для сталих $K$, $L$, $M$ і $N$ систему чотирьох лінійних однорідних рівнянь

$$\begin{gathered}
K + M = 0, \\
L\sqrt[4]{\lambda} + N\sqrt[4]{\lambda} = 0, \\
K \operatorname{ch} \sqrt[4]{\lambda} l + L \operatorname{sh} \sqrt[4]{\lambda} l + M \cos \sqrt[4]{\lambda} l + N \sin \sqrt[4]{\lambda} l = 0, \\
K\sqrt[4]{\lambda} \operatorname{sh} \sqrt[4]{\lambda} l + L\sqrt[4]{\lambda} \operatorname{ch} \sqrt[4]{\lambda} l - M\sqrt[4]{\lambda} \sin \sqrt[4]{\lambda} l + N\sqrt[4]{\lambda} \cos \sqrt[4]{\lambda} l = 0.
\end{gathered} \qquad (4.313)$$

Нетривіальний розв'язок цієї системи існує лише за умови, що її визначник дорівнює нулю:

$$\begin{vmatrix}
1 & 0 & 1 & 0 \\
0 & \sqrt[4]{\lambda} & 0 & \sqrt[4]{\lambda} \\
\operatorname{ch} \sqrt[4]{\lambda} l & \operatorname{sh} \sqrt[4]{\lambda} l & \cos \sqrt[4]{\lambda} l & \sin \sqrt[4]{\lambda} l \\
\sqrt[4]{\lambda} \operatorname{sh} \sqrt[4]{\lambda} l & \sqrt[4]{\lambda} \operatorname{ch} \sqrt[4]{\lambda} l & -\sqrt[4]{\lambda} \sin \sqrt[4]{\lambda} l & \sqrt[4]{\lambda} \cos \sqrt[4]{\lambda} l
\end{vmatrix} =$$

$$= 2\sqrt{\lambda}\left(1 - \operatorname{ch} \sqrt[4]{\lambda} l \cos \sqrt[4]{\lambda} l\right) = 0. \qquad (4.314)$$

Звідси випливає, що крайова задача (4.302), (4.303) має нетривіальні розв'язки $X_n(x)$ лише для тих додатних значень $\lambda_n$, $n = 1, 2, ...$, які є розв'язками рівняння

$$\operatorname{ch} \sqrt[4]{\lambda} l \cos \sqrt[4]{\lambda} l = 1. \qquad (4.315)$$

Ці значення зручно подати у вигляді

$$\lambda_n = \left(\frac{\mu_n}{l}\right)^4, \qquad (4.316)$$

де $\mu_n$ — безрозмірні корені рівняння $\operatorname{ch} \mu \cos \mu = 1$. Оскільки заздалегідь відомо, що власні значення (4.316) додатні, а вигляд останнього рівняння не змінюється при замінах $\mu \to \pm i\mu$ і $\mu \to -\mu$, то достатньо шукати лише додатні корені $\mu_n$. Для перших трьох із них числові розрахунки дають: $\mu_1 = 4{,}730$, $\mu_2 = 7{,}853$, $\mu_3 = 10{,}996$. Із зростанням $\mu$ функція $\operatorname{ch} \mu$ зростає за експоненціальним законом, тому для великих значень $\mu$ з точністю до експоненціально малих членів справджується співвідношення $\cos \mu \approx 0$. Звідси дістаємо асимптотичну оцінку $\mu_n \approx \pi(n + 1/2)$, $n >> 1$. Легко перевірити, що насправді з точністю до третього десяткового знака нею можна користуватися для всіх $\mu_n$ з $n \geq 2$.

Згідно з формулою (4.311) величини $\omega_n = c\sqrt{\lambda_n}$ мають зміст власних частот згинальних коливань системи. З огляду на попередній результат, для однорідного стержня з жорстко закріпленими кінцями маємо:

$$\omega_n = \frac{\mu_n^2}{l^2}\sqrt{\frac{EJ}{\rho_V S}}, \quad n = 1,2,.... \tag{4.317}$$

Відповідні власні функції знаходимо, підставляючи значення $\lambda_n$ у систему (4.313) та розв'язуючи її відносно одного з коефіцієнтів, скажімо $K$. Дістаємо:

$$L_n = -\frac{\operatorname{ch}\mu_n - \cos\mu_n}{\operatorname{sh}\mu_n - \sin\mu_n}K_n, \quad M_n = -K_n, \quad N_n = -L_n,$$

де індекс $n$ указує на залежність коефіцієнтів від $\lambda_n$. Для власних функцій отримуємо

$$\begin{gathered} X_n(x) = K_n X_{\lambda_n}(x), \\ X_{\lambda_n}(x) = \left(\operatorname{ch}\frac{\mu_n x}{l} - \cos\frac{\mu_n x}{l}\right) - \frac{\operatorname{ch}\mu_n - \cos\mu_n}{\operatorname{sh}\mu_n - \sin\mu_n}\left(\operatorname{sh}\frac{\mu_n x}{l} - \sin\frac{\mu_n x}{l}\right), \end{gathered} \tag{4.318}$$

або, додатково врахувавши рівняння для $\mu_n$,

$$X_{\lambda_n}(x) = \left(\operatorname{ch}\frac{\mu_n x}{l} - \cos\frac{\mu_n x}{l}\right) - \operatorname{cth}\frac{\mu_n}{2}\left(\operatorname{sh}\frac{\mu_n x}{l} - \sin\frac{\mu_n x}{l}\right).$$

Бачимо, що кожному власному значенню (4.316) задачі (4.302), (4.303) відповідає лише одна (з точністю до сталого множника) власна функція, тобто ці власні значення *невироджені*.

Оскільки функції $X_{\lambda_n}(x)$ дійсні, сталу $K_n$ також зручно вибрати дійсною. Її значення фіксуємо з умови нормування для функції (4.318). У такий спосіб знаходимо:

$$K_n^{-2} = \int_0^l X_{\lambda_n}^2(x)\,dx = l. \tag{4.319}$$

Результат (4.319) можна отримати двома способами: узявши інтеграл (4.319) безпосередньо або виразивши його через значення власної функції $X_{\lambda_n}(x)$ та її похідних у точці $x = l$ за допомогою такого загального прийому. Нехай $X_\lambda(x)$ — розв'язок рівняння (4.302) при довільному значенні параметра $\lambda$, яке, взагалі кажучи, не збігається з яким-небудь власним значенням $\lambda_n$ крайових задач (4.302)–(4.305); останньому відповідає власна функція $X_{\lambda_n}(x)$. Очевидно, що для функцій $X_\lambda(x)$ і $X_{\lambda_n}(x)$ також справджується співвідношення виду (4.307), оскільки при його виведенні ми скористалися лише тим фактом, що обидві функції задовольняють рівняння (4.302). Зробивши заміни $\lambda_\mu \to \lambda$, $\lambda_\nu \to \lambda_n$, $X_\mu \to X_\lambda$, $X_\nu \to X_{\lambda_n}$ та взявши до уваги дійсність функцій $X_\lambda(x)$ і $X_{\lambda_n}(x)$, подамо його у вигляді

$$\int_0^l X_\lambda(x) X_{\lambda_n}(x)\,dx =$$

$$= \frac{\left[ X_\lambda'''(x) X_{\lambda_n}(x) - X_\lambda(x) X_{\lambda_n}'''(x) + X_\lambda'(x) X_{\lambda_n}''(x) - X_\lambda''(x) X_{\lambda_n}'(x) \right]\Big|_0^l}{\lambda - \lambda_n}. \quad (4.320)$$

Бачимо, що значення інтеграла (4.319) можна знайти, здійснивши граничний перехід $\lambda \to \lambda_n$ та розкривши за допомогою правила Лопіталя невизначеність $0/0$, яка виникає при такому переході у правій частині формули (4.320). Зокрема, для стержня із закріпленими кінцями дістаємо

$$\int_0^l X_{\lambda_n}^2(x)\,dx = \left\{ \lim_{\lambda \to \lambda_n} \left[ -\frac{\partial X_\lambda(x)}{\partial \lambda} X_{\lambda_n}'''(x) + \frac{\partial X_\lambda'(x)}{\partial \lambda} X_{\lambda_n}''(x) \right] \right\} \Bigg|_0^l, \quad (4.321)$$

де ми врахували крайові умови (4.303) для функції $X_{\lambda_n}(x)$.

Нагадаємо тепер, що штрихи в попередніх формулах означають диференціювання за змінною $x$. Оскільки функція $X_\lambda(x)$ має структуру (див. формулу (4.312))

$$X_\lambda(x) = f\left(\sqrt[4]{\lambda}\, x\right),$$

то

$$X_\lambda'(x) = \sqrt[4]{\lambda} f'\left(\sqrt[4]{\lambda}\, x\right), \quad X_\lambda''(x) = \sqrt{\lambda} f''\left(\sqrt[4]{\lambda}\, x\right),$$

де штрихи при функції $f$ означають диференціювання $f$ за її аргументом. Відповідно,

$$\frac{\partial X_\lambda(x)}{\partial \lambda} = \frac{x}{4\left(\sqrt[4]{\lambda}\right)^3} f'\left(\sqrt[4]{\lambda}\, x\right) = \frac{x}{4\lambda} X_\lambda'(x),$$

$$\frac{\partial X_\lambda'(x)}{\partial \lambda} = \frac{1}{4\left(\sqrt[4]{\lambda}\right)^3} f'\left(\sqrt[4]{\lambda}\, x\right) + \frac{x}{4\sqrt{\lambda}} f''\left(\sqrt[4]{\lambda}\, x\right) = \frac{1}{4\lambda} X_\lambda'(x) + \frac{x}{4\lambda} X_\lambda''(x).$$

Користуючись цими співвідношеннями, перейдемо у формулі (4.321) до границі $\lambda \to \lambda_n$, підставимо верхню і нижню межі інтегрування $x = 0$ і $x = l$, та знову врахуємо краєві умови (4.303). Дістанемо

$$\int_0^l X_{\lambda_n}^2(x)\,dx = \frac{l}{4\lambda_n}\left[X_{\lambda_n}''(x)\right]^2\Big|_{x=l}, \qquad (4.322)$$

звідки, з огляду на формули (4.315), (4.316) і (4.318) для власних значень і власних функцій, приходимо до співвідношення (4.319).

Зауважимо, що формулу (4.322) зручно переписати у вигляді

$$\int_0^l X_n^2(x)\,dx = \frac{l}{4}\left[\frac{d^2X_n(z)}{dz^2}\right]^2\Bigg|_{z=\mu_n}, \qquad (4.323)$$

якщо під $z$ розуміти аргумент $\mu_n x/l$ власної функції $X_n(x)$, визначеної формулою (4.318).

З формул (4.311), (4.317)–(4.319) випливає, що власні згинальні коливання однорідного стержня (чи однорідної балки) з жорстко закріпленими кінцями мають вигляд

$$T_n(t)X_n(x) = \left(A\cos\omega_n t + B\sin\omega_n t\right)X_n(x). \qquad (4.324)$$

Кожне окреме коливання (4.324) відповідає такому поперечному руху точок стержня (балки), при якому вони гармонічно коливаються з власними частотами (4.317) та амплітудами, пропорційними власним функціям (4.318). Довільне вільне згинальне коливання стержня є суперпозицією цих власних коливань і описується виразом

$$u(x,t) = \sum_{n=1}^{\infty} q_n(t)X_n(x). \qquad (4.325)$$

Якщо таке коливання виникає внаслідок початкового збудження (3.34), то, скориставшись ортонормованістю системи власних функцій, стандартним чином остаточно знаходимо:

$$q_n(t) = a_n\cos\omega_n t + \frac{b_n}{\omega_n}\sin\omega_n t, \qquad (4.326)$$

$$a_n = \int_0^l u_0(x)X_n(x)\,dx, \quad b_n = \int_0^l v_0(x)X_n(x)\,dx. \qquad (4.327)$$

Розв’язки крайових задач про вільні згинальні коливання стержнів (балок) при інших крайових умовах мають таку саму структуру. У спеціальному випадку, коли $u_0(x)=0$, формула (4.325) набирає вигляду

$$u(x,t) = \rho_V S\int_0^l G(x,x';t)v_0(x')\,dx', \qquad (4.328)$$

де функція

$$G(x,x';t)=\frac{1}{\rho_V S}\sum_{n=1}^{\infty}\frac{\sin\omega_n t}{\omega_n}X_n(x)X_n(x') \qquad (4.329)$$

називається *(часовою) функцією Гріна* для задачі (4.296)–(4.299), (3.34). Ці формули повністю аналогічні відповідним формулам (4.240)–(4.244) для задачі про вільні коливання струни з пружно закріпленими кінцями. Замінивши у формулі (4.245) $\rho$ на $\rho_V S$, дістанемо (часову) функцію Гріна для задачі про «вільні» згинальні коливання стержня (балки) у в'язкому середовищі.

**Завдання 4.12.1.** Знайдіть власні частоти та власні функції задачі про згинальні коливання однорідного стержня, лівий кінець якого закріплений жорстко, а правий вільний (камертон, консольна балка).

*Відповідь*. Власні частоти $\omega_n = c\mu_n^2/l^2$, де $\mu_n$ — додатні корені рівняння $\operatorname{ch}\mu\cos\mu=-1$. Для перших трьох $\mu_n$ числові розрахунки дають: $\mu_1=1{,}875$, $\mu_2=4{,}694$, $\mu_3=7{,}854$. При $n>3$ з точністю до трьох десяткових знаків $\mu_n\approx\pi(n-1/2)$. Якщо основний тон камертона має лінійну частоту $\nu_1=600\,\text{Гц}$, то два наступні обертони мають частоти $\nu_2=3760\,\text{Гц}$, $\nu_3=10530\,\text{Гц}$, а частоти решти обертонів виходять за поріг слухового сприйняття ($\approx 20000$ Гц). Основна частка енергії збудженого камертона припадає на основний тон, вищі гармоніки з часом затухають швидше (унаслідок більшого опору повітря при більших значеннях швидкості руху), і камертон породжує чистий звук на основній частоті.

Власні функції (з точністю до нормувального множника $C_n$)

$$X_n(x)=C_n\left[\left(\operatorname{ch}\frac{\mu_n x}{l}-\cos\frac{\mu_n x}{l}\right)-\frac{\operatorname{ch}\mu_n+\cos\mu_n}{\operatorname{sh}\mu_n+\sin\mu_n}\left(\operatorname{sh}\frac{\mu_n x}{l}-\sin\frac{\mu_n x}{l}\right)\right].$$

**Завдання 4.12.2.** Те саме для однорідної балки, якщо: а) її кінці опираються на тверді стінки; б) її лівий кінець закріплено жорстко, а правий — шарнірно; в) обидва кінці вільні.

*Відповіді*: а) $\omega_n=c\pi^2n^2/l^2$, $X_n(x)=C_n\sin(\pi nx/l)$, де $n=1,2,\ldots$. Власні функції такої балки за своєю структурою збігаються з власними функціями однорідної струни із закріпленими кінцями;

б) $\omega_n=c\mu_n^2/l^2$, де $\mu_n$ — додатні корені рівняння $\operatorname{tg}\mu=\operatorname{th}\mu$,

$$X_n(x)=C_n\left[\left(\operatorname{ch}\frac{\mu_n x}{l}-\cos\frac{\mu_n x}{l}\right)-\frac{\operatorname{ch}\mu_n-\cos\mu_n}{\operatorname{sh}\mu_n-\sin\mu_n}\left(\operatorname{sh}\frac{\mu_n x}{l}-\sin\frac{\mu_n x}{l}\right)\right];$$

в) $\omega_n=c\mu_n^2/l^2$, де $\mu_n$ — невід'ємні корені рівняння $\operatorname{ch}\mu\cos\mu=1$,

$$X_n(x)=C_n\left[\left(\mathrm{ch}\frac{\mu_n x}{l}+\cos\frac{\mu_n x}{l}\right)-\frac{\mathrm{ch}\mu_n-\cos\mu_n}{\mathrm{sh}\mu_n-\sin\mu_n}\left(\mathrm{sh}\frac{\mu_n x}{l}+\sin\frac{\mu_n x}{l}\right)\right].$$

**Зауваження 4.12.1.** Власна функція крайової задачі (4.302), (4.305) (на обох кінцях) при $\lambda=0$ є двопараметричною: $X_0(x)=C_1x+C_2$. Це означає, що для згинальних коливань однорідної балки з незакріпленими кінцями нульова власна частота, яка пов'язана з нульовим коренем $\mu_0=0$ рівняння $\mathrm{ch}\mu\cos\mu=1$, є двічі виродженою. Дві незалежні власні функції $X_{01}(x)$ і $X_{02}(x)$, що відповідають цій частоті (і мають указану структуру), можна вибрати так, щоб вони були нормованими та взаємно ортогональними, наприклад, у вигляді

$$X_{01}(x)=\frac{1}{\sqrt{l}},\quad X_{02}(x)=\frac{2\sqrt{3}}{\sqrt{l}}\left(\frac{x}{l}-\frac{1}{2}\right).$$

При $\lambda=0$ рівняння (4.301) має розв'язок $T_0(t)=a+bt$, $a$ і $b$ — сталі, і тому при $\lambda=0$ для балки з незакріпленими кінцями існують два спеціальні розв'язки рівняння (4.296)

$$u_{01}(x,t)=X_{01}(x)(a_1+b_1t),\quad u_{02}(x,t)=X_{02}(x)(a_2+b_2t).$$

Перший із цих розв'язків описує рівномірний поступальний рух балки як твердого тіла, а другий при малих $t>0$ можна інтерпретувати як обертання балки як цілого зі сталою кутовою швидкістю навколо свого центра мас (при великих $t>0$ ця інтерпретація втрачає зміст, оскільки порушується умова мализни відносних зміщень поперечних перерізів балки, за якої було виведено рівняння (4.296)).

**Завдання 4.12.3.** Доведіть, що власні функції крайових задач (4.302)–(4.305) задовольняють співвідношення

$$\int_0^l X_n^2(x)\,dx=\frac{l}{4}\left[X_n^2(z)+X_n''^2(z)-2X_n'(z)X_n'''(z)\right]\Big|_{z=\mu_n},\qquad (4.330)$$

де тепер штрихи означають диференціювання за змінною $z$, та обчисліть нормувальні множники для власних функцій із завдань 4.12.1 і 4.12.2.

Перейдемо тепер до конспективного аналізу вимушених згинальних коливань однорідного стержня під дією поперечної сили з погонною густиною $F(x,t)$. Задача полягає в тому, щоб знайти розв'язок рівняння (4.296), який задовольняє крайові умови (4.297)–(4.299) та початкові умови (3.34), при цьому достатньо обмежитися випадком, коли початкові функції $u_0(x)=0$, $v_0(x)=0$. При наявності сили тер-тя у праву частину рівняння (4.296) треба додати її погонну густину; останню моделюємо у вигляді $-2\rho_V S\eta u_t(x,t)$.

Користуючись формулами (4.328), (4.245) та принципом Дюамеля, переконуємося, що розв'язок цієї задачі можна подати у вигляді (4.249), якщо під $\rho$ розуміти величину $\rho_V S$, а під $X_n(x)$ та $\Omega_n$ — власні функції крайових задач (4.302)–(4.305) та відповідні частоти.

Якщо ж зовнішня сила $F(x,t)$ — гармонічна, то, подавши її в комплекснозначному вигляді $\tilde{F}(x,t)=\tilde{F}_0(x)e^{-i\omega t}$, знаходимо, що при $t\to\infty$ під її дією встановлюються поперечні коливання стержня, при яких усі його точки рухаються за гармонічним законом $\tilde{A}_\omega(x)e^{-i\omega t}$ з частотою $\omega$ і комплекснозначною амплітудою $\tilde{A}_\omega(x)$. Ця амплітуда описується формулою (4.259), де частотна функція Гріна стержня $G_\omega(x,x')$ має вигляд (4.260) (з $\rho_V S$ замість $\rho$ і відповідними $X_n(x)$, $\Omega_n$).

При $\eta\to 0$ амплітуда $\tilde{A}_\omega(x)$ є розв'язком диференціального рівняння

$$y^{(4)}(x)-\frac{\omega^2}{c^2}y(x)=\frac{1}{\rho_V Sc^2}\tilde{F}_0(x),\quad 0<x<l, \tag{4.331}$$

який задовольняє задані крайові умови типу (4.303)–(4.305), а функція $G_\omega(x,x')$ — узагальненим розв'язком цієї ж задачі для випадку зосередженої сили $\tilde{F}_0(x)=\delta(x-x')$:

$$G_\omega^{(4)}(x,x')-\frac{\omega^2}{c^2}G_\omega(x,x')=\frac{1}{\rho_V Sc^2}\delta(x-x'),\quad 0<x,x'<l. \tag{4.332}$$

Алгоритм побудови функції $G_\omega(x,x')$ наступний. Спершу розбиваємо відрізок $[0,l]$ на дві області, $[0,x')$ (область I) і $(x',l]$ (область II), де права частина рівняння (4.332) обертається в нуль. Далі будуємо загальні розв'язки цього однорідного рівняння для областей I і II та відновлюємо частину сталих інтегрування за заданими крайовими умовами (двома для кожного кінця стержня). Ще чотири сталі інтегрування знаходимо за допомогою чотирьох умов зшивання отриманих розв'язків у точці $x=x'$:

$$G_\omega(x'-0,x')=G_\omega(x'+0,x'), \tag{4.333}$$

$$G'_\omega(x'-0,x')=G'_\omega(x'+0,x'), \tag{4.334}$$

$$G''_\omega(x'-0,x')=G''_\omega(x'+0,x'), \tag{4.335}$$

$$G'''_\omega(x'-0,x')-G'''_\omega(x'+0,x')=-\frac{1}{\rho_V Sc^2}. \tag{4.336}$$

Умови (4.333)–(4.335) випливають з вимоги, щоб форма стержня при вимушених коливаннях під дією гармонічної сили, зосередженої в точці $x'$, описувалася б неперервною функцією з неперервною кривиною, тобто була б у кожний момент часу двічі неперервно диференційовною функцією за координатою $x$. Умова ж (4.336) є наслідком рівняння (4.332). Справді, зінтегрувавши обидві частини рівняння (4.332) за змінною $x$ у межах від $x'-\varepsilon$ до $x'+\varepsilon$, $\varepsilon>0$, та перейшовши в отриманій рівності до границі $\varepsilon\to 0$, з огляду на неперервність $G_\omega(x,x')$ у точці $x=x'$ отримуємо (4.336). Пропонуємо читачеві виконати відповідні перетворення (див. підрозділ 4.9).

**Завдання 4.12.4.** Знайдіть частотні функції Гріна для однорідної балки з жорстко закріпленими кінцями та для систем, описаних у завданнях 4.12.1 і 4.12.2.

Зупинимося коротко на циклі задач про стійкість пружних систем, що безпосередньо примикає до задач на відшукання власних частот згинальних коливань стержнів і балок. Уважається, що механічна система втрачає стійкість, якщо при зміні її параметрів найнижча власна частота досягає *нульового* значення.

Розглянемо, наприклад, умови стійкості до згинання для стиснутого однорідного стержня. У випадку малих деформацій його потенціальну енергію можна записати у вигляді

$$\Pi \approx \frac{1}{2}EJ\int_0^l u_{xx}^2\,dx - \frac{1}{2}F\int_0^l u_x^2\,dx, \qquad (4.337)$$

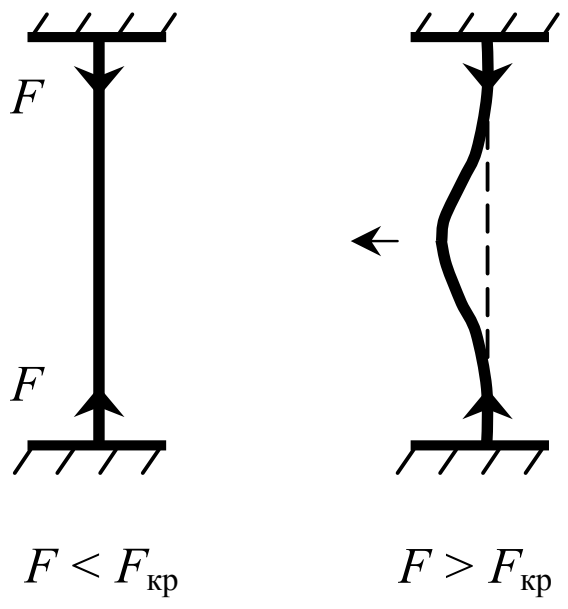

Рис. 4.9. Спонтанне згинання стержня під дією стискаючої сили $F$

де перший член описує потенціальну енергію пружної деформації стержня, а другий зумовлений стискаючою силою $F$ (він подібний до енергії пружної деформації струни, але має протилежний знак). Формула (4.337) відображає той факт, що в стиснутому стержні разом із силами жорсткості, що протидіють згинанню, також виникають сили, що намагаються його зігнути. Якщо величина стискаючої сили залишається меншою від певного порогового значення $F_{кр}$, то мінімуму енергії (4.337) відповідає прямолінійний профіль $u(x)=0$ (див. рис. 4.9). Однак при $F>F_{кр}$ прямолінійна форма стержня відповідає стану нестійкої рівноваги. Досить незначного згинання внаслідок випадкових причин, щоб ця рівновага порушилася і стержень далі сильно зігнувся[1]. Сильно стиснутий стержень узагалі не може реально залишатися в незігнутому стані.

Рівняння для рівноважного профілю $u = u(x)$ стиснутого стержня знаходимо з умови мінімуму потенціальної енергії (4.337):

$$\frac{d^4u}{dx^4} + q^2\frac{d^2u}{dx^2} = 0, \quad q^2 \equiv \frac{F}{EJ}. \tag{4.338}$$

Його загальний розв'язок має вигляд

$$u(x) = C_1\cos qx + C_2\sin qx + C_3x + C_4, \tag{4.339}$$

де сталі інтегрування визначаються крайовими умовами на кінцях стержня.

Нехай, наприклад, кінці стержня жорстко закріплені: $u(0) = u'(0) = = u(l) = u'(l) = 0$. Тоді для сталих інтегрування дістаємо систему лінійних однорідних рівнянь, нетривіальний розв'язок якої існує, якщо справджується рівняння $(\sigma \equiv ql > 0)$

$$2(1-\cos\sigma) - \sigma\sin\sigma = 0.$$

Переписавши його у вигляді

$$2\sin\frac{\sigma}{2}\left(2\sin\frac{\sigma}{2} - \sigma\cos\frac{\sigma}{2}\right) = 0,$$

бачимо, що перший ненульовий корінь збігається з першим ненульовим коренем $\sigma_1 = 2\pi$ першого множника, оскільки на інтервалі $(0, 2\pi]$ другий множник (вираз у дужках) нулів не має. Отже, $F_{кр} = 4\pi^2 EJ/l^2$, при цьому $u(x) = C\sin^2(\pi x/l)$.

**Завдання 4.12.5.** Знайдіть критичну силу $F_{кр}$, при якій однорідний стержень втрачає стійкість, якщо: а) кінці стержня закріплено на шарнірах; б) один кінець стержня замуровано в стінку, а другий вільний (задача Ейлера).

*Вказівка.* У випадку б) для опису рівноважного профілю стиснутого стержня зручно скористатися (малим) кутом $\alpha = \alpha(x)$ між віссю стержня та дотичною до нього в точці з координатою $x$ (рис. 4.10). Тоді функціонал (4.337) набирає вигляду (див. завдання 2.1.2)

$$\Pi \approx \frac{1}{2}\int_0^l dx\left[EJ\alpha'^2(x) - F\alpha^2(x)\right], \qquad (4.340)$$

і задача зводиться до визначення найменшого значення сили $F$, при якому екстремаль фунціонала (4.340) у класі $C^1[0,l]$ кривих із лівим закріпленим і правим вільним кінцями стає відмінною від нуля.

---

[1] Це явище є прикладом спонтанного порушення симетрії: унаслідок згинання порушується циліндрична симетрія стержня, яку він мав до згинання, при цьому передбачити заздалегідь, у якому саме напрямі (перпендикулярному до власної осі) він зігнеться, неможливо.

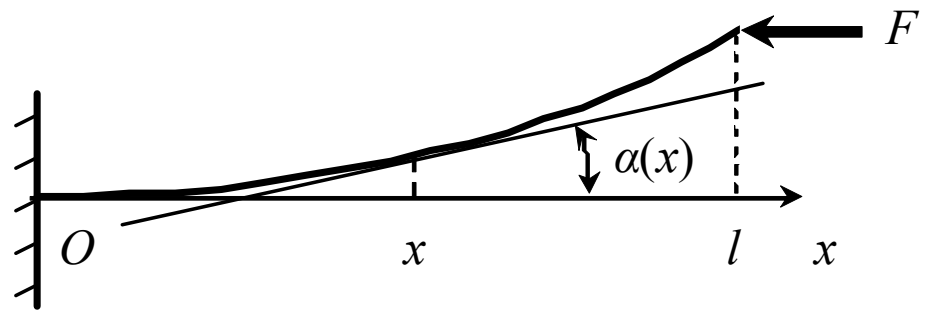

Рис. 4.10. Опис рівноважного профілю стержня за допомогою кута $\alpha(x)$

*Відповіді*: а) $F_{\text{кр}} = \pi^2 EJ/l^2$; б) $F_{\text{кр}} = \pi^2 EJ/4l^2$.

**Завдання 4.12.6.** Виведіть рівняння, яке описує малі згинальні коливання стиснутого однорідного стержня, та сформулюйте відповідну крайову задачу.

## *КОНТРОЛЬНІ ПИТАННЯ ДО РОЗДІЛУ 4*

1. *Як безпосередньо виразити розв'язок задачі Коші для вільних коливань однорідної струни через початкові функції? Для яких початкових умов він має вигляд усамітненої хвилі, що рухається вправо, не змінюючи своєї форми?*
2. *У чому полягає принцип Дюамеля? Як за його допомогою знайти розв'язок задачі Коші для вимушених коливань необмеженої однорідної струни при нульових початкових умовах?*
3. *У чому полягає метод продовження? Як за його допомогою знайти загальний розв'язок задачі Коші для вільних коливань напівобмеженої однорідної струни, кінець якої закріплений жорстко? Вільний?*
4. *Як на основі методу продовження побудувати алгоритм розв'язку крайової задачі для вільних коливань однорідної струни із закріпленими кінцями?*

5. *Як отримати загальний розв'язок крайової задачі для вільних коливань однорідної струни із закріпленими кінцями, виходячи із зображення гладкої періодичної функції на осі у вигляді ряду Фур'є?*
6. *Чи може розв'язок задачі Коші для хвильового рівняння стати з часом неперервним, якщо початкові функції мають розриви першого роду? Чи мають фізичний зміст розривні розв'язки такої задачі?*
7. *У чому полягає явище Гіббса?*
8. *Які рухи механічної системи називаються власними коливаннями? Опишіть частоти власних коливань одновимірної обмеженої струни із закріпленими кінцями. Яка характерна особливість послідовності цих частот втрачається при послабленні жорсткості закріплення кінців струни?*
9. *Знаючи початкові функції, як знайти частку енергії, що припадає на кожне власне коливання однорідної струни?*
10. *Як трансформуються власні коливання однорідної струни під дією тертя? Чому дорівнюють частоти згасаючих власних коливань струни із закріпленими кінцями при наявності тертя?*
11. *Як будується частинний розв'язок крайової задачі про вимушені коливання обмеженої струни під дією зовнішньої сили при нульових початкових умовах?*
12. *Запишіть розклад за власними функціями часової функції Гріна для задачі про вимушені коливання обмеженої струни під дією зовнішньої сили? Який її фізичний зміст?*
13. *Які властивості і зміст має частотна функція Гріна для задачі про усталені вимушені коливання обмеженої струни під дією гармонічної сили? Який вигляд має її розклад за власними функціями відповідної крайової задачі?*
14. *Яке диференціальне рівняння і які крайові умови задовольняє частотна функція Гріна для задачі про вимушені коливання струни під дією гармонічної сили? Як її знайти, користуючись цими властивостями?*
15. *Які частинні розв'язки рівняння коливань є основою методу відокремлення змінних (методу Фур'є)?*
16. *Як будуються розв'язки задач про вільні і вимушені коливання однорідної струни з пружно закріпленими кінцями на основі методу Фур'є?*
17. *Як зростають при* $n \to \infty$ *члени впорядкованої за величинами послідовності власних частот* $\omega_n$ *однорідної струни з пружно закріпленими кінцями?*
18. *Які співвідношення є ключовими при доведенні попарної ортогональності власних функцій крайової задачі про вільні коливання однорід-*

*ної струни з пружно закріпленими кінцями, що відповідають різним власним значенням?*

19. *Як і якими властивостями частотної функції Гріна для однорідної струни з пружно закріпленими кінцями можна скористатися, щоб пересвідчитися в повноті системи власних функцій відповідної крайової задачі?*
20. *Як зростають при $n \to \infty$ члени впорядкованої за величинами послідовності власних частот $\omega_n$ згинальних коливань однорідного стержня з жорстко закріпленими (вільними) кінцями? Порівняйте порядки зростання цих частот і власних частот поперечних коливань однорідної струни.*

## Розділ 5

# КОЛИВАННЯ ОБМЕЖЕНИХ НЕОДНОРІДНИХ СИСТЕМ

### 5.1. КРАЙОВА ЗАДАЧА ШТУРМА — ЛІУВІЛЛЯ

До цього часу ми розглядали одновимірні коливання однорідних систем, механічні і геометричні характеристики яких однакові в усіх їх точках. Рівняння малих коливань для таких систем зводяться до лінійних диференціальних рівнянь у частинних похідних зі сталими коефіцієнтами, а відповідні крайові задачі можуть бути ефективно проаналізовані в рамках підходу, що ґрунтується на методі Фур'є та включає: визначення власних частот і власних коливань системи; відшукання функцій, що описують вільні коливання системи при довільних початкових умовах, у вигляді суперпозиції власних коливань; використання апарату функцій Гріна як універсального інструменту, що містить усю інформацію про динамічні властивості системи; зведення за його допомогою до квадратури алгоритму розв'язування крайових задач про вимушені коливання системи під дією довільних зовнішніх сил.

Слід підкреслити, що для лінійних систем, що допускають відокремлення часової змінної від координат, такий підхід є універсальним. Перш за все, його можна поширити на задачі про одновимірні коливання *систем із розподіленими параметрами*, тобто систем, параметри яких змінюються від точки до точки (неоднорідна струна, неоднорідний стержень тощо). Коло досліджуваних задач можна розширити й далі, припустивши, що коливання систем відбуваються в пружному середовищі, а на їх кінцях реалізуються умови пружного закріплення.

Достатньо загальна математична модель для вивчення коливального руху таких систем була докладно описана нами в підрозділі 3.2. Нагадаємо, що в цій моделі параметри системи з розподіленими параметрами описуються за допомогою функцій $\rho(x) > 0$, $p(x) > 0$ і $q(x) \geq 0$, визначених на відрізку $[0,l]$, $l$ — довжина системи, та сталих $h_1, h_2 \geq 0$. Функції $\rho(x)$ і $q(x)$ уважаються неперервними, $p(x)$ — неперервно диференційовною на $[0,l]$. Коливальний рух системи описується за допомогою скалярної функції $u(x,t)$, визначеної на півсмузі $D:\{0 \le x \le l, t \ge 0\}$. У кожний момент часу $t > 0$ функція $u(x,t)$ дорівнює зміщенню з положення рівноваги поперечного перерізу системи з координатою $x$; у недеформованому стані системи $u(x,t) = 0$. Вимагається, щоб в області $D$ функція $u(x,t)$ була неперервною разом зі своїми першими і другими похідними.

Подальший аналіз цієї моделі дозволяє стверджувати наступне.

1. Якщо сили пружного зв'язку є єдиними зовнішніми силами, що діють на коливальну систему, то функцію $u(x,t)$ можна знайти як розв'язок (з відповідними аналітичними властивостями) крайової задачі, що включає однорідне диференціальне рівняння

$$\rho(x)\frac{\partial^2 u}{\partial t^2} = \frac{\partial}{\partial x}\left(p(x)\frac{\partial u}{\partial x}\right) - q(x)u, \quad 0 < x < l, \quad t > 0, \tag{3.32}$$

крайові умови

$$u_x(0,t) - h_1 u(0,t) = 0, \quad u_x(l,t) + h_2 u(l,t) = 0, \quad h_1, h_2 \ge 0, \quad t \ge 0, \tag{3.33}$$

та початкові умови

$$u(x,0) = u_0(x), \quad u_t(x,0) = v_0(x), \quad 0 \le x \le l, \tag{3.34}$$

де $u_0(x)$ і $v_0(x)$ — принаймні двічі та один раз неперервно диференційовні початкові функції.

2. При наявності зовнішніх сил, що не зводяться до сил пружного зв'язку системи з середовищем і мають погонну густину $F(x,t)$, крайова задача для функції $u(x,t)$ складається з неоднорідного диференціального рівняння

$$\rho(x)\frac{\partial^2 u}{\partial t^2} = \frac{\partial}{\partial x}\left(p(x)\frac{\partial u}{\partial x}\right) - q(x)u + F(x,t), \quad 0 < x < l, \quad t > 0, \tag{3.35}$$

крайових умов (3.33) та початкових умов (3.34)[1].

3. Розв'язки крайових задач (3.32)–(3.34) і (3.33)–(3.35) (при вказаних властивостях функцій $\rho(x)$, $q(x)$, $p(x)$, $u_0(x)$, $v_0(x)$, сталих $h_1$, $h_2$ і для неперервної функції $F(x,t)$) єдині (теорема 3.2.1).

Таким чином, користуючись указаною моделлю, приходимо до крайових задач (3.32)–(3.34) і (3.33)–(3.35), що описують коливальний рух систем із розподіленими параметрами при достатньо загальних припущеннях відносно цих параметрів та умов, у яких зна-ходяться системи. Зокрема, задача (3.32)–(3.34) при $q(x) \neq 0$ описує *коливання системи з розподіленими параметрами в пружному середовищі*, а при $q(x) \equiv 0$ — *вільні коливання цієї системи*. Задача (3.33)–(3.35) описує *коливання системи з розподіленими параметрами в пружному середовищі під дією сили* $F(x,t)$, а при $q(x) \equiv 0$ — *вимушені коливання цієї системи під дією лише сили* $F(x,t)$. В усіх цих випадках на кінцях системи реалізуються *умови пружного закріплення (3.33)*; відповідними граничними переходами можна перейти й до задач, що описують коливання систем із жорстко закріпленими чи вільними кінцями (див. формули (3.36), (3.37)). Аналізу задач типу (3.32)–(3.34) та (3.33)–(3.35) за допомогою методу Фур'є і присвячено цей розділ.

[1] У силу лінійності задач (3.32)–(3.34) і (3.33)–(3.35), що уможливлює їх редукцію, при вивчення задачі (3.33)–(3.35) достатньо обмежитися випадком, коли початкові функції дорівнюють нулю.

Оскільки єдиність розв'язків задач (3.32)–(3.34) і (3.33)–(3.35) уже була доведена, можемо перейти безпосередньо до побудови алгоритму розв'язування цих задач методом Фур'є. Слід, очевидно, почати з розгляду задачі (3.32)–(3.34) та аналізу її власних значень і власних функцій.

Рівняння (3.32), незважаючи на те, що його коефіцієнти залежать від координат, допускає відокремлення часової і просторової змінних; крайові умови (3.33) цьому ніяк не перешкоджають. Це твердження стає очевидним, якщо, користуючись загальною схемою відокремлення змінних, переписати рівняння (3.32) у вигляді

$$\left[\Gamma(t)u\right](x,t) = \left[\Lambda(x)u\right](x,t), \tag{5.1}$$

де

$$\Gamma(t) = \frac{\partial^2}{\partial t^2}, \quad \Lambda(x) = \frac{1}{\rho(x)}\left(\frac{\partial}{\partial x}p(x)\frac{\partial}{\partial x} - q(x)\right). \tag{5.2}$$

Згідно з формулами (5.1) і (5.2), відшукання нетривіальних розв'язків рівняння (5.1) у формі власних коливань $u(x,t) = X(x)\sin(\omega t + \delta)$, що задовольняють крайові умови (3.33), зводиться до відшукання нетривіальних розв'язків лінійної однорідної системи

$$-\frac{d}{dx}\left(p(x)\frac{dX}{dx}\right) + q(x)X = \lambda\rho(x)X, \quad 0 < x < l, \quad \lambda = \omega^2, \tag{5.3}$$

$$X'(0) - h_1 X(0) = 0, \quad X'(l) + h_2 X(l) = 0. \tag{5.4}$$

Лінійне однорідне диференціальне рівняння (5.3) називається *рівнянням Штурма — Ліувілля*, а крайова задача (5.3), (5.4) — *крайовою задачею Штурма — Ліувілля* (КЗШЛ).

При довільних комплексних значеннях параметра $\lambda$ система (5.3), (5.4) має тривіальний розв'язок $X(x) = 0$. Однак можуть існувати й певні значення $\lambda_n$, при яких ця система має нетривіальні розв'язки $X_n(x)$. Ці особливі значення $\lambda_n$ називаються *власними значеннями* КЗШЛ, а самі нетривіальні розв'язки $X_n(x)$ — *власними функціями* КЗШЛ, що відповідають власним значенням $\lambda_n$. Зазначимо, що власні значення КЗШЛ мають зміст квадратів власних частот систем, динаміка яких визначається рівнянням (3.32) і крайовими умовами (3.33).

Таким чином, вихідним пунктом при побудові на основі методу Фур'є алгоритму розв'язування крайових задач (3.32)–(3.34) про коливання систем із розподіленими параметрами стає вивчення загальних властивостей власних значень і власних функцій крайової задачі Штурма — Ліувілля (5.3), (5.4).

## 5.2. ВЛАСНІ ЗНАЧЕННЯ КРАЙОВОЇ ЗАДАЧІ ШТУРМА — ЛІУВІЛЛЯ

При подальшому вивченні властивостей власних значень та власних функцій КЗШЛ ми неодноразово звертатимемося до функціоналів

$$\Pi[y] = \int_0^l dx\, p(x)|y'(x)|^2 + \int_0^l dx\, q(x)|y(x)|^2 + h_1 p(0)|y(0)|^2 + h_2 p(l)|y(l)|^2, \quad (5.5)$$

$$K[y] = \int_0^l dx\, \rho(x)|y(x)|^2, \quad (5.6)$$

визначених на множині функцій, що є неперервно диференційовними та в загальному випадку комплекснозначними на відрізку $[0,l]$. Якщо не обумовлено супротивне, функції $\rho(x)$, $p(x)$ і $q(x)$ уважатимуться неперервними на $[0,l]$.

Функціонали $\Pi[y]$ і $K[y]$ називаються функціоналами відповідно потенціальної і кінетичної енергій. Походження цих назв стає зрозумілим, якщо розглянути власне коливання струни виду $u(x,t) = X(x)\sin(\omega t + \delta)$. Легко переконатися, що потенціальна та кінетична енергії, які припадають на таке коливання, виражаються через функціонали (5.5) і (5.6) співвідношеннями

$$\Pi(t) = \frac{1}{2}\Pi[X]\sin^2(\omega t + \delta), \quad (5.7)$$

$$K(t) = \frac{1}{2}K[X]\,\omega^2\cos^2(\omega t + \delta). \quad (5.8)$$

**Теорема 5.2.1.** При $\rho(x)>0,\ \ p(x)>0$ і $q(x)\geq 0$ власні значення КЗШЛ *дійсні та невід'ємні:* $\lambda \geq 0$ (іншими словами, якщо тертя відсутнє, то власні частоти $\omega$ — дійсні).

*Доведення.* Нехай $X(x)$ — власна функція КЗШЛ, що відповідає власному значенню $\lambda$, тобто є розв'язком рівняння (5.3). Помножимо обидві частини цього рівняння на комплексно спряжену функцію $\overline{X(x)}$ та зінтегруємо обидві частини здобутої рівності за змінною $x$ у межах від 0 до $l$. Дістанемо:

$$-\int_0^l dx\,\overline{X(x)}\frac{d}{dx}\left(p(x)\frac{dX(x)}{dx}\right)+\int_0^l dx\,q(x)\left|X(x)\right|^2=\lambda\int_0^l dx\,\rho(x)\left|X(x)\right|^2.$$

Інтегруючи частинами та користуючись крайовими умовами (5.4), для першого інтеграла зліва маємо:

$$-\int_0^l dx\,\overline{X(x)}\frac{d}{dx}\left(p(x)\frac{dX(x)}{dx}\right)=\left[-p(x)\overline{X(x)}X'(x)\right]\Big|_0^l+\int_0^l dx\,p(x)\left|X'(x)\right|^2=$$

$$=h_1 p(0)\left|X(0)\right|^2+h_2 p(l)\left|X(l)\right|^2+\int_0^l dx\,p(x)\left|X'(x)\right|^2.$$

З огляду на означення (5.5) і (5.6), для власного значення $\lambda$ знаходимо:

$$\lambda=\frac{\Pi[X]}{K[X]}. \qquad (5.9)$$

Оскільки при $\rho(x)>0,\ \ p(x)>0$ і $q(x)\geq 0$ виконуються нерівності $\Pi[X]\geq 0$ і $K[X]>0,$ відразу бачимо, що $\lambda\geq 0.$ Знак рівності справджується лише за умов $q(x)=0,\ h_1=0,\ h_2=0,$ тобто для струни (стержня) з вільними кінцями (при цьому також потрібно, щоб $X=\text{const}$).

З теореми 5.2.1, однак, не випливає, що на півосі $[0,\infty)$ в загальному випадку є хоча б одне власне значення КЗШЛ. Щоб переконатися, що на $[0,\infty)$ справді існує зростаюча послідовність власних значень КЗШЛ, яка прямує до нескінченності, розглянемо докладніше аналітичні властивості розв'язків рівняння Штурма — Ліувілля.

**Теорема 5.2.2.** Якщо функції $\rho(x),\ \ p(x)$ і $q(x)$ задовольняють умови попередньої теореми, а функція $p(x)$ додатково має неперервну похідну на $[0,l]$, то для будь-яких комплексних чисел $a$ і $b$ існує єдиний двічі неперервно диференційовний розв'язок $\theta(x;\lambda)$ рівняння Штурма — Ліувілля

$$-\frac{d}{dx}\left(p(x)\frac{df}{dx}\right)+q(x)f=\lambda\rho(x)f,\ \ 0\le x\le l, \tag{5.10}$$

такий, що

$$\theta(0;\lambda)=a,\ \ \theta'(0;\lambda)=b. \tag{5.11}$$

Для кожного значення змінної $x$ розв'язок $\theta(x;\lambda)$ є цілою функцією параметра $\lambda$.

*Доведення*. Почнемо з того, що покажемо, що розв'язок лінійної системи (5.10), (5.11) задовольняє *інтегральне рівняння*

$$f(x)=a+b\int_0^x\frac{p(0)}{p(x')}dx'+\int_0^x K_\lambda(x,x')f(x')dx',\ \ 0\le x\le l, \tag{5.12}$$

де

$$K_\lambda(x,x')=\left[q(x')-\lambda\rho(x')\right]\int_{x'}^x\frac{dx''}{p(x'')},\ \ 0\le x\le l. \tag{5.13}$$

Для цього спочатку інтегруємо обидві частини рівняння (5.10) за змінною $x$ від нуля до $x$ та враховуємо другу умову (5.11). Дістаємо:

$$p(x)\frac{df}{dx}=p(0)b+\int_0^x\left[q(x')-\lambda\rho(x')\right]f(x')dx'.$$

Далі ділимо обидві частини цієї рівності на $p(x)$, інтегруємо нову рівність за змінною $x$ від нуля до $x$ та враховуємо першу умову (5.11). Маємо

$$f(x)=a+b\int_0^x\frac{p(0)}{p(x')}dx'+\int_0^x\frac{dx''}{p(x'')}\int_0^{x''}\left[q(x')-\lambda\rho(x')\right]f(x')dx',$$

де область інтегрування $D$ в повторному (другому справа) інтегралі визначається як множина точок, що задовольняють співвідношення $0\le x'\le x''$, $0\le x''\le x$ (рис. 5.1). І, нарешті, змінюючи порядок інтегрування в цьому інтегралі та враховуючи, що область $D$ можна розглядати і як множину точок, що задовольняють співвідношення $x'\le x''\le x$, $0\le x'\le x$, остаточно знаходимо

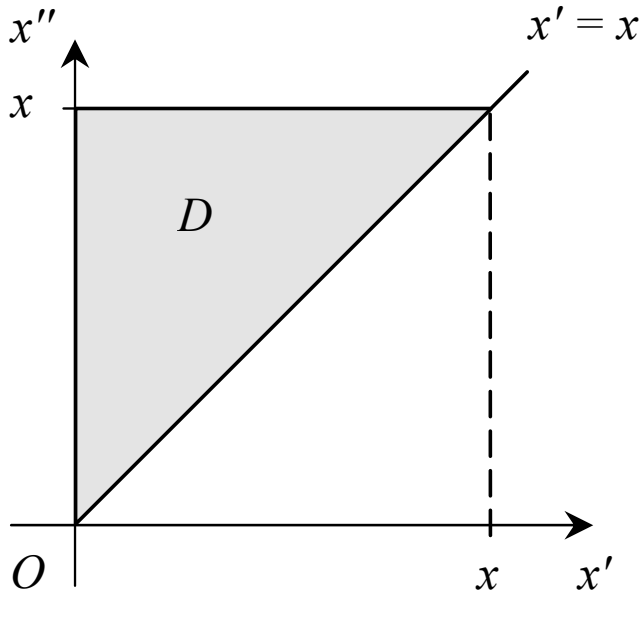

Рис. 5.1. Область інтегрування $D$ в повторному інтегралі

$$f(x)=a+b\int_0^x\frac{p(0)}{p(x')}dx'+$$

$$+\int_0^x\left\{\left[q(x')-\lambda\rho(x')\right]\int_{x'}^{x}\frac{dx''}{p(x'')}\right\}f(x')\,dx',$$

тобто співвідношення (5.12) і (5.13).

Інтегральне рівняння (5.12) і лінійна система (5.10), (5.11) еквівалентні в тому розумінні, що розв'язок рівняння (5.10), який задовольняє умови (5.11), є неперервним розв'язком рівняння (5.12), а неперервний розв'язок рівняння (5.12) є двічі неперервно диференційовною функцією, яка задовольняє рівняння (5.10) й умови (5.11).

Справді, узявши до уваги рівність $K_\lambda(x,x)=0$, бачимо, що для будь-якої неперервної функції $f(x)$ на $[0,l]$ права частина рівняння (5.12) є неперервно диференційовною функцією, похідна якої

$$f'(x)=b\frac{p(0)}{p(x)}+\int_0^x\frac{\partial K_\lambda(x,x')}{\partial x}f(x')\,dx'=$$

$$=b\frac{p(0)}{p(x)}+\frac{1}{p(x)}\int_0^x\left[q(x')-\lambda\rho(x')\right]f(x')\,dx'. \qquad (5.14)$$

Більше того, з рівності (5.14) випливає, що функція $f(x)$ має й другу неперервну похідну, і при цьому справджується співвідношення

$$\frac{d}{dx}\left[p(x)\frac{df(x)}{dx}\right]=\left[q(x)-\lambda\rho(x)\right]f(x). \qquad (5.15)$$

Отже, якщо деяка неперервна функція $f(x)$ задовольняє інтегральне рівняння (5.12), то вона є двічі неперервно диференційовною за змінною $x$ і задовольняє, згідно з формулою (5.15), рівняння Штурма — Ліувілля (5.10). Крім того, така функція задовольняє і крайові умови (5.11), що легко бачити з формул (5.12) і (5.14).

Таким чином, щоб довести теорему 5.2.2, достатньо показати, що існує неперервний розв'язок інтегрального рівняння (5.12) і що він має вказані аналітичні властивості. Для цього розглянемо нескінченну послідовність функцій $f_n(x)$, $n=0,1,2,...$, визначених на відрізку $0\le x\le l$ співвідношеннями

$$f_0(x)=a+b\int_0^x\frac{p(0)}{p(x')}dx',\quad n=0, \qquad (5.16)$$

$$f_n(x)=f_0(x)+\int_0^x K_\lambda(x,x')f_{n-1}(x')\,dx',\quad n=1,2,.... \qquad (5.17)$$

Також позначимо:

$$p_m = \min_{0\le x\le l} p(x), \quad p_M = \max_{0\le x\le l} p(x),$$
$$\rho_m = \min_{0\le x\le l} \rho(x), \quad \rho_M = \max_{0\le x\le l} \rho(x), \quad q_M = \max_{0\le x\le l} q(x). \tag{5.18}$$

Припустимо, що $|\lambda| < N$. Тоді, згідно з формулами (5.16) і (5.13),

$$\left|f_0(x)\right| \le C, \ \ C = |a| + \frac{p_M}{p_m}|b|l, \tag{5.19}$$

$$\left|K_\lambda(x,x')\right| \le D(x-x'), \ \ D = \left(q_M + N\rho_M\right)\frac{1}{p_m}, \ \ 0 \le x' \le x \le l. \tag{5.20}$$

Користуючись оцінками (5.19) і (5.20), маємо:

$$\left|f_1(x) - f_0(x)\right| \le \int_0^x \left|K_\lambda(x,x')\right|\left|f_0(x')\right|dx' \le \int_0^x CD(x-x')dx' = CD\frac{x^2}{2}.$$

При $n \ge 2$ знаходимо:

$$\left|f_n(x) - f_{n-1}(x)\right| \le \int_0^x \left|K_\lambda(x,x')\right|\left|f_{n-1}(x') - f_{n-2}(x')\right|dx' \le$$
$$\le D\int_0^x (x-x')\left|f_{n-1}(x') - f_{n-2}(x')\right|dx',$$

звідки

$$\left|f_2(x) - f_1(x)\right| \le D\int_0^x (x-x')\left|f_1(x') - f_0(x')\right|dx' \le$$
$$\le C\frac{D^2}{2}\int_0^x (x-x')x'^2dx' = C\frac{D^2x^4}{4!}$$

і далі за індукцією

$$\left|f_n(x) - f_{n-1}(x)\right| \le D\int_0^x (x-x')CD^{n-1}\frac{x'^{2(n-1)}}{(2n-2)!}dx' =$$
$$= C\frac{D^n x^{2n}}{(2n)!} \le C\frac{\left(l\sqrt{D}\right)^{2n}}{(2n)!}. \tag{5.21}$$

Звідси випливає, що при $n \to \infty$ послідовність функцій

$$f_0(x) + \sum_{k=1}^{n}\left[f_k(x) - f_{k-1}(x)\right] \tag{5.22}$$

збігається абсолютно та рівномірно відносно $\lambda$ при $|\lambda| < N$ і рівномірно відносно $x$ при $0 \le x \le l$. Тому для таких значень $\lambda$ і $x$ послідовність функцій $\left\{f_n(x)\right\}$ збігається рівномірно до деякої функції $\theta(x;\lambda)$:

$$\lim_{n\to\infty} f_n(x) = f_0(x) + \sum_{n=1}^{\infty}\left[f_n(x) - f_{n-1}(x)\right] = \theta(x;\lambda). \qquad (5.23)$$

Згідно з умовами теореми та означеннями (5.13), (5.16) і (5.17), кожний член ряду (5.23) є неперервною функцією від змінної $x$ і поліномом за параметром $\lambda$. Беручи до уваги щойно встановлені властивості рівномірної збіжності цього ряду, на підставі відомих теорем аналізу можемо стверджувати, що гранична функція $\theta(x;\lambda)$ при кожному скінченному за модулем значенні $\lambda$ є неперервною функцією змінної $x$ на відрізку $0 \le x \le l$, а для кожного значення $x$ із цього відрізка — однозначною аналітичною функцією параметра $\lambda$ в кожному крузі скінченного радіуса комплексної площини, тобто *цілою функцією*. З рівномірної збіжності послідовності $\{f_n(x)\}$ до $\theta(x;\lambda)$ та співвідношень (5.17) також випливає, що

$$\theta(x;\lambda) = \lim_{n\to\infty} f_n(x) = f_0(x) + \lim_{n\to\infty}\int_0^x K_\lambda(x,x')f_{n-1}(x')dx' =$$

$$= f_0(x) + \int_0^x K_\lambda(x,x')\theta(x';\lambda)dx'.$$

Згадавши означення (5.16) функції $f_0(x)$, бачимо, що функція $\theta(x;\lambda)$ є неперервним розв'язком інтегрального рівняння (5.12), а тому й системи (5.10), (5.11).

Підкреслимо, що за умов, які задовольняє функція $K_\lambda(x,x')$, лінійне інтегральне рівняння (5.12) має *єдиний* неперервний розв'язок. Справді, припустимо, що разом з $\theta(x;\lambda)$ існує ще один неперервний розв'язок $\varphi(x;\lambda)$ цього рівняння. Тоді різниця $h(x;\lambda) = \varphi(x;\lambda) - \theta(x;\lambda)$ є неперервним розв'язком однорідного інтегрального рівняння

$$h(x;\lambda) = \int_0^x K_\lambda(x,x')h(x';\lambda)dx', \quad 0 \le x \le l. \qquad (5.24)$$

Позначивши $\max\limits_{0\le x\le l}|h(x;\lambda)| = M$, з формули (5.24) дістаємо оцінку

$$|h(x;\lambda)| \le \int_0^x |K_\lambda(x,x')||h(x';\lambda)|dx' \le MD\int_0^x (x-x')dx' = MD\frac{x^2}{2!}, \quad 0 \le x \le l.$$

Застосувавши її знову до рівняння (5.24), отримуємо нову оцінку, яку потім теж застосовуємо до рівняння (5.24). Багаторазово повторюючи цю процедуру, послідовно знаходимо:

$$|h(x;\lambda)| \le \int_0^x |K_\lambda(x,x')||h(x';\lambda)|dx' \le M\frac{D^2}{2!}\int_0^x (x-x')x'^2dx' = MD^2\frac{x^4}{4!}, \quad 0 \le x \le l,$$

$$\ldots\ldots\ldots\ldots\ldots\ldots,$$

$$\left|h(x;\lambda)\right| \le \int_0^x \left|K_\lambda(x,x')\right|\left|h(x';\lambda)\right|dx' \le$$

$$\le M\frac{D^n}{(2n-2)!}\int_0^x (x-x')x'^{2n-2}dx' = MD^n\frac{x^{2n}}{(2n)!}, \quad 0 \le x \le l.$$

Ліва частина останньої нерівності не залежить від $n$, тоді як права при будь-якому $D > 0$ спадає до нуля при $n \to \infty$. Це означає, що $h(x;\lambda) \equiv 0$. З еквівалентності лінійної системи (5.10), (5.11) та інтегрального рівняння (5.12) далі випливає, що й система (5.10), (5.11) має єдиний розв'язок. Теорему 5.2.2 повністю доведено.

**Завдання 5.2.1.** Покажіть, що для розв'язку $\theta(x;\lambda)$ системи (5.10), (5.11) справджується оцінка

$$\left|\theta(x;\lambda)\right| \le \left(|a| + \frac{p_M}{p_m}|b|l\right)\operatorname{ch}\left(\sqrt{\frac{|\lambda|\rho_M + q_M}{p_m}}\,x\right), \quad 0 \le x \le l. \tag{5.25}$$

*Вказівка.* Скористайтеся оцінками (5.21) і зображенням функції $\theta(x;\lambda)$ у вигляді ряду (5.23).

**Завдання 5.2.2**. Покажіть, що похідна $\theta'(x;\lambda)$ за змінною $x$ розв'язку $\theta(x;\lambda)$ системи (5.10), (5.11) при кожному значенні $x$ із відрізка $0 \le x \le l$ є однозначною аналітичною функцією параметра $\lambda$ в кожному крузі скінченного радіуса комплексної площини, і що справджується оцінка

$$\left|\theta'(x;\lambda)\right| \le \frac{p_M}{p_m}|b| + \left(|a| + \frac{p_M}{p_m}|b|l\right)\sqrt{\frac{|\lambda|\rho_M + q_M}{p_m}}\operatorname{sh}\left(\sqrt{\frac{|\lambda|\rho_M + q_M}{p_m}}\,x\right). \tag{5.26}$$

*Вказівка.* Скористайтеся формулою (5.14), що виражає $\theta'(x;\lambda)$ через $\theta(x;\lambda)$, установленими вище властивостями функції $\theta(x;\lambda)$ та оцінкою (5.25).

**Завдання 5.2.3**. Інтегральні рівняння відносно функції $f(x)$, $-\infty < x_1 \le x \le x_2 < \infty$, що мають вигляд

$$f(x) = g(x) + \int_{x_1}^{x} K(x,x')f(x')dx', \tag{5.27}$$

де функції $g(x)$ і $K(x,x')$ — відомі, називаються *інтегральними рівняннями Вольтерра (другого роду)*; функція $K(x,x')$ називається *ядром інтегрального рівняння (5.27)*.

Нехай $g(x)$ і $K(x,x')$ — неперервні, і

$$\max_{x_1 \le x \le x_2} |g(x)| = C, \quad \max_{x_1 \le x' \le x \le x_2} |K(x,x')| = D.$$

Покажіть, що тоді рівняння (5.27) має єдиний неперервний розв'язок $f(x)$, який можна знайти як рівномірну границю при $n \to \infty$ послідовних наближень

$$f_0(x) = g(x), \;\; f_n(x) = g(x) + \int_{x_1}^{x} K(x,x') f_{n-1}(x')\,dx', \;\; n = 1,2,...,$$

і що для цього розв'язку справджується оцінка $|f(x)| \le Ce^{D(x-x_1)}$, $x_1 \le x \le x_2$.

*Вказівка*. Застосуйте ті ж самі аргументи, що були використані, щоб довести існування та єдиність розв'язку рівняння (Вольтерра) (5.12).

**Завдання 5.2.4**. Скориставшись методом послідовних наближень, знайдіть неперервний розв'язок рівняння Вольтерра

$$f(x) = A + \int_0^x \varphi(x') f(x')\,dx', \quad x > 0,$$

де $\varphi(x)$ — інтегровна функція на кожному скінченному інтервалі півосі $x \ge 0$, $A$ — стала.

*Вказівка*: $\int_0^x dx_1 \int_0^{x_1} dx_2 ... \int_0^{x_{n-1}} dx_n\, \varphi(x_1)\varphi(x_2)...\varphi(x_n) = \frac{1}{n!}\left(\int_0^x dx' \varphi(x')\right)^n$.

*Відповідь*: $f(x) = A\exp\left(\int_0^x \varphi(x')\,dx'\right)$.

**Завдання 5.2.5**. Знайдіть неперервний розв'язок рівняння Вольтерра

$$f(x) = A\varphi(x) + B\varphi(x)\int_0^x \varphi(x') f(x')\,dx', \quad x > 0,$$

де $\varphi(x)$ — неперервна додатна функція на півосі $x \ge 0$, $A$ і $B$ — сталі.

*Вказівка*. Зробіть підстановку $f(x) = \varphi(x)u(x)$ та скористайтеся результатом попереднього завдання.

*Відповідь*: $f(x) = A\varphi(x)\exp\left(B\int_0^x \varphi^2(x')\,dx'\right)$.

**Завдання 5.2.6**. Нехай виконуються умови теореми 5.2.2 і

$$\mu = \min_{x\in[0,l]} \left| a + b\int_0^x \frac{p(0)}{p(x')}dx' \right| > 0. \qquad (5.28)$$

Установивши, що ядро $K_\lambda(x,x')$, визначене формулою (5.13), при $\lambda < 0$ задовольняє нерівність

$$K_\lambda(x,x') \geq \frac{|\lambda|\rho_m}{p_M}(x-x'), \;\; 0 \leq x' \leq x \leq l,$$

покажіть, що для єдиного розв'язку системи (5.10), (5.11) при $\lambda < 0$ справджується оцінка

$$|\theta(x;\lambda)| \geq \mu \operatorname{ch}\left(\sqrt{\frac{|\lambda|\rho_m}{p_M}}\,x\right), \qquad (5.29)$$

тобто $\theta(x;\lambda)$ не є поліноміальною функцією.

*Вказівка.* Зверніть увагу, що за умов (5.28) і $\lambda < 0$ послідовні наближення $f_n(x)$ до розв'язку рівняння Вольтерра (5.12) мають однакові знаки.

**Зауваження 5.2.1**. З єдиності розв'язку системи (5.10), (5.11) випливає, що $\theta(x;\lambda) \equiv 0$, якщо у формулі (5.11) $a = 0$ і $b = 0$.

Позначимо через $\varphi(x;\lambda)$ і $\psi(x;\lambda)$ частинні розв'язки рівняння (5.10), що задовольняють умови

$$\begin{aligned} &\varphi(0;\lambda) = 1, \quad \varphi'(0;\lambda) = 0,\\ &\psi(0;\lambda) = 0, \quad \psi'(0;\lambda) = 1, \end{aligned} \qquad (5.30)$$

тобто частинні розв'язки, для яких відповідно $a = 1$, $b = 0$ і $a = 0$, $b = 1$. Унаслідок умов (5.30) ці розв'язки лінійно незалежні: тотожність $\alpha\varphi(x;\lambda) + \beta\psi(x;\lambda) = 0$ виконується лише за умови, що сталі $\alpha = 0$ і $\beta = 0$.

Розглянемо тепер розв'язок рівняння (5.10) виду $\chi(x;\lambda) = \varphi(x;\lambda) + h_1\psi(x;\lambda)$, який, очевидно, задовольняє першу крайову умову (5.4):

$$\chi'(0;\lambda) - h_1\chi(0;\lambda) = 0. \qquad (5.31)$$

Для ненульової комбінації $\alpha\varphi(x;\lambda) + \beta\psi(x;\lambda)$ умова (5.31) зводиться до співвідношення $\beta - \alpha h_1 = 0$, тому будь-який розв'язок $\theta(x;\lambda)$ рівняння (5.10), який задовольняє умову (5.31), можна подати у вигляді $\alpha\chi(x;\lambda)$, де $\alpha = \theta(0;\lambda)$. Зокрема, якщо $\lambda$ — власне значення КЗШЛ, то відповідна власна функція $X(x)$ із точністю до множника має збігатися з $\chi(x;\lambda)$. Звідси робимо висновок, що *з точністю до сталого* *множника кожному власному значенню крайової задачі Штурма — Ліувілля відповідає лише одна власна функція.*

У лінійних задачах математичної фізики власне значення, якому відповідає лише одна (з точністю до сталого множника) власна функція, називається *невиродженим*. Тому можемо сказати, що *всі власні значення крайової задачі Штурма — Ліувілля невироджені.*

Зауважимо, що невиродженість власних значень КЗШЛ можна довести і більш простим способом. Для цього подамо рівняння Штурма — Ліувілля (5.3) у вигляді

$$-\frac{1}{X}\frac{d}{dx}\left(p(x)\frac{dX}{dx}\right)=\lambda\rho(x)-q(x)$$

та припустимо, що одному й тому самому власному значенню $\lambda$ відповідають щонайменше дві власні функції $X_1(x)$ та $X_2(x)$. Очевидно, що кожна з них задовольняє наведене рівняння. Випишемо відповідні рівняння окремо для $X_1(x)$ та $X_2(x)$. Ураховуючи, що праві частини в цих рівняннях однакові, можемо прирівняти їх ліві частини. Дістаємо рівняння

$$-\frac{1}{X_1}\frac{d}{dx}\left(p(x)\frac{dX_1}{dx}\right)=-\frac{1}{X_2}\frac{d}{dx}\left(p(x)\frac{dX_2}{dx}\right),$$

або

$$X_2\frac{d}{dx}\left(p(x)\frac{dX_1}{dx}\right)-X_1\frac{d}{dx}\left(p(x)\frac{dX_2}{dx}\right)=\frac{d}{dx}\left(p(x)X_2\frac{dX_1}{dx}-p(x)X_1\frac{dX_2}{dx}\right)=0.$$

Бачимо, що

$$p(x)X_2\frac{dX_1}{dx}-p(x)X_1\frac{dX_2}{dx}=C.$$

За допомогою крайових умов (5.4) знаходимо, що стала інтегрування $C=0.$ Отже, дістаємо рівняння

$$X_2\frac{dX_1}{dx}-X_1\frac{dX_2}{dx}=0,$$

яке легко інтегрується:

$$\frac{1}{X_1}\frac{dX_1}{dx}=\frac{1}{X_2}\frac{dX_2}{dx},$$

$$\ln X_1=\ln X_2+\ln A,$$

$$X_1(x)=AX_2(x),$$

де $A$ — нова стала. Таким чином, власні функції $X_1(x)$ та $X_2(x)$ збігаються з точністю до сталої, тобто власне значення $\lambda$, яке їм відповідає, невироджене.

Число $\lambda \geq 0$ є власним значенням КЗШЛ, якщо разом з умовою (5.31) розв'язок $\chi(x;\lambda)$ задовольняє другу крайову умову (5.4):

$$\chi'(l;\lambda) + h_2\chi(l;\lambda) = 0. \tag{5.32}$$

Звідси бачимо, що власні значення КЗШЛ є нулями цілої функції

$$m(\lambda) = \varphi'(l;\lambda) + h_1\psi'(l;\lambda) + h_2\varphi(l;\lambda) + h_1h_2\psi(l;\lambda). \tag{5.33}$$

З формул (5.25) і (5.26) випливає, що при $|\lambda| \to \infty$ для $m(\lambda)$ справджується оцінка

$$\left|m(\lambda)\right| \leq Ql\sqrt{\frac{|\lambda|\rho_M}{p_m}}\exp\left(l\sqrt{\frac{|\lambda|\rho_M}{p_m}}\right), \tag{5.34}$$

де $Q < \infty$ — деяка додатна стала, конкретне значення якої зараз несуттєве.

Будемо розглядати $m(\lambda)$ як функцію $M(w)$ безрозмірного параметра $w = l^2\lambda\rho_M / p_m$. Згідно з оцінкою (5.29) $M(w)$ є цілою функцією, але не є поліномом. З іншого боку, з оцінки (5.34) випливає, що при $|w| \to \infty$ і для будь-якого $p > 1/2$ для $M(w)$ справджується оцінка

$$M(w) \leq C\exp\left(|w|^p\right),\ C < \infty. \tag{5.35}$$

За теоремами Бореля та Адамара[1] теорії функцій комплексного змінного така ціла функція $M(w)$ має нескінченну кількість нулів $w_1, ..., w_n, ... \neq 0,$ таких, що має місце розклад

$$M(w) = Aw^{n_0}\prod_{n=1}^{\infty}\left(1 - \frac{w}{w_n}\right),$$

де $A$ — стала, $n_0$ — ціле додатне число або нуль, і для будь-якого $p > 1/2$

$$\sum_{n=1}^{\infty}\frac{1}{|w_n|^p} < \infty.$$

З останнього співвідношення робимо висновок, що при виконанні умов теореми 5.2.2 *власні значення крайової задачі Штурма — Ліувіл-*

[1] Див. Маркушевич А. И. Теория аналитических функций. Т. 2: Дальнейшее построение теории, гл. 7. — М.: Наука, 1968. — 387 с.

*ля (5.3), (5.4) утворюють нескінченну послідовність невід'ємних чисел* $0 \le \lambda_1 < \lambda_2 < \ldots$, *таку, що для будь-якого* $p > 1/2$

$$\sum_{n=2}^{\infty} \frac{1}{\lambda_n^p} < \infty. \tag{5.36}$$

**Завдання 5.2.7**. Припустивши, що послідовні власні значення крайової задачі Штурма — Ліувілля (5.3), (5.4) зростають із номером $n$ за степеневим законом $\lambda_n = Cn^{\alpha}\left[1 + o(1)\right]$, $C$ — стала, покажіть, що $\alpha = 2$.

## 5.3. ЕКСТРЕМАЛЬНІ ВЛАСТИВОСТІ ВЛАСНИХ ЗНАЧЕНЬ КЗШЛ

З попереднього аналізу випливає, що власні значення $\{\lambda_n\}$ КЗШЛ утворюють нескінченну сукупність точок на півосі $[0,\infty)$. Якщо нумерувати $\lambda_n$ у порядку їх зростання,

$$0 \le \lambda_1 < \lambda_2 < \lambda_3 < \ldots < \lambda_n < \ldots,$$

то єдиною граничною точкою цієї сукупності є $+\infty$.

Ми вже бачили, що найменше власне значення $\lambda_1$ КЗШЛ збігається з мінімумом відношення функціоналів потенціальної $\Pi[X]$ і кінетичної $K[X]$ енергій на множині неперервно диференційовних функцій $C^1([0,l])$ [1]:

$$\lambda_1 = \min_{X \in C^1([0,l])} \frac{\Pi[X]}{K[X]}. \tag{5.37}$$

При цьому, згідно з формулою (5.9),

$$\lambda_1 = \frac{\Pi[X_1]}{K[X_1]}, \tag{5.38}$$

де $X_1$ — власна функція, яка відповідає $\lambda_1$.

При множенні довільної функції $X \in C^1([0,l])$ на ненульову сталу $C$ обидва функціонали $\Pi[X]$ і $K[X]$ множаться на однакову сталу $|C|^2$, а їх відношення не змінюється. Тому мінімум у формулі (5.37) збігається з мінімумом функціонала $\Pi[X]$ на підмножині функцій $X \in C^1([0,l])$, що *задовольняють умову нормування (з вагою* $\rho(x) > 0$*)*

---

[1] Якщо в крайових умовах (5.4) параметр $h_1 \to +\infty$ ($h_2 \to +\infty$), тоді додатково треба вимагати, щоб гладкі функції $X(x)$, серед яких шукається мінімум відношення функціоналів $\Pi[X]$ і $K[X]$, задовольняли умову $X(0) = 0$ ($X(l) = 0$).

$$\|X\|^2 = \int_0^l \rho(x)|X(x)|^2 dx = 1, \tag{5.39}$$

і цей мінімум $\Pi[X]$ досягається саме на власній функції $X_1(x)$, що задовольняє умову (5.39).

Отже, задача про відшукання найменшого власного значення $\lambda_1$ КЗШЛ і відповідної власної функції $X_1(x)$, підкореної умові (5.39), еквівалентна ізопериметричній задачі про відшукання гладкої екстремалі, що надає мінімум функціоналу потенціальної енергії $\Pi[X]$ за умови, що функціонал кінетичної енергії $K[X]=1$.

Задачу про відшукання інших власних значень $\{\lambda_n\}\big|_{n=2}^{\infty}$ КЗШЛ і відповідних власних функцій $\{X_n\}\big|_{n=2}^{\infty}$ також можна звести до ізопериметричних задач для функціонала $\Pi[X]$, якщо скористатися властивістю *ортогональності* власних функцій КЗШЛ.

Нехай $F(x)$ і $G(x)$ — будь-які, взагалі кажучи, комплекснозначні функції, інтегровні з квадратом на відрізку $[0,l]$:

$$\int_0^l \rho(x)|F(x)|^2 dx < \infty, \quad \int_0^l \rho(x)|G(x)|^2 dx < \infty.$$

За означенням, *скалярним добутком* $\langle F\,|\,G\rangle$ функцій $F(x)$ і $G(x)$ (з вагою $\rho(x)$) називається функціонал

$$\langle F\,|\,G\rangle = \int_0^l \rho(x)\overline{F(x)}G(x)dx, \tag{5.40}$$

а скалярним квадратом $\|F\|^2 \equiv \langle F\,|\,F\rangle$ функції $F(x)$ — функціонал

$$\|F\|^2 \equiv \langle F\,|\,F\rangle = \int_0^l \rho(x)|F(x)|^2 dx. \tag{5.41}$$

Якщо скалярний добуток функцій $F(x)$ і $G(x)$ дорівнює нулю, $\langle F\,|\,G\rangle = 0$, то вони називаються ортогональними.

**Теорема 5.3.1.** Власні функції $X_n(x)$ і $X_m(x)$ КЗШЛ, що відповідають її різним власним значенням $\lambda_n \neq \lambda_m$, ортогональні з вагою $\rho(x)$: $\langle X_n\,|\,X_m\rangle = 0$.

*Доведення.* Беручи до уваги, що всі власні значення КЗШЛ невироджені, а кожна власна функція $X_n(x)$ КЗШЛ є розв'язком крайової задачі (5.3), (5.4) з дійсними коефіцієнтами, можна вважати, не втрачаючи загальності, що всі власні функції $X_n(x)$ є дійсними. Запишемо рівняння (5.3) окремо для функцій $X_n(x)$ і $X_m(x)$, помножи-мо перше рівняння на функцію $X_m(x)$, друге — на функцію $X_n(x)$, віднімемо отримані рівняння та зінтегруємо обидві частини здобутої рівності за змінною $x$ у межах від 0 до $l$. Дістанемо:

$$\int_0^l \left[ X_n \frac{d}{dx}\left(p(x)X'_m\right) - X_m \frac{d}{dx}\left(p(x)X'_n\right) \right] dx = \left(\lambda_n - \lambda_m\right) \int_0^l \rho(x) X_n X_m dx.$$

Беручи інтеграл зліва частинами та користуючись крайовими умовами (5.4), далі знаходимо:

$$\int_0^l \left[ X_n \frac{d}{dx}\left(p(x)X'_m\right) - X_m \frac{d}{dx}\left(p(x)X'_n\right) \right] dx = \left[ p(x)X_n X'_m - p(x)X_m X'_n \right]\Big|_0^l = 0.$$

Отже, отримуємо співвідношення

$$\left(\lambda_n - \lambda_m\right) \int_0^l \rho(x) X_n(x) X_m(x)\, dx = 0,$$

з якого бачимо, що при $\lambda_n \neq \lambda_m$

$$\int_0^l \rho(x) X_n(x) X_m(x)\, dx \equiv \left\langle X_n \mid X_m \right\rangle = 0. \tag{5.42}$$

**Теорема 5.3.2.** Нехай функція $\tilde{X}(x)$ надає екстремум функціоналу $П[X]$ у класі $C^1([0,l])$ гладких функцій, для яких $K[X]=1$ і які, крім цього, задовольняють додаткові умови

$$\left\langle X_1 \mid \tilde{X} \right\rangle = \left\langle X_2 \mid \tilde{X} \right\rangle = \ldots = \left\langle X_{n-1} \mid \tilde{X} \right\rangle = 0, \quad n \geq 2. \tag{5.43}$$

Тоді $\tilde{X}(x)$ є власною функцією, яка відповідає $n$-му за величиною власному значенню $\lambda_n$, при цьому $П[\tilde{X}] = \lambda_n$.

*Доведення.* Згідно з теоремою Ейлера, функція $\tilde{X}(x)$ має (як розв'язок ізопериметричної задачі, сформульованої в умовах теореми) надавати екстремум функціоналу

$$П^*[X] = П[X] - \tilde{\lambda} K[X] - \mu_1 \left\langle X_1 \mid X \right\rangle - \mu_2 \left\langle X_2 \mid X \right\rangle - \ldots - \mu_{n-1} \left\langle X_{n-1} \mid X \right\rangle \tag{5.44}$$

при певних значеннях невизначених множників Лагранжа $\tilde{\lambda}$, $\mu_1$, $\mu_2$, ..., $\mu_{n-1}$, а тому задовольняє рівняння Ейлера — Лагранжа

$$-\frac{d}{dx}\left(p(x)\tilde{X}'\right) + q(x)\tilde{X} =$$

$$= \tilde{\lambda} \rho(x) \tilde{X} + \frac{\mu_1}{2} \rho(x) X_1 + \frac{\mu_2}{2} \rho(x) X_2 + \ldots + \frac{\mu_{n-1}}{2} \rho(x) X_{n-1} \tag{5.45}$$

та крайові умови

$$\tilde{X}'(0) - h_1\tilde{X}(0) = 0, \quad \tilde{X}'(l) + h_2\tilde{X}(l) = 0. \qquad (5.46)$$

Помножимо обидві частини рівняння (5.45) на власну функцію $X_m(x)$, $1 \le m \le n-1$, та зінтегруємо їх за змінною $x$ у межах від 0 до $l$. Двічі інтегруючи частинами, користуючись рівнянням (5.3), записаним для функції $X_m(x)$, а також крайовими умовами (5.4) й (5.46), ліву частину здобутої рівності перетворюємо наступним чином:

$$\int_0^l X_m \left[ -\frac{d}{dx}\left(p(x)\tilde{X}'\right) + q(x)\tilde{X} \right] dx =$$

$$= -p(x)X_m\tilde{X}'\Big|_0^l + \int_0^l \left[ p(x)\tilde{X}'X_m' + q(x)\tilde{X}X_m \right] dx =$$

$$= -p(x)X_m\tilde{X}'\Big|_0^l + p(x)X_m'\tilde{X}\Big|_0^l + \int_0^l \left[ -\frac{d}{dx}\left(p(x)X_m'\right) + q(x)X_m \right] \tilde{X}\, dx =$$

$$= \lambda_m \int_0^l \rho(x)X_m\tilde{X}\, dx = \lambda_m \left\langle X_m \mid \tilde{X} \right\rangle.$$

Бачимо, що згідно з умовами (5.43) вона дорівнює нулю. Права ж частина розглядуваної рівності має вигляд

$$\tilde{\lambda}\left\langle X_m \mid \tilde{X} \right\rangle + \sum_{l=1}^{n-1} \frac{\mu_l}{2} \left\langle X_m \mid X_l \right\rangle.$$

З огляду на умови (5.43) та ортогональність власних функцій, що відповідають різним власним значенням, тобто співвідношення $\left\langle X_m \mid X_l \right\rangle = 0$ при $m \ne l$, вона зводиться до виразу $\frac{\mu_m}{2}\left\langle X_m \mid X_m \right\rangle$.

Таким чином, дістаємо співвідношення

$$\mu_m \left\langle X_m \mid X_m \right\rangle = 0.$$

Оскільки $\left\langle X_m \mid X_m \right\rangle \ne 0$, приходимо до висновку, що невизначені множники Лагранжа $\mu_l$, $1 \le l \le n-1$, у виразі (5.44) для функціонала $\Pi^*[X]$ та в рівнянні (5.45) для функції $\tilde{X}(x)$ дорівнюють нулю. Отже, згідно з рівнянням (5.45) та крайовими умовами (5.46), функція $\tilde{X}(x)$, що задовольняє умови теореми, є однією із власних функцій КЗШЛ. Покажемо, що відповідне їй власне значення (див. формулу (5.9))

$$\tilde{\lambda} = \frac{\Pi[\tilde{X}]}{K[\tilde{X}]},$$

яке за умови $K[\tilde{X}] = 1$ збігається з шуканим мінімальним значенням $\Pi[\tilde{X}]$ функціонала $\Pi[X]$, дорівнює власному значенню $\lambda_n$.

Справді, припустивши спершу, що $\tilde{\lambda} > \lambda_n$, візьмемо власну функцію $X_n(x)$, нормовану умовою $K[X_n] = 1$. Очевидно, що функція $X_n(x)$ задовольняє всі вимоги теореми, надаючи при цьому функціоналу $\Pi[X]$ значення $\tilde{\lambda} = \lambda_n$. Приходимо до суперечності. Якщо ж припустити, що $\tilde{\lambda} < \lambda_n$, то, очевидно, $\tilde{\lambda}$ збігається з одним із власних значень $\lambda_m < \lambda_n$, якому відповідає деяка власна функція $X_m(x)$. Оскільки власні значення КЗШЛ невироджені, має справджуватися рівність $\tilde{X}(x) = CX_m(x)$, де $C \neq 0$ — стала. Але для ненульової функції $X_m(x)$ це суперечить умові ортогональності, згідно з якою було би:

$$\left\langle X_m \,|\, \tilde{X} \right\rangle = C\left\langle X_m \,|\, X_m \right\rangle = C\|X_m\|^2 = 0.$$

Отже, мінімум функціонала $\Pi[X]$ дорівнює $\tilde{\lambda} = \lambda_n$ і реалізується на функції $\tilde{X}(x)$, яка збігається із власною функцією $X_n(x)$, нормованою умовою $K[X_n] = 1$.

**Зауваження 5.3.1.** Екстремаль відношення фунціоналів $\Pi[X]/K[X]$ за додаткових умов

$$\left\langle X_1 \,|\, X \right\rangle = \left\langle X_2 \,|\, X \right\rangle = \ldots = \left\langle X_{n-1} \,|\, X \right\rangle = 0$$

визначається з точністю до числового множника. Цей множник завжди можна вибрати так, щоб виконувалася рівність $K[X] = 1$. У загальному ж випадку

$$\min_{\substack{X \in C^1([0,l]),\\ \langle X_1|X\rangle = \ldots = \langle X_{n-1}|X\rangle = 0}} \frac{\Pi[X]}{K[X]} = \lambda_n, \tag{5.47}$$

і цей мінімум досягається на функціях $CX_n(x)$.

Безпосереднім наслідком теореми 5.3.2 та зауваження 5.3.1 є наступна лема.

**Лема 5.3.1.** Якщо в крайових умовах (5.4) $0 \le h_1, h_2 \le \infty$, то для будь-якої гладкої функції $F(x)$ на відрізку $[0,l]$, яка задовольняє умови (5.4) та умови ортогональності

$$\left\langle X_1 \,|\, F \right\rangle = \left\langle X_2 \,|\, F \right\rangle = \ldots = \left\langle X_{n-1} \,|\, F \right\rangle = 0, \tag{5.48}$$

справджується нерівність

$$\|F\|^2 \le \frac{\Pi[F]}{\lambda_n}. \tag{5.49}$$

Скориставшись нерівністю (5.49), уже не важко довести *повноту* системи власних функцій КЗШЛ.

**Теорема 5.3.3.** Нехай функція $F(x)$ — інтегровна з квадратом на відрізку $[0,l]$, яка є ортогональною до всіх власних функцій $\{X_n(x)\}$ КЗШЛ. Тоді

$$\|F\|^2 = \int_0^l \rho(x)|F(x)|^2 dx = 0$$

і, отже, $F(x)=0$ майже скрізь на $[0,l]$.

*Доведення*. Уважатимемо спочатку, що функція $F(x)$, яка ортогональна до всіх власних функцій КЗШЛ, є гладкою. Тоді, згідно з лемою 5.3.1, для будь-якого $n$ справджується нерівність (5.49). Але, як випливає із співвідношення (5.36), власні значення $\{\lambda_n\}$ КЗШЛ утворюють зростаючу послідовність, яка прямує до $+\infty$. З цього факту та нерівності (5.49) випливає, що $\|F\|^2=0$, а, отже, гладка функція $F(x)\equiv 0$ на $[0,l]$.

У випадку, коли функція $F(x)$ не є гладкою, розглянемо замість неї розв'язок $\tilde{F}(x)$ неоднорідної крайової задачі

$$-\frac{d}{dx}\left(p(x)\tilde{F}'\right)+q(x)\tilde{F}-\lambda\rho(x)\tilde{F}=\rho(x)F(x),\ \ 0<x<l, \tag{5.50}$$

$$\tilde{F}'(0)-h_1\tilde{F}(0)=0,\ \ \tilde{F}'(l)+h_2\tilde{F}(l)=0, \tag{5.51}$$

де $\lambda<0$. Навіть якщо $F(x)$ у рівнянні (5.50) не є неперервною, а лише інтегровною з квадратом[1], розв'язок системи (5.50), (5.51) є неперервною функцією, яка до того ж має неперервну першу похідну. У цьому можна переконатися, скориставшись наступним алгоритмом побудови $\tilde{F}(x)$. Спершу знаходимо розв'язок $f(x)$ інтегрального рівняння Вольтерра

$$f(x)=\int_0^x K_\lambda(x,x')f(x')dx'-\int_0^x \frac{dx'}{p(x')}\int_0^{x'}\rho(x'')F(x'')dx'',$$

де функція $K_\lambda(x,x')$ дається формулою (5.13). Цей розв'язок є гладкою функцією, яка задовольняє рівняння (5.50) та умови $f(0)=0$,

[1] Нагадаємо, що функції, які інтегровні з квадратом на скінченному інтервалі, є також інтегровними на цьому інтервалі.

$f'(0) = 0$ (див. теорему 5.2.2 та її доведення), а, отже, і першу крайову умову (5.51). Далі, згадавши, що гладка функція $\chi(x;\lambda)$, яка була введена в попередньому підрозділі, задовольняє однорідне (з $F(x) = 0$) рівняння (5.50) та теж першу крайову умову (5.51), будуємо лінійну комбінацію $f(x) + C\chi(x;\lambda)$. І, нарешті, вибравши сталу $C$ таким чином, щоб ця комбінація задовольняла другу крайову умову (5.51), дістаємо гладкий розв'язок диференціальної системи (5.50), (5.51).

Помножимо тепер обидві частини рівняння (5.50) на власну функцію $X_n(x)$ КЗШЛ та зінтегруємо отриману рівність за змінною $x$ у межах від 0 до $l$. Інтеграл, який виникає в лівій частині, після інтегрування частинами й урахування крайових умов (5.4) та (5.51) зводиться до виразу $(\lambda_n - \lambda)\langle X_n \,|\, \tilde{F}\rangle$. Права ж частина розглядуваної рівності має вигляд $\langle X_n \,|\, F\rangle$ і згідно з умовами теореми дорівнює нулю. Оскільки $\lambda_n \geq 0$, а $\lambda < 0$, приходимо до висновку, що $\langle X_n \,|\, \tilde{F}\rangle = 0$.

Отже, разом із функцією $F(x)$ ортогональною до всіх власних функцій КЗШЛ є й гладка функція $\tilde{F}(x)$. На основі вже наведених міркувань робимо висновок, що $\tilde{F}(x) \equiv 0$. Звідси та з рівняння (5.50) випливає, що майже скрізь на $[0,l]$ $F(x) = 0$.

Щоб відшукати власні значення $\lambda_2, \lambda_3, ..., \lambda_n, ...$ КЗШЛ як відносні мінімуми відношення функціоналів, узагалі не обов'язково попередньо знати власні функції $X_1(x), X_2(x), ..., X_{n-1}(x), ....$ Візьмемо, наприклад, довільну функцію $\Phi(x)$, що є інтегровною з квадратом на відрізку $[0,l]$, та розглянемо функціонал $\Lambda[\Phi]$, визначений як мінімум відношення функціоналів $\Pi[X]/K[X]$ на підмножині функцій із $C^1([0,l])$, ортогональних до $\Phi(x)$:

$$\Lambda[\Phi] = \min_{\substack{X \in C^1([0,l]),\\ \langle \Phi | X \rangle = 0}} \frac{\Pi[X]}{K[X]}.$$

Покажемо, що завжди $\Lambda[\Phi] \leq \lambda_2$.

Для доведення припустимо спершу, що $\langle \Phi \,|\, X_1 \rangle = 0$. Тоді множина функцій, на якій шукається мінімум відношення $\Pi[X]/K[X]$, включає і власну функцію $X_1$. Оскільки, як ми вже знаємо, $\Pi[X_1]/K[X_1] = \lambda_1$ і для будь-якої функції $X \in C^1([0,l])$ відношення $\Pi[X]/K[X]$ не може бути меншим за $\lambda_1$, то $\Lambda[\Phi] = \lambda_1 < \lambda_2$.

Тепер припустимо, що $\langle \Phi \,|\, X_2 \rangle = 0$. Тоді множина функцій, ортогональних до $\Phi(x)$, включає і власну функцію $X_2$. Але, згідно з формулою (5.47), $\Pi[X_2]/K[X_2] = \lambda_2$, тому і в цьому випадку мінімальне значення $\Lambda[\Phi]$ не може бути більшим за $\lambda_2$.

І, нарешті, нехай $\langle \Phi \mid X_1 \rangle \neq 0$ і $\langle \Phi \mid X_2 \rangle \neq 0$. Тоді завжди можна знайти такі сталі $C_1$ і $C_2$, щоб виконувалася умова

$$\langle \Phi \mid C_1 X_1 + C_2 X_2 \rangle = \int_0^l \rho(x)\overline{\Phi(x)}\left[C_1 X_1(x) + C_2 X_2\right]dx =$$

$$= C_1 \langle \Phi \mid X_1 \rangle + C_2 \langle \Phi \mid X_2 \rangle = 0.$$

Справді, останню рівність можна розглядати як лінійне однорідне алгебраїчне рівняння з ненульовими коефіцієнтами відносно невідомих $C_1$ і $C_2$. Оскільки це одне рівняння пов'язує дві невідомі, воно має нескінченну кількість розв'язків. Зокрема, можна покласти

$$C_1 = \langle \Phi \mid X_2 \rangle, \quad C_2 = -\langle \Phi \mid X_1 \rangle.$$

Обчислимо значення функціоналів $\Pi[X]$ і $K[X]$ на лінійній комбінації $C_1 X_1 + C_2 X_2$ власних функцій $X_1$ і $X_2$ із зазначеними значеннями сталих $C_1$ і $C_2$. Застосувавши процедуру інтегрування частинами, рівняння (5.3) і крайові умови (5.4), що їх задовольняють функції $X_1$ і $X_2$, знаходимо:

$$\Pi[C_1 X_1 + C_2 X_2] = |C_1|^2\,\Pi[X_1] + |C_2|^2\,\Pi[X_2] + \left(\overline{C_1}C_2 + C_1\overline{C_2}\right)\times$$

$$\times\left\{\int_0^l \left[p(x)X_1'(x)X_2'(x) + q(x)X_1(x)X_2(x)\right]dx +\right.$$

$$\left. + h_1 p(0)X_1(0)X_2(0) + h_2 p(l)X_1(l)X_2(l)\right\} =$$

$$= |C_1|^2\,\Pi[X_1] + |C_2|^2\,\Pi[X_2] + \left(\overline{C_1}C_2 + C_1\overline{C_2}\right)\times$$

$$\times\int_0^l \left\{-\frac{d}{dx}\left[p(x)X_1'(x)\right] + q(x)X_1(x)\right\}X_2(x)\,dx =$$

$$= |C_1|^2\,\Pi[X_1] + |C_2|^2\,\Pi[X_2] + \left(\overline{C_1}C_2 + C_1\overline{C_2}\right)\lambda_1\int_0^l \rho(x)X_1(x)X_2(x)\,dx.$$

Скориставшись формулою (5.38), формулою (5.47) для $X_2$ і $\lambda_2$ та умовою ортогональності (5.42) для функцій $X_1$ і $X_2$, звідси дістаємо:

$$\Pi[C_1 X_1 + C_2 X_2] = |C_1|^2\,\lambda_1 K[X_1] + |C_2|^2\,\lambda_2 K[X_2]. \tag{5.52}$$

З іншого боку, з ортогональності функцій $X_1$ і $X_2$ випливає, що

$$K[C_1 X_1 + C_2 X_2] = |C_1|^2\,K[X_1] + |C_2|^2\,K[X_2]. \tag{5.53}$$

Беручи до уваги, що $\lambda_2 > \lambda_1$, з формул (5.52) і (5.53) знаходимо:

$$\frac{П[C_1X_1 + C_2X_2]}{K[C_1X_1 + C_2X_2]} = \frac{\lambda_1|C_1|^2 K[X_1] + \lambda_2|C_2|^2 K[X_2]}{|C_1|^2 K[X_1] + |C_2|^2 K[X_2]} \le$$

$$\le \frac{\lambda_2\left(|C_1|^2 K[X_1] + |C_2|^2 K[X_2]\right)}{|C_1|^2 K[X_1] + |C_2|^2 K[X_2]} = \lambda_2. \qquad (5.54)$$

Отже, і тоді, коли не виконується жодна з умов $\langle\Phi \mid X_1\rangle = 0,\ \langle\Phi \mid X_2\rangle = 0,$ знайдеться ненульова функція $Y \in C^1([0,l])$, така, що $\langle\Phi \mid Y\rangle = 0$; при цьому $П[Y]/K[Y] \le \lambda_2$.

Позначимо через $L^2_\rho([0,l])$ множину всіх функцій $\Phi(x)$, інтегровних із квадратом з вагою $\rho(x)$ на відрізку $[0,l]$,

$$L^2_\rho([0,l]) = \left\{\Phi:\ \int_0^l \rho(x)\,|\Phi|^2\,dx < \infty\right\},$$

та розглянемо функціонал $\Lambda[\Phi]$, визначений на цій множині, який кожній функції $\Phi(x)$ ставить у відповідність мінімум відношення функціоналів $П[X]/K[X]$ на підмножині функцій $X(x)$ із $C^1([0,l])$, ортогональних до $\Phi(x)$. Як ми щойно встановили, значення такого функціонала не перевищують власного значення $\lambda_2$ КЗШЛ. З другого боку, згідно з теоремою 5.3.2 значення функціонала $\Lambda[\Phi]$ у випадку, коли $\Phi$ дорівнює власній функції $X_1$, збігається з $\lambda_2$. Отже, ми довели, що

$$\max_{\Phi\in L^2_\rho([0,l])}\left\{\min_{\substack{X\in C^1([0,l]),\\ \langle\Phi|X\rangle=0}} \frac{П[X]}{K[X]}\right\} = \lambda_2. \qquad (5.55)$$

Повторюючи майже дослівно попередні міркування, співвідношення (5.55) можна узагальнити наступним чином.

**Теорема 5.3.4 (про мінімакс)**. Нехай $\Lambda[\Phi_1,\ldots,\Phi_{n-1}]$, $n \ge 2$, — функціонал, який для заданих $n-1$ функцій $\Phi_1,\ldots,\Phi_{n-1}$ із $L^2_\rho([0,l])$ визначається як

$$\Lambda[\Phi_1,\ldots,\Phi_{n-1}] = \min_{\substack{X\in C^1([0,l]),\\ \langle\Phi_1|X\rangle=\ldots=\langle\Phi_{n-1}|X\rangle=0}} \frac{П[X]}{K[X]}.$$

Тоді

$$\max_{\Phi_1,\ldots,\Phi_{n-1}\in L^2_\rho([0,l])} \Lambda[\Phi_1,\ldots,\Phi_{n-1}] = \lambda_n,$$

де $\lambda_n$ — $n$-те за величиною (у порядку зростання) власне значення КЗШЛ.

Теорема про мінімакс дозволяє простежити за рухом власних значень КЗШЛ при змінах функцій $p(x)$, $q(x)$, $\rho(x)$ та параметрів $h_1$, $h_2$, тобто встановити, як змінюються власні частоти коливальної системи при змінах її параметрів.

**Наслідок 5.3.1.** Нехай для двох КЗШЛ коефіцієнти $p^{(1)}(x)$, $q^{(1)}(x)$ та $p^{(2)}(x)$, $q^{(2)}(x)$ у відповідних рівняннях Штурма — Ліувілля (5.3), а також параметри $h_1^{(1)}$, $h_2^{(1)}$ та $h_1^{(2)}$, $h_2^{(2)}$ у відповідних крайових умовах (5.4) пов'язані нерівностями

$$p^{(1)}(x) \le p^{(2)}(x),\ \ q^{(1)}(x) \le q^{(2)}(x),\ \ x \in [0,l],$$

$$h_1^{(1)} \le h_1^{(2)},\ \ h_2^{(1)} \le h_2^{(2)},$$

тоді як коефіцієнти $\rho^{(1)}(x)$ та $\rho^{(2)}(x)$ у зазначених рівняннях збігаються. Тоді для послідовних власних значень цих крайових задач справджуються нерівності

$$\lambda_n^{(1)} \le \lambda_n^{(2)},\ \ n = 1, 2, .... \tag{5.56}$$

З огляду на фізичний зміст власних значень КЗШЛ наслідок 5.3.1 можна сформулювати й так:

**Наслідок 5.3.1а.** При збільшенні (зменшенні) жорсткості коливальної системи її послідовні власні частоти коливань можуть лише збільшуватися (зменшуватися).

Щоб довести ці наслідки, досить помітити, що для будь-якої функції $X \in C^1([0,l])$ значення функціоналів потенціальної енергії для першої і другої КЗШЛ задовольняють нерівності $\Pi^{(1)}[X] \le \Pi^{(2)}[X]$, а відповідні значення функціоналів кінетичної енергії збігаються: $K^{(1)}[X] = K^{(2)}[X]$. Звідси відразу випливає, що для будь-яких функцій $\Phi_1, ..., \Phi_{n-1}$ із $L_\rho^2([0,l])$ значення функціоналів $\Lambda^{(1)}[\Phi_1, ..., \Phi_{n-1}]$ і $\Lambda^{(2)}[\Phi_1, ..., \Phi_{n-1}]$, породжуваних цими КЗШЛ, задовольняють нерівність $\Lambda^{(1)}[\Phi_1, ..., \Phi_{n-1}] \le \Lambda^{(2)}[\Phi_1, ..., \Phi_{n-1}]$. Відповідно, така сама нерівність має справджуватися і для максимумів цих функціоналів. А це і є нерівність (5.56).

**Наслідок 5.3.2.** Нехай для двох КЗШЛ коефіцієнти $p^{(1)}(x)$, $q^{(1)}(x)$ та $p^{(2)}(x)$, $q^{(2)}(x)$ у відповідних рівняннях Штурма — Ліувілля (5.3), а також параметри $h_1^{(1)}$, $h_2^{(1)}$ та $h_1^{(2)}$, $h_2^{(2)}$ у відповідних крайових умовах (5.4) попарно збігаються: $p^{(1)}(x) = p^{(2)}(x)$, $q^{(1)}(x) = q^{(2)}(x)$ тощо, а коефіцієнти $\rho^{(1)}(x)$ та $\rho^{(2)}(x)$ у зазначених рівняннях задовольняють співвідношення $\rho^{(1)}(x) \le \rho^{(2)}(x)$. Тоді для послідовних власних значень цих крайових задач справджуються нерівності

$$\lambda_n^{(1)} \ge \lambda_n^{(2)}, \quad n = 1, 2, \dots . \tag{5.57}$$

**Наслідок 5.3.2а.** При збільшенні (зменшенні) мас складових частин коливальної системи її послідовні власні частоти коливань можуть лише зменшуватися (збільшуватися).

**Завдання 5.3.1.** Доведіть наслідок 5.3.2а.

*Вказівка.* Для будь-якої функції $X \in C^1([0,l])$ значення функціоналів кінетичної енергії для першої та другої КЗШЛ задовольняють нерівність $K^{(1)}[X] \le K^{(2)}[X]$, а значення відповідних функціоналів потенціальної енергії збігаються.

**Завдання 5.3.2.** Нехай значення коефіцієнтів у рівнянні Штурма — Ліувілля (5.3) задовольняють нерівності

$$0 < p_m \le p(x) \le p_M, \; 0 \le q_m \le q(x) \le q_M, \; 0 < \rho_m \le \rho(x) \le \rho_M, \; x \in [0, l].$$

Доведіть, що при будь-яких невід'ємних значеннях параметрів $h_1$, $h_2$ у крайових умовах (5.4) для послідовних власних значень такої КЗШЛ справджуються оцінки

$$\frac{p_m}{\rho_M}\frac{\pi^2 (n-1)^2}{l^2} + \frac{q_m}{\rho_M} \le \lambda_n \le \frac{p_M}{\rho_m}\frac{\pi^2 n^2}{l^2} + \frac{q_M}{\rho_m}, \quad n = 1, 2, \dots . \tag{5.58}$$

*Вказівка.* Проаналізуйте власні значення двох КЗШЛ, що складаються з рівняння (5.3) зі сталими коефіцієнтами $p_m$, $q_m$, $\rho_M$ і $p_M$, $q_M$, $\rho_m$ та крайових умов (5.4) з параметрами, відповідно, $h_1 = h_2 = 0$ і $h_1, h_2 \to \infty$. Потім скористайтеся наслідками 5.3.1 та 5.3.2.

## 5.4. НУЛІ ВЛАСНИХ ФУНКЦІЙ

У подальшому *нулями* неперервної функції називатимемо ті точки її області визначення, де вона дорівнює нулю[1].

Важливим наслідком екстремальних властивостей найменшого власного значення КЗШЛ є наступна теорема.

---

[1] Ці точки також називають *вузлами* функції.

**Теорема 5.4.1.** Власна функція $X_1(x)$ КЗШЛ (5.3), (5.4), яка відповідає найменшому власному значенню $\lambda_1$, не має нулів в інтервалі $(0,l)$.

*Доведення.* Припустимо, усупереч твердженню теореми, що існує точка $x_0 \in (0,l)$, у якій $X_1(x_0)=0$. Очевидно, що при цьому $X_1'(x_0)=C \neq 0$, оскільки єдиним розв'язком однорідного рівняння Штурма — Ліувілля, який задовольняє умови $X(x_0)=0$, $X'(x_0)=0$, $x_0 \in (0,l)$, є тотожний нуль. Отже, в околі точки $x_0$ графік функції $X_1(x)$ є схожим на графік, зображений на рис. 5.2, а).

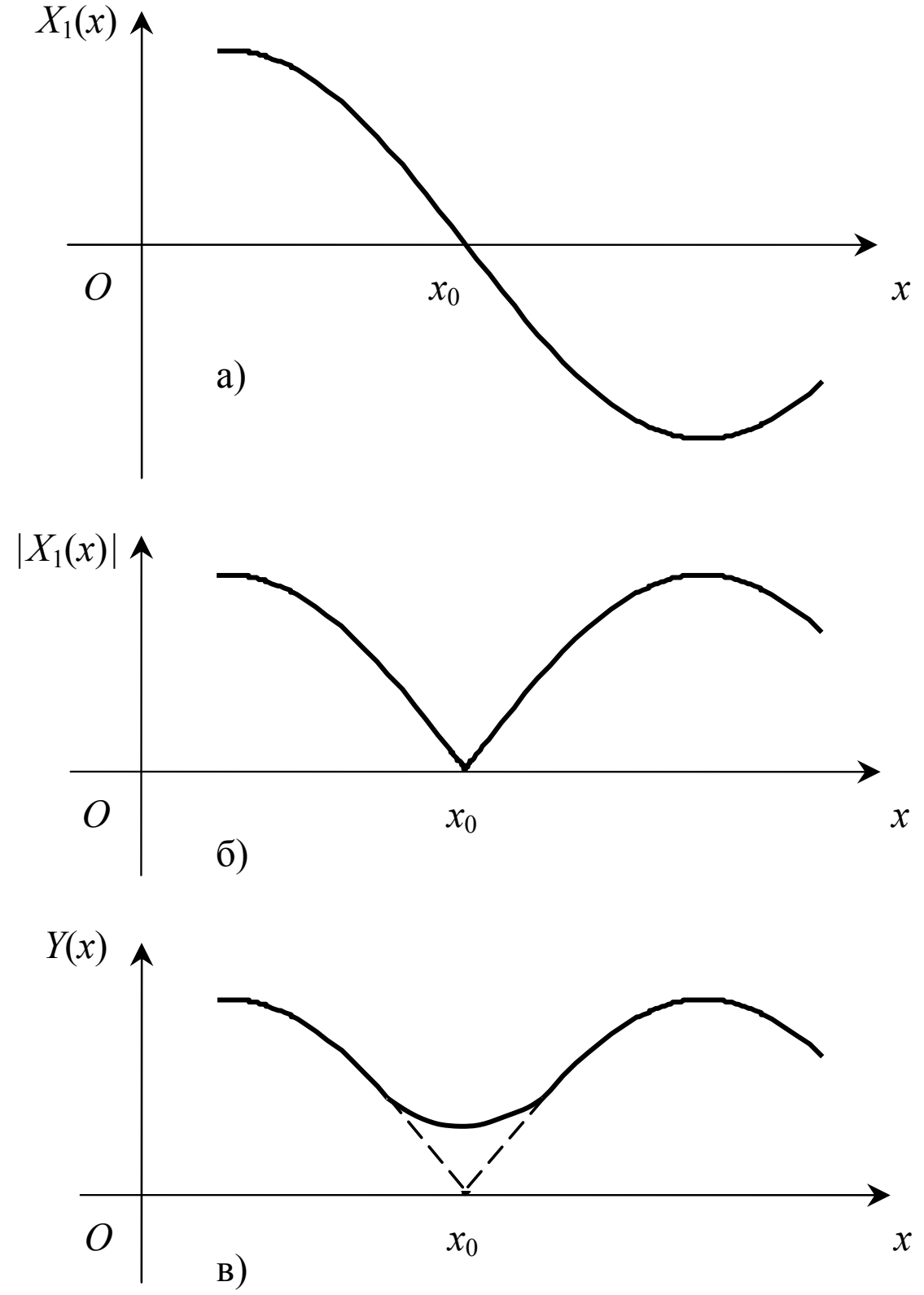

Рис. 5.2. Поведінка а) функції $X_1(x)$, б) функції $|X_1(x)|$ та в) згладженої функції $Y(x)$ в околі можливого нуля $x_0$ функції $X_1(x)$

Звернемо тепер увагу на той факт, що відношення $\Pi[X]/K[X]$ функціоналів потенціальної та кінетичної енергій залежить лише від квадратів $X^2(x)$ і $X'^2(x)$. Тому разом із функцією $X_1(x)$ найменшого власного значення цьому відношенню надає й неперервна кусково-гладка функція $|X_1(x)|$, графік якої має злам у точці $x_0$ (див. рис. 5.2, б)). Згладжуючи функцію $|X_1(x)|$ у малому околі точки $x_0$ таким чином, щоб суттєво зменшити значення $|X_1'(x)|^2$, не дуже збільшивши при цьому значення $X^2(x)$ (рис. 5.2, в)), дістанемо гладку функцію $Y(x)$, для якої мало б справджуватися співвідношення

$$\frac{\Pi[Y]}{K[Y]} < \frac{\Pi[X_1]}{K[X_1]} = \lambda_1.$$

Але це неможливо, оскільки $\lambda_1$ є мінімумом відношення $\Pi[X]/K[X]$ на множині функцій, гладких на відрізку $[0,l]$. Отже, припущення, що $X_1(x_0)=0$, $x_0 \in (0,l)$, веде до суперечності.

**Зауваження 5.4.1.** Наведене доведення теореми 5.4.1 належить Р. Фейнману. Повторюючи ті ж самі аргументи, можна довести, що власні функції, які відповідають найменшим власним значенням крайових задач для багатьох рівнянь із частинними похідними, зокрема, для рівнянь виду

$$-\sum_{n=1}^{N} \frac{\partial^2 \psi}{\partial x_n^2} + V(x_1, \ldots, x_N)\psi = \lambda\psi,$$

не мають нулів. Більше того, з факту відсутності нулів у власної функції, що відповідає в таких задачах найменшому власному значенню, випливає, що це власне значення невироджене. Справді, якби воно було вироджене, то з двох власних функцій $\psi_1$ і $\psi_2$, що йому б відповідали, завжди можна було би побудувати ненульову власну функцію у вигляді лінійної комбінації $C_1\psi_1 + C_2\psi_2$, яка мала б нулі в області визначення, а, отже, не відповідала б найменшому власному значенню.

Для подальшого вивчення питання про наявність і кількість нулів у послідовних власних функцій КЗШЛ уже треба скористатися специфічними властивостями розв'язків рівняння Штурма — Ліувілля

$$-\frac{d}{dx}\big(p(x)y'\big) + q(x)y = \lambda\rho(x)y. \tag{5.59}$$

При цьому без обмеження загальності можна вважати, що функція $p(x)$ у рівнянні (5.59) тотожно дорівнює одиниці: $p(x)\equiv 1$. Справді, заміна змінної

$$x' = \varphi(x) = \int_0^x \frac{ds}{p(s)} \tag{5.60}$$

взаємно однозначно відображає відрізок $[0,l]$ на відрізок $[0,l']$, де $l' = \int_0^l \frac{ds}{p(s)}$. Позначимо через $\psi(x')$, $x' \in [0,l']$, функцію, обернену до функції $\varphi(x)$. Оскільки $p\frac{dy}{dx} = \frac{dy}{dx'}$, то при заміні (5.60) рівняння (5.59) переходить у рівняння

$$-\frac{d^2y}{dx'^2} + \tilde{q}(x')y = \lambda\tilde{\rho}(x')y, \tag{5.61}$$

де $\tilde{q}(x') = p(\psi(x'))q(\psi(x'))$, $\tilde{\rho}(x') = p(\psi(x'))\rho(\psi(x'))$. Також зрозуміло, що при заміні (5.60) будь-яка неперервна функція $f(x)$ на $[0,l]$, яка має $n$ нулів у точках $x_i \in [0,l]$, $i = 1,\dots,n$, переходить у неперервну функцію $\tilde{f}(x') = f(\psi(x'))$, яка має $n$ нулів у точках

$$x_i' = \int_0^{x_i} \frac{ds}{p(s)}, \quad i = 1,\dots,n.$$

Опустивши в рівнянні (5.61) штрихи та тильди, далі розглядатимемо рівняння

$$y'' + \left[\lambda\rho(x) - q(x)\right]y = 0. \tag{5.62}$$

З ним пов'язана така фундаментальна теорема.

**Теорема 5.4.2 (Штурма).** Нехай $u(x)$ — розв'язок рівняння

$$u'' + g(x)u = 0, \tag{5.63}$$

а $v(x)$ — розв'язок рівняння

$$v'' + h(x)v = 0, \tag{5.64}$$

де $g(x) < h(x)$ скрізь на $(0,l)$. Тоді між двома послідовними нулями розв'язку $u(x)$ знаходиться хоча б один нуль розв'язку $v(x)$.

*Доведення.* Помножимо рівняння (5.63) на функцію $v(x)$, рівняння (5.64) — на функцію $u(x)$, та віднімемо здобуті рівності. Знаходимо:

$$u''v - v''u = \frac{d}{dx}\left(u'v - v'u\right) = \left[h(x) - g(x)\right]uv. \tag{5.65}$$

Нехай $x_1$ і $x_2$ — два послідовні нулі розв'язку $u(x)$. Зінтегрувавши рівність (5.65) за змінною $x$ у межах від $x_1$ до $x_2$, дістаємо:

$$u'(x_2)v(x_2)-u'(x_1)v(x_1)=\int\limits_{x_1}^{x_2}\left[h(x)-g(x)\right]u(x)v(x)dx. \qquad (5.66)$$

Припустимо, що розв’язок $v(x)$ не має нулів в інтервалі $(x_1,x_2)$. Не обмежуючи загальності, можемо вважати, що $u(x)>0$ і $v(x)>0$ в $(x_1,x_2)$. Для таких функцій $u(x)$ і $v(x)$ права частина рівності (5.66) додатна. З другого боку, при зроблених нами припущеннях функція $u(x)$ в околі точки $x_1$ зростає, а в околі точки $x_2$ спадає. Тому $u'(x_1)>0$ і $u'(x_2)<0$. Оскільки $v(x_1)\geq 0$, $v(x_2)\geq 0$, то ліва частина рівності (5.66) недодатна. Приходимо до суперечності.

**Наслідок 5.4.1.** Нехай $u(x)$ — нетривіальний розв’язок рівняння (5.63), який задовольняє умову

$$u'(0)-h_1u(0)=0,\;\; h_1\geq 0, \qquad (5.67)$$

а $v(x)$ — нетривіальний розв’язок рівняння (5.64), який задовольняє таку саму умову

$$v'(0)-h_1v(0)=0,\;\; h_1\geq 0. \qquad (5.68)$$

Якщо $u(x)$ має $m$ нулів в інтервалі $0<x\leq l$, то й $v(x)$ має в цьому інтервалі принаймні $m$ нулів, при цьому $s$-й за величиною (у порядку зростання) нуль функції $v(x)$ менший від $s$-го за величиною (у порядку зростання) нуля функції $u(x)$.

*Доведення.* Згідно з теоремою 5.4.2, достатньо показати, що $v(x)$ має принаймні один нуль в інтервалі $(0,x_1)$, де $x_1$ — найменший за величиною нуль функції $u(x)$ в інтервалі $0<x\leq l$. Припустимо спершу, що число $h_1$ — обмежене зверху. Зінтегрувавши рівність (5.65) за змінною $x$ у межах від нуля до $x_1$ та скориставшись умовами (5.67), (5.68), дістаємо:

$$u'(x_1)v(x_1)=\int\limits_{0}^{x_1}\left[h(x)-g(x)\right]u(x)v(x)dx.$$

Якщо $v(x)$ не має жодних нулів в інтервалі $(0,x_1)$, то, уважаючи, як і раніше, що $u(x)>0$ та $v(x)>0$ в $(0,x_1)$, бачимо, що права частина цієї рівності додатна. Однак її ліва частина недодатна, оскільки $u'(x_1)<0$ і $v(x_1)\geq 0$. Приходимо до суперечності.

У випадку, коли $h_1\to\infty$, умови (5.67), (5.68) набирають вигляду $u(0)=0$, $v(0)=0$, і теорема доводиться аналогічно.

Розглянемо тепер власні функції $X(x,\lambda)$ такої КЗШЛ:

$$X'' + \left[\lambda\rho(x) - q(x)\right] X = 0, \tag{5.69}$$

$$X'(0) - h_1 X(0) = 0, \quad h_1 \geq 0, \tag{5.70}$$

$$X'(l) + h_2 X(l) = 0, \quad h_2 \geq 0. \tag{5.71}$$

Позначимо через $n(\lambda)$, $-\infty < \lambda < \infty$, кількість нулів в інтервалі $(0,l)$ нетривіального розв'язку $\tilde{X}(x,\lambda)$ рівняння (5.69), який задовольняє лише умову (5.70). Пам'ятаючи, що $\rho(x) > 0$, $q(x) \geq 0$, за допомогою наслідку 5.4.1 покажемо, що $n(\lambda)$ — це неспадна функція від $\lambda$. З цього факту та теореми 5.4.2 випливає, що кожний нуль $\tilde{X}(x,\lambda)$ зміщується вліво при зростанні $\lambda$.

Нехай, як і раніше,

$$\min_{0\leq x\leq l} \rho(x) = \rho_m > 0, \quad \max_{0\leq x\leq l} \rho(x) = \rho_M < \infty,$$

$$\min_{0\leq x\leq l} q(x) = q_m \geq 0, \quad \max_{0\leq x\leq l} q(x) = q_M < \infty.$$

Спершу порівняємо функцію $\tilde{X}(x,\lambda)$ та розв'язок рівняння

$$Y'' + \lambda\rho_m Y = 0$$

при $\lambda \leq 0$, який задовольняє умову (5.70). З точністю до сталого множника цей розв'язок має вигляд

$$Y(x,\lambda) = \begin{cases} \operatorname{ch}\sqrt{|\lambda|\rho_m}\,x + \dfrac{h_1}{\sqrt{|\lambda|\rho_m}} \operatorname{sh}\sqrt{|\lambda|\rho_m}\,x, & \lambda < 0, \\ 1 + h_1 x, \quad \lambda = 0. \end{cases}$$

Легко бачити, що він взагалі не має нулів на відрізку $[0,l]$. Оскільки при $\lambda \leq 0$ $\lambda\rho(x) - q(x) \leq \lambda\rho_m$, то на основі наслідку 5.4.1 робимо висновок, що при $\lambda \leq 0$ функція $n(\lambda) = 0$.

У випадку, коли $\lambda > 0$, аналізуємо розв'язки рівнянь

$$Y_1'' + \left(\lambda\rho_m - q_M\right) Y_1 = 0, \quad Y_2'' + \lambda\rho_M Y_2 = 0,$$

які задовольняють умову (5.70). Вони мають вигляд

$$Y_1(x,\lambda) = \cos\sqrt{\lambda\rho_m - q_M}\,x + \frac{h_1}{\sqrt{\lambda\rho_m - q_M}} \sin\sqrt{\lambda\rho_m - q_M}\,x,$$

$$Y_2(x,\lambda) = \cos\sqrt{\lambda\rho_M}\,x + \frac{h_1}{\sqrt{\lambda\rho_M}} \sin\sqrt{\lambda\rho_M}\,x.$$

Беручи до уваги, що при $\lambda > 0$

$$\lambda\rho_m - q_M \leq \lambda\rho(x) - q(x) \leq \lambda\rho_M,$$

на підставі наслідку 5.4.1 робимо висновок, що при $\lambda \to \infty$ кількість нулів $n(\lambda)$ функції $\tilde{X}(x,\lambda)$ в інтервалі $(0,l)$ не перевищує кількість нулів функції $Y_2(x,\lambda)$ і не є меншою за кількість нулів функції $Y_1(x,\lambda)$ у цьому ж інтервалі. Звідси випливає, що $n(\lambda)$ — неспадна функція, яка необмежено зростає при $\lambda \to \infty$, залишаючись при цьому скінченною при скінченних значеннях $\lambda$.

Іншим аргументом, який указує на скінченність кількості нулів функції $\tilde{X}(x,\lambda)$ на відрізку $[0,l]$, є наступна теорема.

**Теорема 5.4.3 (Ляпунова).** Нехай функція $g(x)$ — дійсна та неперервна на відрізку $[a,b]$ і $g_+(x) = \max\{g(x),0\}$. Для того, щоб нетривіальний розв'язок рівняння

$$u''(x) + g(x)u(x) = 0 \tag{5.72}$$

мав два нулі на відрізку $[a,b]$, необхідно, щоб виконувалася умова

$$\int_a^b g_+(x)dx \geq \frac{4}{b-a}. \tag{5.73}$$

*Доведення.* Припустимо, що диференціальне рівняння (5.72) має нетривіальний розв'язок із двома нулями $x = \alpha, \beta$ на $[a,b]$, $a \leq \alpha < \beta \leq b$. Тоді на відрізку $[\alpha,\beta]$ воно еквівалентне однорідному інтегральному рівнянню

$$u(x) = \int_\alpha^\beta K(x,x')u(x')dx', \tag{5.74}$$

де

$$K(x,x') = \frac{1}{\beta-\alpha}\begin{cases}(\beta-x)(x'-\alpha)g(x'), & \alpha < x' < x,\\ (x-\alpha)(\beta-x')g(x'), & x < x' < \beta.\end{cases} \tag{5.75}$$

Уведемо функцію

$$K_+(x,x') = \frac{1}{\beta-\alpha}\begin{cases}(\beta-x)(x'-\alpha)g_+(x'), & \alpha < x' < x,\\ (x-\alpha)(\beta-x')g_+(x'), & x < x' < \beta,\end{cases} \tag{5.76}$$

та припустимо, що $\alpha$ і $\beta$ — сусідні нулі функції $u(x)$ і що $u(x) > 0$ при $\alpha < x < \beta$. Нехай також $x = x_0$ — така точка, що $u(x_0) = \max u(x)$ при $x \in (\alpha,\beta)$. Поклавши в рівнянні (5.74) $x = x_0$, дістанемо

$$u(x_0) = \int_\alpha^\beta K(x_0,x')u(x')dx' \leq \int_\alpha^\beta K_+(x_0,x')dx' \cdot u(x_0),$$

звідки

$$1 \le \int_{\alpha}^{\beta} K_+(x_0, x')dx' = \frac{1}{\beta-\alpha}\int_{\alpha}^{x_0}(\beta - x_0)(x'-\alpha)g_+(x')dx' +$$

$$+\frac{1}{\beta-\alpha}\int_{x_0}^{\beta}(x_0-\alpha)(\beta-x')g_+(x')dx'. \qquad (5.77)$$

Оскільки $\beta - x_0 \le \beta - x'$ при $x' \le x_0$ і $x_0 - \alpha \le x' - \alpha$ при $x' \ge x_0$, із співвідношення (5.77) далі знаходимо:

$$1 \le \frac{1}{\beta-\alpha}\int_{\alpha}^{x_0}(\beta - x')(x'-\alpha)g_+(x')dx' +$$

$$+\frac{1}{\beta-\alpha}\int_{x_0}^{\beta}(x'-\alpha)(\beta-x')g_+(x')dx' =$$

$$= \frac{1}{\beta-\alpha}\int_{\alpha}^{\beta}(x'-\alpha)(\beta-x')g_+(x')dx'. \qquad (5.78)$$

Замінивши функцію $(x'-\alpha)(\beta-x')$ в останньому інтегралі формули (5.78) її максимальним значенням $(\beta-\alpha)^2/4$ на проміжку $(\alpha,\beta)$, отримуємо нерівність

$$1 \le \frac{\beta-\alpha}{4}\int_{\alpha}^{\beta} g_+(x')dx',$$

яка ще більше посилиться та перейде в нерівність (5.73), якщо в ній замінити $\alpha$ на $a$ і $\beta$ на $b$.

**Завдання 5.4.1.** Переконайтеся, що рівняння (5.72) і (5.74) еквівалентні. Для цього:

а) перевірте, що розв'язок крайової задачі

$$u''(x) = -f(x), \ \ \alpha < x < \beta, \qquad (5.79)$$

$$u(\alpha) = 0, \ u(\beta) = 0, \qquad (5.80)$$

можна подати у вигляді

$$u(x) = \int_{\alpha}^{\beta} G(x, x') f(x')\, dx', \qquad (5.81)$$

де $G(x,x')$ — неперервний розв'язок крайової задачі

$$\frac{\partial^2 G(x,x')}{\partial x^2} = -\delta(x-x'), \ \ \alpha < x, x' < \beta, \qquad (5.82)$$

$$G(\alpha, x') = 0, \quad G(\beta, x') = 0, \tag{5.83}$$

$\delta(x - x')$ — дельта-функція Дірака. Функція $G(x, x')$ називається функцією Гріна задачі (5.79), (5.80);

б) доведіть, що в точці $x = x'$ разом з умовою неперервності

$$G(x' - 0, x') = G(x' + 0, x') \tag{5.84}$$

функція $G(x, x')$ задовольняє умову

$$\left.\frac{\partial G(x, x')}{\partial x}\right|_{x = x' - 0} - \left.\frac{\partial G(x, x')}{\partial x}\right|_{x = x' + 0} = 1; \tag{5.85}$$

в) покажіть, що розв'язок задачі (5.82), (5.83) має вигляд

$$G(x, x') = \frac{1}{\beta - \alpha} \begin{cases} (x - \alpha)(\beta - x'), & \alpha < x < x', \\ (\beta - x)(x' - \alpha), & x' < x < \beta; \end{cases} \tag{5.86}$$

г) перейдіть від рівняння (5.72) до рівняння (5.74) та доведіть формулу (5.75).

*Вказівки*: а) знайдіть другу похідну від обох частин рівності (5.81) та скористайтеся означенням функції $G(x, x')$; б) зінтегруйте обидві частини рівняння (5.82) за змінною $x$ у межах від $x' - \varepsilon$ до $x' + \varepsilon$, де $\varepsilon > 0$ — мале додатне число, потім перейдіть до границі $\varepsilon \to 0$; в) знайдіть розв'язки задачі (5.82), (5.83) для проміжків $x \in [\alpha, x')$ і $x \in (x', \beta]$, потім зшийте їх за допомогою умов (5.84) і (5.85); г) перепишіть рівняння (5.72) у вигляді $u''(x) = -g(x)u(x)$, скористайтеся результатами (5.81), (5.86) та симетричністю функції $G(x, x')$ відносно перестановки аргументів: $G(x, x') = G(x', x)$. Див. також підрозділ 4.9.

З теореми Ляпунова, зокрема, випливає, що функція $\tilde{X}(x, \lambda)$ може мати два нулі на відрізку $[a, b] \subset (0, l)$, якщо його довжина $b - a$ така, що

$$\frac{4}{\lambda \int_0^l \rho(x)\,dx} \le \frac{4}{\int_a^b \left[\lambda \rho(x) - q(x)\right]_+ dx} \le b - a. \tag{5.87}$$

При будь-якому скінченному (хоч і як завгодно великому) $\lambda$ ця довжина обмежена знизу. Отже, розщеплення нулів функції $\tilde{X}(x, \lambda)$ із зростанням $\lambda$ неможливе.

Для подальшого вивчення властивостей функції $n(\lambda)$ введемо функцію

$$\Theta(x,\lambda)=\operatorname{arctg}\frac{\tilde{X}(x,\lambda)}{\tilde{X}'(x,\lambda)}. \tag{5.88}$$

Оскільки $\tilde{X}(x,\lambda)$ і $\tilde{X}'(x,\lambda)$ не можуть дорівнювати нулю одночасно, то фіксуючи значення

$$\Theta(0,\lambda)=\operatorname{arctg}\frac{1}{h_1} \tag{5.89}$$

у проміжку $[0,\pi/2]$, можемо однозначно визначити $\Theta(x,\lambda)$ як дійсну неперервну функцію змінної $x$, що задовольняє на відрізку $[0,l]$ диференціальне рівняння

$$\Theta'(x,\lambda)=\frac{1}{1+\left[\dfrac{\tilde{X}(x,\lambda)}{\tilde{X}'(x,\lambda)}\right]^2}\left[1-\frac{\tilde{X}(x,\lambda)}{\tilde{X}'^2(x,\lambda)}\tilde{X}''(x,\lambda)\right]=$$

$$=\frac{1}{1+\operatorname{tg}^2\Theta(x,\lambda)}\left\{1+\operatorname{tg}^2\Theta(x,\lambda)\left[\lambda\rho(x)-q(x)\right]\right\}= \tag{5.90}$$

$$=\cos^2\Theta(x,\lambda)+\left[\lambda\rho(x)-q(x)\right]\sin^2\Theta(x,\lambda)$$

та умову (5.89).

**Лема 5.4.1**. Нехай розв’язок $\tilde{X}(x,\lambda)$ має точно $n\geq 1$ нулів $x_j$, $0<x_1<x_2<\ldots<x_n<l$, в інтервалі $(0,l)$. Тоді $\Theta(x_j,\lambda)=j\pi$ та $\Theta(x,\lambda)>j\pi$, якщо $x_j<x<l$, $j=1,\ldots,n$.

*Доведення*. Згідно з формулою (5.88), у точках, де $\tilde{X}(x,\lambda)=0$, значення $\Theta(x,\lambda)$ є цілими кратними $\pi$. Крім того, з рівняння (5.90) видно, що в цих точках $\Theta'(x,\lambda)=1$. Звідси випливає, що в околах нулів розв’язку $\tilde{X}(x,\lambda)$, тобто в околах точок $x_j$, $0<x_j<l$, функція $\Theta(x,\lambda)$ зростає, при цьому $\Theta(x_j,\lambda)=j\pi$. Справді, набувши значення $\Theta(x_j,\lambda)=j\pi$ в точці $x_j$, $0<x_j<l$, та зрісши безпосередньо після цього, функція $\Theta(x,\lambda)$ далі, можливо, навіть спадає, але в жодній точці $x_j^*$, $x_j<x_j^*<l$, уже не повертається до значення $j\pi$, бо інакше виходило би, що $\Theta'(x_j^*,\lambda)\leq 0$, а це суперечить рівнянню (5.90). Як наслідок, $\Theta(x,\lambda)>j\pi$ при $x_j<x<l$.

**Наслідок 5.4.2**. Якщо виконуються умови леми 5.4.1, то

$$n(\lambda)=\left[\frac{1}{\pi}\Theta(l-0,\lambda)\right], \tag{5.91}$$

де $[\xi]$ — ціла частина числа $\xi$.

*Доведення*. При виконанні умов леми 5.4.1 функція $\tilde{X}(x,\lambda)$ вже не має нулів в інтервалі $(x_n,l)$, а тому функція $\Theta(x,\lambda)$, що є неперервною за змінною $x$ на відрізку $[0,l]$, задовольняє на $(x_n,l)$ нерівність $n\pi \le \Theta(x,\lambda)$. При $x > x_n$ ця функція може досягати значення $(n+1)\pi$ лише в точці $x = l$. Беручи до уваги ці властивості $\Theta(x,\lambda)$, отримуємо рівність (5.91).

З доведених теорем і наслідку 5.4.2 випливає, що існує зростаюча послідовність чисел $0 \le \lambda_1 < \lambda_2 < \ldots < \lambda_n < \ldots$, така, що $\tilde{X}(l,\lambda_n) = 0$ і $\tilde{X}(x,\lambda_n)$ має точно $n-1$ нулів в інтервалі $(0,l)$. Якщо в крайовій умові (5.71) $h_2 \to +\infty$, тобто вона має вигляд $X(l) = 0$, то числа $\lambda_1, \lambda_2, \ldots, \lambda_n, \ldots$ є послідовними власними значеннями КЗШЛ.

У випадку, коли $0 \le h_2 < \infty$, розглянемо розв'язки $\tilde{X}(x,\lambda)$ і $\tilde{X}(x,\lambda')$ при фіксованих значеннях $\lambda' > \lambda$. Помітимо, що

$$\frac{\partial}{\partial x}\left\{\frac{\tilde{X}(x,\lambda)}{\tilde{X}(x,\lambda')}\left[\tilde{X}'(x,\lambda)\tilde{X}(x,\lambda') - \tilde{X}(x,\lambda)\tilde{X}'(x,\lambda')\right]\right\} =$$

$$= \frac{\left[\tilde{X}'(x,\lambda)\tilde{X}(x,\lambda') - \tilde{X}(x,\lambda)\tilde{X}'(x,\lambda')\right]^2}{\tilde{X}^2(x,\lambda')} +$$

$$+ \frac{\tilde{X}(x,\lambda)}{\tilde{X}(x,\lambda')}\left[\tilde{X}''(x,\lambda)\tilde{X}(x,\lambda') - \tilde{X}(x,\lambda)\tilde{X}''(x,\lambda')\right] =$$

$$= \frac{\left[\tilde{X}'(x,\lambda)\tilde{X}(x,\lambda') - \tilde{X}(x,\lambda)\tilde{X}'(x,\lambda')\right]^2}{\tilde{X}^2(x,\lambda')} + \tilde{X}^2(x,\lambda)\left[\frac{\tilde{X}''(x,\lambda)}{\tilde{X}(x,\lambda)} - \frac{\tilde{X}''(x,\lambda')}{\tilde{X}(x,\lambda')}\right] =$$

$$= \frac{\left[\tilde{X}'(x,\lambda)\tilde{X}(x,\lambda') - \tilde{X}(x,\lambda)\tilde{X}'(x,\lambda')\right]^2}{\tilde{X}^2(x,\lambda')} + (\lambda' - \lambda)\rho(x)\tilde{X}^2(x,\lambda) > 0.$$

Отже, функція

$$Q(x,\lambda,\lambda') \equiv \frac{\tilde{X}(x,\lambda)}{\tilde{X}(x,\lambda')}\left[\tilde{X}'(x,\lambda)\tilde{X}(x,\lambda') - \tilde{X}(x,\lambda)\tilde{X}'(x,\lambda')\right] =$$

$$= \tilde{X}^2(x,\lambda)\left[\frac{\tilde{X}'(x,\lambda)}{\tilde{X}(x,\lambda)} - \frac{\tilde{X}'(x,\lambda')}{\tilde{X}(x,\lambda')}\right]$$

монотонно зростає за змінною $x$.

Припустимо, що розв'язки $\tilde{X}(x,\lambda)$ і $\tilde{X}(x,\lambda')$ мають однакову кількість $\nu$ нулів в інтервалі $(0,l)$. Найближчий до $l$ нуль $x_\nu$ функції $\tilde{X}(x,\lambda)$ не є нулем функції $\tilde{X}(x,\lambda')$, оскільки між точками 0 і $x_\nu$ знаходяться принаймні $\nu$ нулів функції $\tilde{X}(x,\lambda')$. Звідси, беручи до уваги монотонність функції $Q(x,\lambda,\lambda')$, маємо

$$\tilde{X}^2(l,\lambda)\left[\frac{\tilde{X}'(l,\lambda)}{\tilde{X}(l,\lambda)}-\frac{\tilde{X}'(l,\lambda')}{\tilde{X}(l,\lambda')}\right] > \tilde{X}^2(x_\nu,\lambda)\left[\frac{\tilde{X}'(x_\nu,\lambda)}{\tilde{X}(x_\nu,\lambda')}-\frac{\tilde{X}'(x_\nu,\lambda')}{\tilde{X}(x_\nu,\lambda')}\right]=0$$

і тому

$$\frac{\tilde{X}'(l,\lambda)}{\tilde{X}(l,\lambda)} > \frac{\tilde{X}'(l,\lambda')}{\tilde{X}(l,\lambda')}.$$

З цієї нерівності випливає, що відношення $\tilde{X}'(l,\lambda)/\tilde{X}(l,\lambda)$ як функція параметра $\lambda$ монотонно спадає в кожному інтервалі $(\tilde{\lambda}_s,\tilde{\lambda}_{s+1})$, де $\tilde{\lambda}_s$ і $\tilde{\lambda}_{s+1}$ — послідовні нулі функції $\tilde{X}(l,\lambda)$. Значення цього відношення пробігають усю вісь $(-\infty,\infty)$, спадаючи від $+\infty$ до $-\infty$, оскільки $\tilde{X}(l,\lambda)$ дорівнює нулю на обох кінцях інтервалу $(\tilde{\lambda}_s,\tilde{\lambda}_{s+1})$ і при цьому $\tilde{X}'(l,\lambda)\neq 0$. Отже, в інтервалі $(\tilde{\lambda}_s,\tilde{\lambda}_{s+1})$ існує, і при цьому єдине, число $\lambda_{s+1}$, таке, що

$$\frac{\tilde{X}'(l,\lambda_{s+1})}{\tilde{X}(l,\lambda_{s+1})}=-h_2.$$

Таким чином, для зростаючої послідовності $0\le\lambda_1<\lambda_2<\ldots<\lambda_n<\ldots$ власних значень КЗШЛ (5.69)–(5.71) *кожна власна функція* $X(x,\lambda_m)$, *що відповідає власному значенню* $\lambda_m$, *має точно* $m-1$ *нулів в інтервалі* $(0,l)$.

### *КОНТРОЛЬНІ ПИТАННЯ ДО РОЗДІЛУ 5*

1. *Якому інтегральному рівнянню еквівалентне рівняння Штурма — Ліувілля на відрізку* $[0,l]$ *з параметром* $\lambda$ *і крайовою умовою* $\theta(0;\lambda)=a$, $\theta'(0;\lambda)=b$? *Який загальний алгоритм побудови розв'язку подібних рівнянь?*
2. *Якому класу аналітичних функцій змінної* $\lambda$ *належать єдиний розв'язок* $\theta(x;\lambda)$ *рівняння Штурма — Ліувілля з параметром* $\lambda$ *та його похідна* $\theta'(x;\lambda)$ *за* $x$ *для кожного* $x\in[0,l]$ *за умови, що* $a$ *та* $b$ *у крайових умовах* $\theta(0;\lambda)=a$, $\theta'(0;\lambda)=b$ *не залежать від* $\lambda$? *Які оцінки при цьому справджуються для* $|\theta(x;\lambda)|$ *та* $|\theta'(x;\lambda)|$, *коли* $|\lambda|\to\infty$?

3. *Який висновок відносно зростання послідовності власних значень* $\lambda_1 < \lambda_2 < \ldots < \lambda_n < \ldots$ *крайової задачі Штурма — Ліувілля можна зробити, спираючись на аналітичні властивості вказаного вище розв'язку* $\theta(x;\lambda)$ *та його похідної?*
4. *Який екстремальний зміст мають найменше власне значення крайової задачі Штурма — Ліувілля і відповідна власна функція?*
5. *Який екстремальний зміст мають друге та наступні за величиною власні значення крайової задачі Штурма — Ліувілля?*
6. *Які наслідки випливають з теореми про мінімакс щодо руху власних частот коливальної системи при змінах її жорсткості та мас? Як зростають послідовні власні частоти* $\omega_n$ *неоднорідної струни із зростанням номера* $n$*?*
7. *Як довести повноту системи власних функцій крайової задачі Штурма — Ліувілля, спираючись лише на екстремальні властивості послідовних власних значень та необмежене зростання їх послідовності?*
8. *Чому власна функція, яка відповідає найменшому власному значенню крайової задачі Штурма — Ліувілля, не може мати вузлів у внутрішніх точках відрізка* $[0,l]$*?*
9. *Як змінюється кількість нулів розв'язку рівняння Штурма — Ліувілля* $\theta(x;\lambda)$ *на відрізку* $[0,l]$ *при збільшенні дійсного параметра* $\lambda$*? Чи можуть окремі нулі розщеплюватися при неперервних змінах* $\lambda$*?*
10. *Скільки нулів в інтервалі* $(0,l)$ *має власна функція, що відповідає* $n$*-му за величиною власному значенню крайової задачі Штурма — Ліувілля?*

# Розділ 6
# ЗАДАЧІ ТЕПЛОПРОВІДНОСТІ ТА ДИФУЗІЇ

## 6.1. РІВНЯННЯ ТЕПЛОПРОВІДНОСТІ НА ПРЯМІЙ. МЕТОД ПЕРЕТВОРЕННЯ ФУР'Є

Аналіз задач теплопровідності (дифузії) проведемо для просторово однорідних систем. Почнемо з випадку, коли поширення тепла (частинок домішки) в системі відбувається вздовж необмеженої прямої $-\infty < x < \infty$. Відповідна *задача Коші для рівняння теплопровідності (дифузії)* складається з рівняння

$$\frac{\partial u}{\partial t} = a^2 \frac{\partial^2 u}{\partial x^2} + f(x,t), \quad -\infty < x < \infty, \quad t > 0, \tag{6.1}$$

і початкової умови

$$u(x,0) = u_0(x). \tag{6.2}$$

Побудуємо її розв'язок та з'ясуємо, які властивості повинна мати початкова функція $u_0(x)$, щоб він був неперервно диференційовним за часом і двічі неперервно диференційовним за координатою[1].

Щоб розв'язати поставлену задачу, перейдемо до двох більш простих задач: для однорідного рівняння теплопровідності із заданою початковою умовою та неоднорідного рівняння теплопровідності з нульовою початковою умовою. Вони мають вигляд відповідно

---

[1] Рівняння (6.1) випливає із загального рівняння теплопровідності для просторово неоднорідних систем, у яких діють теплові джерела (стоки). Останнє можна вивести на основі співвідношення теплового балансу, записаного для довільної області системи. Докладні викладки, обговорення фізичних умов, за яких рівняння теплопровідності справджується, та відповідних математичних вимог до функцій, що до нього входять, а також доведення єдиності розв'язку лінійної задачі Коші для рівняння теплопровідності в необмеженому просторі (окремим випадком якої є задача (6.1), (6.2)) можна знайти в §§ 1.1, 1.2 посібника авторів: Вступ до математичної фізики. Introduction to Mathematical Physics. — Одеса: Астропринт, 2003. — 320 с. Питанням обґрунтування рівняння дифузії для просторово неоднорідних систем присвячено §§ 2.1–2.3 цього ж посібника. Також нагадаємо, що для задач теплопровідності параметр $a^2 \equiv \kappa/(c\rho)$ і називається коефіцієнтом температуропровідності ($\kappa$, $c$ та $\rho$ — відповідно коефіцієнт теплопровідності, питома теплоємність та густина речовини тіла). У задачах про дифузію частинок у середовищі $a^2$ має зміст їх коефіцієнта дифузії $D$.

$$\frac{\partial v}{\partial t} = a^2 \frac{\partial^2 v}{\partial x^2}, \quad -\infty < x < \infty, \quad t > 0, \tag{6.3}$$

$$v(x,0) = u_0(x) \tag{6.4}$$

та

$$\frac{\partial w}{\partial t} = a^2 \frac{\partial^2 w}{\partial x^2} + f(x,t), \quad -\infty < x < \infty, \quad t > 0, \tag{6.5}$$

$$w(x,0) = 0. \tag{6.6}$$

Якщо розв'язки $v(x,t)$ і $w(x,t)$ цих задач відомі, то, очевидно, розв'язок задачі (6.1), (6.2) дорівнює їх сумі (принцип суперпозиції)[1]:

$$u(x,t) = v(x,t) + w(x,t). \tag{6.7}$$

Отже, задача зводиться до побудови розв'язків задач (6.3), (6.4) та (6.5), (6.6).

Для розв'язання задачі (6.3), (6.4) припустимо спочатку, що функція $u_0(x)$ — інтегровна, та скористаємося *інтегральним перетворенням Фур'є*. Нагадаємо[2], що перетворенням Фур'є або Фур'є-образом довільної інтегровної функції $g(x)$, визначеної на всій осі $-\infty < x < \infty$, називається функція $\hat{g}(k)$, яка визначена на всій осі $-\infty < k < \infty$ і задається інтегралом

$$\hat{g}(k) = \frac{1}{\sqrt{2\pi}} \int_{-\infty}^{\infty} e^{-ikx} g(x)\,dx, \tag{6.8}$$

де $i$ — уявна одиниця. З цього означення випливає, що Фур'є-образ інтегровної функції — це функція, яка неперервна та рівномірно обмежена на всій осі. Також можна довести, що якщо Фур'є-образ $\hat{g}(k)$ — інтегровна функція, то функція $g(x)$ — неперервна і відновлюється з $\hat{g}(k)$ за *формулою обернення*

---

[1] Такий підхід, який називається *редукцією* вихідної задачі, уможливлюється лінійністю задачі і вже неодноразово використовувався при розгляді малих коливань струни та інших систем, де він мав просту фізичну інтерпретацію: довільне мале коливання системи є суперпозицією її вільних коливань, спричинених ненульовими початковими умовами, та вимушених коливань, викликаних зовнішньою силою. У випадку задач теплопровідності (дифузії) інтерпретація дуже схожа: розподіл температури (домішок) у системі в довільний момент часу є суперпозицією двох розподілів: спричиненого початковим розподілом температури (домішок) та генерованого джерелами, що діють у системі.

[2] Див., наприклад: Адамян В. М., Сушко М. Я. Вступ до математичної фізики. Introduction to Mathematical Physics, § 4.4.

$$g(x)=\frac{1}{\sqrt{2\pi}}\int\limits_{-\infty}^{\infty} e^{ikx}\,\hat{g}(k)\,dk. \qquad (6.9)$$

Для подальшого аналізу суттєву роль відіграє той факт, що при застосуванні перетворення Фур'є операція диференціювання функції зводиться до операції множення Фур'є-образу цієї функції на множник $ik$. Справді, нехай функція $g(x)$ має неперервну та інтегровну першу похідну $g'(x)$. Інтегруючи частинами та враховуючи, що функція $g(x)$ із переліченими властивостями спадає до нуля при $x\to\pm\infty$, з означення (6.8), записаного для Фур'є-образу похідної $g'(x)$, дістаємо:

$$\widehat{g'}(k)=\frac{1}{\sqrt{2\pi}}\int\limits_{-\infty}^{\infty} e^{-ikx}g'(x)\,dx=\frac{1}{\sqrt{2\pi}}e^{-ikx}g(x)\Big|_{-\infty}^{\infty}+\frac{ik}{\sqrt{2\pi}}\int\limits_{-\infty}^{\infty} e^{-ikx}g(x)\,dx=ik\hat{g}(k).$$

Аналогічним чином переконуємося, що якщо функція $g(x)$ додатково має неперервну та інтегровну другу похідну, то Фур'є-образ останньої

$$\widehat{g''}(k)=(ik)^2\hat{g}(k)=-k^2\hat{g}(k). \qquad (6.10)$$

Застосуємо перетворення Фур'є за змінною $x$ до обох частин диференціального рівняння (6.3) та початкової умови (6.4), уважаючи, що шуканий розв'язок є інтегровною функцією, яка має інтегровні перші похідні та інтегровну другу похідну за змінною $x$. Оскільки змінні $x$ та $t$ незалежні, а також справджується формула (6.10), для Фур'є-образу

$$\hat{v}(k,t)=\frac{1}{\sqrt{2\pi}}\int\limits_{-\infty}^{\infty} e^{-ikx}v(x,t)\,dx$$

функції $v(x,t)$ за змінною $x$ маємо такі співвідношення:

$$\widehat{\frac{\partial v}{\partial t}}(k,t)=\frac{1}{\sqrt{2\pi}}\int\limits_{-\infty}^{\infty} e^{-ikx}\frac{\partial v(x,t)}{\partial t}dx=\frac{\partial}{\partial t}\frac{1}{\sqrt{2\pi}}\int\limits_{-\infty}^{\infty} e^{-ikx}v(x,t)\,dx=\frac{\partial\hat{v}(k,t)}{\partial t},$$

$$\widehat{\frac{\partial^2 v}{\partial x^2}}(k,t)=\frac{1}{\sqrt{2\pi}}\int\limits_{-\infty}^{\infty} e^{-ikx}\frac{\partial^2 v(x,t)}{\partial x^2}dx=-k^2\hat{v}(k,t),$$

$$\hat{v}(k,0)=\frac{1}{\sqrt{2\pi}}\int\limits_{-\infty}^{\infty} e^{-ikx}v(x,0)\,dx=\frac{1}{\sqrt{2\pi}}\int\limits_{-\infty}^{\infty} e^{-ikx}u_0(x)\,dx=\hat{u}_0(k). \qquad (6.11)$$

З перших двох формул бачимо, що рівняння (6.3) переходить у диференціальне рівняння першого порядку за змінною $t$ відносно функції $\hat{v}(k,t)$:

$$\frac{\partial \hat{v}(k,t)}{\partial t} = -a^2 k^2 \hat{v}(k,t).$$

Його загальний розв'язок дається виразом

$$\hat{v}(k,t) = C(k) e^{-a^2 k^2 t}.$$

Третя формула має зміст початкової умови для функції $\hat{v}(k,t)$:

$$\hat{v}(k,0) = \hat{u}_0(k).$$

З неї знаходимо сталу інтегрування $C(k)$:

$$C(k) = \hat{u}_0(k).$$

Отже,

$$\hat{v}(k,t) = \hat{u}_0(k) e^{-a^2 k^2 t}.$$

Застосувавши до цієї функції обернене перетворення Фур'є, дістаємо шукану функцію $v(x,t)$ у вигляді

$$v(x,t) = \frac{1}{\sqrt{2\pi}} \int_{-\infty}^{\infty} e^{ikx} \hat{u}_0(k) e^{-a^2 k^2 t} \, dk. \qquad (6.12)$$

Неважко переконатися, що функція (6.12) існує для всіх значень $x \in (-\infty, \infty)$ і $t \in [\tau, \infty)$, де $\tau > 0$ — довільне число, а також диференційовна за $x$ і $t$ на цих інтервалах довільну кількість разів. Для цього достатньо показати, що для вказаних значень $x$ і $t$ інтеграл (6.12) є збіжним, а інтеграли, отримувані з нього диференціюванням за $x$ і $t$ під знаком інтеграла, — рівномірно збіжними. Задача розв'язується просто, якщо згадати, що достатньою умовою рівномірної збіжності інтеграла є існування додатної функції, незалежної від $x$ і $t$, яка мажорує підінтегральний вираз, й інтеграл від якої збігається.

У силу рівномірної обмеженості Фур'є-образу інтегровної функції для довільного $k$ маємо співвідношення $|\hat{u}_0(k)| \le M$, де $M$ — додатна стала. Беручи також до уваги, що $|e^{ikx}| = 1$, бачимо, що для довільних значень $x$ і $t$ із указаних інтервалів інтеграл (6.12) мажорується збіжним інтегралом:

$$|v(x,t)| \le \frac{1}{\sqrt{2\pi}} \int_{-\infty}^{\infty} |e^{ikx}| \left|\hat{u}_0(k)\right| e^{-a^2 k^2 t} \, dk \le \frac{M}{\sqrt{2\pi}} \int_{-\infty}^{\infty} e^{-a^2 k^2 \tau} \, dk = \frac{M}{a\sqrt{2\tau}}.$$

Аналогічно для довільних натуральних $n,m \geq 1$ маємо:

$$\left|\frac{\partial^{n+m} v(x,t)}{\partial x^n \partial t^m}\right| \leq \frac{Ma^{2m}}{\sqrt{2\pi}} 2\int_0^\infty k^{n+2m} e^{-a^2k^2\tau}\, dk = \frac{Ma^{2m}}{\sqrt{2\pi}\left(a\sqrt{\tau}\right)^{n+2m+1}} 2\int_0^\infty x^{n+2m} e^{-x^2}\, dx =$$

$$= \frac{M}{a^{n+1}\sqrt{2\pi}\left(\sqrt{\tau}\right)^{n+2m+1}} \Gamma\left(\frac{n+2m+1}{2}\right),$$

де $\Gamma(z)$ — гамма-функція Ейлера, $\Gamma(z) = \int_0^\infty t^{z-1} e^{-t} dt,\ \operatorname{Re} z > 0.$

Залишається розглянути початкову умову (6.4). Вона буде справджуватися, якщо існує граничне значення інтеграла (6.12) при $t = 0$ $(\tau \downarrow 0)$. Для цього достатньо додатково припустити, що Фур'є-образ $\hat{u}_0(k)$ — інтегровна функція.

Отже, якщо початкова функція $u_0(x)$ є інтегровною та має інтегровний Фур'є-образ, то формула (6.12) дає неперервно диференційовний розв'язок задачі (6.3), (6.4), який при $t = 0$ неперервно примикає до початкової функції в усіх точках неперервності останньої.

Варто зазначити, що формулу (6.12), разом із формулою (6.11), можна розглядати як «неперервний» аналог розкладу за власними функціями, що був раніше застосований та обґрунтований при побудові розв'язків задачі Коші для коливань однорідних струн і стержнів скінченної довжини. Однак тепер у ролі власних функцій, за якими ведеться розклад, виступають функції

$$X_k(x) = \frac{1}{\sqrt{2\pi}} e^{ikx}, \quad -\infty < k < \infty. \tag{6.13}$$

Неважко бачити, що незалежні функції (6.13) задовольняють диференціальне рівняння

$$X_k''(x) = -\lambda X_k(x), \quad \lambda = k^2, \tag{6.14}$$

та інтегральне співвідношення

$$\int_{-\infty}^{\infty} \overline{X_k(x)} X_{k'}(x)\, dx = \lim_{N\to\infty} \int_{-N}^{N} \overline{X_k(x)} X_{k'}(x)\, dx =$$

$$= \lim_{N\to\infty} \frac{\sin(k-k')N}{\pi(k-k')} = \delta(k-k'), \tag{6.15}$$

тобто є попарно ортогональними в тому розумінні, що

$$\int_{-\infty}^{\infty} \overline{X_k(x)} X_{k'}(x)\,dx = 0 \text{ при } k \neq k',$$

але не є інтегровними з квадратом на всій осі:

$$\int_{-\infty}^{\infty} \left|X_k(x)\right|^2 dx = \infty \text{ при } k = k'.$$

Множник $1/\sqrt{2\pi}$ у формулі (6.13) вибирається таким чином, щоб права частина співвідношення (6.15) дорівнювала δ-функції. Це співвідношення називається *умовою нормування на δ-функцію*.

## 6.2. ВЛАСНІ ЗНАЧЕННЯ І СПЕКТР ДИФЕРЕНЦІАЛЬНИХ ОПЕРАТОРІВ НА ОСІ

Щоб зрозуміти особливість ситуації, яка виникає при спробі поширити на випадок задачі Коші (6.3), (6.4) для необмеженої прямої поняття про власні значення, власні функції та ортогональні розклади за власними функціями, що були введені при розгляді крайових задач для обмежених систем, згадаємо, як ці поняття вводяться в лінійній алгебрі. Для цього розглянемо систему лінійних рівнянь

$$\begin{aligned} (a_{11} - \lambda)x_1 + a_{12}x_2 + \ldots + a_{1N}x_N &= y_1, \\ a_{21}x_1 + (a_{22} - \lambda)x_2 + \ldots + a_{2N}x_N &= y_2, \\ &\ldots \\ a_{N1}x_1 + a_{N2}x_2 + \ldots + (a_{NN} - \lambda)x_N &= y_N \end{aligned} \tag{6.16}$$

з параметром $\lambda$. За допомогою матриці $A = \left(a_{jk}\right)\Big|_{j,k=1}^{N}$ та векторів $X = \left(x_k\right)\Big|_{k=1}^{N}$, $Y = \left(y_j\right)\Big|_{j=1}^{N}$ цю систему можемо також записати у вигляді

$$AX - \lambda X = Y. \tag{6.17}$$

Як відомо з лінійної алгебри, при розв'язуванні системи (6.16) із заданим $\lambda$ маємо альтернативу: або система має єдиний розв'язок при довільних числах $y_1$, $y_2$, ..., $y_N$, або однорідна система (усі числа $y_1$, $y_2$, ..., $y_N$ дорівнюють нулю) має нетривіальні розв'язки. Нагадаємо, що однорідна система

$$AX - \lambda X = 0 \tag{6.18}$$

має нетривіальні розв'язки $X$ при тих значеннях $\lambda$, для яких її детермінант дорівнює нулю:

$$|A-\lambda I|=0, \tag{6.19}$$

де $I=\left(\delta_{jk}\right)\Big|_{j,k=1}^{N}$ ( $\delta_{jk}$ — символ Кронекера) — одинична матриця. Ці значення $\lambda$ та відповідні їм розв'язки $X$ називаються власними значеннями та власними векторами матриці $A$, а сукупність власних значень матриці $A$ — її *спектром*. Отже, можемо коротко сказати: або система (6.17) має єдиний розв'язок при *довільній* правій частині, або параметр $\lambda$ належить спектру матриці $A$.

При переході до диференціальних операторів, визначених на певних класах функцій на дійсній осі, півосі чи системі інтервалів із нескінченною загальною довжиною, поняття спектра крайової задачі або, коротко, оператора суттєво збагачується. Справа в тому, що при окресленні множини функцій, до яких повинні належати розв'язки задач, пов'язаних з диференціальними операторами, виникає потреба, з огляду на фізичний зміст цих задач, висунути певні обмеження на характер допустимої поведінки шуканих розв'язків при прямуванні їх просторових аргументів до нескінченності. У переважній більшості фізичних задач ці обмеження покриваються вимогою квадратичної інтегровності за просторовими змінними. Зокрема, для диференціального оператора $D_2\equiv-\partial^2/\partial x^2$ на дійсній осі, який ми зараз розглядаємо, клас функцій, які утворюють його область визначення, зводиться до двічі диференційовних функцій $y(x)$ на осі, що з точністю до лінійної функції відновлюються за своєю другою похідною, тобто

$$y(x)=C_0+C_1x+\int_0^x(x-s)y''(s)\,ds, \tag{6.20}$$

та задовольняють умови[1]

$$\int_{-\infty}^{\infty}|y(x)|^2\,dx<\infty,\quad \int_{-\infty}^{\infty}|y'(x)|^2\,dx<\infty,\quad \int_{-\infty}^{\infty}|y''(x)|^2\,dx<\infty. \tag{6.21}$$

За аналогією з лінійною алгеброю, ми можемо означити *спектр лінійного диференціального оператора* $H$ *на осі з областю визначення (6.20), (6.21),* зокрема, оператора $D_2$, *як множину чисел* $\lambda$ *комплексної площини, для яких рівняння*

$$\left(Hy\right)(x)-\lambda y(x)=f(x) \tag{6.22}$$

[1] Ми опускаємо строге обґрунтування цих умов, оскільки воно виходить за межі цього курсу. Укажемо лише, що в квантовій механіці друга та третя умови забезпечують скінченність середнього значення спостережуваної енергії та скінченність її дисперсії в стані з хвильовою функцією $y(x)$. Також зазначимо, що друга умова в (6.21) є наслідком першої і третьої.

*хоча би при якійсь одній квадратично інтегровній функції $f(x)$ або не має жодного розв'язку з області (6.20), (6.21), або, навпаки, має не єдиний такий розв'язок.* У випадках, коли рівняння (6.22) має не єдиний розв'язок, тобто *коли однорідне рівняння*

$$\big(Hy\big)(x)-\lambda y(x)=0 \qquad (6.23)$$

*має нетривіальні розв'язки, що задовольняють умови (6.20) та (6.21), число $\lambda$ називається (як і для диференціального оператора в крайовій задачі Штурма — Ліувілля) власним значенням оператора $H$, а самі ці розв'язки — власними функціями оператора $H$, що відповідають власному значенню $\lambda$.*

Для диференціальних операторів математичної фізики в необмежених областях спектр оператора і сукупність власних значень можуть не збігатися. Більш того, такі оператори можуть взагалі не мати жодного власного значення. Продемонструємо це на прикладі оператора $D_2$.

**Завдання 6.2.1.** Покажіть, що для довільної функції $f(x)$ із простору квадратично інтегровних функцій $L_2(-\infty,\infty)$ та довільного від'ємного або комплексного значення $\lambda$ рівняння

$$-y''(x)-\lambda y(x)=f(x) \qquad (6.24)$$

має єдиний двічі диференційовний розв'язок, який разом зі своїми першою та другою похідними належить простору $L_2(-\infty,\infty)$.

*Розв'язання.* Якщо значення $\lambda$ — від'ємне або комплексне, то функція

$$y_0(x)=\frac{i}{2\sqrt{\lambda}}\int_{-\infty}^{x}e^{i\sqrt{\lambda}(x-x')}f(x')dx'+\frac{i}{2\sqrt{\lambda}}\int_{x}^{\infty}e^{-i\sqrt{\lambda}(x-x')}f(x')dx'=$$

$$=\frac{i}{2\sqrt{\lambda}}\int_{-\infty}^{\infty}e^{i\sqrt{\lambda}|x-x'|}f(x')dx', \qquad (6.25)$$

де береться гілка $\sqrt{\lambda}$ з додатною уявною частиною, задовольняє рівняння (6.24)[1]. Згідно з нерівністю Коші — Буняковського,

---

[1] Вираз (6.25) є безпосереднім продовженням на комплексні значення $\omega=\sqrt{\lambda}$ формул (4.38) і (4.39) для спеціальних розв'язків одновимірного рівняння Гельмгольца, що прямують до нуля при $x\to\pm\infty$. Втім, його неважко знайти з таких незалежних міркувань. Загальний розв'язок однорідного рівняння (6.24) має вигляд $y(x)=Ae^{i\sqrt{\lambda}x}+Be^{-i\sqrt{\lambda}x}$. Загальний розв'язок неоднорідного рівняння (6.24) слід шукати в такому ж вигляді, уважаючи при цьому коефіцієнти $A$ та $B$ функціями координат, що задовольняють співвідношення $A'(x)e^{i\sqrt{\lambda}x}+B'(x)e^{-i\sqrt{\lambda}x}=0$ (метод варіації сталих). Друге співвідношення для цих функцій дістаємо, підставивши $y(x)$ у рівняння

$$|y_0(x)|^2 \le \frac{1}{4\left|\sqrt{\lambda}\right|^2}\left|\int_{-\infty}^{\infty} e^{-\mathrm{Im}\sqrt{\lambda}|x-x'|} f(x')\,dx'\right|^2 \le$$

$$\le \frac{1}{4\left(\mathrm{Im}\sqrt{\lambda}\right)^2}\left|\int_{-\infty}^{\infty} e^{-\frac{1}{2}\mathrm{Im}\sqrt{\lambda}|x-x'|} e^{-\frac{1}{2}\mathrm{Im}\sqrt{\lambda}|x-x'|} f(x')\,dx'\right|^2 \le$$

$$\le \frac{1}{4\left(\mathrm{Im}\sqrt{\lambda}\right)^2}\int_{-\infty}^{\infty} e^{-\mathrm{Im}\sqrt{\lambda}|x-x'|}dx' \int_{-\infty}^{\infty} e^{-\mathrm{Im}\sqrt{\lambda}|x-x'|}\left|f(x')\right|^2 dx' \le$$

$$\le \frac{1}{2\left(\mathrm{Im}\sqrt{\lambda}\right)^3}\int_{-\infty}^{\infty} e^{-\mathrm{Im}\sqrt{\lambda}|x-x'|}\left|f(x')\right|^2 dx',$$

тому

$$\int_{-\infty}^{\infty} |y_0(x)|^2 dx \le \frac{1}{2\left(\mathrm{Im}\sqrt{\lambda}\right)^3}\int_{-\infty}^{\infty} dx \int_{-\infty}^{\infty} e^{-\mathrm{Im}\sqrt{\lambda}|x-x'|}\left|f(x')\right|^2 dx' =$$

$$= \frac{1}{\left(\mathrm{Im}\sqrt{\lambda}\right)^4}\int_{-\infty}^{\infty} \left|f(x')\right|^2 dx' < \infty.$$

Аналогічним чином доводимо квадратичну інтегровність похідних $y_0'(x)$ і $y_0''(x)$.

Припустимо тепер, що разом з $y_0(x)$ існує ще одна функція $y_1(x)$, яка задовольняє рівняння (6.24) та має ті самі властивості, що й $y_0(x)$. Тоді різниця $w(x) = y_1(x) - y_0(x)$ задовольняє однорідне рівняння

$$-w''(x) - \lambda w(x) = 0 \qquad (6.26)$$

і теж належить простору $L_2(-\infty,\infty)$. Але загальний розв'язок рівняння (6.26) має вигляд

$$w(x) = C_1 e^{i\sqrt{\lambda}x} + C_2 e^{-i\sqrt{\lambda}x},\ \mathrm{Im}\sqrt{\lambda} > 0.$$

---

(6.24): $A'(x)e^{i\sqrt{\lambda}x} - B'(x)e^{-i\sqrt{\lambda}x} = if(x)/\sqrt{\lambda}$. З отриманої системи $A'(x) = \frac{i}{2\sqrt{\lambda}}e^{-i\sqrt{\lambda}x}f(x)$, $B'(x) = -\frac{i}{2\sqrt{\lambda}}e^{i\sqrt{\lambda}x}f(x)$, звідки при $\mathrm{Im}\,\lambda > 0$ знаходимо: $A(x) = \frac{i}{2\sqrt{\lambda}}\int_{-\infty}^{x} e^{-i\sqrt{\lambda}x'}f(x')\,dx' + A_1$, $B(x) = -\frac{i}{2\sqrt{\lambda}}\int_{\infty}^{x} e^{i\sqrt{\lambda}x'}f(x')\,dx' + B_1 = \frac{i}{2\sqrt{\lambda}}\int_{x}^{\infty} e^{i\sqrt{\lambda}x'}f(x')\,dx' + B_1$, де $A_1$, $B_1$ — сталі. Для квадратичної інтегровності розв'язку $y(x)$ ці сталі треба прирівняти нулю.

Загальний розв'язок (6.27) рівняння (6.24) при дійсних $\lambda > 0$ і неперервних фінітних $f(x)$ знаходиться аналогічно. Однак загальний розв'язок рівняння (6.24) при $\lambda = 0$ і тих самих $f(x)$ слід шукати у вигляді $y(x) = A(x)x + B(x)$. Коефіцієнтні функції

Оскільки при $C_1 \neq 0$ модуль першого доданка необмежено зростає при $x \to -\infty$, а при $C_2 \neq 0$ і $x \to \infty$ необмежено зростає модуль другого доданка, то функція $w(x)$ може належати $L_2(-\infty,\infty)$ лише тоді, коли $C_1 = C_2 = 0$. Бачимо, що $y_1(x) = y_0(x)$ для всіх $x \in (-\infty,\infty)$.

Із завдання 6.2.1 та наведеного вище означення випливає, що спектру оператора $D_2$ можуть належати лише точки невід'ємної півосі.

**Теорема 6.2.1.** Усі числа $\lambda \geq 0$ належать спектру оператора $D_2$.

*Доведення.* Достатньо показати, що існують функції $f(x) \in L_2(-\infty,\infty)$, для яких при $\lambda \geq 0$ рівняння (6.24) не має розв'язків, які б належали $L_2(-\infty,\infty)$.

Почнемо з випадку $\lambda > 0$. Тоді для довільної неперервної фінітної функції $f(x)$ загальний розв'язок рівняння (6.24) має вигляд

$$y(x) = C_1 e^{i\sqrt{\lambda}x} + C_2 e^{-i\sqrt{\lambda}x} + \frac{i}{2\sqrt{\lambda}} \int_{-\infty}^{\infty} e^{i\sqrt{\lambda}|x-x'|} f(x')\,dx'. \qquad (6.27)$$

Звідси бачимо, що

$$y(x) \underset{x\to\infty}{=} \left[ C_1 + \frac{i}{2\sqrt{\lambda}} \int_{-\infty}^{\infty} e^{-i\sqrt{\lambda}x'} f(x')\,dx' \right] e^{i\sqrt{\lambda}x} + C_2 e^{-i\sqrt{\lambda}x}, \qquad (6.28)$$

$$y(x) \underset{x\to-\infty}{=} C_1 e^{i\sqrt{\lambda}x} + \left[ C_2 + \frac{i}{2\sqrt{\lambda}} \int_{-\infty}^{\infty} e^{i\sqrt{\lambda}x'} f(x')\,dx' \right] e^{-i\sqrt{\lambda}x}. \qquad (6.29)$$

З формули (6.28) випливає, що розв'язок (6.27) може належати $L_2(-\infty,\infty)$ лише за умови

$$C_1 + \frac{i}{2\sqrt{\lambda}} \int_{-\infty}^{\infty} e^{-i\sqrt{\lambda}x'} f(x')\,dx' = 0, \ \ C_2 = 0. \qquad (6.30)$$

Аналогічно, з формули (6.29) випливає, що розв'язок (6.27) може належати $L_2(-\infty,\infty)$ лише тоді, коли

---

задовольняють систему рівнянь $A'(x)x + B'(x) = 0$, $A'(x) = -f(x)$, звідки $A(x) = -\int_{-\infty}^{x} f(x')\,dx' + A_1$, $B(x) = \int_{-\infty}^{x} x'f(x')\,dx' + B_1$ або $A(x) = -\int_{\infty}^{x} f(x')\,dx' + A_2$, $B(x) = \int_{\infty}^{x} x'f(x')\,dx' + B_2$, де $A_1$, $B_1$, $A_2$, $B_2$ — сталі. Шуканий розв'язок $y(x) = A_1 x + B_1 - \int_{-\infty}^{x} (x-x')f(x')\,dx' = A_1 x + B_1 - \int_{-\infty}^{x} |x-x'| f(x')\,dx'$ або, в іншому вигляді, $y(x) = A_2 x + B_2 + \int_{x}^{\infty} (x-x')f(x')\,dx' = A_2 x + B_2 - \int_{x}^{\infty} |x-x'| f(x')\,dx'$. Половина суми цих виразів дає розв'язок (6.32).

$$C_1 = 0, \quad C_2 + \frac{i}{2\sqrt{\lambda}} \int_{-\infty}^{\infty} e^{i\sqrt{\lambda}x'} f(x')\,dx' = 0. \tag{6.31}$$

Поєднуючи результати (6.30) і (6.31), бачимо, що розв'язок (6.27) може належати $L_2(-\infty,\infty)$ лише якщо Фур'є-образ $\hat{f}(k)$ функції $f(x)$ дорівнює нулю в точках $k = \pm\sqrt{\lambda}$.

При $\lambda = 0$ загальний розв'язок рівняння (6.24) описується виразом

$$y(x) = C_1 + C_2 x - \frac{1}{2} \int_{-\infty}^{\infty} |x - x'| f(x')\,dx'. \tag{6.32}$$

Переходячи в (6.32) до границь $x \to \pm\infty$ та майже повторюючи попередні міркування, знаходимо, що розв'язок (6.32) може належати $L_2(-\infty,\infty)$ лише за умов

$$\int_{-\infty}^{\infty} f(x')\,dx' = 0, \quad \int_{-\infty}^{\infty} x' f(x')\,dx' = 0.$$

Отже, при будь-якій неперервній фінітній функції $f(x) \in L_2(-\infty,\infty)$, що не має перелічених вище властивостей, рівняння (6.24) при $\lambda \geq 0$ не має розв'язків, що належать $L_2(-\infty,\infty)$.

Ще раз підкреслимо, що рівняння

$$\left(D_2 y\right)(x) - \lambda y(x) = 0, \quad -\infty < x < \infty,$$

при будь-якому комплексному, зокрема, дійсному, $\lambda$ не має нетривіальних розв'язків, що належать множині $L_2(-\infty,\infty)$. Тому диференціальний оператор $D_2$ не має жодного власного значення.

Інші ж диференціальні оператори, зокрема, диференціальні оператори $H$ на дійсній осі з областю визначення (6.20), (6.21), які діють за формулою

$$\left(Hy\right)(x) = -y''(x) + q(x)y(x), \quad -\infty < x < \infty, \tag{6.33}$$

де $q(x)$ — неперервна дійсна функція, можуть мати власні значення. Наприклад, диференціальне рівняння

$$-y''(x) - \frac{2a^2}{\operatorname{ch}^2 ax} y(x) = \lambda y(x), \quad a > 0,$$

при $\lambda = -a^2$ має нетривіальний розв'язок $y(x) = A/\operatorname{ch} ax$, який, очевидно, належить множині (6.20), (6.21). При цьому можна показати, що точки півосі $[0,\infty)$ належать спектру оператора (6.33) з $q(x) = -2a^2/\operatorname{ch}^2 ax$.

Зазначимо, що принаймні *при обмежених дійсних функціях* $q(x)$ *оператори (6.33), як і оператори, що асоційовані з крайовою задачею Штурма — Ліувілля, можуть мати лише дійсні власні значення.* Інакше кажучи, рівняння

$$-y''(x)+q(x)y(x)=\lambda y(x),\ \ -\infty<x<\infty, \tag{6.34}$$

при вказаних $q(x)$ може мати нетривіальні розв'язки, які задовольняють умови (6.20), (6.21), лише при дійсних значеннях параметра $\lambda$.

Справді, припустимо, що $y_0(x)$ — саме такий розв'язок рівняння (6.34) при $\lambda=\lambda_0$. Помноживши обидві частини відповідного рівняння (6.34) на спряжену функцію $\overline{y_0(x)}$ та зінтегрувавши їх за змінною $x$ у межах від $-\infty$ до $\infty$, отримуємо рівність

$$-\int_{-\infty}^{\infty}\overline{y_0(x)}y_0''(x)\,dx+\int_{-\infty}^{\infty}q(x)\left|y_0(x)\right|^2dx=\lambda_0\int_{-\infty}^{\infty}\left|y_0(x)\right|^2dx. \tag{6.35}$$

Оскільки

$$\frac{d}{dx}\left[\overline{y_0(x)}y_0'(x)\right]=\overline{y_0(x)}y_0''(x)+\left|y_0'(x)\right|^2, \tag{6.36}$$

то в силу умов (6.20), (6.21) функція $\overline{y_0(x)}y_0'(x)$ інтегровна та має інтегровну першу похідну. Для таких функцій, як відомо, справджуються співвідношення

$$\lim_{x\to\pm\infty}\overline{y_0(x)}y_0'(x)=0. \tag{6.37}$$

Тому на підставі формул (6.36) та (6.37) маємо:

$$-\int_{-\infty}^{\infty}\overline{y_0(x)}y_0''(x)\,dx=-\overline{y_0(x)}y_0'(x)\Big|_{-\infty}^{\infty}+\int_{-\infty}^{\infty}\left|y_0'(x)\right|^2dx=\int_{-\infty}^{\infty}\left|y_0'(x)\right|^2dx. \tag{6.38}$$

Бачимо з формул (6.35) та (6.38), що параметр $\lambda_0$, якому відповідає нетривіальний розв'язок $y_0(x)$, дорівнює відношенню двох дійсних чисел:

$$\lambda_0=\frac{\int_{-\infty}^{\infty}\left|y_0'(x)\right|^2dx+\int_{-\infty}^{\infty}q(x)\left|y_0(x)\right|^2dx}{\int_{-\infty}^{\infty}\left|y_0(x)\right|^2dx}.$$

Підкреслимо, що для *кожного власного значення оператора* $H$ *існують такі функції* $f(x)$, *що належать* $L_2(-\infty,\infty)$, *для яких неоднорідне рівняння (6.22) взагалі не має розв'язків з області визначення* $H$.

Справді, нехай $y_0(x)$ є власною функцією оператора $H$, яка відповідає власному значенню $\lambda_0$, тобто є нетривіальним розв'язком рівняння

$$-y_0''(x)+q(x)y_0(x)-\lambda_0 y_0(x)=0, \qquad (6.39)$$

що належить множині (6.20), (6.21). Без обмеження загальності можемо вважати, що $y_0(x)$ набуває дійсних значень. Розглянемо при тому ж $\lambda_0$ неоднорідне рівняння

$$-y''(x)+q(x)y(x)-\lambda_0 y(x)=y_0(x) \qquad (6.40)$$

та припустимо, що воно має розв'язок $y(x)$, який належить $L_2(-\infty,\infty)$ та, більше того, множині (6.20), (6.21). Помножимо обидві частини рівняння (6.39) на функцію $y(x)$, обидві частини рівняння (6.40) — на функцію $y_0(x)$, та зінтегруємо отримані рівності за $x$ у межах від $-\infty$ до $\infty$. Віднявши далі перший результат із другого, дістанемо

$$\int_{-\infty}^{\infty}\left[y(x)y_0''(x)-y_0(x)y''(x)\right]dx=\int_{-\infty}^{\infty}y_0^2(x)dx,$$

або

$$\int_{-\infty}^{\infty}y_0^2(x)dx=\left[y(x)y_0'(x)-y_0(x)y'(x)\right]\Big|_{-\infty}^{\infty}=0.$$

Але це співвідношення суперечить припущенню про нетривіальність розв'язку $y_0(x)$.

Більш повне доведення твердження про належність власних значень оператора $H$ до його спектра ми опускаємо.

Ми пов'язали поняття спектра лінійних диференціальних операторів із розв'язністю або однозначною розв'язністю неоднорідних лінійних диференціальних рівнянь у певних класах функцій. Однак можна дати наступне еквівалентне означення спектра оператора $H$.

**Означення 6.2.1**. Точка $\lambda$ належить до спектра оператора $H$ тоді й лише тоді, коли існує послідовність нормованих функцій $y_n(x)$,

$$\int_{-\infty}^{\infty}\left|y_n(x)\right|^2 dx=1,$$

з області визначення (6.20), (6.21), таких, що

$$\int_{-\infty}^{\infty}\left|(Hy_n)(x)-\lambda y_n(x)\right|^2 dx=\int_{-\infty}^{\infty}\left|-y_n''(x)+\left[q(x)-\lambda\right]y_n(x)\right|^2 dx\underset{n\to\infty}{\to}0.$$

Якщо послідовності нормованих функцій $y_n(x)$ в означенні 6.2.1 та їх похідних $y_n'(x)$, $y_n''(x)$ збігаються в $L_2(-\infty,\infty)$ до, відповідно, функцій $y_\lambda(x)$, $y_\lambda'(x)$, $y_\lambda''(x)$, то $\lambda$ є звичайним власним значенням оператора $H$, а $y_\lambda(x)$ — власною функцією оператора $H$, яка відповідає цьому власному значенню. Справді, у такому випадку

$$\int_{-\infty}^{\infty} \left|\left(Hy_n\right)(x) - \lambda y_n(x)\right|^2 dx \underset{n\to\infty}{\to} \int_{-\infty}^{\infty} \left|\left(Hy_\lambda\right)(x) - \lambda y_\lambda(x)\right|^2 dx = 0,$$

звідки

$$\left(Hy_\lambda\right)(x) - \lambda y_\lambda(x) = 0.$$

Якщо ж послідовність функцій $y_n(x)$ в означенні 6.2.1 не збігається в $L_2(-\infty,\infty)$ до граничної функції $y_\lambda(x)$, що належить області (6.20), (6.21), то вважається, що $\lambda$ є точкою неперервного спектра $H$.

У загальному випадку можливі ситуації, коли $\lambda$ належить неперервному спектру та одночасно є власним значенням.

**Завдання 6.2.2.** Для оператора $D_2$ наведіть приклад послідовності неперервних функцій $y_n(x)$ з його області визначення, для якої при додатному $\lambda = k_0^2$ справджується співвідношення

$$\int_{-\infty}^{\infty} \left|-y_n''(x) - k_0^2 y_n(x)\right|^2 dx \underset{n\to\infty}{\to} 0.$$

*Відповідь*: $y_n(x) = \dfrac{1}{\sqrt[4]{\pi n}} e^{ik_0 x} e^{-\frac{x^2}{2n}}$.

## 6.3. ФУНКЦІЯ ГРІНА РІВНЯННЯ ТЕПЛОПРОВІДНОСТІ НА ПРЯМІЙ

Покажемо, що результат (6.12) можна подати як згортку певного спеціального розв'язку задачі (6.3), (6.4) і заданої початкової функції, при цьому вимоги до останньої можна суттєво послабити. Для цього у формулі (6.12) запишемо Фур'є-образ $\hat{u}_0(k)$ у вигляді інтеграла (6.11), поміняємо в здобутому виразі порядок інтегрування та скористаємося формулою

$$\int_{-\infty}^{\infty} e^{-\lambda k^2 + ikx} dk = \sqrt{\frac{\pi}{\lambda}} e^{-\frac{x^2}{4\lambda}}, \quad \lambda > 0. \tag{6.41}$$

Маємо $(\lambda = a^2 t)$:

$$v(x,t)=\frac{1}{\sqrt{2\pi}}\int_{-\infty}^{\infty}e^{ikx}\left(\frac{1}{\sqrt{2\pi}}\int_{-\infty}^{\infty}e^{-ikx'}u_0(x')dx'\right)e^{-a^2k^2t}\,dk=$$

$$=\frac{1}{2\pi}\int_{-\infty}^{\infty}\left(\int_{-\infty}^{\infty}e^{ik(x-x')}e^{-a^2k^2t}\,dk\right)u_0(x')dx'=\frac{1}{2\pi}\int_{-\infty}^{\infty}\sqrt{\frac{\pi}{a^2t}}\,e^{-\frac{(x-x')^2}{4a^2t}}u_0(x')dx',$$

або

$$v(x,t)=\int_{-\infty}^{\infty}G(x,x';t)u_0(x')dx', \qquad (6.42)$$

де

$$G(x,x';t)=G(x',x;t)=G(|x-x'|;t)=\frac{1}{\sqrt{4\pi a^2t}}e^{-\frac{(x-x')^2}{4a^2t}}. \qquad (6.43)$$

Диференціюючи функцію $G(x,x';t)$ за змінними $x$ і $t$ при $x\neq x'$ і $t>0,$ знаходимо:

$$\frac{\partial G(x,x';t)}{\partial x}=-\frac{(x-x')}{2a^2t}G(x,x';t),$$

$$\frac{\partial^2 G(x,x';t)}{\partial x^2}=\left[-\frac{1}{2a^2t}+\frac{(x-x')^2}{4a^4t^2}\right]G(x,x';t),$$

$$\frac{\partial G(x,x';t)}{\partial t}=\left[-\frac{1}{2t}+\frac{(x-x')^2}{4a^2t^2}\right]G(x,x';t).$$

Бачимо, що функція $G(x,x',t)$ задовольняє рівняння (6.3):

$$\frac{\partial G(x,x';t)}{\partial t}=a^2\frac{\partial^2 G(x,x';t)}{\partial x^2},\quad -\infty<x<\infty,\quad t>0. \qquad (6.44)$$

Тепер розглянемо послідовність функцій $g_n(x,x')=G(x,x';t_n),$ утворену значеннями функції $G(x,x';t)$ в послідовні моменти часу $t_1>t_2>...>t_n>0,$ узятими в порядку спадання. Очевидно, що функції $g_n(x,x')$ невід'ємні та інтегровні, оскільки при $t>0$ та довільних $x$, $x'$ маємо $G(x,x';t)\geq 0$ і

$$\int_{-\infty}^{\infty}G(x,x';t)dx=\frac{1}{\sqrt{4\pi a^2t}}\int_{-\infty}^{\infty}e^{-\frac{(x-x')^2}{4a^2t}}dx=$$

$$=\left[z=\frac{x-x'}{\sqrt{4a^2t}},dz=\frac{dx}{\sqrt{4a^2t}}\right]=\frac{1}{\sqrt{\pi}}\int_{-\infty}^{\infty}e^{-z^2}dz=1. \qquad (6.45)$$

Графіки функцій $g_n(x,x')$ (див. рис. 6.1) мають куполоподібну форму з максимумами висотою $1\big/\sqrt{4\pi a^2t_n}$ у точці $x=x'$ та півширинами (на піввисоті) $2\sqrt{\ln 2}\,a\sqrt{t_n}$; згідно з (6.45), площі під усіма цими графіками однакові й дорівнюють одиниці. Звідси випливає, що при $n \to \infty$ $(t_n \to 0)$ графіки функцій $g_n(x,x')$ трансформуються наступним чином: їх висоти зростають, півширини спадають, а площі під ними не змінюються. Як відомо з теорії узагальнених функцій, граничним значенням такої послідовності функцій є дельта-функція Дірака:

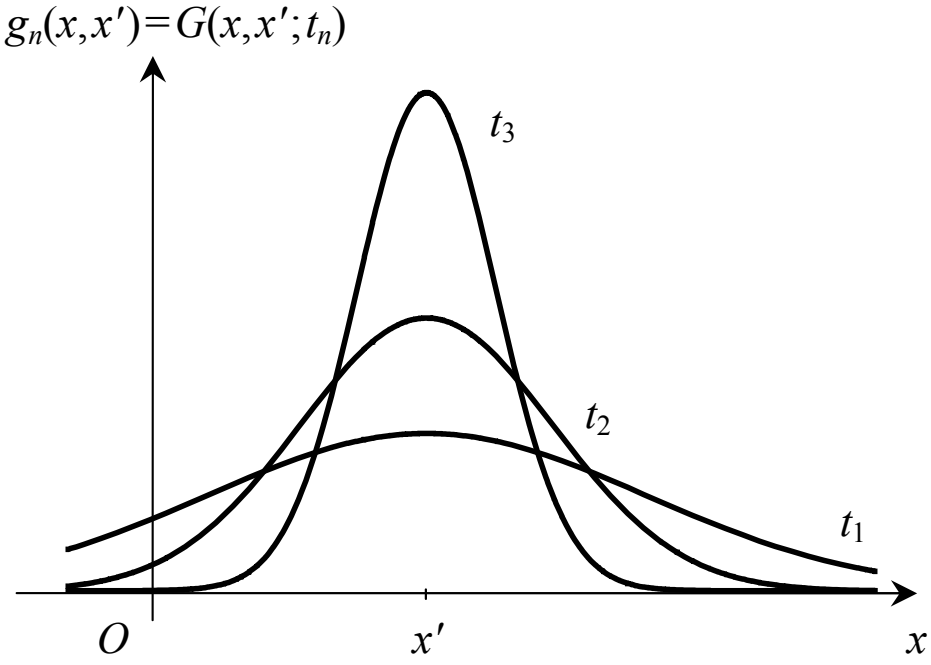

Рис. 6.1. Графіки функції Гріна рівняння теплопровідності для різних моментів часу $t_1 > t_2 > t_3 > 0$

$$\lim_{n\to\infty} g_n(x,x') = \delta(x-x').$$

Повертаючись до функції $G(x,x';t)$, можемо записати:

$$G(x,x';0) = \delta(x-x'). \qquad (6.46)$$

Отже, функцію (6.43) можна розглядати як неперервний, неперервно диференційовний довільну кількість разів скрізь в області $-\infty < x < \infty$, $t > 0$ та інтегровний за координатами розв'язок задачі Коші (6.44), (6.46). Вона називається *функцією Гріна або фундаментальним розв'язком рівняння теплопровідності на нескінченній прямій.*

Аналогічні властивості має й функція Гріна для рівняння дифузії на прямій, яка описується виразом

$$G(x,x';t) = G(x',x;t) = G(|x-x'|;t) = \frac{1}{\sqrt{4\pi Dt}} e^{-\frac{(x-x')^2}{4Dt}}, \quad t > 0. \qquad (6.47)$$

**Завдання 6.3.1.** Доведіть формулу (6.41).

**Завдання 6.3.2.** Доведіть формулу (6.43), розв'язавши крайову задачу (6.44), (6.46) за допомогою інтегрального перетворення Фур'є.

З явного вигляду виразу (6.42) випливає, що він дає розв'язок задачі Коші (6.3), (6.4) не лише для інтегровних, але й обмежених кусково-неперервних початкових функцій, чи навіть початкових функцій, що необмежено зростають при $|x| \to \infty$, наприклад, виду $u_0(x) = Ce^{\beta x}$, $C$ і $\beta$ — сталі.

**Завдання 6.3.3.** Безпосередньою підстановкою доведіть, що вираз (6.42) дає розв'язок задачі Коші (6.3), (6.4) з початковими функціями із щойно зазначених класів.

**Завдання 6.3.4.** Знайдіть розподіл температури в довгому тонкому однорідному стержні з теплоізольованою бічною поверхнею в довільний момент часу $t > 0,$ якщо в початковий момент $t = 0$ він був нерівномірно нагрітий до температури $T(x,0) = T_0 e^{-x^2/l^2}$. Зобразіть графіки цього розподілу для кількох послідовних моментів часу.

*Відповідь*: $v(x,t) = T_0 \dfrac{l}{\sqrt{l^2 + 4a^2 t}} e^{-\frac{x^2}{l^2+4a^2 t}}.$

**Завдання 6.3.5.** У момент часу $t = 0$ у правій половині довгої тонкої скляної трубки, повністю заповненої рідиною, містяться частинки домішки, які розподілені рівномірно з концентрацією $c.$ Знайдіть розподіл концентрації частинок домішки, що встановлюється в трубці внаслідок дифузії до моменту часу $t.$ Проаналізуйте граничні випадки $t \downarrow 0$ та $t \to \infty.$

*Вказівки.* Скориставшись функцією Гріна (6.47), знайдіть розв'язок задачі Коші для однорідного рівняння дифузії на прямій з початковою умовою $u_0(x) = \begin{cases} 0, & x < 0, \\ c, & x > 0. \end{cases}$ Результат виразіть через функцію помилок $\Phi(z) = \dfrac{2}{\sqrt{\pi}} \int_0^z e^{-y^2} dy;$ $\Phi(0) = 0,$ $\Phi(\infty) = 1,$ $\Phi(-z) = -\Phi(z).$

*Відповідь*: $v(x,t) = \dfrac{c}{2}\left[1 + \Phi\left(\dfrac{x}{\sqrt{4Dt}}\right)\right].$

Проаналізуємо окремо ситуацію, коли в момент часу $t_0 = 0$ до малої ділянки $(x_0 - \varepsilon, x_0 + \varepsilon)$ тонкого стержня з площею поперечного перерізу $S$ та теплоізольованою бічною поверхнею, який має нульову температуру, миттєво підводять кількість тепла $Q_0$; тут $\varepsilon > 0$ — інфінітезимальне число. У результаті цього температура ділянки практично миттєво (унаслідок малості ділянки) зростає до значення $T_0 = Q_0/(2\varepsilon\rho c S)$, а температура інших точок стержня все ще залишиться нульовою[1]. Починаючи з цього моменту, температура

[1] Звернемо увагу, що згідно з молекулярно-кінетичною теорією речовини процес поширення тепла в стержні відбувається з обмеженою швидкістю, тоді як формули (6.42) та (6.48) формально показують, що локалізовані теплові неоднорідності чи теплові джерела впливають на температуру будь-яких точок стержня за нескінченно малий проміжок часу. Ця обставина вказує на наближений характер макроскопічного

стержня змінюється за законом (6.42). Розбиваючи інтеграл (6.42) на три інтеграли по інтервалах $(-\infty, x_0 - \varepsilon)$, $(x_0 - \varepsilon, x_0 + \varepsilon)$ і $(x_0 + \varepsilon, \infty)$, знаходимо, що в момент часу $t > 0$ в точці з координатою $x$ вона матиме значення

$$v(x,t) = \int_{x_0-\varepsilon}^{x_0+\varepsilon} G(x,x';t)\frac{Q_0}{2\varepsilon\rho cS}dx' \approx G(x,x_0;t)\frac{Q_0}{2\varepsilon\rho cS}\int_{x_0-\varepsilon}^{x_0+\varepsilon} dx' = \frac{Q_0}{\rho cS}G(x,x_0;t),$$

де ми врахували неперервність функції $G(x,x';t)$ при $t > 0$ та скористалися теоремою про середнє для визначених інтегралів. Отже, з точністю до множника $Q_0/(\rho cS)$ значення функції Гріна $G(x,x_0;t)$ дорівнює температурі, яка встановлюється в точці $x$ стержня в момент часу $t > 0$ внаслідок того, що в початковий момент $t_0 = 0$ в точку $x_0$ стержня, що мав нульову температуру, було миттєво внесено кількість тепла $Q_0$.

Якщо ж тепло $Q_0$ вноситься в точку $x_0$ стержня в момент $t_0 \neq 0$, і стержень до цього мав нульову температуру, то його температура в точці $x$ у момент часу $t > t_0$ дорівнюватиме значенню

$$v(x,t) = \frac{Q_0}{\rho cS}G(x,x_0;t-t_0). \tag{6.48}$$

Рівністю (6.48) визначається фізичний зміст функції Гріна (6.43) в термінах температури. Йому можна надати й іншого звучання, якщо покласти $Q_0 = 1$ та переписати (6.48) у вигляді $G(x,x_0;t-t_0) = c\rho Sv(x,t)$. Згадавши, що добуток $c\rho S$ дорівнює теплоємності одиниці довжини однорідного стержня, можемо стверджувати: *величина* $G(x,x_0;t-t_0)$ *дорівнює погонній густині кількості тепла, яке міститиметься в околі точки* $x$ *однорідного стержня в момент часу* $t > t_0$ *після того, як у його точку* $x_0$ *у момент часу* $t_0$, *коли він мав нульову температуру, було миттєво введено одиницю кількості тепла.*

Зауважимо, що перерозподіл тепла в стержні внаслідок процесу теплопровідності відбувається таким чином, що загальна кількість тепла, зосередженого в стержні, зберігається. Математично цей факт виражається співвідношенням[1]

---

рівняння теплопровідності. Тим не менше, теоретичні результати, отримані на основі цього рівняння, у багатьох випадках можуть уважатися задовільними, оскільки узгоджуються з результатами вимірювань у межах експериментальної чи прийнятної похибки. Легко, наприклад, перевірити, що зміни температури стержня в точках, відносно далеких від теплових джерел, можуть стати експериментально помітними лише через обмежений проміжок часу після того, як ці джерела починають діяти.

$$\int_{-\infty}^{\infty} G(x,x_0;t-t_0)\,dx = 1, \quad t \ge t_0, \tag{6.49}$$

яке доводиться за аналогією з формулою (6.45).

**Завдання 6.3.6.** Нехай у момент часу $t_0 \ne 0$ в точку $x_0$ довгої тонкої трубки з площею поперечного перерізу $S$, заповненої деякою речовиною, миттєво вносять $N_0$ частинок домішки. Покажіть, що концентрація частинок домішки в точці $x$ трубки в момент часу $t > t_0$ виражається через функцію Гріна (6.47) формулою

$$v(x,t) = \frac{N_0}{S} G(x,x_0;t-t_0),$$

тобто величина $G(x,x_0;t-t_0)$ дорівнює густині ймовірності переходу частинки домішки в трубці з перерізу $x_0$ в окіл перерізу $x$ за проміжок часу $t-t_0$. Який зміст має умова (6.49) для цієї $G(x,x_0;t-t_0)$?

**Завдання 6.3.7.** Знайдіть концентрацію частинок домішки в точці $x$ довгої тонкої трубки з одиничною площею поперечного перерізу, якщо в момент часу $t_1$ у точках $x_1$ та $x_2$ трубки в неї вводять відповідно $N_1$ та $N_2$ частинок домішки. Трубку заповнено газом, який вступає в хімічну реакцію з домішкою, у ході котрої частинки домішки поглинаються зі швидкістю $-qu(x,t)$, пропорційною їхній концентрації $u(x,t)$.

*Вказівка*. Треба побудувати функцію Гріна рівняння $u_t = Du_{xx} - qu$, $-\infty < x < \infty$, $t > 0$, яке заміною $u(x,t) = v(x,t)e^{-qt}$ зводиться до рівняння (6.3).

*Відповідь*:

$$u(x,t) = \begin{cases} 0, & t < t_1, \\ \dfrac{e^{-q(t-t_1)}}{\sqrt{4\pi D(t-t_1)}} \left[ N_1 e^{-\frac{(x-x_1)^2}{4D(t-t_1)}} + N_2 e^{-\frac{(x-x_2)^2}{4D(t-t_1)}} \right], & t > t_1. \end{cases}$$

Перейдемо до побудови розв'язку задачі Коші (6.5), (6.6) для неоднорідного рівняння теплопровідності на нескінченній прямій. Знову скористаємося інтегральним перетворенням Фур'є за змінною $x$, застосувавши його тепер до диференціального рівняння (6.5) та початкової умови (6.6). Уважаючи спершу, що функція джерел $f(x,t)$ та її Фур'є-образ $\hat{f}(k,t)$ — інтегровні за своїми змінними, для Фур'є-образу $\hat{w}(k,t)$ шуканої функції $w(x,t)$ дістаємо неоднорідне диференціальне рівняння першого порядку за змінною $t$ та нульову початкову умову:

$$\frac{\partial \widehat{w}(k,t)}{\partial t} + a^2k^2\widehat{w}(k,t) = \widehat{f}(k,t), \tag{6.50}$$

$$\widehat{w}(k,0) = 0. \tag{6.51}$$

Розв'язуємо рівняння (6.50) методом варіації довільної сталої. А саме, шукаємо його розв'язок у вигляді

$$\widehat{w}(k,t) = C(k,t)e^{-a^2k^2t},$$

де $C(k,t)$ — невідома функція. Підставивши цей вираз у рівняння (6.50), для похідної функції $C(k,t)$ дістаємо

$$\frac{\partial C(k,t)}{\partial t} = e^{a^2k^2t}\widehat{f}(k,t),$$

звідки

$$C(k,t) = \int_0^t e^{a^2k^2\tau}\widehat{f}(k,\tau)d\tau + C_1(k),$$

де $C_1(k)$ — стала інтегрування, яка, у принципі, може залежати від $k$. Знаходимо її, скориставшись умовою (6.51), з якої маємо $C(k,0) = 0$ і, отже, $C_1(k) = 0$.

Таким чином,

$$\widehat{w}(k,t) = \int_0^t e^{-a^2k^2(t-\tau)}\widehat{f}(k,\tau)d\tau.$$

Застосувавши до цієї функції обернене перетворення Фур'є, знаходимо:

$$w(x,t) = \frac{1}{\sqrt{2\pi}}\int_0^t d\tau \int_{-\infty}^{\infty} e^{ikx}\, e^{-a^2k^2(t-\tau)}\widehat{f}(k,\tau)dk. \tag{6.52}$$

Не зупиняючись уже докладно на дослідженні властивостей функції (6.52) (для цього можна скористатися тими ж самими методами, що застосовувалися при вивченні функції (6.12)), перейдемо в (6.52) від Фур'є-образу $\widehat{f}(k,\tau)$ функції джерел до самої $f(x,t)$. Оскільки в показнику експоненти $t \geq \tau$, то, з огляду на формулу (6.41), маємо:

$$w(x,t) = \frac{1}{\sqrt{2\pi}}\int_0^t d\tau \int_{-\infty}^{\infty} e^{ikx}\, e^{-a^2k^2(t-\tau)}\left(\frac{1}{\sqrt{2\pi}}\int_{-\infty}^{\infty} e^{-ikx'} f(x',\tau)dx'\right)dk =$$

$$= \frac{1}{2\pi}\int_0^t d\tau \int_{-\infty}^{\infty}\left(\int_{-\infty}^{\infty} e^{ik(x-x')}\, e^{-a^2k^2(t-\tau)}\,dk\right) f(x',\tau)dx' =$$

$$= \frac{1}{2\pi}\int_0^t d\tau \int_{-\infty}^{\infty} \sqrt{\frac{\pi}{a^2(t-\tau)}}\, e^{-\frac{(x-x')^2}{4a^2(t-\tau)}} f(x',\tau)\,dx'.$$

Отже,

$$w(x,t) = \int_0^t d\tau \int_{-\infty}^{\infty} G(x,x';t-\tau) f(x',\tau)\,dx', \tag{6.53}$$

де $G(x,x';t)$ — функція Гріна (6.43).

Структуру формули (6.53) легко зрозуміти, виходячи з фізичного змісту функції Гріна (6.43). Справді, кількість тепла, що генерується джерелами тепла з густиною потужності $F(x,t) = c\rho f(x,t)$ на ділянці стержня $(x', x'+dx')$ за проміжок часу $(\tau, \tau+d\tau)$, дорівнює $dQ(x',\tau) = F(x',\tau)Sdx'd\tau = c\rho f(x',\tau)Sdx'd\tau.$ Згідно з формулою (6.48), це тепло спричиняє зміну температури стержня в точці $x$ у момент часу $t > \tau$ на величину

$$\frac{c\rho f(x',\tau)S\,dx'd\tau}{\rho c S} G(x,x';t-\tau) = G(x,x';t-\tau) f(x',\tau)\,dx'd\tau.$$

Додавши всі такі внески, що генеруються джерелами тепла на всіх ділянках стержня за проміжок часу $[0,t]$, та здійснивши граничний перехід $|dx'| \to 0$ і $|d\tau| \to 0$, дістаємо температуру стержня в точці $x$ у момент часу $t$.

Зазначимо, що формулу (6.53) можна було б записати відразу, пославшись, як і у випадку хвильового рівняння, на принцип Дюамеля (див. підрозділ 4.2, зокрема, завдання 4.2.3, та підрозділ 4.8).

З огляду на формули (6.7), (6.42) і (6.53) розв'язок задачі Коші (6.1), (6.2) має вигляд

$$u(x,t) = \int_{-\infty}^{\infty} G(x,x';t) u_0(x')\,dx' + \int_0^t d\tau \int_{-\infty}^{\infty} G(x,x';t-\tau) f(x',\tau)\,dx'. \tag{6.54}$$

На завершення зазначимо, що використаний у підрозділах 6.1, 6.3 метод інтегрального перетворення Фур'є можна застосувати й до задач Коші для інших лінійних диференціальних рівнянь зі сталими коефіцієнтами на всій прямій, зокрема, для хвильового рівняння.

**Завдання 6.3.8.** Функція Гріна хвильового рівняння на нескінченній прямій означається як узагальнений розв'язок $G(x,x';t) = G(|x-x'|;t)$ задачі Коші

$$\frac{\partial^2 G(x,x';t)}{\partial t^2} = a^2 \frac{\partial^2 G(x,x';t)}{\partial x^2}, \quad -\infty < x < \infty, \quad t > 0, \tag{6.55}$$

$$G(x,x';0)=0,\ \ G_t(x,x';0)=\delta(x-x'). \tag{6.56}$$

Переконайтеся, що в термінах цієї функції Гріна:

а) розв'язок задачі Коші (4.5), (4.6) для однорідного хвильового рівняння на прямій $-\infty<x<\infty$ має при $t>0$ вигляд

$$u(x,t)=\frac{\partial}{\partial t}\int_{-\infty}^{\infty}G(|x-x'|;t)u_0(x')dx'+\int_{-\infty}^{\infty}G(|x-x'|;t)v_0(x')dx'; \tag{6.57}$$

б) розв'язок задачі Коші для неоднорідного хвильового рівняння (4.3) на прямій $-\infty<x<\infty$, який задовольняє нульові початкові умови $u(x,0)=0,\ u_t(x,0)=0,$ має при $t>0$ вигляд

$$u(x,t)=\int_0^t d\tau\int_{-\infty}^{\infty}dx'G(|x-x'|;t-\tau)f(x',\tau). \tag{6.58}$$

*Вказівка.* Узявши до уваги формули (6.55) і (6.56), безпосередньою підстановкою покажіть, що вирази (6.57) і (6.58) формально задовольняють відповідні рівняння та початкові умови.

**Завдання 6.3.9.** Скориставшись перетворенням Фур'є, покажіть, що розв'язок задачі Коші (6.55), (6.56) має вигляд

$$G(x,x';t)=\frac{1}{2\pi}\int_{-\infty}^{\infty}e^{ik(x-x')}\frac{\sin akt}{ak}dk=\frac{1}{2a}\theta\big(at-|x-x'|\big),\ \ t>0, \tag{6.59}$$

де $\theta(z)$ — східчаста функція Хевісайда: $\theta(z)=\begin{cases}0, \text{ якщо } z<0,\\ 1, \text{ якщо } z>0.\end{cases}$

**Завдання 6.3.10.** Покажіть, що вирази (6.57) і (6.58) із функцією Гріна (6.59) зводяться до розв'язків (4.13) і (4.20) відповідних задач Коші для однорідного та неоднорідного хвильових рівнянь.

## 6.4. РІВНЯННЯ ТЕПЛОПРОВІДНОСТІ НА ПІВОСІ. УЗАГАЛЬНЕНЕ ПЕРЕТВОРЕННЯ ФУР'Є

Перейдемо до аналізу *задач Коші для рівняння теплопровідності (дифузії) на півосі* $0\le x<\infty.$ Почнемо з наступної задачі для однорідного рівняння:

$$\frac{\partial v}{\partial t}=a^2\frac{\partial^2 v}{\partial x^2},\ \ x>0,\ \ t>0, \tag{6.60}$$

$$v(0,t) = 0, \tag{6.61}$$

$$v(x,0) = u_0(x), \tag{6.62}$$

де крайова умова (6.61) означає, що на краю $x = 0$ системи підтримується нульова температура (нульова концентрація частинок домішки). Уважаючи, що початкова функція $u_0(x)$ має ті самі властивості, що і в задачі (6.1), (6.2), знайдемо розв'язок задачі (6.60)–(6.62) у класі функцій, неперервно диференційовних за часом та двічі неперервно диференційовних за координатою в області $x > 0, \ t > 0$.

Розв'язуємо задачу (6.60)–(6.62) методом продовження, викладеним у підрозділі 4.3. Для цього спершу продовжуємо початкову функцію $u_0(x)$ на область $x < 0$ непарним чином відносно точки $x = 0$; таке непарне продовження функції $u_0(x)$ має вигляд

$$U_0(x) = \begin{cases} u_0(x), \text{ якщо } x > 0, \\ -u_0(-x), \text{ якщо } x < 0. \end{cases} \tag{6.63}$$

Далі шукаємо розв'язок $V(x,t)$ задачі Коші для рівняння теплопровідності (дифузії) на необмеженій прямій з початковою функцією $U_0(x)$:

$$\frac{\partial V}{\partial t} = a^2 \frac{\partial^2 V}{\partial x^2}, \ -\infty < x < \infty, \ t > 0. \tag{6.64}$$

$$V(x,0) = U_0(x). \tag{6.65}$$

Згідно з результатами попереднього підрозділу, він дається формулою

$$V(x,t) = \int_{-\infty}^{\infty} G(|x - x'|;t) U_0(x') dx'. \tag{6.66}$$

І, нарешті, виокремлюємо ту частину $v(x,t)$ функції (6.66), яка визначена на півосі $x > 0$. Для цього записуємо формулу (6.66) при $x > 0$, ураховуємо, що в цьому випадку $V(x,t) = v(x,t)$, та виконуємо наступні перетворення:

$$v(x,t) = \int_{-\infty}^{0} G(|x - x'|;t) U_0(x') dx' + \int_{0}^{\infty} G(|x - x'|;t) U_0(x') dx' =$$

$$= \left\{ x' \to -x' \text{ у першому інтегралі} \right\} =$$

$$= -\int_{\infty}^{0} G(|x + x'|;t) U_0(-x') dx' + \int_{0}^{\infty} G(|x - x'|;t) U_0(x') dx' = \left\{ U_0(-x') = -U_0(x') \right\} =$$

$$= -\int_0^{\infty} G(|x+x'|;t)U_0(x')\,dx' + \int_0^{\infty} G(|x-x'|;t)U_0(x')\,dx' =$$

$$= \int_0^{\infty} \big[G(|x-x'|;t) - G(|x+x'|;t)\big]u_0(x')\,dx'.$$

Очевидно, що знайдена функція $v(x,t)$ задовольняє рівняння (6.60) і має ті самі властивості, що й розв'язок (6.42) для необмеженої прямої. Неважко перевірити, що вона також задовольняє крайову умову (6.61) та початкову умову (6.62). Справді, при $x=0,\ t>0$ та $x>0,\ t=0$ маємо:

$$v(0,t) = \int_0^{\infty} \big[G(|x'|;t) - G(|x'|;t)\big]u_0(x')\,dx' = 0, \quad t>0,$$

$$v(x,0) = \lim_{t\downarrow 0} \int_0^{\infty} \big[G(|x-x'|;t) - G(|x+x'|;t)\big]u_0(x')\,dx' =$$

$$= \int_0^{\infty} \big[\delta(x-x') - \delta(x+x')\big]u_0(x')\,dx' = u_0(x), \qquad x>0.$$

Отже, розв'язок задачі (6.60)–(6.62) має вигляд

$$v(x,t) = \int_0^{\infty} G_{\infty}(x,x';t)u_0(x')\,dx', \tag{6.67}$$

де

$$G_{\infty}(x,x';t) = G(|x-x'|;t) - G(|x+x'|;t) = \frac{1}{\sqrt{4\pi a^2 t}}\left[e^{-\frac{(x-x')^2}{4a^2t}} - e^{-\frac{(x+x')^2}{4a^2t}}\right]. \tag{6.68}$$

Функція $G_{\infty}(x,x';t)$ має зміст *функції Гріна для рівняння теплопровідності (дифузії) на півосі при крайовій умові (6.61)*. Вона є узагальненим розв'язком спеціальної задачі (6.60)–(6.62) з дельта-подібною початковою умовою:

$$\frac{\partial G_{\infty}(x,x';t)}{\partial t} = a^2 \frac{\partial^2 G_{\infty}(x,x';t)}{\partial x^2}, \quad 0<x,\ x',\ t<\infty, \tag{6.69}$$

$$G_{\infty}(0,x';t) = 0, \tag{6.70}$$

$$G_{\infty}(x,x';0) = \delta(x-x'), \tag{6.71}$$

– і виявляється неперервною та неперервно диференційовною довільну кількість разів скрізь в області $x \ge 0,\ t>0$, а також інтегровною за координатами.

**Завдання 6.4.1.** Напівобмежений стержень $x \geq 0$ з теплоізольованою бічною поверхнею рівномірно нагріто до температури $T_0$. Починаючи з моменту $t = 0$, кінець стержня підтримується при нульовій температурі. Опишіть розподіл температури в стержні в довільний момент часу $t > 0$.

*Відповідь*: $v(x,t) = T_0 \Phi\left(\dfrac{x}{\sqrt{4a^2 t}}\right)$.

**Завдання 6.4.2.** Знайдіть розв'язок задачі для однорідного рівняння дифузії на невід'ємній півосі, який задовольняє початкову умову (6.62) та крайову умову $v_x(0,t) = 0$.

*Відповідь.* Розв'язок дається формулою (6.67) із функцією Гріна

$$G_0(x,x';t) = G(|x-x'|;t) + G(|x+x'|;t) = \frac{1}{\sqrt{4\pi Dt}}\left[e^{-\frac{(x-x')^2}{4Dt}} + e^{-\frac{(x+x')^2}{4Dt}}\right]. \quad (6.72)$$

Знайдемо тепер розв'язок більш загальної задачі, що складається з рівняння (6.60), крайової умови[1]

$$v_x(0,t) - hv(0,t) = 0, \quad 0 \leq h < \infty, \quad (6.73)$$

та початкової умови (6.62). Для побудови відповідної функції Гріна $G_h(x,x';t)$ ми вже не можемо скористатися методом продовження, але, з огляду на отримані ним результати, подамо її у вигляді

$$G_h(x,x';t) = G(x-x';t) + G(x+x';t) - \int_0^\infty G(x+x'+u;t)\varphi(u)du, \quad (6.74)$$

де $\varphi(x)$ — якась функція, яка є обмеженою й неперервно диференційовною на півосі $[0,\infty)$. Для будь-якої такої функції вираз у правій

---

[1] Нагадаємо, що для рівняння теплопровідності крайова умова (6.73) є наслідком *закону Ньютона* $\left[-\kappa\dfrac{\partial T}{\partial x} + \alpha(T - T_0)\right]\Bigg|_{x=0} = 0$, який описує теплообмін між середовищем $x > 0$ і термостатом при температурі $T_0$ через контактну поверхню $x = 0$; тут $\kappa$ — коефіцієнт теплопровідності середовища, $\alpha$ — емпіричний коефіцієнт, який характеризує теплопроникність контактної поверхні. У граничному випадку $h \equiv \alpha/\kappa \to 0$ приходимо до умови на теплоізольованій поверхні: $T_x(0,t) = 0$. Другий граничний випадок $h \to \infty$ відповідає ситуації, коли контактна поверхня підтримується при заданій температурі: $T(0,t) = T_0$.

При наявності у крайових умовах неоднорідних членів $(T_0 \neq 0)$ розв'язок лінійної крайової задачі для рівняння теплопровідності шукаємо у вигляді $T(x,t) = u(x,t) + \chi(x,t)$, де функцію $\chi(x,t)$ підбираємо таким чином, щоб вона задовольняла задані неоднорідні крайові умови (для наведених вище умов достатньо покласти $\chi(x,t) = T_0$).

частині (6.74) задовольняє при фіксованих значеннях $x' > 0$ і $t > 0$ рівняння (6.60) й умову

$$\lim_{t\downarrow 0} G_h(x,x';t) = \lim_{t\downarrow 0} G(x-x';t) = \delta(x-x'). \tag{6.75}$$

Залишається так підібрати функцію $\varphi(x)$, щоб при довільних $x' > 0$ і $t > 0$ функція (6.74) задовольняла умову (6.73). Оскільки

$$G_h(0,x';t) = G(-x';t) + G(x';t) - \int_0^\infty G(x'+u;t)\varphi(u)\,du =$$

$$= 2G(x';t) - \int_0^\infty G(x'+u;t)\varphi(u)\,du \tag{6.76}$$

та

$$\left.\frac{\partial G_h(x,x';t)}{\partial x}\right|_{x=0} =$$

$$= \left.\left[\frac{\partial G(x-x';t)}{\partial x} + \frac{\partial G(x+x';t)}{\partial x} - \int_0^\infty \frac{\partial G(x+x'+u;t)}{\partial x}\varphi(u)\,du\right]\right|_{x=0} =$$

$$= -\int_0^\infty \frac{\partial G(x'+u;t)}{\partial u}\varphi(u)\,du = \varphi(0)G(x';t) + \int_0^\infty G(x'+u;t)\varphi'(u)\,du, \tag{6.77}$$

то після підстановки виразів (6.76) і (6.77) у ліву частину (6.73) дістаємо рівність

$$\left[\varphi(0) - 2h\right]G(x';t) + \int_0^\infty G(x'+u;t)\left[\varphi'(u) + h\varphi(u)\right]du = 0.$$

Вона справджується при довільних $x' > 0$ і $t > 0$ лише за умов

$$\varphi'(u) + h\varphi(u) = 0, \quad \varphi(0) = 2h.$$

Звідси знаходимо, що

---

Після цього для функції $u(x,t)$ формулюємо та розв'язуємо лінійну крайову задачу з уже однорідними крайовими умовами та, взагалі кажучи, переозначеними початковими умовами й функцією джерел.

Для задач дифузії аналогом закону Ньютона виступає умова $\left.\left[-D\frac{\partial n}{\partial x} + \beta\left(n - n_0\right)\right]\right|_{x=0} = 0$, що описує процес дифузії частинок домішки між середовищем, у якому їх концентрація дається функцією $n(x,t)$, і резервуаром, де вона має певне значення $n_0$, через поверхню розділу $x = 0$. У цій умові $D$ — коефіцієнт дифузії частинок домішки в середовищі, $\beta$ — емпіричний коефіцієнт, який характеризує проникність поверхні розділу по відношенню до цих частинок. Відповідно, тепер $h \equiv \beta/D$, граничний випадок $h \to 0$ відповідає умові $n_x(0,t) = 0$ на поверхні, що є непроникною для частинок домішки, а граничний випадок $h \to \infty$ відповідає ситуації, коли на поверхні розділу їх концентрація підтримується на значенні $n(0,t) = n_0$.

$$\varphi(u) = 2he^{-hu}$$

і, отже,

$$G_h(x,x';t) = G(x-x';t) + G(x+x';t) - 2h\int_0^\infty G(x+x'+u;t)e^{-hu}\,du. \qquad (6.78)$$

**Завдання 6.4.3.** Покажіть, що у граничних випадках $h \to 0$ та $h \to \infty$ функція Гріна (6.78) переходить, відповідно, у функції Гріна (6.72) та (6.68).

Безпосередня перевірка показує, що функції

$$v_k(x,t) = Ce^{-a^2k^2t}X_k(x), \qquad (6.79)$$

де

$$X_k(x) = \sqrt{\frac{2}{\pi}}\frac{1}{\sqrt{1+\dfrac{h^2}{k^2}}}\left(\cos kx + \frac{h}{k}\sin kx\right),\ \ 0 \le k < \infty, \qquad (6.80)$$

є частинними розв'язками рівняння (6.60) при $x > 0,\ t > 0,$ які задовольняють крайову умову (6.73). Вибір множника перед дужками в (6.80) стане зрозумілим з подальшого розгляду (див. доведення теореми 6.4.1); зокрема, він забезпечує виконання при $h \ge 0$ *умови повноти*

$$\int_0^\infty X_k(x)X_k(x')\,dk = \delta(x-x') \qquad (6.81)$$

та *узагальненої умови ортогональності*

$$\int_0^\infty X_k(x)X_{k'}(x)\,dx = \delta(k-k'). \qquad (6.82)$$

**Теорема 6.4.1.** Для будь-якої початкової функції $u_0(x),$ інтегровної на півосі $(0,x),$ розв'язок задачі (6.60), (6.73), (6.62) при $t > 0$ допускає зображення

$$v(x,t) = \int_0^\infty e^{-a^2k^2t}X_k(x)\hat{u}_0(k)\,dk, \qquad (6.83)$$

де

$$\hat{u}_0(k) = \int_0^\infty X_k(x)u_0(x)\,dx. \qquad (6.84)$$

При цьому в усіх точках неперервності функції $u_0(x)$ справджується співвідношення

$$\lim_{t\downarrow 0} v(x,t)=u_0(x). \qquad (6.85)$$

*Доведення.* Скориставшись виразом

$$G(x,t)=\frac{1}{2\pi}\int_{-\infty}^{\infty} e^{ikx}e^{-a^2k^2t}dk=\frac{1}{\pi}\int_0^{\infty}\cos kx\, e^{-a^2k^2t}dk$$

для функції Гріна на осі (див. підрозділ 6.3) та формулами

$$\int_0^{\infty}\cos ku\, e^{-hu}du=\frac{h}{h^2+k^2},\quad \int_0^{\infty}\sin ku\, e^{-hu}du=\frac{k}{h^2+k^2},$$

перетворимо функцію $G_h(x,x';t)$ наступним чином:

$$G_h(x,x';t)=G(x-x';t)+G(x+x';t)-2h\int_0^{\infty}G(x+x'+u;t)e^{-hu}\,du=$$

$$=\frac{1}{\pi}\int_0^{\infty}dk\, e^{-a^2k^2t}\left[\cos k(x-x')+\cos k(x+x')-2h\int_0^{\infty}\cos k(x+x'+u)e^{-hu}du\right]=$$

$$=\frac{1}{\pi}\int_0^{\infty}dk\, e^{-a^2k^2t}\left[2\cos kx\cos kx'-2h\cos k(x+x')\int_0^{\infty}\cos ku\, e^{-hu}du+\right.$$

$$\left.+2h\sin k(x+x')\int_0^{\infty}\sin ku\, e^{-hu}du\right]=$$

$$=\frac{2}{\pi}\int_0^{\infty}dk\, e^{-a^2k^2t}\left(\cos kx\cos kx'-\frac{h^2}{k^2+h^2}\cos kx\cos kx'+\right.$$

$$\left.+\frac{h^2}{k^2+h^2}\sin kx\sin kx'+\frac{hk}{k^2+h^2}\sin kx\cos kx'+\frac{hk}{k^2+h^2}\cos kx\sin kx'\right)=$$

$$=\frac{2}{\pi}\int_0^{\infty}dk\frac{e^{-a^2k^2t}}{1+\frac{h^2}{k^2}}\left(\cos kx+\frac{h}{k}\sin kx\right)\left(\cos kx'+\frac{h}{k}\sin kx'\right).$$

Згадавши означення (6.80), бачимо, що функцію $G_h(x,x';t)$ можна подати у вигляді

$$G_h(x,x';t)=\int_0^{\infty}e^{-a^2k^2t}X_k(x)X_k(x')dk. \qquad (6.86)$$

Звідси для інтегровної початкової функції $u_0(x)$ і $t>0$ маємо:

$$v(x,t)=\int_0^{\infty}G_h(x,x';t)u_0(x')dx'=$$

$$=\int_0^\infty\left(\int_0^\infty e^{-a^2k^2t}X_k(x)X_k(x')\,dk\right)u_0(x')\,dx'=\int_0^\infty e^{-a^2k^2t}X_k(x)\hat{u}_0(k)\,dk. \quad (6.87)$$

Останнє твердження теореми випливає із співвідношення (6.75) для функції Гріна $G_h(x,x';t)$ та відомої властивості функції Гріна $G(x,x';t)$ задачі на осі, згідно з якою для будь-якої інтегровної функції на осі, зокрема, функції $u_0(x)$, відмінної від нуля лише на півосі $(0,\infty)$, в усіх точках неперервності $u_0(x)$ виконується співвідношення

$$\lim_{t\downarrow 0}\int_{-\infty}^{\infty}G(x-x';t)u_0(x')\,dx'=\lim_{t\downarrow 0}\int_0^\infty G(x-x';t)u_0(x')\,dx'=u_0(x).$$

Інтегральне перетворення (6.84) визначено на множині функцій $u_0(x)$, принаймні інтегровних на півосі $(0,\infty)$. З теореми 6.4.1 випливає, що для цього перетворення справджується формула обернення, згідно з якою в точках неперервності інтегровної функції $u_0(x)$ виконується співвідношення

$$u_0(x)=\lim_{t\downarrow 0}\int_0^\infty e^{-a^2k^2t}X_k(x)\hat{u}_0(k)\,dk. \quad (6.88)$$

У випадку, коли функція $\hat{u}_0(k)$ виявляється інтегровною, формула обернення для *узагальненого* перетворення Фур'є, визначеного формулою (6.84), зводиться до застосування до $\hat{u}_0(k)$ перетворення, аналогічного (6.84):

$$u_0(x)=\int_0^\infty X_k(x)\hat{u}_0(k)\,dk. \quad (6.89)$$

Зауважимо, що формули (6.84) і (6.89) для, відповідно, узагальненого перетворення Фур'є та оберненого перетворення еквівалентні співвідношенням (6.81) і (6.82).

Зазначимо, що узагальнене перетворення Фур'є $\hat{u}_0(k)$ інтегровної функції $u_0(x)$ само є інтегровною та інтегровною з квадратом функцією, якщо, наприклад, $u_0(x)$ — двічі диференційовна в кожній точці, має інтегровні першу й другу похідні та задовольняє умову

$$u_0'(0)-hu_0(0)=0.$$

Справді, функції $X_k(x)$ задовольняють рівняння

$$-X_k''(x)=k^2X_k(x),\quad k>0, \quad (6.90)$$

і крайову умову (6.73). На підставі цього маємо:

$$\hat{u}_0(k)=\int_0^\infty X_k(x)u_0(x)\,dx=-\frac{1}{k^2}\int_0^\infty X_k''(x)u_0(x)\,dx=$$

$$=-\frac{1}{k^2}\int_0^\infty\left\{\frac{d}{dx}\left[X_k'(x)u_0(x)\right]-X_k'(x)u_0'(x)\right\}dx=\frac{1}{k^2}\left[X_k'(0)u_0(0)-X_k(0)u_0'(0)\right]-$$

$$-\frac{1}{k^2}\int_0^\infty X_k(x)u_0''(x)\,dx=-\frac{1}{k^2}\int_0^\infty X_k(x)u_0''(x)\,dx.$$

У силу інтегровності функції $u_0''(x)$ останній інтеграл є обмеженою неперервною функцією $k$, яка до того ж прямує до нуля при $k\to\infty$ (згідно з теоремою Рімана — Лебега). Отже,

$$\hat{u}_0(k)\underset{k\to\infty}{=}-\frac{1}{k^2}o(1),$$

тобто функція $\hat{u}_0(k)$ — інтегровна та інтегровна з квадратом.

Для вказаних функцій $u_0(x)$ і $\hat{u}_0(k)$, з огляду на формулу (6.88), дістаємо:

$$\int_0^\infty\left|u_0(x)\right|^2dx=\lim_{t\downarrow 0}\int_0^\infty dx\,\overline{u_0(x)}\int_0^\infty dk\,e^{-a^2k^2t}X_k(x)\hat{u}_0(k)=$$

$$=\lim_{t\downarrow 0}\int_0^\infty dk\,e^{-a^2k^2t}\left(\int_0^\infty dx\,\overline{u_0(x)}X_k(x)\right)\hat{u}_0(k)=\lim_{t\downarrow 0}\int_0^\infty\left|\hat{u}_0(k)\right|^2e^{-a^2k^2t}dk=\int_0^\infty\left|\hat{u}_0(k)\right|^2dk.$$

Рівність

$$\int_0^\infty\left|u_0(x)\right|^2dx=\int_0^\infty\left|\hat{u}_0(k)\right|^2dk \tag{6.91}$$

називається *рівністю Парсеваля*. Ми вивели її при певних обмеженнях на функції $u_0(x)$ та $\hat{u}_0(k)$, які насправді можна значно послабити. Зокрема, можна показати, що якщо функція $u_0(x)$ — інтегровна з квадратом на півосі $(0,\infty)$, то тоді майже скрізь на півосі $(0,\infty)$ існує і її узагальнене Фур'є-перетворення $\hat{u}_0(k)$, яке є інтегровним з квадратом, і при цьому $u_0(x)$ та $\hat{u}_0(k)$ задовольняють рівність (6.91).

**Завдання 6.4.4.** Запишіть перетворення (6.84) та формули обернення (6.88), (6.89) для граничних випадків $h=0$ і $h\to\infty$ (відповідними формулами означаються так звані *косинус- і синус-перетворення Фур'є*).

Позначимо, як звичайно, через $L_2(0,\infty)$ множину функцій $f(x)$, квадратично інтегровних на півосі $(0,\infty)$:

$$\|f\|^2 \equiv \int_0^\infty |f(x)|^2\, dx < \infty.$$

Не торкаючись суто технічних питань збіжності, на підставі попередніх результатів можемо стверджувати, що будь-яка така функція допускає зображення

$$f(x) = \int_0^\infty X_k(x)\widehat{f}(k)\,dk,$$

де

$$\widehat{f}(k) = \int_0^\infty X_k(x) f(x)\,dx.$$

Серед функцій $f \in L_2(0,\infty)$ виділимо підмножину функцій, що задовольняють умову

$$\int_0^\infty (1+k^4)\left|\widehat{f}(k)\right|^2 dk < \infty, \tag{6.92}$$

та означимо на ній оператор $D_{2,h}$, який діє за формулою

$$\left(D_{2,h} f\right)(x) = \int_0^\infty X_k(x) k^2 \widehat{f}(k)\,dk. \tag{6.93}$$

Якщо інтегровна з квадратом функція $k^2\widehat{f}(k)$ є ще й просто інтегровною на півосі $(0,\infty)$, то з рівняння (6.90) випливає, що

$$\left(D_{2,h} f\right)(x) = -\frac{\partial^2}{\partial x^2} f(x).$$

При цьому

$$\frac{\partial}{\partial x} f(x) = \int_0^\infty X_k'(x)\widehat{f}(k)\,dk,$$

а тому

$$f'(0) - hf(0) = \int_0^\infty \left[X_k'(0) - hX_k(0)\right]\widehat{f}(k)\,dk = 0.$$

Спираючись на ці факти, можемо довести, що оператор $D_{2,h}$ збігається з оператором другої похідної $-\partial^2/\partial x^2$ на множині абсолютно неперервних функцій $f(x)$ із $L_2(0,\infty)$, що мають абсолютно неперервну першу похідну та задовольняють умови

$$\int_0^\infty |f'(x)|^2\,dx<\infty,\quad \int_0^\infty |f''(x)|^2\,dx<\infty,$$

$$f'(0)-hf(0)=0.$$

Покажемо, що спектр оператора $D_{2,h}$ збігається з піввіссю $[0,\infty)$ комплексної площини і є чисто неперервним, тобто жодна точка півосі $[0,\infty)$ не є власним значенням оператора $D_{2,h}$.

Дійсно, припустимо спершу, що комплексне число $\lambda=\mu+i\tau$, $\tau\neq 0$, належить спектру оператора $D_{2,h}$. Тоді, за означенням, існує така послідовність нормованих функцій $g_n(x)$ ($\|g_n\|^2=1$) з області визначення $D_{2,h}$, що

$$\lim_{n\to\infty}\int_0^\infty \left|\left(D_{2,h}g_n\right)(x)-\lambda g_n(x)\right|^2 dx=0. \qquad (6.94)$$

З другого боку, з формул (6.92), (6.93) та рівності Парсеваля (6.91) випливає, що

$$\int_0^\infty \left|\left(D_{2,h}g_n\right)(x)-\lambda g_n(x)\right|^2 dx=\int_0^\infty \left|k^2\hat{g}_n(k)-\lambda\hat{g}_n(k)\right|^2 dk=$$

$$=\int_0^\infty \left[(k^2-\mu)^2+\tau^2\right]\left|\hat{g}_n(k)\right|^2 dk\geq \tau^2\int_0^\infty \left|\hat{g}_n(k)\right|^2 dk=\tau^2\int_0^\infty |g_n(x)|^2\,dx=\tau^2>0,$$

а це суперечить співвідношенню (6.94).

Далі припустимо, що спектру оператора $D_{2,h}$ належить дійсне число $\lambda<0$. Скориставшись співвідношенням (6.94) для відповідної послідовності нормованих функцій $g_n(x)$ з області визначення $D_{2,h}$, формулами (6.92) і (6.93), рівністю Парсеваля (6.91) та нерівністю Коші — Буняковського, знову приходимо до суперечності:

$$0<|\lambda|=-\lambda\int_0^\infty |g_n(x)|^2\,dx=\int_0^\infty (-\lambda)\left|\hat{g}_n(k)\right|^2 dk\leq\int_0^\infty (k^2-\lambda)\left|\hat{g}_n(k)\right|^2 dk\leq$$

$$\leq\left\{\int_0^\infty (k^2-\lambda)^2\left|\hat{g}_n(k)\right|^2 dk\right\}^{1/2}\left\{\int_0^\infty \left|\hat{g}_n(k)\right|^2 dk\right\}^{1/2}\underset{n\to\infty}{\to}0.$$

Нехай тепер $\lambda=k_0^2\geq 0$ — невід’ємне число. Для будь-якого такого $\lambda$ можна ввести, наприклад, послідовність функцій

$$g_n(x)=\int_{k_0}^{\sqrt{k_0^2+\frac{\delta^2}{n}}}\left[\sqrt{k_0^2+\frac{\delta^2}{n}}-k_0\right]^{-\frac{1}{2}} X_k(x)\,dk,$$

які є оберненими перетвореннями Фур'є кусково-неперервних функцій

$$\hat{g}_n(k)=\begin{cases}\left[\sqrt{k_0^2+\dfrac{\delta^2}{n}}-k_0\right]^{-\frac{1}{2}}, & \text{якщо } k\in\left(k_0,\sqrt{k_0^2+\dfrac{\delta^2}{n}}\right),\\ 0, & \text{якщо } k\notin\left(k_0,\sqrt{k_0^2+\dfrac{\delta^2}{n}}\right),\end{cases}$$

заданих на півосі $k\in[0,\infty)$. Функції $g_n(x)$ нормовані, що легко бачити з рівності Парсеваля

$$\int_0^\infty |g_n(x)|^2\,dx=\int_0^\infty \left|\hat{g}_n(k)\right|^2 dk=\frac{1}{\sqrt{k_0^2+\dfrac{\delta^2}{n}}-k_0}\int_{k_0}^{\sqrt{k_0^2+\frac{\delta^2}{n}}} dk=1,$$

та задовольняють умову (6.92). Знову скориставшись рівністю Парсеваля (6.91), для послідовності цих функцій $g_n(x)$ дістаємо:

$$\int_0^\infty\left|\left(D_{2,h}g_n\right)(x)-k_0^2g_n(x)\right|^2dx=\frac{1}{\sqrt{k_0^2+\dfrac{\delta^2}{n}}-k_0}\int_{k_0}^{\sqrt{k_0^2+\frac{\delta^2}{n}}}\left(k^2-k_0^2\right)^2dk\le$$

$$\le\frac{1}{\sqrt{k_0^2+\dfrac{\delta^2}{n}}-k_0}\int_{k_0}^{\sqrt{k_0^2+\frac{\delta^2}{n}}}\left(k_0^2+\frac{\delta^2}{n}-k_0^2\right)^2dk=\frac{\delta^4}{n^2}.$$

Звідси бачимо, що $k_0^2$ належить спектру оператора $D_{2,h}$. При цьому $k_0^2\ge 0$ не може бути власним значенням оператора $D_{2,h}$, оскільки тоді б для відповідної ненульової власної функції $g_{k_0^2}(x)$ ми би мали:

$$\int_0^\infty\left|\left(D_{2,h}\,g_{k_0^2}\right)(x)-k_0^2\,g_{k_0^2}(x)\right|^2dx=\int_0^\infty\left(k^2-k_0^2\right)^2\left|\hat{g}_{k_0^2}(k)\right|^2dk=0.$$

Виходило б, що функція $\left(k^2-k_0^2\right)\hat{g}_{k_0^2}(k)$ майже скрізь дорівнює нулю, а тому $\hat{g}_{k_0^2}(k)=0$ (майже скрізь) і, отже, $g_{k_0^2}(x)=0$ (майже скрізь).

## 6.5. НЕОДНОРІДНЕ РІВНЯННЯ ТЕПЛОПРОВІДНОСТІ НА ПІВОСІ

Тепер розглянемо задачу Коші для неоднорідного рівняння теплопровідності (дифузії) на півосі, яку запишемо у вигляді

$$\frac{\partial w}{\partial t} = a^2 \frac{\partial^2 w}{\partial x^2} + f(x,t), \quad 0 < x,\ t < \infty, \tag{6.95}$$

$$w_x(0,t) - hw(0,t) = 0, \quad 0 \le h < \infty, \tag{6.96}$$

$$w(x,0) = 0. \tag{6.97}$$

Функцію $f(x,t)$ вважаємо обмеженою в області $x \ge 0,\ \ t \ge 0,$ а початкову функцію вибираємо, без обмеження загальності, рівною нулю.

Розв’язуємо задачу, скориставшись принципом Дюамеля. А саме, уведемо функцію $\varphi(x,t \,|\, \tau),$ яка для фіксованого значення параметра $\tau$ задовольняє однорідне рівняння

$$\frac{\partial \varphi}{\partial t} = a^2 \frac{\partial^2 \varphi}{\partial x^2}, \quad 0 < x < \infty, \ \ t > \tau, \tag{6.98}$$

крайову умову

$$\varphi_x(0,t \,|\, \tau) - h\varphi(0,t \,|\, \tau) = 0 \tag{6.99}$$

та початкову умову

$$\varphi(x,t \,|\, \tau)\Big|_{t=\tau} = f(x,\tau). \tag{6.100}$$

Далі будуємо функцію

$$w(x,t) = \int_0^t \varphi(x,t \,|\, \tau)\, d\tau. \tag{6.101}$$

Вона і є розв’язком задачі (6.95)–(6.97).

Справді, з огляду на формули (6.98), (6.100) і (6.101) маємо:

$$\frac{\partial w(x,t)}{\partial t} = \varphi(x,t \,|\, \tau)\Big|_{\tau=t} + \int_0^t \frac{\partial \varphi(x,t \,|\, \tau)}{\partial t} d\tau = f(x,t) + \int_0^t \frac{\partial \varphi(x,t \,|\, \tau)}{\partial t} d\tau,$$

$$\frac{\partial^2 w(x,t)}{\partial x^2} = \int_0^t \frac{\partial^2 \varphi(x,t \,|\, \tau)}{\partial x^2} d\tau,$$

$$\frac{\partial w(x,t)}{\partial t} - a^2 \frac{\partial^2 w(x,t)}{\partial x^2} = f(x,t) + \int_0^t \left[ \frac{\partial \varphi(x,t \,|\, \tau)}{\partial t} - a^2 \frac{\partial^2 \varphi(x,t \,|\, \tau)}{\partial x^2} \right] d\tau = f(x,t).$$

Отже, рівняння (6.95) справджується. З формул (6.99) і (6.101) бачимо, що

$$w_x(0,t) - hw(0,t) = \int_0^t \left[\varphi_x(0,t\,|\,\tau) - h\varphi(0,t\,|\,\tau)\right] d\tau = 0,$$

тобто виконується й крайова умова (6.96). І, нарешті, з формули (6.101) при $t = 0$ приходимо до початкової умови (6.97):

$$w(x,0) = \int_0^0 \varphi(x,0\,|\,\tau)\, d\tau = 0.$$

Залишається виписати функцію $\varphi(x,t\,|\,\tau)$ в явному вигляді. Для цього порівнюємо задачі (6.60), (6.73), (6.62) та (6.98)–(6.100). Крім початкових функцій, вони відрізняються вибором початків відліку часу, які дорівнюють відповідно $t = 0$ і $t = \tau$. Тому в першій задачі аргумент $t$ має зміст проміжку часу, що пройшов з початкового моменту, а в другій — означає, що з початкового моменту пройшов проміжок часу $t - \tau$. Звідси випливає, що значення функції $\varphi(x,t\,|\,\tau)$ (розв'язку задачі (6.98)–(6.100)) у момент часу $t$ збігається із значенням функції (6.87) (розв'язку задачі (6.60), (6.73), (6.62)) у момент часу $t - \tau$, записаної для початкової функції $f(x,\tau)$:

$$\varphi(x,t\,|\,\tau) = v(x,t-\tau)\,\left\{\text{із заміною } u_0(x') \to f(x',\tau)\right\} =$$

$$= \int_0^\infty G_h(x,x';t-\tau)\, f(x',\tau)\, dx'. \qquad (6.102)$$

Підставивши вираз (6.102) у формулу (6.101), дістаємо шуканий розв'язок задачі (6.95)–(6.97):

$$w(x,t) = \int_0^t d\tau \int_0^\infty dx'\, G_h(x,x';t-\tau)\, f(x',\tau), \qquad (6.103)$$

де функція $G_h(x,x';t)$ дається формулою (6.86).

**Завдання 6.5.1.** Доведіть формулу (6.103) методом узагальненого перетворення Фур'є.

**Завдання 6.5.2.** Методом продовження знайдіть розв'язок задачі Коші для неоднорідного рівняння дифузії на невід'ємній півосі, який задовольняє початкову умову (6.97) та крайову умову: а) $w(0,t) = 0$; б) $w_x(0,t) = 0$.

У загальному випадку задача теплопровідності (дифузії) на півосі $0 \le x < \infty$ складається як з неоднорідного рівняння (6.95), так і з ненульової початкової умови (6.62). При виконанні крайової умови (6.73), (6.96) вона є лінійною, тому редукується на задачі (6.60), (6.73), (6.62) і (6.95)–(6.97), а її розв'язок є суперпозицією розв'язків цих двох задач:

$$u(x,t)=\int_0^\infty dx'\,G_h(x,x';t)u_0(x')+\int_0^t d\tau\int_0^\infty dx'\,G_h(x,x';t-\tau)\,f(x',\tau). \qquad (6.104)$$

У граничних випадках $h\to\infty$ та $h\to 0$ звідси отримуємо розв'язки для задач теплопровідності (дифузії) на півосі з крайовими умовами відповідно $u(0,t)=0$ та $u_x(0,t)=0$. Функції Гріна для цих задач даються формулами (6.68) та (6.72).

## 6.6. РІВНЯННЯ ТЕПЛОПРОВІДНОСТІ НА ВІДРІЗКУ

Перейдемо до розгляду *крайових задач для рівняння теплопровідності (дифузії) на відрізку* $[0,l]$ дійсної осі. Такі задачі описують, наприклад, процеси теплопровідності у скінченних тонких стержнях та процеси дифузії частинок домішки в тонких пробірках. Якщо скористатися методом відокремлення змінних, то їх аналіз значною мірою схожий на аналіз крайових задач про коливання струни, однорідної чи неоднорідної. Основна відмінність між розв'язками цих задач полягає в характері їх часової залежності.

Щоб продемонструвати сказане, розглянемо процес теплопровідності в однорідному стержні $[0,l]$, бічна поверхня якого теплоізольована, а на кінцях відбувається теплообмін за законом Ньютона з навколишнім середовищем, що має нульову температуру. Відповідна крайова задача записується у вигляді

$$\frac{\partial u}{\partial t}=a^2\frac{\partial^2 u}{\partial x^2}+f(x,t),\ \ 0<x<l,\ \ t>0, \qquad (6.105)$$

$$u_x(0,t)-h_1u(0,t)=0,\ \ u_x(l,t)+h_2u(l,t)=0,\ \ t\ge 0, \qquad (6.106)$$

$$u(x,0)=u_0(x),\ \ 0\le x\le l, \qquad (6.107)$$

де сталі $h_1,h_2\ge 0$, а функція $u_0(x)$ має зміст початкової температури точок стержня; для простоти, уважатимемо її кусково-неперервною на $[0,l]$.

Розв'язок задачі (6.105)–(6.107) дорівнює сумі

$$u(x,t)=v(x,t)+w(x,t), \qquad (6.108)$$

де $v(x,t)$ — розв'язок крайової задачі для однорідного рівняння теплопровідності з крайовими умовами (6.106) і початковою умовою (6.107), а $w(x,t)$ — розв'язок крайової задачі для неоднорідного рівняння теплопровідності з тими самими крайовими умовами та нульовою початковою умовою.

Розв'язуємо першу задачу:

$$\frac{\partial v}{\partial t} = a^2 \frac{\partial^2 v}{\partial x^2}, \quad 0 < x < l, \quad t > 0, \tag{6.109}$$

$$v_x(0,t) - h_1 v(0,t) = 0, \quad v_x(l,t) + h_2 v(l,t) = 0, \quad t \geq 0, \tag{6.110}$$

$$v(x,0) = u_0(x), \quad 0 \leq x \leq l. \tag{6.111}$$

Згідно із загальною схемою відокремлення змінних, частинні розв'язки рівняння (6.109) шукаємо у вигляді добутку $v(x,t) = T(t)X(x)$. Підставивши його спершу в рівняння (6.109), дістаємо співвідношення

$$\frac{T'(t)}{a^2 T(t)} = \frac{X''(x)}{X(x)}.$$

Воно справджується для довільних $x$ і $t$ за умови, що обидві його частини дорівнюють деякій сталій $-\lambda$. Звідси випливає, що функції $T(t)$ та $X(x)$ задовольняють звичайні диференціальні рівняння

$$T'(t) + \lambda a^2 T(t) = 0, \quad t > 0, \tag{6.112}$$

та

$$X''(x) + \lambda X(x) = 0, \quad 0 < x < l. \tag{6.113}$$

Підставляючи далі частинні розв'язки $v(x,t) = T(t)X(x)$ у крайові умови (6.110), знаходимо крайові умови для функції $X(x)$:

$$X'(0) - h_1 X(0) = 0, \quad X'(l) + h_2 X(l) = 0. \tag{6.114}$$

Отже, для координатної частини шуканого розв'язку $v(x,t)$ дістаємо крайову задачу Штурма — Ліувілля (6.113), (6.114). Вона повністю збігається з відповідною крайовою задачею Штурма — Ліувілля (4.210), (4.211) для коливань струни з пружно закріпленими кінцями, а тому, як було показано в підрозділі 4.10, за умови, що значення параметра $\lambda$ збігаються з певними — власними — значеннями $\lambda_n$, що даються формулою (4.225), існують її нетривіальні нормовані розв'язки $X_n(x)$, які описуються формулами (4.226) і (4.233).

Загальний розв'язок рівняння (6.112) при $\lambda = \lambda_n$ дається виразом

$$T_n(t) = A_n e^{-\lambda_n a^2 t}, \qquad (6.115)$$

де $A_n$ — стала. Нагадаємо, що власні значення $\lambda_n$ — дійсні та невід'ємні, а відповідні розв'язки (власні функції) $X_n(x)$ — попарно ортогональні.

Таким чином, нетривіальні частинні розв'язки рівняння (6.109), що задовольняють крайові умови (6.110), описуються виразом

$$T_n(x)X_n(x) = A_n e^{-\lambda_n a^2 t} X_n(x) = A_n e^{-\lambda_n a^2 t} C_n \left( \cos\frac{\xi_n x}{l} + \frac{h_1 l}{\xi_n}\sin\frac{\xi_n x}{l} \right), \qquad (6.116)$$

де $\lambda_n = \xi_n{}^2 / l^2$, $\xi_n$ — невід'ємні дійсні корені рівняння (4.220). За аналогією із задачею про коливання обмеженої струни, розв'язок крайової задачі (6.109)–(6.111) можна подати у вигляді суперпозиції розв'язків (6.116):

$$v(x,t) = \sum_{n=1}^{\infty} a_n e^{-\lambda_n a^2 t} X_n(x). \qquad (6.117)$$

Коефіцієнти $a_n$ у цій формулі визначаються з початкової умови (6.111):

$$v(x,0) = \sum_{n=1}^{\infty} a_n X_n(x) = u_0(x),$$

звідки

$$a_n = \int_0^l u_0(x') X_n(x') dx'. \qquad (6.118)$$

Ряд (6.117) має прозорий фізичний зміст. Кожний його член описує певну *теплову моду*, тобто такий розподіл температури по довжині стержня, який у будь-який момент часу має вигляд стоячої хвилі з профілем, що визначається функцією $X_n(x)$, та з довжиною $l_n = 2\pi / \sqrt{\lambda_n}$. Цей розподіл формується внаслідок накладання теплових потоків, що йдуть до та від країв стержня при заданих крайових умовах. Але, на відміну від обмежених пружних систем, у яких при збудженні певної гармоніки всі точки коливаються (при відсутності тертя) за гармонічним законом зі сталими амплітудами, амплітудні значення температури в тепловій моді спадають в усіх внутрішніх точках стержня за експоненціальним законом $a_n e^{-t/\tau_n}$, де $\tau_n = 1/(a^2\lambda_n)$ — характерний час релаксації цієї теплової моди.

Довільний розподіл температури в стержні в будь-який момент часу дорівнює суперпозиції (6.117) теплових мод, що можуть виникати в ньому. Оскільки час $\tau_n$ швидко спадає з номером моди $n$, то фактично вже через проміжок часу порядку кількох $\tau_n$ $n$-та теплова мода та моди з вищими номерами згасають і перестають впливати на еволюцію температури стержня.

**Зауваження 6.6.1.** Якщо кінці стержня теплоізольовані, то найменшому власному значенню $\lambda_0 = 0$ відповідає нормована власна функція $X_0 = 1/\sqrt{l}$. У цьому спеціальному випадку існує частинний розв'язок рівняння теплопровідності виду $T_0 X_0 = A_0/\sqrt{l}$, який має зміст рівноважного значення температури, що встановлюється в стержні з часом.

Аналогічну інтерпретацію допускають і частинні розв'язки виду (6.116) крайової задачі для однорідного рівняння дифузії. А саме, вони описують *дифузійні моди*, що характеризують просторовий розподіл та часову еволюцію концентрації частинок домішки. Характерні довжини дифузійних мод $l_n = 2\pi/\sqrt{\lambda_n}$, характерні часи релаксації $\tau_n = 1/(D\lambda_n)$. Якщо краї системи непроникні для частинок домішки, то останні в кінцевому підсумку розподіляються рівномірно по всій такій замкненій системі.

З огляду на формулу (6.118), розв'язок (6.117) можна подати у вигляді

$$v(x,t) = \int_0^l G(x,x';t)u_0(x')\,dx', \tag{6.119}$$

де

$$G(x,x';t) = \sum_{n=1}^{\infty} e^{-\lambda_n a^2 t} X_n(x) X_n(x'), \; t > 0, \tag{6.120}$$

*— часова функція Гріна рівняння теплопровідності (6.109) на відрізку при крайових умовах (6.110).*

Як і розв'язок (6.117), функція Гріна (6.120) є при $t > 0$ неперервною та неперервно диференційовною довільну кількість разів за своїми аргументами. Це випливає з того факту, що ряд (6.120) та ряди, отримувані з нього почленним диференціюванням за змінними $x$ та $t$, є рівномірно збіжними, оскільки мажоруються абсолютно збіжними (при $t > 0$) числовими рядами. Справді, для власних значень задачі (6.113), (6.114) справджується оцінка $c_1(n-1)^2 \le \lambda_n \le c_2 n^2$, де сталі $c_1, c_2 > 0$ (див. формулу (5.58)). Згідно ж із формулами (4.226) і (4.233), відповідні власні функції $X_n(x)$ та їх похідні $X_n^{(k)}(x)$, $k \geq 1$, рівномірно обмежені на $[0,l]$:

$$\left|X_n(x)\right| < M < \infty, \quad \left|X_n^{(k)}(x)\right| < M\lambda_n^{k/2}.$$

Тому

$$\left|\frac{\partial^{k+l}G(x,x';t)}{\partial x^k \partial t^l}\right| \leq \sum_{n=1}^{\infty}\left|\frac{\partial^l e^{-\lambda_n a^2 t}}{\partial t^l}\right|\left|X_n^{(k)}(x)X_n(x')\right| \leq$$

$$\leq M^2 a^{2l}\sum_{n=1}^{\infty}\lambda_n^{l+k/2}e^{-\lambda_n a^2 t} \leq M^2 a^{2l}c_2^{l+k/2}\sum_{n=1}^{\infty}n^{2l+k}\, e^{-c_1(n-1)^2 a^2 t}.$$

Оскільки загальний член останнього ряду $b_n = n^{2l+k}\, e^{-c_1(n-1)^2 a^2 t}$, то при $t > 0$ маємо:

$$\lim_{n\to\infty}\frac{b_{n+1}}{b_n} = \lim_{n\to\infty}\left(1+\frac{1}{n}\right)^{2l+k} e^{-c_1(2n-1)a^2 t} = 0.$$

Отже, за ознакою Д'Аламбера, цей ряд збігається.

Аналогічним способом переконуємося, що ряд (6.120) можна також почленно інтегрувати, тобто функція Гріна (6.120) інтегровна за своїми змінними.

**Завдання 6.6.1.** Необмежену плоскопаралельну однорідну пластину товщиною $l$ нагріто до температури $T$. Починаючи з моменту часу $t=0$, поверхні пластини підтримуються при нульовій температурі. Знайдіть розподіл температури у пластині в довільний момент часу. Оцініть проміжок часу, за який середня температура пластини впаде на порядок.

*Розв'язання.* Напрямимо вісь $OX$ перпендикулярно до поверхонь пластини. Для визначення температури у пластині маємо крайову задачу виду (6.109)–(6.111) із крайовими умовами $v(0,t) = v(l,t) = 0$ та початковою умовою $v(x,0) = T$. Застосувавши метод відокремлення змінних, для координатної частини дістаємо крайову задачу Штурма — Ліувілля $X'' + \lambda X = 0$, $X(0) = X(l) = 0$, звідки знаходимо власні значення $\lambda_n = \dfrac{\pi^2 n^2}{l^2}$ та нормовані власні функції $X_n(x) = \sqrt{\dfrac{2}{l}}\sin\dfrac{\pi n x}{l}$, $n = 1,2,\ldots$. Для часової частини маємо рівняння (6.112), розв'язок якого дається формулою (6.115). Розподіл температури у пластині описується рядом (6.117) з коефіцієнтами (див. формулу (6.118))

$$a_n=\int_0^l TX_n(x)\,dx=T\sqrt{\frac{2}{l}}\frac{l}{\pi n}\left[1-(-1)^n\right]=$$

$$=\begin{cases}0, & \text{якщо } n-\text{парне, тобто } n=2m,\\ 2T\sqrt{\frac{2}{l}}\frac{l}{\pi n}, & \text{якщо } n-\text{непарне, } n=2m+1,\end{cases}$$

де $m=0,1,2,\ldots.$

Отже, температура пластини в перерізі $x$ у момент часу $t$

$$v(x,t)=\sum_{n=1}^{\infty}a_n e^{-\lambda_n a^2 t}X_n(x)=\frac{4T}{\pi}\sum_{m=0}^{\infty}\frac{1}{(2m+1)}e^{-\frac{\pi^2(2m+1)^2a^2}{l^2}t}\sin\frac{\pi(2m+1)x}{l},$$

її середня температура

$$\bar{v}(t)=\frac{1}{l}\int_0^l v(x,t)\,dx=\frac{8T}{\pi^2}\sum_{m=0}^{\infty}\frac{1}{(2m+1)^2}e^{-\frac{\pi^2(2m+1)^2a^2}{l^2}t}.$$

Для оцінок достатньо обмежитися першим членом $(m=0)$ останнього ряду. Проміжок часу $\Delta t,$ за який він зменшиться на порядок, знаходимо з умови, що показник експоненти при $t=\Delta t$ дорівнює приблизно 2,3 $(e^{-2,3}\approx 0{,}10).$ Звідси $\Delta t\approx 2{,}3l^2/(\pi^2a^2)\approx 0{,}23l^2/a^2.$

**Завдання 6.6.2.** У лівій частині, що має довжину $h,$ закритої горизонтальної хімічної пробірки довжиною $l$ міститься розчин деякої речовини; концентрація речовини в ньому $c_0.$ Починаючи з моменту $t=0,$ розчинена речовина дифундує в розчинник, яким перед цим заповнюють усю праву частину пробірки. Опишіть процес вирівнювання концентрації речовини в пробірці. Кінці пробірки непроникні для речовини та розчинника.

*Розв'язання*. Для концентрації речовини маємо крайову задачу виду (6.109)–(6.111) із крайовими умовами $v_x(0,t)=v_x(l,t)=0$ і початковою умовою $v(x,0)=\begin{cases}c_0, & 0<x<h,\\ 0, & h<x<l.\end{cases}$ Власні значення та нормовані власні функції відповідної крайової задачі Штурма — Ліувілля наступні:

$$\lambda_0=0,\ \ X_0(x)=\frac{1}{\sqrt{l}},$$

$$\lambda_n=\frac{\pi^2n^2}{l^2},\ \ X_n(x)=\sqrt{\frac{2}{l}}\cos\frac{\pi nx}{l},\ \ n=1,2,\ldots.$$

Шуканий розв'язок дається рядом

$$v(x,t)=a_0X_0+\sum_{n=1}^{\infty}a_ne^{-\lambda_nDt}X_n(x)$$

із коефіцієнтами

$$a_0=\int_0^l v(x,0)X_0(x)dx=\int_0^h c_0\frac{1}{\sqrt{l}}dx+\int_h^l 0\frac{1}{\sqrt{l}}dx=\frac{c_0h}{\sqrt{l}},$$

$$a_n=\int_0^l v(x,0)X_n(x)dx=\int_0^h c_0\sqrt{\frac{2}{l}}\cos\frac{\pi nx}{l}dx=c_0\sqrt{\frac{2}{l}}\frac{l}{\pi n}\sin\frac{\pi nh}{l},\quad n=1,2,\dots .$$

Остаточно знаходимо:

$$v(x,t)=c_0\left[\frac{h}{l}+\frac{2}{\pi}\sum_{n=1}^{\infty}\frac{1}{n}\sin\frac{\pi nh}{l}e^{-\frac{\pi^2n^2D}{l^2}t}\cos\frac{\pi nx}{l}\right].$$

При $t\to\infty$ (а практично вже через проміжок часу $\Delta t\approx 4l^2/(\pi^2D)\approx$ $\approx 0{,}4l^2/D$) речовина рівномірно розподілиться по об'єму пробірки; її рівноважна концентрація становитиме $c_0h/l$.

Перейдемо до розгляду крайової задачі для неоднорідного рівняння теплопровідності:

$$\frac{\partial w}{\partial t}=a^2\frac{\partial^2w}{\partial x^2}+f(x,t),\quad 0<x<l,\ \ t>0, \tag{6.121}$$

$$w_x(0,t)-h_1w(0,t)=0,\ \ w_x(l,t)+h_2w(l,t)=0,\ \ t\ge 0, \tag{6.122}$$

$$w(x,0)=0,\ \ 0\le x\le l. \tag{6.123}$$

Замість того, щоб безпосереднього скористатися розв'язком (6.117) (чи (6.119)) та методом Дюамеля, розв'язуємо її методом відокремлення змінних, щоб продемонструвати деталі застосування останнього до неоднорідних крайових задач.

Як уже зазначалося, розподіл температури в стержні в довільний момент часу можна уявляти собі як лінійну суперпозицію теплових мод, що можуть виникати в ньому при заданих крайових умовах. Якщо цей розподіл є результатом еволюції температурного стану, створеного в стержні в початковий момент часу, то вагові коефіцієнти в цій суперпозиції збігаються з коефіцієнтами розкладу початкової функції в ряд Фур'є за власними функціями відповідної крайової задачі. Оскільки власні функції системи визначаються лише її фізичними властивостями та крайовими умовами, то природно припустити, що розподіл температури в тому ж стержні, генерований джерелами тепла, теж можна подати у вигляді лінійної суперпозиції тих самих теплових мод, але з ваговими коефіцієнтами, що визначаються густиною потужності джерел.

Отже, шукаємо розв'язок крайової задачі (6.121)–(6.123) для неоднорідного рівняння теплопровідності у вигляді ряду Фур'є за власними функціями $X_n(x)$ крайової задачі (6.109)–(6.111):

$$w(x,t) = \sum_n \theta_n(t) X_n(x). \tag{6.124}$$

Додатково припустимо, що функцію джерел (стоків) $f(x,t) = F(x,t)/(c\rho)$ теж можна розвинути в ряд Фур'є за тією самою системою функцій $X_n(x)$:

$$f(x,t) = \sum_n f_n(t) X_n(x), \tag{6.125}$$

де коефіцієнти Фур'є для функції джерел

$$f_n(t) = \int_0^l f(x',t) X_n(x') dx'. \tag{6.126}$$

Підставляючи ряди (6.124) та (6.125) у рівняння (6.121), отримуємо рівність

$$\sum_n \left[ \dot{\theta}_n(t) X_n(x) - a^2 \theta_n(t) X_n''(x) - f_n(t) X_n(x) \right] = 0.$$

Згідно з рівнянням (6.113) $X_n''(x) = -\lambda_n X_n(x)$, тому далі маємо:

$$\sum_n \left[ \dot{\theta}_n(t) + \lambda_n a^2 \theta_n(t) - f_n(t) \right] X_n(x) = 0.$$

Звідси, скориставшись повнотою та ортонормованістю системи функцій $X_n(x)$, для функцій $\theta_n(t)$ дістаємо звичайне неоднорідне диференціальне рівняння

$$\dot{\theta}_n(t) + \lambda_n a^2 \theta_n(t) = f_n(t), \quad t > 0. \tag{6.127}$$

Тепер підставляємо ряд (6.124) у крайові умови (6.122) та початкову умову (6.123). Перші справджуються автоматично, оскільки власні функції $X_n(x)$ задовольняють ті самі крайові умови (див. формули (6.114)):

$$\sum_n \theta_n(t) \left[ X_n'(0) - h_1 X_n(0) \right] = 0, \quad \sum_n \theta_n(t) \left[ X_n'(l) + h_2 X_n(l) \right] = 0.$$

Друга дає співвідношення

$$\sum_n \theta_n(0) X_n(x) = 0,$$

звідки знаходимо початкові умови для функцій $\theta_n(t)$, якими треба доповнити рівняння (6.127):

$$\theta_n(0) = 0. \qquad (6.128)$$

Скориставшись методом варіації довільної сталої, нетривіальний розв'язок рівняння (6.127) шукаємо у вигляді

$$\theta_n(t) = A_n(t) e^{-\lambda_n a^2 t}.$$

Підставивши цей вираз у (6.127), для похідної функції $A_n(t)$ дістаємо:

$$\frac{dA_n(t)}{dt} = e^{\lambda_n a^2 t} f_n(t).$$

Звідси

$$A_n(t) = \int_0^t e^{\lambda_n a^2 \tau} f_n(\tau) d\tau + B_n,$$

де в силу умов (6.128) сталі інтегрування $B_n = 0$. Остаточно маємо:

$$\theta_n(t) = \left( \int_0^t e^{\lambda_n a^2 \tau} f_n(\tau) d\tau \right) e^{-\lambda_n a^2 t}. \qquad (6.129)$$

Таким чином, розв'язок крайової задачі (6.121)–(6.123) дається рядом

$$w(x,t) = \sum_n \left( \int_0^t e^{\lambda_n a^2 \tau} f_n(\tau) d\tau \right) e^{-\lambda_n a^2 t} X_n(x). \qquad (6.130)$$

Згадавши вираз (6.126) для коефіцієнтів $f_n(\tau)$, можемо подати його через функцію Гріна (6.120) у вигляді квадратури:

$$w(x,t) = \int_0^t d\tau \int_0^l dx' G(x,x';t-\tau) f(x',\tau). \qquad (6.131)$$

**Завдання 6.6.3.** Знайдіть температуру, що встановлюється в тонкому однорідному стержні $0 \le x \le l$ з теплоізольованою бічною поверхнею під дією періодичного джерела тепла з густиною потужності $F(x,t) = F_0(x)\sin(\omega t + \varphi(x))$. Кінці стержня підтримуються при нульовій температурі.

*Розв'язання.* Увівши комплекснозначну функцію $\tilde{F}(x,t) = \tilde{F}_0(x) e^{-i\omega t}$, $\tilde{F}_0(x) = F_0(x) e^{-i\varphi(x)}$, та взявши до уваги формули (6.120) і (6.131), шукану температуру знаходимо за формулою $w(x,t) = -\operatorname{Im} \tilde{w}(x,t)$, де

$$\tilde{w}(x,t)=\lim_{t\to\infty}\int_0^t d\tau\int_0^l dx'\,G(x,x';t-\tau)\frac{\tilde{F}(x',\tau)}{c\rho}=\tilde{A}(x)e^{-i\omega t},$$

$$\tilde{A}_\omega(x)=\int_0^l dx'\,G_\omega(x,x')\,\tilde{F}_0(x'),$$

$$G_\omega(x,x')=\frac{2}{c\rho l}\sum_{n=1}^{\infty}\frac{1}{\lambda_n a^2-i\omega}\sin\frac{\pi nx}{l}\sin\frac{\pi nx'}{l},\ \ \lambda_n=\frac{\pi^2 n^2}{l^2}.$$

Функція $\tilde{w}(x,t)$ має зміст усталеної *комплекснозначної температури*, що встановлюється в стержні під дією теплового джерела з густиною потужності $\tilde{F}(x,t)=\tilde{F}_0(x)e^{-i\omega t}$ і коливається з частотою $\omega$ джерела та амплітудою $\tilde{A}_\omega(x)$. Остання задовольняє неоднорідне диференціальне рівняння

$$-\frac{d^2y(x)}{dx^2}-\frac{i\omega}{a^2}y(x)=\frac{1}{c\rho a^2}\tilde{F}_0(x),\qquad 0<x<l,$$

та крайові умови

$$y(0)=y(l)=0.$$

Функція $G_\omega(x,x')$ є спеціальним неперервним розв'язком цієї задачі при $\tilde{F}_0(x)=\delta(x-x')$:

$$-\frac{\partial^2G_\omega(x,x')}{\partial x^2}-\frac{i\omega}{a^2}G_\omega(x,x')=\frac{1}{c\rho a^2}\delta(x-x'),\quad 0<x,x'<l,$$

$$G_\omega(0,x')=0,\ \ G_\omega(l,x')=0.$$

Вона називається *частотною функцією Гріна для рівняння теплопровідності на відрізку* $[0,l]$ (при заданих крайових умовах). Щоб отримати її у скінченному вигляді, повторюємо схему розв'язування задачі (4.183), (4.184): розбивши відрізок $[0,l]$ на два проміжки $0\le x<x'$ і $x'<x\le l,$ знаходимо на них розв'язки однорідного рівняння

$$-\frac{\partial^2G_\omega(x,x')}{\partial x^2}-\frac{i\omega}{a^2}G_\omega(x,x')=0,$$

що задовольняють відповідні крайові умови, та зшиваємо ці розв'язки в точці $x=x'$ за допомогою умов

$$G_\omega(x'-0,x')=G_\omega(x'+0,x'),\ \ G'_\omega(x'-0,x')-G'_\omega(x'+0,x')=\frac{1}{c\rho a^2}.$$

У результаті дістаємо:

$$G_\omega(x,x') = \frac{1}{c\rho a\sqrt{i\omega}\sin\left(\sqrt{i\omega}\frac{l}{a}\right)}\begin{cases}\sin\left(\sqrt{i\omega}\frac{x}{a}\right)\sin\left(\sqrt{i\omega}\frac{l-x'}{a}\right), & 0\le x\le x',\\ \sin\left(\sqrt{i\omega}\frac{l-x}{a}\right)\sin\left(\sqrt{i\omega}\frac{x'}{a}\right), & x'\le x\le l.\end{cases}$$

Варто зазначити, що при $\omega = 0$ функція $G_\omega(x,x')$ особливостей не має:

$$\lim_{\omega\to 0} G_\omega(x,x') = \frac{1}{c\rho a^2 l}\begin{cases}x(l-x'), & 0\le x\le x',\\ (l-x)x', & x'\le x\le l.\end{cases}$$

При достатньо великих частотах $\omega$ у тих внутрішніх точках $x,\ x'$, де справджуються умови

$$\min\left[\sqrt{\frac{\omega}{2}}\frac{x}{a}, \sqrt{\frac{\omega}{2}}\frac{x'}{a}, \sqrt{\frac{\omega}{2}}\frac{l-x}{a}, \sqrt{\frac{\omega}{2}}\frac{l-x'}{a}\right] >> 1,$$

маємо:

$$G_\omega(x,x') = \frac{i}{2c\rho a\sqrt{i\omega}} e^{i\sqrt{i\omega}\frac{|x-x'|}{a}},\ \ \mathrm{Im}\sqrt{i\omega}\ge 0.$$

### *КОНТРОЛЬНІ ПИТАННЯ ДО РОЗДІЛУ 6*

1. *Яка властивість перетворень Фур'є визначає їх роль у задачах для диференціальних рівнянь зі сталими коефіцієнтами на дійсній осі і в необмеженому просторі?*
2. *Як визначається спектр диференціальних операторів математичної фізики? Чи збігається їх спектр із сукупністю їх власних значень?*
3. *Які точки комплексної площини утворюють спектр оператора* $-d^2/dx^2$ *на множині функцій* $y(x)$, *абсолютно неперервних разом зі своїми першими похідними на дійсній осі і таких, що* $\int_{-\infty}^{\infty}\left[|y(x)|^2 + |y''(x)|^2\right]dx < \infty$? *Чи має цей оператор власні значення?*
4. *Нехай* $q(x)$ *— обмежена дійсна функція. Чи може оператор* $-d^2/dx^2 + q(x)$ *на дійсній осі з указаною вище областю визначення мати власні значення з ненульовою уявною частиною?*
5. *Як визначається функція Гріна для рівняння теплопровідності зі сталими коефіцієнтами на всій осі? Який вона має фізичний зміст?*
6. *Як виражається функція Гріна рівняння теплопровідності (дифузії) зі сталими коефіцієнтами на додатній півосі при крайовій умові*

$v_x(0,t) - hv(0,t) = 0, \quad h \geq 0,$ *через функцію Гріна для відповідного рівняння на всій осі?*

7. *Чому дорівнюють обмежені узагальнені власні функції диференціального оператора* $-d^2/dx^2$ *на додатній півосі з указаною вище крайовою умовою?*
8. *Як треба нормувати узагальнені власні функції в попередньому питанні, щоб для узагальненого перетворення Фур'є за цими функціями справджувалася рівність Парсеваля?*
9. *Які точки комплексної площини складають спектр диференціального оператора* $-d^2/dx^2$ *на додатній півосі, визначеного вказаною крайовою умовою в точці* $x = 0$ *на множині функцій, абсолютно неперервних разом зі своїми першими похідними та інтегровних із квадратом разом зі своїми другими похідними?*
10. *Як визначаються теплові (дифузійні) моди для рівняння теплопровідності (дифузії) на обмеженому інтервалі? Який зміст для цих мод мають власні значення асоційованої крайової задачі Штурма — Ліувілля?*

# Розділ 7
# ПРОСТОРОВІ КРАЙОВІ ЗАДАЧІ

## 7.1. ПОПЕРЕЧНІ КОЛИВАННЯ ПРЯМОКУТНОЇ МЕМБРАНИ

Під прямокутною мембраною розумітимемо *натягнену* (що робить її пружинистою) тонку плівку прямокутної форми. Будемо вважати, що в стані рівноваги вона збігається з ділянкою $D:\{0\le x\le l_1,\ 0\le y\le l_2\}$ площини $XOY$; межу цієї ділянки, тобто краї мембрани, позначимо через Г. Під дією зовнішніх факторів (початкової деформації, початкового імпульсу, прикладеної сили) у такій системі можуть збудитися поперечні коливання, при яких точки мембрани зміщуються перпендикулярно до площини $XOY$. Для їх опису введемо функцію $u=u(x,y,t)$, яка має зміст поперечного зміщення точок $(x,y)\in D$ мембрани в момент часу $t$.

Будемо вважати, що мембрана однорідна, а її коливання малі. Тоді функція Лагранжа мембрани дається виразом (див. завдання 3.1.2)

$$L(t)=\iint\limits_D \mathcal{L}(u,u_t,u_x,u_y)\,dxdy, \tag{7.1}$$

де

$$\mathcal{L}=\frac{1}{2}\rho u_t^2(x,y,t)-\frac{1}{2}k\left(u_x^2(x,y,t)+u_y^2(x,y,t)\right)+F(x,y,t)u(x,y,t), \tag{7.1a}$$

$\rho$ — поверхнева густина (маса одиниці площі) мембрани, $k>0$ — коефіцієнт, який характеризує її натяг, $F(x,y,t)$ — поверхнева густина поперечної зовнішньої сили, прикладеної до мембрани. Екстремальне значення функціонала дії мембрани

$$S=\int\limits_{t_1}^{t_2} L(t)\,dt$$

досягається на функції $u(x,y,t)$, що має задовольняти рівняння Ейлера — Остроградського (2.104) для лагранжіана $\mathcal{L}$. Звідси знаходимо рівняння руху мембрани

$$\frac{\partial^2 u}{\partial t^2}=a^2\left(\frac{\partial^2 u}{\partial x^2}+\frac{\partial^2 u}{\partial y^2}\right)+f(x,y,t),\ \ (x,y)\in D,\ \ t>0, \tag{7.2}$$

де $a^2\equiv k/\rho,\ \ f(x,y,t)=F(x,y,t)/\rho$.

Якщо краї мембрани закріплені, то рівняння (7.2) треба доповнити крайовими умовами

$$u(x,y,t)\big|_{\Gamma} = 0. \qquad (7.3)$$

Для прямокутної мембрани ці умови слід деталізувати наступним чином:

$$u(0,y,t)=0,\ \ u(l_1,y,t)=0,\ \ u(x,0,t)=0,\ \ u(x,l_2,t)=0. \qquad (7.3а)$$

Якщо ж деякі з країв мембрани незакріплені і можуть вільно рухатися в напрямі, перпендикулярному до її рівноважної поверхні, то крайові умови на них визначаються як необхідні умови існування екстремуму функціонала дії мембрани у класі функцій з цією властивістю. Ці умови даються формулою (2.105). У випадку лагранжіана (7.1а) для кожного вільного краю мембрани вона набирає вигляду

$$\left.\frac{\partial u}{\partial n}\right|_{\Gamma} = 0, \qquad (7.4)$$

де $n$ указує напрям у площині $XOY$, перпендикулярний краю мембрани. Зокрема, для мембрани, закріпленої лише за край $x=0$, крайові умови мають вигляд

$$u(0,y,t)=0,\ \ u_x(l_1,y,t)=0,\ \ u_y(x,0,t)=0,\ \ u_y(x,l_2,t)=0. \qquad (7.4а)$$

Нарешті, початкові умови для малих поперечних коливань мембрани вибираємо у стандартному вигляді, задавши початкове поперечне зміщення $u_0(x,y)$ та початкову поперечну швидкість $v_0(x,y)$ точок мембрани:

$$u(x,y,0)=u_0(x,y),\ \ u_t(x,y,0)=v_0(x,y). \qquad (7.5)$$

Тепер опишемо алгоритм розв’язування крайових задач про поперечні коливання прямокутних мембран, що базується на методі відокремлення змінних (див. підрозділ 4.10). Почнемо з розгляду вільних коливань прямокутної мембрани із закріпленими краями. Відповідна крайова задача складається з однорідного рівняння (7.2) (з $f(x,y,t)=0$), крайових умов (7.3) та початкових умов (7.5). Її нетривіальні частинні розв’язки шукаємо у вигляді добутку $u(x,y,t)=T(t)\Phi(x,y)$; підставивши його в рівняння (7.2), дістаємо рівність

$$\frac{T''(t)}{a^2T(t)} = \frac{1}{\Phi(x,y)}\left(\frac{\partial^2\Phi(x,y)}{\partial x^2} + \frac{\partial^2\Phi(x,y)}{\partial y^2}\right).$$

Очевидно, що кожна з її сторін дорівнює деякій сталій, яку позначимо через $-\lambda$. Як результат, для часової та координатної частин шуканого розв’язку маємо два рівняння:

$$T''(t) + \lambda a^2 T(t) = 0,\ \ t > 0, \tag{7.6}$$

загальний розв'язок якого

$$T(t) = A\cos\sqrt{\lambda}at + B\sin\sqrt{\lambda}at, \tag{7.7}$$

та

$$\frac{\partial^2 \Phi(x,y)}{\partial x^2} + \frac{\partial^2 \Phi(x,y)}{\partial y^2} + \lambda \Phi(x,y) = 0,\ \ (x,y) \in D. \tag{7.8}$$

Останнє рівняння являє собою *двовимірне рівняння Гельмгольца*. Його треба доповнити крайовими умовами, які отримуємо, підставивши шукані частинні розв'язки в умови (7.3):

$$\Phi(x,y)\big|_{\Gamma} = 0, \tag{7.9}$$

тобто

$$\Phi(0,y) = 0,\ \ \Phi(l_1,y) = 0,\ \ \Phi(x,0) = 0,\ \ \Phi(x,l_2) = 0. \tag{7.9а}$$

Отже, для знаходження координатної функції $\Phi(x,y)$ маємо крайову задачу (7.8), (7.9а). Звернемо увагу, що (7.8) — це рівняння в частинних похідних. Розв'язуємо його, знову застосувавши метод відокремлення змінних. А саме, шукаємо його нетривіальні частинні розв'язки у вигляді $\Phi(x,y) = X(x)Y(y)$, де $X(x)$ і $Y(y)$ — нетривіальні функції. Підставивши цей добуток у рівняння (7.8), дістаємо

$$X''(x)Y(y) + X(x)Y''(y) + \lambda X(x)Y(y) = 0,$$

або, з огляду на нетривіальність $X(x)$ і $Y(y)$,

$$\frac{X''(x)}{X(x)} + \frac{Y''(y)}{Y(y)} + \lambda = 0.$$

Звідси бачимо, що доданки, що залежать від різних змінних, розділяються, якщо покласти

$$\frac{X''(x)}{X(x)} = -\mu,\ \ \frac{Y''(y)}{Y(y)} = -\nu,\ \ \lambda = \mu + \nu,$$

де $\mu$ і $\nu$ — сталі.

Таким чином, функції $X(x)$ і $Y(y)$ задовольняють звичайні диференціальні рівняння

$$X''(x) + \mu X(x) = 0,\ \ 0 < x < l_1, \tag{7.10}$$

$$Y''(y) + \nu Y(y) = 0,\ \ 0 < y < l_2. \tag{7.11}$$

Підставивши добуток $\Phi(x,y) = X(x)Y(y)$ у крайові умови (7.9а), знаходимо крайові умови для цих рівнянь:

$$X(0)=0,\ \ X(l_1)=0, \tag{7.10a}$$

$$Y(0)=0,\ \ Y(l_2)=0. \tag{7.11a}$$

Неважко помітити, що для функцій $X(x)$ і $Y(y)$ ми дістали дві одновимірні крайові задачі Штурма — Ліувілля (7.10), (7.10a) і (7.11), (7.11a). Їх власні значення та відповідні нормовані власні функції відомі:

$$\mu_m=\frac{\pi^2 m^2}{l_1^2},\ \ X_m(x)=\sqrt{\frac{2}{l_1}}\sin\frac{\pi m x}{l_1},\ \ m=1,2,...,$$

$$\nu_n=\frac{\pi^2 n^2}{l_2^2},\ \ Y_n(y)=\sqrt{\frac{2}{l_2}}\sin\frac{\pi n y}{l_2},\ \ n=1,2,....$$

Бачимо, що власні значення крайової задачі (7.8), (7.9a) для функції $\Phi(x,y)$ описуються формулою

$$\lambda_{mn}=\mu_m+\nu_n=\frac{\pi^2 m^2}{l_1^2}+\frac{\pi^2 n^2}{l_2^2},\ \ m,n=1,2,..., \tag{7.12}$$

а відповідні власні функції — формулою

$$\Phi_{mn}(x,y)=\frac{2}{\sqrt{l_1 l_2}}\sin\frac{\pi m x}{l_1}\sin\frac{\pi n y}{l_2}. \tag{7.13}$$

Звернемо увагу, що ці величини залежать від двох індексів $m$ і $n$. Для довільних значень цих індексів власні значення $\lambda_{mn}$ додатні, а власні функції $\Phi_{mn}(x,y)$ задовольняють умову ортонормованості

$$\iint\limits_D \Phi_{mn}(x,y)\Phi_{m'n'}(x,y)\,dxdy=\int_0^{l_1}dx\int_0^{l_2}dy\,\Phi_{mn}(x,y)\Phi_{m'n'}(x,y)=\delta_{mm'}\delta_{nn'}. \tag{7.14}$$

Крім того, *система власних функцій* $\Phi_{mn}(x,y)$ *повна*. Її повнота є прямим наслідком повноти систем функцій $X_m(x)$ і $Y_n(y)$.

Зауважимо, що якщо відношення $l_1^2/l_2^2$ не є раціональним числом, то кожному значенню $\lambda_{mn}$ відповідає лише одна (з точністю до числового множника) власна функція $\Phi_{mn}(x,y)$, тобто всі власні значення задачі (7.8), (7.9a) є невиродженими. У супротивному випадку для пари натуральних чисел $(m,n)$ може існувати інша пара натуральних чисел $(m_1,n_1)$ (не обов'язково одна), така, що $\lambda_{mn}=\lambda_{m_1 n_1}$, і тоді власному значенню $\lambda_{mn}$ відповідають принаймні дві ортогональні власні функції $\Phi_{mn}(x,y)$ і $\Phi_{m_1 n_1}(x,y)$. Іншими словами, при раціональному значенні відношення $l_1^2/l_2^2$ серед власних значень задачі (7.8), (7.9a) можуть бути власні значення, кратність виродження яких дорівнює принаймні двом.

**Завдання 7.1.1**. Переконайтеся в тому, що для квадратної мембрани ($l_1 = l_2$) усі власні значення $\lambda_{mn}$ задачі (7.8), (7.9а) з $m \neq n$ принаймні двократно вироджені. Якщо ж натуральні числа $m, k, l$ з $m > k$ утворюють піфагорову трійку, тобто $m^2 + k^2 = l^2$, то власні значення $\lambda_{m-k, m+k}$ вироджені щонайменше тричі.

*Вказівка*. При $m \neq n$ власні функції $\Phi_{mn}(x, y)$ і $\Phi_{nm}(x, y)$ ортогональні.

З огляду на попередні результати часова частина (7.7) шуканих частинних розв'язків набирає вигляду

$$T_{mn}(t) = A_{mn} \cos \omega_{mn} t + B_{mn} \sin \omega_{mn} t, \quad \omega_{mn} = a\sqrt{\lambda_{mn}}, \qquad (7.15)$$

де $A_{mn}$ і $B_{mn}$ — невідомі коефіцієнти, а $\omega_{mn}$ — додатні величини, що мають зміст власних частот коливань мембрани. Самі ж частинні розв'язки однорідного рівняння (7.2), що задовольняють крайові умови (7.3), описуються формулою

$$u_{mn}(x, y, t) = T_{mn}(t)\Phi_{mn}(x, y) = \left(A_{mn} \cos \omega_{mn} t + B_{mn} \sin \omega_{mn} t\right)\Phi_{mn}(x, y). \quad (7.16)$$

Вони мають зміст власних коливань, тобто гармонік прямокутної мембрани. Лінійна суперпозиція власних коливань

$$u(x, y, t) = \sum_{m,n=1}^{\infty} C_{mn} u_{mn}(x, y, t) =$$

$$= \sum_{m,n=1}^{\infty} \left( a_{mn} \cos \omega_{mn} t + \frac{b_{mn}}{\omega_{mn}} \sin \omega_{mn} t \right) \Phi_{mn}(x, y) \qquad (7.17)$$

визначає загальну форму функції, що описує вільні поперечні коливання прямокутної мембрани. Коефіцієнти $a_{mn}$ і $b_{mn}$ у ній обчислюємо за допомогою початкових умов (7.5):

$$u(x, y, 0) = u_0(x, y) = \sum_{m,n=1}^{\infty} a_{mn}\, \Phi_{mn}(x, y),$$

$$u_t(x, y, 0) = v_0(x, y) = \sum_{m,n=1}^{\infty} b_{mn}\, \Phi_{mn}(x, y),$$

звідки, з огляду на умову ортонормованості (7.14),

$$a_{mn} = \int_0^{l_1} dx' \int_0^{l_2} dy'\, u_0(x', y')\Phi_{mn}(x', y'),$$

$$b_{mn} = \int_0^{l_1} dx' \int_0^{l_2} dy'\, v_0(x', y')\Phi_{mn}(x', y'). \qquad (7.18)$$

Тепер розв’яжемо крайову задачу про вимушені коливання прямокутної мембрани із закріпленими краями, спричинені зовнішньою поперечною силою, тобто побудуємо розв’язок неоднорідного рівняння (7.2), який задовольняє крайові умови (7.3а) та, без втрати загальності, початкові умови (7.5) з нульовими початковими функціями. Знову користуючись методом відокремлення змінних, подаємо шуканий розв’язок та функцію $f(x,y,t)$ у вигляді рядів за власними функціями (7.13):

$$u(x,y,t)=\sum_{m,n=1}^{\infty} q_{mn}(t)\Phi_{mn}(x,y), \tag{7.19}$$

$$f(x,y,t)=\sum_{m,n=1}^{\infty} f_{mn}(t)\Phi_{mn}(x,y), \tag{7.20}$$

де

$$f_{mn}(t)=\int_0^{l_1}dx'\int_0^{l_2}dy'\,f(x',y',t)\,\Phi_{mn}(x',y'). \tag{7.21}$$

Повторюючи майже дослівно схему розв’язування завдання 4.10.4, для функцій $q_{mn}(t)$ дістаємо звичайне неоднорідне диференціальне рівняння

$$\ddot{q}_{mn}(t)+\omega_{mn}^2 q_n(t)=f_{mn}(t),\ \ t>0, \tag{7.22}$$

та початкові умови

$$q_{mn}(0)=0,\quad \dot{q}_{mn}(0)=0. \tag{7.23}$$

Звідси знаходимо:

$$q_{mn}(t)=\int_0^t \frac{\sin\omega_{mn}(t-\tau)}{\omega_{mn}} f_{mn}(\tau)\,d\tau. \tag{7.24}$$

Розв’язок крайової задачі (7.2), (7.3), (7.5) дорівнює сумі розв’язків (7.17) і (7.19). Скориставшись виразами (7.18), (7.21) та (7.24), його можна подати у вигляді

$$\begin{aligned} u(x,y,t)&=\frac{\partial}{\partial t}\rho\iint_D dx'dy'\,G(x,x',y,y';t)\,u_0(x',y')\ +\\ &+\rho\iint_D dx'dy'\,G(x,x',y,y';t)\,v_0(x',y')+\\ &+\int_0^t d\tau\iint_D dx'dy'\,G(x,x',y,y';t-\tau)\,F(x',y',\tau), \end{aligned} \tag{7.25}$$

де

$$G(x,x',y,y';t)=\frac{1}{\rho}\sum_{m,n=1}^{\infty}\frac{\sin\omega_{mn}t}{\omega_{mn}}\Phi_{mn}(x,y)\Phi_{mn}(x',y') \qquad (7.26)$$

– *часова функція Гріна для однорідної мембрани*. За своєю структурою та фізичним змістом вона є двовимірним аналогом часової функції Гріна (4.244) для однорідної струни.

**Завдання 7.1.2**. У стані рівноваги прямокутна мембрана шириною $l_1$ і довжиною $l_2$ збігається з ділянкою $D:\{0\le x\le l_1, 0\le y\le l_2\}$ площини $XOY$. Край $x=l_1$ мембрани вільний, решта країв жорстко закріплені. Знайдіть:

а) власні частоти та нормовані власні функції малих поперечних коливань мембрани;

б) зміщення точок мембрани та енергію її окремих гармонік у довільний момент часу, якщо в початковий момент часу всім її точкам надали однакової поперечної швидкості $v_0$;

в) те саме, якщо в початковий момент часу мембрану здеформували і відпустили; мала початкова деформація мембрани описується формулою $u_0(x,y)=\alpha x^2 y(y-l_2)$, $\alpha=\mathrm{const}$;

г) амплітуду усталених вимушених коливань мембрани під дією поперечної періодичної сили $f(x,y,t)=e^{-\varepsilon t}\sin\omega t\,\delta\left(x-\frac{l_1}{2}\right)\delta\left(y-\frac{l_2}{2}\right)$, $\varepsilon>0$, $t>0$, коли $\varepsilon\downarrow 0$.

## 7.2. ВІСЕСИМЕТРИЧНІ КОЛИВАННЯ КРУГЛОЇ МЕМБРАНИ. ФУНКЦІЯ БЕССЕЛЯ

Як випливає з попереднього аналізу, застосування методу відокремлення змінних до задачі про коливання прямокутної мембрани виявляється особливо ефективним завдяки тому факту, що краї мембрани збігаються з координатними лініями декартової системи координат[1]. Така ситуація є характерною й для багатовимірних крайових задач, що формулюються для тіл з формою поверхні, відмінною від прямокутної. Розв'язуючи їх, слід користуватися тими криволінійними координатами, у яких форма координатних поверхонь збігається з формою поверхні тіл. Вибором криволінійних координат, що відображають геометричну симетрію тіл, вдається суттєво спро-стити математичні викладки за умови, що початкові і крайові умови, а також функції джерел (стоків) не змінюються при перетвореннях координат, при яких тіло суміщається з собою. Урахування цієї симетрії може підказати структуру розв'язку та виключити з розгляду несуттєві змінні.

[1] Координатні лінії плоскої декартової системи координат визначаються рівняннями $x=\mathrm{const}$ і $y=\mathrm{const}$.

Продемонструємо сказане спершу на задачі про малі поперечні коливання однорідної круглої мембрани радіусом $R$ із жорстко закріпленим краєм. Ця задача цікава ще й у тому відношенні, що дозволяє торкнутися питання про спеціальні функції та їх використання для побудови координатних частин розв'язків крайових задач.

Нехай у стані рівноваги мембрана збігається з кругом $D:\{x^2+y^2<R^2\}$ площини $XOY$, обмеженим гладким колом $\Gamma:\{x^2+y^2=R^2\}$, а її вісь симетрії $OZ$ проходить через центр $D$ перпендикулярно до площини $XOY$. Якщо користуватися декартовою системою координат, то крайова задача для поперечного (уздовж осі $OZ$) зміщення $u=u(x,y,t)$ точок мембрани в довільний момент часу складається з рівняння коливань

$$u_{tt}=a^2\Delta u+f(x,y,t),\ (x,y)\in D,\ t>0, \tag{7.27}$$

де $\Delta\equiv\frac{\partial^2}{\partial x^2}+\frac{\partial^2}{\partial y^2}$ — оператор Лапласа у плоскій декартовій системі координат, крайової умови

$$u\big|_\Gamma=0 \tag{7.28}$$

та початкових умов

$$u(x,y,0)=u_0(x,y),\ \ u_t(x,y,0)=v_0(x,y). \tag{7.29}$$

Очевидно, що крайова умова (7.28) не дозволяє розділити декартові координати $x$ та $y$ безпосередньо. Тому, з огляду на кругову симетрію мембрани, перейдемо до полярної системи координат $(r,\alpha)$ з початком у центрі мембрани. Перехід здійснюється за формулами

$$x=r\cos\alpha,\ \ y=r\sin\alpha,\ \ 0\le r\le R,\ \ 0\le\alpha<2\pi, \tag{7.30}$$

його якобіан

$$\frac{D(x,y)}{D(r,\alpha)}=\begin{vmatrix}\frac{\partial x}{\partial r} & \frac{\partial x}{\partial\alpha}\\ \frac{\partial y}{\partial r} & \frac{\partial y}{\partial\alpha}\end{vmatrix}=\begin{vmatrix}\cos\alpha & -r\sin\alpha\\ \sin\alpha & r\cos\alpha\end{vmatrix}=r, \tag{7.31}$$

при цьому оператор $\Delta$ набирає вигляду

$$\Delta = \frac{1}{r}\frac{\partial}{\partial r}\left(r\frac{\partial}{\partial r}\right) + \frac{1}{r^2}\frac{\partial^2}{\partial \alpha^2}. \tag{7.32}$$

Відповідно, у полярних координатах рівняння коливань (7.27) для поперечного зміщення $u = u(r,\alpha,t)$ точок мембрани як функції цих координат і часу має вигляд

$$\frac{\partial^2 u}{\partial t^2} = a^2\left[\frac{1}{r}\frac{\partial}{\partial r}\left(r\frac{\partial u}{\partial r}\right) + \frac{1}{r^2}\frac{\partial^2 u}{\partial \alpha^2}\right] + f(r,\alpha,t). \tag{7.33}$$

Крайова умова (7.28) та початкові умови (7.29) записуються наступним чином:

$$u(R,\alpha,t) = 0, \tag{7.34}$$

$$u(r,\alpha,0) = u_0(r,\alpha), \;\; u_t(r,\alpha,0) = v_0(r,\alpha). \tag{7.35}$$

Звернемо увагу, що при $r = 0$ рівняння (7.33), а тому й шукана функція $u(r,\alpha,t)$ можуть мати сингулярність усередині досліджуваної області $D$, тоді як вихідна функція $u(x,y,t)$ жодних особливостей у цій області не мала. Ця обставина пояснюється тим фактом, що перехід від декартової системи координат до полярної є взаємно-однозначним лише за умови, що якобіан (7.31) перетворень (7.30) відмінний від нуля: $r \neq 0$. У точці $r = 0$, де ця умова порушується, указаний перехід не є взаємно-однозначним, а тому функція $u(r,\alpha,t)$ може мати сингулярність. Остання має фізичний зміст лише у спеціальних випадках, коли точка $r = 0$ якось виділена, наприклад, коли саме в ній зосереджено джерела коливань (тепла, частинок тощо). В інших випадках сингулярність має бути усунена за допомогою додаткової вимоги, щоб функція $u(r,\alpha,t)$ залишалася неперервною в точці $r = 0$, де якобіан (7.31) дорівнює нулю. Умову неперервності достатньо замінити вимогою

$$|u(0,\alpha,t)| < \infty. \tag{7.36}$$

Аналогічні співвідношення виникають і при використанні інших криволінійних координат. *Вони відіграють роль додаткових крайових умов, накладених на шукану функцію у відповідних особливих точках або на відповідних кривих.*

Щоб завершити постановку задачі про відшукання функції $u(r,\alpha,t)$, залишається взяти до уваги ще одну обставину. А саме, треба явно врахувати, що $u(r,\alpha,t)$ як функція змінної $\alpha$ повинна, з одного боку, однозначно визначатися значеннями $\alpha$ з проміжку $[0,2\pi)$, а, з другого, залишатися незмінною — інваріантною — при поворотах мембрани навколо своєї осі $OZ$ на довільні кути, кратні $2\pi$. Для цього достатньо накласти на $u(r,\alpha,t)$ *умову періодичності*

$$u(r,\alpha,t) = u(r,\alpha + 2\pi,t). \tag{7.37}$$

Почнемо вивчення загальної задачі (7.33)–(7.37) з окремого випадку вільних $(f(r,\alpha,t) = 0)$ коливань мембрани, коли початкові умови не залежать від полярного кута $\alpha$. Оскільки мембрана однорідна, зовнішні сили відсутні, а крайові та, тепер, початкові умови є інваріантними відносно поворотів навколо осі мембрани $OZ$, тобто для будь-яких двох значень $\alpha$ і $\alpha + \delta\alpha$ задовольняють співвідношення

$$u(r,\alpha,t)\big|_{\Gamma} = u(r,\alpha+\delta\alpha,t)\big|_{\Gamma}, \; u(r,\alpha,0)\big|_{D} = u(r,\alpha+\delta\alpha,0)\big|_{D},$$

$$u_t(r,\alpha,0)\big|_{D} = u_t(r,\alpha+\delta\alpha,0)\big|_{D},$$

то немає жодних фізичних причин для того, щоб залежність зміщення від $\alpha$, яка відсутня при $t = 0$, усе ж таки почала би проявлятися з часом. Робимо висновок, що змінна $\alpha$ є, на відміну від решти змінних, несуттєвою в розглядуваній задачі, а тому миттєві поперечні зміщення точок мембрани описуються функцією лише $r$ і $t$: $u = u(r,t)$. Такі коливання круглої мембрани називаються *вісесиметричними*. У будь-який момент часу мембрана, яка їх здійснює, має форму поверхні обертання.

Оскільки умова періодичності (7.37) для вісесиметричних коливань справджується автоматично, для визначення функції $u = u(r,t)$ маємо наступну крайову задачу:

$$\frac{\partial^2 u}{\partial t^2} = a^2 \frac{1}{r}\frac{\partial}{\partial r}\left(r\frac{\partial u}{\partial r}\right), \; 0 < r < R, \; t > 0, \tag{7.38}$$

$$u(R,t) = 0, \; |u(0,t)| < \infty, \tag{7.39}$$

$$u(r,0) = u_0(r), \; u_t(r,0) = v_0(r). \tag{7.40}$$

Її аналіз почнемо з пошуку нетривіальних частинних розв'язків у вигляді добутку $u(r,t) = \chi(r)T(t)$.

Відокремивши часову і координатну частини, для першої дістаємо рівняння (7.6) та його загальний розв'язок у вигляді (7.7), а для другої — одновимірну крайову задачу

$$\frac{1}{r}\frac{d}{dr}\left(r\frac{d\chi}{dr}\right) + \lambda\chi = 0, \; 0 < r < R, \tag{7.41}$$

$$\chi(R)=0,\ \ |\chi(0)|<\infty. \tag{7.42}$$

Якщо незалежну змінну $r$ перепозначити через $x$, а її значення $R$ — через $l$, то відразу бачимо, що маємо справу з крайовою задачею типу Штурма — Ліувілля на відрізку $[0,l]$, у якій $p(x)=x$, $\rho(x)=x$ і $q(x)\equiv 0$. Очевидно, що умови $p(x)>0$ й $\rho(x)>0$, для яких КЗШЛ досліджувалася в попередніх підрозділах, тепер порушуються при $x=0$. Покажемо, що ця обставина веде до появи сингулярного внеску в загальному розв'язку рівняння (7.41), однак для задачі (7.41), (7.42) не змінює встановлені вище загальні висновки про невід'ємність власних значень і попарну ортогональність власних функцій КЗШЛ.

За допомогою підстановки $r=x/\sqrt{\lambda}$ перепишемо рівняння (7.41) як

$$\chi''(x)+\frac{1}{x}\chi'(x)+\chi(x)=0 \tag{7.43}$$

і, далі, шукаємо його розв'язки у вигляді степеневого ряду

$$\chi(x)=x^{\sigma}\sum_{s=0}^{\infty}a_s x^s=\sum_{s=0}^{\infty}a_s x^{s+\sigma}. \tag{7.44}$$

З умови $|\chi(0)|<\infty$ випливає, що показник $\sigma$ в (7.44) може бути лише невід'ємним.

Оскільки

$$\frac{1}{x}\chi'(x)=\sum_{s=0}^{\infty}a_s(s+\sigma)x^{s+\sigma-2},$$

$$\chi''(x)=\sum_{s=0}^{\infty}a_s(s+\sigma)(s+\sigma-1)x^{s+\sigma-2},$$

то після підстановки цих рядів у рівняння (7.43) маємо:

$$\sum_{s=0}^{\infty}a_s(s+\sigma)^2x^{s+\sigma-2}+\sum_{s=0}^{\infty}a_s x^{s+\sigma}=0. \tag{7.45}$$

Виокремимо з першого ряду зліва члени з $s=0$ і $s=1$, а в решті його членів перейдемо від індексу підсумовування $s\geq 2$ до індексу підсумовування $p\geq 0$ за правилом $s=p+2$. Маємо:

$$\sum_{s=0}^{\infty}a_s(s+\sigma)^2x^{s+\sigma-2}=a_0\sigma^2x^{\sigma-2}+a_1(1+\sigma)^2x^{\sigma-1}+\sum_{p=0}^{\infty}a_{p+2}(p+2+\sigma)^2x^{p+\sigma}.$$

Підставивши цей вираз у рівність (7.45), одержуємо:

$$a_0\sigma^2x^{\sigma-2}+a_1(1+\sigma)^2x^{\sigma-1}+\sum_{s=0}^{\infty}\left[a_{s+2}(s+2+\sigma)^2+a_s\right]x^{s+\sigma}=0.$$

Остання рівність справджується при довільних $x$ лише за умови, що дорівнюють нулю всі коефіцієнти степеневого ряду зліва. Звідси дістаємо:

$$a_0\sigma^2 = 0,$$

$$a_1(1+\sigma)^2 = 0,$$

$$a_{s+2}(s+2+\sigma)^2 + a_s = 0, \quad s = 0,1,2,....$$

Оскільки $\sigma \geq 0,$ то з другого рівняння випливає, що $a_1 = 0.$ Відтак із третього рівняння знаходимо за індукцією, що всі коефіцієнти з непарними номерами дорівнюють нулю,

$$a_{2s+1} = 0,$$

а коефіцієнти з парними номерами виражаються через $a_0$ за формулами

$$a_2 = -\frac{1}{(2+\sigma)^2}a_0, \quad a_4 = \frac{1}{(2+\sigma)^2(4+\sigma)^2}a_0, \ ...,$$

$$a_{2s} = \frac{(-1)^s}{(2+\sigma)^2(4+\sigma)^2...(2s+\sigma)^2}a_0.$$

При цьому з першого рівняння бачимо, що нетривіальний обмежений розв'язок рівняння (7.43) у вигляді збіжного степеневого ряду (7.44) існує лише при $\sigma = 0.$ Отже, шуканий розв'язок має вигляд

$$\chi(x) = a_0 J_0(x), \tag{7.46}$$

де $J_0(x)$ — функція, яка визначається степеневим рядом

$$J_0(x) = \sum_{s=0}^{\infty}\frac{(-1)^s}{(s!)^2}\left(\frac{x}{2}\right)^{2s}. \tag{7.47}$$

Для будь-якого дійсного або комплексного $x,$ що задовольняє умову $|x| \leq M,$ ряд (7.47) мажорується числовим рядом:

$$\sum_{s=0}^{\infty}\frac{(-1)^s}{2^{2s}(s!)^2}x^{2s} \leq \sum_{s=0}^{\infty}\frac{1}{2^{2s}(s!)^2}M^{2s}.$$

Останній, згідно з ознакою Д'Аламбера, є збіжним для будь-якого додатного $M.$ Звідси випливає, що ряд (7.47) збігається рівномірно при довільних $x,$ а його сума $J_0(x)$ є однозначною аналітичною функцією в будь-якій області комплексної площини, тобто цілою функцією. Вона називається *функцією Бесселя нульового порядку*; її графік для невід'ємних дійсних $x$ зображено на рис. 7.1.

Отже, частинний розв'язок рівняння (7.41), який задовольняє другу умову (7.42), має вигляд

$$\chi(r) = a_0 J_0\left(\sqrt{\lambda}r\right). \tag{7.48}$$

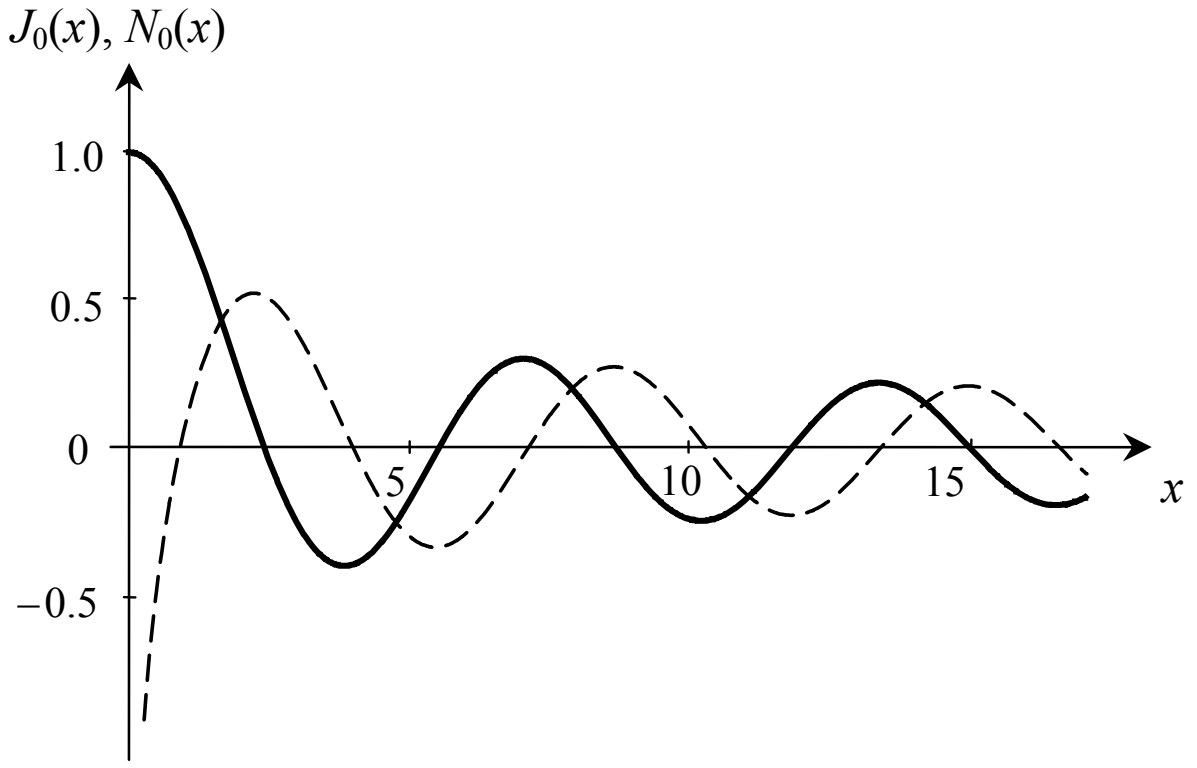

Рис. 7.1. Графіки функцій Бесселя (суцільна лінія) та Неймана (штрихова) нульового порядку для невід'ємних значень аргументу. Перші три нулі цих функцій дорівнюють відповідно 2.405, 5.520, 8.654 та 0.894, 3.958, 7.086

Покажемо, що лише такі частинні розв'язки задовольняють умови (7.42).

Нехай на інтервалі $(a,b]$ відомо нетривіальний розв'язок $u(x)$ рівняння

$$(p(x)y')' + k(x)y = 0 \tag{7.49}$$

з неперервними на $(a,b]$ коефіцієнтами $p(x)$, $k(x)$, при цьому $p(x) \neq 0$ і $u(x) \neq 0$ на $(a,b]$. Тоді, згідно із загальною теорією лінійних диференціальних рівнянь другого порядку, загальний розв'язок рівняння (7.49) на $(a,b)$ має вигляд[1]

$$y(x) = a_0 u(x) + a_1 v(x), \tag{7.50}$$

де $a_0$, $a_1$ — довільні сталі, $u(x)$ та

$$v(x) = u(x)\left[\frac{v(b)}{u(b)} - \int_x^b \frac{dx'}{p(x')u^2(x')}\right] \tag{7.51}$$

– лінійно незалежні розв'язки рівняння (7.49), для яких

$$p(x)\left[u(x)v'(x) - u'(x)v(x)\right] = 1. \tag{7.52}$$

[1] Помножимо рівняння (7.49) на $u(x)$, аналогічне рівняння для $u(x)$ — на $y(x)$, та віднімемо отримані рівності. Дістаємо $u[p(x)y']' - y[p(x)u']' = [p(x)uy' - p(x)u'y]' = 0$, звідки $p(x)uy' - p(x)u'y = a_1$, де стала $a_1 \neq 0$ (у противному разі $y(x)$ та $u(x)$ відрізняються лише числовим множником). Поділивши обидві частини останнього рівняння на $p(x)u^2(x)$, отримуємо $\frac{y'}{u} - \frac{u'}{u^2}y = \frac{a_1}{p(x)u^2(x)}$, тобто $\frac{d}{dx}\left(\frac{y}{u}\right) = \frac{a_1}{p(x)u^2(x)}$, а відтак, після інтегрування, — формули (7.50)–(7.52).

Рівняння (7.43), очевидно, можна подати у вигляді (7.49) з $p(x)=k(x)=x.$ Оскільки, за означенням, $J_0(0)=1,$ виберемо в якості $u(x)$ в околі точки $x=0$ саме функцію $J_0(x).$ Тоді другим лінійно незалежним розв'язком (7.51) при достатньо малих $0<x<b$ виступає функція

$$v(x)=J_0(x)\left[\frac{v(b)}{J_0(b)}-\int_x^b \frac{dx'}{x'J_0^2(x')}\right]=$$

$$=J_0(x)\left[\frac{v(b)}{J_0(b)}+\ln\frac{x}{b}\right]-J_0(x)\int_x^b\left[\frac{1}{J_0^2(x')}-1\right]\frac{dx'}{x'}. \tag{7.53}$$

Оскільки $J_0(x)$ є аналітичною функцією, то в силу умови $J_0(0)=1$ такою ж є й функція

$$\left[\frac{1}{J_0^2(x)}-1\right]\frac{1}{x},$$

а тому і другий доданок у правій частині формули (7.53). Скориставшись цим результатом та порівнявши формули (7.53) і (7.50), бачимо, що в околі точки $x=0$ загальний розв'язок рівняння (7.43) допускає зображення

$$\chi(x)=\varphi(x)+\psi(x)\ln x, \tag{7.54}$$

де $\varphi(x)$ і $\psi(x)$ — аналітичні функції[1]. При цьому розв'язки, що обмежені в околі нуля $(\psi(x)=0),$ мають вигляд (7.46).

Вимагаючи тепер, щоб розв'язок (7.48) задовольняв першу крайову умову (7.42), дістаємо рівняння

$$a_0 J_0\left(\sqrt{\lambda}R\right)=0. \tag{7.55}$$

Бачимо, що нетривіальні розв'язки задачі (7.41), (7.42) існують лише при тих (власних) значеннях параметра $\lambda_k$, для яких числа $\alpha_k=\sqrt{\lambda_k}R$

---

[1] Загальний розв'язок диференціального рівняння Бесселя $\chi''(x)+\frac{1}{x}\chi'(x)+\chi(x)=0$ ($x$ — будь-яке число) часто записують у вигляді лінійної комбінації $\chi(x)=a_0J_0(x)+$ $+b_0N_0(x),$ де

$$N_0(x)=\frac{2}{\pi}J_0(x)\left(\ln\frac{x}{2}+\gamma\right)-\frac{2}{\pi}\sum_{s=1}^{\infty}\frac{(-1)^s}{(s!)^2}\left(\frac{x}{2}\right)^{2s}\left(1+\frac{1}{2}+\frac{1}{3}+\ldots+\frac{1}{s}\right)$$

– частинний розв'язок рівняння Бесселя, який називається *функцією Неймана нульового порядку*, $\gamma=\lim_{s\to\infty}\left(1+\frac{1}{2}+\frac{1}{3}+\ldots+\frac{1}{s}-\ln s\right)\approx 0{,}57722$ — стала Ейлера. Графік функції $N_0(x)$ для додатних значень аргументу зображено на рис. 7.1.

є нулями функції $J_0(x)$. Доведемо, що всі ці нулі дійсні, а відповідні власні значення $\lambda_k$ — додатні.

Спершу згадаємо, що $J_0(x)$ — ціла функція комплексної змінної $x$. Як уже вказувалося (див. підрозділ 1.5), така функція може мати лише ізольовані нулі, що утворюють послідовність точок $\{\alpha_k\}$ з єдиною точкою скупчення на нескінченності. Нехай $\alpha_k = \sqrt{\lambda_k} R$ і $\alpha_n = \sqrt{\lambda_n} R$ — два довільні комплекснозначні нулі $J_0(x)$. Маємо такі рівняння та умови:

$$\frac{1}{r}\frac{d}{dr}\left[r\frac{dJ_0\left(\sqrt{\lambda_k}\, r\right)}{dr}\right] + \lambda_k J_0\left(\sqrt{\lambda_k}\, r\right) = 0, \quad 0 < r < R, \tag{7.56}$$

$$\frac{1}{r}\frac{d}{dr}\left[r\frac{dJ_0\left(\sqrt{\lambda_n}\, r\right)}{dr}\right] + \lambda_n J_0\left(\sqrt{\lambda_n}\, r\right) = 0, \quad 0 < r < R, \tag{7.57}$$

$$J_0\left(\sqrt{\lambda_k} R\right) = 0, \quad J_0\left(\sqrt{\lambda_n} R\right) = 0. \tag{7.58}$$

Помножимо рівняння (7.56) на функцію $J_0\left(\sqrt{\lambda_n}\, r\right)$, рівняння (7.57) — на функцію $J_0\left(\sqrt{\lambda_k}\, r\right)$, віднімемо отримані рівності та зінтегруємо результат з вагою $r$ за змінною $r$ у межах від 0 до $R$. Дістаємо:

$$\int_0^R \left\{ J_0\left(\sqrt{\lambda_n}\, r\right)\frac{d}{dr}\left[r\frac{dJ_0\left(\sqrt{\lambda_k}\, r\right)}{dr}\right] - J_0\left(\sqrt{\lambda_k}\, r\right)\frac{d}{dr}\left[r\frac{dJ_0\left(\sqrt{\lambda_n}\, r\right)}{dr}\right]\right\} dr =$$

$$= \left(\lambda_n - \lambda_k\right)\int_0^R rJ_0\left(\sqrt{\lambda_n}\, r\right)J_0\left(\sqrt{\lambda_k}\, r\right)dr. \tag{7.59}$$

Обчисливши перший інтеграл частинами та скориставшись умовами (7.58), переконуємося, що він дорівнює нулю. Маємо:

$$\left(\lambda_n - \lambda_k\right)\int_0^R rJ_0\left(\sqrt{\lambda_n}\, r\right)J_0\left(\sqrt{\lambda_k}\, r\right)dr = 0, \tag{7.60}$$

або, перейшовши до $\alpha_n$ і $\alpha_k$,

$$\left(\alpha_n^2 - \alpha_k^2\right)\int_0^R rJ_0\left(\frac{\alpha_n}{R} r\right)J_0\left(\frac{\alpha_k}{R} r\right)dr = 0. \tag{7.61}$$

При $k \neq n$ формули (7.60) і (7.61) дають співвідношення

$$\int_0^R rJ_0\left(\frac{\alpha_n}{R}r\right)J_0\left(\frac{\alpha_k}{R}r\right)dr = 0, \text{ якщо } \alpha_n \neq \alpha_k. \tag{7.62}$$

З іншого боку, ураховуючи, що згідно з означенням (7.47) $\overline{J_0(x)} = J_0(\overline{x})$, для двох комплексно спряжених коренів $\alpha_k$ і $\alpha_n = \overline{\alpha}_k$ із співвідношення (7.61) дістаємо:

$$\left(\overline{\alpha}_k^2 - \alpha_k^2\right)\int_0^R rJ_0\left(\frac{\overline{\alpha}_k}{R}r\right)J_0\left(\frac{\alpha_k}{R}r\right)dr = \left(\overline{\alpha}_k^2 - \alpha_k^2\right)\int_0^R r\left|J_0\left(\frac{\alpha_k}{R}r\right)\right|^2 dr = 0.$$

Звідси

$$\overline{\alpha}_k^2 - \alpha_k^2 = 0,$$

тобто

$$\overline{\alpha}_k = \alpha_k \text{ або } \overline{\alpha}_k = -\alpha_k.$$

Бачимо, що нулі функції $J_0(x)$ можуть бути лише дійсними або уявними. Однак для уявних чисел $\alpha_k = \pm i|\alpha_k|$ ряд

$$J_0\left(\pm i|\alpha_k|\right) = \sum_{s=0}^{\infty}\frac{(-1)^s}{2^{2s}(s!)^2}\left(\pm i|\alpha_k|\right)^{2s} = \sum_{s=0}^{\infty}\frac{1}{2^{2s}(s!)^2}|\alpha_k|^{2s}$$

є сумою лише додатних чисел, а тому не дорівнює нулю. Отже, усі нулі функції $J_0(x)$ є дійсними, а всі власні значення $\lambda_k = \alpha_k^2/R^2$, $k = 1,2,3,...,$ задачі (7.41), (7.42) — додатними. Їм відповідають власні функції (див. (7.48))

$$\chi_k(r) = a_{0k}J_0\left(\frac{\alpha_k}{R}r\right), \quad k = 1,2,3,..., \tag{7.63}$$

де в силу парності функції $J_0(x)$ під $\alpha_k$ слід розуміти лише її додатні нулі, розташовані в порядку зростання.

Щоб знайти нормувальний множник $a_{0k}$, повернімося до формули (7.59). Розглядаючи параметр $\lambda_n$ у ній уже не як власне значення, а як деяке число, що прямує до $\lambda_k$ і для якого $J_0\left(\sqrt{\lambda_n}R\right) \neq 0$, після інтегрування частинами її лівої частини дістаємо співвідношення

$$R\sqrt{\lambda_k}J_0\left(\sqrt{\lambda_n}R\right)J_0'\left(\sqrt{\lambda_k}R\right) = \left(\lambda_n - \lambda_k\right)\int_0^R rJ_0\left(\sqrt{\lambda_k}\,r\right)J_0\left(\sqrt{\lambda_n}\,r\right)dr, \tag{7.64}$$

де штрих означає похідну функції Бесселя за аргументом. Шуканий інтеграл знаходимо як границю

$$\int_0^R r\left[J_0\left(\sqrt{\lambda_k}\,r\right)\right]^2 dr = \lim_{\lambda_n \to \lambda_k}\frac{R\sqrt{\lambda_k}J_0\left(\sqrt{\lambda_n}R\right)J_0'\left(\sqrt{\lambda_k}R\right)}{\lambda_n - \lambda_k}.$$

Розкриваючи невизначеність типу 0/0 у правій частині за правилом Лопіталя, маємо:

$$\int_0^R r\left[J_0\left(\sqrt{\lambda_k}\, r\right)\right]^2 dr = \frac{R^2}{2}\left[J_0'\left(\sqrt{\lambda_k} R\right)\right]^2 = \frac{R^2}{2}\left[J_0'(\alpha_k)\right]^2. \quad (7.65)$$

Звідси та з умови нормування

$$\int_0^R r\left[\chi_k(r)\right]^2 dr = 1 \quad (7.66)$$

знаходимо:

$$a_{0k} = \frac{\sqrt{2}}{R\left|J_0'(\alpha_k)\right|}, \ k = 1,2,3,.... \quad (7.67)$$

**Зауваження 7.2.1.** У вираз (7.63) для власних функцій $\chi_k(r)$ можна для зручності внести додатковий множник $1/\sqrt{2\pi}$. Тоді умова ортонормованості для нових власних функцій

$$\Phi_k(r) = \frac{1}{\sqrt{2\pi}}\chi_k(r) \quad (7.68)$$

записується через подвійний інтеграл по всій області $D$, зайнятій вільною мембраною у стані рівноваги:

$$\int_D \Phi_k(r)\Phi_{k'}(r)dS = \int_0^R dr r\int_0^{2\pi} d\alpha\, \Phi_k(r)\Phi_{k'}(r) = \delta_{kk'}. \quad (7.69)$$

Скориставшись здобутими власними функціями крайової задачі (7.41), (7.42) та загальним розв'язком (7.7) для часової частини $T_k(t)$, записуємо частинні і загальний розв'язки задачі (7.38)–(7.40) у вигляді

$$u_k(r,t) = \left(A_k \cos\omega_k t + B_k \sin\omega_k t\right)\Phi_k(r), \quad (7.70)$$

$$u(r,t) = \sum_{k=1}^{\infty} c_k T_k(t)\Phi_k(r) = \sum_{k=1}^{\infty}\left(\tilde{A}_k \cos\omega_k t + \tilde{B}_k \sin\omega_k t\right)\Phi_k(r), \quad (7.71)$$

де

$$\omega_k = \frac{\alpha_k a}{R}, \ k = 1,2,3,..., \quad (7.72)$$

– частоти вісесиметричних коливань (гармонік (7.70)) мембрани.

Залишається знайти коефіцієнти $\tilde{A}_k$ і $\tilde{B}_k$ у формулі (7.71). З початкових умов (7.40) маємо:

$$u(r,0) = \sum_{k=1}^{\infty}\tilde{A}_k\, \Phi_k(r) = u_0(r), \ \ u_t(r,0) = \sum_{k=1}^{\infty}\omega_k \tilde{B}_k\, \Phi_k(r) = v_0(r).$$

Звідси за допомогою умови ортонормованості (7.69) дістаємо:

$$\tilde{A}_k = \int_D u_0(r)\Phi_k(r)\,dS = \sqrt{2\pi}\int_0^R u_0(r)\chi_k(r)rdr =$$

$$= \sqrt{2\pi}\frac{\sqrt{2}}{R|J_0'(\alpha_k)|}\int_0^R u_0(r)J_0\left(\frac{\alpha_k}{R}r\right)rdr,$$

$$\tilde{B}_k = \frac{1}{\omega_k}\int_D v_0(r)\Phi_k(r)\,dS = \frac{\sqrt{2\pi}}{\omega_k}\int_0^R v_0(r)\chi_k(r)rdr =$$

$$= \frac{\sqrt{2\pi}}{a\alpha_k}\frac{\sqrt{2}}{|J_0'(\alpha_k)|}\int_0^R v_0(r)J_0\left(\frac{\alpha_k}{R}r\right)rdr.$$

Отже, остаточно маємо:

$$u(r,t) = \sum_{k=1}^{\infty}\left[\left(\int_0^R u_0(r')\chi_k(r')r'dr'\right)\cos\omega_k t + \right.$$

$$\left. + \left(\frac{1}{\omega_k}\int_0^R v_0(r')\chi_k(r')r'dr'\right)\sin\omega_k t\right]\chi_k(r) =$$

$$= \frac{2}{R^2}\sum_{k=1}^{\infty}\frac{1}{[J_0'(\alpha_k)]^2}\left[\left(\int_0^R u_0(r')J_0\left(\frac{\alpha_k}{R}r'\right)r'dr'\right)\cos\left(\frac{\alpha_k a}{R}t\right) + \right.$$

$$\left. + \left(\frac{R}{a\alpha_k}\int_0^R v_0(r')J_0\left(\frac{\alpha_k}{R}r'\right)r'dr'\right)\sin\left(\frac{\alpha_k a}{R}t\right)\right]J_0\left(\frac{\alpha_k}{R}r\right). \qquad (7.73)$$

**Завдання 7.2.1.** Опишіть вимушені малі вісесиметричні коливання однорідної круглої мембрани із закріпленим краєм під дією поперечної сили з поверхневою густиною $F(r,t)$.

*Вказівка.* Відповідна крайова задача складається з неоднорідного рівняння ($\rho$ — поверхнева густина мембрани)

$$\frac{\partial^2 u}{\partial t^2} = a^2\frac{1}{r}\frac{\partial}{\partial r}\left(r\frac{\partial u}{\partial r}\right) + \frac{1}{\rho}F(r,t), \quad 0 < r < R, \ t > 0, \qquad (7.74)$$

крайових умов (7.39) та початкових умов (7.40) з $u_0(r) = 0$ і $v_0(r) = 0$. З огляду на відомі власні функції (7.63) та розв’язок (7.73) її розв’язок можна знайти за допомогою методу відокремлення змінних або принципу Дюамеля (див. підрозділ 4.10).

*Відповідь*:

$$u(r,t)=\int_0^t d\tau\int_0^R dr'r'G(r,r';t-\tau)F(r',\tau), \tag{7.75}$$

де

$$G(r,r';t)=\frac{1}{\rho}\sum_{k=1}^{\infty}\frac{\sin\omega_k t}{\omega_k}\chi_k(r)\chi_k(r')=$$

$$=\frac{2}{\rho R^2}\sum_{k=1}^{\infty}\frac{\sin\omega_k t}{\omega_k\left[J_0'(\alpha_k)\right]^2}J_0\left(\frac{\alpha_k}{R}r\right)J_0\left(\frac{\alpha_k}{R}r'\right) \tag{7.76}$$

– функція Гріна рівняння малих вісесиметричних коливань круглої мембрани із закріпленим краєм.

Слід зазначити, що техніка обчислення інтегралів від добутку функції Бесселя (Неймана) нульового чи ненульових порядків (останні обговорюються в наступному підрозділі) з багатьма елементарними функціями добре розвинута і часто базується на використанні степеневих рядів, якими ці функції зображуються. Наприклад, функція Бесселя першого порядку визначається як сума ряду

$$J_1(x)=\sum_{s=0}^{\infty}\frac{(-1)^s}{s!(s+1)!}\left(\frac{x}{2}\right)^{2s+1}. \tag{7.77}$$

Безпосереднім диференціюванням легко перевірити (див. формули (7.47) і (7.77)), що справджуються співвідношення

$$J_0'(x)=-J_1(x), \tag{7.78}$$

$$\frac{d}{dx}\left[xJ_1(x)\right]=xJ_0(x) \tag{7.79}$$

і, отже,

$$\int_0^z xJ_0(x)dx=\int_0^z\frac{d}{dx}\left[xJ_1(x)\right]dx=zJ_1(z). \tag{7.80}$$

Зокрема,

$$\int_0^{\delta} r'J_0\left(\frac{\alpha_k}{R}r'\right)dr'=\frac{R^2}{\alpha_k^2}\int_0^{\frac{\alpha_k}{R}\delta}xJ_0(x)dx=\frac{R\delta}{\alpha_k}J_1\left(\frac{\alpha_k}{R}\delta\right). \tag{7.81}$$

**Завдання 7.2.2.** Опишіть малі поперечні коливання однорідної круглої мембрани $r\le R$ із закріпленим краєм, викликані ударним імпульсом $P$, прикладеним перпендикулярно до її поверхні і розподіленим рівномірно по ділянці $r\le\delta$.

*Відповідь:* $u(r,t)=\dfrac{2P}{\pi a\delta\rho}\sum_{k=1}^{\infty}\dfrac{1}{\alpha_k^2 J_1^2(\alpha_k)}J_1\left(\dfrac{\alpha_k}{R}\delta\right)J_0\left(\dfrac{\alpha_k}{R}r\right)\sin\left(\dfrac{\alpha_k a}{R}t\right).$

**Завдання 7.2.3.** Опишіть малі поперечні коливання однорідної круглої мембрани $r\le R$ із закріпленим краєм, викликані періодичними коливаннями тиску $P=P_0\sin\omega t$, прикладеного до одного боку мембрани і рівномірно розподіленого по її поверхні. Частота коливань тиску не дорівнює жодній власній частоті мембрани.

*Вказівки.* Скористайтеся розв'язком завдання 7.2.1, замінивши в ньому $F(r,t)$ на $P_0\sin\omega t$. Тоді

$$u(r,t)=-\frac{2P_0}{\rho}\sum_{k=1}^{\infty}\frac{1}{\alpha_k J_1(\alpha_k)}\frac{1}{\omega^2-\omega_k^2}J_0\left(\frac{\alpha_k}{R}r\right)\sin\omega t+$$
$$+\frac{2P_0\omega}{\rho}\sum_{k=1}^{\infty}\frac{1}{\alpha_k J_1(\alpha_k)}\frac{1}{\omega_k\left(\omega^2-\omega_k^2\right)}J_0\left(\frac{\alpha_k}{R}r\right)\sin\omega_k t.$$

Перший доданок у цій формулі описує вимушені коливання мембрани виду $A(r)\sin\omega t$, де амплітуда $A(r)$ (сума першого ряду) задовольняє рівняння

$$-\omega^2 A=a^2\frac{1}{r}\frac{d}{dr}\left(r\frac{dA}{dr}\right)+\frac{1}{\rho}P_0,\quad 0<r<R,$$

та умови $|A(0)|<\infty$, $A(R)=0$. Його обмежений загальний розв'язок має вигляд $A(r)=-\dfrac{P_0}{\rho\omega^2}+CJ_0\left(\dfrac{\omega r}{a}\right)$. Остаточно матимемо:

$$u(r,t)=\frac{P_0}{\rho\omega^2}\left[J_0\left(\frac{\omega}{a}r\right)\Big/J_0\left(\frac{\omega}{a}R\right)-1\right]\sin\omega t+$$
$$+\frac{2P_0\omega R^3}{\rho a}\sum_{k=1}^{\infty}\frac{1}{\alpha_k^2 J_1(\alpha_k)\left(\omega^2R^2-a^2\alpha_k^2\right)}J_0\left(\frac{\alpha_k}{R}r\right)\sin\left(\frac{\alpha_k a}{R}t\right).$$

Покажіть, що другий доданок у цій формулі є розв'язком задачі (7.38)–(7.40) з початковими функціями $u_0(r)=0$ і $v_0(r)=-\omega A(r)$. При обчисленнях скористайтеся формулами (7.64) та (7.78)–(7.81).

Поведінку отриманих вище рядів можна дослідити (пропонуємо читачеві зробити це самостійно), скориставшись асимптотичними властивостями функції Бесселя та її нулів.

**Завдання 7.2.4.** Здійснивши в рівнянні (7.43) підстановку $\chi(x) = x^{-1/2}\eta(x)$, покажіть, що функція $\eta(x)$ задовольняє рівняння

$$\eta'' + \left(1 + \frac{1}{4x^2}\right)\eta = 0,$$

тобто асимптотично функції $J_0(x)$ та $N_0(x)$ при $x \to \infty$ мають вигляд спадних гармонік

$$\chi(x) = \frac{C}{\sqrt{x}}\sin(x + \delta) + O\left(x^{-3/2}\right), \tag{7.82}$$

де $C$ і $\delta$ — сталі.

За допомогою методів теорії функцій комплексної змінної асимптотики (7.82) можна уточнити:

$$\begin{aligned} J_0(x) \underset{x\to\infty}{=} \sqrt{\frac{2}{\pi x}}\sin\left(x + \frac{\pi}{4}\right) + O\left(x^{-3/2}\right), \\ N_0(x) \underset{x\to\infty}{=} \sqrt{\frac{2}{\pi x}}\sin\left(x - \frac{\pi}{4}\right) + O\left(x^{-3/2}\right). \end{aligned} \tag{7.83}$$

Перша з них відразу дає асимптотичну оцінку для додатних нулів функції $J_0(x)$:

$$\alpha_k \approx \frac{3\pi}{4} + \pi(k-1), \quad k >> 1. \tag{7.84}$$

Ця оцінка тим точніша, чим більший номер $k$, і вже при $k = 7$ відрізняється від точного значення менше, ніж на 0,006.

Узагалі, ряд виду

$$f(r) = \sum_{k=1}^{\infty} c_k J_0\left(\frac{\alpha_k}{R}r\right), \quad 0 \le r \le R, \tag{7.85}$$

з коефіцієнтами, що визначаються за формулою

$$c_k = \frac{2}{R^2\left[J_0'(\alpha_k)\right]^2}\int_0^R r f(r) J_0\left(\frac{\alpha_k}{R}r\right)dr, \tag{7.86}$$

називається *рядом Фур'є — Бесселя* для функції $f(r)$ за функціями Бесселя $J_0\left(\frac{\alpha_k}{R}r\right)$. Можна довести, що для кусково-гладкої $f(r)$ ряд (7.85), (7.86) збігається до значення $\frac{1}{2}\left[f(r-0) + f(r+0)\right]$ у кожній точці $r \in (0, R)$, дорівнює нулю при $r = R$ та (якщо $r = 0$ — точка неперервності $f(r)$) $f(0)$ при $r = 0$.

## 7.3. КОЛИВАННЯ КРУГЛОЇ МЕМБРАНИ. ЗАГАЛЬНИЙ ВИПАДОК

Повернімося до загальної задачі (7.33)–(7.37) про малі поперечні коливання однорідної круглої мембрани. У силу лінійності задачі її загальний розв'язок $u = u(r,\alpha,t)$ дорівнює суперпозиції розв'язку задачі про вільні коливання мембрани $(f(r,\alpha,t) = 0)$ та розв'язку задачі про вимушені коливання мембрани $(f(r,\alpha,t) \neq 0)$ під дією поперечної сили і при нульових початкових умовах $(u_0(r,\alpha) = 0,\ v_0(r,\alpha) = 0)$.

Почнемо з розгляду задачі про вільні коливання. Користуючись методом відокремлення змінних, нетривіальні частинні розв'язки однорідного рівняння (7.33) шукаємо у вигляді добутку $u(r,\alpha,t) = T(t)\Phi(r,\alpha)$. Підставивши його в однорідне рівняння (7.33) та поділивши обидві частини здобутого співвідношення на $a^2T(t)\Phi(r,\alpha)$, дістаємо рівність

$$\frac{T''(t)}{a^2T(t)} = \frac{1}{\Phi(r,\alpha)}\left[\frac{1}{r}\frac{\partial}{\partial r}\left(r\frac{\partial\Phi(r,\alpha)}{\partial r}\right) + \frac{1}{r^2}\frac{\partial^2\Phi(r,\alpha)}{\partial\alpha^2}\right] = -\lambda,$$

де, як завжди, ми врахували, що функції різних змінних дорівнюють одна одній лише за умови, що вони дорівнюють деякій сталій $-\lambda$. Бачимо, що для часової частини маємо рівняння (7.6), загальний розв'язок якого дається формулою (7.7), а для координатної частини — рівняння

$$\frac{1}{r}\frac{\partial}{\partial r}\left(r\frac{\partial\Phi(r,\alpha)}{\partial r}\right) + \frac{1}{r^2}\frac{\partial^2\Phi(r,\alpha)}{\partial\alpha^2} + \lambda\Phi(r,\alpha) = 0,\quad (r,\alpha)\in D. \qquad (7.87)$$

Підставивши далі добуток $u(r,\alpha,t) = T(t)\Phi(r,\alpha)$ у співвідношення (7.34), (7.36) і (7.37), знаходимо, що рівняння (7.87) треба доповнити крайовими умовами

$$\Phi(R,\alpha) = 0,\quad \left|\Phi(0,\alpha)\right| < \infty \qquad (7.88)$$

та умовою періодичності

$$\Phi(r,\alpha) = \Phi(r,\alpha + 2\pi). \qquad (7.89)$$

Отже, для координатної частини $\Phi(r,\alpha)$ маємо крайову задачу (7.87)–(7.89). Її нетривіальні частинні розв'язки шукаємо знову методом відокремлення змінних, тобто у вигляді добутку $\Phi(r,\alpha) = \chi(r)\psi(\alpha)$, який підставляємо спершу в рівняння (7.87), а потім — в умови (7.88) і (7.89). Зокрема, рівняння (7.87) набирає вигляду

$$\frac{1}{r}\frac{d}{dr}\left(r\frac{d\chi(r)}{\partial r}\right)\psi(\alpha) + \frac{1}{r^2}\chi(r)\frac{d^2\psi(\alpha)}{d\alpha^2} + \lambda\chi(r)\psi(\alpha) = 0.$$

Поділивши обидві частини цього рівняння на $\chi(r)\psi(\alpha)$, отримуємо:

$$\frac{1}{r\chi(r)}\frac{d}{dr}\left(r\frac{d\chi(r)}{\partial r}\right)+\frac{1}{r^2}\frac{1}{\psi(\alpha)}\frac{d^2\psi(\alpha)}{d\alpha^2}+\lambda=0. \qquad (7.90)$$

Бачимо, що доданки, які залежать від змінних $r$ і $\alpha$, можна відокремити один від одного, якщо покласти

$$\frac{1}{\psi(\alpha)}\frac{d^2\psi(\alpha)}{d\alpha^2}=-\mu,$$

де $\mu$ — деяка стала. Звідси та з умови періодичності (7.89) дістаємо задачу для кутової частини:

$$\frac{d^2\psi}{d\alpha^2}+\mu\psi=0,\ \ 0\le\alpha<2\pi, \qquad (7.91)$$

$$\psi(\alpha)=\psi(\alpha+2\pi), \qquad (7.92)$$

а далі — з рівності (7.90) та крайових умов (7.88) — крайову задачу для радіальної частини:

$$\frac{1}{r}\frac{d}{dr}\left(r\frac{d\chi}{\partial r}\right)+\left(\lambda-\frac{\mu}{r^2}\right)\chi=0,\ \ 0<r<R, \qquad (7.93)$$

$$\chi(R)=0,\ \ |\chi(0)|<\infty. \qquad (7.94)$$

Частинні розв’язки рівняння (7.91) мають вигляд

$$\psi^{(1)}(\alpha)=A\cos\left(\sqrt{\mu}\,\alpha\right),\ \ \psi^{(2)}(\alpha)=B\sin\left(\sqrt{\mu}\,\alpha\right)$$

і, згідно з умовою періодичності (7.92), мають задовольняти співвідношення

$$\cos\left(\sqrt{\mu}\,\alpha\right)=\cos\left(\sqrt{\mu}\,\alpha+2\pi\sqrt{\mu}\right),\ \ \sin\left(\sqrt{\mu}\,\alpha\right)=\sin\left(\sqrt{\mu}\,\alpha+2\pi\sqrt{\mu}\right).$$

Звідси випливає, що $\sqrt{\mu}$ — ціле число. Більше того, ураховуючи парність функції $\psi^{(1)}$ та непарність функції $\psi^{(2)}$, лінійно незалежні власні функції задачі (7.91), (7.92) отримаємо, обмежившись невід’ємними цілими значеннями $\sqrt{\mu}$. Отже, власні значення цієї задачі

$$\mu_m=m^2,\ \ m=0,1,2,.... \qquad (7.95)$$

Їм відповідають нормовані власні функції

$$\psi_0^{(1)}=\frac{1}{\sqrt{2\pi}},\quad m=0,$$

$$\psi_m^{(1)}(\alpha)=\frac{1}{\sqrt{\pi}}\cos m\alpha,\quad \psi_m^{(2)}(\alpha)=\frac{1}{\sqrt{\pi}}\sin m\alpha,\quad m=1,2,..., \tag{7.96}$$

для яких справджується умова ортонормованості

$$\int_0^{2\pi}\psi_m^{(i)}(\alpha)\psi_{m'}^{(i')}(\alpha)d\alpha=\delta_{ii'}\,\delta_{mm'},$$
$$i,i'=1,2,\quad m,m'=i-1,i'-1;i,i';i+1,i'+1;.... \tag{7.97}$$

Власні значення і нормовані власні функції радіальної задачі (7.93), (7.94) шукаємо в той самий спосіб, що був використаний при розв'язуванні задачі (7.41), (7.42). Зазначимо його основні моменти, уже не зупиняючись на подробицях обчислень. Після підстановки $r=x/\sqrt{\lambda}$ рівняння (7.93) набирає вигляду

$$\chi''(x)+\frac{1}{x}\chi'(x)+\left(1-\frac{m^2}{x^2}\right)\chi(x)=0. \tag{7.98}$$

Будуючи його обмежений ($|\chi(0)|<\infty$) частинний розв'язок у формі степеневого ряду (7.44), знаходимо, що кожному значенню $m$ відповідає розв'язок

$$\chi_m(x)=a_m J_m(x), \tag{7.99}$$

де

$$J_m(x)=\sum_{s=0}^{\infty}\frac{(-1)^s}{s!(s+m)!}\left(\frac{x}{2}\right)^{2s+m} \tag{7.100}$$

– *функція Бесселя порядку m*. Для цілих $m$ вона визначена скрізь у комплексній площині, однозначна та ціла, тобто має лише ізольовані нулі. Позначивши через $\alpha_k^{(m)}$ додатні нулі функції $J_m(x)$ [1], пронумеровані в порядку зростання, та врахувавши першу крайову умову (7.94), знаходимо нетривіальні власні значення та нормовані власні функції крайової задачі (7.93), (7.94) ($m=0,1,2,...,\ k=1,2,3,...$):

$$\lambda_{mk}=\frac{\left(\alpha_k^{(m)}\right)^2}{R^2}, \tag{7.101}$$

$$\chi_{mk}(r)=a_{mk}J_m\left(\frac{\alpha_k^{(m)}}{R}r\right), \tag{7.102}$$

де нормувальний множник

---

[1] У попередньому підрозділі додатні нулі $\alpha_k^{(0)}$ функції $J_0(x)$ позначалися як $\alpha_k$.

$$a_{mk}=\frac{\sqrt{2}}{R\left|J'_m\left(\alpha_k^{(m)}\right)\right|} \qquad (7.103)$$

і власні функції (7.102) задовольняють умову ортонормованості

$$\int_0^R r\chi_m\left(\frac{\alpha_k^{(m)}}{R}r\right)\chi_m\left(\frac{\alpha_{k'}^{(m)}}{R}r\right)dr=\delta_{kk'}. \qquad (7.104)$$

**Завдання 7.3.1.** Нехай $\gamma_k^{(m)}$ — корені рівняння $\gamma J'_m(\gamma)+hJ_m(\gamma)=0$, $h\geq 0$. Доведіть співвідношення ортогональності

$$\int_0^R rJ_m\left(\frac{\gamma_k^{(m)}}{R}r\right)J_m\left(\frac{\gamma_{k'}^{(m)}}{R}r\right)dr=0,\ \text{якщо}\ k\neq k', \qquad (7.105)$$

та обчисліть норму

$$\int_0^R rJ_m^2\left(\frac{\gamma_k^{(m)}}{R}r\right)dr=\frac{R^2}{2}\left\{\left[J'_m\left(\gamma_k^{(m)}\right)\right]^2+\left(1-\frac{m^2}{\gamma_k^{(m)2}}\right)\left[J_m\left(\gamma_k^{(m)}\right)\right]^2\right\}. \qquad (7.106)$$

Як наслідки, отримайте формули (7.103) і (7.104).

**Завдання 7.3.2.** Виходячи з означення (7.100), доведіть співвідношення

$$\frac{d}{dx}\left[x^mJ_m(x)\right]=x^mJ_{m-1}(x),\quad \frac{d}{dx}\left[x^{-m}J_m(x)\right]=-x^{-m}J_{m+1}(x). \qquad (7.107)$$

На їх основі знайдіть рекурентні формули

$$J_{m+1}(x)=-J_{m-1}(x)+\frac{2m}{x}J_m(x),\quad J'_m(x)=-J_{m+1}(x)+\frac{m}{x}J_m(x) \qquad (7.108)$$

та невизначені інтеграли

$$\int x^mJ_{m-1}(x)dx=x^mJ_m(x)+C,\quad \int x^{-m}J_{m+1}(x)dx=-x^{-m}J_m(x)+C, \qquad (7.109)$$

$$\int xJ_1(x)dx=-xJ_0(x)+\int J_0(x)dx+C, \qquad (7.110)$$

$$\int x^2J_2(x)dx=-x^2J_1(x)-3xJ_0(x)+3\int J_0(x)dx+C, \qquad (7.111)$$

$$\int J_3(x)dx=J_0(x)-\frac{4}{x}J_1(x)+C, \qquad (7.112)$$

де $C$ — довільна стала.

*Вказівка.* При обчисленні інтегралів (7.110) і (7.111) скористайтеся інтегруванням частинами та інтегралами (7.109), які випливають із співвідношень (7.107). Інтеграл (7.112) знаходиться за допомогою співвідношень (7.108) й інтегралів (7.109). Інтеграл $\int_0^w J_0(x)dx$ не виражається в замкненій формі, однак його значення добре протабульовані.

**Завдання 7.3.3.** Доведіть, що другий лінійно незалежний розв'язок[1] рівняння Бесселя (7.98) при $m \geq 1$ має в точці $x = 0$ сингулярність виду $1/x^m$.

**Завдання 7.3.4.** Доведіть, що при $x \to \infty$ функції Бесселя $J_m(x)$ і Неймана $N_m(x)$ ведуть себе як спадні гармоніки виду (7.82)[2].

З огляду на результати (7.95)–(7.97) і (7.101)–(7.104) можемо стверджувати, що власні значення $\lambda_{mk}$ $(m = 0,1,2,...,\ k = 1,2,3,...)$ крайової задачі (7.87)–(7.89) даються формулою (7.101), і їм відповідають нормовані власні функції

$$\begin{gathered}
\Phi_{0k}^{(1)}(r) = \chi_{0k}(r)\psi_0^{(1)} = \frac{1}{\sqrt{\pi}R\left|J_0'\left(\alpha_k^{(0)}\right)\right|} J_0\left(\frac{\alpha_k^{(0)}}{R}r\right), \\
\Phi_{mk}^{(1)}(r,\alpha) = \chi_{mk}(r)\psi_m^{(1)}(\alpha) = \\
= \frac{\sqrt{2}}{\sqrt{\pi}R\left|J_m'\left(\alpha_k^{(m)}\right)\right|} J_m\left(\frac{\alpha_k^{(m)}}{R}r\right)\cos m\alpha, \quad m \geq 1, \\
\Phi_{mk}^{(2)}(r,\alpha) = \chi_{mk}(r)\psi_m^{(2)}(\alpha) = \\
= \frac{\sqrt{2}}{\sqrt{\pi}R\left|J_m'\left(\alpha_k^{(m)}\right)\right|} J_m\left(\frac{\alpha_k^{(m)}}{R}r\right)\sin m\alpha, \quad m \geq 1,
\end{gathered} \tag{7.113}$$

що задовольняють умову ортонормованості

$$\int_D \Phi_{mk}^{(i)}(r,\alpha)\Phi_{m'k'}^{(i')}(r,\alpha)dS = \int_0^R drr\int_0^{2\pi} d\alpha\,\Phi_{mk}^{(i)}(r,\alpha)\Phi_{m'k'}^{(i')}(r,\alpha) = \delta_{ii'}\,\delta_{mm'}\,\delta_{kk'}. \tag{7.114}$$

---

[1] Він пропорціональний *функції Неймана порядку m*, яка для цілих *m* означається як границя $N_m(x) = \lim_{\mu\to m} N_\mu(x)$, де $N_\mu(x) = \dfrac{J_\mu(x)\cos\pi\mu - J_{-\mu}(x)}{\sin\pi\mu}$ та $J_\mu(x) = \sum_{s=0}^{\infty}\dfrac{(-1)^s}{\Gamma(s+1)\Gamma(s+\mu+1)}\left(\dfrac{x}{2}\right)^{2s+\mu}$ — функції Неймана та Бесселя, доозначені на довільне неціле число $\mu$, $\Gamma(x)$ — гамма-функція Ейлера.

[2] Більш докладні вирази для асимптотик мають вигляд:

$$J_m(x) \underset{x\to\infty}{=} \sqrt{\frac{2}{\pi x}}\cos\left(x - \frac{\pi}{4} - \frac{\pi}{2}m\right) + O\left(x^{-3/2}\right), \quad N_m(x) \underset{x\to\infty}{=} \sqrt{\frac{2}{\pi x}}\sin\left(x - \frac{\pi}{4} - \frac{\pi}{2}m\right) + O\left(x^{-3/2}\right).$$

Скориставшись далі виразом (7.7) для часової частини $T_{mk}(t)$, загальний розв’язок крайової задачі (7.33)–(7.37) про вільні коливання круглої мембрани із закріпленим краєм можемо записати у вигляді

$$u(r,\alpha,t)=\sum_{k=1}^{\infty}\left(A_{0k}^{(1)}\cos\omega_{0k}t+B_{0k}^{(1)}\sin\omega_{0k}t\right)\Phi_{0k}^{(1)}(r)+ \\ +\sum_{i=1}^{2}\sum_{m=1}^{\infty}\sum_{k=1}^{\infty}\left(A_{mk}^{(i)}\cos\omega_{mk}t+B_{mk}^{(i)}\sin\omega_{mk}t\right)\Phi_{mk}^{(i)}(r,\alpha), \tag{7.115}$$

де

$$\omega_{mk}=\frac{\alpha_k^{(m)}a}{R} \tag{7.116}$$

– частоти коливань мембрани. Коефіцієнти $A_{mk}^{(i)}$ і $B_{mk}^{(i)}$ визначаємо з початкових умов (7.35):

$$\sum_{k=1}^{\infty}A_{0k}^{(1)}\Phi_{0k}^{(1)}(r)+\sum_{i=1}^{2}\sum_{m=1}^{\infty}\sum_{k=1}^{\infty}A_{mk}^{(i)}\Phi_{mk}^{(i)}(r,\alpha)=u_0(r,\alpha), \tag{7.117}$$

$$\sum_{k=1}^{\infty}\omega_{0k}B_{0k}^{(1)}\Phi_{0k}^{(1)}(r)+\sum_{i=1}^{2}\sum_{m=1}^{\infty}\sum_{k=1}^{\infty}\omega_{mk}B_{mk}^{(i)}\Phi_{mk}^{(i)}(r,\alpha)=v_0(r,\alpha), \tag{7.118}$$

звідки, у силу умови ортонормованості (7.114),

$$A_{0k}^{(1)}=\int_D u_0(r,\alpha)\Phi_{0k}^{(1)}(r)dS,\quad B_{0k}^{(1)}=\frac{1}{\omega_{0k}}\int_D v_0(r,\alpha)\Phi_{0k}^{(1)}(r)dS,\quad k\ge 1, \tag{7.119}$$

$$A_{mk}^{(i)}=\int_D u_0(r,\alpha)\Phi_{mk}^{(i)}(r,\alpha)dS, \\ B_{mk}^{(i)}=\frac{1}{\omega_{mk}}\int_D v_0(r,\alpha)\Phi_{mk}^{(i)}(r,\alpha)dS,\quad i=1,2,\quad m,k\ge 1. \tag{7.120}$$

Знайдені коефіцієнти є фактично коефіцієнтами рядів Фур’є — Бесселя за функціями Бесселя $J_m\left(\frac{\alpha_k^{(m)}}{R}r\right)$, тобто рядів виду

$$f(r)=\sum_{k=1}^{\infty}c_kJ_m\left(\frac{\alpha_k^{(m)}}{R}r\right),\quad 0\le r\le R, \tag{7.121}$$

з коефіцієнтами

$$c_k=\frac{2}{R^2\left[J'_m\left(\alpha_k^{(m)}\right)\right]^2}\int_0^R rf(r)J_m\left(\frac{\alpha_k^{(m)}}{R}r\right)dr. \tag{7.122}$$

(У цьому легко переконатися, помноживши обидві частини рівностей (7.117) і (7.118) на $\cos m\alpha$ чи $\sin m\alpha$ та зінтегрувавши нові рівності за $\alpha$ в межах від 0 до $2\pi$.) Для кожної кусково-гладкої функції $f(r)$ ряди (7.121), (7.122) з $m>0$ збігаються до значення $\frac{1}{2}\left[f(r-0)+f(r+0)\right]$ у кожній точці $r\in(0,R)$ та дорівнюють нулю при $r=R$ і $r=0$.

Зауважимо, що перша сума в розв'язку (7.115) описує вісесиметричні коливання мембрани при початкових умовах (7.35). Якщо ці умови не залежать явно від кута $\alpha$, то всі коефіцієнти (7.120) дорівнюють нулю, а вказана сума збігається з розв'язком (7.73).

**Завдання 7.3.5.** Опишіть малі коливання однорідної круглої мембрани із закріпленим краєм під дією поперечної сили з поверхневою густиною $F(r,\alpha,t)$ і при нульових початкових умовах.

*Відповідь*: $u(r,\alpha,t)=\int\limits_0^t d\tau\int\limits_D dS'G(r,\alpha,r',\alpha';t-\tau)F(r',\alpha',\tau)$,

де

$$G(r,\alpha,r',\alpha';t)=\frac{1}{\rho}\sum_{k=1}^{\infty}\frac{\sin\omega_{0k}t}{\omega_{0k}}\Phi_{0k}^{(1)}(r)\Phi_{0k}^{(1)}(r')+$$

$$+\frac{1}{\rho}\sum_{i=1}^{2}\sum_{m=1}^{\infty}\sum_{k=1}^{\infty}\frac{\sin\omega_{mk}t}{\omega_{mk}}\Phi_{mk}^{(i)}(r,\alpha)\Phi_{mk}^{(i)}(r',\alpha')$$

– функція Гріна рівняння малих поперечних коливань круглої мембрани із закріпленим краєм.

## 7.4. КРАЙОВІ ЗАДАЧІ ДЛЯ ЦИЛІНДРИЧНИХ ОБЛАСТЕЙ

Функції Бесселя, Неймана та їх лінійні комбінації також називають циліндричними, оскільки їх поява є типовою при розв'язуванні крайових задач, поставлених для тіл чи областей із циліндричною геометрією. Проілюструємо сказане кількома простими прикладами.

Почнемо із задачі про остигання однорідного круглого стержня радіусом $R$ і висотою $H$, початкова температура якого описується радіальним розподілом $T_0(r)$; стержень остигає з моменту часу $t=0$, з якого його поверхня підтримується при нульовій температурі.

Нехай початок декартової системи координат знаходиться в центрі стержня, координатна вісь $OZ$ напрямлена по осі симетрії стержня,

а дві інші координатні осі $OX$ і $OY$ лежать у площині його поперечного перерізу. Ураховуючи геометричну симетрію стержня, перейдемо до циліндричної системи координат з початком у центрі стержня і полярною віссю, напрямленою по його осі. Формули переходу від декартових координат $(x,y,z)$ до циліндричних $(r,\alpha,z)$ мають вигляд

$$x = r\cos\alpha, \quad y = r\sin\alpha, \quad z = z, \tag{7.123}$$

при цьому точкам просторової області $C$, зайнятої стержнем, відповідають значення координат, що задовольняють нерівності

$$0 \le r \le R, \quad 0 \le \alpha < 2\pi, \quad -H/2 \le z \le H/2; \tag{7.124}$$

якобіан цього переходу

$$\frac{D(x,y,z)}{D(r,\alpha,z)} = \begin{vmatrix} \dfrac{\partial x}{\partial r} & \dfrac{\partial x}{\partial \alpha} & \dfrac{\partial x}{\partial z} \\ \dfrac{\partial y}{\partial r} & \dfrac{\partial y}{\partial \alpha} & \dfrac{\partial y}{\partial z} \\ \dfrac{\partial z}{\partial r} & \dfrac{\partial z}{\partial \alpha} & \dfrac{\partial z}{\partial z} \end{vmatrix} = \begin{vmatrix} \cos\alpha & -r\sin\alpha & 0 \\ \sin\alpha & r\cos\alpha & 0 \\ 0 & 0 & 1 \end{vmatrix} = r. \tag{7.125}$$

У циліндричній системі координат оператор Лапласа

$$\Delta = \frac{1}{r}\frac{\partial}{\partial r}\left(r\frac{\partial}{\partial r}\right) + \frac{1}{r^2}\frac{\partial^2}{\partial \alpha^2} + \frac{\partial^2}{\partial z^2},$$

тому рівняння теплопровідності для однорідного середовища має вигляд

$$\frac{\partial u}{\partial t} = a^2\left[\frac{1}{r}\frac{\partial}{\partial r}\left(r\frac{\partial u}{\partial r}\right) + \frac{1}{r^2}\frac{\partial^2 u}{\partial \alpha^2} + \frac{\partial^2 u}{\partial z^2}\right] + f(r,\alpha,z,t), \tag{7.126}$$

де $u = u(r,\alpha,z,t)$ — температура стержня як функція координат і часу, $f(r,\alpha,z,t)$ — функція теплових джерел і стоків (у нашому випадку вона дорівнює нулю).

У точках $r = 0$, де якобіан переходу (7.125) обертається в нуль, порушується взаємна однозначність між декартовими і циліндричними координатами, а тому формальний розв'язок рівняння (7.126) може мати нефізичну сингулярність. Як уже зазначалося, її можна усунути, наклавши вимогу, щоб функція $u(r,\alpha,z,t)$ була обмеженою при $r = 0$:

$$|u(0,\alpha,z,t)| < \infty. \tag{7.127}$$

Крім того, на поверхні стержня $\Gamma$ (його основах та бічній поверхні) функція $u(r,\alpha,z,t)$ повинна задовольняти крайові умови

$$u(r,\alpha,z,t)\big|_{\Gamma}=0, \tag{7.128}$$

або, докладніше,

$$u(r,\alpha,-H/2,t)=0,\ \ u(r,\alpha,H/2,t)=0,\ \ u(R,\alpha,z,t)=0, \tag{7.128a}$$

а в усіх внутрішніх точках $D$ стержня — початкову умову

$$u(r,\alpha,z,0)\big|_{D}=T_0(r). \tag{7.129}$$

І, нарешті, повинна справджуватися умова періодичності

$$u(r,\alpha,z,t)=u(r,\alpha+2\pi,z,t). \tag{7.130}$$

Візьмемо тепер до уваги, що стержень однорідний, крайові та початкові умови *аксіально симетричні* (не залежать явно від кута $\alpha$), а теплові джерела відсутні. Тому немає жодних фізичних причин, які б з часом порушували початкову інваріантність задачі відносно поворотів навколо полярної осі $OZ$. Слід очікувати, що в кожний фіксований момент часу $t$ температура стержня є функцією лише $r$ і $z$: $u=u(r,z,t)$. Для її визначення дістаємо таку крайову задачу:

$$\frac{\partial u}{\partial t}=a^2\left[\frac{1}{r}\frac{\partial}{\partial r}\left(r\frac{\partial u}{\partial r}\right)+\frac{\partial^2 u}{\partial z^2}\right],\ \ 0<r<R,\ \ -H/2<z<H/2,\ \ t>0, \tag{7.131}$$

$$u(R,z,t)=0,\ \ u(r,-H/2,t)=0,\ \ u(r,H/2,t)=0,\ \ \left|u(0,z,t)\right|<\infty, \tag{7.132}$$

$$u(r,z,0)=T_0(r). \tag{7.133}$$

Шукаємо нетривіальні частинні розв’язки задачі (7.131)–(7.133) у вигляді добутку $\Phi(r,z)T(t)$. Відокремивши часову і координатну частини, дістаємо рівняння

$$T'(t)+\lambda a^2 T(t)=0,\ \ t>0, \tag{7.134}$$

$$\frac{1}{r}\frac{\partial}{\partial r}\left(r\frac{\partial \Phi}{\partial r}\right)+\frac{\partial^2 \Phi}{\partial z^2}+\lambda\Phi=0,\ \ 0<r<R,\ \ -H/2<z<H/2, \tag{7.135}$$

та крайові умови для функції $\Phi(r,z)$:

$$\Phi(R,z)=0,\ \ \left|\Phi(0,z)\right|<\infty,\ \ \Phi(r,-H/2)=0,\ \ \Phi(r,H/2)=0. \tag{7.136}$$

Крайову задачу (7.135), (7.136) розв’язуємо, знову скориставшись методом відокремлення змінних. Шукаючи її частинні розв’язки у вигляді $\Phi(r,z)=\chi(r)Z(z)$, для функцій $\chi(r)$ і $Z(z)$ дістаємо дві одновимірні крайові задачі:

$$\frac{1}{r}\frac{d}{dr}\left(r\frac{d\chi}{dr}\right)+\mu\chi=0,\ \ 0<r<R, \tag{7.137}$$

$$\chi(R)=0,\ \ |\chi(0)|<\infty, \tag{7.138}$$

$$\frac{d^2Z}{dz^2}+\nu Z=0,\ \ -H/2<z<H/2, \tag{7.139}$$

$$Z(-H/2)=0,\ \ Z(H/2)=0, \tag{7.140}$$

при цьому

$$\lambda=\mu+\nu. \tag{7.141}$$

Власні значення $\nu_n$ та нормовані власні функції $Z_n(z)$ задачі (7.139), (7.140) знаходимо, підставивши загальний розв'язок

$$Z(z)=A\cos\left(\sqrt{\nu}\,z\right)+B\sin\left(\sqrt{\nu}\,z\right)$$

рівняння (7.139) у крайові умови (7.140) та проаналізувавши умови існування нетривіальних розв'язків системи однорідних лінійних рівнянь, отримуваних для коефіцієнтів $A$ і $B$. Після нормування знайдених власних функцій дістаємо:

$$\nu_n=\frac{\pi^2n^2}{H^2},\quad n=1,2,...,\quad Z_n(z)=\sqrt{\frac{2}{H}}\begin{cases}\cos\dfrac{\pi nz}{H}, & \text{якщо } n=1,3,5,...,\\ \sin\dfrac{\pi nz}{H}, & \text{якщо } n=2,4,6,....\end{cases} \tag{7.142}$$

Задача (7.137), (7.138) уже аналізувалася при розгляді вісесиметричних коливань круглої мембрани із закріпленим краєм — див. задачу (7.41), (7.42), у якій $\lambda$ треба замінити на $\mu$. Її власні значення і нормовані власні функції даються формулами

$$\mu_k=\frac{\alpha_k^2}{R^2},\ \ \chi_k(r)=\frac{\sqrt{2}}{R\left|J_0'(\alpha_k)\right|}J_0\left(\frac{\alpha_k}{R}r\right),\ \ k=1,2,3,..., \tag{7.143}$$

де $\alpha_k$ — додатні нулі функції $J_0(x)$, розташовані в порядку зростання.

На основі формул (7.142) і (7.143) можемо вже виписати власні значення $\lambda_{kn}$ та власні функції $\Phi_{kn}(r,z)$ крайової задачі (7.135), (7.136):

$$\lambda_{kn}=\mu_k+\nu_n,\ \ \Phi_{kn}(r,z)=\frac{1}{\sqrt{2\pi}}\chi_k(r)Z_n(z),\ \ k,n=1,2,3,.... \tag{7.144}$$

Зауважимо, що множник $1/\sqrt{2\pi}$ у виразі (7.144) для $\Phi_{kn}(r,z)$ не обов'язковий — його зручно ввести для того, щоб умова ортонормованості цих функцій записувалася через об'ємний інтеграл по всій циліндричній області $C$, зайнятій стержнем:

$$\int\limits_C \Phi_{kn}(r,z)\Phi_{k'n'}(r,z)dV = \int\limits_0^R drr\int\limits_0^{2\pi} d\alpha \int\limits_{-H/2}^{H/2} dz\,\Phi_{kn}(r,z)\Phi_{k'n'}(r,z) = \delta_{kk'}\,\delta_{nn'}. \quad (7.145)$$

З огляду на розв'язки рівняння (7.134) частинні та загальний розв'язки задачі (7.131)–(7.133) мають вигляд

$$u_{kn}(r,z,t) = A_{kn}e^{-\lambda_{kn}a^2t}\Phi_{kn}(r,z),$$

$$u(r,z,t) = \sum_{k=1}^{\infty}\sum_{n=1}^{\infty} c_{kn}u_{kn}(r,z,t) = \sum_{k=1}^{\infty}\sum_{n=1}^{\infty} a_{kn}e^{-\lambda_{kn}a^2t}\Phi_{kn}(r,z). \quad (7.146)$$

Залишається знайти коефіцієнти $a_{kn}$ в останній формулі. З початкової умови (7.133) маємо:

$$u(r,z,0) = \sum_{k=1}^{\infty}\sum_{n=1}^{\infty} a_{kn}\Phi_{kn}(r,z) = T_0(r).$$

Звідси, скориставшись умовою ортонормованості (7.145), дістаємо:

$$a_{kn} = \int\limits_C T_0(r)\Phi_{kn}(r,z)dV = \int\limits_0^R drrT_0(r)\int\limits_0^{2\pi} d\alpha \int\limits_{-H/2}^{H/2} dz\frac{1}{\sqrt{2\pi}}\chi_k(r)Z_n(z) =$$

$$= \sqrt{2\pi}\int\limits_0^R drrT_0(r)\chi_k(r)\int\limits_{-H/2}^{H/2} dz\,Z_n(z).$$

Для функцій (7.142) далі маємо

$$\int\limits_{-H/2}^{H/2} Z_n(z)dz = \sqrt{\frac{2}{H}}\int\limits_{-H/2}^{H/2}\cos\frac{\pi nz}{H}dz = \sqrt{2H}\,\frac{2}{\pi n}\sin\frac{\pi n}{2}, \quad n = 1,3,5,...,$$

$$\int\limits_{-H/2}^{H/2} Z_n(z)dz = \sqrt{\frac{2}{H}}\int\limits_{-H/2}^{H/2}\sin\frac{\pi nz}{H}dz = 0, \quad n = 2,4,6,...,$$

тобто відмінними від нуля є коефіцієнти $a_{kn}$ з непарними значеннями $n = 2p+1$, де $p = 0,1,2,...$:

$$a_{kp} = \sqrt{2\pi}\frac{\sqrt{2}}{R|J_0'(\alpha_k)|}\left(\int\limits_0^R rT_0(r)J_0\left(\frac{\alpha_k}{R}r\right)dr\right)\sqrt{2H}\,\frac{2}{\pi(2p+1)}(-1)^p,$$

$$k = 1,2,..., \quad p = 0,1,2,....$$

У підсумку ряд (7.146) набирає вигляду

$$u(r,z,t)=\frac{8}{\pi R^2}\sum_{k=1}^{\infty}\sum_{p=0}^{\infty}\frac{(-1)^p}{(2p+1)\left[J_0'(\alpha_k)\right]^2}\left[\int_0^R r'T_0(r')J_0\left(\frac{\alpha_k}{R}r'\right)dr'\right]\times$$
$$\times e^{-\left[\frac{\alpha_k^2}{R^2}+\frac{\pi^2(2p+1)^2}{H^2}\right]a^2t}J_0\left(\frac{\alpha_k}{R}r\right)\cos\frac{\pi(2p+1)z}{H}. \tag{7.147}$$

Зокрема, ураховуючи, що

$$\sum_{p=0}^{\infty}\frac{(-1)^p}{(2p+1)}=\frac{\pi}{4},$$

у граничному випадку $H\to\infty$ знаходимо:

$$u(r,t)=\frac{2}{R^2}\sum_{k=1}^{\infty}\frac{1}{\left[J_0'(\alpha_k)\right]^2}\left[\int_0^R r'T_0(r')J_0\left(\frac{\alpha_k}{R}r'\right)dr'\right]e^{-\frac{\alpha_k^2}{R^2}a^2t}J_0\left(\frac{\alpha_k}{R}r\right). \tag{7.148}$$

Цей перехід відповідає ситуації, коли краї стержня знаходяться настільки далеко від точки спостереження, що їх впливом можна знехтувати. У системі з’являється ще один елемент симетрії — однорідність уздовж осі стержня, тобто неможливість знайти відмінності між фізичними умовами в точках, що відрізняються лише значеннями координати $z$. Як результат, залежність від цієї координати зникає.

Якщо стержень у початковий момент часу був нагрітий рівномірно, $T_0(r)=T_0$, то інтеграл у рядах (7.147) і (7.148) зводиться до інтеграла (7.81), і ці ряди набирають простішого вигляду. Ураховуючи ще й співвідношення (7.78), відповідно маємо:

$$u(r,z,t)=\frac{8T_0}{\pi}\sum_{k=1}^{\infty}\sum_{p=0}^{\infty}\frac{(-1)^p}{(2p+1)\alpha_k J_1(\alpha_k)}e^{-\left[\frac{\alpha_k^2}{R^2}+\frac{\pi^2(2p+1)^2}{H^2}\right]a^2t}J_0\left(\frac{\alpha_k}{R}r\right)\cos\frac{\pi(2p+1)z}{H},$$

$$u(r,t)=2T_0\sum_{k=1}^{\infty}\frac{1}{\alpha_k J_1(\alpha_k)}e^{-\frac{\alpha_k^2}{R^2}a^2t}J_0\left(\frac{\alpha_k}{R}r\right).$$

За аналогічною схемою розв’язуються й інші задачі з аксіальною симетрією для циліндричних областей.

**Завдання 7.4.1.** Узагальніть попередню задачу та її розв’язок на випадок, коли початкова температура стержня описується функцією $T_0(r,z)$.

**Завдання 7.4.2.** Знайдіть власні частоти радіальних коливань газу в довгому порожнистому циліндрі радіусом $R$.

*Вказівка*. Крайова задача для потенціалу швидкості $u = u(r,t)$ газу включає рівняння

$$\frac{\partial^2 u}{\partial t^2} = a^2 \frac{1}{r}\frac{\partial}{\partial r}\left(r\frac{\partial u}{\partial r}\right), \quad 0 < r < R, \quad t > 0,$$

крайові умови

$$\left.\frac{\partial u(r,t)}{\partial r}\right|_{r=R} = 0, \quad |u(0,t)| < \infty$$

та початкові умови

$$u(r,0) = u_0(r), \quad \left.\frac{\partial u(r,t)}{\partial t}\right|_{t=0} = v_0(r).$$

*Відповідь*: $\omega_k = a\beta_k / R$, де $\beta_k$ — послідовні додатні корені рівняння $J_0'(\beta) = 0$ (додатні нулі функції Бесселя $J_1(x)$). Звернемо увагу, що вказана крайова задача включає й нульове власне значення, для якого відповідний частинний розв'язок $A + Bt$ описує стан спокою газу при певному тиску.

**Завдання 7.4.3.** Знайдіть закон остигання довгої однорідної циліндричної труби, нагрітої до температури $T_0(r)$, якщо з моменту часу $t = 0$ на її внутрішній (радіусом $R_1$) і зовнішній (радіусом $R_2$) поверхнях підтримується нульова температура.

*Вказівки*. Крайова задача для визначення температури труби $u = u(r,t)$ в довільний момент часу має вигляд

$$\frac{\partial u}{\partial t} = a^2 \frac{1}{r}\frac{\partial}{\partial r}\left(r\frac{\partial u}{\partial r}\right), \quad R_1 < r < R_2, \quad t > 0,$$

$$u(R_1,t) = 0, \quad u(R_2,t) = 0,$$

$$u(r,0) = T_0(r).$$

Зауважимо, що умова $|u(0,t)| < \infty$ тут не використовується, оскільки вісь труби $r = 0$ лежить поза межами досліджуваної області.

Для радіальної частини $\Phi(r)$ частинного розв'язку $u(r,t) = \Phi(r)T(t)$ дістаємо одновимірну крайову задачу

$$\frac{1}{r}\frac{d}{dr}\left(r\frac{d\Phi}{dr}\right) + \lambda\Phi = 0, \quad R_1 < r < R_2, \qquad (7.149)$$

$$\Phi(R_1) = 0, \quad \Phi(R_2) = 0. \qquad (7.150)$$

Загальний розв'язок рівняння (7.149)

$$\Phi(r) = AJ_0\left(\sqrt{\lambda}\, r\right) + BN_0\left(\sqrt{\lambda}\, r\right)$$

( $A$ і $B$ – сталі), тому крайові умови (7.150) дають однорідну алгебраїчну систему

$$AJ_0\left(\sqrt{\lambda} R_1\right) + BN_0\left(\sqrt{\lambda} R_1\right) = 0,$$
$$AJ_0\left(\sqrt{\lambda} R_2\right) + BN_0\left(\sqrt{\lambda} R_2\right) = 0.$$

Звідси

$$\Phi_k(r) = C_k F_0\left(\sqrt{\lambda_k}\, r\right) \equiv$$
$$\equiv C_k\left[J_0\left(\sqrt{\lambda_k} R_1\right)N_0\left(\sqrt{\lambda_k}\, r\right) - N_0\left(\sqrt{\lambda_k} R_1\right)J_0\left(\sqrt{\lambda_k}\, r\right)\right] = \qquad (7.151)$$
$$= C_k \frac{J_0\left(\sqrt{\lambda_k} R_1\right)}{J_0\left(\sqrt{\lambda_k} R_2\right)}\left[J_0\left(\sqrt{\lambda_k} R_2\right)N_0\left(\sqrt{\lambda_k}\, r\right) - N_0\left(\sqrt{\lambda_k} R_2\right)J_0\left(\sqrt{\lambda_k}\, r\right)\right],$$

де $C_k$ — нові сталі, $\lambda_k = \gamma_k^2 / R_2^2$ $(k = 1,2,...)$, $\gamma_k$ — послідовні додатні корені рівняння

$$J_0\left(\gamma \frac{R_1}{R_2}\right)N_0(\gamma) - N_0\left(\gamma \frac{R_1}{R_2}\right)J_0(\gamma) = 0.$$

Скориставшись рівнянням (7.149), крайовими умовами (7.150) та значенням вронскіана[1]

$$W[J_0(x), N_0(x)] = J_0(x)N_0'(x) - J_0'(x)N_0(x) = \frac{2}{\pi x},$$

---

[1] Нехай $\chi_1$ і $\chi_2$ — два лінійно незалежні розв'язки рівняння Бесселя $\chi''(x) + \frac{1}{x}\chi'(x) + \chi(x) = 0,$ $W\left(\chi_1, \chi_2\right)(x) = \begin{vmatrix} \chi_1 & \chi_2 \\ \chi_1' & \chi_2' \end{vmatrix} = \chi_1\chi_2' - \chi_2\chi_1'$ — їх вронскіан. Користуючись властивостями детермінантів та рівняннями для $\chi_1$ і $\chi_2$, знаходимо диференціальне рівняння для $W\left(\chi_1, \chi_2\right)$:

$$\frac{dW\left(\chi_1, \chi_2\right)}{dx} = \begin{vmatrix} \chi_1' & \chi_2' \\ \chi_1' & \chi_2' \end{vmatrix} + \begin{vmatrix} \chi_1 & \chi_2 \\ \chi_1'' & \chi_2'' \end{vmatrix} = \begin{vmatrix} \chi_1 & \chi_2 \\ \chi_1'' & \chi_2'' \end{vmatrix} = \begin{vmatrix} \chi_1 & \chi_2 \\ -\frac{1}{x}\chi_1' - \chi_1 & -\frac{1}{x}\chi_2' - \chi_2 \end{vmatrix} =$$
$$= -\frac{1}{x}\begin{vmatrix} \chi_1 & \chi_2 \\ \chi_1' & \chi_2' \end{vmatrix} - \begin{vmatrix} \chi_1 & \chi_2 \\ \chi_1 & \chi_2 \end{vmatrix} = -\frac{1}{x}W\left(\chi_1, \chi_2\right).$$

Звідси $W\left(\chi_1, \chi_2\right)(x) = \frac{C}{x}$, де $C \neq 0$ — стала. Фактично ми отримали формулу (7.52). Сталу $C$ обчислюємо, наприклад, за допомогою асимптотик (7.83).

переконайтеся, що функції (7.151) попарно ортогональні з вагою $r$ та задовольняють співвідношення

$$\int_{R_1}^{R_2} r\left[\Phi_k(r)\right]^2 dr = C_k^2\left\{\frac{R_2^2}{2}\left[F_0'\left(\sqrt{\lambda_k}R_2\right)\right]^2 - \frac{R_1^2}{2}\left[F_0'\left(\sqrt{\lambda_k}R_1\right)\right]^2\right\},$$

тобто при $C_k^2 = \dfrac{\pi^2}{2}\dfrac{\lambda_k J_0^2\left(\sqrt{\lambda_k}R_2\right)}{J_0^2\left(\sqrt{\lambda_k}R_1\right) - J_0^2\left(\sqrt{\lambda_k}R_2\right)}$ вони ортонормовані з вагою $r$:

$$\int_{R_1}^{R_2} r\Phi_n(r)\Phi_k(r)\,dr = \delta_{nk}.$$

*Відповідь*:

$$u(r,t) = \frac{\pi^2}{2}\sum_{k=1}^{\infty}\frac{\lambda_k J_0^2\left(\sqrt{\lambda_k}R_2\right)}{J_0^2\left(\sqrt{\lambda_k}R_1\right) - J_0^2\left(\sqrt{\lambda_k}R_2\right)}\left[\int_{R_1}^{R_2} r'T_0(r')F_0\left(\sqrt{\lambda_k}\,r'\right)dr'\right]e^{-\lambda_k a^2 t}F_0\left(\sqrt{\lambda_k}\,r\right),$$

де $F_0\left(\sqrt{\lambda_k}\,r\right) \equiv J_0\left(\sqrt{\lambda_k}R_1\right)N_0\left(\sqrt{\lambda_k}r\right) - N_0\left(\sqrt{\lambda_k}R_1\right)J_0\left(\sqrt{\lambda_k}\,r\right).$

Як приклад задачі, поставленої для циліндричної області, але з порушеною аксіальною симетрією, розглянемо остигання довгого однорідного стержня з початковою температурою $T_0(r,\alpha)$. Уважаючи, як і раніше, що поверхня стержня підтримується при нульовій температурі, а вісь стержня збігається з полярною віссю $OZ$, для визначення його температури $u = u(r,\alpha,t)$ маємо таку крайову задачу:

$$\frac{\partial u}{\partial t} = a^2\left[\frac{1}{r}\frac{\partial}{\partial r}\left(r\frac{\partial u}{\partial r}\right) + \frac{1}{r^2}\frac{\partial^2 u}{\partial \alpha^2}\right],\ \ 0 < r < R,\ \ 0 \le \alpha < 2\pi,\ \ t > 0, \qquad (7.152)$$

$$u(R,\alpha,t) = 0,\ \ \left|u(0,\alpha,t)\right| < \infty, \qquad (7.153)$$

$$u(r,\alpha,0) = T_0(r,\alpha), \qquad (7.154)$$

$$u(r,\alpha,t) = u(r,\alpha + 2\pi,t). \qquad (7.155)$$

Шукаємо нетривіальні розв’язки рівняння (7.152) у вигляді добутку $u(r,\alpha,t) = T(t)\Phi(r,\alpha)$. Відокремивши часову і координатні змінні, для функції $T(t)$ дістаємо рівняння (7.134), а для функції $\Phi(r,\alpha)$ — крайову задачу (7.87)–(7.89). З огляду на результати підрозділу 7.3 відразу можемо стверджувати, що загальний розв’язок крайової задачі (7.152)–(7.155) дається рядом

$$u(r,\alpha,t) = \sum_{k=1}^{\infty} A_{0k}^{(1)}\, e^{-\lambda_{0k}a^2 t}\Phi_{0k}^{(1)}(r) + \sum_{i=1}^{2}\sum_{m=1}^{\infty}\sum_{k=1}^{\infty} A_{mk}^{(i)}\, e^{-\lambda_{mk}a^2 t}\Phi_{mk}^{(i)}(r,\alpha), \qquad (7.156)$$

де (див. формулу (7.101)) власні значення $\lambda_{mk} = \left(\alpha_k^{(m)}\right)^2 \Big/ R^2$, $\alpha_k^{(m)}$ — додатні нулі функції $J_m(x)$, розташовані в порядку зростання, і власні функції $\Phi_{mk}^{(i)}(r,\alpha)$ даються формулою (7.113). Коефіцієнти в цих рядах знаходимо з початкової умови

$$\sum_{k=1}^{\infty} A_{0k}^{(1)} \Phi_{0k}^{(1)}(r) + \sum_{i=1}^{2}\sum_{m=1}^{\infty}\sum_{k=1}^{\infty} A_{mk}^{(i)} \Phi_{mk}^{(i)}(r,\alpha) = T_0(r,\alpha),$$

скориставшись умовою ортонормованості (7.114) для функцій $\Phi_{mk}^{(i)}(r,\alpha)$:

$$A_{0k}^{(1)} = \int_D T_0(r,\alpha)\Phi_{0k}^{(1)}(r)\,dS, \quad k \ge 1,$$

$$A_{mk}^{(i)} = \int_D T_0(r,\alpha)\Phi_{mk}^{(i)}(r,\alpha)\,dS, \quad i = 1,2,\ m,k \ge 1.$$

Розписавши ці коефіцієнти докладно за допомогою формул (7.113), після незначних перетворень ряд (7.156) можемо подати у вигляді

$$u(r,\alpha,t) = \sum_{k=1}^{\infty} B_k\, e^{-\lambda_{0k} a^2 t} J_0\left(\frac{\alpha_k^{(0)}}{R} r\right) +$$
$$+ \sum_{m=1}^{\infty}\sum_{k=1}^{\infty} \left[C_{mk} \cos m\alpha + D_{mk} \sin m\alpha\right] e^{-\lambda_{mk} a^2 t} J_m\left(\frac{\alpha_k^{(m)}}{R} r\right), \qquad (7.157)$$

де

$$B_k = \frac{1}{\pi R^2 \left[J_0'\left(\alpha_k^{(0)}\right)\right]^2} \int_0^R dr\, r \int_0^{2\pi} d\alpha\, T_0(r,\alpha) J_0\left(\frac{\alpha_k^{(0)}}{R} r\right), \ k \ge 1,$$

$$C_{mk} = \frac{2}{\pi R^2 \left[J_m'\left(\alpha_k^{(m)}\right)\right]^2} \int_0^R dr\, r \int_0^{2\pi} d\alpha\, T_0(r,\alpha) \cos m\alpha\, J_m\left(\frac{\alpha_k^{(m)}}{R} r\right), \quad m,k \ge 1,$$

$$D_{mk} = \frac{2}{\pi R^2 \left[J_m'\left(\alpha_k^{(m)}\right)\right]^2} \int_0^R dr\, r \int_0^{2\pi} d\alpha\, T_0(r,\alpha) \sin m\alpha\, J_m\left(\frac{\alpha_k^{(m)}}{R} r\right), \quad m,k \ge 1.$$

Якщо початкова температура не залежить від кута $\alpha$, то $C_{mk} = D_{mk} = 0$, і ми від (7.157) повертаємося до розв'язку (7.148).

**Завдання 7.4.4.** Знайдіть закон остигання довгого однорідного бруса з напівкруглим поперечним перерізом $\tilde{D}: \left\{0 \le r \le R, 0 \le \alpha \le \pi\right\}$, якщо з моменту $t = 0$, коли його температура дорівнює $T_0(r,\alpha)$, поверхня бруса підтримується при нульовій температурі.

*Вказівка*. Крайові умови (7.153) треба доповнити умовами $u(r,0,t)=u(r,\pi,t)=0$.

*Відповідь*: $u(r,\alpha,t)=\sum_{m=1}^{\infty}\sum_{k=1}^{\infty} D_{mk}\, e^{-\lambda_{mk}a^2t}\sin m\alpha\, J_m\left(\frac{\alpha_k^{(m)}}{R}r\right)$,

де $\lambda_{mk}=\left(\alpha_k^{(m)}\right)^2\Big/R^2$, $\alpha_k^{(m)}$ — послідовні додатні нулі функції $J_m(x)$,

$$D_{mk}=\frac{4}{\pi R^2\left[J_m'\left(\alpha_k^{(m)}\right)\right]^2}\int_0^R dr r\int_0^\pi d\alpha T_0(r,\alpha)\sin m\alpha\, J_m\left(\frac{\alpha_k^{(m)}}{R}r\right).$$

## 7.5. ЗАДАЧІ ЗІ СФЕРИЧНОЮ СИМЕТРІЄЮ

Перейдемо до розгляду крайових задач теплопровідності і коливань, що характеризуються сферичною симетрією. Як приклад, розглянемо задачу про остигання однорідної кулі радіусом $R$ і з радіально-симетричним початковим розподілом температури $T_0(r)$, яке відбувається внаслідок того, що з моменту часу $t=0$ поверхня кулі підтримується при нульовій температурі. Якщо користуватися тривимірною декартовою системою координат, то крайова задача для миттєвої температури $u(x,y,z,t)$ точок кулі складається з однорідного рівняння теплопровідності

$$u_t=a^2\Delta u, \tag{7.158}$$

заданого у просторовій області $S:\left\{x^2+y^2+z^2<R^2\right\}$ при $t>0$, крайової умови

$$u\big|_\Gamma=0 \tag{7.159}$$

на поверхні кулі $\Gamma:\left\{x^2+y^2+z^2=R^2\right\}$ та початкової умови

$$u(x,y,z,0)=T_0\left(\sqrt{x^2+y^2+z^2}\right). \tag{7.160}$$

У рівнянні (7.158) $\Delta\equiv\frac{\partial^2}{\partial x^2}+\frac{\partial^2}{\partial y^2}+\frac{\partial^2}{\partial z^2}$ — оператор Лапласа у тривимірній декартовій системі координат.

Очевидно, що крайова умова (7.159) не дозволяє розділити декартові координати $x$, $y$ та $z$ безпосередньо. Тому, з огляду на симетрію кулі, перейдемо до сферичної системи координат з початком у центрі кулі. Перехід здійснюється за формулами

$$x = r\sin\theta\cos\alpha, \quad y = r\sin\theta\sin\alpha, \quad z = r\cos\theta,$$
$$0 \le r \le R, \quad 0 \le \theta \le \pi, \quad 0 \le \alpha < 2\pi, \tag{7.161}$$

його якобіан

$$\frac{D(x,y,z)}{D(r,\theta,\alpha)} = \begin{vmatrix} \frac{\partial x}{\partial r} & \frac{\partial x}{\partial \theta} & \frac{\partial x}{\partial \alpha} \\ \frac{\partial y}{\partial r} & \frac{\partial y}{\partial \theta} & \frac{\partial y}{\partial \alpha} \\ \frac{\partial z}{\partial r} & \frac{\partial z}{\partial \theta} & \frac{\partial z}{\partial \alpha} \end{vmatrix} =$$

$$= \begin{vmatrix} \sin\theta\cos\alpha & r\cos\theta\cos\alpha & -r\sin\theta\sin\alpha \\ \sin\theta\sin\alpha & r\cos\theta\sin\alpha & r\sin\theta\cos\alpha \\ \cos\theta & -r\sin\theta & 0 \end{vmatrix} = r^2\sin\theta. \tag{7.162}$$

Оператор $\Delta$ набирає вигляду

$$\Delta = \Delta_r + \frac{1}{r^2}\Delta_{\theta\alpha}, \tag{7.163}$$

де його радіальна $\Delta_r$ та кутова $\Delta_{\theta\alpha}$ частини даються виразами

$$\Delta_r = \frac{1}{r^2}\frac{\partial}{\partial r}\left(r^2\frac{\partial}{\partial r}\right), \tag{7.164}$$

$$\Delta_{\theta\alpha} = \frac{1}{\sin\theta}\frac{\partial}{\partial\theta}\left(\sin\theta\frac{\partial}{\partial\theta}\right) + \frac{1}{\sin^2\theta}\frac{\partial^2}{\partial\alpha^2}. \tag{7.165}$$

Отже, у сферичних координатах однорідне рівняння теплопровідності (7.158) має вигляд

$$\frac{\partial u}{\partial t} = a^2\left[\frac{1}{r^2}\frac{\partial}{\partial r}\left(r^2\frac{\partial u}{\partial r}\right) + \frac{1}{r^2\sin\theta}\frac{\partial}{\partial\theta}\left(\sin\theta\frac{\partial u}{\partial\theta}\right) + \frac{1}{r^2\sin^2\theta}\frac{\partial^2 u}{\partial\alpha^2}\right], \tag{7.166}$$

де $u = u(r,\theta,\alpha,t)$ — температура кулі як функція цих координат і часу. Крайова умова (7.159) та початкова умова (7.160) записуються в цих координатах наступним чином:

$$u(R,\theta,\alpha,t) = 0, \tag{7.167}$$

$$u(r,\theta,\alpha,0) = T_0(r). \tag{7.168}$$

Візьмемо тепер до уваги той факт, що в точках, де $r=0$ і (або) $\theta=0,\pi,$ якобіан (7.162) перетворень (7.161) дорівнює нулю, тобто порушується взаємно-однозначна відповідність між декартовою та сферичною системами координат, а тому шукана функція $u(r,\theta,\alpha,t)$ може мати сингулярності. Раніше вже зазначалося (див. підрозділ 7.2), що якщо такі точки фізично ніяк не виділені (у них не зосереджуються джерела тепла, частинок або коливань), то функція $u(r,\theta,\alpha,t)$ має залишатися в них неперервною, і щоб добитися виконання цієї умови, достатньо вимагати, щоб у цих точках $u(r,\theta,\alpha,t)$ була обмеженою:

$$|u(0,\theta,\alpha,t)|<\infty,\ |u(r,0,\alpha,t)|<\infty,\ |u(r,\pi,\alpha,t)|<\infty. \qquad (7.169)$$

У загальному випадку також повинна справджуватися умова періодичності

$$u(r,\theta,\alpha,t)=u(r,\theta,\alpha+2\pi,t). \qquad (7.170)$$

Перейдемо до розв'язування крайової задачі (7.166)–(7.170). Побудова розв'язку суттєво спрощується, якщо тепер узяти до уваги симетрію крайової та початкової умов. Оскільки останні, як і сама куля, ізотропні (не залежать від кутових змінних), джерела тепла відсутні, то немає жодних фізичних причин, які б порушували початкову ізотропність задачі. Природно виснувати, що кутові змінні в розглядуваній задачі є несуттєвими, а шукана температура залежить лише від радіальної змінної $r$ та часу $t$: $u=u(r,t)$. У цьому випадку $\Delta_{\theta\alpha}u=0$ і для функції $u(r,t)$ дістаємо таку крайову задачу на відрізку:

$$\frac{\partial u}{\partial t}=a^2\frac{1}{r^2}\frac{\partial}{\partial r}\left(r^2\frac{\partial u}{\partial r}\right),\ 0<r<R,\ t>0, \qquad (7.171)$$

$$u(R,t)=0,\ |u(0,t)|<\infty, \qquad (7.172)$$

$$u(r,0)=T_0(r). \qquad (7.173)$$

Шукаємо нетривіальні частинні розв'язки отриманої задачі у вигляді $u(r,t)=\Phi(r)T(t)$. Щоб відокремити змінні, підставляємо цей добуток у рівняння (7.171) та ділимо обидві частини одержуваної рівності на $a^2\Phi(r)T(t)$. Приходимо до співвідношення

$$\frac{T'(t)}{a^2T(t)}=\frac{1}{r^2\Phi(r)}\frac{d}{dr}\left[r^2\frac{d\Phi(r)}{dr}\right]=-\lambda,$$

звідки дістаємо два рівняння для часової та координатної частин:

$$T'(t) + \lambda a^2 T(t) = 0, \tag{7.174}$$

$$\frac{1}{r^2}\frac{d}{dr}\left[r^2\frac{d\Phi(r)}{dr}\right] + \lambda\Phi(r) = 0. \tag{7.175}$$

Далі з умов (7.172) знаходимо крайові умови для функції $\Phi(r)$:

$$\Phi(R) = 0, \quad |\Phi(0)| < \infty. \tag{7.176}$$

Розв’язок крайової задачі (7.175), (7.176) легко знайти, якщо скористатися підстановкою

$$\Phi(r) = \frac{\varphi(r)}{r}, \tag{7.177}$$

де $\varphi(r)$ — нова шукана функція. Оскільки

$$\frac{d\Phi(r)}{dr} = \frac{\varphi'(r)}{r} - \frac{\varphi(r)}{r^2},$$

$$\frac{d}{dr}\left[r^2\frac{d\Phi(r)}{dr}\right] = \frac{d}{dr}\left[r\varphi'(r) - \varphi(r)\right] = r\varphi''(r),$$

то з рівняння (7.175) та формули (7.177) знаходимо рівняння для $\varphi(r)$:

$$\varphi''(r) + \lambda\varphi(r) = 0. \tag{7.178}$$

З умови обмеженості функції $\Phi(r)$ у нулі випливає (див. другу формулу (7.176) та формулу (7.177)), що $\varphi(r) \to 0$ при $r \to 0.$ Беручи також до уваги першу формулу (7.176), для функції $\varphi(r)$ дістаємо крайові умови

$$\varphi(0) = 0, \quad \varphi(R) = 0. \tag{7.179}$$

З точністю до позначень задача (7.178), (7.179) повністю збігається з крайовою задачею Штурма — Ліувілля для закріпленої однорідної струни довжиною $R$ (див. завдання 4.10.2). Відразу можемо стверджувати, що власні значення задачі (7.178), (7.179) невироджені та даються формулою

$$\lambda_k = \frac{\pi^2 k^2}{R^2}, \quad k = 1, 2, .... \tag{7.180}$$

Їм відповідають власні функції

$$\varphi_k(r) = C_k \sin\frac{\pi k r}{R}, \tag{7.181}$$

де $C_k$ — сталі.

Повертаючись до крайової задачі (7.175), (7.176), бачимо, що її власні значення теж даються формулою (7.180), а відповідні власні функції мають вигляд

$$\Phi_k(r) = \frac{C_k}{r}\sin\frac{\pi k r}{R}. \tag{7.182}$$

Нормуємо ці функції умовою

$$\int\limits_S \Phi_k^2(r)\,dV = 1, \tag{7.183}$$

де інтеграл береться по всій досліджуваній області $S$. Маємо:

$$\int\limits_0^R dr\,r^2\int\limits_0^\pi d\theta\sin\theta\int\limits_0^{2\pi} d\alpha\frac{C_k^2}{r^2}\sin^2\frac{\pi k r}{R} = 4\pi C_k^2\int\limits_0^R dr\sin^2\frac{\pi k r}{R} = 2\pi R C_k^2 = 1,$$

тобто

$$C_k = \frac{1}{\sqrt{2\pi R}}. \tag{7.184}$$

Очевидно, що власні функції (7.182), (7.184) задовольняють умову ортонормованості

$$\int\limits_S \Phi_n(r)\Phi_k(r)\,dV = \delta_{nk} \tag{7.185}$$

та утворюють повну систему функцій.

З рівняння (7.174) з точністю до сталих коефіцієнтів $A_k$ знаходимо часові функції:

$$T_k(t) = A_k e^{-\frac{\pi^2 k^2 a^2}{R^2}t},\quad k = 1,2,.... \tag{7.186}$$

Отже, частинні розв’язки рівняння теплопровідності (7.171), що задовольняють крайові умови (7.172), описуються виразами

$$u_k(r,t) = T_k(t)\Phi_k(r) = \frac{A_k}{\sqrt{2\pi R}}e^{-\frac{\pi^2 k^2 a^2}{R^2}t}\frac{1}{r}\sin\frac{\pi k r}{R}. \tag{7.187}$$

Розв’язок задачі (7.171)–(7.173) про остигання радіально-симетрично нагрітої кулі будуємо у вигляді їх лінійної суперпозиції:

$$u(r,t) = \sum_{k=1}^{\infty} c_k T_k(t)\Phi_k(r) = \sum_{k=1}^{\infty} a_k e^{-\frac{\pi^2 k^2 a^2}{R^2}t}\Phi_k(r). \tag{7.188}$$

Коефіцієнти $a_k$ у цій формулі знаходимо з початкової умови (7.173). Маємо:

$$u(r,0)=\sum_{k=1}^{\infty}a_k\Phi_k(r)=T_0(r),$$

звідки, з огляду на умову ортонормованості (7.185),

$$a_k=\int_S T_0(r)\Phi_k(r)dV=\int_0^R drr^2T_0(r)\int_0^{\pi}d\theta\sin\theta\int_0^{2\pi}d\alpha\frac{1}{r\sqrt{2\pi R}}\sin\frac{\pi kr}{R}=$$

$$=\frac{4\pi}{\sqrt{2\pi R}}\int_0^R drrT_0(r)\sin\frac{\pi kr}{R}.$$

Підставивши коефіцієнти $a_k$ у формулу (7.188), скориставшись явним виглядом (7.182), (7.184) функцій $\Phi_k(r)$ та виконавши необхідні скорочення, остаточно знаходимо:

$$u(r,t)=\frac{2}{rR}\sum_{k=1}^{\infty}\left[\int_0^R dr'r'T_0(r')\sin\frac{\pi kr'}{R}\right]e^{-\frac{\pi^2k^2a^2}{R^2}t}\sin\frac{\pi kr}{R}. \qquad (7.189)$$

Зокрема, якщо в початковий момент часу куля була нагріта рівномірно, $T_0(r)=T_0$, то

$$a_k=T_0\frac{4\pi}{\sqrt{2\pi R}}\int_0^R drr\sin\frac{\pi kr}{R}=T_0\frac{4\pi}{\sqrt{2\pi R}}\frac{R^2}{\pi k}(-1)^{k+1}$$

і, відповідно,

$$u(r,t)=\frac{2T_0R}{\pi r}\sum_{k=1}^{\infty}\frac{(-1)^{k+1}}{k}e^{-\frac{\pi^2k^2a^2}{R^2}t}\sin\frac{\pi kr}{R}.$$

При дії всередині кулі теплових джерел зі сферично симетричною густиною потужності $F(r,t)$ природно виникає крайова задача для неоднорідного рівняння теплопровідності $(f(r,t)=F(r,t)/(c\rho))$

$$\frac{\partial u}{\partial t}=a^2\frac{1}{r^2}\frac{\partial}{\partial r}\left(r^2\frac{\partial u}{\partial r}\right)+f(r,t),\ 0<r<R,\ t>0, \qquad (7.190)$$

при крайових умовах (7.172) та, без обмеження загальності, нульовій початковій умові $u(r,0)=0$. Розв’язуємо її методом відокремлення змінних за схемою, повністю аналогічною схемі розв’язування неоднорідних крайових задач в одновимірному випадку: подаємо шуканий розв’язок та неоднорідний член $f(r,t)$ у вигляді рядів Фур’є

$$u(r,t)=\sum_{k=1}^{\infty}\theta_k(t)\Phi_k(r),$$

$$f(r,t)=\sum_{k=1}^{\infty}f_k(t)\Phi_k(r),\ \ f_k(t)=\int_S f(r,t)\Phi_k(r)dV=4\pi\int_0^R f(r,t)\Phi_k(r)r^2dr,$$

за власними функціями $\Phi_k(r)$ крайової задачі (7.175), (7.176), підставляємо ці ряди в рівняння (7.190) і (нульову) початкову умову, отримуємо для коефіцієнтів $\theta_k(t)$ звичайне диференціальне рівняння (6.127) та початкову умову (6.128), а звідси і явні вирази (6.129) для цих коефіцієнтів.

Наведений вище аналіз крайових задач про теплопровідність (дифузію) у системах зі сферичною симетрією легко узагальнюється і на крайові задачі про радіальні коливання газів (рідин) усередині порожнистих сферично-симетричних об'ємів. З математичного погляду розв'язки цих задач відрізняються лише структурою часових частин.

**Завдання 7.5.1.** Знайдіть власні частоти, власні функції і загальний розв'язок задачі про радіальні коливання газу всередині порожнистої сфери з твердою оболонкою.

*Вказівка.* Нехай $u(r,t)$ — потенціал радіальної швидкості газу в момент часу $t$ в точках, розташованих на відстані $r$ від центра сфери радіусом $R$. Крайова задача для функції $u(r,t)$ має вигляд

$$\frac{\partial^2 u}{\partial t^2} = a^2 \frac{1}{r^2}\frac{\partial}{\partial r}\left(r^2 \frac{\partial u}{\partial r}\right), \quad 0 < r < R, \quad t > 0,$$

$$\left.\frac{\partial u(r,t)}{\partial r}\right|_{r=R} = 0, \quad |u(0,t)| < \infty,$$

$$u(r,0) = u_0(r), \quad \left.\frac{\partial u(r,t)}{\partial t}\right|_{t=0} = v_0(r),$$

де $u_0(r)$ і $v_0(r)$ — початкові значення потенціалу та його часової похідної.

*Відповідь.* Власні частоти $\omega_k = a\sqrt{\lambda_k} = a\gamma_k / R$, де $\lambda_k = \gamma_k^2 / R^2$ — власні значення, $\gamma_k$ — послідовні додатні корені рівняння $\operatorname{tg}\gamma = \gamma$; при $k >> 1$ $\gamma_k \approx \frac{\pi}{2} + \pi k$. Нормовані власні функції $\Phi_k(r) = \frac{1}{\sqrt{2\pi R}} \frac{\sqrt{1+\gamma_k^2}}{\gamma_k} \frac{1}{r} \sin\frac{\gamma_k r}{R}$. Розв'язок крайової задачі

$$u(r,t) = \frac{2}{rR}\sum_{k=1}^{\infty}\left(1 + \frac{1}{\gamma_k^2}\right)\left[\left(\int_0^R r' u_0(r') \sin\frac{\gamma_k r'}{R} dr'\right)\cos\frac{a\gamma_k t}{R} + \right.$$

$$\left. + \frac{R}{a\gamma_k}\left(\int_0^R r' v_0(r') \sin\frac{\gamma_k r'}{R} dr'\right)\sin\frac{a\gamma_k t}{R}\right]\sin\frac{\gamma_k r}{R}$$

(член виду $A + Bt$ опущено).

**Завдання 7.5.2.** Знайдіть власні частоти радіальних коливань газу між твердими концентричними сферичними оболонками, якщо зовнішній радіус меншої оболонки становить $R_1$, а внутрішній радіус більшої — $R_2$.

*Вказівка.* Крайові умови для потенціалу швидкості мають вигляд

$$\left.\frac{\partial u(r,t)}{\partial r}\right|_{r=R_1}=0,\quad \left.\frac{\partial u(r,t)}{\partial r}\right|_{r=R_2}=0.$$

*Відповідь:* $\omega_k = a\gamma_k/(R_2 - R_1)$, де $\gamma_k$ — послідовні додатні корені рівняння $\left[1+\frac{R_1R_2}{(R_2-R_1)^2}\gamma^2\right]\sin\gamma=\gamma\cos\gamma.$

**Завдання 7.5.3.** Однорідна куля радіусом $R$ має в усіх точках однакову температуру $T_0$. Знайдіть закон остигання кулі, якщо з моменту часу $t=0$ на поверхні кулі відбувається теплообмін за законом Ньютона з навколишнім середовищем, яке має нульову температуру.

*Вказівки.* Крайова умова на поверхні кулі: $\left[\frac{\partial u(r,t)}{\partial r}+hu(r,t)\right]_{r=R}=0.$ Власні значення $\lambda_k=\gamma_k^2/R^2$, де $\gamma_k$ — послідовні додатні корені рівняння $\gamma\cos\gamma+(hR-1)\sin\gamma=0.$ Власні функції $\Phi_k(r)=$ $=\frac{1}{\sqrt{2\pi R}}\left[\frac{\gamma_k^2+(hR-1)^2}{\gamma_k^2+(hR-1)hR}\right]^{1/2}\frac{1}{r}\sin\frac{\gamma_k r}{R}.$

*Відповідь*:

$$u(r,t)=\frac{2T_0hR^2}{r}\sum_{k=1}^{\infty}(-1)^{k+1}\frac{\sqrt{\gamma_k^2+(hR-1)^2}}{\gamma_k[\gamma_k^2+(hR-1)hR]}e^{-\frac{\gamma_k^2}{R^2}a^2t}\sin\frac{\gamma_k r}{R}.$$

**Завдання 7.5.4.** Усередині однорідної кулі з радіусом $R$ та нульовою початковою температурою діють джерела тепла, що мають сталу густину потужності $q$. Знайдіть температуру кулі в довільний момент часу, якщо на поверхні кулі: а) підтримується нульова температура; б) відбувається теплообмін за законом Ньютона з навколишнім середовищем, яке має нульову температуру. Яка температура встановлюється всередині кулі з часом?

*Вказівки*:

а) $f_k(t)=(-1)^{k+1}\frac{4qR^2}{kc\rho\sqrt{2\pi R}},\quad \theta_k(t)=(-1)^{k+1}\frac{4qR^4}{\pi^2k^3\kappa\sqrt{2\pi R}}\left(1-e^{-\frac{\pi^2k^2a^2}{R^2}t}\right),$

де $\kappa = a^2 c\rho$ — коефіцієнт теплопровідності речовини кулі. Температура кулі в довільний момент часу

$$u(r,t) = \frac{2qR^3}{\pi^3 \kappa r} \sum_{k=1}^{\infty} \frac{(-1)^{k+1}}{k^3} \left( 1 - e^{-\frac{\pi^2 k^2 a^2}{R^2} t} \right) \sin \frac{\pi k r}{R}.$$

Оскільки при $0 \le x \le \pi$ $\sum_{k=1}^{\infty} \frac{(-1)^{k+1}}{k^3} \sin kx = \frac{x}{12} (\pi^2 - x^2)$, то стаціонарна температура кулі при $t \to \infty$

$$u(r) = \frac{2qR^3}{\pi^3 \kappa r} \sum_{k=1}^{\infty} \frac{(-1)^{k+1}}{k^3} \sin \frac{\pi k r}{R} = \frac{qR^2}{6\kappa} \left( 1 - \frac{r^2}{R^2} \right).$$

Цю температуру можна знайти і як розв'язок стаціонарної задачі $\kappa \Delta_r u(r) + q = 0, \ |u(0)| < \infty, \ u(R) = 0.$

*Відповідь*:

б) $$u(r,t) = \frac{2qhR^4}{\kappa r} \sum_{k=1}^{\infty} (-1)^{k+1} \frac{\sqrt{\gamma_k^2 + (hR-1)^2}}{\gamma_k^3 [\gamma_k^2 + (hR-1)hR]} \left( 1 - e^{-\frac{\gamma_k^2}{R^2} a^2 t} \right) \sin \frac{\gamma_k r}{R}$$

(див. $\gamma_k$ у завданні 7.5.3). Стаціонарна температура $u(r) = \frac{qR^2}{6\kappa} \left( 1 - \frac{r^2}{R^2} \right) + \frac{qR}{3\kappa h}.$

## 7.6. КРАЙОВІ ЗАДАЧІ ДЛЯ СФЕРИЧНИХ ОБЛАСТЕЙ. ПОЛІНОМИ ЛЕЖАНДРА, СФЕРИЧНІ ФУНКЦІЇ БЕССЕЛЯ, ПРИЄДНАНІ ФУНКЦІЇ ЛЕЖАНДРА

Припустимо тепер, що в задачі про остигання кулі початкову умову (7.168) замінено умовою більш загального виду $u(r,\theta,\alpha,0) = T_0(r,\theta)$, де $T_0(r,\theta)$ — кусково-неперервна функція у сферичній області $S: \{0 \le r \le R, 0 \le \theta \le \pi, 0 \le \alpha < 2\pi\}$. При такій початковій умові розподіл температури в кулі в будь-який пізніший момент часу $t > 0$ характеризується вже не сферичною симетрією, а аксіальною, тобто описується функцією $u = u(r,\theta,t)$. Останню знаходимо як неперервний при $t > 0$ розв'язок наступної крайової задачі, поставленої для області $S$:

$$\frac{\partial u}{\partial t} = a^2 \left[ \frac{1}{r^2} \frac{\partial}{\partial r} \left( r^2 \frac{\partial u}{\partial r} \right) + \frac{1}{r^2 \sin\theta} \frac{\partial}{\partial \theta} \left( \sin\theta \frac{\partial u}{\partial \theta} \right) \right], \quad t > 0, \qquad (7.191)$$

$$u(R,\theta,t) = 0, \quad |u(0,\theta,t)| < \infty, \qquad (7.192)$$

$$|u(r,0,t)|<\infty,\quad |u(r,\pi,t)|<\infty, \tag{7.193}$$

$$u(r,\theta,0)=T_0(r,\theta). \tag{7.194}$$

Будуємо нетривіальні частинні розв'язки задачі (7.191)–(7.194) у вигляді добутку $u(r,t)=T(t)\Phi(r,\theta)$. Стандартним способом відокремивши часову і просторові змінні, для функції $\Phi(r,\theta)$ дістаємо крайову задачу

$$\frac{1}{r^2}\frac{\partial}{\partial r}\left(r^2\frac{\partial\Phi}{\partial r}\right)+\frac{1}{r^2\sin\theta}\frac{\partial}{\partial\theta}\left(\sin\theta\frac{\partial\Phi}{\partial\theta}\right)+\lambda\Phi=0,\quad 0<r<R,\quad 0<\theta<\pi, \tag{7.195}$$

$$\Phi(R,\theta)=0,\quad |\Phi(0,\theta)|<\infty, \tag{7.196}$$

$$|\Phi(r,0)|<\infty,\quad |\Phi(r,\pi)|<\infty, \tag{7.197}$$

де $\lambda$ — невідома стала. Подавши в рівнянні (7.195) $\Phi(r,\theta)$ у вигляді добутку $\Phi(r,\theta)=\chi(r)\vartheta(\theta)$ та поділивши отриману рівність на цей добуток, далі дістаємо:

$$\frac{1}{r^2\chi(r)}\frac{d}{dr}\left(r^2\frac{d\chi(r)}{dr}\right)+\frac{1}{r^2\vartheta(\theta)\sin\theta}\frac{d}{d\theta}\left(\sin\theta\frac{d\vartheta(\theta)}{d\theta}\right)+\lambda=0. \tag{7.198}$$

Змінні $r$ і $\theta$ легко відокремити, якщо покласти

$$\frac{1}{\vartheta(\theta)\sin\theta}\frac{d}{d\theta}\left(\sin\theta\frac{d\vartheta(\theta)}{d\theta}\right)=-\nu,$$

$\nu$ — нова стала. Звідси та з крайових умов (7.197) для кутової частини $\vartheta(\theta)$ дістаємо крайову задачу

$$\frac{1}{\sin\theta}\frac{d}{d\theta}\left(\sin\theta\frac{d\vartheta(\theta)}{d\theta}\right)+\nu\vartheta(\theta)=0,\quad 0<\theta<\pi, \tag{7.199}$$

$$|\vartheta(0)|<\infty,\quad |\vartheta(\pi)|<\infty, \tag{7.200}$$

а з рівності (7.198) і крайових умов (7.196) — крайову задачу для радіальної частини $\chi(r)$:

$$\frac{1}{r^2}\frac{d}{dr}\left(r^2\frac{d\chi}{\partial r}\right)+\left(\lambda-\frac{\nu}{r^2}\right)\chi=0,\quad 0<r<R, \tag{7.201}$$

$$\chi(R)=0,\quad |\chi(0)|<\infty. \tag{7.202}$$

Після підстановки $x=\cos\theta$, $\frac{d}{dx}=-\frac{1}{\sin\theta}\frac{d}{d\theta}$, задача (7.199), (7.200) набирає вигляду

$$\frac{d}{dx}\left[(1-x^2)\frac{d\vartheta(x)}{dx}\right]+\nu\vartheta(x)=0,\ \ -1<x<1, \qquad (7.203)$$

$$\left|\vartheta(-1)\right|<\infty,\ \ \left|\vartheta(1)\right|<\infty, \qquad (7.204)$$

тобто зводиться до відшукання нетривіальної функції, що є неперервною в інтервалі $(-1, 1)$, задовольняє рівняння Штурма — Ліувілля (7.203) з $p(x)=1-x^2$, $\rho(x)=1$, $q(x)\equiv 0$ (це рівняння також називають *рівнянням Лежандра*), та є обмеженою на кінцях цього інтервалу, де $p(x)=0$, а тому обмеженою на всьому відрізку $[-1, 1]$. Ситуація аналогічна до тої, з якою ми зустрілися при аналізі крайової задачі (7.41), (7.42), тож можна очікувати, що загальний розв'язок рівняння (7.203) включає внесок, сингулярний у точках $x=\pm 1$, власні значення задачі (7.203), (7.204) є невід'ємними, а відповідні власні функції — попарно ортогональними.

Шукаємо розв'язок задачі (7.203), (7.204) у вигляді степеневого ряду

$$\vartheta(x)=\sum_{s=0}^{\infty}a_s x^s. \qquad (7.205)$$

Підставивши його в рівняння (7.203), дістаємо рівність

$$\sum_{s=2}^{\infty}a_s s(s-1)x^{s-2}+\sum_{s=0}^{\infty}a_s\left[-s(s+1)+\nu\right]x^s=0,$$

або, після переходу в першій сумі до нового індексу підсумовування за правилом $s\to s+2$,

$$\sum_{s=0}^{\infty}\left\{a_{s+2}(s+1)(s+2)+a_s\left[-s(s+1)+\nu\right]\right\}x^s=0.$$

Очевидно, що всі коефіцієнти степеневого ряду зліва повинні дорівнювати нулю. Звідси дістаємо рекурентне співвідношення між коефіцієнтами ряду (7.205):

$$a_{s+2}=-\frac{\nu-s(s+1)}{(s+1)(s+2)}a_s,\ \ s=0,1,2,.... \qquad (7.206)$$

Якщо при будь-якому значенні $s$ чисельник у формулі (7.206) відмінний від нуля, то для довільного обмеженого числа $\nu$ існує такий індекс $s_0$, починаючи з якого всі коефіцієнти $a_s$ мають однаковий знак, а за модулем наближаються при $s\to\infty$ до модуля першого ненульового коефіцієнта ряду (7.205). Такий ряд є розбіжним при $x=1$, тому залишається припустити, що при деякому невід'ємному цілому $s=n$ ряд (7.205) обривається (коефіцієнт $a_{n+2}=0$), тобто зводиться до полінома степеня $n$. З рекурентного співвідношення (7.206) далі випливає, що якщо $n$ — парне число, то відмінними від нуля в цьому поліномі є лише коефіцієнти з парними номерами $(s = 0, 2, ..., n)$, а якщо непарне — то лише коефіцієнти з непарними номерами $(s = 1, 3, ..., n)$. Існує, очевидно, нескінченна сукупність таких поліномів. Без обмеження загальності перші ненульові коефіцієнти в них можна було би прирівняти до одиниці. Однак за традицією значення цих коефіцієнтів фіксуються умовою, щоб у точці $x = 1$ усі ці поліноми дорівнювали одиниці.

Таким чином, крайова задача (7.203), (7.204) має нетривіальні неперервні і, отже, обмежені розв'язки на $[-1, 1]$ лише за умови, що $\nu$ набуває значень

$$\nu_n = n(n+1), \quad n = 0, 1, 2, .... \tag{7.207}$$

Цим власним значенням відповідають власні функції

$$\vartheta_n(x) = c_n P_n(x), \tag{7.208}$$

де $c_n$ — сталі,

$$P_n(x) = \begin{cases} \sum_{s=0}^{n/2} a_{2s} x^{2s}, & \text{якщо } n \text{ — парне,} \\ \sum_{s=0}^{[n/2]} a_{2s+1} x^{2s+1}, & \text{якщо } n \text{ — непарне,} \end{cases} \tag{7.209}$$

– поліноми степенів $n \geq 0$[1] з коефіцієнтами, що визначаються співвідношеннями

$$a_0 \neq 0, \quad a_1 \neq 0, \quad a_{s+2} = -\frac{n(n+1) - s(s+1)}{(s+1)(s+2)} a_s, \quad 0 \leq s \leq n-2, \tag{7.210}$$

та нормовані умовою

$$P_n(1) = 1. \tag{7.211}$$

Ці поліноми називаються *поліномами Лежандра*. Графіки кількох із них зображено на рис. 7.2.

З формул (7.209)–(7.211) випливає, що поліноми Лежандра з парними та непарними номерами можна подати у вигляді $(n = 0, 1, 2, ...)$

$$P_{2n}(x) = \frac{(-1)^n}{2^{2n-1}} \sum_{s=0}^{n} \frac{(-1)^s (2n+2s-1)!}{(2s)!(n+s-1)!(n-s)!} x^{2s}, \tag{7.212}$$

$$P_{2n+1}(x) = \frac{(-1)^n}{2^{2n}} \sum_{s=0}^{n} \frac{(-1)^s (2n+2s+1)!}{(2s+1)!(n+s)!(n-s)!} x^{2s+1}. \tag{7.213}$$

---

[1] Нагадаємо, що символ $[n/2]$ позначає цілу частину числа $n/2$.

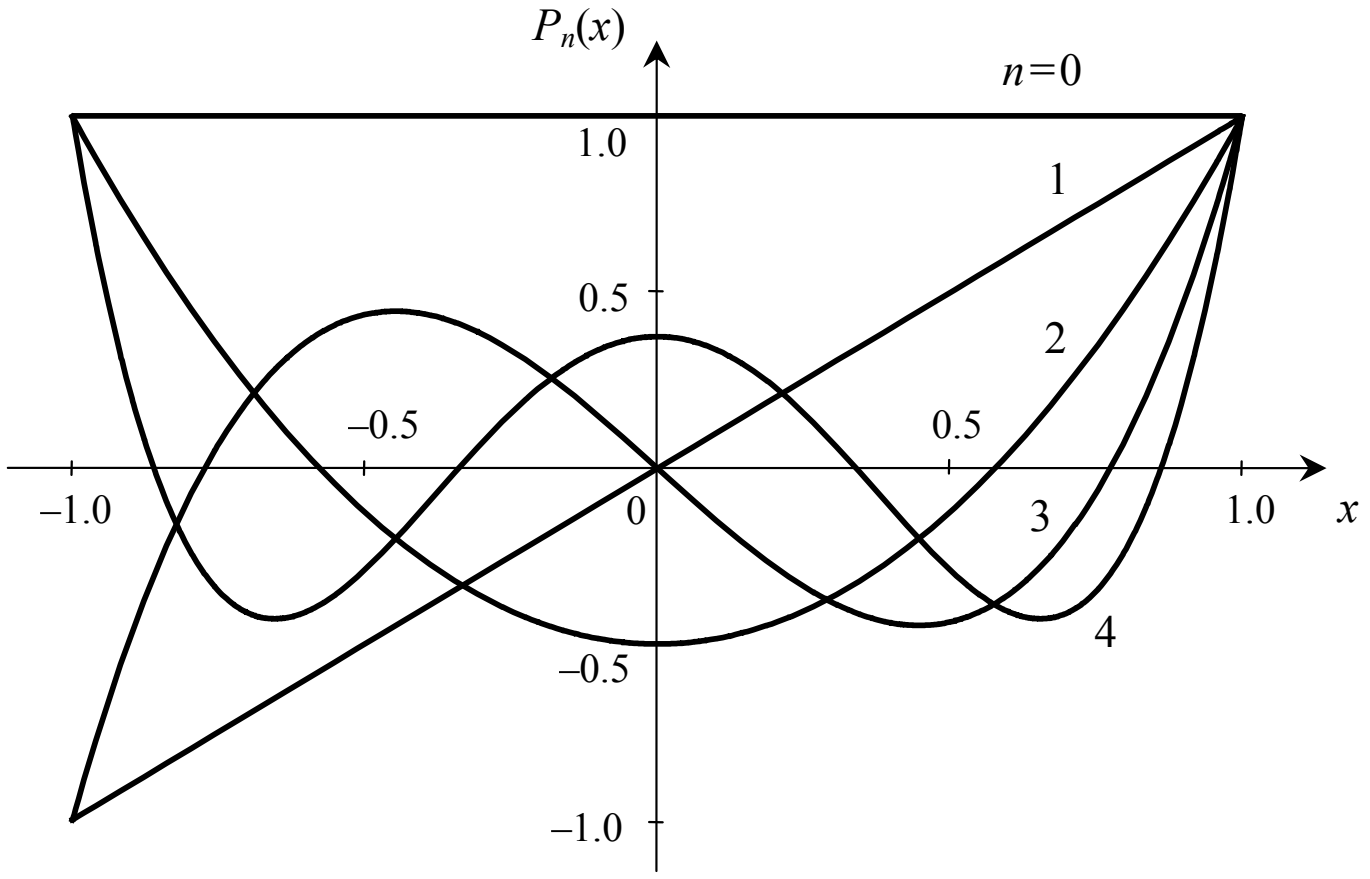

Рис. 7.2. Графіки поліномів Лежандра $P_n(x)$ ($n = 0 \div 4$) на відрізку $[-1,1]$

Ці дві формули можна об’єднати:

$$P_n(x) = \frac{1}{2^n} \sum_{s=0}^{[n/2]} \frac{(-1)^s (2n-2s)!}{s!(n-s)!(n-2s)!} x^{n-2s}. \qquad (7.214)$$

**Завдання 7.6.1.** Покажіть, що перші шість поліномів $P_n(x)$ описуються виразами

$$P_0(x) = 1, \quad P_1(x) = x, \quad P_2(x) = \frac{1}{2}(3x^2 - 1), \quad P_3(x) = \frac{1}{2}(5x^3 - 3x),$$
$$P_4(x) = \frac{1}{8}(35x^4 - 30x^2 + 3), \quad P_5(x) = \frac{1}{8}(63x^5 - 70x^3 + 15x). \qquad (7.215)$$

**Завдання 7.6.2.** Здиференціювавши $n$ разів біномний розклад $(x^2-1)^n = \sum_{s=0}^{n} \frac{(-1)^s n!}{s!(n-s)!} x^{2n-2s}$ та порівнявши результат з формулою (7.214), доведіть *формулу Родріга*

$$P_n(x) = \frac{1}{2^n n!} \frac{d^n}{dx^n}\left[(x^2-1)^n\right], \quad n = 0, 1, 2, \ldots. \qquad (7.216)$$

**Завдання 7.6.3.** Доведіть рекурентні співвідношення $(n = 1, 2, 3, \ldots)$

$$P_{n+1}(x) = \frac{2n+1}{n+1} x P_n(x) - \frac{n}{n+1} P_{n-1}(x), \qquad (7.217)$$

$$P_n(x) = \frac{1}{2n+1}\left[P'_{n+1}(x) - P'_{n-1}(x)\right]. \qquad (7.218)$$

*Вказівка.* Скористайтеся формулами (7.212) і (7.213), або формулою (7.216).

**Завдання 7.6.4.** Користуючись рівнянням (7.203), доведіть попарну ортогональність поліномів Лежандра на відрізку $[-1, 1]$:

$$\int_{-1}^{1} P_k(x)P_n(x)dx = 0, \text{ якщо } k \neq n. \qquad (7.219)$$

**Завдання 7.6.5.** Доведіть, що квадрат норми поліномів Лежандра

$$\left\|P_n(x)\right\|^2 \equiv \int_{-1}^{1}\left[P_n(x)\right]^2 dx = \frac{2}{2n+1}, \quad n = 0, 1, 2, .... \qquad (7.220)$$

*Вказівка.* З рекурентної формули (7.217) при $n = k-1$ та співвідношення ортогональності (7.219) випливає, що $\left\|P_k(x)\right\|^2 = \frac{2k-1}{k}\int_{-1}^{1} xP_k(x)P_{k-1}(x)dx.$ Виразивши $xP_k(x)$ знову з формули (7.217), записаної тепер для $n = k,$ для квадратів норми поліномів Лежандра виведіть рекурентне співвідношення $\left\|P_k(x)\right\|^2 = \frac{2k-1}{2k+1}\left\|P_{k-1}(x)\right\|^2,$ $k = 2, 3, 4, ....$ Безпосередньо обчисливши $\left\|P_0(x)\right\|^2 = 2$ та $\left\|P_1(x)\right\|^2 = \frac{2}{3},$ далі за індукцією доведіть (7.220).

Можна також в інтегралі (7.220) замінити $P_n(x)$ виразом (7.216) та застосувати інтегрування частинами $n$ разів. Тоді: $\left\|P_n(x)\right\|^2 = \frac{2(2n)!}{2^{2n}(n!)^2}\int_{0}^{1}(1-x^2)^n dx = \frac{2}{2n+1}.$

**Зауваження 7.6.1.** Для практичних обчислень із використанням сферичної системи координат співвідношення (7.219) і (7.220) зручно об'єднати та записати як

$$\int_{0}^{\pi} P_k(\cos\theta)P_n(\cos\theta)\sin\theta d\theta = \frac{2}{2n+1}\delta_{kn}. \qquad (7.221)$$

Говорячи про інші властивості поліномів Лежандра, зазначимо, що всі нулі поліномів $P_n(x)$ (з $n \geq 1$) та їх похідних $d^m P_n(x)/dx^m$ (при $m < n$) лежать усередині відрізка $[-1, 1]$ та є простими й дійсними.

Сукупність поліномів $P_n(x)$ утворює повну систему функцій, тобто будь-яка інша неперервна функція, яка визначена на $[-1,1]$ й ортогональна всім $P_n(x)$ $(n=0,1,2,...)$, тотожно дорівнює нулю. І, нарешті, довільній функції $f(x)$, визначеній щонайменше на відрізку $[-1,1]$, можна поставити у відповідність ряд

$$f(x)=\sum_{n=0}^{\infty}c_n P_n(x),\ -1\le x\le 1, \tag{7.222}$$

з коефіцієнтами

$$c_n=\frac{2n+1}{2}\int_{-1}^{1}f(x)P_n(x)dx, \tag{7.223}$$

який називається *рядом Лежандра* для функції $f(x)$ і є окремим прикладом узагальнених рядів Фур'є. Якщо $f(x)$ — кусково-гладка, то в інтервалі $(-1,1)$ її ряд Лежандра збігається до $f(x)$ у кожній точці неперервності $f(x)$ та до $\frac{1}{2}\left[f(x-0)+f(x+0)\right]$ у точках розриву $f(x)$, а на краях цього інтервалу — до значень $f(-1+0)$ при $x=-1$ та $f(1-0)$ при $x=1$.

Поліномами Лежандра $P_n(x)$ вичерпується множина неперервних та обмежених розв'язків крайової задачі (7.203), (7.204). Саме ж рівняння Лежандра (7.203) при $\nu=n(n+1)$ має, згідно із загальною теорією диференціальних рівнянь другого порядку, ще одну множину розв'язків $\tilde{\vartheta}_n(x)=\tilde{c}_n Q_n(x)$, лінійно незалежних від $P_n(x)$. За допомогою формул (7.51), (7.52) та явних виразів (7.215) для $P_n(x)$ легко знаходимо кілька перших функцій $Q_n(x)$:

$$\begin{aligned}
Q_0(x)&=\frac{1}{2}\ln\frac{1+x}{1-x},\\
Q_1(x)&=\frac{1}{2}x\ln\frac{1+x}{1-x}-1=xQ_0(x)-1=P_1(x)Q_0(x)-1,\\
Q_2(x)&=\frac{1}{4}(3x^2-1)\ln\frac{1+x}{1-x}-\frac{3}{2}x=\\
&=\frac{3}{2}xQ_1(x)-\frac{1}{2}Q_0(x)=P_2(x)Q_0(x)-\frac{3}{2}x.
\end{aligned} \tag{7.224}$$

Узагальнюючи ці результати, можна довести, що відповідним чином нормовані функції $Q_n(x)$ задовольняють такі самі рекурентні співвідношення, що й функції $P_n(x)$, зокрема

$$Q_{n+1}(x)=\frac{2n+1}{n+1}xQ_n(x)-\frac{n}{n+1}Q_{n-1}(x), \tag{7.225}$$

і що для всіх таких $Q_n(x)$ $(n=1,2,...)$ справджується формула

$$Q_n(x)=P_n(x)Q_0(x)-\sum_{s=0}^{[(n-1)/2]}\frac{(2n-4s-1)}{(2s+1)(n-s)}P_{n-2s-1}(x),\ |x|<1. \quad (7.226)$$

Функції $Q_n(x)$ називаються *функціями Лежандра другого роду*. Усі вони мають логарифмічні розбіжності при $x=\pm 1$[1].

**Завдання 7.6.6.** Доведіть формули (7.224).

Перейдемо тепер до розгляду крайової задачі (7.201), (7.202) для радіальної функції $\chi(r)$, де додатково врахуємо, що $\nu=n(n+1)$. За допомогою підстановки $\chi(r)=r^{-1/2}\varphi(r)$ вона зводиться до крайової задачі

$$\varphi''(r)+\frac{1}{r}\varphi'(r)+\left[\lambda-\frac{(n+1/2)^2}{r^2}\right]\varphi(r)=0,\quad 0<r<R, \quad (7.227)$$

$$\varphi(R)=0,\ \ \varphi(0)=0. \quad (7.228)$$

Рівняння (7.227) є узагальненням рівняння Бесселя (7.93) на випадок нецілих значень чисельника дробу, що стоїть усередині квадратних дужок, і аналізується методами, розглянутими в підрозділах 7.2, 7.3. Після заміни змінної $x=\sqrt{\lambda}\,r$ воно набирає вигляду

$$\varphi''(x)+\frac{1}{x}\varphi'(x)+\left(1-\frac{\mu^2}{x^2}\right)\varphi(x)=0, \quad (7.229)$$

де в нашому випадку $\mu^2=\left(n+1/2\right)^2$. Для довільного дійсного числа $\mu\geq 0$ загальний розв'язок рівняння (7.229) має вигляд

$$\varphi_\mu(x)=aJ_\mu(x)+bN_\mu(x), \quad (7.230)$$

де $a$ і $b$ — сталі, $J_\mu(x)$ і $N_\mu(x)$ — функції Бесселя і Неймана порядку $\mu$ (див. їх явні вирази у виносці до завдання 7.3.3). Цей розв'язок є обмеженим (і, зокрема, дорівнює нулю) при $x=0$ лише за умови $b=0$. Ураховуючи ще й першу крайову умову (7.228), робимо висновок, що власні значення і власні функції задачі (7.201), (7.202) при $\nu=n(n+1)$ даються формулами $(n=0,1,2,...,\ k=1,2,3,...)$

$$\lambda_{nk}=\frac{\left[\alpha_k^{(n+1/2)}\right]^2}{R^2}, \quad (7.231)$$

[1] Варто зазначити, що функції $Q_n(x)$, як і поліноми $P_n(x)$, можна продовжити на інтервал $x>1$, де вони, на відміну від $P_n(x)$, спадають до нуля при $x\to\infty$.

$$\chi_{nk}(r)=\frac{a_{nk}}{\sqrt{r}}J_{n+\frac{1}{2}}\left(\frac{\alpha_k^{(n+1/2)}}{R}r\right), \tag{7.232}$$

де $\alpha_k^{(n+1/2)}$ — додатні нулі функції Бесселя $J_{n+1/2}(x)$, розташовані в порядку зростання.

Функції виду (7.232) прийнято записувати в термінах так званих *сферичних функцій Бесселя* $j_n(x)$ *і Неймана* $y_n(x)$. Останні означаються співвідношеннями

$$j_n(x)=\sqrt{\frac{\pi}{2x}}J_{n+\frac{1}{2}}(x),\;\; y_n(x)=\sqrt{\frac{\pi}{2x}}Y_{n+\frac{1}{2}}(x). \tag{7.233}$$

Відповідно, загальний розв'язок рівняння (7.201) при $\nu=n(n+1)$, $n=0,1,2,...,$ записується у вигляді

$$\chi_n(r)=aj_n\left(\sqrt{\lambda}\,r\right)+by_n\left(\sqrt{\lambda}\,r\right), \tag{7.234}$$

а власні функції (7.232) крайової задачі (7.201), (7.202), які відповідають власним значенням (7.231), — у вигляді (з іншими сталими $a_{nk}$)

$$\chi_{nk}(r)=a_{nk}j_n\left(\frac{\alpha_k^{(n+1/2)}}{R}r\right),\;\; n=0,1,2,...,\;\; k=1,2,3,.... \tag{7.235}$$

**Завдання 7.6.7.** Виходячи з означення (7.233) та степеневих рядів для $J_{1/2}(x)$ і $N_{1/2}(x)$, покажіть, що сферичні функції $j_0(x)$ і $y_0(x)$ виражаються через елементарні функції:

$$j_0(x)=\frac{\sin x}{x},\;\; y_0(x)=-\frac{\cos x}{x}.$$

*Вказівка*. Для довільних додатних чисел $z$ справджуються рівності $\Gamma(z)\Gamma\left(z+\frac{1}{2}\right)=2^{1-2z}\sqrt{\pi}\,\Gamma(2z)$, $\Gamma(z+1)=z\Gamma(z)$.

**Завдання 7.6.8.** Доведіть співвідношення

$$\frac{d}{dx}\left[x^{n+1}j_n(x)\right]=x^{n+1}j_{n-1}(x),\;\; \frac{d}{dx}\left[x^{-n}j_n(x)\right]=-x^{-n}j_{n+1}(x),$$

$$j_{n+1}(x)=-j_{n-1}(x)+\frac{2n+1}{x}j_n(x),\;\; j_n'(x)=-j_{n+1}(x)+\frac{n}{x}j_n(x)$$

та такі самі співвідношення для $y_n(x)$. За їх допомогою покажіть, що

$$j_1(x)=\frac{\sin x}{x^2}-\frac{\cos x}{x},\;\; j_2(x)=\left(\frac{3}{x^3}-\frac{1}{x}\right)\sin x-\frac{3}{x^2}\cos x,$$

$$y_1(x)=-\frac{\cos x}{x^2}-\frac{\sin x}{x},\quad y_2(x)=-\left(\frac{3}{x^3}-\frac{1}{x}\right)\cos x-\frac{3}{x^2}\sin x.$$

**Завдання 7.6.9.** Доведіть співвідношення

$$\int_0^R j_n\left(\frac{\alpha_k^{(n+1/2)}}{R}r\right)j_n\left(\frac{\alpha_{k'}^{(n+1/2)}}{R}r\right)r^2dr=\delta_{kk'}\frac{\pi R^3}{4\alpha_k^{(n+1/2)}}\left[J'_{n+\frac{1}{2}}\left(\alpha_k^{(n+1/2)}\right)\right]^2. \quad (7.236)$$

Тепер ми спроможні виписати повну систему нормованих власних функцій крайової задачі (7.195)–(7.197):

$$\Phi_{nk}(r,\theta)=\frac{\left[(2n+1)\alpha_k^{(n+1/2)}\right]^{1/2}}{\pi R^{3/2}\left|J'_{n+1/2}\left(\alpha_k^{(n+1/2)}\right)\right|}j_n\left(\frac{\alpha_k^{(n+1/2)}}{R}r\right)P_n(\cos\theta),$$
$$n=0,1,2,\ldots,\quad k=1,2,3,\ldots. \quad (7.237)$$

Ці функції відповідають власним значенням (7.231) та задовольняють співвідношення

$$\int_S \Phi_{nk}(r,\theta)\Phi_{n'k'}(r,\theta)dV=\delta_{nn'}\delta_{kk'}. \quad (7.238)$$

За допомогою функцій (7.237) будуємо загальний розв'язок рівняння (7.191), який задовольняє крайові умови (7.192) і (7.193):

$$u(r,\theta,t)=\sum_{n=0}^{\infty}\sum_{k=1}^{\infty}c_{nk}e^{-\lambda_{nk}a^2t}\Phi_{nk}(r,\theta). \quad (7.239)$$

Коефіцієнти $c_{nk}$ у цьому ряді відновлюємо, скориставшись початковою умовою (7.194) та умовою ортонормованості (7.238):

$$c_{nk}=\int_S T_0(r,\theta)\Phi_{nk}(r,\theta)dV. \quad (7.240)$$

Після незначних обчислень і спрощень для розв'язку крайової задачі (7.191)–(7.194) остаточно маємо:

$$u(r,\theta,t)=\sum_{n=0}^{\infty}\sum_{k=1}^{\infty}A_{nk}e^{-\lambda_{nk}a^2t}j_n\left(\frac{\alpha_k^{(n+1/2)}}{R}r\right)P_n(\cos\theta), \quad (7.241)$$

де

$$A_{nk}=\frac{2(2n+1)\alpha_k^{(n+1/2)}}{\pi R^3\left[J'_{n+1/2}\left(\alpha_k^{(n+1/2)}\right)\right]^2}\times$$
$$\times\int_0^R dr'r'^2\int_0^\pi d\theta'\sin\theta' T_0(r',\theta')j_n\left(\frac{\alpha_k^{(n+1/2)}}{R}r'\right)P_n(\cos\theta'). \quad (7.242)$$

**Завдання 7.6.10.** Покажіть, що для радіально-симетричного розподілу початкової температури формули (7.241) і (7.242) зводяться до формули (7.189).

**Завдання 7.6.11.** Знайдіть стаціонарний розподіл температури всередині кулі радіусом $R$, поверхня якої підтримується при заданій температурі $T(\theta)$.

*Вказівки.* В області $0<r<R,\ \ 0<\theta<\pi$ треба знайти розв'язок $u=u(r,\theta)$ рівняння

$$\frac{1}{r^2}\frac{\partial}{\partial r}\left(r^2\frac{\partial u}{\partial r}\right)+\frac{1}{r^2\sin\theta}\frac{\partial}{\partial\theta}\left(\sin\theta\frac{\partial u}{\partial\theta}\right)=0, \qquad (7.243)$$

який задовольняє крайові умови

$$u(R,\theta)=T(\theta),\ \ |u(0,\theta)|<\infty,\ \ |u(r,0)|<\infty,\ \ |u(r,\pi)|<\infty.$$

Після підстановки $u(r,\theta)=\chi(r)\vartheta(\theta)$ і відокремлення змінних кутова частина $\vartheta(\theta)$ знаходиться як розв'язок крайової задачі (7.199), (7.200), а радіальна $\chi(r)$ — як розв'язок рівняння $\frac{d}{dr}\left(r^2\frac{d\chi(r)}{dr}\right)-n(n+1)\chi(r)=0$, обмежений при $r=0$. Лінійно незалежні розв'язки цього рівняння шукаються у вигляді $\chi(r)=Ar^{\alpha}$ ($A$, $\alpha$ — сталі) і дорівнюють $\chi_1(r)=A_1r^n$, $\chi_2(r)=A_2r^{-(n+1)}$, а тому найбільш загальний вираз для розв'язку рівняння (7.243), придатний і для зовнішньої області $r>R$, $0<\theta<\pi$, має структуру $u(r,\theta)=\sum_{n=0}^{\infty}\left(a_nr^n+\frac{b_n}{r^{n+1}}\right)P_n(\cos\theta)$.

*Відповідь*: $u(r,\theta)=\sum_{n=0}^{\infty}a_n\left(\frac{r}{R}\right)^n P_n(\cos\theta)$,

де $a_n=\left(n+\frac{1}{2}\right)\int_0^{\pi}T(\theta')P_n(\cos\theta')\sin\theta'd\theta'$.

На завершення проаналізуємо стисло ситуацію, коли внаслідок різних причин (асиметрії початкових умов, крайових умов, чи теплових джерел) у задачі про остигання кулі відсутні будь-які елементи симетрії. Тоді миттєвий розподіл температури всередині кулі описується функцією, що залежить від усіх просторових змінних: $u=u(r,\theta,\alpha,t)$. Природно виникає питання про побудову власних функцій відповідної крайової задачі, для чого спершу треба знайти частинні розв'язки рівняння (7.166), які задовольняють умови обмеженості (7.169) та умову періодичності (7.170). Знову подаючи шукані розв'язки у вигляді добутку $u(r,t)=T(t)\Phi(r,\theta,\alpha)$ і застосовуючи метод відокремлення змінних, для координатної частини дістаємо рівняння $(0 < r < R,\ \ 0 < \theta < \pi,\ \ 0 \le \alpha < 2\pi)$

$$\frac{1}{r^2}\frac{\partial}{\partial r}\left(r^2\frac{\partial\Phi}{\partial r}\right)+\frac{1}{r^2\sin\theta}\frac{\partial}{\partial\theta}\left(\sin\theta\frac{\partial\Phi}{\partial\theta}\right)+\frac{1}{r^2\sin^2\theta}\frac{\partial^2\Phi}{\partial\alpha^2}+\lambda\Phi=0, \qquad (7.244)$$

умови обмеженості[1]

$$\left|\Phi(0,\theta,\alpha)\right|<\infty,\ \ \left|\Phi(r,0,\alpha)\right|<\infty,\ \ \left|\Phi(r,\pi,\alpha)\right|<\infty \qquad (7.245)$$

та умову періодичності

$$\Phi(r,\theta,\alpha)=\Phi(r,\theta,\alpha+2\pi), \qquad (7.246)$$

де $\lambda$ — стала (для незалежних від часу розв'язків $\lambda = 0$). Шукаючи частинні розв'язки цієї задачі у вигляді добутку $\Phi(r,\theta,\alpha)=\chi(r)\vartheta(\theta)\psi(\alpha)$ та відокремлюючи змінні, дістаємо три звичайні диференціальні рівняння з відповідними додатковими умовами:

$$\frac{d^2\psi}{d\alpha^2}+\mu\psi=0,\ \ 0\le\alpha<2\pi, \qquad (7.247)$$

$$\psi(\alpha)=\psi(\alpha+2\pi); \qquad (7.248)$$

$$\frac{1}{\sin\theta}\frac{d}{d\theta}\left(\sin\theta\frac{d\vartheta(\theta)}{d\theta}\right)+\left(\nu-\frac{\mu}{\sin^2\theta}\right)\vartheta(\theta)=0,\ \ 0<\theta<\pi, \qquad (7.249)$$

$$\left|\vartheta(0)\right|<\infty,\ \ \left|\vartheta(\pi)\right|<\infty; \qquad (7.250)$$

$$\frac{1}{r^2}\frac{d}{dr}\left(r^2\frac{d\chi}{\partial r}\right)+\left(\lambda-\frac{\nu}{r^2}\right)\chi=0,\ \ 0<r<R, \qquad (7.251)$$

$$\left|\chi(0)\right|<\infty. \qquad (7.252)$$

Власні значення та нормовані власні функції задачі (7.247), (7.248) були знайдені в підрозділі 7.3 і даються формулами

$$\mu_m=m^2,\ \ m=0,1,2,..., \qquad (7.95)$$

$$\psi_0^{(1)}=\frac{1}{\sqrt{2\pi}},\quad m=0,$$
$$\psi_m^{(1)}(\alpha)=\frac{1}{\sqrt{\pi}}\cos m\alpha,\quad \psi_m^{(2)}(\alpha)=\frac{1}{\sqrt{\pi}}\sin m\alpha,\ \ m=1,2,.... \qquad (7.96)$$

---

[1] При постановці зовнішніх крайових задач ($r > R$) перша умова (7.245), зрозуміло, замінюється відповідною умовою на поверхні $r = R$.

Стандартною заміною змінної $x = \cos\theta$ задача (7.249), (7.250) зводиться до вигляду

$$\frac{d}{dx}\left[(1-x^2)\frac{d\vartheta(x)}{dx}\right]+\left(\nu-\frac{m^2}{1-x^2}\right)\vartheta(x)=0,\ \ -1<x<1, \tag{7.253}$$

$$\left|\vartheta(-1)\right|<\infty,\ \ \left|\vartheta(1)\right|<\infty. \tag{7.254}$$

Рівняння (7.253) узагальнює рівняння Лежандра (7.203) на випадок $m \neq 0$ і називається *приєднаним рівнянням Лежандра*. Враховуючи, що похідні $d^m P_n(x)/dx^m \equiv w(x)$ поліномів $P_n(x)$ задовольняють рівняння

$$\begin{gathered}(1-x^2)\frac{d^2w(x)}{dx^2}-2(m+1)x\frac{dw(x)}{dx}+\\ +\left[n(n+1)-m(m+1)\right]w(x)=0,\end{gathered} \tag{7.255}$$

безпосередньою підстановкою $\vartheta(x)=(1-x^2)^{m/2}w(x)$ далі переконуємося, що рівняння (7.253) перетворюється на тотожність за умови, що $\nu$ пробігає значення (7.207):

$$\nu_n = n(n+1),\ \ n = 0, 1, 2,.... \tag{7.207}$$

Відповідні лінійно незалежні розв'язки цього рівняння з точністю до сталої виражаються через функції

$$\begin{gathered}P_n^m(x)=(1-x^2)^{m/2}\frac{d^m P_n(x)}{dx^m},\\ Q_n^m(x)=(1-x^2)^{m/2}\frac{d^m Q_n(x)}{dx^m},\ \ m=0,1,2,...,n,\end{gathered} \tag{7.256}$$

які називаються *приєднаними функціями Лежандра першого та другого родів*.

Обмежені на відрізку $[-1,1]$ розв'язки крайової задачі (7.253), (7.254), що відповідають власним значенням (7.207), є неперервними на цьому відрізку і даються формулою

$$\vartheta_n^m(x)=c_{nm}P_n^m(x),\ \ n=0,1,2,...,\ \ m=0,1,2,...,n, \tag{7.257}$$

де $c_{nm}$ — сталі.

**Завдання 7.6.12.** Доведіть рівняння (7.255) та подальші твердження до формули (7.257) включно.

**Завдання 7.6.13.** Покажіть, що $P_n^0(x) = P_n(x)$, а кілька перших функцій $P_n^m(x)$ з $m > 0$ [1] мають вигляд:

$$P_1^1(x) = (1-x^2)^{1/2},\quad P_2^1(x) = 3x(1-x^2)^{1/2},\quad P_2^2(x) = 3(1-x^2),$$

$$P_3^1(x) = \frac{3}{2}(5x^2-1)(1-x^2)^{1/2},\quad P_3^2(x) = 15x(1-x^2),\quad P_3^3(x) = 15(1-x^2)^{3/2}.$$

**Завдання 7.6.14.** Доведіть рекурентні співвідношення:

$$P_{n+1}^m(x) = \frac{2n+1}{n-m+1} x P_n^m(x) - \frac{n+m}{n-m+1} P_{n-1}^m(x),$$

$$P_n^{m+1}(x) = 2mx(1-x^2)^{-1/2} P_n^m(x) - \left[n(n+1) - m(m-1)\right] P_n^{m-1}(x).$$

**Завдання 7.6.15.** Обчисліть інтеграли:

$$\int_{-1}^{1} P_k^m(x) P_n^m(x)\,dx = \delta_{kn} \frac{2(n+m)!}{(2n+1)(n-m)!},$$

$$\int_{0}^{\pi} P_k^m(\cos\theta) P_n^m(\cos\theta)\sin\theta\,d\theta = \delta_{kn} \frac{2(n+m)!}{(2n+1)(n-m)!}, \qquad (7.258)$$

$$\int_{-1}^{1} (1-x^2)^{-1} P_n^k(x) P_n^m(x)\,dx = \delta_{km} \frac{(n+m)!}{m(n-m)!}.$$

Повертаючись до задачі (7.249), (7.250) та беручи до уваги результати (7.207), (7.256)–(7.258), можемо вибрати значення коефіцієнтів $c_{nm}$ таким чином, щоб її власні функції (які відповідають власним значенням (7.207)) були нормовані на одиницю:

$$\vartheta_n^m(\theta) = \sqrt{\frac{(2n+1)(n-m)!}{2(n+m)!}}\, P_n^m(\cos\theta),\quad n = 0, 1, 2, \ldots,\quad m = 0, 1, 2, \ldots, n. \qquad (7.259)$$

Залишається згадати, що розв'язки задачі (7.251), (7.252) при $\nu = n(n+1)$, $n$ — невід'ємне ціле число, та $\lambda > 0$ були знайдені при розв'язуванні крайової задачі (7.201), (7.202) і мають вигляд

$$\chi_n(r) = a_n j_n\left(\sqrt{\lambda}\, r\right),\quad n = 0, 1, 2, \ldots. \qquad (7.260)$$

---

[1] Функції $P_n^m(x)$ можна продовжити і на від'ємні цілі $m = -n, -n+1, \ldots, -1$. Продовжені функції легко відновити через $P_n^m(x)$ з $m > 0$, скориставшись співвідношенням $P_n^{-|m|}(x) = (-1)^{|m|} \frac{(n-|m|)!}{(n+|m|)!} P_n^{|m|}(x)$.

Отже, усі лінійно незалежні обмежені (фактично — неперервні) періодичні розв'язки *рівняння Гельмгольца*

$$\Delta\Phi + \lambda\Phi = 0, \ \ \lambda > 0, \tag{7.244а}$$

для сферичної області $S$ вичерпуються функціями $\Phi_{nm}^{(i)}(r,\theta,\alpha) = \chi_n(r)\vartheta_n^m(\theta)\psi_m^{(i)}(\alpha), \ \ i = 1,2.$ У більш докладній формі $(n = 0, 1, 2,...)$:

$$\begin{aligned}
\Phi_{n0}^{(1)}(r,\theta) &= A_{n0} j_n\left(\sqrt{\lambda}\, r\right) P_n(\cos\theta), \qquad && m = 0,\\
\Phi_{nm}^{(1)}(r,\theta,\alpha) &= A_{nm} j_n\left(\sqrt{\lambda}\, r\right) P_n^m(\cos\theta)\cos m\alpha, && m = 1,2,...,n,\\
\Phi_{nm}^{(2)}(r,\theta,\alpha) &= B_{nm} j_n\left(\sqrt{\lambda}\, r\right) P_n^m(\cos\theta)\sin m\alpha, && m = 1,2,...,n,
\end{aligned} \tag{7.261}$$

де $A_{nm}$ і $B_{nm}$ — сталі.

Згадавши результати завдання 7.6.11, можемо також виписати розв'язки задачі (7.251), (7.252) при $\nu = n(n+1)$, $n$ — невід'ємне ціле число, та $\lambda = 0$:

$$\chi_n(r) = a_n r^n, \ \ n = 0, 1, 2,.... \tag{7.262}$$

Відповідно, усі лінійно незалежні обмежені (неперервні) періодичні розв'язки *рівняння Лапласа*

$$\Delta\Phi = 0 \tag{7.244б}$$

для сферичної області $S$ даються функціями $(n = 0, 1, 2,...)$

$$\begin{aligned}
\Phi_{n0}^{(1)}(r,\theta) &= A_{n0} r^n P_n(\cos\theta), \qquad && m = 0,\\
\Phi_{nm}^{(1)}(r,\theta,\alpha) &= A_{nm} r^n P_n^m(\cos\theta)\cos m\alpha, && m = 1,2,...,n,\\
\Phi_{nm}^{(2)}(r,\theta,\alpha) &= B_{nm} r^n P_n^m(\cos\theta)\sin m\alpha, && m = 1,2,...,n,
\end{aligned} \tag{7.263}$$

$A_{nm}$ і $B_{nm}$ — сталі.

Загальний розв'язок $u = u(r,\theta,\alpha,t)$ задачі про остигання кулі при асиметричному розподілі початкової температури (як і інших подібних лінійних крайових задач для рівнянь теплопровідності і коливань у сферичної області $S$) будується як лінійна комбінація функцій (7.261). Значення параметра $\lambda$ та числових коефіцієнтів у цій комбінації визначаються за допомогою крайових та початкових умов. При відшуканні незалежних від часу розв'язків $u = u(r,\theta,\alpha)$ задач для області $S$ використовується система функцій (7.263), а числові коефіці-єнти у відповідних лінійних комбінаціях визначаються за допомогою крайових умов.

**Завдання 7.6.16.** Побудуйте всі лінійно незалежні обмежені періодичні розв'язки рівнянь Гельмгольца й Лапласа для області $\tilde{S}:\{r \geq R, 0 \leq \theta \leq \pi, 0 \leq \alpha < 2\pi\}$, що спадають до нуля при $r \to \infty$.

*Відповідь.* У формулах (7.261) $A_{nm} j_n\left(\sqrt{\lambda}\, r\right)$ треба замінити на лінійну комбінацію $A_{nm} j_n\left(\sqrt{\lambda}\, r\right) + C_{mn} y_n\left(\sqrt{\lambda}\, r\right)$, а у формулах (7.263) $A_{nm} r^n$ — на $A_{nm} r^{-(n+1)}$. Аналогічно для $B_{nm} j_n\left(\sqrt{\lambda}\, r\right)$ і $B_{nm} r^n$.

**Завдання 7.6.17.** У момент часу $t = 0$ температура однорідної кулі радіусом $R$ описується функцією $T_0(r,\theta,\alpha)$. Знайдіть температуру кулі в довільний момент часу $t > 0$, якщо: а) поверхня кулі підтримується при нульовій температурі; б) поверхня кулі теплоізольована; в) на поверхні кулі відбувається теплообмін за законом Ньютона з навколишнім середовищем, яке має нульову температуру.

**Завдання 7.6.18.** Знайдіть стаціонарний розподіл температури всередині однорідної кулі радіусом $R$, поверхня якої підтримується при температурі $T(\theta,\alpha)$.

**Завдання 7.6.19.** Оцініть критичний радіус сферичного реактора з поглинаючою поверхнею.

*Відповідь*: $R_c = \pi\sqrt{D/\beta}$, $D$ — коефіцієнт дифузії нейтронів, $\beta$ — стала швидкості народження нейтронів.

**Зауваження 7.6.2.** Взаємно ортогональні функції

$$\begin{aligned} Y_{nm}^{(1)}(\theta,\alpha) &= P_n^m(\cos\theta)\cos m\alpha, \quad m = 0,1,2,\dots,n, \\ Y_{nm}^{(2)}(\theta,\alpha) &= P_n^m(\cos\theta)\sin m\alpha, \quad m = 1,2,\dots,n, \end{aligned} \tag{7.264}$$

називаються *фундаментальними сферичними функціями n-го порядку*. Згідно з попередніми результатами, вони визначають (з точністю до сталої) усі дійсні, обмежені (неперервні) та періодичні розв'язки рівняння

$$\Delta_{\theta\alpha} Y + \nu Y = 0, \;\; 0 \leq \theta \leq \pi, \;\; 0 \leq \alpha < 2\pi, \tag{7.265}$$

які відповідають власним значенням $\nu_n = n(n+1)$, $n = 0, 1, 2, \dots$. Ці власні значення вироджені з кратністю $2n+1$, тобто загальна кількість власних функцій (7.264), що відповідають заданому $\nu_n$, дорівнює $2n+1$.

У квантовій механіці природно виникає задача про відшукання обмежених та періодичних розв'язків рівняння (7.265) на множині комплекснозначних функцій. Доведіть самостійно, що власні значення цієї узагальненої задачі теж описуються формулою $\nu_n = n(n+1)$ і вироджені з кратністю $2n+1$, а відповідними власними функціями, нормованими на одиничній сфері, виступають так звані *сферичні гармоніки*

$$Y_n^m(\theta,\alpha) = \sqrt{\frac{2n+1}{4\pi}\frac{(n-m)!}{(n+m)!}}\,P_n^m(\cos\theta)e^{im\alpha},$$
$$n = 0, 1, 2, \ldots,\quad m = -n, -n+1, \ldots, n, \tag{7.266}$$

$$\int_0^{\pi} d\theta \sin\theta \int_0^{2\pi} d\alpha\, \overline{Y_n^m(\theta,\alpha)}\, Y_{n'}^{m'}(\theta,\alpha) = \delta_{nn'}\delta_{mm'}, \tag{7.267}$$

де риска означає комплексне спряження. Очевидно, що сферичні функції (7.264) можна розглядати як компоненти сферичних гармонік (7.266).

## 7.7. АСИМПТОТИЧНИЙ РОЗПОДІЛ ВЛАСНИХ ЗНАЧЕНЬ

У крайових задачах на власні значення для рівняння Штурма — Ліувілля на скінченних інтервалах або рівняння Гельмгольца в обмежених областях на площині чи у тривимірному просторі, які розглядалися в попередніх підрозділах, послідовні власні значення $\{\lambda_n\}$ утворювали обмежені знизу ($-\infty < \xi \le \lambda_n$) дискретні множини точок на дійсній осі, які скупчувалися лише на $+\infty$. Для подібних задач уводять неспадну функцію $\mathcal{N}(\lambda)$ на дійсній осі, означену як суму кратностей $m_n$ усіх власних значень $\lambda_n < \lambda$:

$$\mathcal{N}(\lambda) = \sum_{\lambda_n < \lambda} m_n. \tag{7.268}$$

Функція $\mathcal{N}(\lambda)$ називається *функцією розподілу власних значень*. За означенням (7.268) вона:

1) тотожно дорівнює нулю при $\lambda < \xi$;

2) є сталою у проміжках між послідовними власними значеннями;

3) має в точках $\lambda_n$ стрибки величиною $m_n$.

Звідси випливає, що для розглянутих нами крайових задач ($\xi \ge 0$) її можна подати у вигляді

$$\mathcal{N}(\lambda) = \int_0^{\lambda} \rho(\lambda')\,d\lambda', \tag{7.269}$$

де

$$\rho(\lambda) = \sum_n m_n \delta(\lambda - \lambda_n) \tag{7.270}$$

— так звана *спектральна густина*.

Оскільки вказані крайові задачі мають необмежену кількість власних значень із граничною точкою $+\infty$, то для них $\mathcal{N}(\lambda) \to \infty$ при $\lambda \to \infty$. Для низки важливих задач теоретичної фізики треба знати, як саме швидко зростає $\mathcal{N}(\lambda)$ з $\lambda$. Дослідженню асимптотичної поведінки $\mathcal{N}(\lambda)$ на нескінченності і присвячено цей підрозділ.

Почнемо з розгляду власних значень крайової задачі про коливання квадратної мембрани з вільними краями. Нагадаємо, що сукупність цих власних значень описується формулою[1]

$$\lambda_{jk} = \frac{\pi^2 a^2}{l^2}(j^2 + k^2), \tag{7.271}$$

де $l$ — довжина сторони мембрани, $j$ і $k$ — будь-які невід'ємні цілі числа. У цьому випадку функція $\mathcal{N}(\lambda) \equiv \mathcal{N}_{\text{в}}(\lambda)$ дорівнює, згідно з формулою (7.271), кількості різних пар невід'ємних цілих чисел $(j, k)$, що задовольняють умову

$$j^2 + k^2 < \frac{\lambda l^2}{\pi^2 a^2}.$$

Геометрична множина точок на площині з координатами $(x, y)$, такими, що

$$x^2 + y^2 < \frac{\lambda l^2}{\pi^2 a^2},$$

утворює круг із центром у початку координат і радіусом $l\sqrt{\lambda}/(\pi a)$. Тому $\mathcal{N}_{\text{в}}(\lambda)$ збігається з кількістю точок, які лежать у чверті цього круга, розташованій у першому квадранті декартової площини, та мають цілочислові, включаючи й нульові, координати.

Проведемо через усі точки декартової площини з цілочисловими координатами прямі, паралельні координатним осям (див. рис. 7.3). Тоді площина покриється сіткою, елементарними комірками якої є квадратики з одиничною стороною й, відповідно, одиничною площею. Оскільки кожний окремий вузол сітки є вершиною для чоти-рьох різних елементарних комірок, а вершинами кожної елементарної комірки є чотири різні вузли сітки, то можна кожному вузлу сітки поставити у взаємно-однозначну відповідність одну елементарну комірку, покривши всіма комірками всю площину. Зрозуміло, що при цьому мінімальна сума площ елементарних комірок, які цілком покривають довільну область $\Omega$ площини, є не меншою за кількість точок із цілочисловими координатами в області $\Omega$. Звідси робимо висновок, що число $\mathcal{N}_в(\lambda)$ є не меншим за площу чверті круга радіусом $l\sqrt{\lambda}/(\pi a)$, тобто

$$\frac{l^2\lambda}{4\pi a^2} \le \mathcal{N}_в(\lambda). \tag{7.272}$$

[1] У цьому підрозділі параметр $a^2$ включено у власні значення $\lambda_{jk}$. Останні, таким чином, набувають змісту квадратів власних частот коливань мембрани.

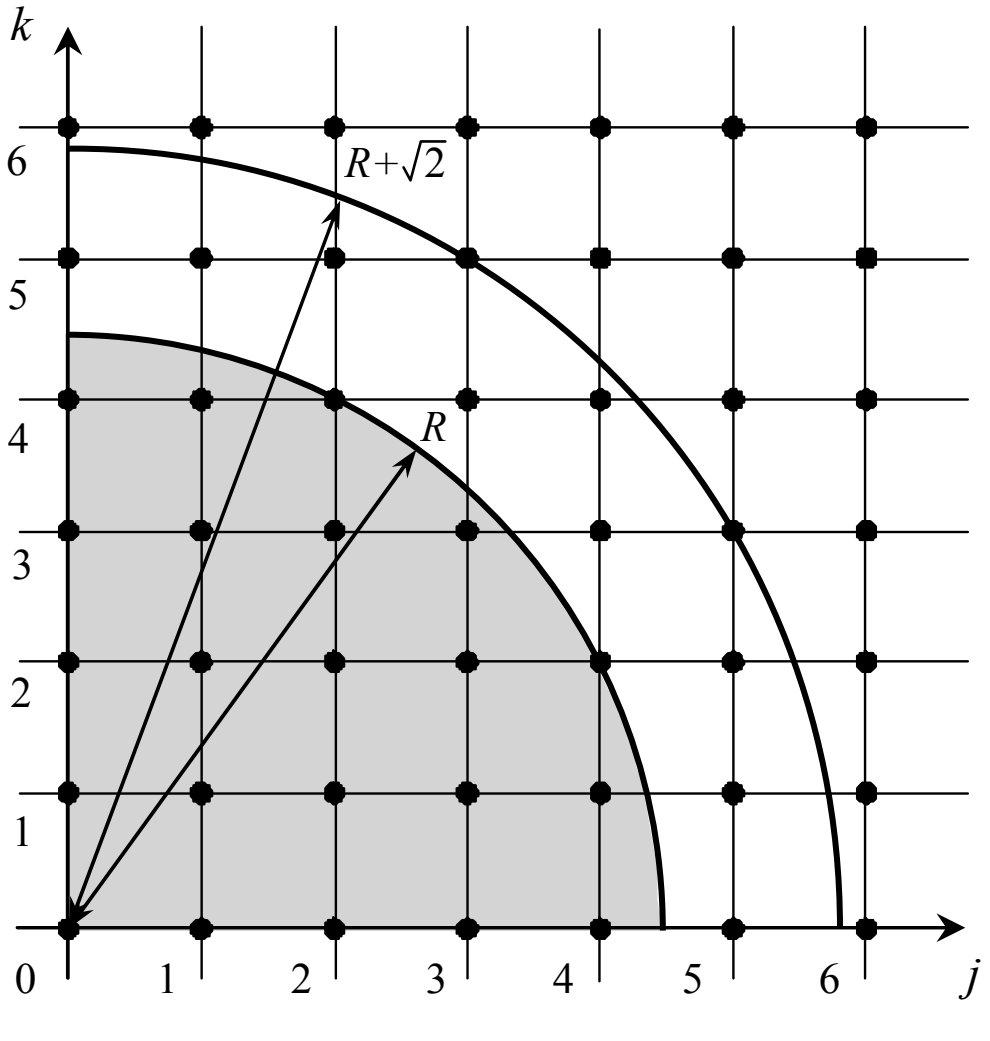

Рис. 7.3. До оцінки функції розподілу власних значень квадратної мембрани

З іншого боку, з елементарних геометричних міркувань випливає, що якщо координати $(j,k)$ хоча б однієї з вершин якоїсь елементарної комірки задовольняють умову $j^2+k^2 \le R^2$, то для координат $(j',k')$ будь-якої з трьох її інших вершин справджується нерівність $j'^2+k'^2 \le \left(R+\sqrt{2}\right)^2$. Отже, усі комірки, у яких хоча б одна з вершин лежить у чверті круга $x^2+y^2 \le R^2$, цілком лежать у чверті круга $x^2+y^2 \le \left(R+\sqrt{2}\right)^2$ $(x,y \ge 0)$. Маємо нерівність

$$\mathcal{N}_{\text{в}}(\lambda) \le \frac{\pi}{4}\left(\frac{l\sqrt{\lambda}}{\pi a}+\sqrt{2}\right)^2 = \frac{l^2\lambda}{4\pi a^2}+\frac{l\sqrt{\lambda}}{\sqrt{2}\,a}+\frac{\pi}{2}. \qquad (7.273)$$

Порівнюючи співвідношення (7.272) і (7.273), знаходимо:

$$\mathcal{N}_{\text{в}}(\lambda) \underset{\lambda\to\infty}{=} \frac{l^2\lambda}{4\pi a^2}+O\left(\sqrt{\lambda}\right). \qquad (7.274)$$

Аналогічними аргументами можна скористатися для обчислення асимптотики функції $\mathcal{N}(\lambda)\equiv\mathcal{N}_{\text{з}}(\lambda)$ для задачі про коливання квадратної мембрани з жорстко закріпленими краями. Власні значення цієї задачі теж описуються формулою (7.271), однак тепер $j$ і $k$ — лише натуральні числа (нульові значення виключаються). Зводячи, як і раніше, відшукання $\mathcal{N}_{\text{з}}(\lambda)$ до обчислення кількості точок із цілочисловими координатами в замкненій чверті круга радіусом $l\sqrt{\lambda}/(\pi a)$, маємо не враховувати ті з них, які лежать на координатних осях. Кількість цих точок, зважаючи на те, що довжини відрізків, на яких вони лежать, збігаються з радіусом указаного круга, дорівнює $2\left[l\sqrt{\lambda}/(\pi a)\right]+1$, де $[x]$ — ціла частина числа $x$. Тому $\mathcal{N}_{\text{з}}(\lambda)$ у порівнянні з $\mathcal{N}_{\text{в}}(\lambda)$ є меншою на величину $2\left[l\sqrt{\lambda}/(\pi a)\right]+1$:

$$\mathcal{N}_{\text{з}}(\lambda)=\mathcal{N}_{\text{в}}(\lambda)-2\left[\frac{l}{\pi a}\sqrt{\lambda}\right]-1,$$

отже

$$\mathcal{N}_{\text{з}}(\lambda) \underset{\lambda\to\infty}{=} \mathcal{N}_{\text{в}}(\lambda)+O\left(\sqrt{\lambda}\right)=\frac{l^2\lambda}{4\pi a^2}+O\left(\sqrt{\lambda}\right). \qquad (7.275)$$

Рівність головних членів асимптотик (7.274) і (7.275) наводить на думку, що головний член асимптотичного розкладу $\mathcal{N}(\lambda)$ при $\lambda\to\infty$ не залежить, принаймні в задачах про коливання квадратної мембрани, від виду крайових умов. Більше того, зараз ми покажемо, що для обмеженої мембрани він не залежить і від її (достатньо довільної) форми, а визначається лише площею мембрани $S$:

$$\mathcal{N}(\lambda) \underset{\lambda\to\infty}{=} \frac{S\lambda}{4\pi a^2}+o(\lambda).\text{[1]} \qquad (7.276)$$

Нехай $\Omega$ — обмежена область на площині з кусково-гладкою межею $\Sigma$ і площею $S$. Позначимо через $L_2(\Omega)$ множину (лінійний простір) функцій $f(x,y)\equiv f(\mathbf{r})$, квадратично інтегровних на $\Omega$,

---

[1] Це твердження вперше сформулював Г. А. Лоренц (1908 р.) і згодом довів Г. Вейль (1911 р.). Наведене нижче доведення запропонував М. Кац — див.: The American Mathematical Monthly, Vol. 73. No 4, Part 2: Papers in Analysis (1966), pp. 1–23.

$$\|f\|^2 = \iint_{\Omega} |f(x,y)|^2\, dxdy \equiv \int_{\Omega} |f(\mathbf{r})|^2\, d\mathbf{r} < \infty,$$

а через $D_\Delta$ — підмножину функцій, двічі неперервно диференційовних усередині $\Omega$ та неперервних у замкненій області $\bar{\Omega} = \Omega \bigcup \Sigma$. Для будь-якої пари функцій $f, g \in L_2(\Omega)$ функціонал

$$\langle f \mid g \rangle = \int_{\Omega} \overline{f(\mathbf{r})} g(\mathbf{r})\, d\mathbf{r}$$

називатимемо, як і раніше, їх скалярним добутком, а самі функції $f$ і $g$ — ортогональними в $L_2(\Omega)$, якщо $\langle f \mid g \rangle = 0$. Послідовність $\{\varphi_\alpha\}$ з $L_2(\Omega)$ називатимемо повною, якщо ортогональність функції $f \in L_2(\Omega)$ до всіх функцій $\varphi_\alpha$ справджується тоді й лише тоді, коли $f = 0$.

Розглянемо крайову задачу на відшукання власних значень і власних функцій з підмножини $D_\Delta$ для однорідної системи

$$\begin{gathered} -a^2 \Delta\varphi = \lambda\varphi, \\ \left.\left( \frac{\partial \varphi}{\partial n} + \sigma\varphi \right)\right|_{\Sigma} = 0, \end{gathered} \tag{7.277}$$

де $\partial\varphi/\partial n$ — складова градієнта $\nabla\varphi$ в напрямі зовнішньої нормалі до межі $\Sigma$, а $\sigma$ — невід'ємна функція на $\Sigma$. Уважатимемо, що $\sigma$ може набувати й значення $+\infty$ на кусково-гладкій частині $\Sigma'$ межі $\Sigma$, і на цій частині $\Sigma$ крайова умова в (7.277) зводиться до вигляду $\varphi|_{\Sigma'} = 0$; на другій частині $\Sigma$ функція $\sigma$ неперервна.

**Завдання 7.7.1.** Доведіть, що власні значення $\lambda$ крайової задачі (7.277) невід'ємні, а власні функції $\varphi_\lambda(\mathbf{r})$ і $\varphi_{\lambda'}(\mathbf{r})$ з $D_\Delta$, які відповідають різним власним значенням $\lambda$ й $\lambda'$, ортогональні в $L_2(\Omega)$.

*Вказівка*. Скористайтеся теоремою Остроградського — Гаусса та тими ж самими прийомами, що були застосовані при доведенні аналогічних результатів для крайової задачі Штурма — Ліувілля.

При вказаних вище умовах справджується наступна теорема, доведення якої досить складне й тут не наводиться.

**Теорема 7.7.1.** Множина власних значень крайової задачі (7.277) утворює послідовність точок півосі $[0,\infty)$ з єдиною граничною точкою $+\infty$, при цьому кожне власне значення має скінченну кратність. Власні функції крайової задачі (7.277) утворюють повну систему функцій в $L_2(\Omega)$.

Розташуємо власні значення крайової задачі (7.277) у порядку зростання,

$$0 \le \lambda_1 \le \lambda_2 \le \ldots \le \lambda_n \le \ldots,\ \lambda_n \to \infty,\ n \to \infty, \tag{7.278}$$

повторюючи в цій послідовності кожне власне значення стільки разів, як і його кратність. Кожному власному значенню $\lambda_k$ послідовності (7.278) можна поставити у відповідність власну функцію $\varphi_k(\mathbf{r})$ задачі (7.277) таким чином[1], щоби функції послідовності $\{\varphi_k\}$ були дійсними й утворювали ортонормовану систему в $L_2(\Omega)$:

$$\langle \varphi_k \mid \varphi_l \rangle = \int_{\Omega} \varphi_k(\mathbf{r}) \varphi_l(\mathbf{r}) d\mathbf{r} = \delta_{kl}.$$

З повноти системи власних функцій задачі (7.277) випливає, що кожну функцію $f(\mathbf{r})$ із $L_2(\Omega)$ можна подати у вигляді ряду Фур'є

$$f(\mathbf{r}) = \sum_{k=1}^{\infty} \langle \varphi_k \mid f \rangle \varphi_k(\mathbf{r})$$

за ортонормованою системою $\{\varphi_k\}$, який збігається в середньому квадратичному.

На підставі наведених фактів загальний розв'язок крайової задачі

$$\begin{gathered} \frac{\partial T}{\partial t} = a^2 \Delta T,\ \ t > 0, \\ \left. \left( \frac{\partial T}{\partial n} + \sigma u \right) \right|_{\Sigma} = 0, \\ T(\mathbf{r},0) = f(\mathbf{r}),\ \ f \in L_2(\Omega), \end{gathered} \tag{7.279}$$

для рівняння теплопровідності в області $\Omega$ можемо записати у вигляді ряду

$$T(\mathbf{r},t) = \sum_{k=1}^{\infty} e^{-\lambda_k t} \langle \varphi_k \mid f \rangle \varphi_k(\mathbf{r})$$

---

[1] Важливо підкреслити, що довільно взяті незалежні власні функції $\varphi_\alpha(\mathbf{r})$, які відповідають одному й тому самому виродженому (з кратністю *m*) власному значенню крайової задачі, узагалі кажучи, не будуть ортогональними між собою. Однак від функцій $\varphi_\alpha(\mathbf{r})$ завжди можна перейти до *m* їх лінійних комбінацій $\psi_\alpha(\mathbf{r})$, які теж є власними функціями, що відповідають заданому власному значенню, але вже взаємно ортогональними. Такий перехід від системи функцій $\varphi_\alpha(\mathbf{r})$ до системи функцій $\psi_\alpha(\mathbf{r})$ називається ортогоналізацією і може бути здійснений різними способами. Зазначимо також, що як функції $\varphi_\alpha(\mathbf{r})$, так і функції $\psi_\alpha(\mathbf{r})$ є ортогональними до всіх власних функцій, що відповідають іншим власним значенням розглядуваної крайової задачі.

і, далі, у вигляді інтеграла

$$T(\mathbf{r},t)=\int_{\Omega}G(\mathbf{r},\mathbf{r}';t)f(\mathbf{r}')d\mathbf{r}',$$

де

$$G(\mathbf{r},\mathbf{r}';t)=\sum_{k=1}^{\infty}e^{-\lambda_k t}\varphi_k(\mathbf{r})\varphi_k(\mathbf{r}')$$

– функція Гріна задачі (7.279).

Нехай $\mathbf{r}$ — внутрішня точка області $\Omega$. Розглянемо значення функції

$$G(\mathbf{r},\mathbf{r};t)=\sum_{k=1}^{\infty}e^{-\lambda_k t}\varphi_k^2(\mathbf{r}) \qquad (7.280)$$

при $t>0$. Нагадаємо, що з фізичного погляду функція (7.280) має зміст густини теплової енергії середовища в точці $\mathbf{r}$ у момент часу $t>0$ за умов, що в момент часу $t=0$ у цю ж саму точку було введено одиницю кількості тепла, до цього моменту температура середовища скрізь дорівнювала нулю, і в замкненій області $\overline{\Omega}$ не було й немає джерел тепла. З огляду на природу явища теплопровідності як результату багатьох міжчастинкових зіткнень зрозуміло, що наявність десь межі області $\Omega$ та умови на ній не можуть при $t\to 0$ вплинути на розподіл тепла в малому околі внутрішньої точки $\mathbf{r}\in\Omega$, і він спочатку буде таким самим, яким би був у необмеженому однорідному просторі. Розподіл тепла в останньому при вказаних вище умовах визначається за допомогою функції Гріна

$$G_0(\mathbf{r},\mathbf{r}';t)=G_0(|\mathbf{r}-\mathbf{r}'|;t)=\frac{1}{4\pi a^2 t}e^{-\frac{|\mathbf{r}-\mathbf{r}'|^2}{4a^2t}},\quad t>0,$$

рівняння теплопровідності в однорідному двовимірному просторі. Бачимо, що

$$G(\mathbf{r},\mathbf{r};t)\underset{t\to 0}{=}G_0(0;t)\left[1+o(1)\right]=\frac{1}{4\pi a^2 t}\left[1+o(1)\right]. \qquad (7.281)$$

Спираючись на вираз (7.281) й опускаючи всі технічні деталі, приходимо, зокрема, до висновку, що

$$\int_{\Omega}G(\mathbf{r},\mathbf{r};t)d\mathbf{r}\underset{t\to 0}{\approx}\frac{S}{4\pi a^2 t}\left[1+o(1)\right]. \qquad (7.282)$$

З іншого боку, беручи до уваги, що всі члени ряду у правій частині формули (7.280) невідємні, а функції $\varphi_k(\mathbf{r})$ нормовані, знаходимо:

$$\int_{\Omega} G(\mathbf{r},\mathbf{r};t)\,d\mathbf{r} = \sum_{k=1}^{\infty} e^{-\lambda_k t} = \sum_{n=1}^{\infty} e^{-\lambda'_n t} m_n, \qquad (7.283)$$

де $0 \le \lambda'_1 < \lambda'_2 < \ldots < \lambda'_n < \ldots$ — різні власні значення крайової задачі (7.279), занумеровані в порядку зростання, $m_n$ — їх кратності. Згадавши означення функції розподілу спектра $\mathcal{N}(\lambda)$ крайової задачі (7.279) та формули (7.269) і (7.270), згідно з якими

$$d\mathcal{N}(\lambda) = \sum_n m_n \delta(\lambda - \lambda'_n)\,d\lambda,$$

останню суму у формулі (7.283) можемо записати у вигляді інтеграла:

$$\sum_{n=1}^{\infty} e^{-\lambda'_n t} m_n = \int_0^{\infty} e^{-\lambda t} d\mathcal{N}(\lambda). \qquad (7.284)$$

З формул (7.282)–(7.284) випливає, що функція $\mathcal{N}(\lambda)$ є такою, що

$$\int_0^{\infty} e^{-\lambda t} d\mathcal{N}(\lambda) \underset{t\downarrow 0}{\approx} \frac{S}{4\pi a^2 t}\left[1 + o(1)\right]. \qquad (7.285)$$

Щоб із формули (7.285) отримати бажане співвідношення (7.276), залишається скористатися відомим тонким результатом аналізу, а саме, так званою *тауберовою теоремою Харді — Літтлвуда*[1] в наступному формулюванні.

**Теорема 7.7.2.** Нехай $\mathcal{N}(\lambda)$ — неспадна функція на півосі $[0,\infty)$. Тоді для $\mathcal{N}(\lambda)$ і деякого $\rho \ge 0$ співвідношення

$$\int_0^{\infty} e^{-\lambda t} d\mathcal{N}(\lambda) \underset{t\downarrow 0}{=} C t^{-\rho}\left[1 + o(1)\right]$$

справджується тоді й лише тоді, коли

$$\mathcal{N}(\lambda) \underset{\lambda\to\infty}{=} \frac{C}{\Gamma(1+\rho)} \lambda^{\rho}\left[1 + o(1)\right],$$

де $C$ — стала, $\Gamma(x)$ — гамма-функція Ейлера.

Нехай тепер $\Omega$ — обмежена область $n$-вимірного простору з кусково-гладкою межею $\Sigma$ та об'ємом $V_\Omega$, $\Delta = \frac{\partial^2}{\partial x_1^2} + \frac{\partial^2}{\partial x_2^2} + \ldots + \frac{\partial^2}{\partial x_n^2}$ — оператор Лапласа в $\Omega$. Можна довести, що й у цьому випадку для крайової задачі на відшукання власних значень і власних функцій з відповідного класу $D_\Delta$, яка виникає при розв'язуванні крайової задачі виду (7.279) для рівняння теплопровідності, дослівно справджується теорема 7.7.1, а асимптотичний вираз для відповідної функції $\mathcal{N}(\lambda)$ має вигляд

$$\mathcal{N}(\lambda) \underset{\lambda\to\infty}{\approx} \frac{V_\Omega \lambda^{n/2}}{(4\pi a^2)^{n/2}\Gamma(1+n/2)}. \tag{7.286}$$

[1] Див.: Феллер В. Введение в теорию вероятностей и ее приложения. Т. 2, гл. XIII. — М.: Мир, 1984. — 752 с.

**Завдання 7.7.2.** Доведіть, що для функції розподілу власних значень крайової задачі (7.279) для обмеженої тривимірної області об'ємом $V$ справджується асимптотичний вираз

$$\mathcal{N}(\lambda) \underset{\lambda\to\infty}{\approx} \frac{V\lambda^{3/2}}{6\pi^2 a^3}. \tag{7.287}$$

*Вказівка.* Згадавши, що функція Гріна для рівняння теплопровідності в необмеженому тривимірному просторі має вигляд

$$G_0(r,t) = \frac{1}{(4\pi a^2 t)^{3/2}} e^{-\frac{r^2}{4a^2 t}},$$

а

$$\Gamma(1+3/2) = \int_0^\infty x^{3/2} e^{-x} dx = \frac{3}{4}\sqrt{\pi},$$

повторіть міркування, які привели нас до формули (7.276) для двовимірного простору.

Наведемо простий приклад з електронної теорії твердого тіла на застосування асимптотики (7.287). Згідно зі спрощеною (але досить ефективною) моделлю А. Зоммерфельда валентні електрони в куску металу $\Omega$ об'ємом $V$ можна розглядати як газ однакових невзаємодіючих частинок. Енергії, які можуть мати частинки такого газу, вичерпуються послідовними власними значеннями $\varepsilon_1 < \varepsilon_2 < \ldots$ крайової задачі типу (7.277) з параметром $a^2 = \hbar^2/2\mu$, де $\hbar$ — стала Планка, $\mu$ — маса частинки (електрона). При цьому, згідно з принципом Паулі, кількість частинок такого газу $N_k$, що можуть одночасно мати енергію $\varepsilon_k$, не перевищує $2m_k$, де $m_k$ — кратність власного значення $\varepsilon_k$ $(0 \le N_k \le 2m_k)$. Крайові умови на поверхні $\Sigma$ куска металу $\Omega$ у відповідній крайовій задачі (7.277), узагалі кажучи, не конкретизуються, але, у відповідності з фундаментальними положеннями квантової теорії, уважаються такими, що справджуються твердження теореми 7.7.1.

На числа $N_k$ додатково накладається очевидне обмеження

$$\sum_k N_k = N, \tag{7.288}$$

де $N$ — загальна кількість частинок електронного газу, що розглядається.

З урахуванням обмежень $0 \le N_k \le 2m_k$ та (7.288) мінімальне значення повної енергії електронного газу

$$\sum_k \varepsilon_k N_k = E$$

отримуємо у випадку, коли для певної допустимої енергії $\varepsilon_F$ числа $N_k$ визначаються умовами

$$N_k = \begin{cases} 2m_k, & \varepsilon_k < \varepsilon_F, \\ 0, & \varepsilon_k > \varepsilon_F, \end{cases} \tag{7.289}$$

та

$$2\sum_{\varepsilon_k < \varepsilon_F} m_k + N_F = N, \tag{7.290}$$

де $N_F$ — кількість частинок з енергією $\varepsilon_F$.

Енергія $\varepsilon_F$ називається *енергією Фермі* і є фундаментальною характеристикою металу. Беручи до уваги, що для реального макроскопічного куска металу число $N$ надзвичайно велике ($N/V \sim 10^{28}$ м$^{-3}$), можемо визначити її за допомогою формул (7.289), (7.290) та співвідношення

$$2\mathcal{N}(\varepsilon_F) = N,$$

де $\mathcal{N}(\varepsilon)$ — функція розподілу спектра відповідної крайової задачі (7.277). Знову ж таки, ураховуючи, що число $N$ дуже велике, можемо скористатися для $\mathcal{N}(\varepsilon_F)$ асимптотичним виразом (7.287). Дістаємо:

$$\left(\frac{2\mu}{\hbar^2}\right)^{3/2} \frac{\varepsilon_F^{3/2}}{3\pi^2} V = N.$$

Бачимо, що незалежно від форми макроскопічного куска металу та крайових умов на його поверхні

$$\varepsilon_F = \frac{\hbar^2}{2\mu}(3\pi^2 n)^{2/3}, \tag{7.291}$$

де $n = N/V$ — густина вільних електронів металу.

## *КОНТРОЛЬНІ ПИТАННЯ ДО РОЗДІЛУ 7*

1. *У яких випадках крайова задача на власні значення для рівняння Гельмгольца в обмеженій області допускає відокремлення змінних? Як воно здійснюється в задачах про коливання прямокутної мембрани з жорстко закріпленими та вільними краями?*
2. *Звідки випливає повнота системи незалежних власних функцій у задачах про коливання прямокутної мембрани з жорстко закріпленими та вільними краями?*
3. *Яка особливість виникає у крайовій задачі Штурма — Ліувілля, до якої зводиться відшукання вісесиметричних власних коливань круглої мембрани з жорстко закріпленим краєм? Як ця особливість відбивається на незалежних розв'язках відповідного рівняння Штурма — Ліувілля?*
4. *Як здійснюється відокремлення змінних у задачах про коливання круглої мембрани? Чому радіальні частини власних функцій крайової задачі для рівняння Гельмгольца у цих випадках зводяться лише до функцій Бесселя?*
5. *Як здійснюється відокремлення змінних у крайових задачах для сферичних областей?*
6. *Чому дорівнюють кутові частини власних функцій у крайових задачах для рівнянь теплопровідності й дифузії у сферичних областях зі сферично-симетричними крайовими умовами?*
7. *Які кратності виродження мають власні значення крайових задач для рівнянь теплопровідності й дифузії у сферичних областях зі сферично-симетричними крайовими умовами?*
8. *Чому дорівнює асимптотика функції розподілу квадратів частот прямокутної мембрани з площею $S$ і жорстко закріпленими краями? Чи буде змінюватися ця асимптотика при неперервних змінах форми мембрани при незмінній площі $S$ і при змінах умов закріплення на краях?*
9. *Як узагальнюється попередній результат для асимптотики розподілу власних значень аналогічної крайової задачі, поставленої для рівняння Гельмгольца в обмеженій області $n$-вимірного простору з об'ємом $V$ і кусково-гладкою межею?*

# Література

# Предметний покажчик

У посібнику розглядаються постановка й методи розв'язування простих лінійних задач класичної математичної фізики. Коло розглянутих питань охоплює принципи варіаційного числення, одновимірні крайові задачі теорії коливань і теплопровідності з докладним аналізом крайової задачі Штурма — Ліувілля та обґрунтуванням методу Фур'є, приклади розв'язування відповідних задач у двох і трьох вимірах із необхідними елементами теорії спеціальних функцій.

Для студентів фізичних, інженерно-фізичних та математичних спеціальностей університетів.

This book (in Ukrainian) considers posing and the methods of solving simple linear boundary-value problems in classical mathematical physics. The questions encompassed include: the fundamentals of calculus of variations; one-dimensional boundary-value problems in the oscillation and heat conduction theories, with a detailed analysis of the Sturm—Liouville boundary-value problem and substantiation of the Fourier method; sample solutions of the corresponding problems in two and three dimensions, with essential elements of the special function theory.

The text is designed for Physics, Engineering, and Mathematics majors.

**АДАМЯН Вадим Мовсесович**
**СУШКО Мирослав Ярославович**

# ВСТУП ДО МАТЕМАТИЧНОЇ ФІЗИКИ. ВАРІАЦІЙНЕ ЧИСЛЕННЯ ТА КРАЙОВІ ЗАДАЧІ

Навчальний посібник
для студентів фізичних
та інженерно-фізичних спеціальностей
вищих навчальних закладів